# The State of Food Systems Worldwide: Counting Down to 2030

Last updated: March 23, 2023

Kate Schneider[1], Jessica Fanzo[1], Lawrence Haddad[2], Mario Herrero*[3], Jose Rosero Moncayo[4], Anna Herforth*[5], Roseline Remans*[6], Alejandro Guarin*[7], Danielle Resnick*[8], Namukolo Covic*[9], Christophe Béné*[6,10], Andrea Cattaneo*[4], Nancy Aburto[4], Ramya Ambikapathi[3], Destan Aytekin[1], Simon Barquera[11], Jane Battersby-Lennard[12], Ty Beal[2], Paulina Bizzoto Molina[13], Carlo Cafiero[4], Christine Campeau[14], Patrick Caron[15], Piero Conforti[4], Kerstin Damerau[3], Michael DiGirolamo[1], Fabrice DeClerck[16], Deviana Dewi[1], Ismahane Elouafi[4], Carola Fabi[4], Pat Foley[17], Ty Frazier[18], Jessica Gephart[19], Christopher Golden[5], Carlos Gonzalez Fischer[3], Sheryl Hendriks[20], Maddalena Honorati[21], Jikun Huang[22], Gina Kennedy[2], Amos Laar[23], Rattan Lal[24], Preetmoninder Lidder[4], Brent Loken[25], Quinn Marshall[26], Yuta Masuda[27], Rebecca McLaren[1], Lais Miachon[1], Hernán Muñoz[4], Stella Nordhagen[2], Naina Qayyum[28], Michaela Saisana[29], Diana Suhardiman[30], Rashid Sumaila[31], Maximo Torrero Cullen[4], Francesco Tubiello[4], Jose-Luis Vivero-Pol[17], Patrick Webb[28], Keith Wiebe[26]

[*] Denotes working group lead
[1] Johns Hopkins University
[2] Global Alliance for Improved Nutrition (GAIN)
[3] Cornell University
[4] Food and Agriculture Organization of the United Nations (FAO)
[5] Harvard University
[6] Alliance of CIAT-Bioversity
[7] International Institute for Environment & Development (IIED)
[8] Brookings Institution
[9] International Livestock Research Institute (ILRI)
[10] Wageningen Economic Research Group
[11] Instituto Nacional de Salud Pública (INSP), México
[12] University of Cape Town
[13] European Centre for Development Policy Management
[14] CARE
[15] University of Montpellier, Cirad, ART-DEV
[16] EAT Forum
[17] United Nations World Food Programme (WFP)
[18] Oakridge National Laboratory
[19] American University
[20] University of Greenwich
[21] World Bank
[22] Peking University
[23] University of Ghana
[24] Ohio State University
[25] World Wildlife Fund (WWF)
[26] International Food Policy Research Institute
[27] Vulcan
[28] Tufts University
[29] Joint Research Centre (JRC) of the European Commission
[30] KIT Royal Tropical Institute (Netherlands)
[31] University of British Columbia

# The State of Food Systems Worldwide: Counting Down to 2030

## Summary

Transforming food systems is essential to bring about a healthier, equitable, sustainable, and resilient future, including achieving global development and sustainability goals.[1–3] To date, no comprehensive framework exists to track food systems transformation and their contributions to global goals. In 2021, the Food Systems Countdown to 2030 Initiative (FSCI) articulated an architecture to monitor food systems across five themes: (1) diets, nutrition, and health; (2) environment, natural resources, and production; (3) livelihoods, poverty, and equity; (4) governance; and (5) resilience and sustainability.[1] Each theme comprises three-to-five indicator domains. This paper builds on that architecture, presenting the inclusive, consultative process used to select indicators and an application of the indicator framework using the latest available data, constructing the first global food systems baseline to track transformation. While data are available to cover most themes and domains, critical indicator gaps exist such as off-farm livelihoods, food loss and waste, and governance. Baseline results demonstrate every region or country can claim positive outcomes in some parts of food systems, but none are optimal across all domains, and some indicators are independent of national income. These results underscore the need for dedicated monitoring and transformation agendas specific to food systems. Tracking these indicators to 2030 and beyond will allow for data-driven food systems governance at all scales and increase accountability for urgently needed progress toward achieving global goals.

## Main

Food systems fundamentally shape lives, wellbeing, and human and planetary health, and they are central to tackling some of the most pressing global challenges of our time. The United Nations (UN) held its first-ever Food Systems Summit (UNFSS) in 2021, which demonstrated the interconnectedness of food systems with the Sustainable Development Goals (SDGs) and provided a space for countries to develop national pathways towards food systems transformation. Food systems also featured prominently at the 26th and 27th UN Climate Change Conference (COP26/7),[4] and in the Kunming-Montreal Global Biodiversity Framework targets.[5] This context offers growing momentum to influence public policy, private sector, and civil society actions to transform food systems from their current unsustainable and inequitable trajectories.[6–8] Yet while their contributions to other global goals are recognized and the clear need for monitoring has been articulated,[9] no indicator framework has been defined to track food systems. Decision-makers across sectors thus lack a means to assess their food systems, guide action, or evaluate progress.

In 2021, the Food Systems Countdown to 2030 Initiative (FSCI) emerged from the UNFSS as an interdisciplinary collaboration of dozens of scientists. This paper uses the term "food systems" throughout, in line with the UNFSS language. However, the indicator framework presented here takes an expanded concept of agrifood systems given that many indicators cannot distinguish between food and non-food components of production and value addition, although such non-food components greatly influence the environment, social outcomes, and the food people ultimately eat. Hence, food systems as used here encompass activities and processes around non-food agricultural products (e.g., forestry, fibers, biofuels, etc.) that are interconnected with food for human consumption.[1] As a first step, in 2021, the FSCI published an architecture to monitor food systems comprising five thematic areas each with three to five indicator domains.[1] Next, the FSCI undertook a consultative process with additional scientific experts and policy stakeholders to select a set of existing indicators (or modifications thereof). The consultative process selected 50 indicators, a list as comprehensive as possible given available indicators and data, and which constitutes the indicator framework applied in this paper.

Descriptive analysis at the global, regional, and income group provides a baseline of the world's food systems; an essential first step in a global food systems research agenda and the starting point to track change. For the next seven years (2023-2030), the FSCI will publish annual updates and incorporate new indicators to fill the remaining gaps. Recognizing the descriptive summaries only scratch the surface of potential knowledge this rich dataset can offer, the FSCI also plans deeper analyses, beginning with assessments of interactions between different indicators, establishing benchmarks, and analyzing relative performance over the next two years.

The fundamental contributions of this paper are (1) an application of the global architecture previously developed;[1] (2) the comprehensive yet actionable set of indicators legitimated through consultative process; and (3) a baseline dataset to track progress on food systems transformation to 2030. Many of these data have long time series available, while some indicators are new[1] but expected to be collected/computed globally going forward. Government officials responsible for developing food system transformation pathways coming out of the UNFSS have expressed clear demand for guidance on indicators[10–12] and the UNFSS 2023 "stock-taking" provides an entry point to support decision-makers in steering food systems.[13] The selected indicators address topics that appear in these pathways, offering a menu of indicators that may be most relevant for accountability to stated commitments. At the global level, the framework enables policymakers, advisors, private sector, and civil society actors to monitor food systems worldwide.[2] The indicators can be used to establish local-to-global targets, track progress against those targets, provide accountability for commitments, and drive progress towards desired outcomes, while recognizing the diversity of pathways at national and subnational levels that can be compatible with global goals.

---

[1] Specifically diet quality indicators, sustainable nitrogen management, landholdings by gender, proportion of the urban population living in a municipality signed onto the Milan Urban Food Policy Pact (MUFPP), and the share of agricultural land with minimum species richness.
[2] See **Supplementary Data A (SD-A) Figure A.1** for the full theory of transformation.

## Results

### Data coverage

The indicator framework used here follows the architecture defined in Fanzo et al. (2021)[1] consisting of five themes: (1) diets, nutrition, and health; (2) environment, natural resources, and production; (3) livelihoods, poverty, and equity; (4) governance; and (5) resilience and sustainability. Within each theme there are several indicator domains that specific indicators are selected to reflect. **Table 1** presents the themes, indicator domains, and the global distribution of the selected indicators (henceforth "the baseline"), and the country level data are provided as **Supplementary Data (SD)-F**.[3] **Figure 1** presents a data coverage heatmap from 2000 forward showing that the indicators with greatest country coverage and longest time series are those associated with agricultural development such as yields and the share of agriculture in GDP. Data coverage for other food system indicators has been steadily increasing over time, making it possible to compile this set of indicators covering all domains.

There are, however, notable data gaps that emerged through the indicator selection process. These include the economic value of food systems, food safety, the true cost of food, the magnitude and composition of populations working in food systems, productivity in the sector (e.g., value-added as a share of GDP and per worker), policy coherence (alignment across policy areas) for food systems transformation, budgetary allocations to food systems, food loss and waste at the country level, and livelihood indicators that can capture the welfare of food system workers beyond agriculture (especially measures of decent work, gender equity, and violations of human rights in food systems), and food production and supply indicators inclusive of aquatic and wild foods. For other indicators – adult diet quality, biodiversity, and agro- and food diversity – the country and year coverage remain sparse. There are no adult diet quality data for Oceania, a priority gap, given that region's high burden of diet-related disease.[14,15] Furthermore, environmental indicators predominantly relate to production and largely exclude loss and waste as well as pollution related to processes (e.g., packaging) further down the value chain. Additional indicators of governance and resilience specific to food systems are also lacking.

[3] See **Table 4** in the data and methods section for the full description, data sources, and rationale for each indicator.

**Figure 1. Data coverage, number of years per country-indicator, 2000-2021**

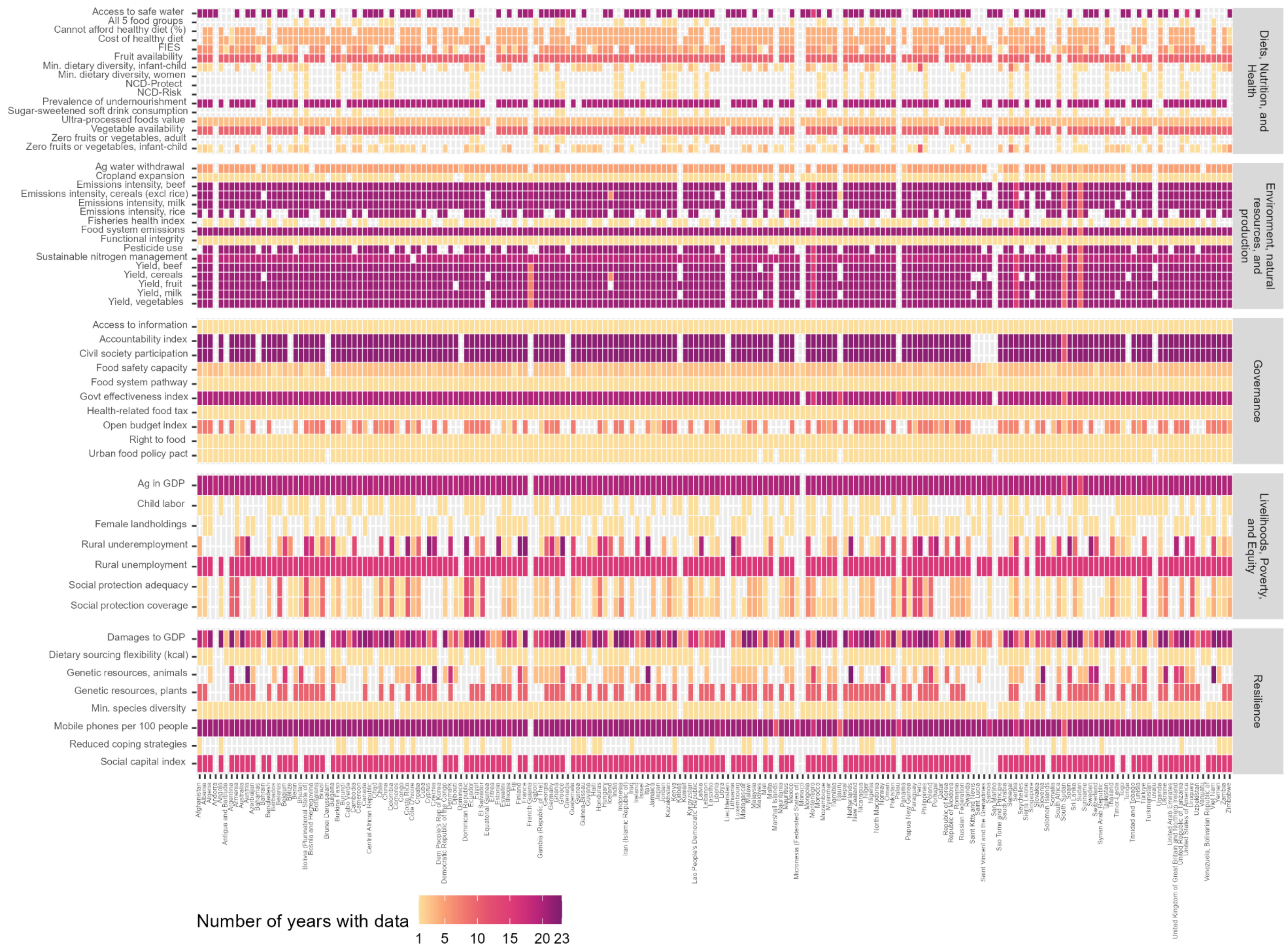

Notes: Heatmaps showing the indicator-country time series by region are available in **SD-A Figures A.3 – A.11**. Gray cells indicate zero observations (no data). Maximum country coverage is all UN member states but differs per indicator depending on data availability. Differences in indicator coverage largely drive observed differences across countries. Specifically, the indicators with most heterogeneous coverage are the six indicators of diet quality sourced from the Global Diet Quality project currently available for only 41 mostly low- and lower middle-income countries, the livelihoods indicators of employment, social protection, child labor, and landholdings, and resilience indicators of genetic resources, coping strategies (available for countries with high prevalence of food insecurity). Looking across countries within each indicator, countries with the indicator typically have similar duration time series available. Yield and emissions intensity for additional products are provided in **SD-A** and the baseline dataset.

## Table 1. Indicator list & global baseline‡ distributions (see Table 4 for data sources; year is the latest data point per country per indicator)

| Domain | | Indicator | | Unit | Min. | 25th | Median | 75th | Max | Weighted Mean | Weighted SD | Weighted by |
|---|---|---|---|---|---|---|---|---|---|---|---|---|
| ***Diets, nutrition, and health*** | | | | | | | | | | | | |
| Food environments | **1** | Cost of a healthy diet | | current PPP US$/person/day | 1.9 | 3.0 | 3.4 | 3.9 | 6.7 | 3.3 | 0.6 | Population |
| | **2** | Availability of fruits and vegetables | Fruits | grams/capita/day | 14.3 | 134.2 | 201.3 | 280.5 | 999.1 | 223.8 | 145.8 | (unweighted) |
| | | | Vegetables | grams/capita/day | 17.7 | 119.0 | 210.0 | 297.9 | 1,059.9 | 246.8 | 186.5 | (unweighted) |
| | **3** | Retail value of ultra-processed foods | | current (nominal) US$/capita | 10.8 | 44.2 | 163.7 | 365.4 | 1,465.5 | 204.0 | 293.1 | Population |
| | **4** | % Population using safely managed drinking water services (SDG 6.1.1) | | % population | 5.6 | 47.5 | 85.7 | 98.3 | 99.9 | 66.3 | 30.9 | Population |
| Food security | **5** | Prevalence of Undernourishment (SDG 2.1.1) | | % population | 2.5 | 2.5 | 5.6 | 15.3 | 52.2 | 9.4 | 8.9 | Population |
| | **6** | % Population experiencing moderate or severe food insecurity (SDG 2.1.2) | | % population | 2.2 | 9.9 | 26.5 | 50.4 | 88.7 | 29.5 | 23.0 | Population |
| | **7** | % Population who cannot afford a healthy diet | | % population | 0.0 | 2.0 | 21.4 | 70.9 | 97.2 | 42.3 | 33.9 | Population |
| Diet quality | **8** | MDD-W: minimum dietary diversity for women | | % population, women 15-49 | 35.9 | 53.9 | 71.7 | 79.9 | 88.7 | 65.7 | 20.3 | Population |
| | **9** | MDD (IYCF): minimum dietary diversity for infants and young children | | % population, 6-23 months | 8.1 | 22.4 | 34.4 | 53.5 | 85.9 | 31.8 | 15.9 | Population |
| | **10** | All-5: consumption of all 5 food groups | | % adult population (≥15 y) | 15.9 | 23.5 | 30.5 | 44.3 | 63.4 | 39.0 | 13.7 | Population |
| | **11** | Zero fruit or vegetable consumption | Adults | % adult population (≥15 y) | 1.9 | 4.6 | 8.4 | 12.2 | 22.3 | 10.8 | 7.9 | Population |
| | **12** | | Children 6-23 months | % population 6-23 months | 2.2 | 18.3 | 31.5 | 47.9 | 69.2 | 39.1 | 15.8 | Population |
| | **13** | NCD-Protect | | Score (points out of 9) | 2.5 | 3.1 | 3.5 | 4.0 | 4.9 | 3.8 | 0.7 | Population |
| | **13** | NCD-Risk | | Score (points out of 9) | 0.9 | 1.5 | 2.0 | 2.8 | 3.9 | 2.1 | 0.7 | Population |
| | **14** | Sugar-sweetened soft drink consumption | | % adult population (≥15 y) | 6.7 | 16.7 | 24.1 | 33.9 | 51.4 | 18.9 | 10.6 | Population |

| Domain | | Indicator | | Unit | Min. | 25th | Median | 75th | Max | Weighted Mean | Weighted SD | Weighted by |
|---|---|---|---|---|---|---|---|---|---|---|---|---|
| ***Environment, production, and natural resources*** | | | | | | | | | | | | |
| Greenhouse gas emissions | **15** | Food systems greenhouse gas emissions | | kt $CO_2$eq (AR5) | 4.6 | 5,010.1 | 18,626.2 | 61,612.6 | 1,862,042.1 | 82,463.9 | 226,713.0 | (unweighted) |
| | **16** | Greenhouse gas emissions intensity, by product group[$] | Cereals (excl. rice)[†] | kg $CO_2$eq/kg product | 0.0 | 0.1 | 0.2 | 0.3 | 100.6 | 0.2 | 0.1 | Area harvested |
| | | | Beef | kg $CO_2$eq/kg product | 0.4 | 16.4 | 37.3 | 66.5 | 271.7 | 30.3 | 28.2 | Animals slaughtered |
| | | | Cow's milk | kg $CO_2$eq/kg product | 0.2 | 0.7 | 1.4 | 3.4 | 40.1 | 1.0 | 1.0 | Producing animals |
| | | | Rice | kg $CO_2$eq/kg product | 0.2 | 0.9 | 1.6 | 2.7 | 6,671.0 | 1.1 | 0.6 | Area harvested |
| Production | **17** | Food product yield, by food group[$] | Cereals[†] | tonnes/ha | 0.2 | 16.5 | 32.6 | 48.6 | 292.4 | 40.7 | 20.7 | Area harvested |
| | | | Fruit[†] | tonnes/ha | 4.9 | 65.5 | 103.0 | 151.7 | 358.3 | 136.7 | 50.9 | Area harvested |
| | | | Beef | kg/animal | 71.6 | 138.9 | 188.4 | 251.2 | 450.0 | 231.5 | 95.1 | Animals slaughtered |
| | | | Cow's milk | kg/animal | 100.9 | 621.1 | 1,537.4 | 4,841.7 | 12,700.1 | 2,676.6 | 2,713.3 | Producing animals |
| | | | Vegetables[†] | kg/ha | 11.4 | 80.5 | 139.3 | 250.0 | 755.0 | 197.0 | 90.9 | Area harvested |
| Land | **18** | Cropland expansion (relative change 2003-2019) | | % | -78.0 | 2.9 | 16.0 | 51.3 | 1,800.0 | 19.1 | 39.2 | Cropland° |
| Water | **19** | Agriculture water withdrawal as % of total renewable water resources | | % total renewable | 0.0 | 0.4 | 1.9 | 13.0 | 3,892.0 | 16.9 | 52.6 | Cropland |
| Biosphere integrity | **20** | Functional integrity: % agricultural land with minimum level of natural habitat | | % agricultural land | 19.6 | 78.7 | 93.4 | 98.6 | 100.0 | 88.3 | 13.9 | Agricultural land[œ] |
| | **21** | Fishery health index progress score | | index | 0.1 | 11.4 | 22.3 | 31.0 | 64.0 | 21.4 | 12.8 | Population |
| Pollution | **22** | Total pesticides per unit of cropland | | kg/ha | 0.0 | 0.3 | 1.4 | 4.0 | 20.5 | 1.8 | 1.9 | Cropland |
| | **23** | Sustainable nitrogen management index | | Index | 0.2 | 0.7 | 0.9 | 1.1 | 1.4 | 0.7 | 0.2 | Cropland |
| ***Livelihoods, poverty, and equity*** | | | | | | | | | | | | |
| Poverty and income | **24** | Share of agriculture in GDP | | % GDP | 0.0 | 2.5 | 7.9 | 17.9 | 61.3 | 4.4 | 5.2 | GDP |
| Employment | **25** | Unemployment, rural | | % working age population | 0.1 | 2.6 | 4.9 | 8.7 | 34.4 | 5.7 | 4.1 | Population |
| | **26** | Underemployment rate, rural | | % working age population | 0.2 | 2.2 | 4.4 | 7.7 | 36.1 | 7.3 | 8.2 | Population |

| Domain | | Indicator | Unit | Min. | 25th | Median | 75th | Max | Weighted Mean | Weighted SD | Weighted by |
|---|---|---|---|---|---|---|---|---|---|---|---|
| Social protection | **27** | Social protection coverage | % population | 0.9 | 16.8 | 40.8 | 63.6 | 94.0 | 55.8 | 28.0 | Population |
| | **28** | Social protection adequacy | % welfare of beneficiary households | 0.5 | 11.3 | 23.3 | 32.5 | 67.0 | 21.0 | 15.1 | Population |
| Rights | **29** | % Children 5-17 engaged in child labor | % children 5-17 (sex specific is % children 5-17 of each sex) | 0.3 | 3.4 | 9.0 | 17.5 | 40.5 | 9.4 | 9.6 | Population |
| | **30** | Female share of landholdings | % landholdings by sex of operator | 1.7 | 12.8 | 18.7 | 27.7 | 50.5 | 16.8 | 8.3 | Land area |
| ***Governance*** | | | | | | | | | | | |
| Shared vision and strategic planning | **31** | Civil society participation index | index | 0.0 | 0.5 | 0.7 | 0.8 | 1.0 | 0.6 | 0.2 | Population |
| | **32** | % Urban population living in cities signed onto the Milan Urban Food Policy Pact ⁺ | % urban population | 0.0 | 0.0 | 0.0 | 13.2 | 70.8 | 7.2 | 10.4 | Urban population |
| | **33** | Degree of legal recognition of the Right to Food (1 = Explicit protection or directive principle of state policy 2= Other implicit or national codification of international obligations or relevant provisions 3 = None) | categorical | 1.0 | 2.0 | 2.0 | 2.0 | 3.0 | 1.9 | 0.6 | (unweighted) |
| | **34** | Presence of a national food system transformation pathway (0 = No, 1 = yes) | binary | 0.0 | 0.0 | 1.0 | 1.0 | 1.0 | 0.6 | 0.5 | (unweighted) |
| Effective implementation | **35** | Government effectiveness index | index | -2.3 | -0.7 | -0.1 | 0.5 | 2.3 | 0.1 | 0.8 | Population |
| | **36** | International Health Regulations State Party Assessment report (IHR SPAR), Food safety capacity | score | 0.0 | 40.0 | 80.0 | 80.0 | 100.0 | 69.4 | 21.6 | Population |
| | **37** | Presence of health-related food taxes ⁺ | binary | 0.0 | 0.0 | 0.0 | 0.0 | 1.0 | 0.3 | 0.5 | Population |
| Accountability | **38** | V-Dem Accountability index | index | -1.7 | 0.0 | 0.7 | 1.3 | 2.0 | 0.3 | 0.9 | Population |
| | **39** | Open Budget Index Score | index | 0.0 | 31.0 | 46.0 | 61.5 | 87.0 | 43.1 | 21.3 | Population |

| Domain | | Indicator | | Unit | Min. | 25th | Median | 75th | Max | Weighted Mean | Weighted SD | Weighted by |
|---|---|---|---|---|---|---|---|---|---|---|---|---|
| | **40** | Guarantees for public access to information (SDG 16.10.2) | | binary | 0.0 | 0.0 | 1.0 | 1.0 | 1.0 | 1.9 | 0.6 | Population |
| ***Resilience & Sustainability*** | | | | | | | | | | | | |
| Exposure to shocks | **41** | Ratio of total damages of all disasters to GDP | | ratio | 0.0 | 0.0 | 0.0 | 0.0 | 279.2 | 0.3 | 0.8 | GDP |
| Resilience capacities | **42** | Dietary sourcing flexibility index | | index | 0.1 | 0.6 | 0.7 | 0.8 | 1.0 | 0.7 | 0.1 | Population |
| | **43** | Mobile cellular subscriptions (per 100 people) | | number per 100 people | 12.0 | 85.7 | 108.8 | 130.2 | 187.9 | 105.5 | 35.0 | (unweighted) |
| | **44** | Social capital index | | index | 0.1 | 0.3 | 0.4 | 0.5 | 0.9 | 0.5 | 0.2 | Population |
| Agro- and Food Diversity | **45** | Proportion of agricultural land with minimum level of species diversity (crop and pasture) ⁺ | | % agricultural land | 0.0 | 0.0 | 14.1 | 48.5 | 100.0 | 22.5 | 23.6 | Agricultural land § |
| | **46** | Number of (a) plant and (b) animal genetic resources for food and agriculture secured in either medium- or long-term conservation facilities (SDG 2.5.1) | Plants | thousands | 0.0 | 2.1 | 7.0 | 33.0 | 846.3 | 161.4 | 174.5 | Land area |
| | **47** | | Animals | number | 0.0 | 0.0 | 0.0 | 2.0 | 37.0 | 4.4 | 8.8 | Land area |
| Resilience responses/ strategies | **48** | Coping strategies index | | % population | 12.0 | 31.5 | 39.0 | 50.1 | 59.9 | 38.5 | 12.7 | Population |
| Long-term outcomes | **49** | Food price volatility ⁺ | | unitless | 0.0 | 0.6 | 0.7 | 0.9 | 1.4 | 0.7 | 0.3 | (unweighted) |
| | **50** | Food supply variability | | kcal/capita/day | 6.0 | 19.0 | 27.0 | 37.0 | 114.0 | 29.9 | 17.2 | (unweighted) |

‡ Baseline data are comprised of the latest available data point per country-indicator. Latest data point per country-indicator differs given data availability and is reported **SD-E Metadata and Codebook.** 92.5% of data points are from 2017-2022, 6.5% from 2010-2016, and only 1% are from 2000-2009.

† Product mix varies across countries.

° Cropland variable used for weighted means comes from the FAOSTAT database and adheres to the definition of croplands as described in **Table 4.**

⁺ Indicates FSCI value-added to existing data.

$ Additional products included in **SD-A** and in the baseline dataset (**SD-F**)

œ Weighted by agricultural land in 2015 in concordance with the only available year of data for this indicator.

§ Weighted by agricultural land in 2010 in concordance with the only available year of data for this indicator.

Sources: Author's calculations based on data sources listed in **Table 4**.

Notes: Indicator data sources and definitions are in **Table 4.** Each indicator includes a maximum of all UN member states as of August 2022, country list differs per indicator given data availability (see **SD-A Figures A.3-A.11**).

## Global baseline, by theme

**Diets, nutrition, and health.**[iv] Supporting human health is one of three fundamental goals of food systems. The three indicator domains in this theme are food environments (the interface between individuals and the food system), food security, and diet quality.

Indicators of food environments include the availability of fruits and vegetables, per capita sales of ultra-processed[16] foods (UPFs), and access to clean water, essential for avoiding foodborne and waterborne illness. The cost of a healthy diet is the cost of purchasing the least expensive locally available foods to meet requirements for energy and food-based dietary guidelines. The affordability of that diet (cost relative to income) is one of three food security indicators alongside the prevalence of undernourishment and the percentage of the population experiencing moderate or severe food insecurity. Different aspects of diet quality are measured for sub-populations, with parallel indicators of dietary diversity for women and children. Additional indicators for adult populations include "All-5", which measures consumption of the five food groups typically recommended for daily consumption in food-based dietary guidelines (fruits; vegetables; pulses, nuts, or seeds; animal-source foods; and starchy staples)[17], dietary factors that either protect against or increase risk for non-communicable diseases (NCDs), and unhealthy dietary practices over the lifecycle, aligned with international guidance.[17,18]

All food environment indicators suggest inequalities across countries: availability of fruits and vegetables is generally a challenge in low- and middle-income countries (LMICs), while high-income countries (HICs) generally have widespread availability of ultra-processed foods (UPFs). The cost of healthy diets is similar across most countries, but given wide differences in purchasing power, that cost is largely unaffordable across LMICs. GDP per capita is only modestly associated with dietary factors protective against NCDs but strongly associated with those that increase risk for NCDs, though large variation across countries and regions exists.[17]

---

[iv] **SD-A Figures S1.1-S1.24** visualize each indicator in this theme.

**Environment, food production, and natural resources.**[v] Food systems are a major contributor to environmental degradation, but they can also protect and restore environmental outcomes if managed appropriately. The six domains of environmental indicators address the multiple environmental impacts of food systems: greenhouse gas (GHG) emissions, land, biosphere integrity, water, pollution (conceptually including nutrient runoff, chemical exposure, and solid waste), and agricultural production, which interacts with all other domains.

Indicators of GHG emissions include total emissions (from production through consumption and waste disposal) and emissions intensities (emissions per unit of primary product) of major foods. Land use change, measured by cropland expansion and water use, expressed by how much agricultural water withdrawals place pressure on renewable freshwater resources. Overuse of pesticides and sustainable nitrogen management capture pollution. Functional integrity – the capacity for biodiversity to support sustainable food production and other ecosystem services – and the integrity of fishery stocks capture biosphere integrity.[vi] Yields interact with all other domains; increases are directly tied to the observed declining trends in emissions intensities.

Reductions in nitrogen pollution are observed in many places while water withdrawals are stable or modestly decreasing everywhere. Northern Africa and Western and Southern Asia remain at greatest risk of exhausting available water resources. Despite some improving trajectories, total food system emissions are increasing and remain high in HICs. Pesticide application has increased in many countries, highlighting a potential trade-off with increasing yields. Only 88% of agricultural lands have the minimum of 10% functional integrity needed to support food production, meaning over one-tenth of the world's agricultural lands lack foundational ecosystem services such as crop pollination, pest regulation, and soil protection, and other research suggests that the 10% threshold may be insufficient.[19]

---

[v] **SD-A Figures S2.1-S2.21** visualize each indicator in this theme.
[vi] Yields for cereals, vegetables, and cow's milk and emissions intensities for rice and beef are reported in the main analysis as the most relevant to track for food system transformation. Additional products are included in the SD-A and SD-F (the baseline dataset).

**Livelihoods, poverty, and equity.**[vii] Poverty is most prevalent in rural areas where people earn significant income shares from agriculture (including marginalized groups such as Indigenous Peoples and female-headed households).[20–22] Food systems provide employment for 1.23 billion people and including household members support over 3.83 billion livelihoods, in all stages of the value chain across rural and urban areas.[23] Four indicator domains capture their wellbeing: income and poverty, employment, social protection, and rights. Compared to other themes, the available data are more limited due in large part to lack of disaggregation to identify food system livelihoods from others.

Lacking a rural poverty indicator with sufficient coverage, income and rural (monetary) poverty are captured by the share of GDP from agriculture, as a proxy for a country's overall level of development.[24] Unemployment and underemployment capture employment, though not 'decent' work.[25] Though lacking sectoral disaggregation, the rural rates proxy the status of agricultural and farm-related labor markets.[26] Social protection systems increase access to food quantity and quality, reduce producers' risk, and incentivize productive investment.[27,28] Social protection programs may be particularly impactful in breaking the cycle of poverty for small-scale food producers and informal workers who face chronic food insecurity and vulnerability to shocks.[28] Finally, among the many rights related to livelihoods, indicators currently available capture women's access to land and the specific human rights violation of child labor, of which an estimated 60% occurs in agriculture.[29]

Available data provide only a partial view of food system-based livelihoods, but even the incomplete picture suggests deep inequalities. The share of agriculture in GDP remains high in most low-income countries (LICs), indicating limited opportunity for income diversification out of agriculture. Important differences in unemployment and underemployment between rural and urban areas show unemployment prevalent in urban areas while underemployment is more prevalent in rural areas. Other evidence shows a

---

[vii] **SD-A Figures S3.1-S3.13** visualize each indicator in this theme.

larger gender gap in labor force participation in rural areas.[30] Even where there is adequate coverage of social protection programs, the level of benefits provided may be insufficient to produce meaningful impacts, and informal and seasonal workers are often excluded.[31–34] Finally, access to land shows a stark gender disparity with no country approaching gender equality in landholdings.

**Governance.**[viii] Governance is foundational for inclusive food system transformation, encompassing not only the political commitment to adopt supportive policies but also promoting participatory processes and accountability to ensure that policies have legitimacy and reach the intended target group. Furthermore, governance involves strengthening capacities for implementation across sectors to ensure aspirational goals are technically feasible. Three indicator domains collectively capture these dimensions of governance: shared vision and strategic planning, effective implementation, and accountability. There are few indicators of governance specific to food systems, but broad indices of the governance landscape may have significant impacts on food system choices and outcomes. Further research is especially needed in this area to develop more direct indicators of food systems governance.

Indicators of shared vision and strategic planning include one broad indicator beyond food systems and three others reflecting intentionality by governments to pursue food systems objectives. The Civil Society Participation Index captures whether civil society organizations (e.g., non-governmental organizations (NGOs), unions, social movements) have opportunities to convey their views to policymakers. Food system-specific indicators are the presence of a legal recognition of the Right to Food; the existence of a food system transformation pathway; and the share of the urban population living in cities that have signed onto the Milan Urban Food Policy Pact (MUFPP). The MUFPP is an innovative policy mechanism that has rapidly become the leading international tool for urban food policy governance (37 recommended actions and specific indicators) as well as a platform for cooperation, organizing, and political influence.[35]

---

[viii] **SD-A Figures S4.1-S4.17** visualize the indicators in this theme.

Effective implementation is also measured by a combination of indicators that are contextual (broader than the food system but establish the governance regime within which food system actors can operate) and specific to food systems. The Government Effectiveness Index reflects the quality of public services, civil service, policy formulation, implementation, and credibility. Public tracking of investments for food systems requires transparency over budgets and guarantees for public information access, reflected in the Open Budget Index score and guarantees for public access to information, as well as the overall Accountability Index, which encompasses the existence of mechanisms to keep officials responsive to the public (e.g., checks and balances, elections, press freedoms). Specific to food systems, available data can monitor two policy tools for achieving healthy food systems: health-related food taxes and food safety capacity (the number of specific mechanisms in place to detect and respond to foodborne disease and contamination).

The data show that indicators of overall governance track country income, while those more closely related to food systems show more heterogeneity across regions and income groups. For example, only 29 countries explicitly recognize the Right to Food, while the US, Canada, the UK, and Australia notably have no degree of legal recognition. And health-related food taxes exist in 38 countries spread across all continents.

**Resilience and sustainability.**[ix] The COVID-19 pandemic and conflict in Ukraine both demonstrated the imperative to better understand and strengthen the resilience of local and global food systems to a numerous shocks and stressors – not just climate change. Assessing resilience requires a combination of indicators related to two domains: (i) the contextual elements of resilience – the level of exposure of the system to adverse events, and the capacities of that system to anticipate, absorb, or adapt to those events –

---

[ix] **SD-A Figures S5.1-S5.17** visualize the indicators in this theme.

and (ii) the short- and longer-term outcomes of resilience – generally measured through individual and system wellbeing, ideally considered at multiple scales.[36]

A range of indicators are necessary to capture these different components of resilience and to better understand how to establish more efficient, inclusive, and sustainable food systems in the face of increasingly complex and intertwined shocks. As such, indicators of resilience cover five domains: exposure to shocks, resilience capacities, agro- and food diversity, short-term resilience responses, and long-term outcomes.

Exposure to shocks depends on the intensity, nature, and frequency of shocks and stressors and can be proxied by the cumulative costs of those events relative to GDP. Resilience capacities are the different elements that can be used to buffer and respond to adverse events. Those capacities take many forms. In food systems, diversity and redundancy of food sources, national infrastructure – proxied by mobile phone coverage – and social capital are some of the key elements that constitute resilience capacities. Also critical to food systems resilience is the level of biodiversity on which food production relies, captured by the number of plant and animal genetic resources conserved for use. Understanding how actors react and respond in the short term to the impact of shocks is also a foundational element of resilience analysis. This element can be measured using the coping strategies index, while longer-term outcomes of food system resilience can be captured by the ability of the system to maintain low price volatility and low food supply variability.

Looking across resilience indicators for a sub-group of countries **(SD-A Figure S5.17)**, the data reveal some notable results. The Philippines, Nicaragua, and Indonesia, for instance, demonstrate relatively higher food price volatility or food supply variability than, e.g., the Netherlands, Thailand, or even India. This empirical observation suggests that countries affected by higher exposure and/or lower resilience capacities (e.g., Nicaragua, Ecuador) are also faring worse in their food system outcomes than those less exposed to shocks and/or characterized by higher social capital and dietary sourcing flexibility (Thailand, the Netherlands). This trend however displays important variability, reflecting the specificity in how shocks propagate through a country’s food system, and calls for more in-depth analyses as well as close monitoring to better understand individual and combined dynamics. Future research aims to develop a rigorous and robust way to assess and monitor food system resilience, and eventually aim at coupling this with further work on food system sustainability, recognizing that resilience is a prerequisite for long-term sustainability.

### Regional baselines

Regional[x] patterns relative to global means present a broad picture of where each region stands per indicator, using current global means and the desirable direction of change (defined in **Tables 2**, **3**, and **4**) to provide broad characterizations of relative status.[xi] The terms "better" and "worse" are used throughout to communicate the directionality of the region relative to the global average given that the desirable direction of change varies per indicator, however, the global mean may well be far from established thresholds or benchmarks that characterize positive food system transformation. Thus, the characterization is meant to be descriptive of the relative baseline starting points and not intended as a performance assessment, which is a subsequent research agenda of the FSCI in the next two years.

**Figure 2** illustrates the distribution of regional weighted means relative to the global weighted mean for each indicator, aligned to the desirable direction of change such that the regions to the left are performing "worse" than the global mean and vice versa. **Table 2** complements the figure to show whether differences observed are meaningful by indicating their magnitude (percent difference), direction (sign aligned to desirable direction), and whether statistically significantly different from zero.[xii]. For brevity, the discussion concentrates on indicators significant at the 1% level. No region demonstrates desirable performance on all indicators. Only two indicators do not differ across regions: the presence of a food system transformation pathway and coping strategies.

---

[x] Modified M-49 groupings are illustrated in **SD-A Figure 1**. Country classification is provided in **SD-E Metadata and Codebook.**
[xi] Furthermore, country level data mask heterogeneity within countries. To better manage food systems and achieve 2030 goals, countries ought to monitor these indicators at sub-national level. Sub-national monitoring is beyond the scope of what this global initiative can do.
[xii] For the level of regional weighted means, see **SD-A Table A.1**. Medians are provided in **SD-A Table A.2**.

**Figure 2. Indicator regional means relative to global mean, by thematic area and desirable direction of change**

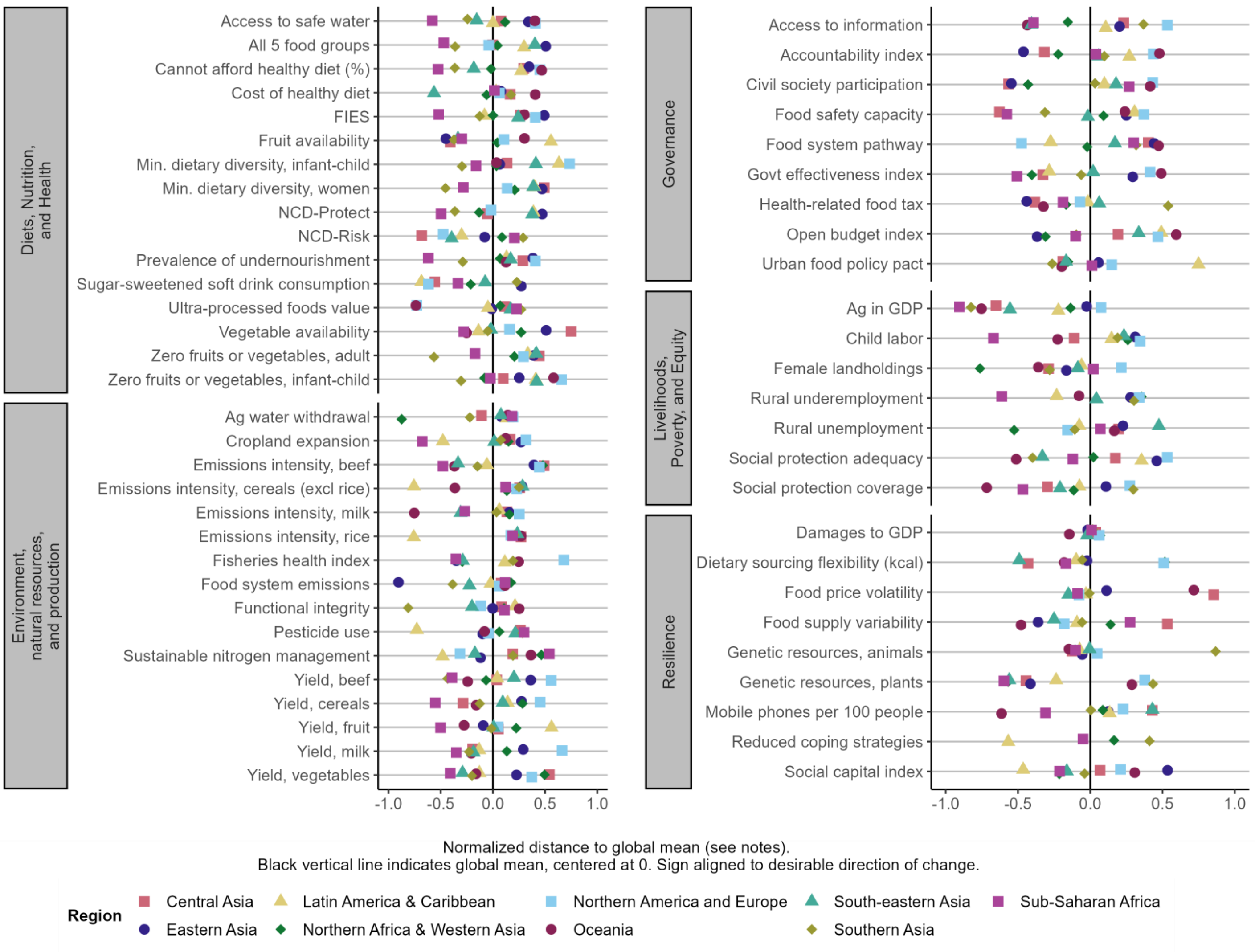

Sources: Author's calculations based on data sources listed in **Table 4**.

Notes: Normalized distance to global mean (weighted means following weights defined in Table 1) is calculated relative to the global mean and scaled to the minimum and maximum of regional mean, per indicator (global mean centered at 0). The sign of the normalized distance has been reversed for all indicators where a lower value is more desirable, such that -1 can be interpreted as "worse than" and 1 can be interpreted as "better than" the global mean. Degree of legal recognition of the Right to Food not shown. Product mix in aggregate categories of emissions intensities (cereals) and yields (cereals, fruit, and vegetables) differ across countries. Yield and emissions intensity for additional products included in **SD-A** and in the baseline dataset.

No region stands out for better than average performance on all indicators, but each has certain indicators that do. For **Latin America and the Caribbean** these are minimum dietary diversity and fruit and vegetable consumption, functional integrity, social protection adequacy, child labor, municipalities signed onto the MUFPP, and government accountability. **North America and Europe** unsurprisingly performs better than average on food security and indicators of dietary diversity and fruit and vegetable consumption, as well as yields, emissions intensities (beef and milk), agricultural water withdrawal, underemployment, social protection coverage, child labor, government effectiveness, open budget index, and accountability. **Oceania** performs well on access to safe drinking water, the percent of people who can afford a healthy diet, consumption of fruits and vegetables by infants and young children, total food system emissions, beef emissions intensities, milk yields, agricultural water withdrawal, functional integrity, fisheries health index, government accountability, and open budgets. For **North Africa and Western Asia,** the better-than-average indicators are total food system emissions, functional integrity, underemployment, and dietary sourcing flexibility. **Central Asia** displays better than average performance on vegetable availability, prevalence of undernourishment, food insecurity experience, minimum dietary diversity for women, adult fruit and vegetable consumption, total food systems emissions, beef emissions intensities, functional integrity, and food price volatility. In **Eastern Asia**, nearly all indicators of diets, nutrition, and health are above global averages except NCD-Risk. It performs well on beef and rice emissions intensities, cereal, fruit, and vegetable yields, nearly all indicators of livelihoods, government effectiveness, food safety capacity, and social capital. **South-Eastern Asia** shows better than average UPF sales, prevalence of undernourishment, dietary diversity, fruit and vegetable consumption, and use of pesticides and rural unemployment. In **Southern Asia**, the notable indicators currently better than the global average include UPF sales and NCD-Risk, rice emissions intensity, fisheries health, and pesticide use, rural underemployment, child labor, and conservation of animal genetic resources. **Sub-Saharan Africa** has better than average UPF sales and civil society participation as well as indicators of input use (water withdrawal, pesticide use, sustainable nitrogen management), though given low yields, this level of input use is likely insufficient for sustainable production.

## Table 2. Percent deviation from global mean, by region and desirable direction of change[‡]

| Domain | | Indicator | | Global weighted mean | Dir. of Δ | % Deviation of regional weighted mean from global weighted mean, sign aligned to desirable direction of change | | | | | | | | | |
|---|---|---|---|---|---|---|---|---|---|---|---|---|---|---|---|
| | | | | | | **Latin America & Caribbean** | **Northern America & Europe** | **Oceania** | **Northern Africa & Western Asia** | **Central Asia** | **Eastern Asia** | **South-eastern Asia** | **Southern Asia** | **Sub-Saharan Africa** | **Joint signif. (F-test)[§]** |
| ***Diets, Nutrition, & Health*** | | | | | | | | | | | | | | | |
| Food environments | **1** | Cost of a healthy diet | | US PPP $3.3 per person/day | ↓ | 1.2 | 4.9 | 21.1*** | -3.1 | 10.3 | 1.3 | -27.4*** | 6.2 | -1.1 | *** |
| | **2** | Availability of fruits and vegetables | Fruits | 223.8 grams per capita/ day | ↑ | 37.6* | 7.9 | 20.4 | 4 | -30.1 | -36.3* | -22.3 | -29.9 | -24.3* | ** |
| | | | Vegetables | 246.8 grams/ capita/ day | ↑ | -25.2* | 32.6** | -37.4** | 50.1** | 133.2*** | 91.9 | -9.6 | -13.7 | -50*** | *** |
| | **3** | Retail value of ultra-processed foods | | US PPP $204.0 per capita | ↓ | -19.6 | -245.7*** | -256.7* | 26.9 | 41.4** | -10.8 | 52.1*** | 88.1*** | 79.8*** | *** |
| | **4** | % Population using safely managed drinking water services (SDG 6.1.1) | | 66.3% population | ↑ | 4.2 | 42.2*** | 42.9*** | 13.9* | 5.3 | 41.4*** | -16 | -24.3 | -69.3*** | *** |
| Food security | **5** | Prevalence of Undernourishment (SDG 2.1.1) | | 9.4% population | ↓ | 17.5 | 73.1*** | 25.5 | 11.3 | 61.9*** | 66.3*** | 37*** | -62.9*** | -116.7*** | *** |
| | **6** | % Population experiencing moderate or severe food insecurity (SDG 2.1.2) | | 29.5% population | ↓ | -10.8 | 74.4*** | 56.5*** | 1.7 | 48.9* | 84.3*** | 36.4 | -21.2 | -104.1*** | *** |
| | **7** | % Population who cannot afford a healthy diet | | 42.3% population | ↓ | 46.5*** | 95.6*** | 93.6*** | 2.9 | 48.9 | 74*** | -27.5 | -65.6*** | -100.8*** | *** |
| Diet quality | **8** | MDD-W: minimum dietary diversity for women | | 65.7% population, women 15-49 | ↑ | 24.9*** | 9.3*** | | 14.3*** | 33.6*** | 31.3*** | 28.5*** | -31.6*** | -20.2** | -- |
| | **9** | MDD (IYCF): minimum dietary diversity for infants and young children | | 31.8% population, 6-23 months | ↑ | 95.4*** | 122.9*** | 1.1 | 10.7 | 14.7 | 16.7*** | 66.5*** | -40.2*** | -27.4*** | *** |
| | **10** | All-5: consumption of all 5 food groups | | 39.0% adult population (≥15 y) | ↑ | 20.7 | -1.8 | | 6.9 | -3.2 | 39*** | 27.4 | -29.2*** | -35.8*** | -- |
| | **11** | Zero fruit or vegetable consumption | Adults | 10.8% adult population (≥15 y) | ↓ | 53.1*** | 49.3*** | | 38.7*** | 71.1*** | 65.3*** | 59.6*** | -84.7*** | -26.1 | -- |
| | | | Children 6-23 months | 39.1% population, 6-23 months | ↓ | 50*** | 81.7*** | 66.4*** | -9.3 | 13.1 | 24.8*** | 45.1*** | -38*** | -8.9 | *** |
| | **12** | NCD-Protect | | 3.8 points (out of 9) | ↑ | 15.2 | -2 | | -5.7 | -2.6 | 20.4*** | 14.4* | -14*** | -19.7*** | -- |
| | **13** | NCD-Risk | | 2.1 points (out of 9) | ↓ | -29.1*** | -46.6*** | | 12.4 | -68*** | -12.1*** | -40.9*** | 30.6*** | 17.1 | -- |

| Domain | | Indicator | | Global weighted mean | Dir. of Δ | % Deviation of regional weighted mean from global weighted mean, sign aligned to desirable direction of change | | | | | | | | | |
|---|---|---|---|---|---|---|---|---|---|---|---|---|---|---|---|
| | | | | | | Latin America & Caribbean | Northern America & Europe | Oceania | Northern Africa & Western Asia | Central Asia | Eastern Asia | South-eastern Asia | Southern Asia | Sub-Saharan Africa | Joint signif. (F-test)§ |
| | **14** | Sugar-sweetened soft drink consumption | | 18.9% adult population (≥15 y) | ↓ | -91.4*** | -76.9* | | -31** | -77*** | 38.9*** | -9.8 | 24.6** | -44* | -- |
| ***Environment, natural resources, and production*** | | | | | | | | | | | | | | | |
| Greenhouse gas emissions | **16** | Food systems greenhouse gas emissions | | 82,463.9 kt $CO_2$eq (AR5) | ↓ | 3.6 | 6.4 | 72.6*** | 63.3*** | 62.5*** | -436.5 | -102 | -215.6 | 40.9* | *** |
| | **16** | Greenhouse gas emissions intensity, by product group$ | Cereals (excl. rice)† | 0.2 kg $CO_2$eq/ kg product | ↓ | 10 | 8.2 | -51.2*** | -30.9** | 15 | -2.7*** | 4.5 | -37.6*** | 17.3** | *** |
| | | | Beef | 30.3 kg $CO_2$eq/ kg product | ↓ | -34.4*** | 49.9*** | 30.2*** | 30.9 | 44.6*** | 49.6*** | -75.6*** | -108.8 | -147* | *** |
| | | | Cow's milk | 1.0 kg $CO_2$eq/ kg product | ↓ | -4.5 | 38.1*** | 17.7** | -6 | -18.7 | 13.5 | -185.3 | -35.1** | -288*** | *** |
| | | | Rice | 1.1 kg $CO_2$eq/ kg product | ↓ | 12.5 | -75.9 | -33.3 | -6.7 | -166.8 | 16.3*** | -33.9* | 20.4*** | -46.8* | *** |
| Production | **17** | Food product yield, by food group$ | Cereals† | 40.7 tonnes/ha | ↑ | 16.3* | 32.5 | -58.2*** | -43.8** | -59*** | 53.2*** | 6.2 | -19*** | -59.8*** | *** |
| | | | Fruit† | 136.7 tonnes/ha | ↑ | 24.9** | -5.6 | -2.5 | 1.8 | -1.8 | 18.7*** | 3.1 | 2.3 | -43.5*** | *** |
| | | | Beef | 231.5 kg/animal | ↑ | 18.2 | 37.3** | 0.7 | -17.7 | -19.9*** | -31.9*** | -8.7 | -44.2*** | -35.9*** | *** |
| | | | Cow's milk | 2676.6 kg/animal | ↑ | -6.9 | 184.1*** | 82.6*** | -29.9 | -15.8 | 14.3 | -59.5*** | -41.1*** | -81.4*** | *** |
| | | | Vegetables† | 197.0 kg/ha | ↑ | -5.2 | 45.9** | 2.9 | 27.1 | 74.2** | 29.9*** | -40.5*** | -21.7*** | -71*** | *** |
| Land | **18** | Cropland expansion (relative change 2003-2019) | | 19.1% | ↓ | -150.9* | 98*** | 42.1 | 32.5 | 60*** | 71.5*** | 17.5 | 28.3** | -209.1** | *** |
| Water | **19** | Agriculture water withdrawal as % of total renewable water resources | | 16.9% total renewable | ↓ | 78.6*** | 80.3*** | 86.2*** | -474.1* | -86.8 | 20.5*** | 49.5*** | -139.8*** | 75*** | *** |
| Biosphere integrity | **20** | Functional integrity: % agricultural land with minimum level of natural habitat | | 88.3% agricultural land | ↑ | 7.2*** | -3.4 | 9*** | 5.8 | 3.1*** | -0.4 | -7.1 | -30.1* | 2.6 | *** |
| | **21** | Fishery health index progress score | | 21.4 | ↑ | 13.2 | 79.4** | 27.9*** | -36* | | -42.7*** | -35.9* | 27.4*** | -42.3* | *** |
| Pollution | **22** | Total pesticides per unit of cropland | | 1.8 kg/ha | ↓ | -195.9*** | -10.7 | -20.1* | 29 | 58.5** | -30.9 | 46.7 | 76.1*** | 77.4*** | *** |
| | **23** | Sustainable nitrogen management index | | 0.7 | ↑ | -20 | -12.5 | 16.1*** | 22.1 | 9.7** | -5.4** | -6.4 | 10.7* | 24.3*** | *** |
| ***Livelihoods, Poverty, & Equity*** | | | | | | | | | | | | | | | |
| Poverty and income | **24** | Share of agriculture in GDP | | 4.4% GDP | ↓ | -32.1** | 69.1*** | 36.8 | -19.2 | -151.1 | -30.2 | -143.9*** | -305*** | -315.3** | *** |
| Employment | **25** | Unemployment, rural | | 5.7% working age population | ↓ | -11.7 | -15.5 | 29.3* | -77.4*** | 21.6** | 34.1*** | 63.4*** | -20.1 | 4.4 | *** |

| Domain | | Indicator | Global weighted mean | Dir. of Δ | % Deviation of regional weighted mean from global weighted mean, sign aligned to desirable direction of change: Latin America & Caribbean | Northern America & Europe | Oceania | Northern Africa & Western Asia | Central Asia | Eastern Asia | South-eastern Asia | Southern Asia | Sub-Saharan Africa | Joint signif. (F-test)[§] |
|---|---|---|---|---|---|---|---|---|---|---|---|---|---|---|
| | **26** | Underemployment rate, rural | 7.3% working age population | ↓ | -45.1*** | 60.2*** | -16.9*** | 71.9*** | 90 | 59.9*** | 6.8 | 61.8*** | -116* | *** |
| Social protection | **27** | Social protection coverage | 55.8% population | ↑ | -14.1* | 32*** | -86.7*** | -18.8 | -34.2** | 13*** | -23.3** | 38.4 | -59.7*** | *** |
| | **28** | Social protection adequacy | 21.0% welfare of beneficiary households | ↑ | 55.6** | 79.9* | -81.9*** | -0.3 | 32.5 | 75.3*** | -46.1* | -60.7*** | -20 | *** |
| Rights | **29** | % Children 5-17 engaged in child labor | 9.4% children 5-17 | ↓ | 39.4** | 65.1*** | -52.6*** | 54.4** | -13.8 | 52.5*** | 45.3* | 43.4*** | -134.9*** | *** |
| | **30** | Female share of landholdings | 16.8% landholdings by sex of operator | ↑ | -11.5 | 13.2 | -99.4*** | -3.4 | -47.4*** | -15.9*** | 9.7 | -59.5** | -13.3 | -- |
| ***Governance*** | | | | | | | | | | | | | | |
| Shared vision and strategic planning | **31** | Civil society participation index | 0.6 | ↑ | 7.9 | 37.6** | 29.5*** | -37.6*** | -42.4*** | -41.3*** | 14.5 | 5.8 | 18.4*** | *** |
| | **32** | % Urban population living in cities signed onto the Milan Urban Food Policy Pact | 7.2% urban population | ↑ | 250.7** | 66.9* | -76.4*** | -65.5*** | -58.4* | 16 | -70.9*** | -91.3*** | 16.4 | *** |
| | **33** | Degree of legal recognition of the Right to Food (1 = Explicit protection or directive principle of state policy 2= Other implicit or national codification of international obligations or relevant provisions 3 = None) | 1.9 | ↓ | 4.5 | -4.7 | -33.1** | -9.3* | -4.7*** | 5.7 | -4.7 | 34.3*** | 8.3* | -- |
| | **34** | Presence of a national food system transformation pathway (0 = No, 1 = yes) | 0.6 | ↑ | -18.2 | -35.5** | 34.6 | -0.2 | 31.6 | 31.6 | 15 | 23.3 | 23.3* | |
| Effective implemen-tation | **35** | Government effectiveness index | 0.1 | ↑ | -380.3*** | 637.4*** | 683.9 | -549.3*** | -438** | 467.3*** | 51.5 | -55 | -756.9*** | *** |
| | **36** | International Health Regulations State Party Assessment report (IHR SPAR), Food safety capacity | 69.4 | ↑ | 22.3** | 27.7*** | 18.8 | 4 | -38.9 | 17.9*** | -0.4 | -17.6*** | -35.3*** | *** |

| Domain | | Indicator | | Global weighted mean | Dir. of Δ | % Deviation of regional weighted mean from global weighted mean, sign aligned to desirable direction of change: Latin America & Caribbean | Northern America & Europe | Oceania | Northern Africa & Western Asia | Central Asia | Eastern Asia | South-eastern Asia | Southern Asia | Sub-Saharan Africa | Joint signif. (F-test)[§] |
|---|---|---|---|---|---|---|---|---|---|---|---|---|---|---|---|
| | 37 | Presence of health-related food taxes | | 0.3 | ↑ | 1.1 | -18.7 | -87.3*** | -47.6 | -100*** | -100*** | 4.2 | 137 | -47.5 | -- |
| Accountability | 38 | V-Dem Accountability index | | 0.3 | ↑ | 229.5*** | 337*** | 427.5*** | -187.1*** | -259.1*** | -389.7** | 31.5 | 59 | 62.5 | *** |
| | 39 | Open Budget Index Score | | 43.1 | ↑ | 53** | 55.8*** | 68.1*** | -28.8 | 15.9 | -41.8** | 37.3** | -12.8*** | -12.7 | *** |
| | 40 | Guarantees for public access to information (SDG 16.10.2) | | 1.9 | ↑ | 3.3 | 15.6*** | -12.8 | -4.8 | 5.2 | 5.2 | -10.4 | 9.1 | -11.5** | -- |
| ***Resilience & Sustainability*** | | | | | | | | | | | | | | | |
| Exposure to shocks | 41 | Ratio of total damages of all disasters to GDP | | 0.3 | ↓ | 19.5 | -43.4 | 27.3*** | 88.2*** | 99*** | 47.4*** | 56.4*** | 15.9 | 74.8*** | *** |
| Resilience capacities | 42 | Dietary sourcing flexibility index | | 0.7 | ↑ | -2.1 | 9.8 | -3.9 | 9.9*** | -8.2*** | -1.2 | -9.6 | -1.6 | -3.6* | *** |
| | 43 | Mobile cellular subscriptions (per 100 people) | | 105.5 per 100 people | ↑ | 4.4 | 13.4*** | -28.1** | 3.1 | 20.2* | 7.1 | 21.5* | -0.1 | -17.4** | *** |
| | 44 | Social capital index | | 0.5 | ↑ | -46.2*** | 16.3 | 28.9* | -17.9*** | 3 | 46.1*** | -16.4* | -8.1* | -23.6*** | *** |
| Agro- and food diversity | 45 | Proportion of agricultural land with minimum level of species diversity (crop and pasture) | | 22.5 % agricultural land | ↑ | -60.3*** | -12.9 | -33*** | -59* | -71.2*** | 45.7 | 92.1* | 96.9 | 40.1* | -- |
| | 46 | Number of (a) plant and (b) animal genetic resources for food and agriculture secured in either medium- or long-term conservation facilities (SDG 2.5.1) | Plants | 161.4 (thousands) | ↑ | -35 | 55.9 | 41.9** | -88.4*** | -74.9*** | -63.8* | -93.1*** | 62.7 | -92.1*** | *** |
| | 47 | | Animals | 4.4 | ↑ | -86.9*** | 30.2 | -100*** | -95.1*** | -100*** | -71.4 | 7.2 | 731.2*** | -70.2*** | -- |
| Resilience responses/ strategies | 48 | Coping strategies index | | 38.5% population | ↓ | -19.9 | | | 5.4 | | | | 15 | -2.4 | |
| Long-term outcomes | 49 | Food price volatility | | 0.7 | ↓ | -1.8 | -1.9 | 19.1 | -2.1 | 21.2*** | 4 | -4 | -0.6 | -1.9 | *** |
| | 50 | Food supply variability | | 29.9 kcal per capita/ day | ↑ | -3.2 | -8.2 | -22.4 | 6.8 | 28 | -16.1 | -12 | -3.2 | 12.5 | -- |

*** p < 0.001 ** p < 0.01 * p < 0.05

§ Reflects p-value of joint significance tests (F-test), '--' indicates insufficient observations in one or more regions to compute the F-test with cluster robust standard errors, required due to unequal variances by region.

‡ See **SD-A Table A.1** for regional means and **SD-A Table A.2** for regional medians.

† Product mix varies across countries.

$ Additional products included in **SD-A Figures S2.3-S2.10** and in the baseline dataset.

Sources: Author's calculations based on data sources listed in **Table 4**.

**Relationship to country income**

Patterns across income groups can also be informative to guide priorities and action. Many aspects of food systems are associated with country income level,[24] begging the questions of whether this indicator framework is redundant with income and, if not, where there are inflection points by income that might help countries set priorities. Disaggregating by country income level (**Figure 3** and **Table 3**) proceeds as above and then the relationship between the indicators and GDP, by region, is illustrated (**Figure 4**). The results demonstrate that only some food system indicators show clear patterns of association with GDP, underscoring the potential for transformation across income contexts as well as the need for dedicated monitoring of food systems.

Several indicators show no statistically significant difference ($p>0.05$) by country income group at all (**Table 3** joint significance column). Of those that do differ by income, HICs generally have the advantage, but perform worse than average on UPF sales, NCD-Risk, and soft drink consumption. Similarly, LICs show certain disadvantages but perform better than the global average for NCD-Risk and pesticide use. Lower middle-income countries perform largely in line with the global average for most indicators with better performance on UPF sales and pesticide use and worse than average on the affordability of healthy diets, beef and milk yields, social protection adequacy, and food safety capacity. Upper middle-income countries perform in line with or better than the global average on nearly all indicators where there are differences by income, excepting NCD-Risk and conservation of animal genetic resources.

Beyond country income level, understanding each indicator's relationship to GDP per capita is useful for hypothesis generation. **Figure 4** shows the relationship between two illustrative indicators per theme and GDP per capita, selected purposively to illustrate one indicator in each theme that does and does not show some clear relationship to GDP.[13] These findings underscore the potential for policymakers and other actors to influence more desirable outcomes on at least some indicators of food systems even in LICs, and to identify where income seems to be a necessary driver (though alone likely insufficient) of more desirable outcomes.

[13] **SD-A Figures A.12-A.16** show all continuous indicators.

**Figure 3. Income group means relative to global mean, by thematic area**

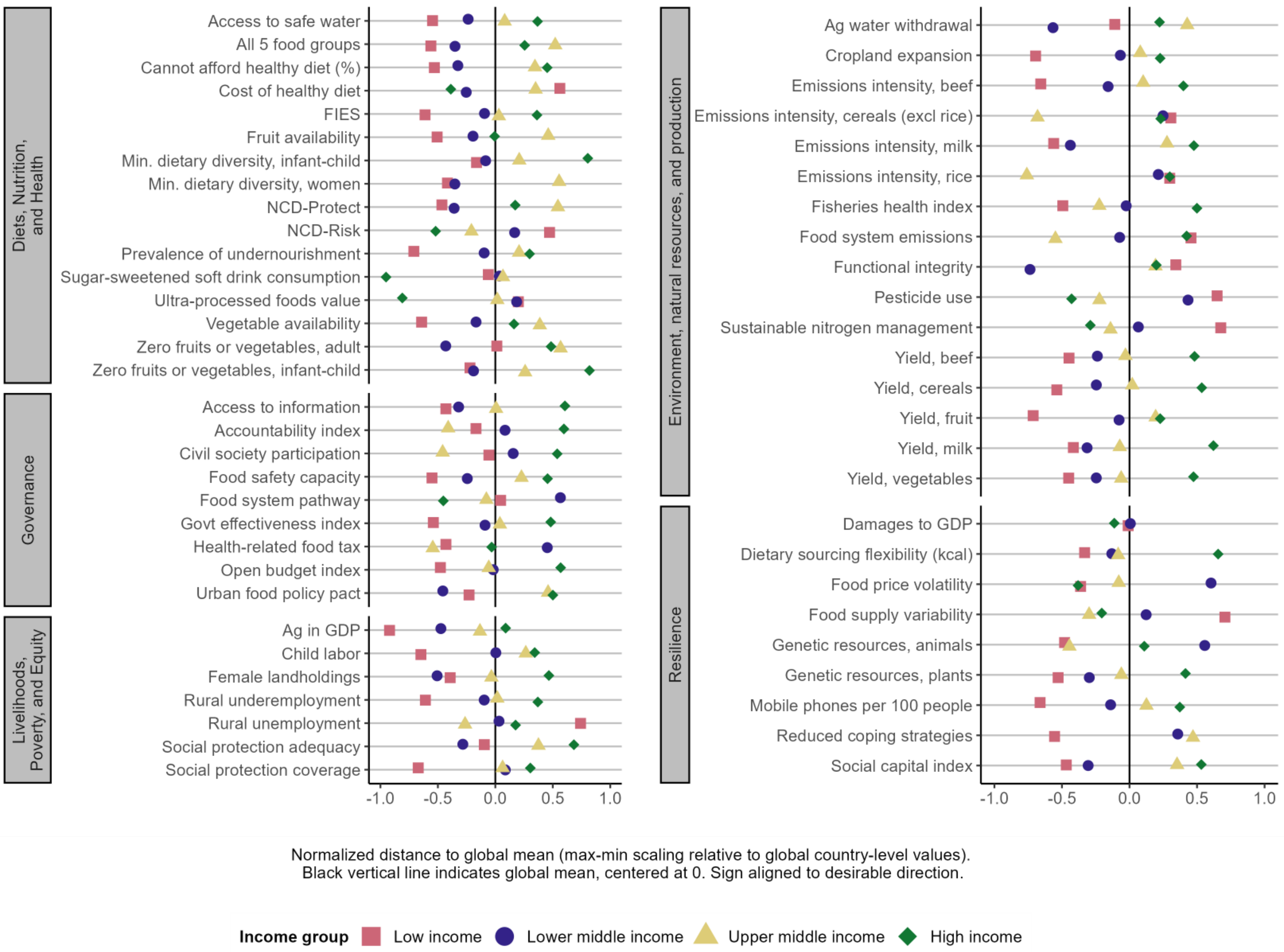

Sources: Author's calculations based on data sources listed in **Table 4**.
Notes: Normalized distance to global mean (weighted means following weights defined in Table 1) is calculated relative to the global mean and scaled to the minimum and maximum of income group mean, per indicator (global mean centered at 0). The sign of the normalized distance has been reversed for all indicators where a lower value is more desirable, such that -1 can be interpreted as "worse than" and 1 can be interpreted as "better than" the global mean. Number of people who cannot afford a healthy diet and Degree of legal recognition of the right to food not shown. Product mix in aggregate categories of emissions intensities (cereals) and yields (cereals, citrus, fruit, pulses, roots and tubers, and vegetables) differ across countries. Yield and emissions intensity for additional products are included in the **SD-A** and baseline dataset.

## Table 3. Income group deviation from global weighted means, by desirable direction of change[‡]

| Domain | | Indicator | | Global weighted mean | Dir. of Δ | % Deviation of income group weighted mean from global weighted mean, sign aligned to desirable direction of change | | | | |
|---|---|---|---|---|---|---|---|---|---|---|
| | | | | | | Low income | Lower middle income | Upper middle income | High income | Joint signif. (F-test)[§] |
| ***Diets, Nutrition, & Health*** | | | | | | | | | | |
| Food environments | **1** | Cost of a healthy diet | | US PPP $3.3 per person/day | ↓ | 7.5 | -3.1 | 5 | -5.2 | |
| | **2** | Availability of fruits and vegetables | Fruits | 223.8 grams per capita/ day | ↑ | -27.3* | -9.5 | 25.7* | -0.6 | |
| | | | Vegetables | 246.8 grams/ capita/ day | ↑ | -48.4*** | -9.5 | 26 | 11 | |
| | **3** | Retail value of ultra-processed foods | | US PPP $204.0 per capita | ↓ | 88.1*** | 77.8*** | 11.1 | -293*** | *** |
| | **4** | % Population using safely managed drinking water services (SDG 6.1.1) | | 66.3% population | ↑ | -70*** | -26.9* | 9 | 47.8*** | *** |
| Food security | **5** | Prevalence of Undernourishment (SDG 2.1.1) | | 9.4% population | ↓ | -211.3*** | -33.4 | 58.6*** | 71.5*** | *** |
| | **6** | % Population experiencing moderate or severe food insecurity (SDG 2.1.2) | | 29.5% population | ↓ | -123.1*** | -16.9 | 15.4 | 76.5*** | *** |
| | **7** | % Population who cannot afford a healthy diet | | 42.3% population | ↓ | -108.9*** | -63.9*** | 63.8*** | 96.3*** | *** |
| Diet quality | **8** | MDD-W: minimum dietary diversity for women[++] | | 65.7% population, women 15-49 | ↑ | -22.3** | -17.2 | 27.4*** | | |
| | **9** | MDD (IYCF): minimum dietary diversity for infants and young children | | 31.8% population, 6-23 months | ↑ | -32.5*** | -14.7 | 34.6* | 119.7*** | *** |
| | **10** | All-5: consumption of all 5 food groups | | 39.0% adult population (≥15 y) | ↑ | -29.7** | -18.5 | 27.2* | 13.5*** | *** |
| | **11** | Zero fruit or vegetable consumption | Adults | 10.8% adult population (≥15 y) | ↓ | 5 | -45.2 | 58.8*** | 50.3*** | |
| | | | Children 6-23 months | 39.1% population, 6-23 months | ↓ | -19.1 | -15.4 | 30.9*** | 85.9*** | *** |
| | **12** | NCD-Protect | | 3.8 points (out of 9) | ↑ | -13.1* | -10* | 15.3** | 3.9*** | *** |
| | **13** | NCD-Risk | | 2.1 points (out of 9) | ↓ | 40.7*** | 15.8 | -16.5*** | -51.3*** | *** |
| | **14** | Sugar-sweetened soft drink consumption | | 18.9% adult population (≥15 y) | ↓ | -2.7 | 6.6 | 10.2 | -105*** | |
| ***Environment, natural resources, and production*** | | | | | | | | | | |
| Greenhouse gas emissions | **16** | Food systems greenhouse gas emissions | | 82,463.9 kt $CO_2$eq (AR5) | ↓ | 27.1 | -7.1 | -38.2 | 25.6 | |
| | **16** | Greenhouse gas emissions intensity, by product group[$] | Cereals (excl. rice)[†] | 0.2 kg $CO_2$eq/ kg product | ↓ | 24.5** | -20.1 | 5.1 | 2.3 | |
| | | | Beef | 30.3 kg $CO_2$eq/ kg product | ↓ | -213.1*** | -62.2 | 0.1 | 45.7*** | *** |
| | | | Cow's milk | 1.0 kg $CO_2$eq/ kg product | ↓ | -354.2*** | -46.3** | 8.8 | 40.5*** | *** |
| | | | Rice | 1.1 kg $CO_2$eq/ kg product | ↓ | -40.7 | -0.4 | 5.9 | -22.8 | |
| Production | **17** | Food product yield, by food group[$] | Cereals[†] | 40.7 tonnes/ha | ↑ | -64.6*** | -20** | 13.5 | 45.6 | *** |

| Domain | | Indicator | | Global weighted mean | Dir. of Δ | % Deviation of income group weighted mean from global weighted mean, sign aligned to desirable direction of change: Low income | Lower middle income | Upper middle income | High income | Joint signif. (F-test)[§] |
|---|---|---|---|---|---|---|---|---|---|---|
| | | | Fruit[†] | 136.7 tonnes/ha | ↑ | -51.4*** | -5.4 | 15.4*** | 5.2 | *** |
| | | | Beef | 231.5 kg/animal | ↑ | -46.9*** | -31.5*** | -2.3 | 38.2*** | *** |
| | | | Cow's milk | 2676.6 kg/animal | ↑ | -83.9*** | -43.9*** | 11.2 | 193.1*** | *** |
| | | | Vegetables[†] | 197.0 kg/ha | ↑ | -52.7*** | -34.3** | 25.7*** | 65.2*** | *** |
| Land | **18** | Cropland expansion (relative change 2003-2019) | | 19.1% | ↓ | -226.5** | -12.1 | 23.6 | 78.8*** | *** |
| Water | **19** | Agriculture water withdrawal as % of total renewable water resources | | 16.9% total renewable | ↓ | -8.3 | -49.3 | 34.2 | 21.5 | |
| Biosphere integrity | **20** | Functional integrity: % agricultural land with minimum level of natural habitat | | 88.3% agricultural land | ↑ | 5.1** | -11.4 | 2.3 | 3.5 | |
| | **21** | Fishery health index progress score | | 21.4 | ↑ | -56.8*** | -1.5 | -22.9 | 67.6* | *** |
| Pollution | **22** | Total pesticides per unit of cropland | | 1.8 kg/ha | ↓ | 89.5*** | 65.5*** | -38.9 | -57.1*** | *** |
| | **23** | Sustainable nitrogen management index | | 0.7 | ↑ | 31.7*** | 4.7 | -5 | -13.2 | *** |
| ***Livelihoods, Poverty, & Equity*** | | | | | | | | | | |
| Poverty and income | **24** | Share of agriculture in GDP | | 4.4% GDP | ↓ | -486.5*** | -278*** | -54.5*** | 70.9*** | *** |
| Employment | **25** | Unemployment, rural | | 5.7% working age population | ↓ | 11.9 | -0.1 | -5 | 2.8 | -- |
| | **26** | Underemployment rate, rural | | 7.3% working age population | ↓ | -105.2 | -10.7 | 5.5 | 55.8*** | |
| Social protection | **27** | Social protection coverage | | 55.8% population | ↑ | -74.9*** | 7.1 | 9.3* | 37.1** | *** |
| | **28** | Social protection adequacy | | 21.0% welfare of beneficiary households | ↑ | -23.5 | -48.6*** | 62.8*** | 125.1** | |
| Rights | **29** | % Children 5-17 engaged in child labor | | 9.4% children 5-17 | ↓ | -123.2* | 10.4 | 55.5*** | 74.7*** | *** |
| | **30** | Female share of landholdings | | 16.8% landholdings by sex of operator | ↑ | -28.7* | -38.4** | 36.8 | 8 | |
| ***Governance*** | | | | | | | | | | |
| Shared vision and strategic planning | **31** | Civil society participation index | | 0.6 | ↑ | -1.6 | 8 | -29.9* | 40.9*** | -- |
| | **32** | % Urban population living in cities signed onto the Milan Urban Food Policy Pact | | 7.2% urban population | ↑ | -38.7* | -68.9*** | 68 | 78.1** | *** |
| | **33** | Degree of legal recognition of the Right to Food (1 = Explicit protection or directive principle of state policy 2= Other implicit or national codification of international obligations or relevant provisions 3 = None) | | 1.9 | ↓ | 12* | 8 | 1.6 | -15.1*** | *** |
| | **34** | Presence of a national food system transformation pathway (0 = No, 1 = yes) | | 0.6 | ↑ | 5.5 | 27.2** | -2.9 | -22.7* | |
| | **35** | Government effectiveness index | | 0.1 | ↑ | -1010.6*** | -186.2 | 138.8 | 856.3*** | *** |

| Domain | | Indicator | | Global weighted mean | Dir. of Δ | % Deviation of income group weighted mean from global weighted mean, sign aligned to desirable direction of change | | | | |
|---|---|---|---|---|---|---|---|---|---|---|
| | | | | | | Low income | Lower middle income | Upper middle income | High income | Joint signif. (F-test)§ |
| Effective implementation | **36** | International Health Regulations State Party Assessment report (IHR SPAR), Food safety capacity | | 69.4 | ↑ | -40.7*** | -16*** | 18.3*** | 30*** | *** |
| | **37** | Presence of health-related food taxes | | 0.3 | ↑ | -44.8 | 53.1 | -54.4 | -6.7 | *** |
| Accountability | **38** | V-Dem Accountability index | | 0.3 | ↑ | -105.9* | 55 | -236.7 | 406.8*** | *** |
| | **39** | Open Budget Index Score | | 43.1 | ↑ | -41.2** | -1.7 | -8.3 | 50.7*** | |
| | **40** | Guarantees for public access to information (SDG 16.10.2) | | 1.9 | ↑ | -6.7 | -6 | -0.3 | 9.9*** | |
| ***Resilience & Sustainability*** | | | | | | | | | | |
| Exposure to shocks | **41** | Ratio of total damages of all disasters to GDP | | 0.3 | ↓ | 30.2 | 36.4* | 53.8*** | -31.6 | *** |
| Resilience capacities | **42** | Dietary sourcing flexibility index | | 0.7 | ↑ | -7** | -2.3 | -2.7 | 13.8* | ** |
| | **43** | Mobile cellular subscriptions (per 100 people) | | 105.5 per 100 people | ↑ | -37.4*** | -6.8 | 6.2 | 20.9*** | |
| | **44** | Social capital index | | 0.5 | ↑ | -20.5*** | -14.6** | 15.2 | 21.9** | *** |
| Agro- and food diversity | **45** | Proportion of agricultural land with minimum level of species diversity (crop and pasture) | | 22.5 % agricultural land | ↑ | 61.5* | 42.9 | -2.5 | -54.2** | *** |
| | **46** | Number of (a) plant and (b) animal genetic resources for food and agriculture secured in either medium- or long-term conservation facilities (SDG 2.5.1) | Plants | 161.4 (thousands) | ↑ | -89.7*** | -47.3 | -17 | 70 | *** |
| | **47** | | Animals | 4.4 | ↑ | -83.3*** | 106.2 | -78.8*** | 27.9 | *** |
| Resilience responses/ strategies | **48** | Coping strategies index | | 38.5% population | ↓ | -7.2 | 4.5 | 5.8 | 0*** | |
| Long-term outcomes | **49** | Food price volatility | | 0.7 | ↓ | -6.5 | 9.5 | -0.7 | -6.1 | *** |
| | **50** | Food supply variability | | 29.9 kcal per capita/ day | ↑ | 12.4 | 0.3 | -7 | -6 | *** |

*** $p < 0.001$ ** $p < 0.01$ * $p < 0.05$

§ Reflects p-value of joint significance tests (F-test), '--' indicates insufficient observations in one or more income groups to compute the F-test with cluster robust standard errors, required due to unequal variances by income group.

‡ See **SD-A Table A.3** for income group weighted means and **SD-A Table A.4** for income group medians.

++ Note that MDD-W has not been validated in HICs.

† Product mix varies across countries.

$ Additional products are included in the **SD-A** and baseline dataset.

+ Indicates FSCI value-added to existing data.

Sources: Author's calculations based on data sources listed in **Table 4**.

## Figure 4. Relation to GDP per capita, selected indicators by thematic area

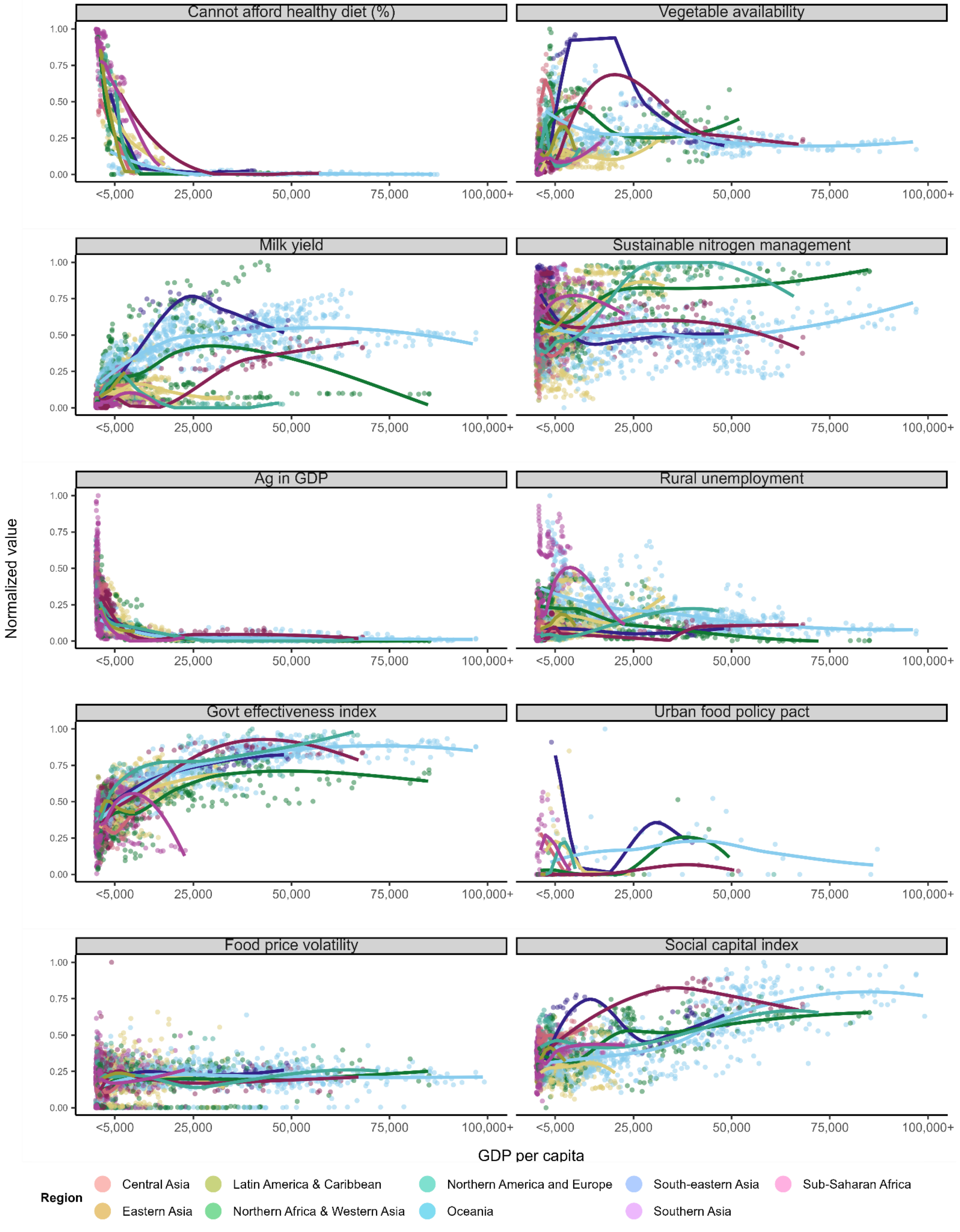

Sources: Author's calculations based on data sources listed in **Table 4**.
Notes: Colored lines reflects the local polynomial regression by region (Loess). GDP per capita is in nominal terms and data come from 2021 for most countries, see **SD-E** for the specific year of data for GDP and each variable. Normalized values are calculated using max-min normalization relative to the global weighted mean. Readers should note that normalizing by indicator range obscures differences in range magnitude across indicators, for which they are directed to regional distribution in **Table 2**. Each dot represents a country data point. Countries with GDP>$100,000 not shown on figure but included in model. Complete figures of all indicators are available in **SD-A Figures A.12-A.16.**

## Conclusion

The indicator framework presented in this paper allows for progress across global food systems to be meaningfully tracked, complementing the SDG's and other indicator frameworks with a scientifically vetted and curated set of indicators to monitor food systems. It provides the foundation for future research to better understand how and where change comes about, and importantly how to identify where improvements in any one domain do not necessarily translate into improvements in others.[37,38] The baseline dataset provides a starting point for tracking and the framework of indicators can be used by policymakers and other food system actors to diagnose their food systems and formulate appropriate responses, including transformation plans, and monitor advances in their countries. The baseline description demonstrates that no country or region shows positive outcomes across all dimensions. In addition, given that some food system outcomes are independent of national income levels, dedicated monitoring, and transformation agendas specific to food systems are needed.

Looking across this baseline, the indicators included offer a trove of information that provide transparency and specificity to the important constructs but do not prescribe obvious or uniform actions. Two examples illustrate tradeoffs in current policy choices. First, low resource use tends to correlate with low yields. Sub-Saharan Africa has among the lowest water withdrawals and pesticide use and better-than-average sustainable nitrogen management index (low nitrogen pollution), but these results reflect low use of external inputs in agriculture and instead a strategy relying on land expansion for production. Southern Asia has similar input use patterns but without the same land expansion, suggesting it may be on a more sustainable path. The goal would be to arrive at an optimal range of resource use that satisfies food needs with the lowest possible environmental impacts while contributing to regeneration. Second, several regions (Latin America and Caribbean, Oceania, South-eastern Asia, and Sub-Saharan Africa) have better-than-average (lower) rural unemployment but high underemployment, highlighting a tension between the need to work and low labor demand in low productivity activities. Two regions – Latin America and the Caribbean – have better social protection adequacy than the global mean, but not coverage, demonstrating a policy choice to provide higher benefits to fewer people. These examples illustrate the interconnectedness of food systems and need for alignment across goals and actions.

This indicator framework was developed with usefulness to countries and other food system decision-makers as a driving purpose. Following the UNFSS process, over 117 governments have developed national food system transformation pathways. The five domains of the FSCI architecture map closely into these pathways and will allow them to be well-monitored with the indicators selected and presented here.[14] There is utility in tracking national progress relative to goals as well as relative progress within a region, by income peer group, or the world overall. In addition to meeting the information needs at a country level, the indicator framework is also useful in addressing the supranational and transboundary issues within food systems that require alignment, coordination, and goals at higher jurisdictional levels. Decision-makers can use the framework as a starting point to consider what changes in indicators are achievable at different scales and can forge coalitions to drive change. Furthermore, different actors may find certain indicators more useful to guide action than others. For example, donors may be more concerned with cross-country comparisons when deciding how to allocate resources. National policymakers may be more interested to understand how their country is doing over time on indicators under more direct national influence or control.

The process of indicator selection identified key data gaps – the specific information that needs to be collected at scale to achieve the ambitious goal of tracking and informing food systems transformation. The gaps span all themes, for example, livelihood indicators beyond agriculture, food loss and waste, and governance of food systems. Many ongoing initiatives are working to fill some gaps (**Supplementary Information (SI), Appendix D**) with notable achievements already in bringing data together (e.g., the "Food Systems Dashboard"[39]). Ongoing expansion of the FAOSTAT database and the Global Diet Quality Project[17] will also help fill these gaps.[40] Other advances are significantly reducing costs and increasing the quality and granularity of new data collection (e.g., the 50x2030 Initiative).[41–44]

---

[14] Top ten priorities are (% countries with this priority): Zero Hunger (88 %), Sustainable Productivity Growth (80%), Climate and disasters, resilience (80%), Resilient food supply chains (75%), Healthy Diets from Sustainable food systems for all (72%), Decent work and living incomes for all food system workers (69%), Food systems for women and girls (69%), Food Loss and Waste (68%), Food Quality Safety (68%), Water (68%).

This baseline sets the stage, but future work is needed to close data gaps, assess status relative to benchmarks aligned to transformation, understand how food systems evolve over time including interactions across different indicators, and better understand and take action to support the needs of national and global data users. The FSCI will undertake this research and action agenda in the coming years alongside regularly updated assessments tracking progress from this baseline forward including the addition of new indicators or refinement of the current set of indicators as food systems science progresses. By doing so, the FSCI aims to facilitate and accelerate food systems transformation to deliver a healthier, more equitable, sustainable, and resilient future for all.

## Data and Methods

A rigorous set of prerequisite criteria were established that all indicators had to meet in order to be considered at all for this work, which included: feasibility (having recent data and are planned to be updated within the next 8 years), coverage (at least 70 countries across regions and income levels), and transparency (no modeled indicators with undisclosed or untraceable methodologies). A comprehensive multi-stage, multi-stakeholder process was then conducted to select the list of indicators analyzed in this paper (described in further detail below). Using a quantitative survey, dozens of experts were asked to rate each candidate indicator on its relevance, quality of the data and methods, and its interpretability for policy purposes. Indicators assessed to be relevant, high quality, and interpretable were considered to be useful, and a usefulness criterion was applied to the suite of indicators selected to monitor each domain to ensure sufficient but not redundant information. Finally, crucial input on regional priorities and policy utility provided by policy stakeholders was incorporated. Several indicators come from common sources such as FAOSTAT, Gallup World Poll, and the World Bank, but also, data from many other academic and NGO sources are also included. This replicable protocol including the survey and consultation processes culminated in the co-authors' final selection of indicators presented in this paper. All data and replication code are publicly available.

**Data.** Data used in this paper were sourced from many global, publicly available data sources. **Table 4** provides the data source, description, rationale for inclusion, and coverage metadata for each indicator. **SD-E** provides an Excel spreadsheet containing complete metadata, a codebook, country and year coverage, and the year of the latest data point per country-indicator that comprises the baseline. **SD-F** contains the complete baseline dataset of the latest data point per country per indicator used in the baseline analysis presented herein.

## Table 4. Indicator metadata

| Domain | Indicator | | Data Source | Description | Rationale for inclusion | Country coverage | Years covered | Desirable direction* |
|---|---|---|---|---|---|---|---|---|
| Food environments | 1 | Cost of a healthy diet | FAOSTAT | The per capita cost of the least expensive locally available foods to meet requirements for energy and food-based dietary guidelines, per capita, per day (2017 US$).[6,45] | Food-based dietary guidelines are designed to achieve nutrient adequacy and provide protection of health. This indicator reflects the cost of purchasing a diet aligned to a diet that reflects average amounts in guidelines. | 157 | 2017-2020 | ↓ |
| | 2 | Availability of fruits and vegetables<br>Fruits<br>Vegetables | FAOSTAT | Amounts of fruits and vegetables available in a country's food supply at the national level (expressed as grams per person per day). | Availability of fruits and vegetables is an essential precondition, yet not a guarantee, for their consumption. Consumption of abundant fruits and vegetables is universally recommended in global and national dietary guidance. | 174 | 2010-2019 | ↑ |
| | 3 | Retail value of ultra-processed foods | Euromonitor | Total sales of ultra-processed foods in the calendar year per person (USD/person). | This indicator proxies the availability of UPFs, defined as foods made of mostly industrial ingredients and additives with minimal amounts of unprocessed foods. These additives are not naturally occurring in the food but are added in the processing phase in order to increase palatability and shelf life. Examples of UPFs include sweet and savory snacks, instant noodles, confectionery, meat substitutes, and soft drinks, among others. These data are not publicly available but have been acquired for use by this Initiative and no comparable public sector data exist to capture this important aspect of food environments. | 187 | 2017-2019 | ↓ |
| | 4 | % Population using safely managed drinking water services (SDG 6.1.1) | WHO/UNICEF Joint Monitoring Programme | Percentage of the population that obtains drinking water from an improved water source, defined as one located on premises, available when needed, and free from fecal and chemical contamination. | Access to clean water is essential for food and nutrition security, to avoid foodborne and waterborne illness. | 117 | 2000-2020 | ↑ |
| Food Security | 5 | Prevalence of Undernourishment (SDG 2.1.1) | FAOSTAT | An estimate of the proportion of the population that lacks enough dietary energy for a healthy, active life. | An indicator used to monitor hunger at the global and regional level. The estimate is obtained with a model that compares the distribution of habitual food consumption levels with the dietary energy requirements for an average individual in the population. | 161 | 2001-2020 | ↓ |
| | 6 | % Population experiencing moderate or severe food insecurity (SDG 2.1.2) | FAOSTAT | Prevalence of the population experiencing moderate or severe food insecurity as measured by the FIES. | The FIES is an experience-based food security scale used to produce a measure of access to food at different levels of severity that can be compared across contexts. It relies on data obtained by asking people, directly in surveys, about the occurrence of conditions and behaviors that are known to reflect constrained access to food. | 120 | 2015-2020 | ↓ |
| | 7 | % Population who cannot afford a healthy diet | FAOSTAT | The share of the population whose food budget is below the cost of a healthy diet.[6,45] | The food budget is defined as 52% of household income, based on the average share of income that households in low-income countries spend on food. Where the minimum cost of a healthy diet (see definition above) exceeds this amount of income, it is considered unaffordable. | 141 | 2017-2020 | ↓ |
| Diet quality | 8 | MDD-W: % adult women meeting minimum dietary diversity | Gallup World Poll | Percentage of women 15-49 years of age who consumed at least five out of ten defined food groups the previous day or night. It is associated with a higher probability of nutrient adequacy for 11 micronutrients.[6] | It is a food group diversity indicator that reflects micronutrient adequacy, summarized across 11 micronutrients. The proportion of women aged 15-49 years who achieve this minimum of five food groups out of ten can be used as a proxy indicator for higher micronutrient adequacy. | 41 | 2021 | ↑ |
| | **9** | MDD (IYCF): % children 6-23 months meeting minimum dietary diversity | UNICEF | Percentage of children 6–23 months of age who consumed foods and beverages from at least five out of eight defined food groups during the previous day.[46] | WHO guiding principles for infant and young child feeding recommend that children aged 6–23 months be fed a variety of foods to ensure that nutrient needs are met. A diet lacking in diversity can increase the risk of micronutrient deficiencies, which may have a damaging effect on children's physical and cognitive development. | 100 | 2005-2020 | ↑ |
| | **10** | All-5: % adult population consuming all 5 food groups | Gallup World Poll | Proportion of the population age 15 years and older consuming all five food groups typically recommended for daily consumption: fruits; vegetables; pulses, nuts, or seeds; animal-source foods; and starchy staples.[17] | This indicator reflects the proportion of the population consuming any non-zero amount of each food group, and therefore may reflect minimal adherence to dietary guidelines. These food groups are aligned with the food groups used in the "Cost of a healthy diet" indicator. | 41 | 2021 | ↑ |
| | **11** | Zero fruit or vegetable consumption<br>Adults | Gallup World Poll | Proportion of the population age 15 years and older who did not consume any vegetables or fruits in the previous day.[17] | Consumption of zero vegetables or fruits is an unhealthy practice, as these food groups are associated with reduced risk of NCDs. It is a general population diet quality indicator aligned with the infant and young child feeding indicator (see next). | 41 | 2021 | ↓ |
| | | Children 6-23 months | UNICEF | Percentage of children 6–23 months of age who did not consume any vegetables or fruits during the previous day.[46] | Consumption of zero vegetables or fruits is an unhealthy practice, as these food groups are recommended for IYC, and are associated with reduced risk of NCDs. | 99 | 2005-2020 | ↓ |
| | **12** | NCD-Protect | Gallup World Poll | The NCD-Protect score is an indicator of dietary factors protective against NCDs, based on consumption during the previous day or night of nine food groups that are associated with meeting WHO recommendations on fruits, vegetables, whole grains, pulses, nuts and seeds, and fiber. The score ranges from zero to nine expressed as an average score for the population age 15 years and older.[17] | Dietary factors protective against NCDs include consumption of whole grains, pulses, nuts or seeds, at least 400g fruits and vegetables per day, and at least 25g of fiber per day. A higher NCD-Protect score indicates inclusion of more health-promoting foods in the diet, and correlates positively with meeting global dietary recommendations. | 41 | 2021 | ↑ |
| | **13** | NCD-Risk | Gallup World Poll | The NCD-Risk score is an indicator of dietary risk factors for NCDs, based on consumption during the previous day or night of eight food groups that are negatively associated with meeting WHO recommendations on free sugar, salt, total and saturated fat, and red and processed meat. The score ranges from zero to nine expressed as an average score for the population age 15 years and older.[17] | Dietary risk factors for NCDs include consumption of >10% of dietary energy from free sugar/day, >5g of salt/day, >30% of total fat/day, >10% of dietary energy from saturated fat/day, >350-500g red meat/week, and any processed meat. A higher NCD-Risk score indicates inclusion of more foods and drinks to limit in the diet, and correlates negatively with meeting global dietary recommendations. The NCD-Risk score is also a proxy for UPF intake; a higher NCD-Risk score indicates higher UPF consumption. | 41 | 2021 | ↓ |
| | **14** | Sugar-sweetened soft drink consumption | Gallup World Poll | Proportion of the population age 15 years and older who consumed a sugar-sweetened soft drink during the previous day or night. Sugar-sweetened soft drinks include soda, energy drinks, and sports drinks. | Sugar-sweetened soft drinks are a large source of free sugars; WHO recommends added sugars be limited to less than 10% of total energy. Excessive consumption may increase the risk of overweight and obesity and diet-related noncommunicable diseases. | 41 | 2021 | ↓ |
| **Greenhouse Gas Emissions** | **15** | Food systems greenhouse gas emissions | FAOSTAT | Production based greenhouse gas emissions (carbon dioxide, methane, nitrous oxide and F-gases) for food systems, expressed in kT CO2eq (AR5). | Food systems account for about 30% of total anthropogenic emissions. Reducing food systems emissions is crucial to reduce the impact of climate change and reach the targets of the Paris Agreement. And it is a sub-indicator of the FAO monitoring progress towards sustainable agriculture (SDG 2.4.1). | 194 | 1990-2020 | ↓ |
| | **16** | Greenhouse gas emissions intensity, by product group<br>Cereals (excl. rice) | FAOSTAT | Greenhouse gas emissions, CO2 equivalents (carbon dioxide, methane, and nitrous oxide) from the production of different crops and livestock and commodities within the farm gate. | Reducing the emissions intensity is a necessary - but not sufficient solution to reduce GHG emissions. Differences in emission intensities across countries reflect differences in environmental conditions, production systems and production efficiency. Changes in emission intensity over | 176 | 1961-2020 | ↓ |
| | | Beef | | | time helps to track improvements in efficiency, adoption of better practices and other changes in production systems. And it is a sub-indicator of the FAO monitoring progress towards sustainable agriculture (SDG 2.4.1). The most informative products for monitoring are shown in the main paper with additional products included in the supplementary data (SD-A) and dataset (SD-F). | 184 | | ↓ |
| | | Cow's milk | | | | 179 | | ↓ |
| | | Rice | | | | 119 | | ↓ |
| **Production** | **17** | Food product yield, by food group<br>Cereals | FAOSTAT | Yield means the harvested production per ha for the area under cultivation, expressed in tonnes per hectare or kg per animal. | Yields measure the efficiency with which inputs are used to produce agricultural output. Improving production efficiency is considered a critical path to meet food and nutrition security needs of current and future generations. The yields data, subtracted from consumptive need can serve as an indicator of coherence between sustainable production and healthy consumption targets. The most informative products for monitoring are shown in the main paper with additional products included in the supplementary data (SD-A) and dataset (SD-F). | 184 | 1961-2020 | ↑ |
| | | Fruits | | | | 186 | | ↑ |
| | | Beef | | | | 179 | | ↑ |
| | | Cow's milk | | | | 167 | | ↑ |
| | | Vegetables | | | | 187 | | ↑ |
| **Land** | **18** | Cropland expansion (relative change 2003-2019) | Potapov et al. (2021)[47] | Net change in cropland area from all sources, relative to the baseline. The measure of cropland used in this dataset excludes permanent crops. | The increasing demand for food can be met by increasing yields or expanding agricultural lands. Land use change is a major driver of biodiversity loss and climate change. Even with improved production efficiency, the total impact of the food system can still increase. Halting or reducing cropland expansion is another crucial step towards reducing the impact of the food system. | 173 | 2019 | ↓ |
| **Water** | **19** | Agriculture water withdrawal as % of total renewable water resources | AQUASTAT | Water withdrawn for irrigation in a given year, expressed in percent of the total renewable water resources. | Water is often a limiting factor for agricultural production and increasing irrigation is one of the main proposed ways to increase food production. At the same time, water scarcity is already a serious issue in many regions of the world and the situation is expected to get worse under climate change and increasing demand. This is also one of the eight proxy sub-indicators currently proposed for monitoring SDG 2.4.1 "sustainable and productive agriculture". | 175 | 1967-2018 | ↓ |
| **Biosphere Integrity** | **20** | Functional integrity: % agricultural land with minimum level of natural habitat | DeClerck et al (2021)[48] | Measures the proportion of semi-natural or natural habitat per $km^2$ of cropland or rangeland. The threshold is set at 10%, based on evidence that agroecosystem services are lost below 10%. At national level this indicator is articulated as the proportion of the country's agricultural lands having above or below the integrity threshold. | There is great pressure to expand agricultural area to increase food production. At the same time, there is growing concern about the loss of biodiversity and ecosystems services provided by natural habitats in agriculture. For example, fruit, vegetable, and legume production depend on pollination services, safe food production is dependent on regulations of pest and diseases provided by natural predators. | 194 | 2015 | ↑ |
| | **21** | Fishery health index progress score | Minderoo Foundation | The product of stock data availability and stock sustainability. | Fisheries can provide a substantial contribution to people's diets. Overfishing and environmental degradation have resulted in a drop of catching rates and raised questions about the future contribution that some fisheries could have in the future. | 122 | 2021 | ↑ |
| **Pollution** | **22** | Total pesticides per unit of cropland | FAOSTAT | The use of pesticides per area of cropland (which is the sum of arable land and land under permanent crops) at national level expressed as kg/ha. | Pesticide use in general, and use of hazardous pesticides in particular pollutes the biosphere at all levels, damaging flora and fauna and putting human health at risk. Reducing use and reverting current trends is a fundamental component of sustainable agricultural production and hence food systems. This is also one of the eight proxy sub-indicators currently proposed for monitoring SDG 2.4.1 "sustainable and productive agriculture". | 153 | 1990-2020 | ↓ |
| | **23** | Sustainable nitrogen management index | Zhang et al (2022)[49] | A one-dimensional ranking score that combines two efficiency measures in crop production: Nitrogen use efficiency and land use efficiency (crop yield), to provide a measure of the environmental efficiency of agricultural production. | Overuse of synthetic fertilizers pollutes the biosphere at all levels, with specific risk to aquifers and pollution downstream and into the oceans. Reducing use and reverting current trends is a fundamental component of sustainable agricultural production and hence food systems. This is also one of the eight proxy sub-indicators currently proposed for monitoring SDG 2.4.1 "sustainable and productive agriculture". | 188 | 1961-2018 | ↑ |
| **Poverty & Income** | **24** | Share of agriculture in GDP | FAOSTAT | The share of income derived from agriculture is a key parameter in agricultural transformation and the evolution of food systems. Historically, structural transformation has been characterized by a transition from a low-productivity agriculture-based economy that employs the majority of workers and generates the most output, to one dominated by industry and services and a smaller, more productive agriculture sector. | The indicator is a measure of the stage of food system transformation. It is correlated with economic development: countries with larger share of income coming from agriculture are poorer. | 192 | 2001-2020 | ↓ |
| **Employment** | **25** | Unemployment rate, rural | ILO | Share of people of employment age that are unemployed, disaggregated by total, urban, and rural. | The share of unemployed people in rural areas is an indicator of economic activity and livelihood opportunities for people in areas dominated by agriculture. | 177 | 2005-2020 | ↓ |
| | **26** | Underemployment rate, rural | ILO | Time-related underemployment is defined as people who (during the reference period) are: willing and available to work additional hours and worked less than a relevant nationally determined threshold of working time. | The use of the time-based underemployment rate is recommended to be used in combination with the unemployment rate, as unemployment does not capture the quality of employment. This use of these two indicators together was recommended in the 19th International Conference of Labour Statisticians in 2013. Monitoring the relationship between unemployment and time-related underemployment in rural vs urban areas by region may be useful in tracking shifts in agricultural employment. Disaggregated analysis by sex (SD-A) provides additional understanding of the shifts taking place in rural agricultural employment. | 104 | 1996-2021 | ↓ |
| **Social Protection** | **27** | Social protection coverage | World Bank | The share of individuals in the total population from households where at least one member participates in a social protection and labor market program, including non-contributory social safety nets (e.g. cash transfers, school feeding), contributory social insurance (e.g. old-age pension, health insurance), and labor market programs (e.g. job training, unemployment insurance). | Social protection (SP) can improve both the demand and supply of AFS: SP supports healthier diets through school feeding, nutrition sensitive programs, income support for increased consumption of nutrient-dense foods, dietary diversity, and micronutrient intake. SP (cash transfers especially when combined with skills development interventions and entrepreneurship support programs) may have positive impacts on supply of AFS, increasing the productivity of smallholder farmers and food retailers increasing investments in agricultural inputs, asset holdings, food sale and distribution. | 122 | 2000-2019 | ↑ |
| | **28** | Social protection adequacy | World Bank | The total social protection benefit amount received by beneficiary households (direct and indirect beneficiaries) as a percentage of beneficiaries' post-transfer, household wealth. This includes non-contributory social safety nets and contributory social insurance with monetary transfers but excludes safety nets without a monetary transfer and labor market programs. | The adequacy of social protection programs measures the likelihood that social protection intervention will have desired impacts on households' consumption of healthy diets and on improving the productivity of food production and distribution. | 117 | 2000-2019 | ↑ |
| **Rights** | **29** | % Children 5-17 engaged in child labor | UNICEF | Percent of children 5-17 years classified as engaged in child labor over the total population aged 5-17 years, disaggregated by sex. Criteria for child labor varies by age group:<br>1) Age 5 to 11 years: At least 1 hour of economic work or 21 hours of unpaid household services per week.<br>2) Age 12 to 14 years: At least 14 hours of economic work or 21 hours of unpaid household services per week.<br>3) Age 15 to 17 years: At least 43 hours of economic work per week. | ILO conventions 138 and 182 (which outlaw child labor and exploitation) have been almost universally ratified. It is a regularly monitored indicator with good global coverage. While the data used for this indicator are not disaggregated by economic activity, child labor is predominantly associated with agriculture and rural areas, which makes this indicator informative for food systems monitoring. | 99 | 2010-2021 | ↓ |
| | **30** | Female share of landholdings | FAO | Distribution of land holdings by sex (female %).[50] | The stark inequality in female land ownership occurs in countries of all income levels and belies other types of gender-based discrimination with profound consequences for livelihoods and resource use within and beyond the food system. The data from FAO used here are not currently being updated, and will eventually be replaced by SDG data on landholding by sex but which are not yet available. | 112 | 1988-2016 | ↑ |
| **Shared Vision & Strategic Planning** | **31** | Civil society participation index | Varieties of Democracy | The core civil society index is designed to provide a measure of a robust civil society, understood as one that enjoys autonomy from the state and in which citizens freely and actively pursue their political and civic goals, however conceived. | Captures whether an enabling environment exists for citizens to articulate their preferences over the food system, ensuring that policy goals are broadly representative | 172 | 1960-2021 | ↑ |
| | **32** | % Urban population living in cities signed onto the Milan Urban Food Policy Pact ⁺ | Milan Urban Food Policy Pact / Oakridge National Laboratory / FSCI | Proportion of the total national urban population living in cities that signed on to the Milan Urban Food Policy Pact. There were 221 cities from 73 different countries in 2020. | Indicates intentionality by subnational governments to prioritize food policy in their urban planning. | 187 | 2020 | ↑ |
| | **33** | Degree of legal recognition of the Right to Food ⁺ | FAOLEX / FSCI | Policies are classified based on the FAOLEX database five-group typology as:<br>1) Explicit protection of the right to food or directive of state policy.<br>2) Some other implicit recognition, codification of international statues, or other pertinent provisions.<br>3) None: countries with no policies catalogued in the FAOLEX database. | Indicates a government's recognition that guaranteeing food security to its citizens is one of its main responsibilities. As a human right, governments have duties to protect and fulfil this right, and the citizens have entitlements to be free from hunger and food insecurity. | 194 | 2021 | ↓ |
| | **34** | Presence of a food system transformation pathway (from the UNFSS) | FAO / FSCI | The country has developed a food system transformation pathway, as reported to the FAO UN Food Systems Summit Hub. | Specific strategy that addresses food systems from a systems perspective at the country level. | 194 | 2022 | ↑ |
| **Effective Implementation** | **35** | Government effectiveness index | World Governance Indicators | Perceptions of the quality of public services, the quality of the civil service and the degree of its independence from political pressures, the quality of policy formulation and implementation, and the credibility of the government's commitment to such policies. | A credible government with high levels of bureaucratic quality is more likely to effectively implement complex food systems policies. | 192 | 1996-2020 | ↑ |
| | **36** | International Health Regulations State Party Assessment report (IHR SPAR), Food safety capacity | WHO Global Health Observatory | Mechanisms are established and functioning for detecting and responding to foodborne disease and food contamination. | Indicates whether a government can effectively manage food safety challenges. | 191 | 2018-2020 | ↑ |
| | **37** | Presence of health-related food taxes [+] | World Cancer Research Fund International NOURISHING / FSCI | Reflects the presence of any health-related tax at all in the country (defined as binary given the diversity of policy mechanisms and objectives catalogued). In some countries, however, the tax may not be a federal policy and may only apply to certain subnational areas such as municipalities. | Indicates a government's willingness and ability to implement policy instruments aimed at promoting consumer health. | 194 | 2021 | ↑ |
| **Accountability** | **38** | V-Dem Accountability index | Varieties of Democracy | Combines horizontal (across government ministries), vertical (between voters and leaders), and diagonal (between civil society organizations, media, and government) accountability. It is measured by the following questions:<br>4) To what extent is the ideal of government accountability achieved?<br>5) Does the state's population hold government accountable through elections?<br>6) Are there checks and balances between institutions?<br>7) Is there oversight by civil society organizations and media activity? | Governments that demonstrate higher levels of general accountability are more likely to allow for citizen/legislative oversight of food system commitments and spending. | 172 | 1960-2021 | ↑ |
| | **39** | Open Budget Index Score | International Budget Partnership | A score based on an assessment of the public's access to information on how the central government raises and spends public resources. | Governments that demonstrate higher levels of general accountability are more likely to allow for citizen/legislative oversight of food system commitments and spending. | 120 | 2006-2021 | ↑ |
| | **40** | Guarantees for public access to information (SDG 16.10.2) | Sustainable Development Goals | The number of countries that adopt and implement constitutional, statutory, and/or policy guarantees for public access to information | Citizens are more likely to be able to hold their governments accountable for food system and other policies if they can access information about government activities | 194 | 2021 | ↑ |
| **Exposure to Shocks** | **41** | Ratio of total damages of all disasters to GDP | EM-DAT | Total estimated damages (nominal 000'USD, meaning unadjusted for inflation) divided by GDP (nominal 000'USD) and multiplied by 100 for readability. | An essential element of any resilience analysis is to assess and, if possible, document the intensity, nature, and frequency of the shock and stressors that a particular system is exposed to. | 187 | 1960-2021 | ↓ |
| **Resilience capacities** | **42** | Dietary sourcing flexibility index | FAO (soon to be available in FAOSTAT) | Measures the diversity of pathways through which food reaches consumers. Expresses how difficult it is to disrupt a country's food supply. Considers three possible pathways a unit of food can reach a consumer:<br>1) Food produced domestically;<br>2) Imported food;<br>3) Stocks carried over from the previous year.<br>Total sources for calories is the indicator used, disaggregation by source and for other nutrients and food groups are available. | Diversification (of portfolios, livelihoods, income, source of food, etc.) is an essential risk strategy well established in both theoretically and empirically in the general literature. In the present case the dietary sourcing flexibility index capture the possible diversification of sources of food supply. | 151 | 2018 | ↑ |
| | **43** | Mobile cellular subscriptions (per 100 people) | International Telecommunications Union / World Bank | Subscriptions to a public mobile telephone service using cellular technology, which provides access to the public | The number of mobile cellular telephone subscriptions has been included in the computation of food system resilience as a | 193 | 1991-2020 | **43** |
| | | | | switched telephone network. Includes post-paid and prepaid subscriptions. | proxy for countries' infrastructure level as well as a direct indicator of response capacity. | | | |
| | **44** | Social capital index | Legatum Institute / FSCI | A composite index based on a subset of indicators from the Social Capital pillar of the Legatum Prosperity Index, which assesses social cohesion and engagement, community and family networks, and political participation and institutional trust. The social capital index used here is a composite of the indicators of "Help from family & friends"; "Generalized interpersonal trust"; "Confidence in financial institutions"; and "Public trust in politicians or confidence in national government", based on the relevance of these indicators to resilience. The index is calculated as the geometric mean of the four variables, each measured on a scale that ranges from 0 (low) to 100 (high). | Social capital is another extremely important element of resilience in general. The score on the Social Capital pillar of the Legatum Prosperity Index has been included in that regard. | 165 | 2007-2021 | ↑ |
| **Agro- and Food Diversity** | **45** | Proportion of agricultural land with minimum level of species diversity (crop and pasture) ⁺ | Jones et al. (2021)[51] | The proportion of agricultural land with minimum species diversity is defined by the top global quartile of land with the highest species richness. The threshold number of species at which (and above) covers 25% of total global ag land (the 25% of land with the most diversity) is 24 species. Therefore, the indicator reflects the percentage of agricultural land per country with 24 or more species. | Diversification (of portfolios, livelihoods, income, source of food, etc.) is an essential risk strategy well established in both theoretically and empirically in the general literature. The proportion of agricultural land with minimum level of species diversity (crop and pasture) contributes to landscape and livelihood buffering capacity, with more functionality to cope with for example unexpected environmental changes or shocks and spreading risks to cope with for example market volatility. It is therefore a relevant indicator reflecting the diversity of the agricultural production component of the food system. | 180 | 2010 | ↑ |
| | **46** | Number of (a) plant and (b) animal genetic resources for food and agriculture secured in either medium- or long-term conservation facilities (SDG 2.5.1)<br>Plants | Sustainable Development Goals | The conservation of plant genetic resources for food and agriculture in medium- or long-term conservation facilities (ex situ in gene banks) represents the most trusted means of conserving genetic resources worldwide. Measured as:<br>a) Plant genetic resources accessions stored ex situ (number). | Diversification (of portfolios, livelihoods, income, source of food, etc.) is an essential risk strategy well established in both theoretically and empirically in the general literature. The Number of plant and animal genetic resources for food and agriculture secured in conservation facilities is considered a buffering capacity, providing a back-up in times of crises and shocks and indicating the food and agricultural diversity ex-situ 'stock' of a country | 97 | 2000-2020 | ↑ |
| | **47** | Animals | | b) Number of animal genetic resources represents the number of species with sufficient genetic material stored for reconstitution. | | 97 | 2000-2021 | ↑ |
| **Resilience Responses / Strategies** | **48** | Coping strategies index | WFP | The reduced Coping Strategies Index (rCSI) measures the frequency and severity of household behaviors when faced with shortages of food or financial resources to buy food. It is calculated using five standard food consumption-based strategies and severity weighting, a higher score indicates more frequent and/or extreme negative coping strategies. It measures the proportion of population using extreme coping strategies (with a rCSI >=19). It is the highest observed daily prevalence throughout the year. | Understanding and documenting the type of responses that actors adopt when they are faced with shocks or stressors is an essential part of resilience analysis. In theory the objective of a resilience building intervention is to reduce the propensity of actors to engage in detrimental responses and to increase their ability to engage in more 'positive', adaptive or transformative responses. At the present time the rCSI is only recorded in a limited number of, mostly low-income, countries. | 114 | 2021 | ↓ |
| **Long-term Outcomes** | **49** | Food price volatility ⁺ | FAOSTAT / FSCI | Domestic food price volatility index measures the variation (volatility) in domestic food prices over time, measured as the relative variation in the domestic food price index, a standardized measure of the cost of a basket of goods. High values indicate a higher volatility (more variation) in food prices. | Different indicators can be used to assess the long-term outcomes of a system resilience. In the case of food systems, the ability of the system to maintain a low price-volatility in the face of shocks, is a direct way to assess the system resilience. The lower the volatility, the better. | 42 | 2000-2021 | ↓ |
| | **50** | Food supply variability | FAOSTAT | This indicator uses the data on dietary energy supply from the Food Balance Sheet to measure annual fluctuations in the per capita food supply (kcal), represented as the standard deviation over the previous five years per capita food supply. Food supply variability results from a combination of instability and responses in production, trade, consumption, and storage, in addition to changes in government policies such as trade restrictions, taxes and subsidies, stockholding, and public distribution. | Along with food price volatility (see above), the ability of the system to maintain a low variability in the supply of food products in the face of shocks is a direct way to assess the system long-term resilience. A resilient food system would be able to keep the variability of food supply low despite being hit by shocks. Therefore, the lower the food supply variability, the better. | 181 | 2000-2021 | ↓ |
| **Contextual variables** | GDP | | World Bank | Sum of gross value added by all resident producers in the economy plus any product taxes and minus any subsidies not included in the value of the products. Data are in current U.S. dollars. Dollar figures for GDP are converted from domestic currencies using single year official exchange rates. | Provides the denominator for the exposure to shocks indicator and for analysis of indicator relationship to GDP. Weighting variable for weighted means as specified in **Table 4**. | 192 | 1960-2021 | |
| | Population | | World Bank | Total population, counting all residents regardless of legal status or citizenship at midyear estimates. | Weighting variable for weighted means as specified in **Table 4**. | 193 | 1960-2021 | |
| | Urban population | | Oakridge National Laboratory / Landscan | Total urban population based on GADM administrative unit boundaries. | Weighting variable for weighted means as specified in **Table 4**. | 187 | 2020 | |
| | Land area | | FAOSTAT | Country area excluding area under inland waters and coastal waters. | Weighting variable for weighted means as specified in **Table 4**. | 194 | 1961-2020 | |
| | Cropland | | FAOSTAT | Land used for cultivation of crops. Includes all arable land and permanent crops. | Weighting variable for weighted means as specified in **Table 4**. | 193 | 1961-2020 | |
| | Agricultural land | | FAOSTAT | Land used for cultivation of crops and animal husbandry. Includes cropland and permanent meadow and pasture. | Weighting variable for weighted means as specified in **Table 4**. | 193 | 1961-2020 | |

⁺ Indicates substantial value add by FSCI to existing data.

* Desirable direction: ↑ denotes a higher value is more desirable, ↓ denotes a lower value is more desirable.

Country and year coverage refers to coverage in the FSCI dataset, limited to UN member states and all available years as of November 2022 when the data were last pulled.

**Indicator selection.** The authors employed a multi-stage, multi-stakeholder process to select the list of indicators analyzed in this paper. A preliminary set of criteria was previously published in Fanzo et al. (2021).[1] In the first stage of indicator selection, the authors refined these criteria by deeming three attributes to be essential: feasibility, coverage, and transparency. Next, the authors refined the four criteria established previously: relevance, high quality, interpretable, and useful. **Table 5** details the requirements, criteria definitions, and sub-criteria.

## Table 5. Requirements and criteria used to select indicators.

| | |
|---|---|
| **Requirements** | • **Feasibility.** Recent (within the last 10 years) data exist or are planned to be collected in the coming 1-2 years and will be updated over the next 8 years.<br>• **Coverage.** Data exist for at least 70 countries and the proportions of countries in low-, middle-, and high-income countries approximate the distribution of countries by income level in the World Bank classification (14% LIC; 49% MIC; 37% HIC)<br>• **Transparency.** No indicators calculated with undisclosed modeling, methodology, or assumptions and no composite indicators where change cannot be clearly traced to underlying components. In other words, no "black boxes." |
| **Criteria** | **Relevant.** Indicator measures something meaningful for food systems across a variety of settings and during relevant time periods.<br>• Can be clearly mapped to the food systems framework.<br>• Observing change in the indicator is possible within a decade (meaning that the phenomena can change on that timescale and that the data exist to observe change).<br>**High-Quality.** Best practices in data collection and aggregation (including quality controls) and rigorous statistical methodologies.<br>• Well-documented methodologies and metadata.<br>• Data are nationally representative.<br>**Interpretable.** Clear desirable direction of change, comparable across time and space, and easily communicated.<br>• Change has a clear interpretation.<br>• Data are comparable across countries.<br>**Useful.** Useful for policy, planning, and decision-making.<br>• Useful individual indicators meet all three other criteria: they are relevant, high quality, and interpretable.<br>• Suites of indicators (i.e., per domain) should satisfy the criterion of usefulness, that they are together "useful for policy, planning, and decision-making." |

Note: Criteria definitions describe the overall criteria. Indicators were assessed by the sub-criteria in the bulleted lists used to operationalize each criterion.

Working group members compiled a list of candidate indicators for each domain that met the prerequisite requirements for potential inclusion. **SI-B** contains the indicator catalogue of all candidates, indicator options excluded for failure to meet the prerequisites, and all relevant information that was provided to assess the indicators. This list of candidate indicators was assessed against the first three criteria (relevance, quality, interpretability) using an online survey by all the collaborators and an additional group of over two dozen external experts who were volunteer respondents based on a list of experts generated by all the authors with additional research to reach relevant people unknown to the author group. Everyone assessed indicators in the domain(s) aligned with their expertise. Respondents were asked to choose their level of agreement (from 1 to 5) with the statement that the candidate indicator met each sub-criterion, the elements in the bulleted lists in **Table 5**. In addition, all respondents were also asked to state their agreement that the indicator is important for tracking food system transformation and to share their interpretation of both importance and transformation in that context, providing complementary qualitative data. Finally, external experts were also asked to suggest additional data sources for candidate indicators and to describe any observed gaps in the domains and indicators and how they recommend filling those gaps. For those who assessed governance indicators, an additional question asked what new indicators the respondent deemed necessary and recommendations for their construction. **SI-C** contains the full report of the survey procedures and outcomes.

In parallel, the FAO convened five regional policy stakeholder consultations in Latin America and the Caribbean, sub-Saharan Africa, North Africa and the Middle East, Asia and Pacific, and Europe. Over 500 people participated, averaging 75-100 per region, The consultations included a short overview presentation and breakout discussions of each thematic area. Participants were asked to assess the local pertinence of the architecture and indicator framework and solicit regional priorities, interests, and needs. **SI-C** contains the reports for each regional consultation. The consultation asked experts and stakeholders to suggest alternative indicators and data sources and identify gaps, which resulted in the addition of several indicators to the initial list of candidates.

The authors made a final selection of indicators considering the results of the assessment process, summary statistics and visualizations of the candidate indicators, and results of the policy stakeholder workshops. Specifically, scores from the assessment of indicators against the six sub-criteria of relevance, quality, and interpretability criteria were summed to the indicator level with equal weighting providing a single score per indicator. Usefulness was assessed qualitatively at the level of indicator domains, with emphasis on meeting the needs illuminated by the policy stakeholder workshops. Several indicators were added and ultimately included in the final set after the consultations based on the feedback provided during those consultations and the gaps identified. These indicators are: safe drinking water, agri-food system emissions, yields, share of agriculture in GDP, underemployment, degree of legal recognition of the Right to Food, percent of the urban population living in a signatory municipality to the MUFPP, food safety capacity, health-related food taxes, guarantees for public access to information, proportion of agricultural land with minimum species richness, and the number of animal and plant genetic resources in conservation facilities. Not all gaps identified in the consultations could be filled and are instead described in the data gaps and research agenda discussion, in particular, lack of food loss and waste data was a prominent theme of the consultations.

**Analysis methods.** Analyses were carried out in Stata 17 and R v4.2.2. The data were compiled into a dataset where all years of available data per country and indicator were included. In two instances (EM-DAT and Varieties of Democracy indices) data prior to 1960 were excluded because no other datasets provided data before that year. Initially, all territories classified in the UN Global Administrative Units List dataset[52] and present in any datasets were included (94 areas in total). After compiling the complete dataset with all indicators, the authors investigated whether there was sufficient coverage across all indicators for any territories or areas that are not UN Member States to remain in the dataset. A criterion was applied that the area must have at least 80% of all indicators. In practice, all territories were dropped at a much lower threshold, none having more than the median number of variables present for Member States (40; where certain indicators are represented in the dataset by more than one variable). In sum, the dataset contains all the available data from 1960 to 2021 for all UN Member States, and one indicator (the presence of a food system transformation pathway) defined only in 2022.

The focus of this manuscript is a baseline dataset comprised of the latest data point per country per year. Overall, 92.5% of all data points are from 2017-2022, 6.5% are from 2010-2016, and only 1% are from 2000-2010. A small number of observations (N = 24 across all indicators) are dropped from the dataset because the latest data point for that country-indicator pair comes from prior to 2000. The only indicator where this drops more than a few observations is female share of landholdings which has 13 countries whose data point in that cross-sectional dataset is from the 1990s or before. A new data source will become available through the SDG process for this indicator in future years.

The supplementary data include analysis of the data from 2000 forward wherever time series are available. Countries are grouped into regions based on modified groupings of the M49 classification system of the UN Statistical Commission, using a combination of continental and sub-regional groupings. **SD-A Figure A.2** depicts the alignment of countries to the modified M49 regional grouping used in this paper. Countries are identified by income group using the World Bank country income classification.[53]

Distribution of the indicators by region and income group relative to the global weighted mean (**Figures 3** and **4**, respectively) are presented as the normalized difference from the global weighted mean. The global weighted mean is subtracted from the region (income group) weighted mean and normalized using min-max scaling, which divides the demeaned observation by the total range across all regions (income groups) (i.e., divide by the maximum observed minus the minimum observed). Deviations of region and income group weighted means from the global weighted mean **Tables 2** and **3**) are calculated using weighted least squares regression with heteroskedasticity robust standard errors regressing region (income group) on the demeaned observation. Demeaned observations are calculated by subtracting the global weighted mean from each observation. The sign of the demeaned observation is reversed for all indicators where the desirable direction of change is lower. Regression coefficients are the regional (income group) deviation from the global average with the sign indicating whether the region is performing worse (negative sign) or better (positive sign) than the global average. The signed deviation is then translated into a percentage deviation by dividing by the global average to harmonize the presentation of indicators given the different units and scales of their level measurements.

**Data availability.** Analysis in this paper relies on numerous datasets in the public domain unless otherwise noted (for which permission to include in our dataset was secured). Metadata contains necessary links to access the underlying raw data.

**Code availability.** Replication code for this paper will be available on GitHub upon publication.

## Supplementary Materials

### Supplementary Data – Appendix A

1. Theory of transformation.
2. Countries by regional group (modified M49 grouping)
3. Regional visualizations of data coverage.
4. Supplementary analysis: regional and income group weighted means and medians.
5. Analysis and visualization of indicator relationship with GDP per capita, by theme.
6. Comprehensive visualizations for all indicators over time, and by region, income group, and other relevant disaggregation, where data are available.

### Supplementary Information – Appendix B

The list of candidate indicators that were ultimately excluded.

### Supplementary Information – Appendix C

Reports from the indicator selection process (survey report, reports from regional consultations).

### Supplementary Information – Appendix D

List of related initiatives and how they relate to the FSCI work.

### Supplementary Data – Appendix E

Metadata and codebook Excel workbook.

### Supplementary Data – Appendix F

Baseline dataset (latest data point per country-indicator) in .dta and .csv formats.

## Acknowledgements

The authors would like to acknowledge the essential contribution made by the external experts and participants in the regional stakeholder consultations. Our work would not have been possible without their input. Their names are listed in the Supplementary Information (Appendix C). We would also like to thank Rahul Agrawal Bejarano of American University and Alexa Bellows of the University of Edinburgh for their assistance with R coding as well as Zhongyun (Bella) Zhang for her contributions to the data team. We thank Rihanna Ibrahim of the World Food Programme (WFP), Giulia Martini also of WFP, and the Milan Urban Food Policy Pact team for facilitating access to data underlying their online displays in dataset formats. And we thank Charlotte Pedersen of GAIN for her contributions to the indicator selection process. The views expressed in this paper are the authors' only and cannot be taken to represent those of any organization's (FAO, GAIN, Cornell University, Johns Hopkins University, nor the institutional home of any other author) policies and views on the subject matter.

# Supplementary Material, Appendix A: Supplementary Analysis

**Figure A.1: FSCI Theory of Transformation**

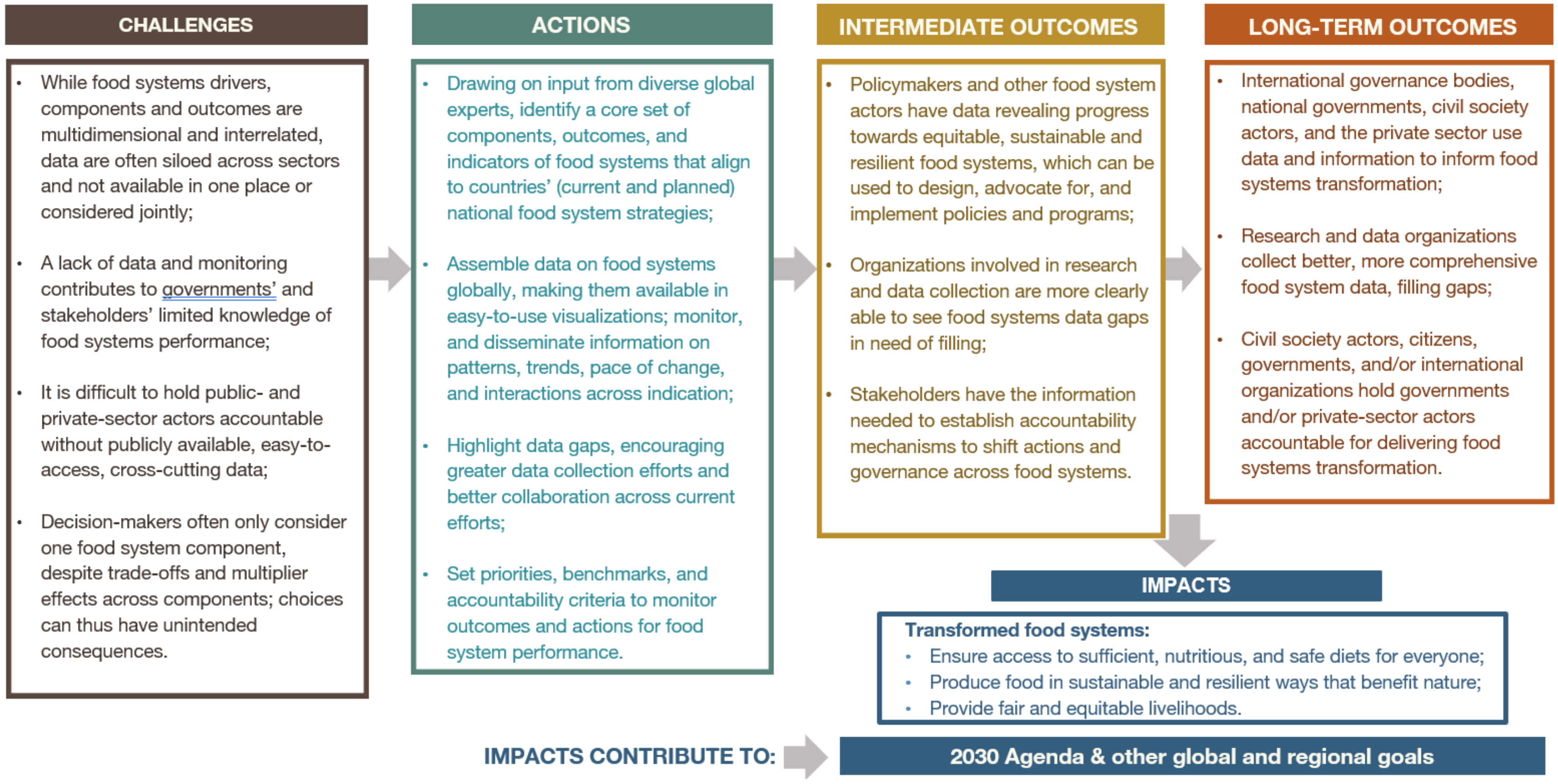

**Figure A.2 Country alignment to modified M49 regions**

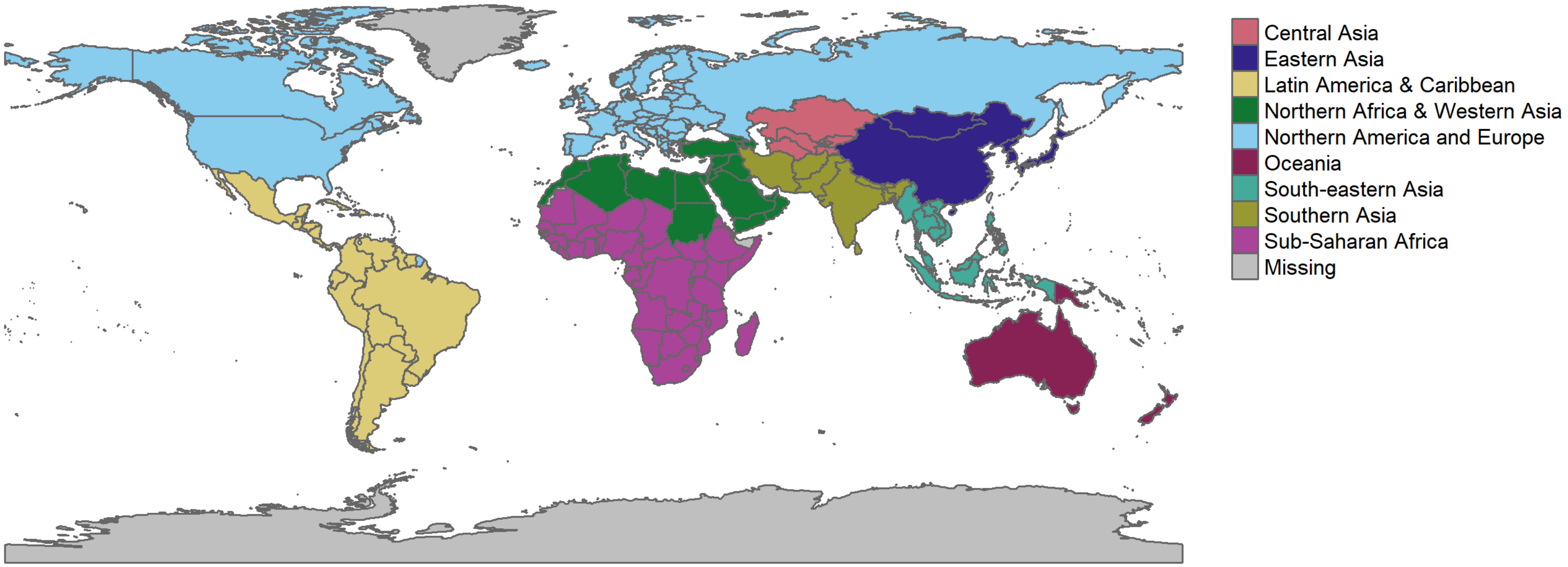

**Figure A.3 Indicator coverage by country and extent of time series**

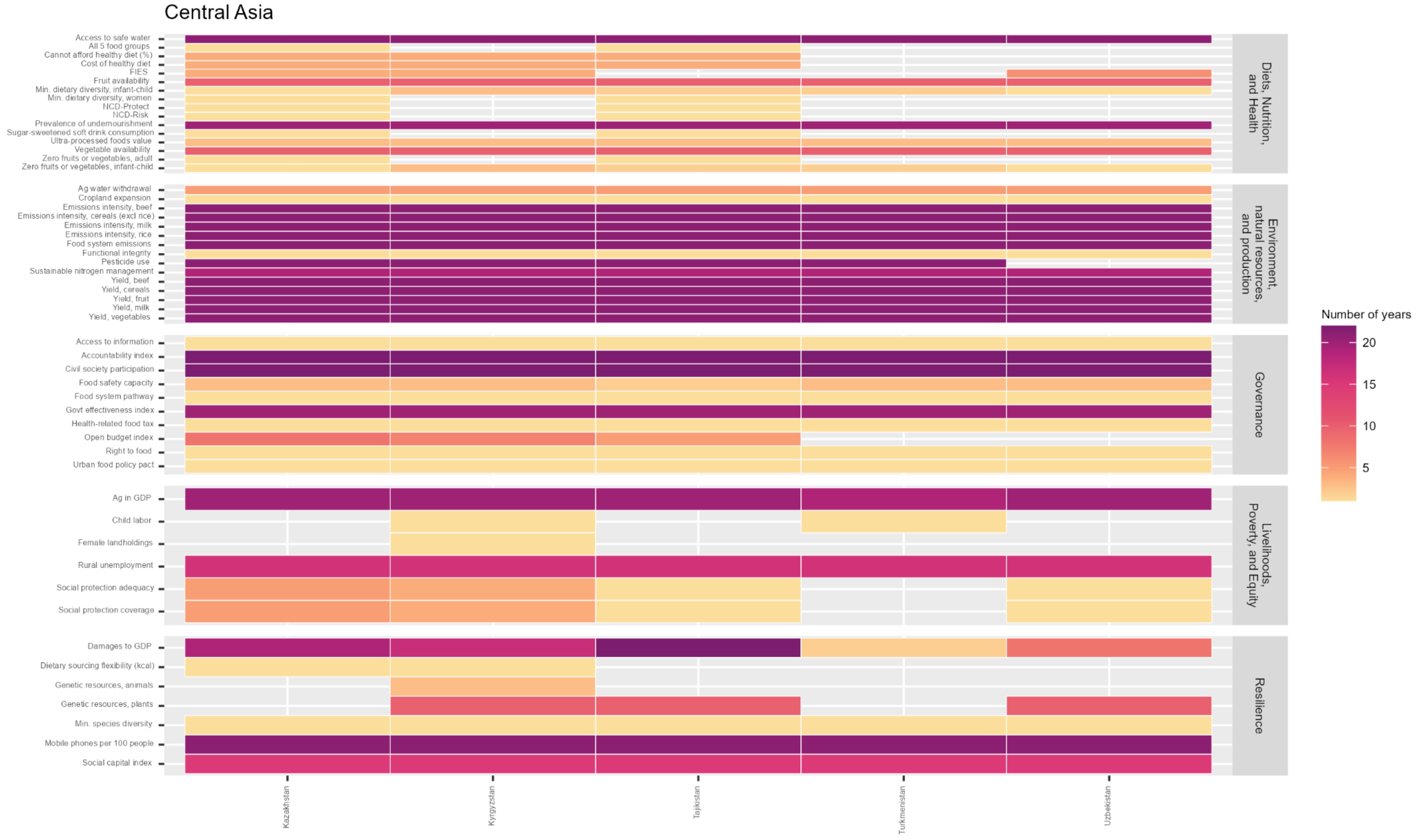

**Figure A.4 Indicator coverage by country and extent of time series**

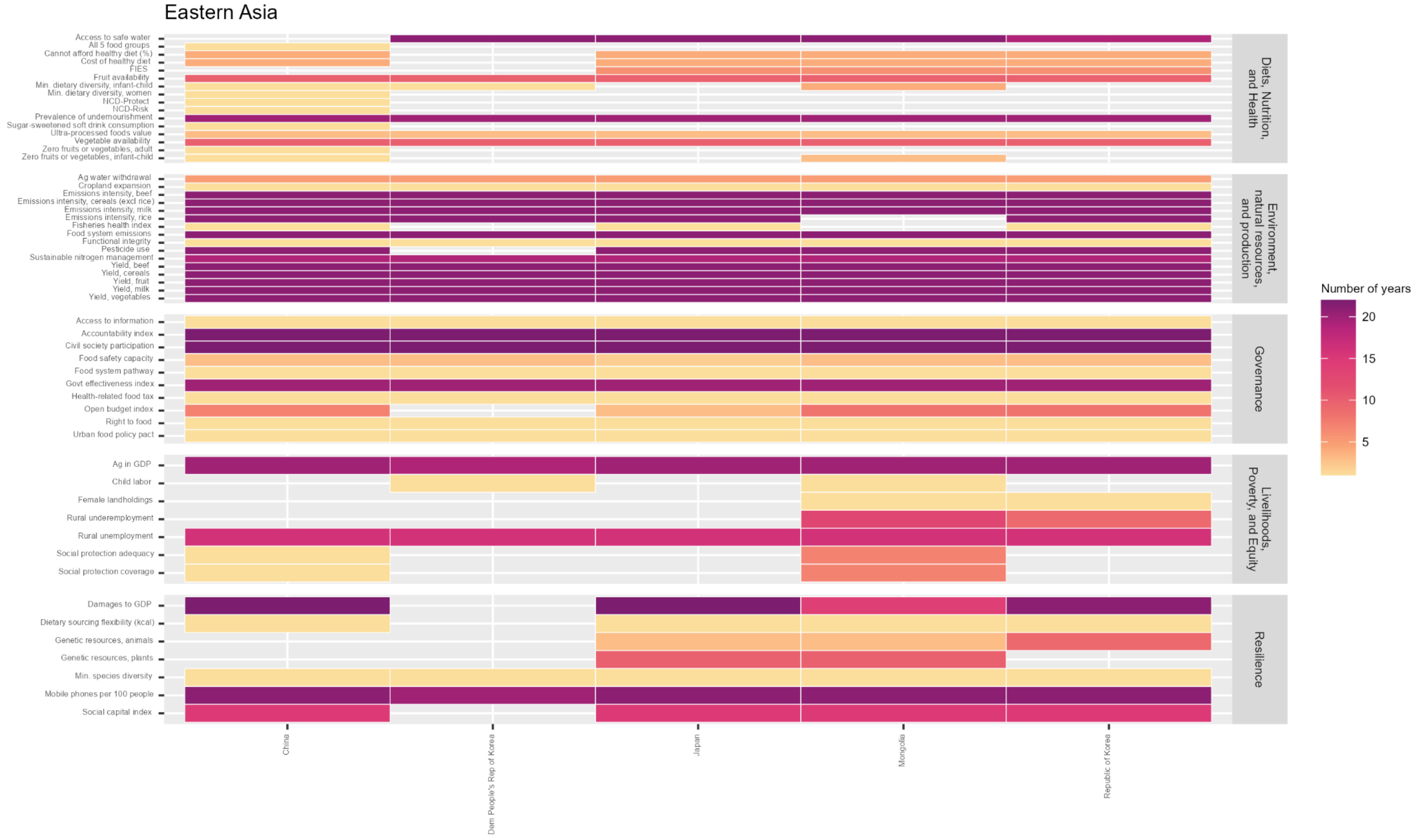

**Figure A.5 Indicator coverage by country and extent of time series**

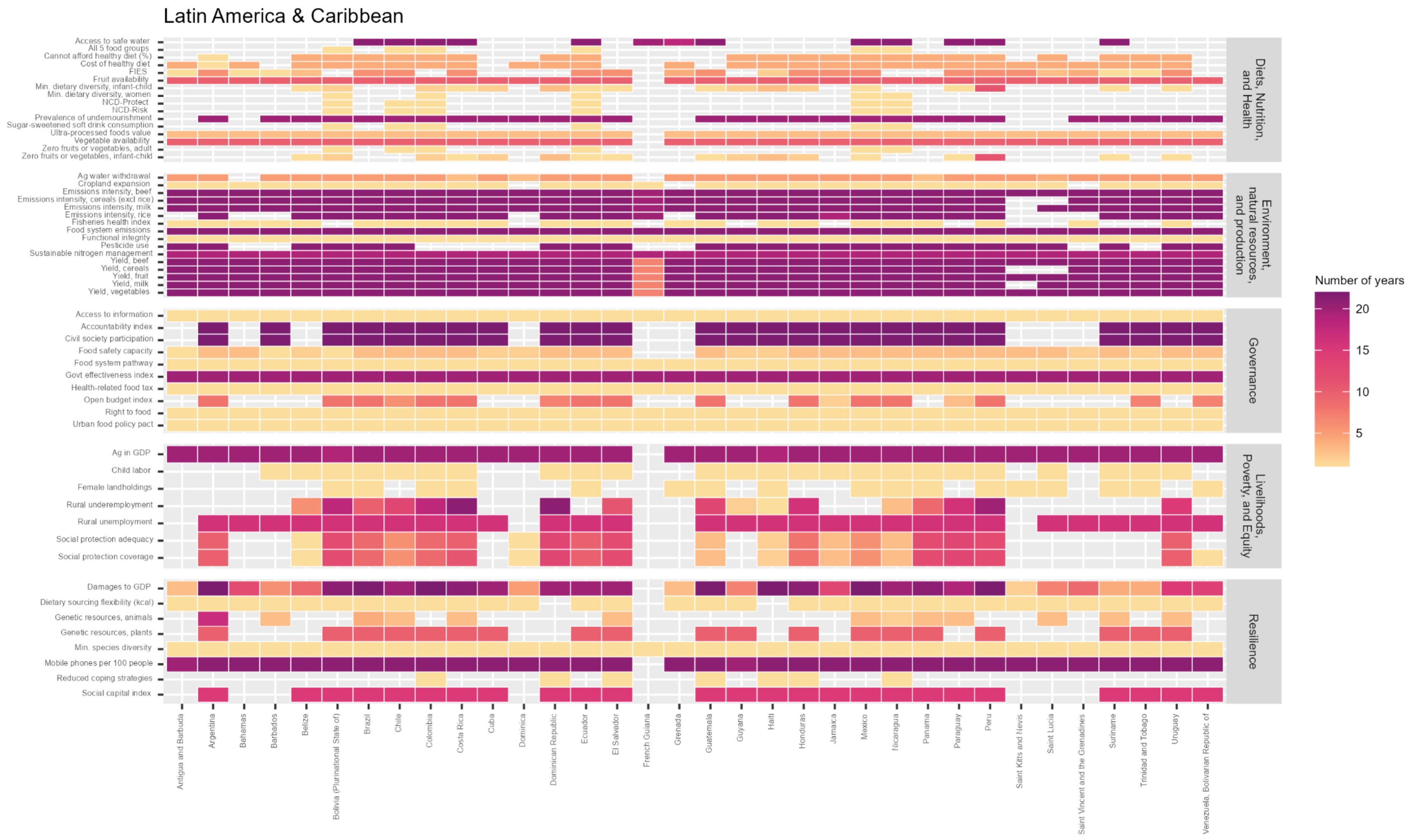

**Figure A.6 Indicator coverage by country and extent of time series**

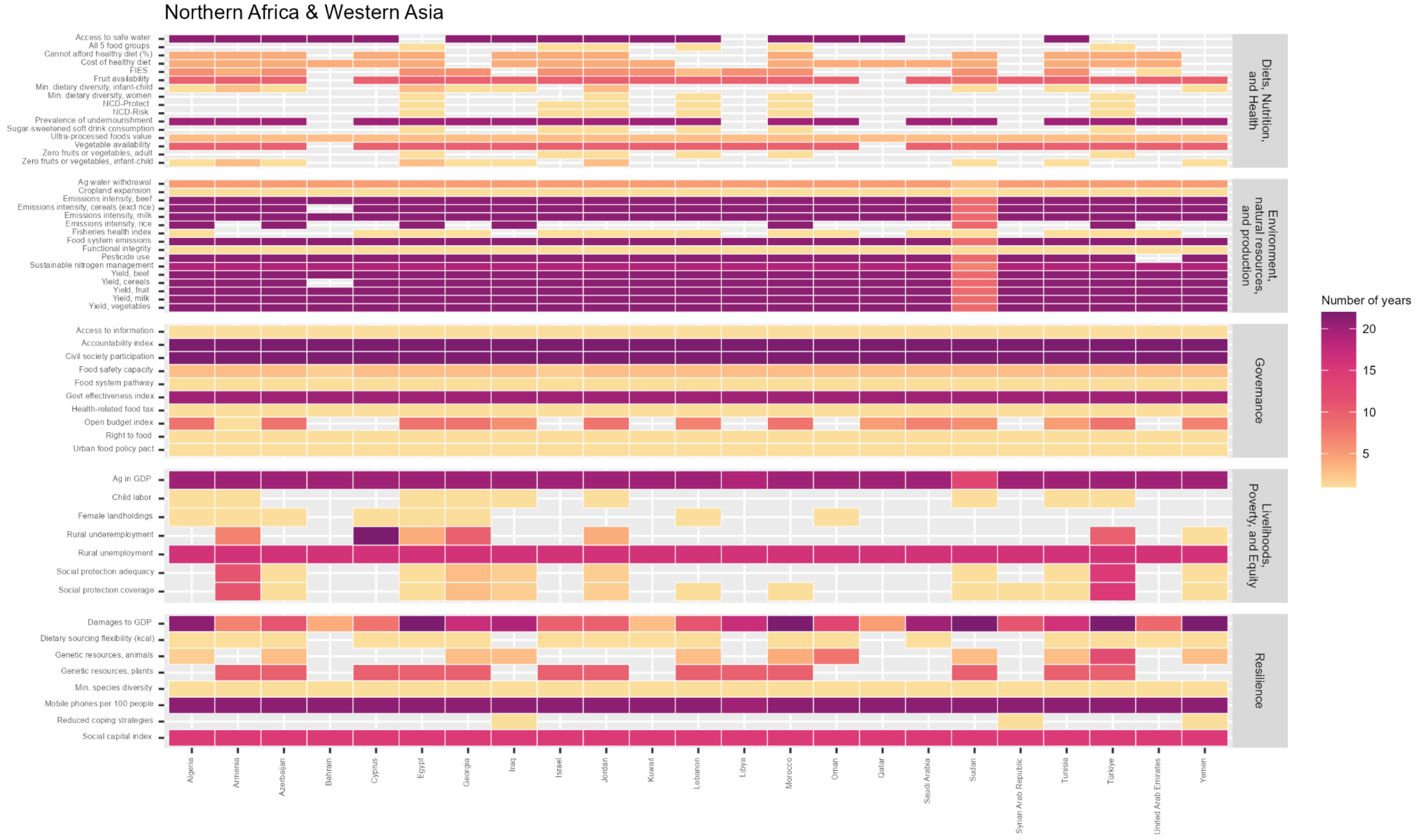

**Figure A.7 Indicator coverage by country and extent of time series**

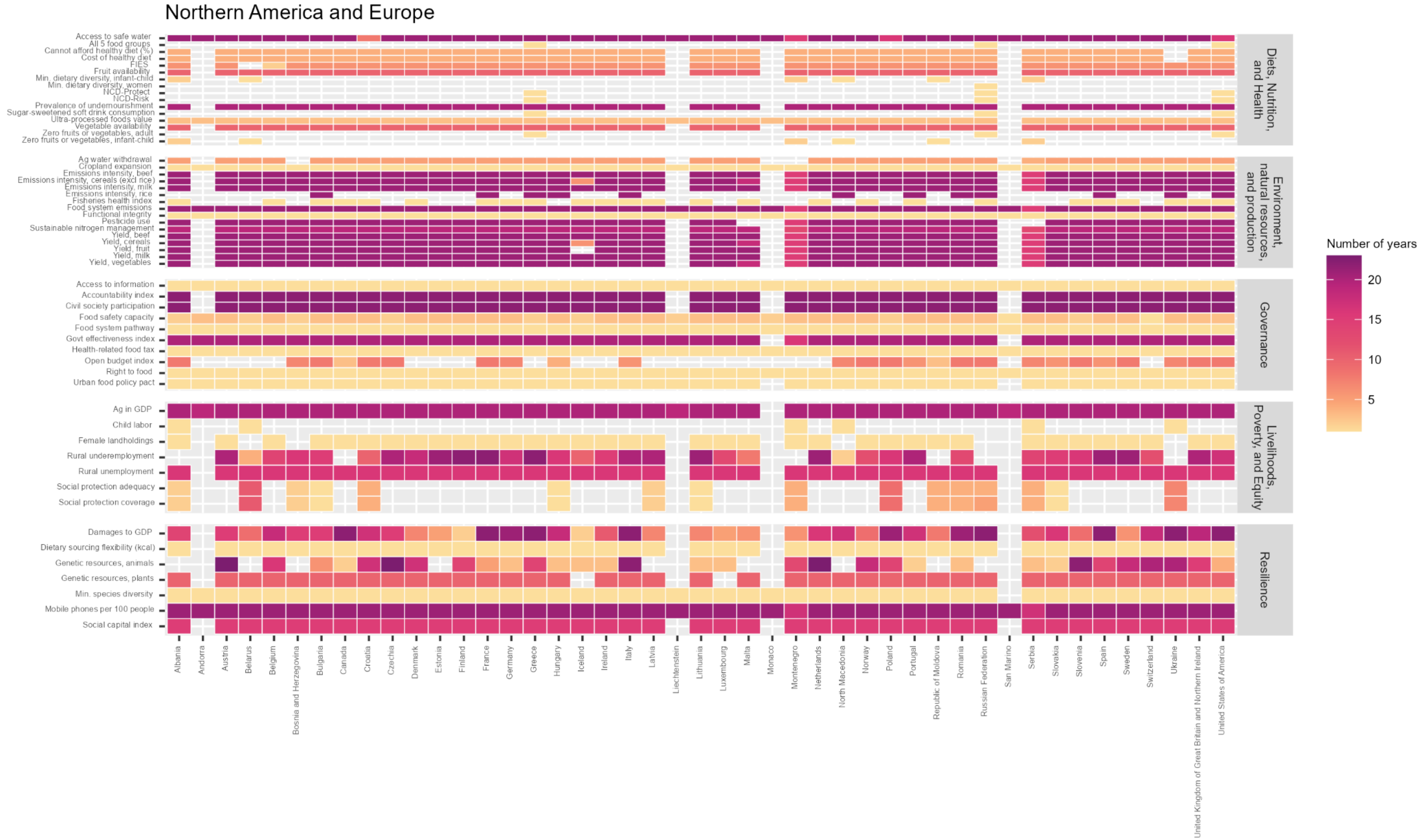

**Figure A.8 Indicator coverage by country and extent of time series**

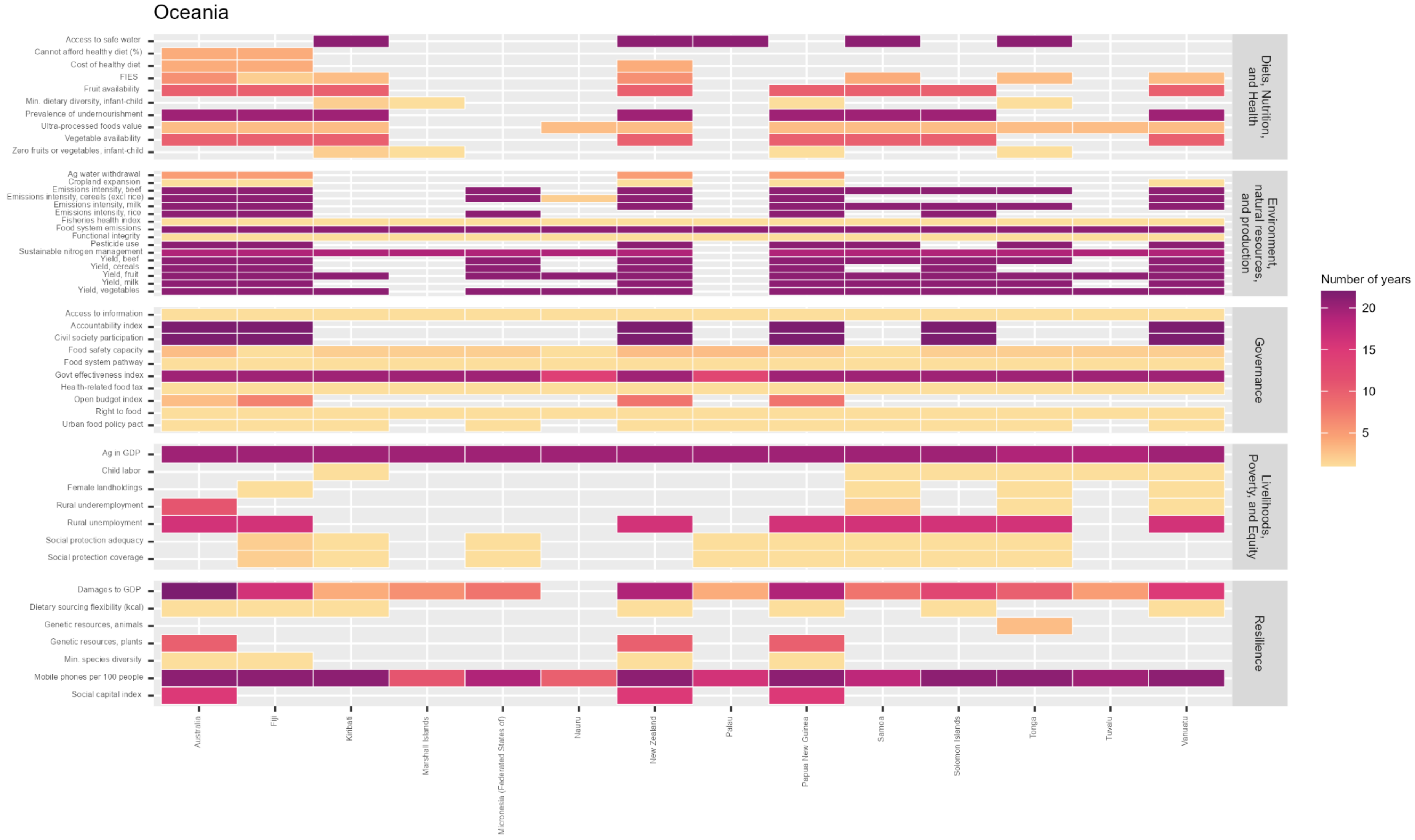

**Figure A.9 Indicator coverage by country and extent of time series**

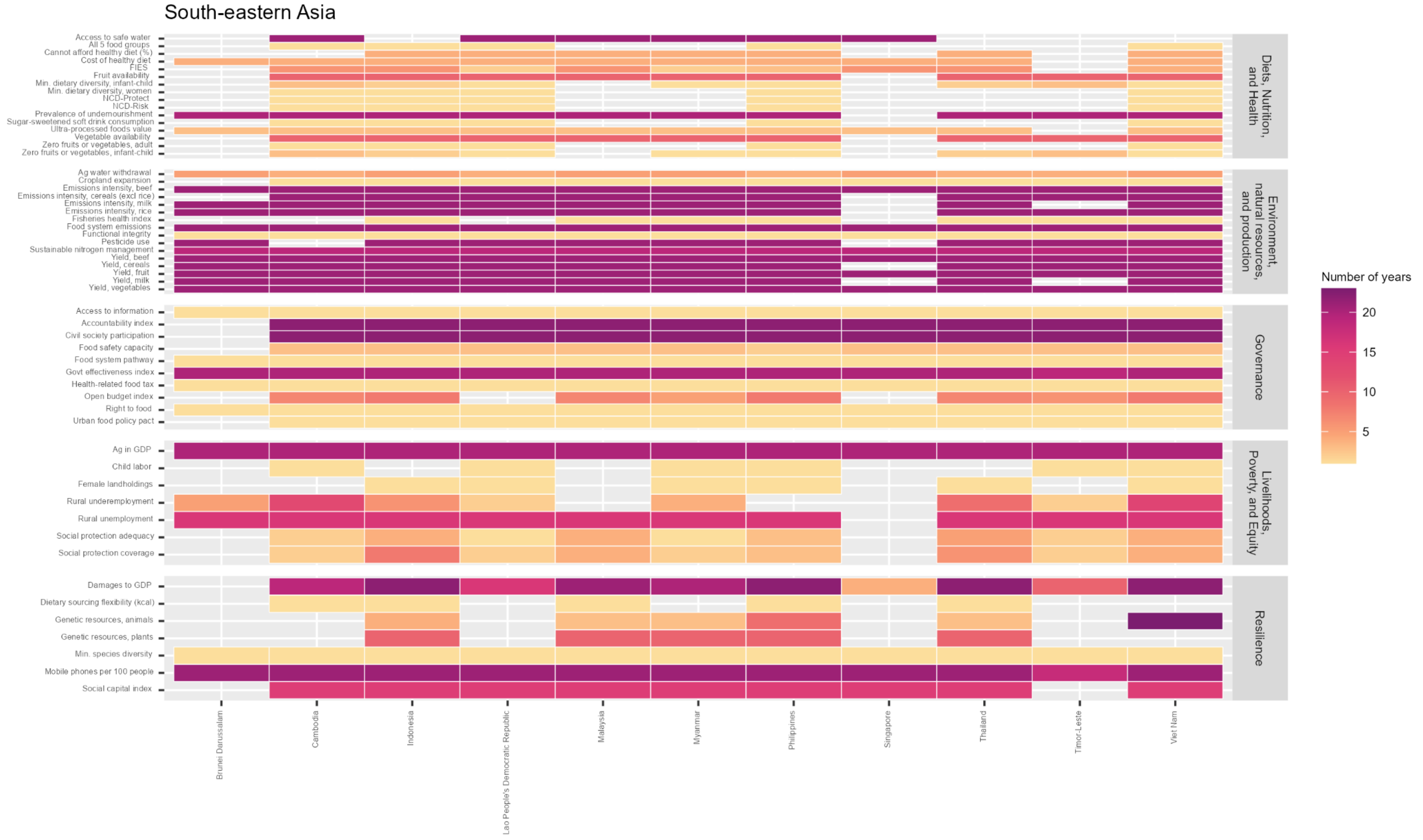

**Figure A.10 Indicator coverage by country and extent of time series**

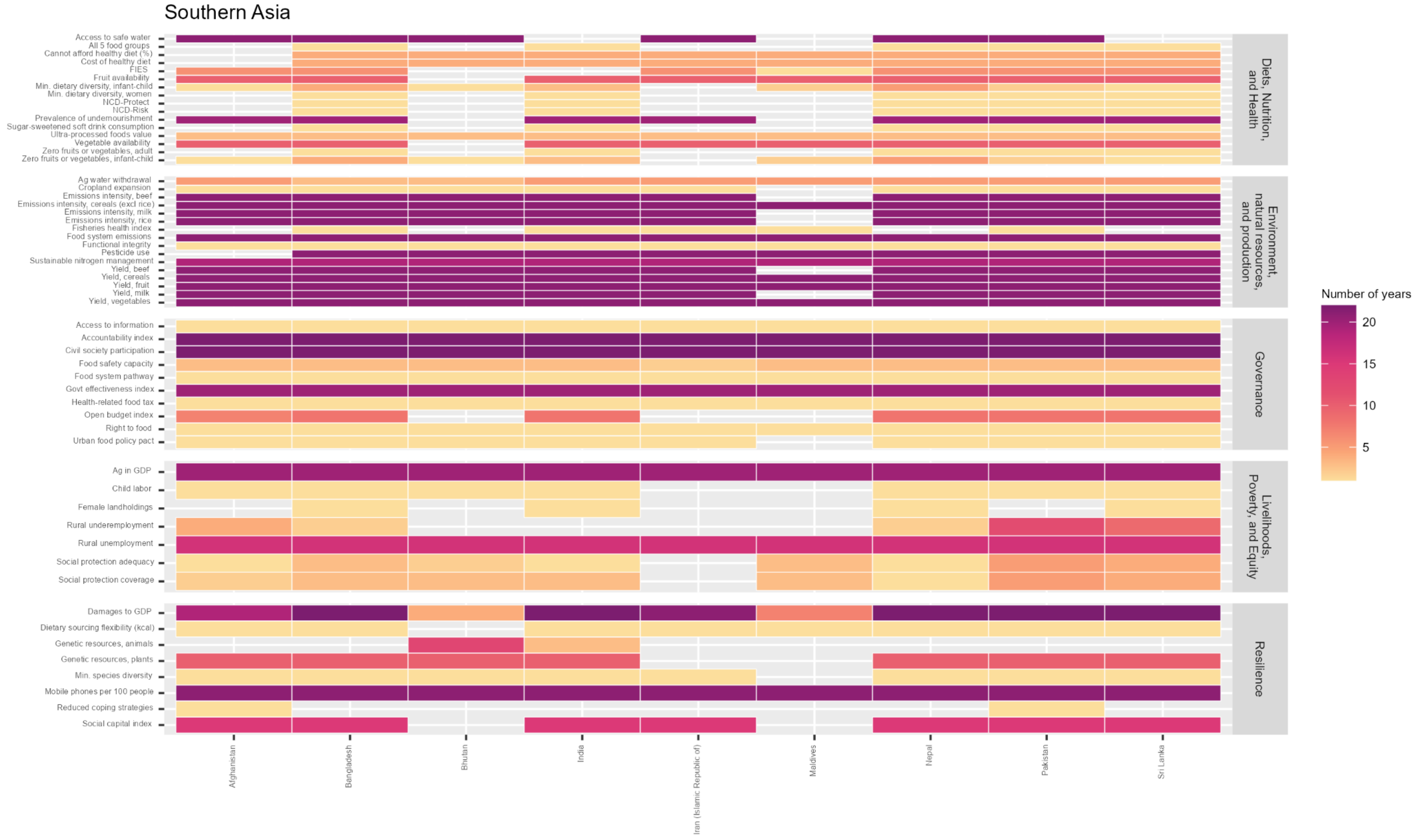

**Figure A.11 Indicator coverage by country and extent of time series**

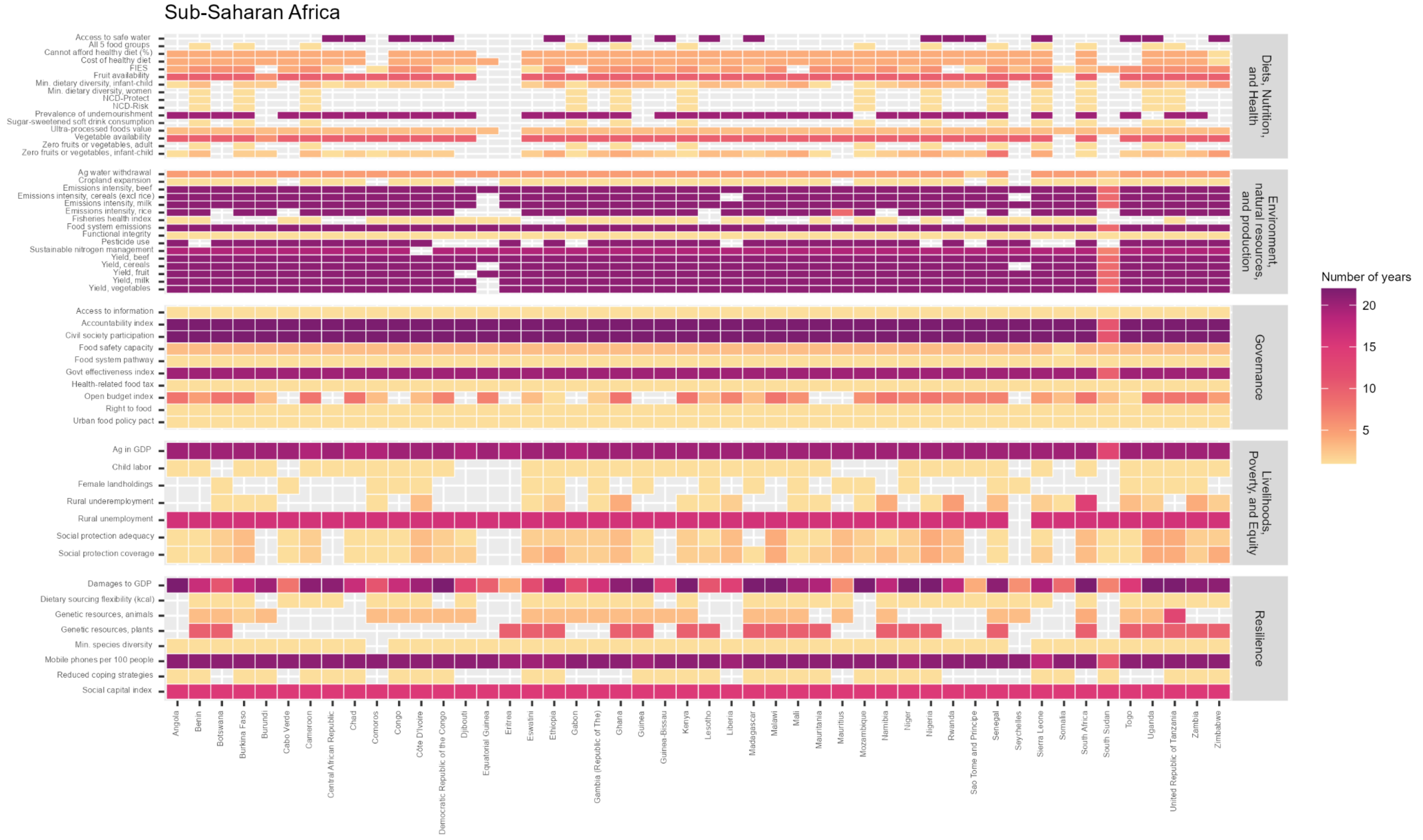

## Table A.1 Regional weighted mean values

| Domain | | Indicator | | Global weighted mean | Dir. of Δ | Latin America & Caribbean | Northern America & Europe | Oceania | Northern Africa & Western Asia | Central Asia | Eastern Asia | South-eastern Asia | Southern Asia | Sub-Saharan Africa |
|---|---|---|---|---|---|---|---|---|---|---|---|---|---|---|
| ***Diets, Nutrition, & Health*** | | | | | | | | | | | | | | |
| Food environments | **1** | Cost of a healthy diet | | US PPP $3.3 per person/day | ↓ | 3.3 | 3.2 | 2.6 | 3.4 | 3.0 | 3.3 | 4.2 | 3.1 | 3.4 |
| | **2** | Availability of fruits and vegetables | Fruits | 223.8 grams per capita/ day | ↑ | 308.1 | 241.5 | 269.5 | 232.9 | 156.6 | 142.6 | 174.1 | 157.0 | 169.5 |
| | | | Vegetables | 246.8 grams/ capita/ day | ↑ | 184.8 | 327.6 | 154.8 | 370.7 | 575.9 | 473.9 | 223.4 | 213.3 | 123.6 |
| | **3** | Retail value of ultra-processed foods | | US PPP $204.0 per capita | ↓ | 243.9 | 705.3 | 727.6 | 149.2 | 119.6 | 226.0 | 97.8 | 24.3 | 41.2 |
| | **4** | % Population using safely managed drinking water services (SDG 6.1.1) | | 66.3% population | ↑ | 69.1 | 94.3 | 94.7 | 75.6 | 69.8 | 93.8 | 55.7 | 50.2 | 20.3 |
| Food security | **5** | Prevalence of Undernourishment (SDG 2.1.1) | | 9.4% population | ↓ | 7.8 | 2.5 | 7.0 | 8.4 | 3.6 | 3.2 | 6.0 | 15.4 | 20.5 |
| | **6** | % Population experiencing moderate or severe food insecurity (SDG 2.1.2) | | 29.5% population | ↓ | 32.7 | 7.6 | 12.8 | 29.0 | 15.1 | 4.6 | 18.8 | 35.8 | 60.3 |
| | **7** | % Population who cannot afford a healthy diet | | 42.3% population | ↓ | 22.6 | 1.9 | 2.7 | 41.1 | 21.6 | 11.0 | 53.9 | 70.0 | 84.9 |
| Diet quality | **8** | MDD-W: minimum dietary diversity for women | | 65.7% population, women 15-49 | ↑ | 82.1 | 71.8 | | 75.1 | 87.7 | 86.2 | 84.4 | 44.9 | 52.4 |
| | **9** | MDD (IYCF): minimum dietary diversity for infants and young children | | 31.8% population, 6-23 months | ↑ | 62.1 | 70.8 | 32.1 | 35.2 | 36.4 | 37.1 | 52.9 | 19.0 | 23.1 |
| | **10** | All-5: consumption of all 5 food groups | | 39.0% adult population (≥15 y) | ↑ | 47.0 | 38.3 | | 41.7 | 37.7 | 54.2 | 49.7 | 27.6 | 25.0 |
| | **11** | Zero fruit or vegetable consumption | Adults | 10.8% adult population (≥15 y) | ↓ | 5.1 | 5.5 | | 6.6 | 3.1 | 3.8 | 4.4 | 20.0 | 13.7 |
| | | | Children 6-23 months | 39.1% population, 6-23 months | ↓ | 19.6 | 7.1 | 13.1 | 42.7 | 34.0 | 29.4 | 21.4 | 54.0 | 42.6 |
| | **12** | NCD-Protect | | 3.8 points (out of 9) | ↑ | 4.4 | 3.8 | | 3.6 | 3.7 | 4.6 | 4.4 | 3.3 | 3.1 |
| | **13** | NCD-Risk | | 2.1 points (out of 9) | ↓ | 2.8 | 3.1 | | 1.9 | 3.6 | 2.4 | 3.0 | 1.5 | 1.8 |
| | **14** | Sugar-sweetened soft drink consumption | | 18.9% adult population (≥15 y) | ↓ | 36.2 | 33.4 | | 24.8 | 33.5 | 11.6 | 20.8 | 14.3 | 27.2 |

| Domain | | Indicator | | Global weighted mean | Dir. of Δ | Latin America & Caribbean | Northern America & Europe | Oceania | Northern Africa & Western Asia | Central Asia | Eastern Asia | South-eastern Asia | Southern Asia | Sub-Saharan Africa |
|---|---|---|---|---|---|---|---|---|---|---|---|---|---|---|
| ***Environment, natural resources, and production*** | | | | | | | | | | | | | | |
| Greenhouse gas emissions | **16** | Food systems greenhouse gas emissions | | 82463.9 kt $CO_2$eq (AR5) | ↓ | 82584.4 | 80290.1 | 25641.1 | 33310.2 | 33971.7 | 445521.9 | 169650.5 | 263351.1 | 51790.6 |
| | **16** | Greenhouse gas emissions intensity, by product group[$] | Cereals (excl. rice)[†] | 0.2 kg $CO_2$eq/kg product | ↓ | 0.2 | 0.2 | 0.3 | 0.2 | 0.2 | 0.2 | 0.2 | 0.2 | 0.1 |
| | | | Beef | 30.3 kg $CO_2$eq/kg product | ↓ | 40.7 | 15.2 | 21.2 | 20.9 | 16.8 | 15.3 | 53.2 | 63.2 | 74.8 |
| | | | Cow's milk | 1.0 kg $CO_2$eq/kg product | ↓ | 1.0 | 0.6 | 0.8 | 1.0 | 1.2 | 0.8 | 2.8 | 1.3 | 3.8 |
| | | | Rice | 1.1 kg $CO_2$eq/kg product | ↓ | 1.0 | 1.9 | 1.5 | 1.2 | 2.9 | 0.9 | 1.5 | 0.9 | 1.6 |
| Production | **17** | Food product yield, by food group[$] | Cereals[†] | 40.7 tonnes/ha | ↑ | 47.4 | 53.9 | 17.0 | 22.9 | 16.7 | 62.4 | 43.2 | 33.0 | 16.4 |
| | | | Fruit[†] | 136.7 tonnes/ha | ↑ | 170.7 | 129.1 | 133.3 | 139.1 | 134.2 | 162.3 | 140.9 | 139.8 | 77.2 |
| | | | Beef | 231.5 kg/animal | ↑ | 273.6 | 317.9 | 233.2 | 190.6 | 185.5 | 157.8 | 211.4 | 129.3 | 148.4 |
| | | | Cow's milk | 2676.6 kg/animal | ↑ | 2491.2 | 7605.2 | 4887.0 | 1876.5 | 2254.9 | 3058.7 | 1083.3 | 1575.6 | 499.0 |
| | | | Vegetables[†] | 197.0 kg/ha | ↑ | 186.7 | 287.4 | 202.8 | 250.3 | 343.2 | 255.9 | 117.2 | 154.2 | 57.1 |
| Land | **18** | Cropland expansion (relative change 2003-2019) | | 19.1% | ↓ | 47.9 | 0.4 | 11.0 | 12.9 | 7.6 | 5.4 | 15.7 | 13.7 | 59.0 |
| Water | **19** | Agriculture water withdrawal as % of total renewable water resources | | 16.9% total renewable | ↓ | 3.6 | 3.3 | 2.3 | 96.9 | 31.5 | 13.4 | 8.5 | 40.5 | 4.2 |
| Biosphere integrity | **20** | Functional integrity: % agricultural land with minimum level of natural habitat | | 88.3% agricultural land | ↑ | 94.6 | 85.2 | 96.2 | 93.4 | 91.0 | 87.9 | 82.0 | 61.7 | 90.6 |
| | **21** | Fishery health index progress score | | 21.4 | ↑ | 24.2 | 38.3 | 27.3 | 13.7 | 0.0 | 12.2 | 13.7 | 27.2 | 12.3 |
| Pollution | **22** | Total pesticides per unit of cropland | | 1.8 kg/ha | ↓ | 5.3 | 2.0 | 2.1 | 1.3 | 0.7 | 2.3 | 0.9 | 0.4 | 0.4 |
| | **23** | Sustainable nitrogen management index | | 0.7 | ↑ | 0.6 | 0.6 | 0.8 | 0.9 | 0.8 | 0.7 | 0.7 | 0.8 | 0.9 |
| ***Livelihoods, Poverty, & Equity*** | | | | | | | | | | | | | | |
| Poverty and income | **24** | Share of agriculture in GDP | | 4.4% GDP | ↓ | 5.8 | 1.4 | 2.8 | 5.2 | 11.0 | 5.7 | 10.7 | 17.7 | 18.2 |
| Employment | **25** | Unemployment, rural | | 5.7% working age population | ↓ | 6.4 | 6.6 | 4.0 | 10.1 | 4.5 | 3.8 | 2.1 | 6.9 | 5.5 |
| | **26** | Underemployment rate, rural | | 7.3% working age population | ↓ | 10.7 | 2.9 | 8.6 | 2.1 | | 2.9 | 6.9 | 2.8 | 15.9 |

| Domain | | Indicator | Global weighted mean | Dir. of Δ | Latin America & Caribbean | Northern America & Europe | Oceania | Northern Africa & Western Asia | Central Asia | Eastern Asia | South-eastern Asia | Southern Asia | Sub-Saharan Africa |
|---|---|---|---|---|---|---|---|---|---|---|---|---|---|
| Social protection | **27** | Social protection coverage | 55.8% population | ↑ | 48.0 | 73.7 | 7.4 | 45.3 | 36.7 | 63.1 | 42.9 | 77.3 | 22.5 |
| | **28** | Social protection adequacy | 21.0% welfare of beneficiary households | ↑ | 32.7 | 37.8 | 3.8 | 20.9 | 27.8 | 36.8 | 11.3 | 8.2 | 16.8 |
| Rights | **29** | % Children 5-17 engaged in child labor | 9.4% children 5-17 | ↓ | 5.7 | 3.3 | 14.3 | 4.3 | 10.6 | 4.4 | 5.1 | 5.3 | 22.0 |
| | **30** | Female share of landholdings | 16.8% landholdings by sex of operator | ↑ | 18.4 | 22.6 | 3.6 | 19.8 | 12.4 | 17.7 | 22.0 | 10.4 | 18.1 |
| ***Governance*** | | | | | | | | | | | | | |
| Shared vision and strategic planning | **31** | Civil society participation index | 0.6 | ↑ | 0.7 | 0.8 | 0.8 | 0.4 | 0.3 | 0.4 | 0.7 | 0.6 | 0.7 |
| | **32** | % Urban population living in cities signed onto the Milan Urban Food Policy Pact | 7.2% urban population | ↑ | 25.4 | 12.1 | 1.7 | 2.5 | 3.0 | 8.4 | 2.1 | 0.6 | 8.4 |
| | **33** | Degree of legal recognition of the Right to Food (1 = Explicit protection or directive principle of state policy 2= Other implicit or national codification of international obligations or relevant provisions 3 = None) | 1.9 | ↑ | 1.8 | 2.0 | 2.5 | 2.1 | 2.0 | 1.8 | 2.0 | 1.3 | 1.8 |
| | **34** | Presence of a national food system transformation pathway (0 = No, 1 = yes) | 0.6 | ↑ | 0.5 | 0.4 | 0.8 | 0.6 | 0.8 | 0.8 | 0.7 | 0.8 | 0.8 |
| Effective implemen-tation | **35** | Government effectiveness index | 0.1 | ↑ | -0.4 | 0.9 | 1.0 | -0.6 | -0.4 | 0.7 | 0.2 | 0.1 | -0.8 |
| | **36** | International Health Regulations State Party Assessment report (IHR SPAR), Food safety capacity | 69.4 | ↑ | 84.9 | 88.7 | 82.5 | 72.2 | 42.4 | 81.9 | 69.2 | 57.2 | 44.9 |
| | **37** | Presence of health-related food taxes | 0.3 | ↑ | 0.3 | 0.2 | 0.0 | 0.2 | 0.0 | 0.0 | 0.3 | 0.7 | 0.2 |
| Accountability | **38** | V-Dem Accountability index | 0.3 | ↑ | 0.9 | 1.2 | 1.5 | -0.2 | -0.4 | -0.8 | 0.4 | 0.4 | 0.5 |

| Domain | | Indicator | | Global weighted mean | Dir. of Δ | Latin America & Caribbean | Northern America & Europe | Oceania | Northern Africa & Western Asia | Central Asia | Eastern Asia | South-eastern Asia | Southern Asia | Sub-Saharan Africa |
|---|---|---|---|---|---|---|---|---|---|---|---|---|---|---|
| | 39 | Open Budget Index Score | | 43.1 | ↑ | 65.9 | 67.1 | 72.4 | 30.7 | 49.9 | 25.1 | 59.1 | 37.5 | 37.6 |
| | 40 | Guarantees for public access to information (SDG 16.10.2) | | 1.9 | ↑ | 0.8 | 1.0 | 0.5 | 0.6 | 0.8 | 0.8 | 0.5 | 0.9 | 0.5 |
| ***Resilience & Sustainability*** | | | | | | | | | | | | | | |
| Exposure to shocks | 41 | Ratio of total damages of all disasters to GDP | | 0.3 | ↓ | 0.2 | 0.4 | 0.2 | 0.0 | 0.0 | 0.1 | 0.1 | 0.2 | 0.1 |
| Resilience capacities | 42 | Dietary sourcing flexibility index | | 0.7 | ↑ | 0.7 | 0.7 | 0.6 | 0.7 | 0.6 | 0.7 | 0.6 | 0.7 | 0.7 |
| | 43 | Mobile cellular subscriptions (per 100 people) | | 105.5 per 100 people | ↑ | 110.6 | 120.2 | 76.3 | 109.3 | 127.3 | 113.5 | 128.7 | 105.9 | 87.6 |
| | 44 | Social capital index | | 0.5 | ↑ | 0.3 | 0.6 | 0.6 | 0.4 | 0.5 | 0.7 | 0.4 | 0.4 | 0.4 |
| Agro- and food diversity | 45 | Proportion of agricultural land with minimum level of species diversity (crop and pasture) | | 22.5 % agricultural land | ↑ | 8.9 | 19.6 | 15.1 | 9.2 | 6.5 | 32.8 | 43.2 | 44.3 | 31.5 |
| | 46 | Number of (a) plant and (b) animal genetic resources for food and agriculture secured in either medium- or long-term conservation facilities (SDG 2.5.1) | Plants | 161.4 (thousands) | ↑ | 105.0 | 251.6 | 229.1 | 18.7 | 40.5 | 58.4 | 11.1 | 262.7 | 12.8 |
| | 47 | | Animals | 4.4 | ↑ | 0.6 | 5.7 | 0.0 | 0.2 | 0.0 | 1.3 | 4.7 | 36.5 | 1.3 |
| Resilience responses/ strategies | 48 | Coping strategies index | | 38.5% population | ↓ | 46.1 | | | 36.4 | | | | 32.7 | 39.4 |
| Long-term outcomes | 49 | Food price volatility | | 0.7 | ↓ | 0.8 | 0.8 | 0.6 | 0.8 | 0.6 | 0.7 | 0.8 | 0.7 | 0.8 |
| | 50 | Food supply variability | | 29.9 kcal per capita/ day | ↑ | 28.8 | 27.4 | 23.1 | 31.9 | 38.2 | 25.0 | 26.2 | 28.9 | 33.6 |

*** $p < 0.001$ ** $p < 0.01$ * $p < 0.05$

§ Reflects p-value of joint significance tests (F-test), '--' indicates insufficient observations in one or more regions to compute the F-test with cluster robust standard errors, required due to unequal variances by region.

† Product mix varies across countries.

$ Additional products included in **SM Figures S2.3-S2.10** and in the baseline dataset.

Sources: Author's calculations based on data sources listed in **Table 4**.

## Table A.2 Regional median values

| Domain | Indicator | | Unit | Latin America & Caribbean | Northern America & Europe | Oceania | Northern Africa & Western Asia | Central Asia | Eastern Asia | South-eastern Asia | Southern Asia | Sub-Saharan Africa |
|---|---|---|---|---|---|---|---|---|---|---|---|---|
| ***Diets, nutrition, and health*** | | | | | | | | | | | | |
| Food environments | Cost of a healthy diet | | current PPP dollar/ person/ day | 3.7 | 3.1 | 2.7 | 3.3 | 3.2 | 5.1 | 4.1 | 3.8 | 3.4 |
| | Availability of fruits and vegetables | Fruits | grams/ capita/ day | 264.7 | 221.9 | 189.6 | 233.6 | 135.1 | 143.0 | 189.0 | 129.0 | 120.4 |
| | | Vegetables | grams/ capita/ day | 159.9 | 264.1 | 146.7 | 331.5 | 630.5 | 334.6 | 173.7 | 179.8 | 93.1 |
| | Retail value of ultra-processed foods | | current (nominal) US$/capita | 196.4 | 496.1 | 125.4 | 176.8 | 90.7 | 156.0 | 107.6 | 21.3 | 31.0 |
| | % Population using safely managed drinking water services (SDG 6.1.1) | | % population | 66.8 | 97.6 | 46.2 | 86.9 | 70.1 | 82.5 | 53.1 | 36.2 | 21.1 |
| Food security | % Population experiencing moderate or severe food insecurity (SDG 2.1.2) | | % population | 7.5 | 2.5 | 5.1 | 5.1 | 3.5 | 3.2 | 5.8 | 11.4 | 19.3 |
| | % Population who cannot afford a healthy diet | | % population | 33.1 | 7.4 | 23.2 | 27.6 | 6.6 | 5.3 | 25.5 | 32.6 | 61.1 |
| | PoU: Prevalence of Undernourishment (SDG 2.1.1) | | % population | 21.0 | 0.9 | 30.6 | 16.7 | 42.1 | 7.3 | 65.1 | 61.8 | 83.5 |
| Diet quality | MDD-W: minimum dietary diversity for women | | % population, women 15-49 | 83.7 | 71.8 | | 75.6 | 87.5 | 86.2 | 79.9 | 58.8 | 50.9 |
| | MDD (IYCF): minimum dietary diversity for infants and young children | | % population, 6-23 months | 59.3 | 67.8 | 33.2 | 35.6 | 49.0 | 45.0 | 47.0 | 28.0 | 22.8 |
| | All-5: consumption of all 5 food groups | | % adult population (≥15 y) | 46.2 | 29.2 | | 38.9 | 44.0 | 54.2 | 36.4 | 28.2 | 23.3 |
| | Zero fruit or vegetable consumption | Adults | % adult population (≥15 y) | 4.6 | 5.3 | | 6.9 | 2.8 | 3.8 | 5.1 | 12.7 | 12.1 |
| | | Children 6-23 months | % population 6-23 months | 18.3 | 10.1 | 29.6 | 32.1 | 21.2 | 37.4 | 27.9 | 49.8 | 42.7 |
| | NCD-Protect | | Score (points out of 9) | 4.1 | 3.3 | | 3.5 | 3.8 | 4.6 | 3.9 | 3.4 | 3.0 |
| | NCD-Risk | | Score (points out of 9) | 2.7 | 2.9 | | 2.0 | 3.4 | 2.4 | 2.8 | 1.5 | 1.5 |
| | Sugar-sweetened soft drink consumption | | % adult population (≥15 y) | 38.0 | 24.1 | | 24.5 | 33.3 | 11.6 | 24.6 | 12.7 | 19.7 |

| Domain | Indicator | | Unit | Latin America & Caribbean | Northern America & Europe | Oceania | Northern Africa & Western Asia | Central Asia | Eastern Asia | South-eastern Asia | Southern Asia | Sub-Saharan Africa |
|---|---|---|---|---|---|---|---|---|---|---|---|---|
| ***Environment, natural resources, and production*** | | | | | | | | | | | | |
| Greenhouse gas emissions | Food systems greenhouse gas emissions | | kt $CO_2$eq (AR5) | 17,024.2 | 20,674.9 | 237.7 | 15,327.3 | 22,024.1 | 87,730.7 | 92,989.9 | 36,144.2 | 21,595.8 |
| | Greenhouse gas emissions intensity, by product group | Cereals (excl. rice) | kg CO2eq/ kg product | 0.3 | 0.2 | 0.2 | 0.3 | 0.2 | 0.2 | 0.2 | 0.1 | 0.2 |
| | | Beef | kg $CO_2$eq/ kg product | 49.7 | 16.7 | 68.5 | 11.6 | 15.8 | 22.5 | 58.7 | 33.2 | 66.9 |
| | | Cow's milk | kg $CO_2$eq/ kg product | 1.7 | 0.6 | 3.5 | 0.8 | 1.3 | 0.9 | 3.2 | 1.6 | 4.2 |
| | | Rice | kg $CO_2$eq/ kg product | 1.2 | 3.2 | 3.6 | 2.6 | 2.4 | 1.0 | 1.5 | 0.9 | 1.4 |
| Production | Food product yield, by food group | Cereals | tonnes/ha | 35.0 | 52.1 | 19.1 | 26.0 | 32.6 | 60.5 | 37.1 | 32.0 | 14.0 |
| | | Fruit | tonnes/ha | 150.6 | 98.0 | 86.1 | 120.5 | 118.3 | 155.0 | 102.3 | 110.8 | 68.0 |
| | | Beef | kg/animal | 193.8 | 271.8 | 157.2 | 200.5 | 180.8 | 150.0 | 199.9 | 116.6 | 145.4 |
| | | Cow's milk | kg/animal | 1,481.0 | 7,040.4 | 1,187.0 | 2,305.5 | 1,950.9 | 2,839.8 | 953.0 | 1,305.3 | 448.5 |
| | | Vegetables | kg/ha | 133.3 | 249.1 | 118.8 | 253.5 | 291.3 | 255.8 | 87.7 | 127.5 | 70.8 |
| Land | Cropland expansion (2003-2019) | | % | 23.6 | 1.8 | 33.4 | 7.5 | 7.3 | 5.2 | 19.3 | 14.2 | 62.7 |
| Water | Agriculture water withdrawal as % of total renewable water resources | | % total renewable | 1.2 | 0.4 | 0.6 | 65.9 | 33.7 | 12.5 | 4.4 | 21.4 | 1.0 |
| Biosphere integrity | Functional integrity: % agricultural land with minimum level of natural habitat | | % agricultural land | 98.6 | 80.1 | 98.9 | 92.3 | 91.6 | 85.4 | 82.4 | 76.7 | 94.1 |
| | Fishery health index progress score | | index | 23.0 | 32.5 | 23.0 | 18.6 | 0.0 | 13.6 | 9.3 | 24.4 | 12.6 |
| Pollution | Total pesticides per unit of croplands | | kg/ha | 4.8 | 2.1 | 2.0 | 2.0 | 0.5 | 6.2 | 0.9 | 0.4 | 0.1 |
| | Sustainable nitrogen management index | | index | 0.8 | 0.8 | 0.9 | 1.0 | 0.7 | 0.8 | 0.7 | 0.8 | 0.9 |
| ***Livelihoods, poverty, and equity*** | | | | | | | | | | | | |
| Poverty and income | Share of agriculture in GDP | | % GDP | 6.7 | 2.1 | 17.6 | 6.1 | 13.5 | 7.4 | 13.7 | 17.2 | 18.5 |
| Employment | Unemployment, rural | | % working age population | 5.5 | 6.3 | 3.2 | 8.4 | 4.8 | 2.9 | 1.9 | 4.9 | 4.0 |
| | Underemployment rate, rural | | % working age population | 8.9 | 3.3 | 7.1 | 4.1 | 0.0 | 1.8 | 2.2 | 3.0 | 6.5 |

| Domain | Indicator | Unit | Latin America & Caribbean | Northern America & Europe | Oceania | Northern Africa & Western Asia | Central Asia | Eastern Asia | South-eastern Asia | Southern Asia | Sub-Saharan Africa |
|---|---|---|---|---|---|---|---|---|---|---|---|
| Social protection | Social protection coverage | % population | 52.0 | 71.5 | 13.4 | 54.6 | 44.2 | 78.5 | 35.3 | 31.3 | 20.7 |
| | Social protection adequacy | % welfare of beneficiary households | 27.7 | 38.4 | 21.6 | 14.0 | 30.3 | 30.0 | 17.7 | 12.6 | 13.6 |
| Rights | % Children 5-17 engaged in child labor | % children 5-17 | 3.6 | 3.4 | 13.4 | 1.9 | 10.2 | 6.0 | 7.7 | 5.9 | 17.5 |
| | Female share of landholdings | % landholdings by sex of operator | 25.2 | 17.3 | 19.4 | 14.6 | 12.4 | 15.8 | 12.9 | 15.8 | 20.3 |
| ***Governance*** | | | | | | | | | | | |
| Shared vision and strategic planning | Civil society participation index | index | 0.7 | 0.8 | 0.7 | 0.5 | 0.3 | 0.6 | 0.5 | 0.7 | 0.7 |
| | % Urban population living in cities signed onto the Milan Urban Food Policy Pact [+] | % urban population | 0.0 | 0.4 | 0.0 | 0.0 | 0.0 | 15.6 | 0.0 | 0.0 | 0.0 |
| | Degree of legal recognition of the Right to Food [+] | Categorical | 2.0 | 2.0 | 3.0 | 2.0 | 2.0 | 2.0 | 2.0 | 1.0 | 2.0 |
| | Presence of a national food system transformation pathway | Binary | 0.5 | 0.0 | 1.0 | 1.0 | 1.0 | 1.0 | 1.0 | 1.0 | 1.0 |
| Effective implementation | Government effectiveness index | index | -0.2 | 1.0 | -0.1 | -0.1 | -0.5 | 0.6 | 0.2 | -0.5 | -0.8 |
| | International Health Regulations State Party Assessment report (IHR SPAR), Food safety capacity | Score | 80.0 | 80.0 | 80.0 | 80.0 | 40.0 | 80.0 | 70.0 | 60.0 | 40.0 |
| | Presence of health-related food taxes [+] | Binary | 0.0 | 0.0 | 0.0 | 0.0 | 0.0 | 0.0 | 0.0 | 0.0 | 0.0 |
| Accountability | V-Dem Accountability index | index | 1.0 | 1.5 | 1.2 | -0.2 | -0.5 | 1.0 | 0.1 | 0.6 | 0.4 |
| | Open Budget Index Score | index | 50.0 | 65.0 | 64.5 | 42.0 | 62.0 | 60.5 | 49.5 | 38.0 | 31.0 |
| | Guarantees for public access to information (SDG 16.10.2) | Binary | 1.0 | 1.0 | 0.0 | 1.0 | 1.0 | 1.0 | 0.0 | 1.0 | 0.0 |

| Domain | Indicator | | Unit | Latin America & Caribbean | Northern America & Europe | Oceania | Northern Africa & Western Asia | Central Asia | Eastern Asia | South-eastern Asia | Southern Asia | Sub-Saharan Africa |
|---|---|---|---|---|---|---|---|---|---|---|---|---|
| ***Resilience and sustainability*** | | | | | | | | | | | | |
| Exposure to shocks | Ratio of total damages of all disasters to GDP | | Ratio | 0.0 | 0.0 | 0.0 | 0.0 | 0.0 | 0.1 | 0.0 | 0.0 | 0.0 |
| Resilience capacities | Dietary sourcing flexibility index | | index | 0.7 | 0.8 | 0.6 | 0.8 | 0.6 | 0.7 | 0.6 | 0.7 | 0.6 |
| | Mobile cellular subscriptions (per 100 people) | | Number per 100 people | 108.8 | 120.2 | 72.7 | 103.9 | 129.4 | 133.1 | 135.1 | 107.0 | 87.5 |
| | Social capital index | | index | 0.3 | 0.5 | 0.7 | 0.4 | 0.5 | 0.6 | 0.5 | 0.4 | 0.4 |
| Agro- and Food Diversity | Proportion of agricultural land with minimum level of species diversity (crop and pasture) ⁺ | | % agricultural land | 13.2 | 0.0 | 12.2 | 0.1 | 12.2 | 39.9 | 30.9 | 26.3 | 22.8 |
| | Number of (a) plant and (b) animal genetic resources for food and agriculture secured in either medium- or long-term conservation facilities (SDG 2.5.1) | Plants | thousands | 6.0 | 12.0 | 36.1 | 13.2 | 4.8 | 122.0 | 12.1 | 12.4 | 3.6 |
| | | Animals | number | 0.0 | 1.5 | 0.0 | 0.0 | 0.0 | 0.0 | 0.5 | 19.0 | 0.0 |
| Resilience responses/ strategies | Coping strategies index | | % population | 45.0 | | | 55.6 | | | | 43.4 | 38.4 |
| Long-term outcomes | Food price volatility ⁺ | | unitless | 0.7 | 0.7 | 0.7 | 0.7 | 0.6 | 0.7 | 0.8 | 0.7 | 0.8 |
| | Food supply variability | | kcal/ capita/ day | 27.0 | 26.0 | 23.5 | 27.0 | 31.0 | 18.0 | 27.5 | 31.0 | 28.5 |

⁺ Indicates FSCI value-added to existing data

Sources: Author’s calculations based on data sources listed in **Table 4**.

## Table 3. Income group weighted means‡

| Domain | | Indicator | | Unit | Low income | Lower middle income | Upper middle income | High income | Global |
|---|---|---|---|---|---|---|---|---|---|
| ***Diets, nutrition & health*** | | | | | | | | | |
| Food environments | **1** | Cost of a healthy diet | | current PPP dollar/ person/ day | 3.1 | 3.4 | 3.2 | 3.5 | 3.3 |
| | **2** | Availability of fruits and vegetables | Fruits | grams/capita/day | 162.8 | 202.6 | 281.5 | 222.5 | 223.8 |
| | | | Vegetables | grams/capita/day | 127.5 | 223.7 | 311.3 | 274.2 | 246.8 |
| | **3** | Retail value of ultra-processed foods | | current (nominal) US$/capita | 24.4 | 45.2 | 181.4 | 801.6 | 204.0 |
| | **4** | % Population using safely managed drinking water services (SDG 6.1.1) | | % population | 19.9 | 48.5 | 72.3 | 98.0 | 66.3 |
| Food security | **5** | Prevalence of Undernourishment (SDG 2.1.1) | | % population | 29.4 | 12.6 | 3.9 | 2.7 | 9.4 |
| | **6** | % Population experiencing moderate or severe food insecurity (SDG 2.1.2) | | % population | 65.9 | 34.6 | 25.0 | 7.0 | 29.5 |
| | **7** | % Population who cannot afford a healthy diet | | % population | 88.3 | 69.3 | 15.3 | 1.6 | 42.3 |
| Diet quality | **8** | MDD-W: minimum dietary diversity for women | | % population, women 15-49 | 51.0 | 54.4 | 83.7 | | 65.7 |
| | **9** | MDD (IYCF): minimum dietary diversity for infants and young children | | % population, 6-23 months | 21.4 | 27.1 | 42.8 | 69.8 | 31.8 |
| | **10** | All-5: consumption of all 5 food groups | | % adult population (≥15 y) | 27.4 | 31.8 | 49.6 | 44.2 | 39.0 |
| | **11** | Zero fruit or vegetable consumption | Adults | % adult population (≥15 y) | 10.3 | 15.7 | 4.5 | 5.4 | 10.8 |
| | | | Children 6-23 months | % population 6-23 months | 46.6 | 45.1 | 27.0 | 5.5 | 39.1 |
| | **12** | NCD-Protect | | Score (points out of 9) | 3.3 | 3.5 | 4.4 | 4.0 | 3.8 |
| | **13** | NCD-Risk | | Score (points out of 9) | 1.3 | 1.8 | 2.5 | 3.2 | 2.1 |
| | **14** | Sugar-sweetened soft drink consumption | | % adult population (≥15 y) | 19.4 | 17.7 | 17.0 | 38.8 | 18.9 |
| ***Environment, natural resources, & production*** | | | | | | | | | |
| Greenhouse gas emissions | **15** | Food systems greenhouse gas emissions | | kt $CO_2$eq (AR5) | 63,201.0 | 91,359.8 | 117,046.8 | 64,454.8 | 82,463.9 |
| | **16** | Greenhouse gas emissions intensity, by product group$ | Cereals (excl. rice)† | kg $CO_2$eq/kg product | 0.1 | 0.2 | 0.2 | 0.2 | 0.2 |
| | | | Beef | kg $CO_2$eq/kg product | 94.8 | 49.1 | 30.3 | 16.4 | 30.3 |
| | | | Cow's milk | kg $CO_2$eq/kg product | 4.4 | 1.4 | 0.9 | 0.6 | 1.0 |
| | | | Rice | kg $CO_2$eq/kg product | 1.6 | 1.1 | 1.0 | 1.4 | 1.1 |
| Production | **17** | Food product yield, by food group$ | Cereals† | tonnes/ha | 14.4 | 32.6 | 46.2 | 59.3 | 40.7 |
| | | | Fruit† | tonnes/ha | 66.5 | 129.3 | 157.8 | 143.7 | 136.7 |
| | | | Beef | kg/animal | 123.0 | 158.6 | 226.2 | 320.0 | 231.5 |

| Domain | | Indicator | | Unit | Low income | Lower middle income | Upper middle income | High income | Global |
|---|---|---|---|---|---|---|---|---|---|
| | | | Cow's milk | kg/animal | 430.1 | 1,502.8 | 2,976.0 | 7,845.7 | 2,676.6 |
| | | | Vegetables† | kg/ha | 93.2 | 129.5 | 247.7 | 325.5 | 197.0 |
| Land | **18** | Cropland expansion (2003-2019) | | % | 62.3 | 21.4 | 14.6 | 4.1 | 19.1 |
| Water | **19** | Agriculture water withdrawal as % of total renewable water resources | | % total renewable | 18.3 | 25.2 | 11.1 | 13.3 | 16.9 |
| Biosphere integrity | **20** | Functional integrity: % agricultural land with minimum level of natural habitat | | % agricultural land | 92.8 | 78.2 | 90.3 | 91.3 | 88.3 |
| | **21** | Fishery health index progress score | | index | 9.2 | 21.0 | 16.5 | 35.8 | 21.4 |
| Pollution | **22** | Total pesticides per unit of land | | kg/ha | 0.2 | 0.6 | 2.5 | 2.8 | 1.8 |
| | **23** | Sustainable nitrogen management index | | index | 0.9 | 0.7 | 0.7 | 0.6 | 0.7 |
| ***Livelihoods, poverty & equity*** | | | | | | | | | |
| Poverty and income | **24** | Share of agriculture in GDP | | % GDP | 25.6 | 16.5 | 6.8 | 1.3 | 4.4 |
| Employment | **25** | Unemployment, rural | | % working age population | 5.0 | 5.7 | 6.0 | 5.6 | 5.7 |
| | **26** | Underemployment rate, rural | | % working age population | 15.1 | 8.1 | 6.9 | 3.3 | 7.3 |
| Social protection | **27** | Social protection coverage | | % population | 14.0 | 59.8 | 61.0 | 76.6 | 55.8 |
| | **28** | Social protection adequacy | | % welfare of beneficiary households | 16.0 | 10.8 | 34.2 | 47.2 | 21.0 |
| Rights | **29** | % Children 5-17 engaged in child labor | | % children 5-17 (sex specific is % children 5-17 of each sex) | 20.9 | 8.4 | 4.2 | 2.4 | 9.4 |
| | **30** | Female share of landholdings | | % landholdings by sex of operator | 14.4 | 13.1 | 18.3 | 18.7 | 16.8 |
| ***Governance*** | | | | | | | | | |
| Shared vision and strategic planning | **31** | Civil society participation index | | index | 0.6 | 0.7 | 0.4 | 0.9 | 0.6 |
| | **32** | % Urban population living in cities signed onto the Milan Urban Food Policy Pact + | | % urban population | 4.4 | 2.2 | 12.2 | 12.9 | 7.2 |
| | **33** | Degree of legal recognition of the Right to Food + | | categorical | 1.7 | 1.8 | 1.9 | 2.2 | 1.9 |
| | **34** | Presence of a national food system transformation pathway | | binary | 0.6 | 0.8 | 0.6 | 0.5 | 0.6 |
| Effective implementation | **35** | Government effectiveness index | | index | -1.1 | -0.1 | 0.3 | 1.2 | 0.1 |
| | **36** | International Health Regulations State Party Assessment report (IHR SPAR), Food safety capacity | | score | 41.2 | 58.3 | 82.2 | 90.3 | 69.4 |
| | **37** | Presence of health-related food taxes + | | binary | 0.2 | 0.5 | 0.1 | 0.3 | 0.3 |
| Accountability | **38** | V-Dem Accountability index | | index | 0.0 | 0.4 | -0.4 | 1.4 | 0.3 |
| | **39** | Open Budget Index Score | | index | 25.3 | 42.3 | 39.5 | 64.9 | 43.1 |
| | **40** | Guarantees for public access to information (SDG 16.10.2) | | binary | 0.6 | 0.6 | 0.7 | 0.9 | 1.9 |

| Domain | | Indicator | | Unit | Low income | Lower middle income | Upper middle income | High income | Global |
|---|---|---|---|---|---|---|---|---|---|
| ***Resilience & sustainability*** | | | | | | | | | |
| Exposure to shocks | **41** | Ratio of total damages of all disasters to GDP | | ratio | 0.2 | 0.2 | 0.1 | 0.4 | 0.3 |
| Resilience capacities | **42** | Dietary sourcing flexibility index | | index | 0.6 | 0.7 | 0.7 | 0.8 | 0.7 |
| | **43** | Mobile cellular subscriptions (per 100 people) | | number per 100 people | 66.5 | 98.8 | 112.5 | 128.1 | 105.5 |
| | **44** | Social capital index | | index | 0.4 | 0.4 | 0.6 | 0.6 | 0.5 |
| Agro- and Food Diversity | **45** | Proportion of agricultural land with minimum level of species diversity (crop and pasture) ⁺ | | % agricultural land | 36.4 | 32.2 | 22.0 | 10.3 | 22.5 |
| | **46** | Number of (a) plant and (b) animal genetic resources for food and agriculture secured in either medium- or long-term conservation facilities (SDG 2.5.1) | Plants | thousands | 16.7 | 85.0 | 134.0 | 274.5 | 161.4 |
| | **47** | | Animals | number | 0.7 | 9.1 | 0.9 | 5.6 | 4.4 |
| Resilience responses/ strategies | **48** | Coping strategies index | | % population | 41.2 | 36.7 | 36.2 | | 38.5 |
| Long-term outcomes | **49** | Food price volatility ⁺ | | unitless | 0.8 | 0.7 | 0.7 | 0.8 | 0.7 |
| | **50** | Food supply variability | | kcal/capita/day | 33.5 | 29.9 | 27.7 | 28.0 | 29.9 |

‡ See **SM Table A.4** for income group medians.
† Product mix varies across countries.
$ Additional products are included in the **SM-A** and baseline dataset.
⁺ Indicates FSCI value-added to existing data.
Sources: Author's calculations based on data sources listed in **Table 4**.

## Table A.4 Income group median values

| Domain | Indicator | | Unit | Low income | Lower middle income | Upper middle income | High income |
|---|---|---|---|---|---|---|---|
| ***Diets, nutrition & health*** | | | | | | | |
| Food environments | Cost of a healthy diet | | current PPP dollar/person/day | 3.2 | 3.6 | 3.6 | 3.1 |
| | Availability of fruits and vegetables | Fruits | grams/capita/day | 117.5 | 177.2 | 240.7 | 208.0 |
| | | Vegetables | grams/capita/day | 93.2 | 149.2 | 217.4 | 245.1 |
| | Retail value of ultra-processed foods | | current (nominal) US$/capita | 21.7 | 50.8 | 178.1 | 485.0 |
| | % Population using safely managed drinking water services (SDG 6.1.1) | | % population | 19.0 | 46.1 | 76.8 | 99.4 |
| Food security | % Population experiencing moderate or severe food insecurity (SDG 2.1.2) | | % population | 30.4 | 8.2 | 6.1 | 2.5 |
| | % Population who cannot afford a healthy diet | | % population | 72.4 | 41.7 | 25.3 | 7.5 |
| | PoU: Prevalence of Undernourishment (SDG 2.1.1) | | % population | 89.0 | 66.8 | 18.0 | 0.8 |
| Diet quality | MDD-W: minimum dietary diversity for women | | % population, women 15-49 | 44.9 | 69.3 | 74.1 | |
| | MDD (IYCF): minimum dietary diversity for infants and young children | | % population, 6-23 months | 22.8 | 30.6 | 53.0 | 69.8 |
| | All-5: consumption of all 5 food groups | | % adult population (≥15 y) | 24.7 | 28.2 | 29.7 | 42.0 |
| | Zero fruit or vegetable consumption | Adults | % adult population (≥15 y) | 11.3 | 9.6 | 6.8 | 3.9 |
| | | Children 6-23 months | % population 6-23 months | 44.1 | 36.3 | 21.2 | 5.5 |
| | NCD-Protect | | Score (points out of 9) | 3.1 | 3.6 | 3.5 | 4.0 |
| | NCD-Risk | | Score (points out of 9) | 1.2 | 1.6 | 2.5 | 3.2 |
| | Sugar-sweetened soft drink consumption | | % adult population (≥15 y) | 15.9 | 21.2 | 31.6 | 42.4 |
| ***Environment, natural resources & production*** | | | | | | | |
| Greenhouse gas emissions | Food systems greenhouse gas emissions | | kt $CO_2$eq (AR5) | 35,806.0 | 23,912.9 | 13,618.9 | 17,965.5 |
| | Greenhouse gas emissions intensity, by product group | Cereals (excl. rice) | kg $CO_2$eq/kg product | 0.1 | 0.2 | 0.3 | 0.2 |
| | | Beef | kg $CO_2$eq/kg product | 78.6 | 57.6 | 32.5 | 17.5 |
| | | Cow's milk | kg $CO_2$eq/kg product | 4.1 | 2.5 | 1.4 | 0.6 |
| | | Rice | kg $CO_2$eq/kg product | 1.3 | 1.4 | 1.7 | 2.2 |
| Production | Food product yield, by food group | Cereals | tonnes/ha | 12.4 | 20.1 | 34.0 | 58.7 |
| | | Fruit | tonnes/ha | 66.1 | 104.7 | 115.1 | 119.9 |
| | | Beef | kg/animal | 125.0 | 165.2 | 199.2 | 269.5 |
| | | Cow's milk | kg/animal | 459.3 | 1,108.2 | 2,022.3 | 7,022.9 |

| Domain | Indicator | Unit | Low income | Lower middle income | Upper middle income | High income |
|---|---|---|---|---|---|---|
| | Vegetables | tonnes/ha | 81.4 | 120.4 | 152.9 | 252.2 |
| Land | Cropland expansion (2003-2019) | % | 54.6 | 22.5 | 15.8 | 2.9 |
| Water | Agriculture water withdrawal as % of total renewable water resources | % total renewable | 1.6 | 5.1 | 1.3 | 0.8 |
| Biosphere integrity | Functional integrity: % agricultural land with minimum level of natural habitat | % agricultural land | 93.2 | 91.6 | 95.1 | 91.5 |
| | Fishery health index progress score | index | 6.8 | 19.2 | 22.8 | 31.0 |
| Pollution | Total pesticides per unit of cropland | kg/ha | 0.1 | 0.4 | 2.2 | 2.7 |
| | Sustainable nitrogen management index | index | 0.9 | 0.8 | 0.8 | 0.9 |
| ***Livelihoods, poverty & equity*** | | | | | | |
| Poverty and income | Share of agriculture in GDP | % GDP | 23.8 | 14.9 | 7.1 | 1.8 |
| Employment | Unemployment, rural | % working age population | 3.7 | 4.6 | 7.9 | 5.0 |
| | Underemployment rate, rural | % working age population | 6.7 | 3.3 | 5.0 | 3.9 |
| Social protection | Social protection coverage | % population | 7.8 | 33.6 | 55.2 | 80.9 |
| | Social protection adequacy | % welfare of beneficiary households | 9.7 | 18.7 | 28.4 | 34.2 |
| Rights | % Children 5-17 engaged in child labor | % children 5-17 (sex specific is % children 5-17 of each sex) | 18.9 | 11.5 | 3.4 | 2.2 |
| | Female share of landholdings | % landholdings by sex of operator | 16.5 | 18.7 | 25.0 | 18.3 |
| ***Governance*** | | | | | | |
| Shared vision and strategic planning | Civil society participation index | index | 0.7 | 0.7 | 0.7 | 0.9 |
| | % Urban population living in cities signed onto the Milan Urban Food Policy Pact [+] | % urban population | 0.0 | 0.0 | 0.0 | 0.0 |
| | Degree of legal recognition of the Right to Food [+] | Categorical | 2.0 | 2.0 | 2.0 | 2.0 |
| | Presence of a national food system transformation pathway | Binary | 1.0 | 1.0 | 1.0 | 0.0 |
| Effective implementation | Government effectiveness index | index | -1.2 | -0.5 | -0.1 | 1.1 |
| | International Health Regulations State Party Assessment report (IHR SPAR), Food safety capacity | Score | 40.0 | 60.0 | 80.0 | 80.0 |
| | Presence of health-related food taxes [+] | Binary | 0.0 | 0.0 | 0.0 | 0.0 |
| Accountability | V-Dem Accountability index | index | 0.1 | 0.4 | 0.8 | 1.5 |
| | Open Budget Index Score | index | 27.0 | 42.0 | 55.0 | 64.0 |
| | Guarantees for public access to information (SDG 16.10.2) | Binary | 1.0 | 1.0 | 1.0 | 1.0 |

| Domain | Indicator | | Unit | Low income | Lower middle income | Upper middle income | High income |
|---|---|---|---|---|---|---|---|
| ***Resilience & sustainability*** | | | | | | | |
| Exposure to shocks | Ratio of total damages of all disasters to GDP | | Ratio | 0.0 | 0.0 | 0.0 | 0.0 |
| Resilience capacities | Dietary sourcing flexibility index | | index | 0.6 | 0.7 | 0.7 | 0.8 |
| | Mobile cellular subscriptions (per 100 people) | | Number per 100 people | 57.8 | 99.1 | 110.6 | 125.2 |
| | Social capital index | | index | 0.4 | 0.4 | 0.4 | 0.6 |
| Agro- and Food Diversity | Proportion of agricultural land with minimum level of species diversity (crop and pasture) [+] | | % agricultural land | 43.8 | 22.1 | 12.6 | 0.0 |
| | Number of (a) plant and (b) animal genetic resources for food and agriculture secured in either medium- or long-term conservation facilities (SDG 2.5.1) | Plants | thousands | 4.7 | 6.5 | 6.0 | 20.8 |
| | | Animals | number | 0.0 | 0.0 | 0.0 | 1.0 |
| Resilience responses/ strategies | Coping strategies index | | % population | 40.9 | 38.3 | 38.6 | |
| Long-term outcomes | Food price volatility [+] | | unitless | 0.8 | 0.7 | 0.7 | 0.8 |
| | Food supply variability | | kcal/capita/day | 28.5 | 26.5 | 26.0 | 26.0 |

[+] Indicates FSCI value-added to existing data

Sources: Author's calculations based on data sources listed in **Table 4**.

## Figure A.12 Relation to GDP per capita: Diets, nutrition, and health

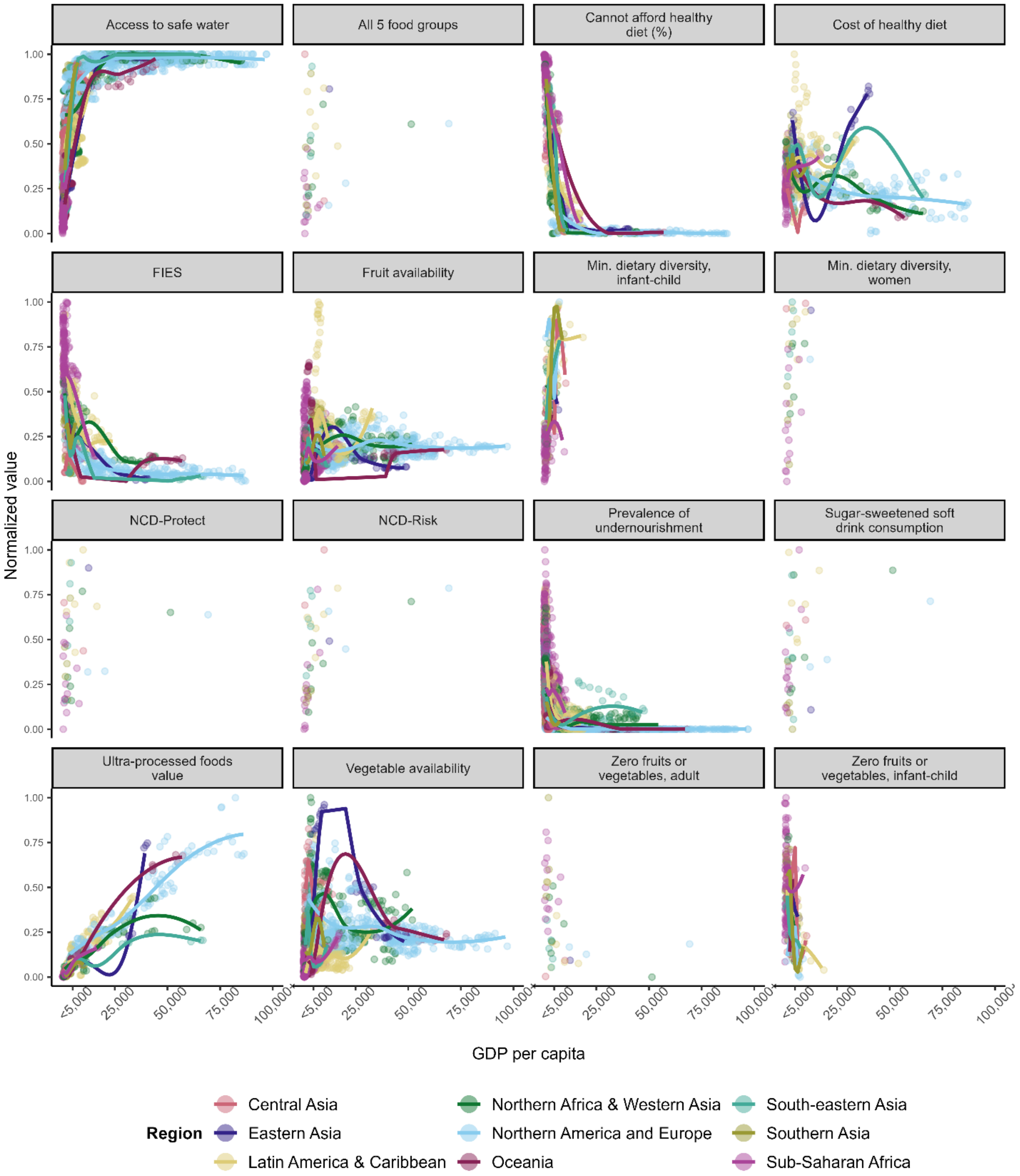

Sources: Author's calculations based on data sources listed in **Table 4**.

Notes: Colored lines reflects the local polynomial regression by region (Loess). Regression lines shown only for indicators with at least three data points in all regions necessary to estimate the Loess model. GDP per capita is in nominal terms and data come from 2021 for most countries, see **SM-E** for the specific year of data for GDP and each variable. Normalized values are calculated using max-min normalization relative to the global weighted mean. Each dot represents a country data point. Countries with GDP>$100,000 not shown on figure but included in model.

## Figure A.13 Relation to GDP per capita: Environment, natural resources, and production

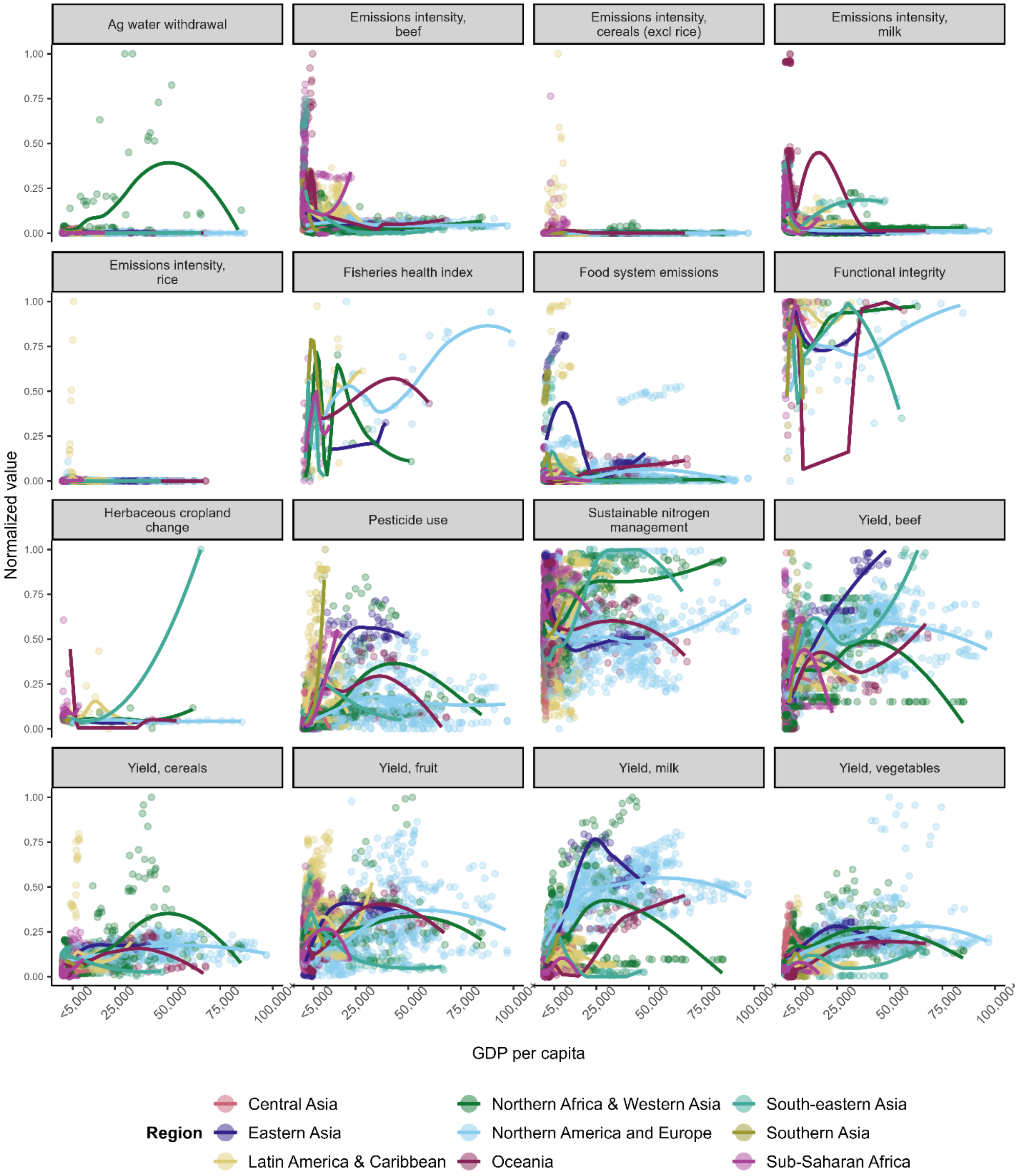

Sources: Author's calculations based on data sources listed in **Table 4**.
Notes: Colored lines reflects the local polynomial regression by region (Loess). GDP per capita is in nominal terms and data come from 2021 for most countries, see **SM-E** for the specific year of data for GDP and each variable. Normalized values are calculated using max-min normalization relative to the global weighted mean. Each dot represents a country data point. Countries with GDP>$100,000 not shown on figure but included in model.

## Figure A.14 Relation to GDP per capita: Livelihoods, poverty, and equity

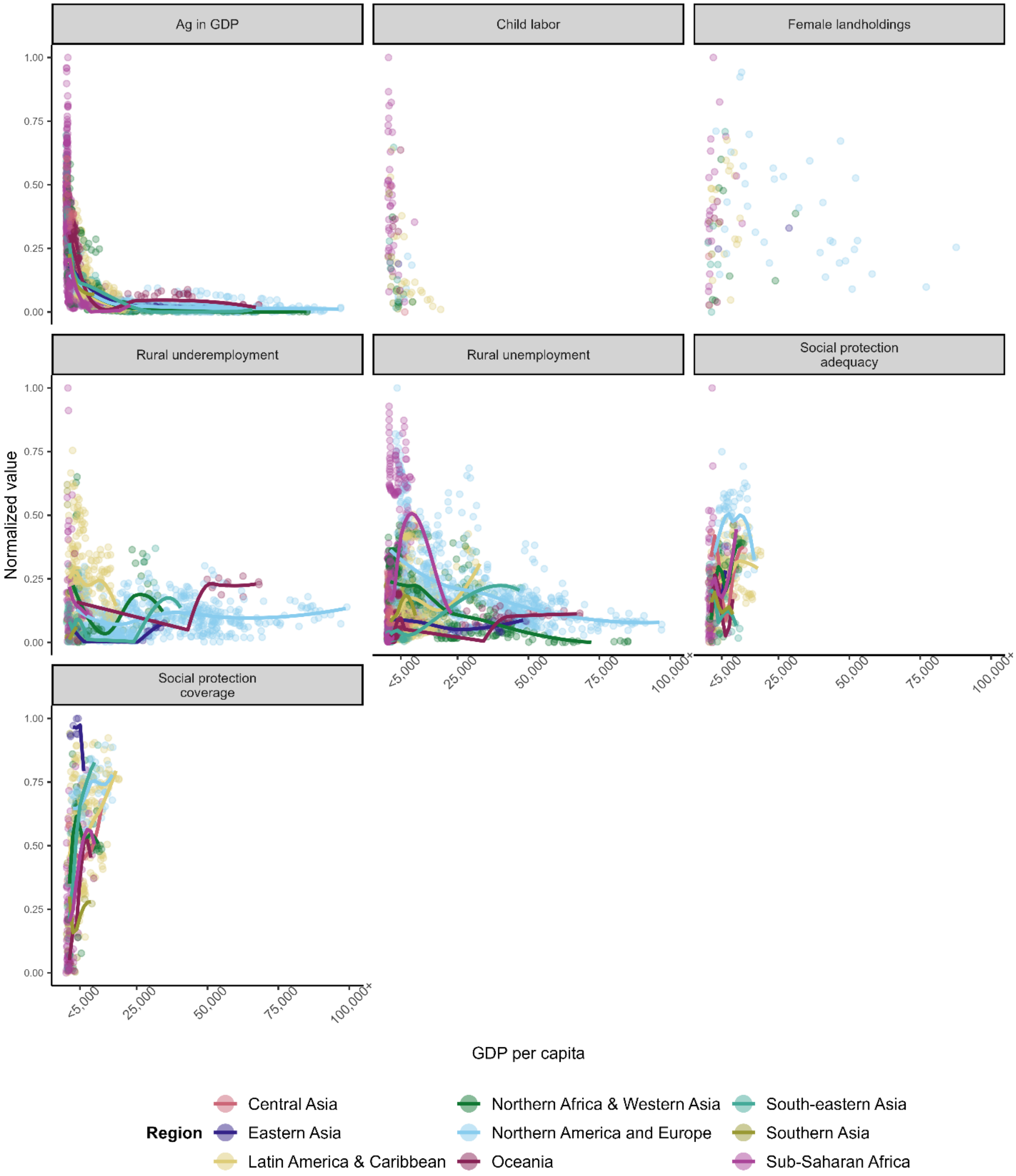

Sources: Author's calculations based on data sources listed in **Table 4**.

Notes: Colored lines reflects the local polynomial regression by region (Loess). Regression lines shown only for indicators with at least three data points in all regions necessary to estimate the Loess model. GDP per capita is in nominal terms and data come from 2021 for most countries, see **SM-E** for the specific year of data for GDP and each variable. Normalized values are calculated using max-min normalization relative to the global weighted mean. Each dot represents a country data point. Countries with GDP>$100,000 not shown on figure but included in model.

## Figure A.15 Relation to GDP per capita: Governance

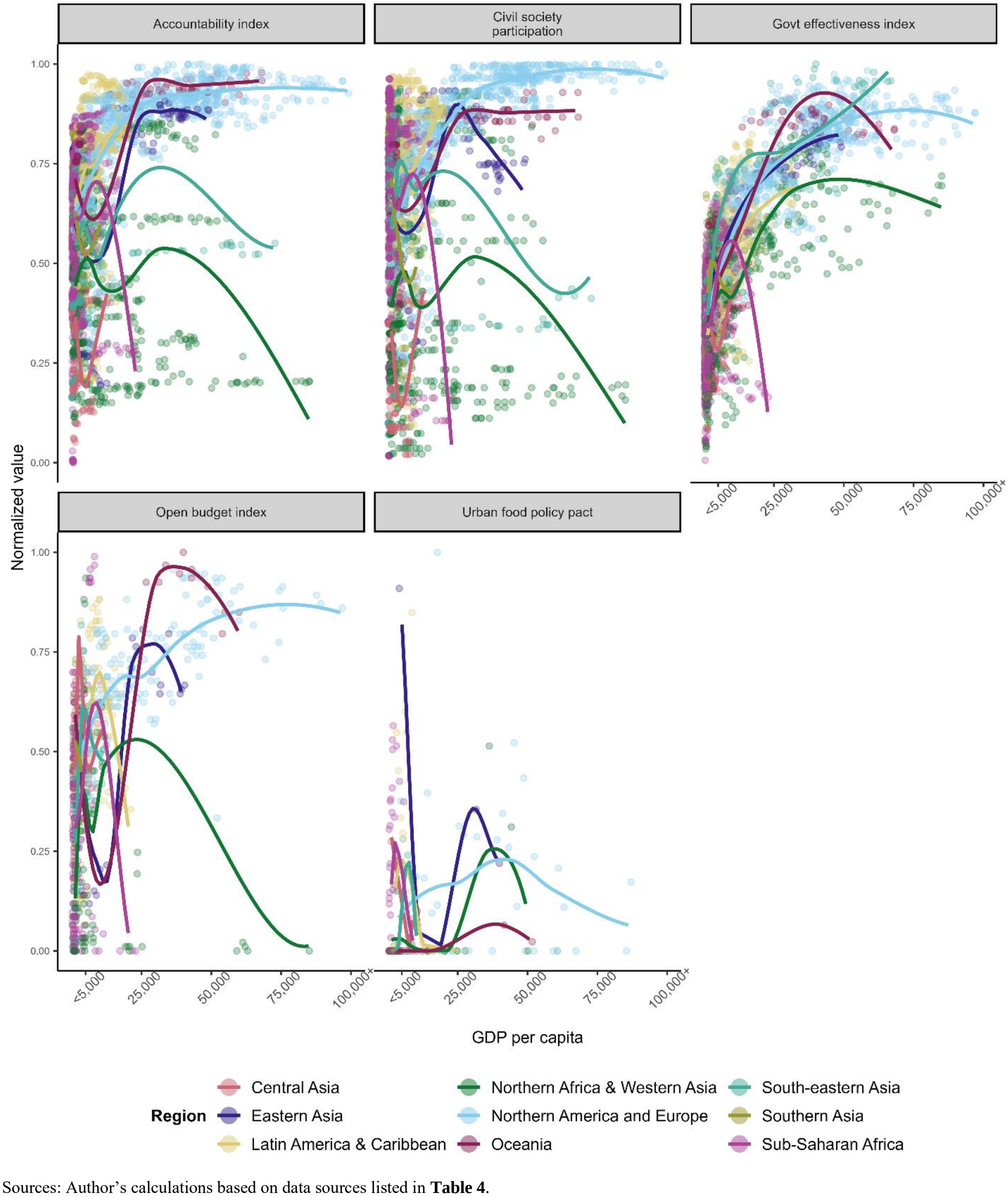

Sources: Author's calculations based on data sources listed in **Table 4**.
Notes: Colored lines reflects the local polynomial regression by region (Loess). Only continuous indicators shown. GDP per capita is in nominal terms and data come from 2021 for most countries, see **SM-E** for the specific year of data for GDP and each variable. Normalized values are calculated using max-min normalization relative to the global weighted mean. Each dot represents a country data point. Countries with GDP>$100,000 not shown on figure but included in model.

## Figure A.16 Relation to GDP per capita: Resilience

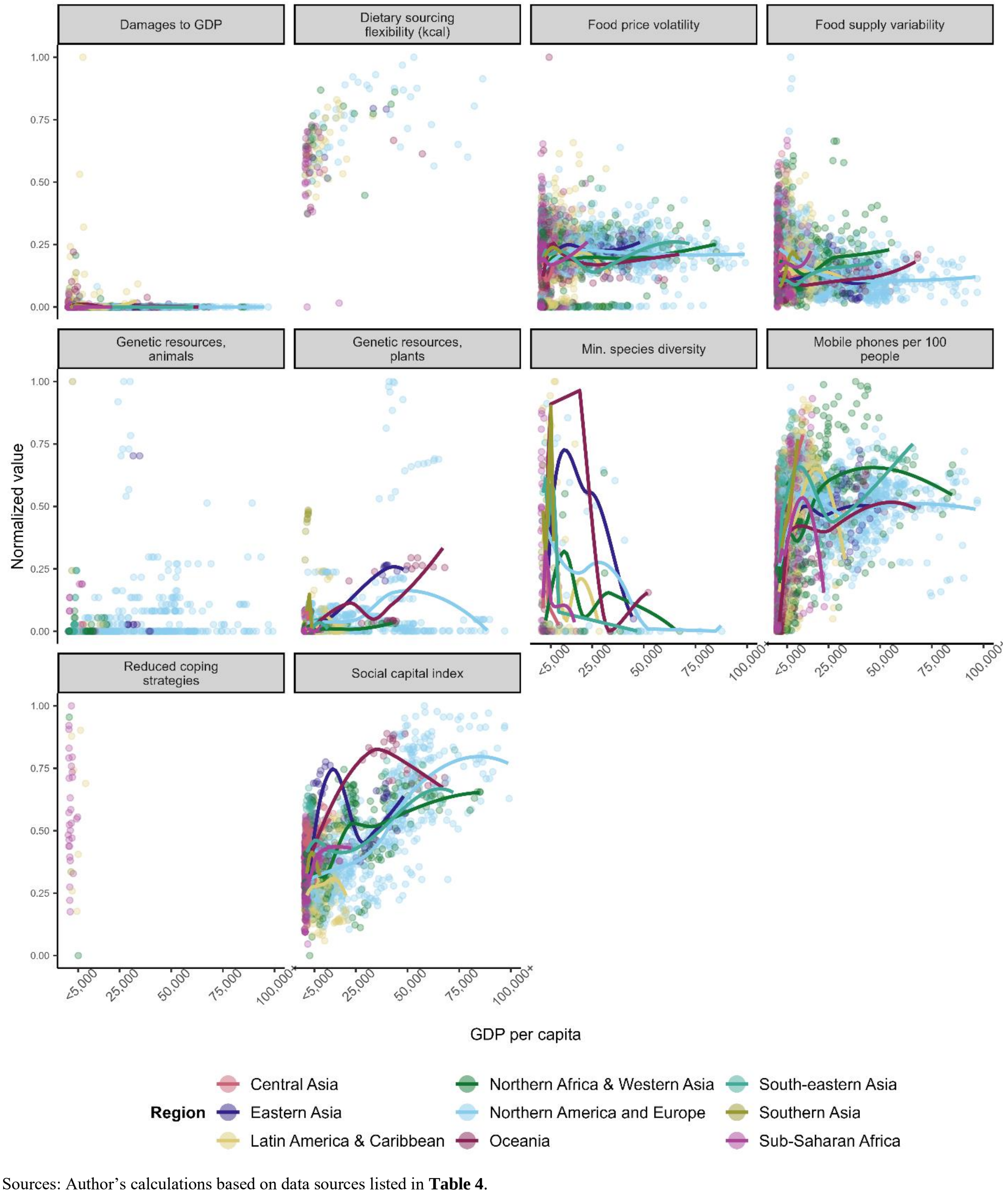

Sources: Author's calculations based on data sources listed in **Table 4**.
Notes: Colored lines reflects the local polynomial regression by region (Loess). Regression lines shown only for indicators with at least three data points in all regions necessary to estimate the Loess model. GDP per capita is in nominal terms and data come from 2021 for most countries, see **SM-E** for the specific year of data for GDP and each variable. Normalized values are calculated using max-min normalization relative to the global weighted mean. Each dot represents a country data point. Countries with GDP>$100,000 not shown on figure but included in model.

## Supplementary Analysis: Figures S1.1 to S5.16, visualization(s) for every indicator, by theme

All sources: Author’s calculations based on data sources listed in **Table 4**.

### S1.1 Distribution of the cost of the least-cost healthy diet ($PPP/capita/day) by region, 2017-2020

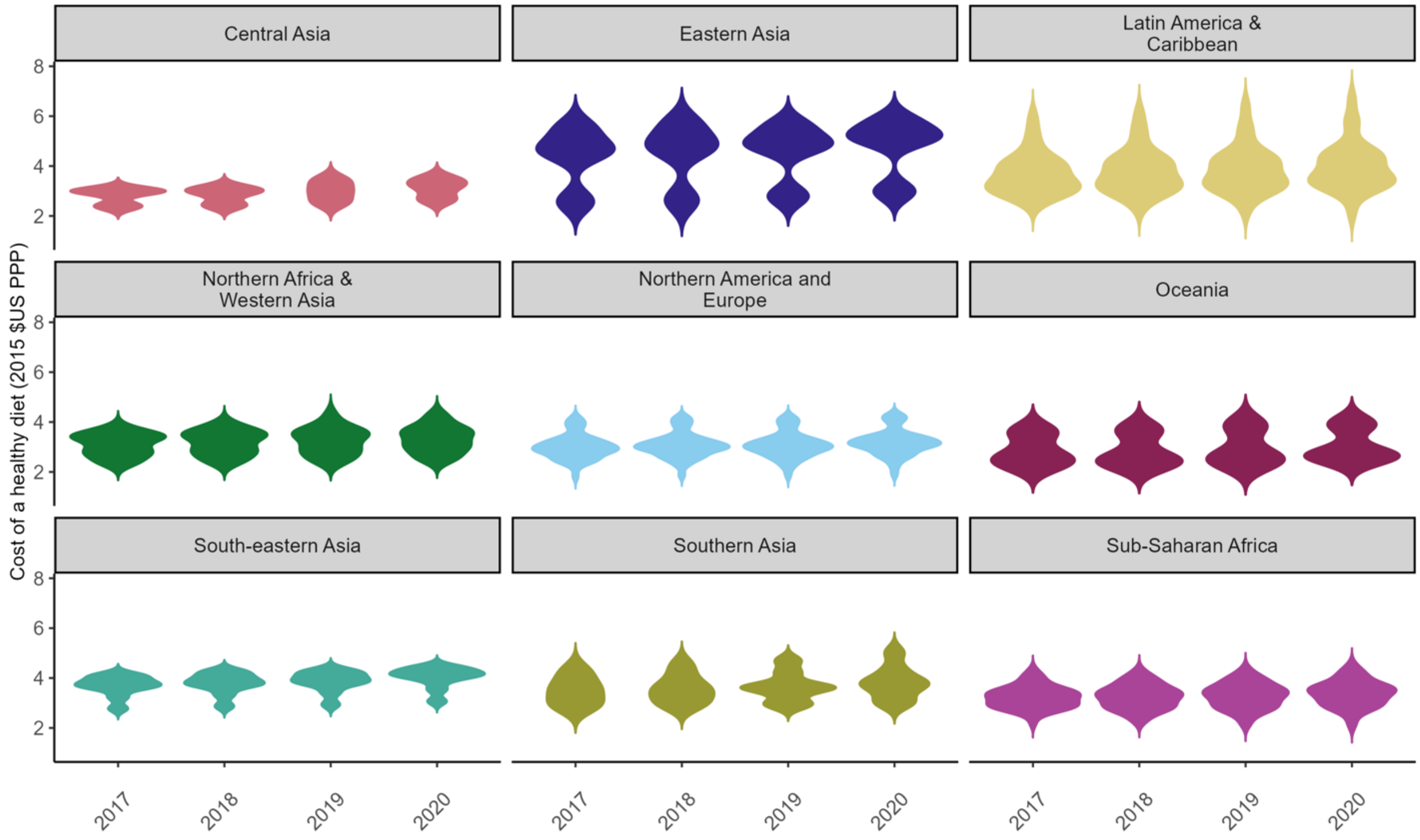

Violin plots reflect the distribution of the observations where the curved line and colored fill illustrate the probability distribution of the data.

## S1.2 Distribution of the cost of the least-cost healthy diet ($PPP/capita/day) by country income groups, 2017-2020

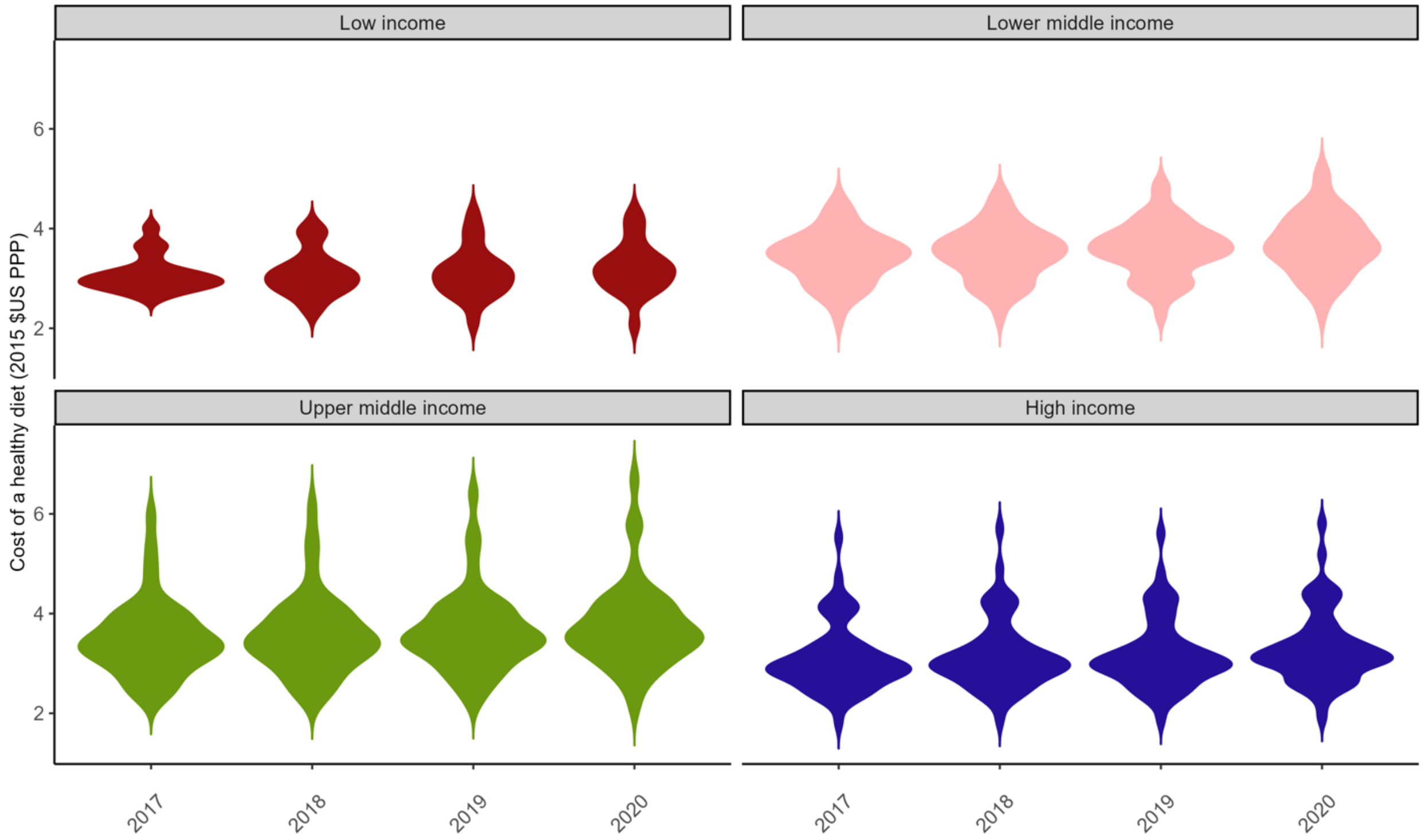

## S1.3 Change in fruit and vegetable availability, 2010-2019, by region

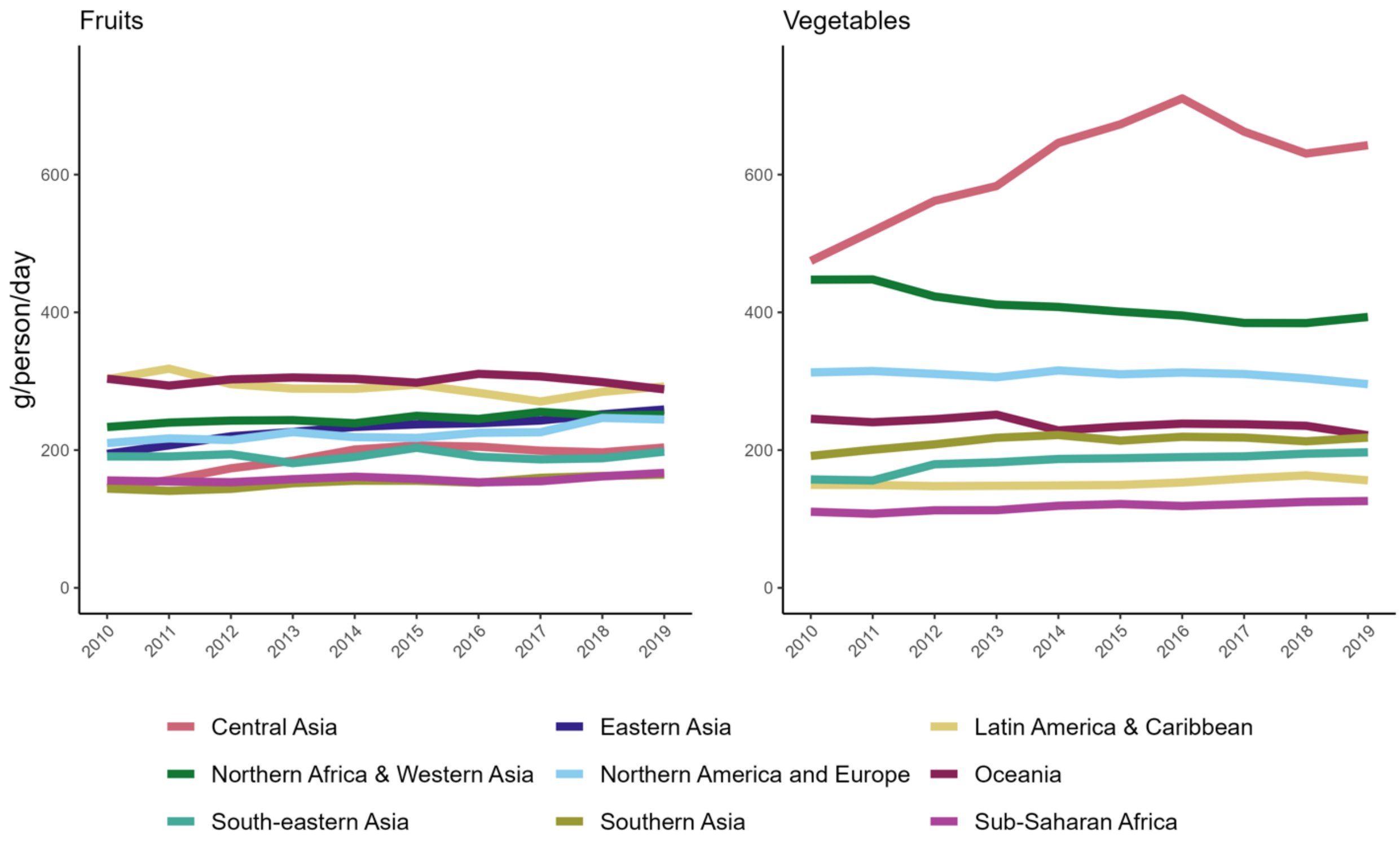

Unweighted regional means.

## S1.4 Change in fruit and vegetable availability, 2010-2019, by income group

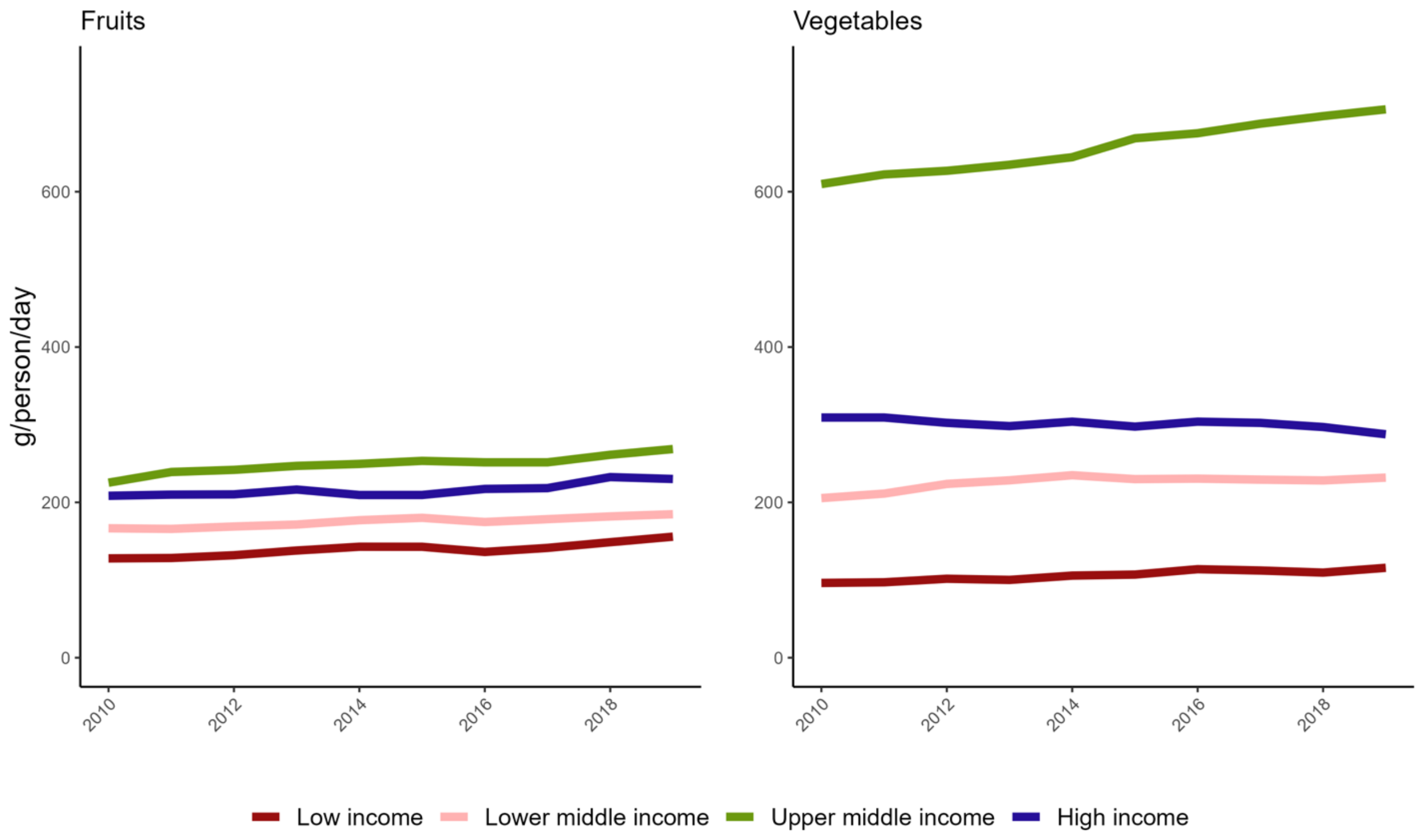

Unweighted income group means.

## S1.5 Retail value of Ultra-Processed Foods per capita, 2017-2019, by region

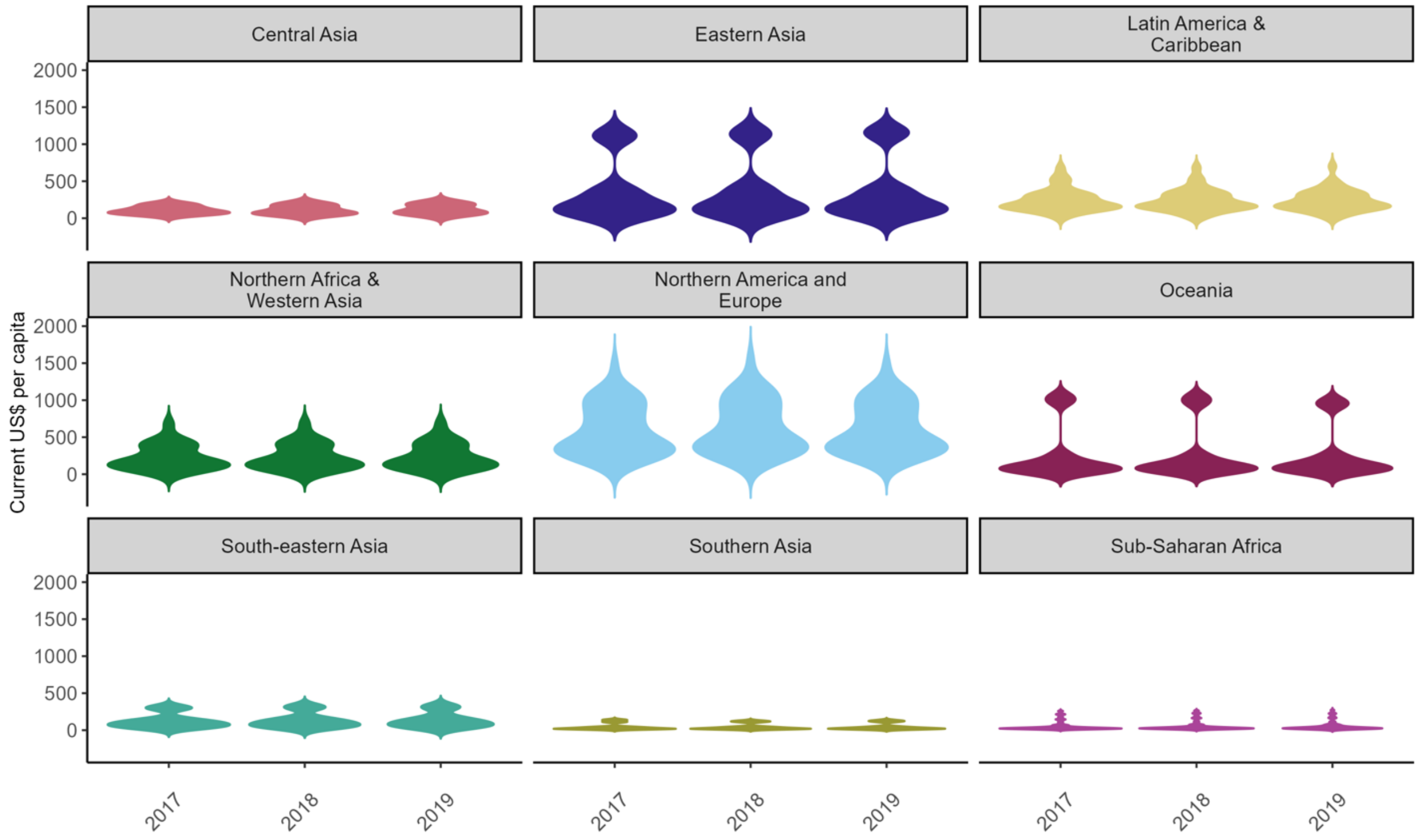

## S1.6 Retail value of Ultra-Processed Foods per capita, 2019

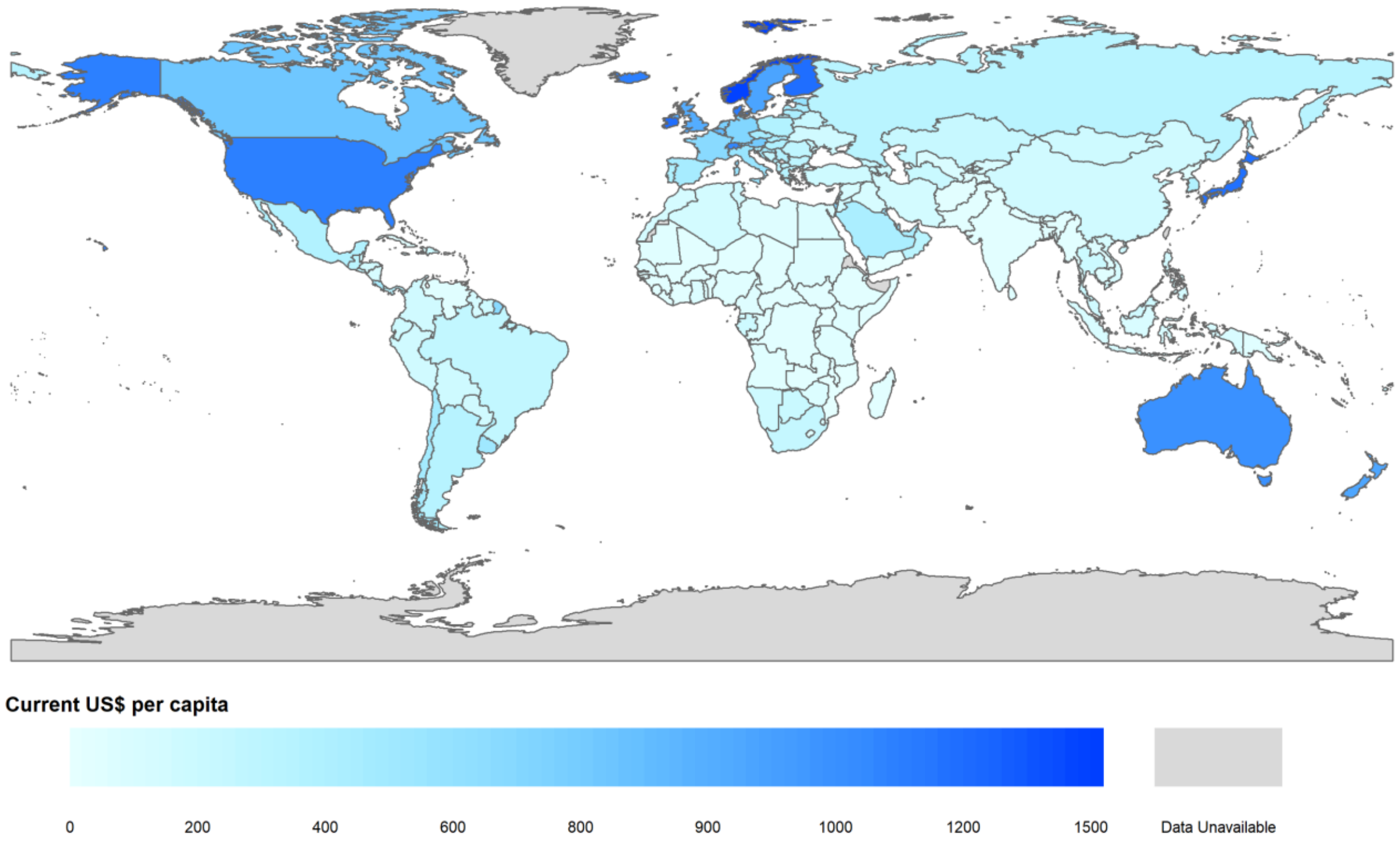

**S1.7 Percent population using safely managed drinking water services (SDG 6.1.1), 2000-2020, by region**

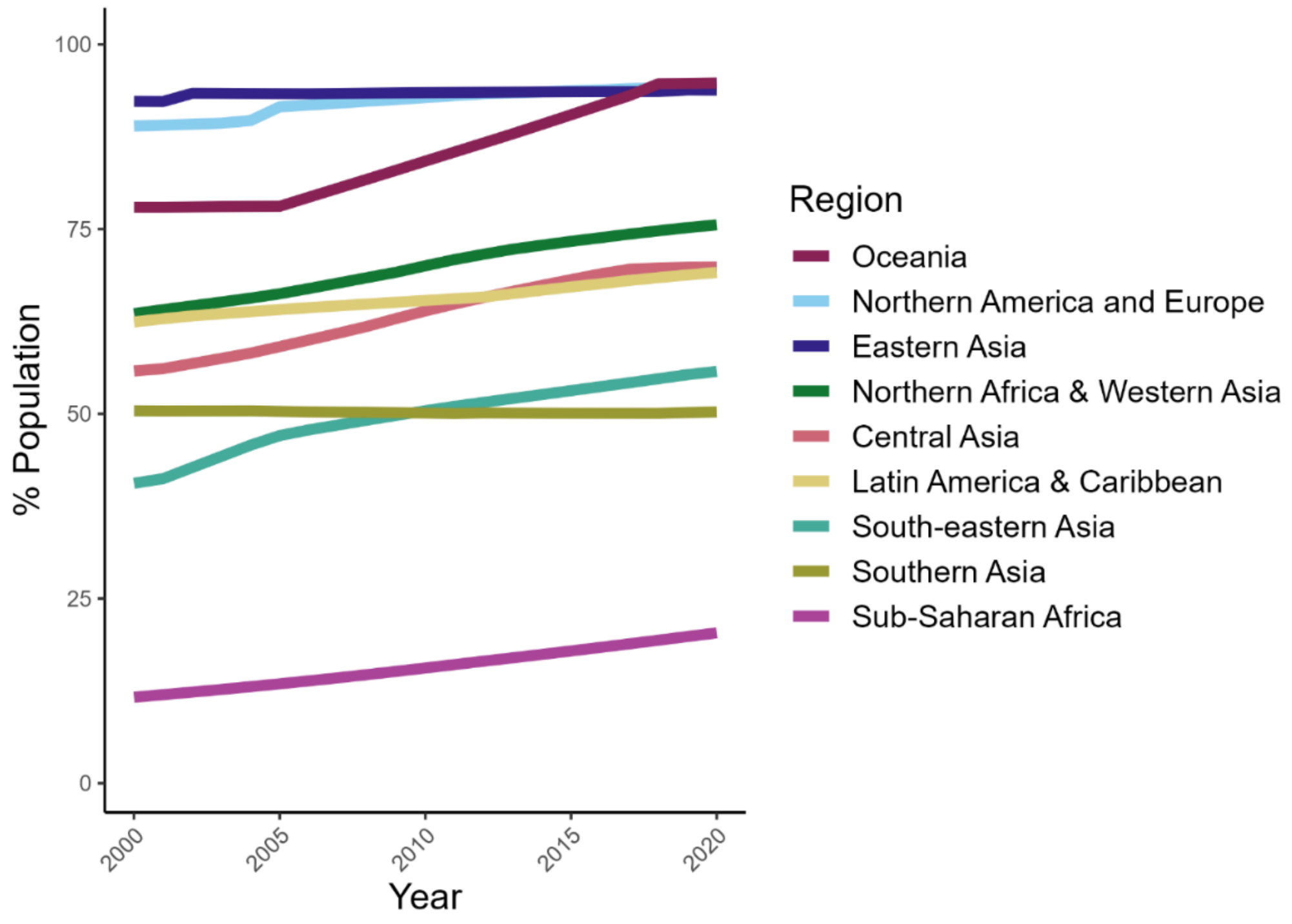

Population-weighted regional means.

**S1.8 Percent population using safely managed drinking water services (SDG 6.1.1), 2000-2020, by income group**

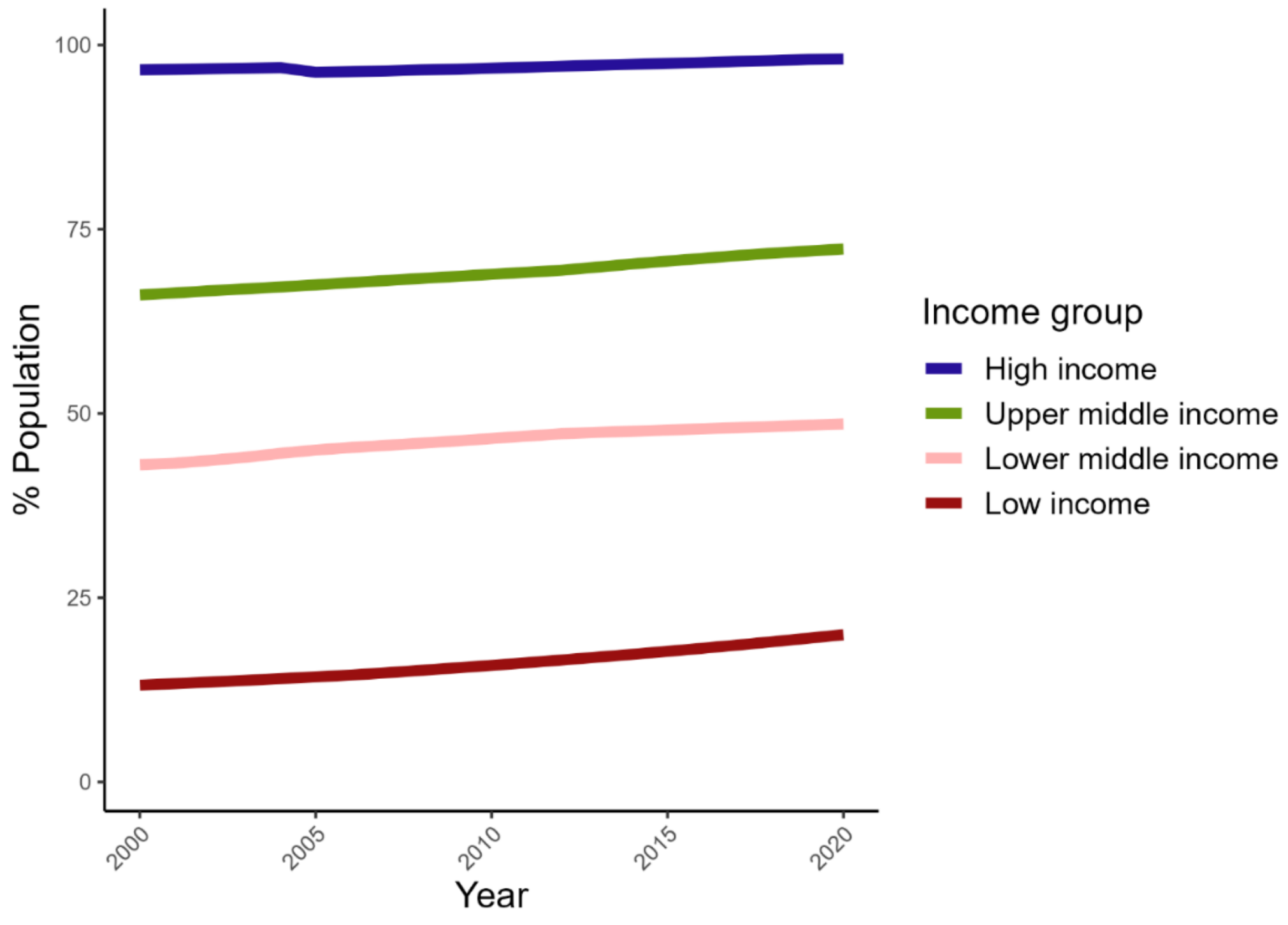

Population-weighted income group means.

## S1.9 Prevalence of Undernourishment, 2001-2020, by region

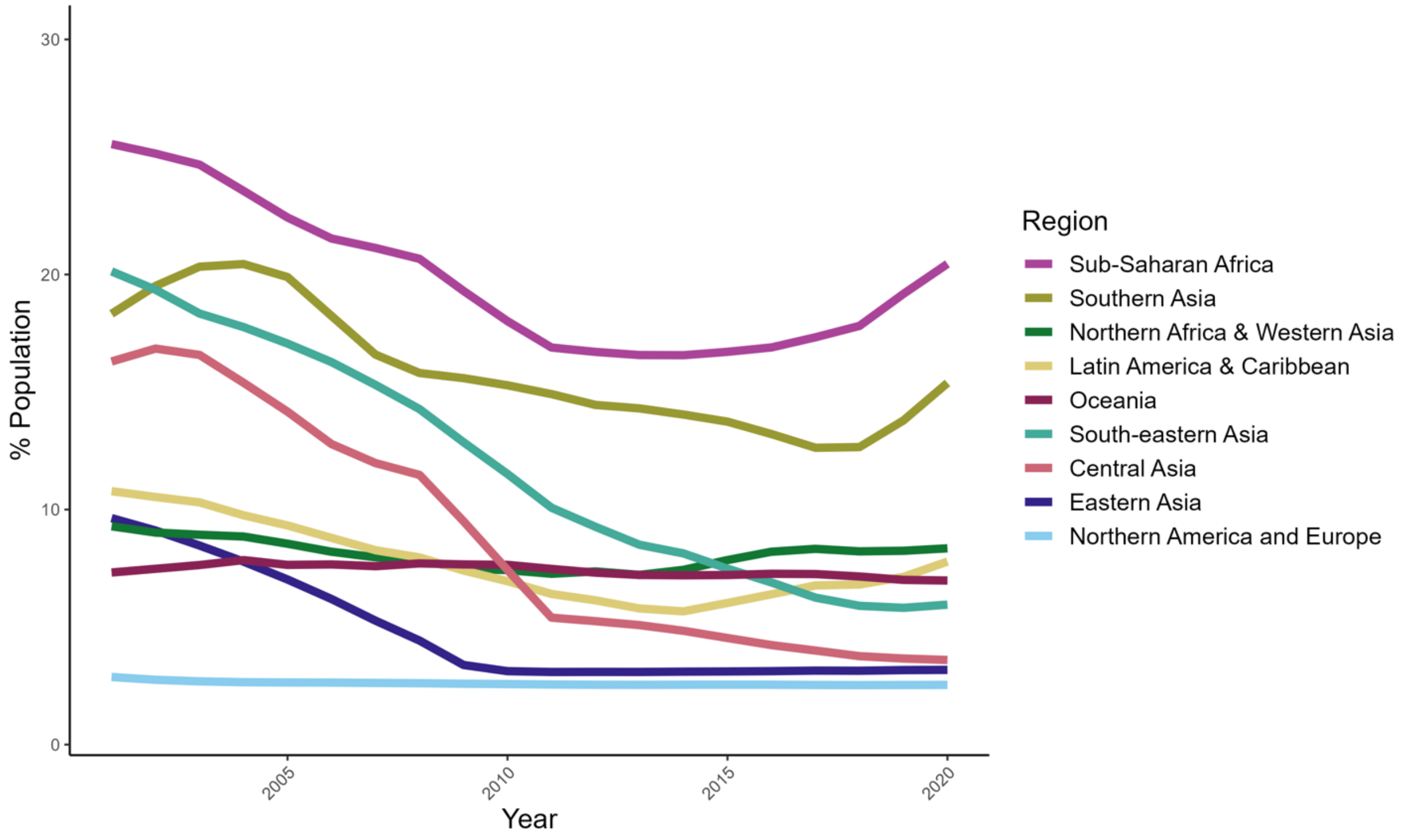

Annual value taken at the midpoint year of 3-year average. Population-weighted regional mean.

## S1.10 Prevalence of Undernourishment, 2001-2020, by income group

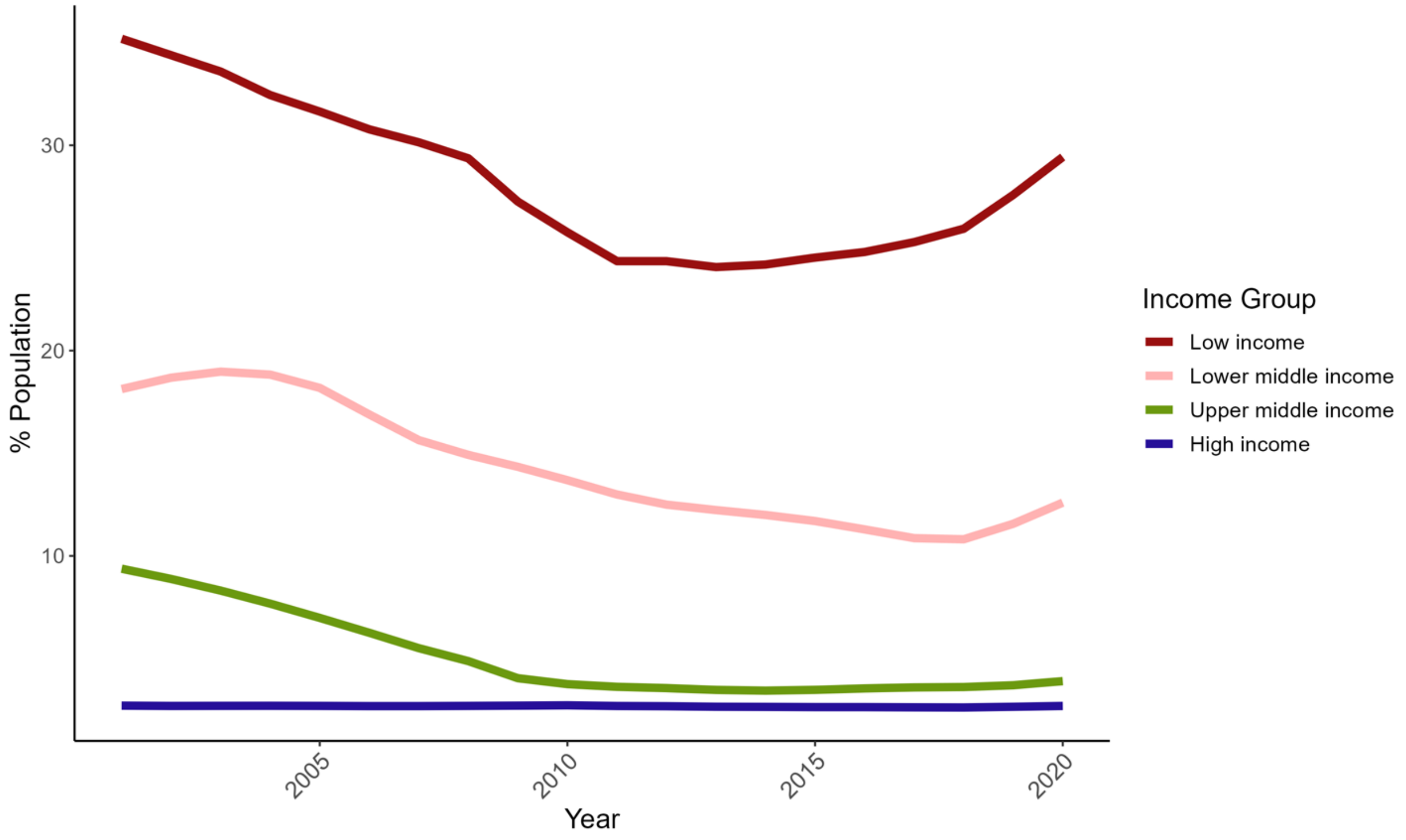

Annual value taken at the midpoint year of 3-year average. Population-weighted income group mean.

## S1.11 Percent population experiencing moderate or severe food insecurity, 2014-2020, by region

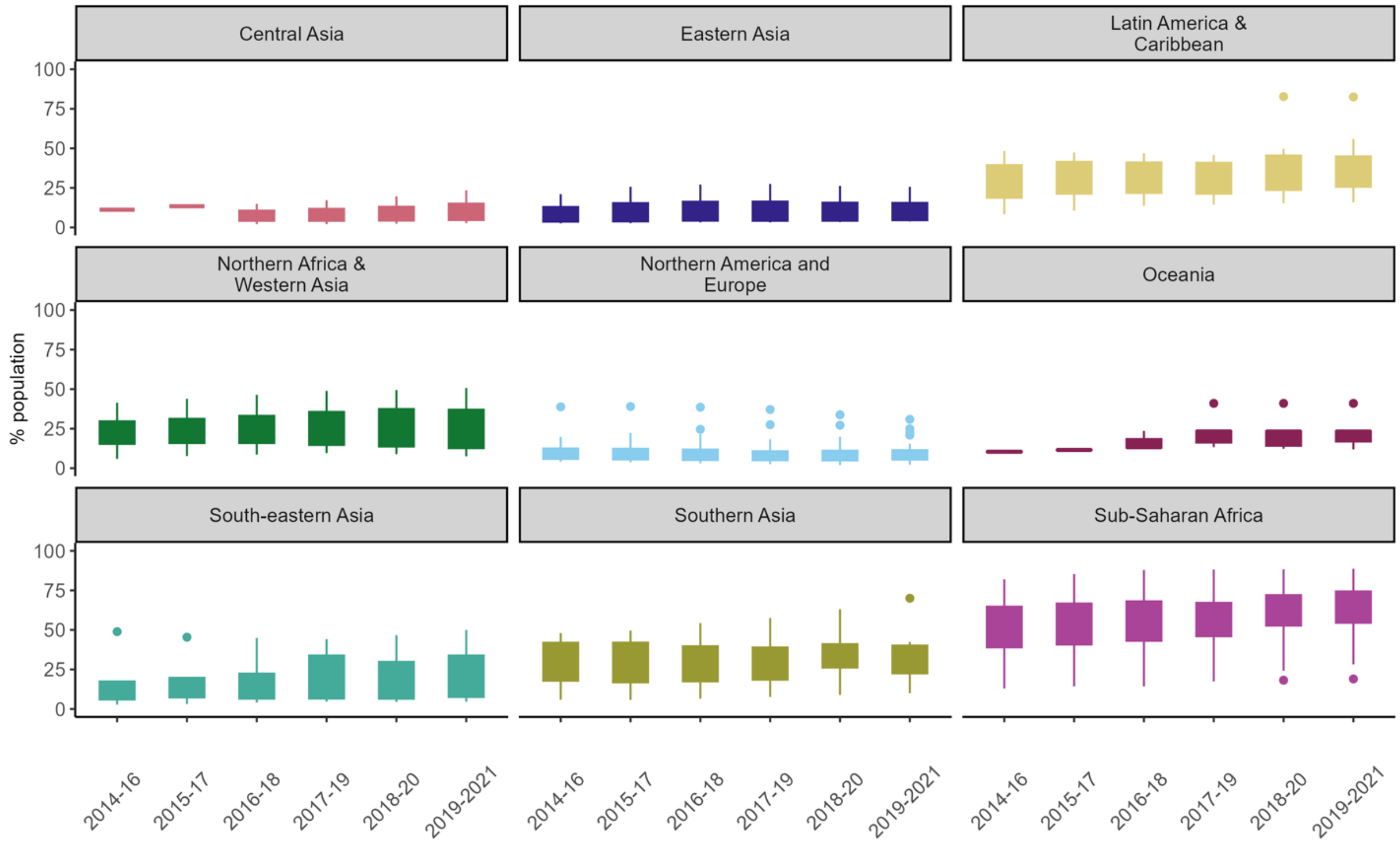

Annual value taken at the midpoint year of 3-year average.

## S1.12 Percent population experiencing moderate or severe food insecurity, 2014-2020, by income group

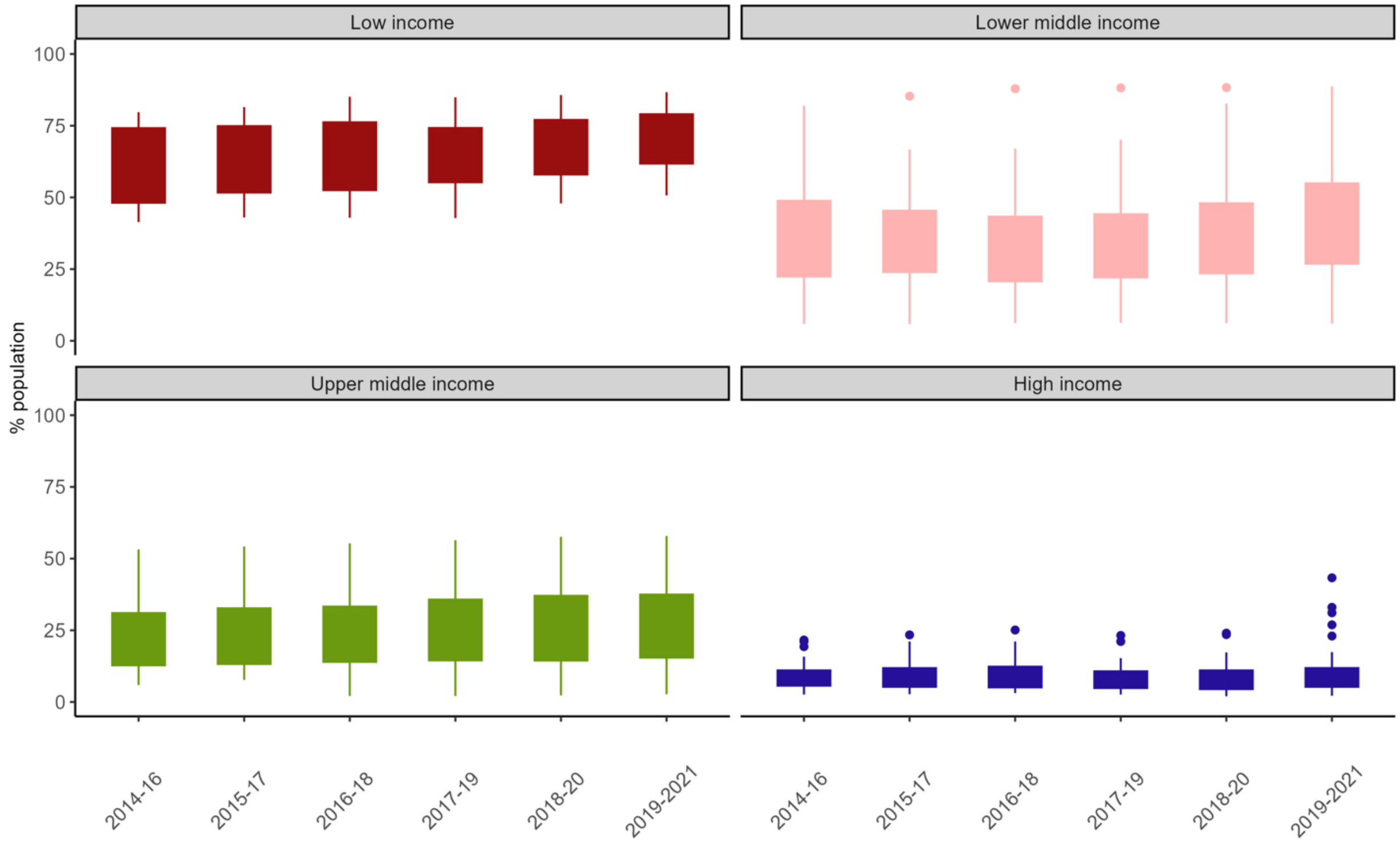

Annual value taken at the midpoint year of 3-year average.

**S1.13 Percent population who cannot afford a healthy diet, 2017-2020, by region**

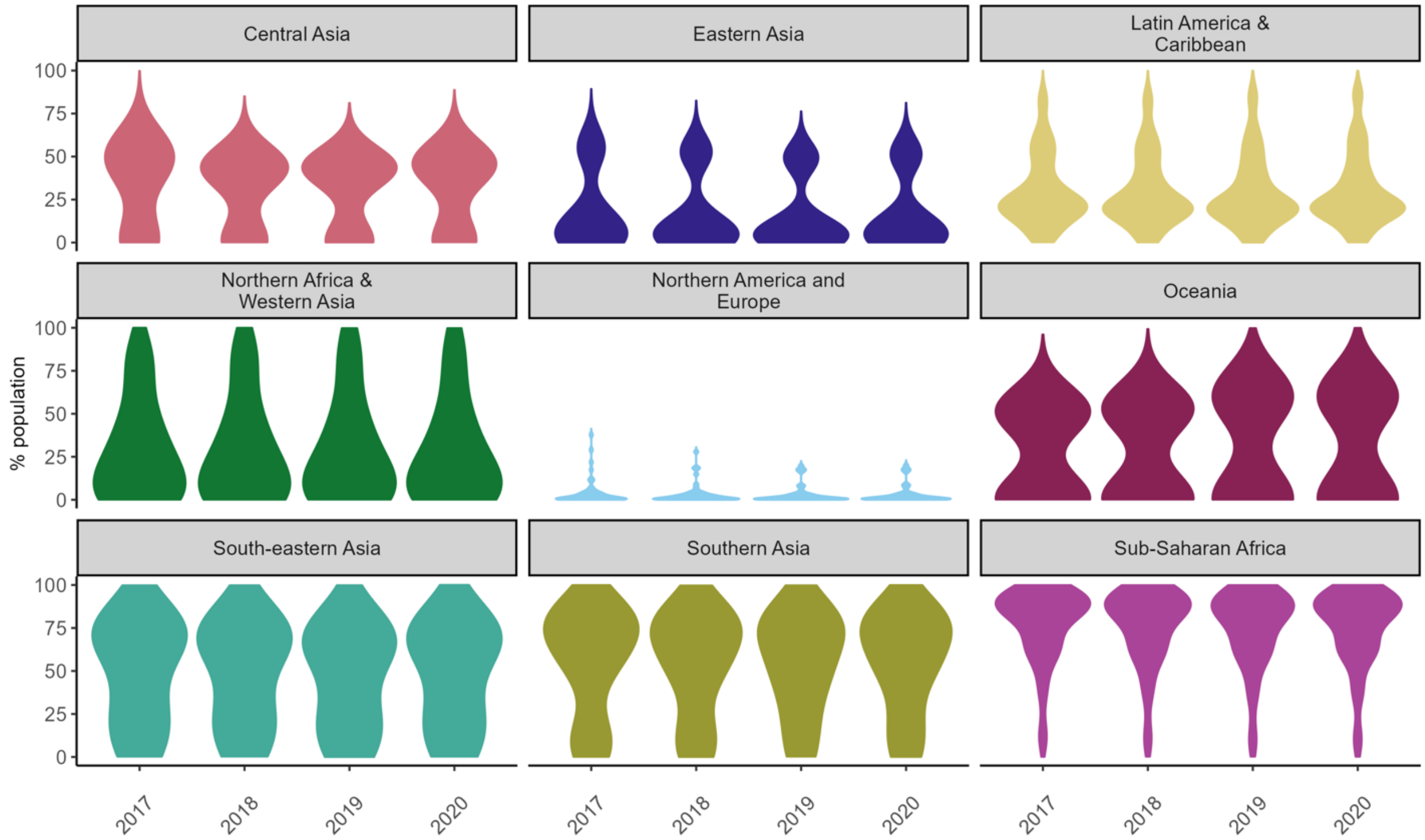

**S1.14 Percent population who cannot afford a healthy diet, 2017-2020, by income group**

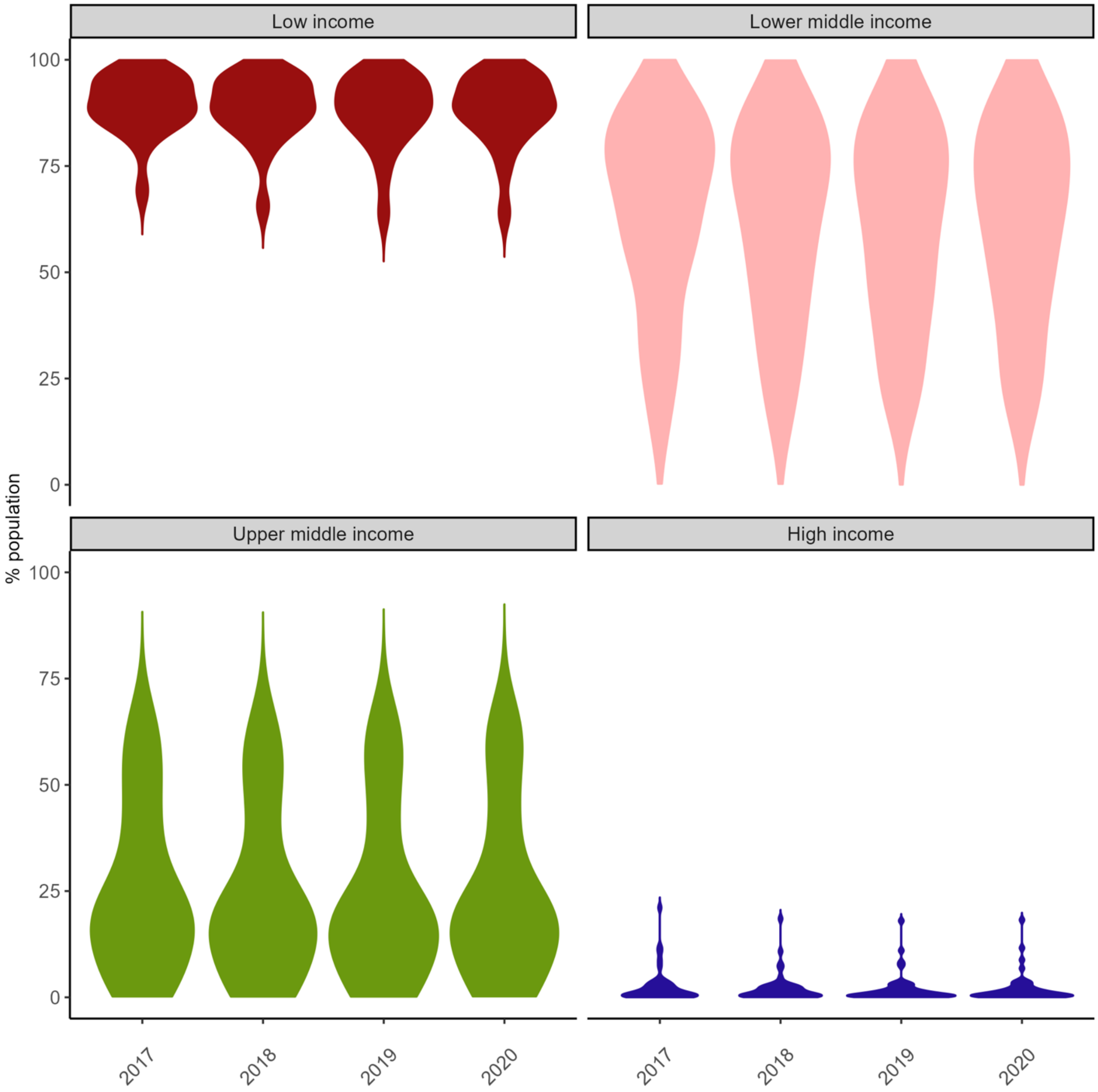

## S1.15 Percent adult women (15-49 y) meeting minimum dietary diversity, 2021

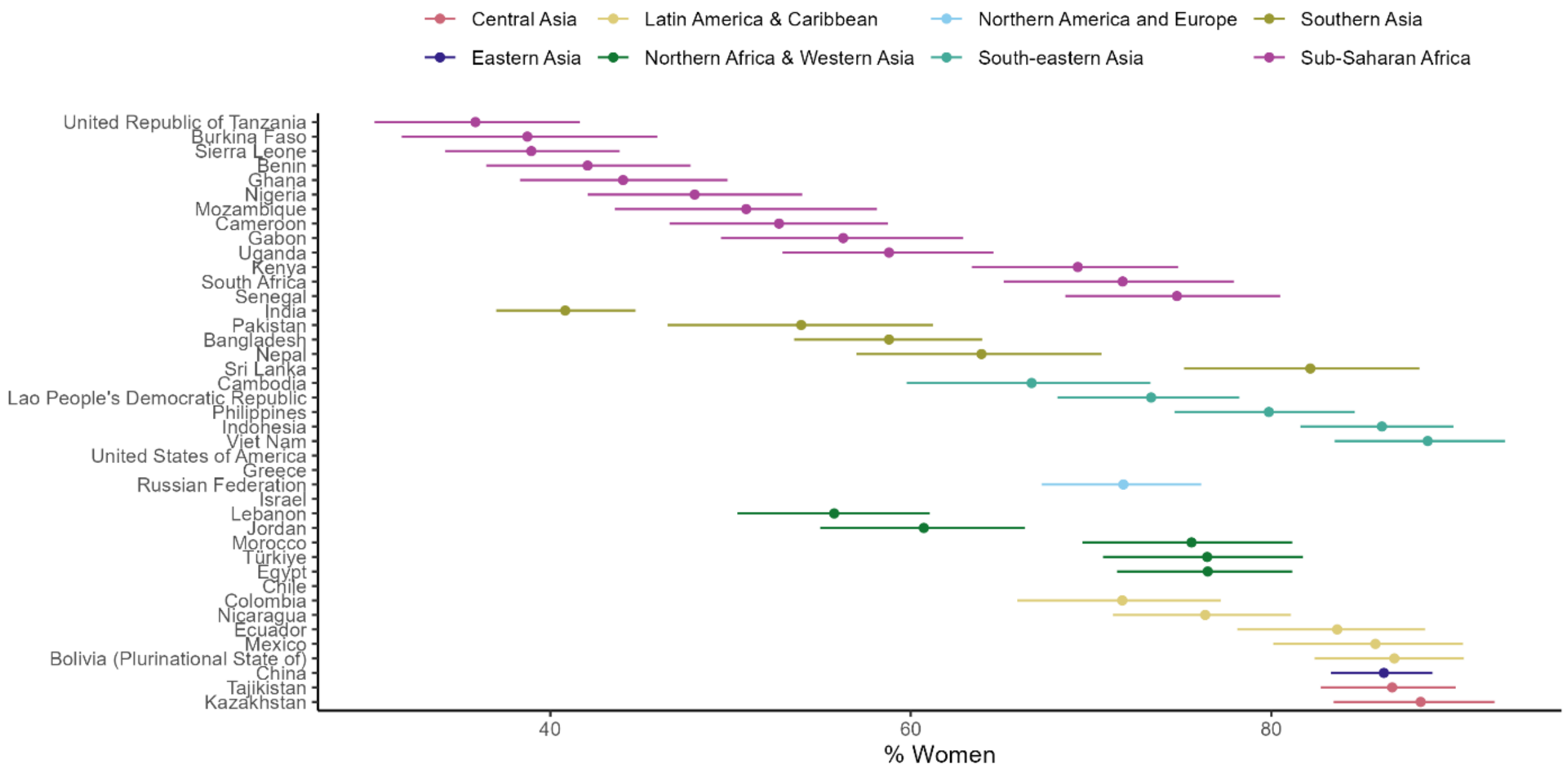

Legend bar shows the 95% confidence interval around the mean point estimate.

## S1.16 Percent of children 6-23 months meeting minimum dietary diversity, by sex

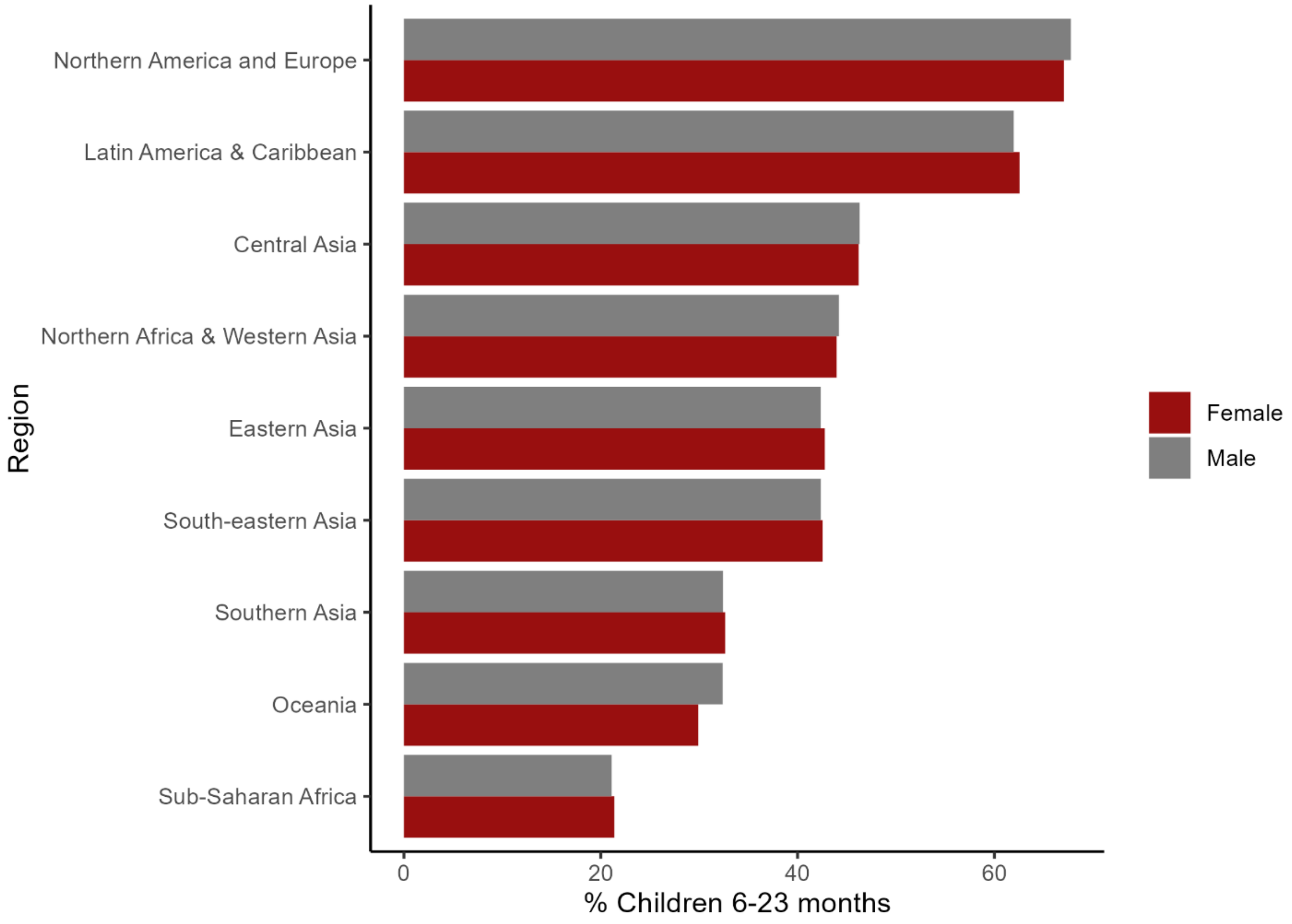

Data source year differs by country. See table A1.2.

### S1.17 Percent of children 6-23 months meeting minimum dietary diversity, by urban/rural

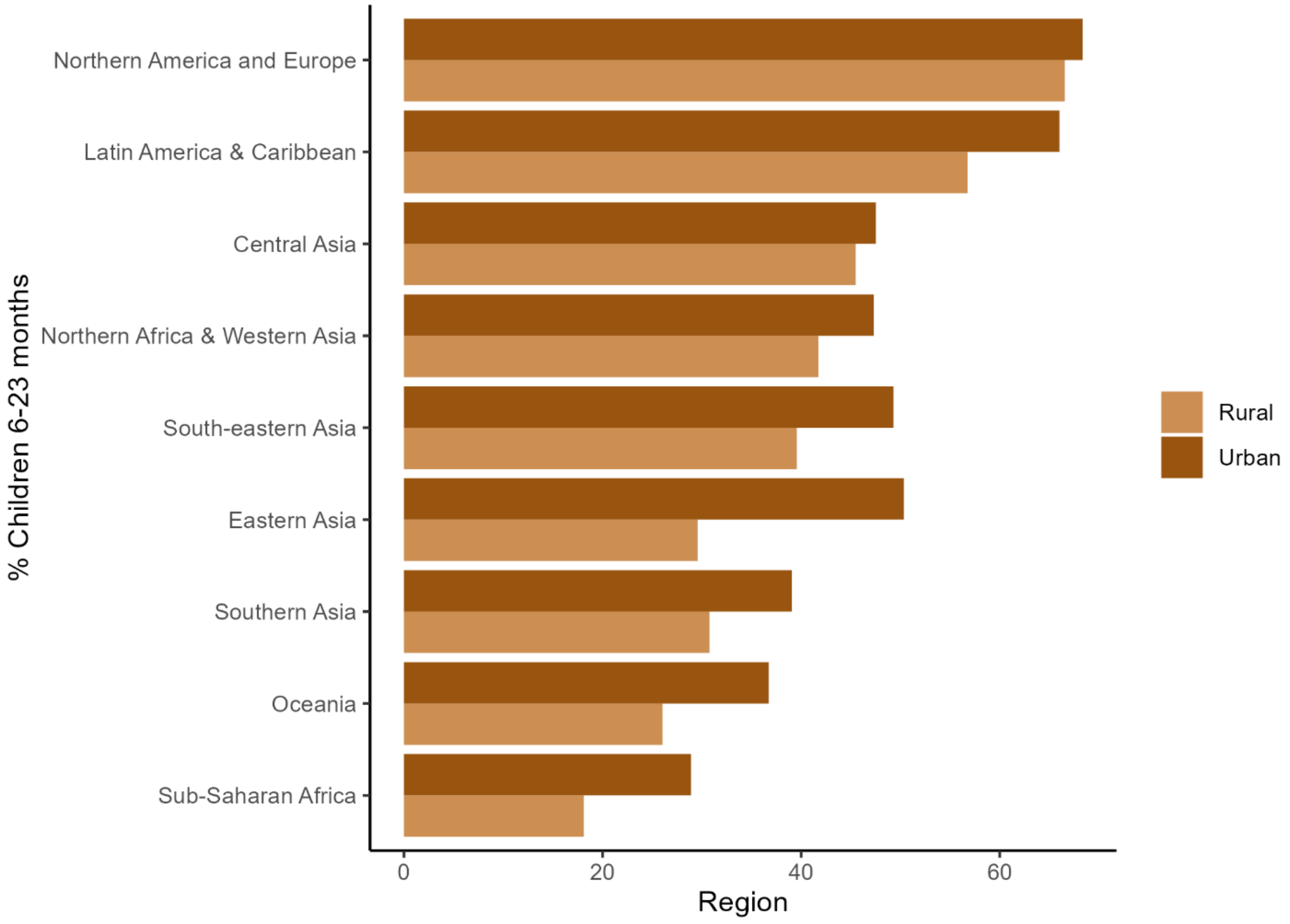

Data source year differs by country. See table A1.2.

## S1.18 Percent adult population (≥15 y) consuming all 5 food groups, 2021

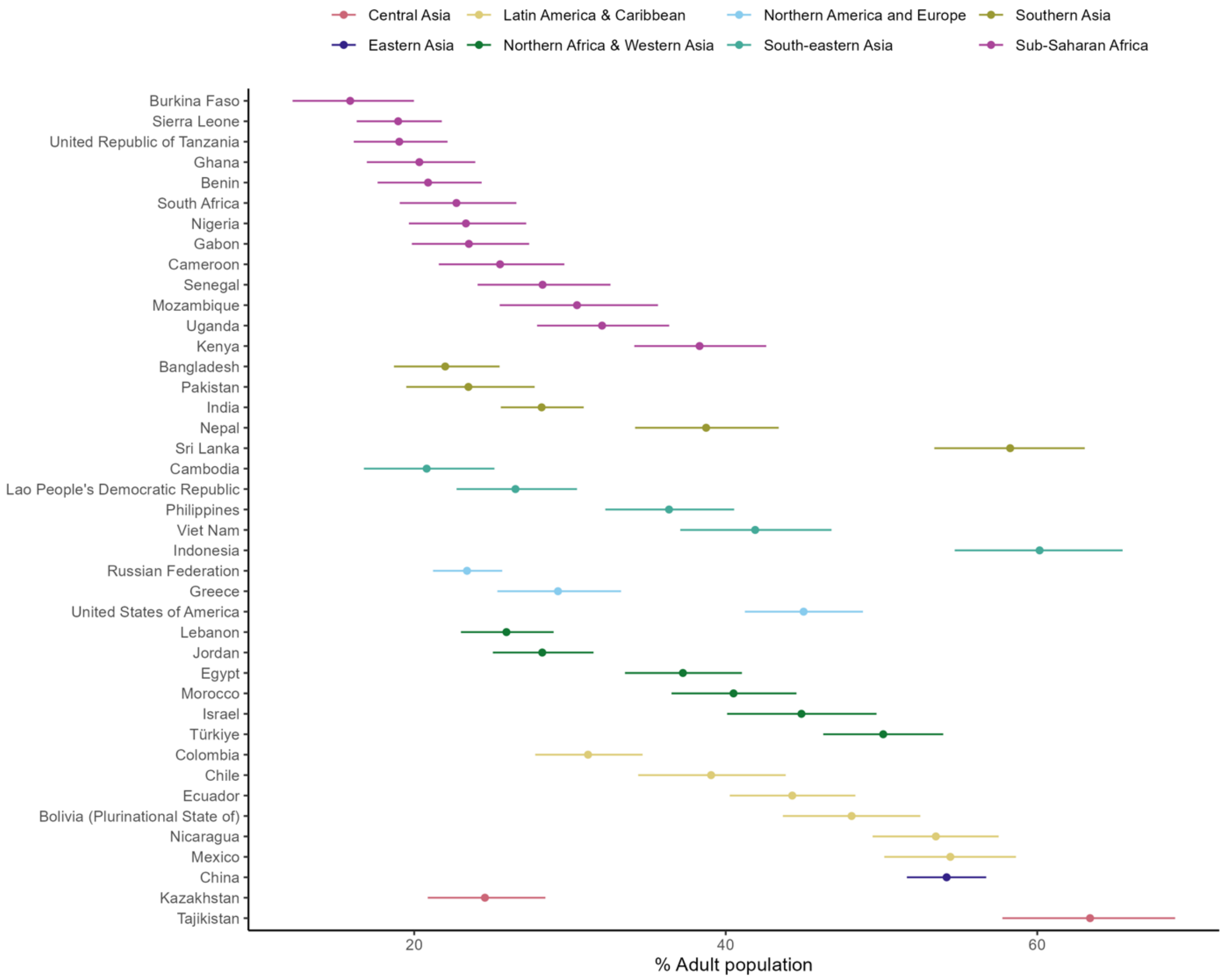

Legend bar shows the 95% confidence interval around the mean point estimate.

## S1.19 Percent adult population (≥15 y) with zero fruits and vegetables consumption, 2021

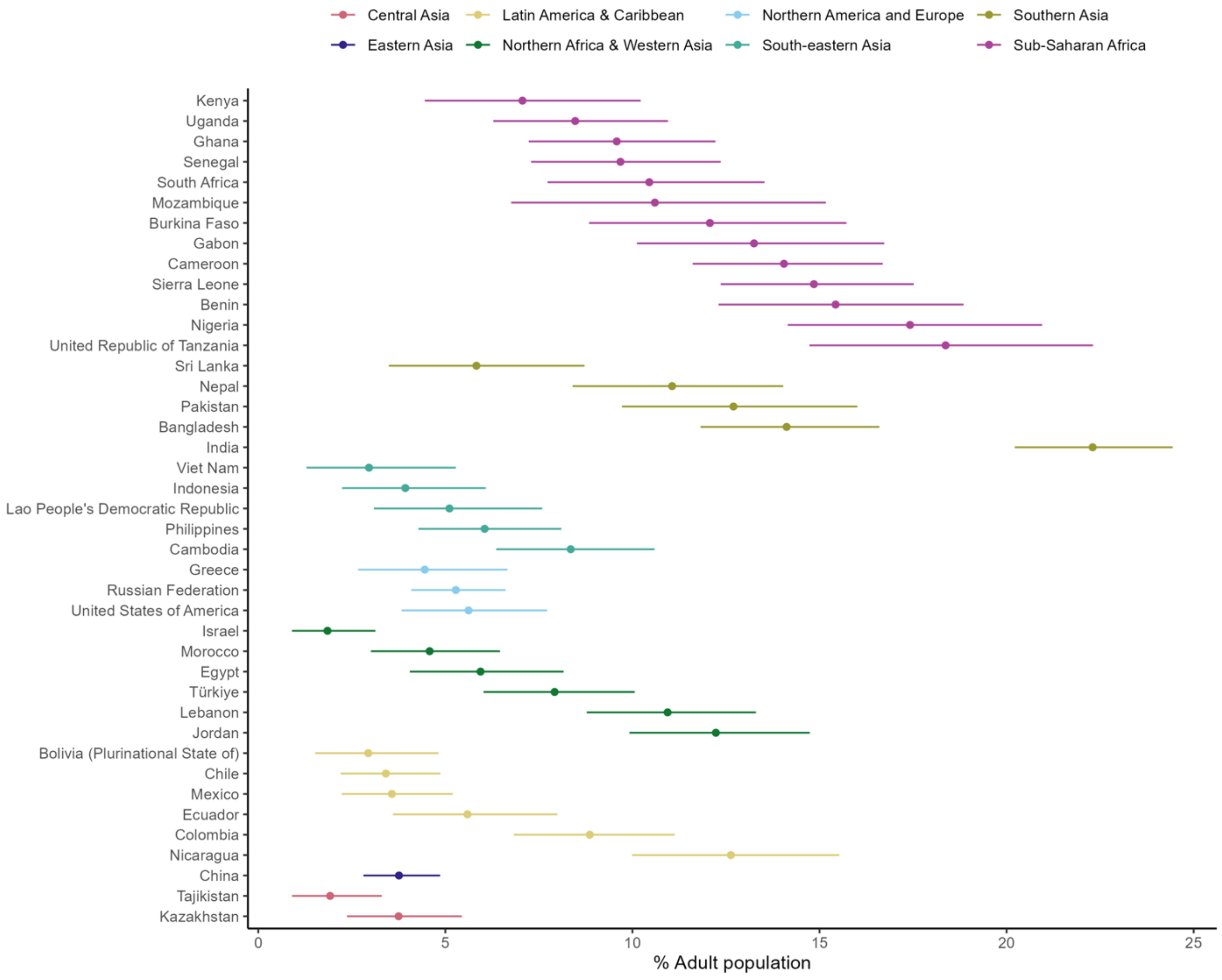

Legend bar shows the 95% confidence interval around the mean point estimate.

**S1.20 Prevalence of zero fruits and vegetables consumption, children 6-23 months, by sex**

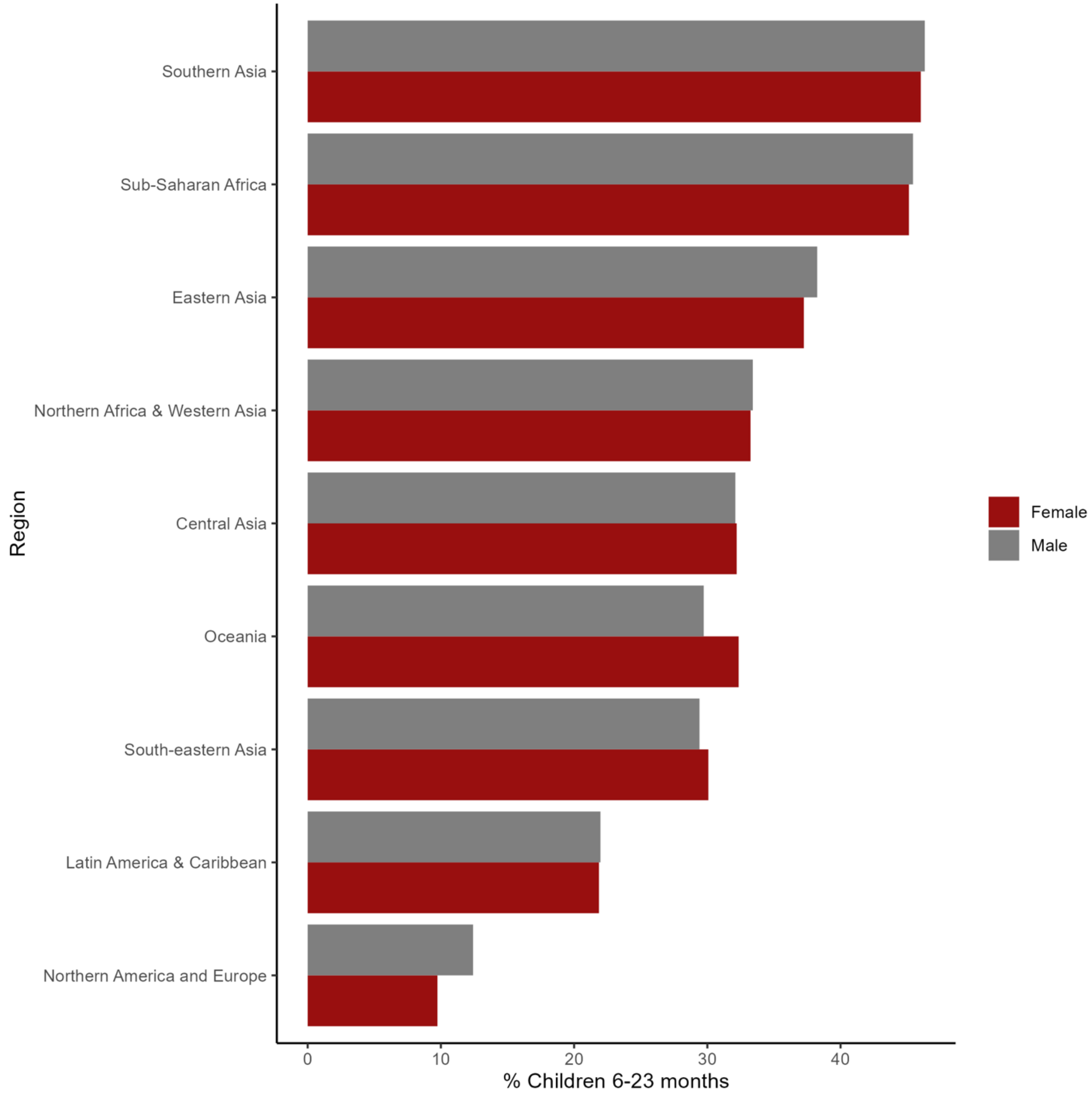

Data source year differs by country. See table A1.2.

**S1.21 Prevalence of zero fruits and vegetables consumption, children 6-23 months, by urban/rural**

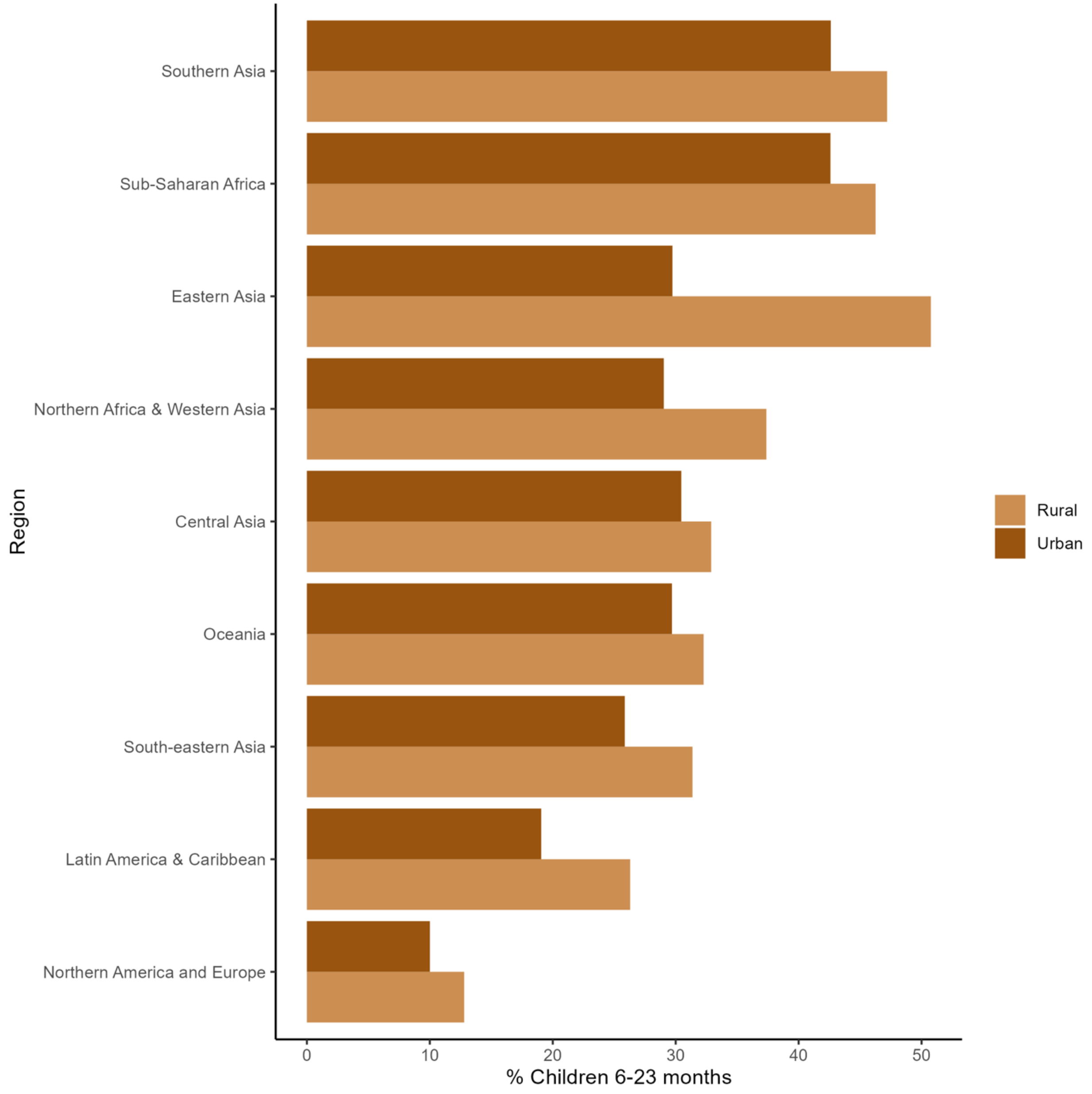

Data source year differs by country. See table A1.2.

## S1.22 NCD-Protect, adults (≥15 y), 2021

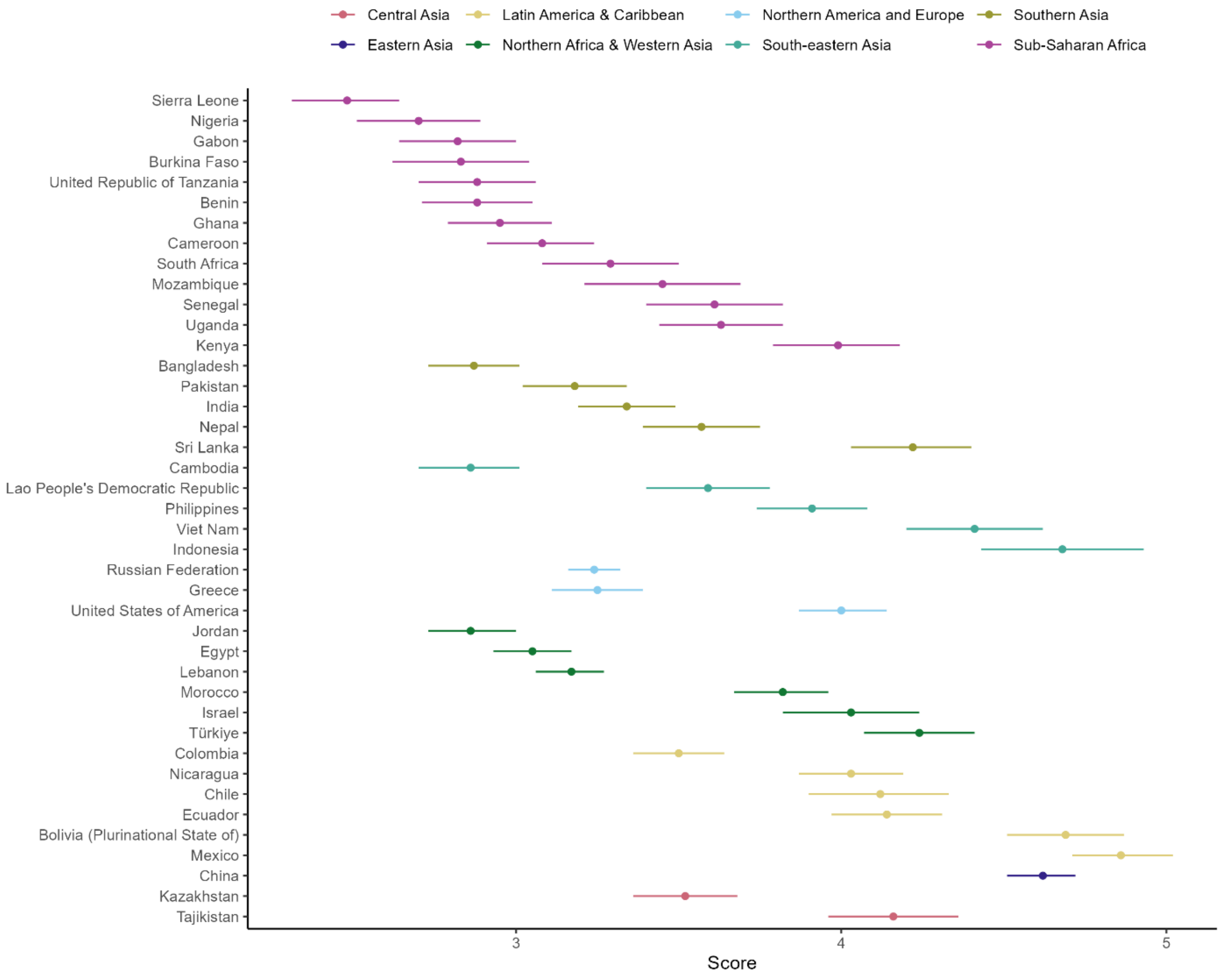

Legend bar shows the 95% confidence interval around the mean point estimate.

## S1.23 NCD-Risk, adults (≥15 y), 2021

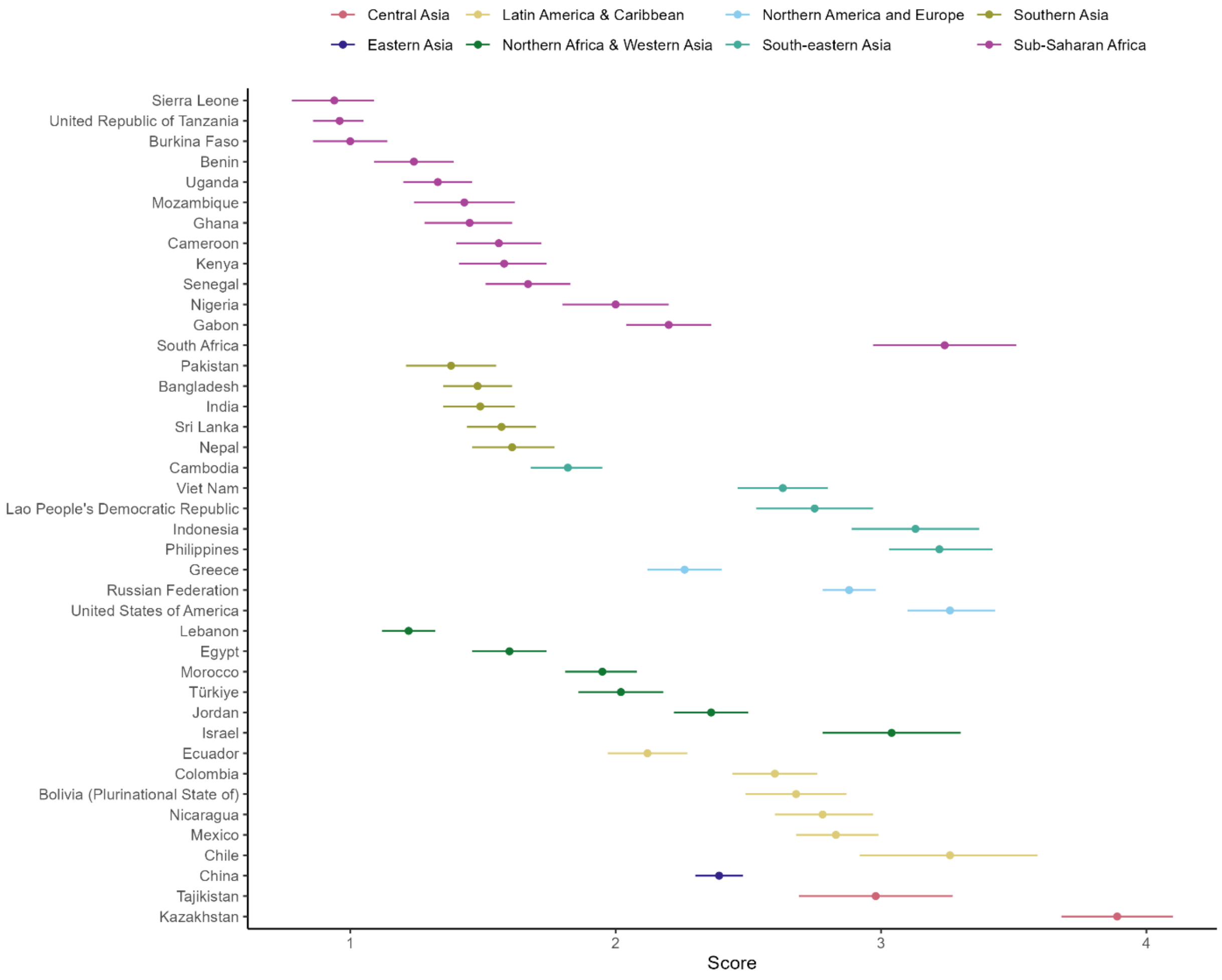

Legend bar shows the 95% confidence interval around the mean point estimate.

## S1.24 Prevalence of sugar-sweetened soft drink consumption, adults (≥15 y), 2021

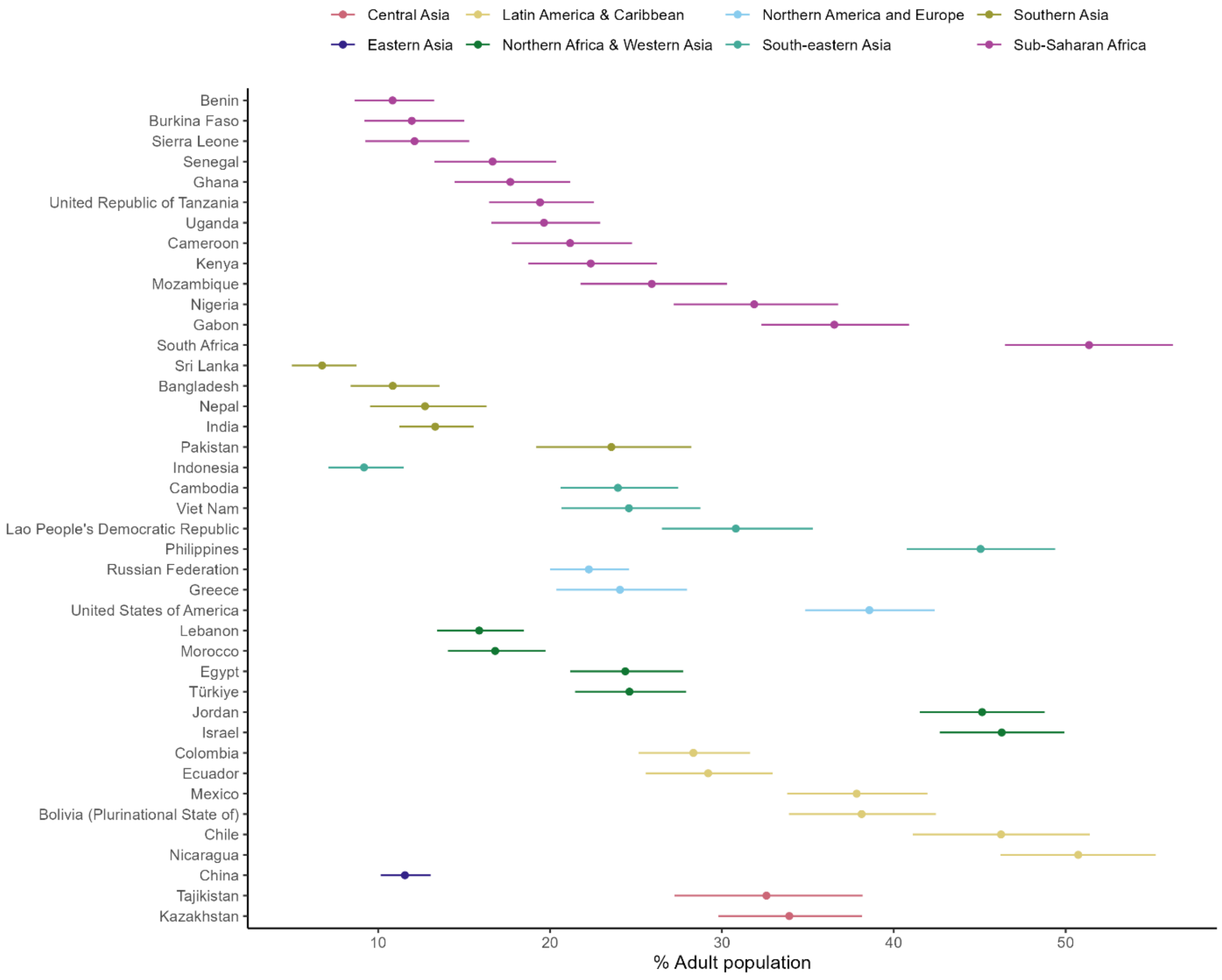

Legend bar shows the 95% confidence interval around the mean point estimate.

## S2.1 Total food system emissions, 2000-2020, by region

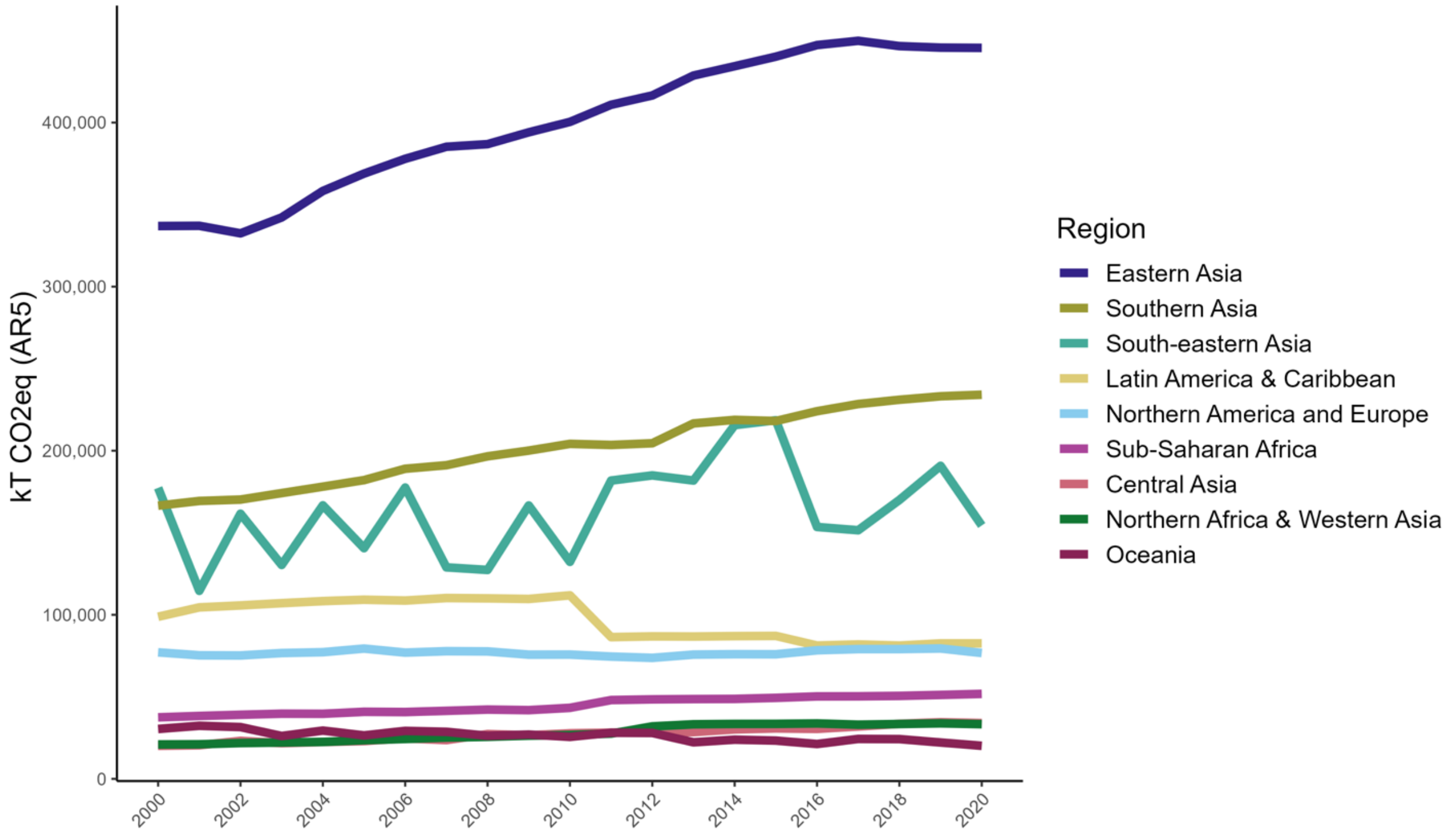

Unweighted mean by region.

## S2.2 Total food system emissions, 2020

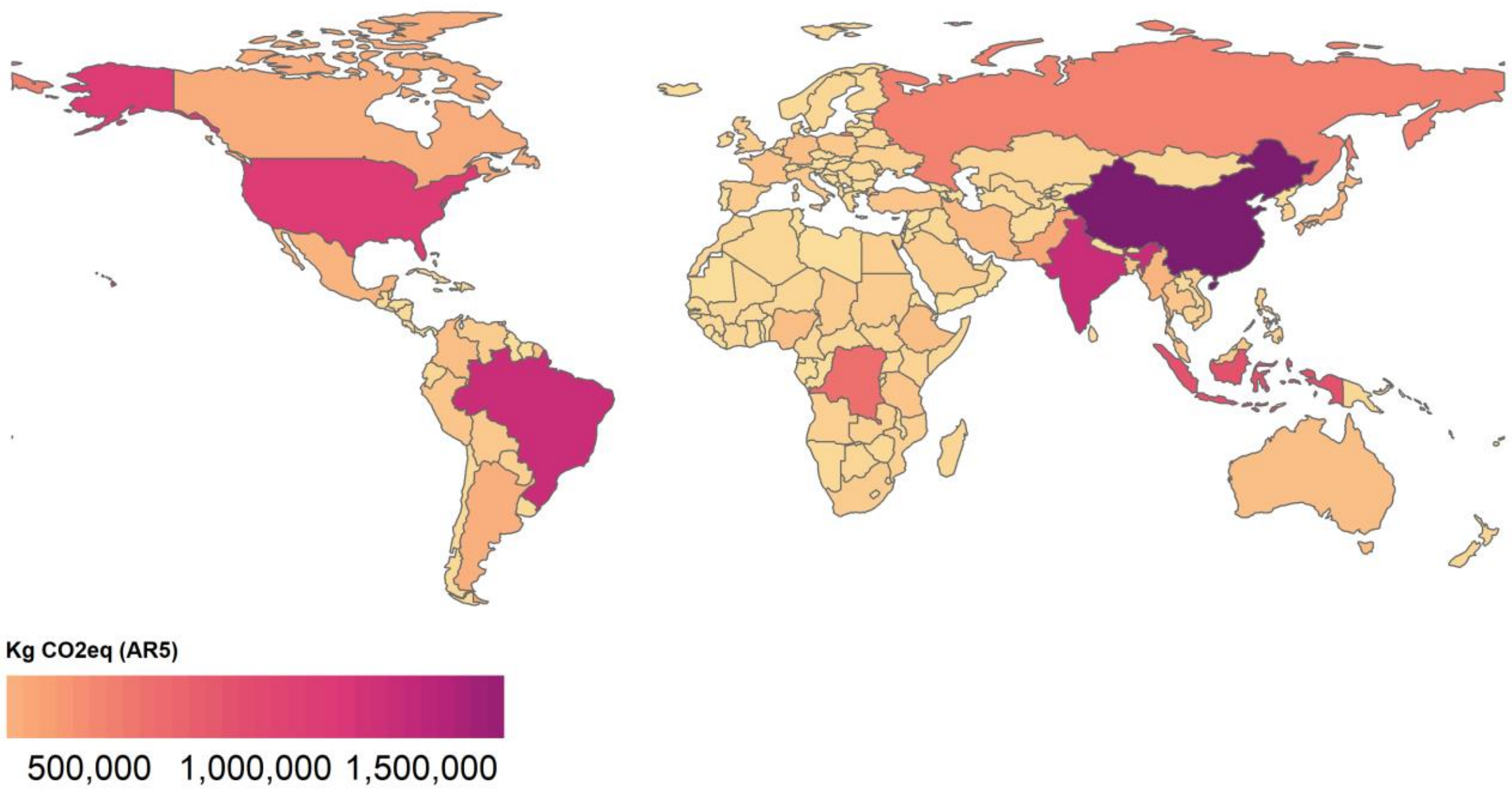

## S2.3 Emissions intensity (kg $CO_2$eq/kg product), Staple foods, 2000-2020, by region

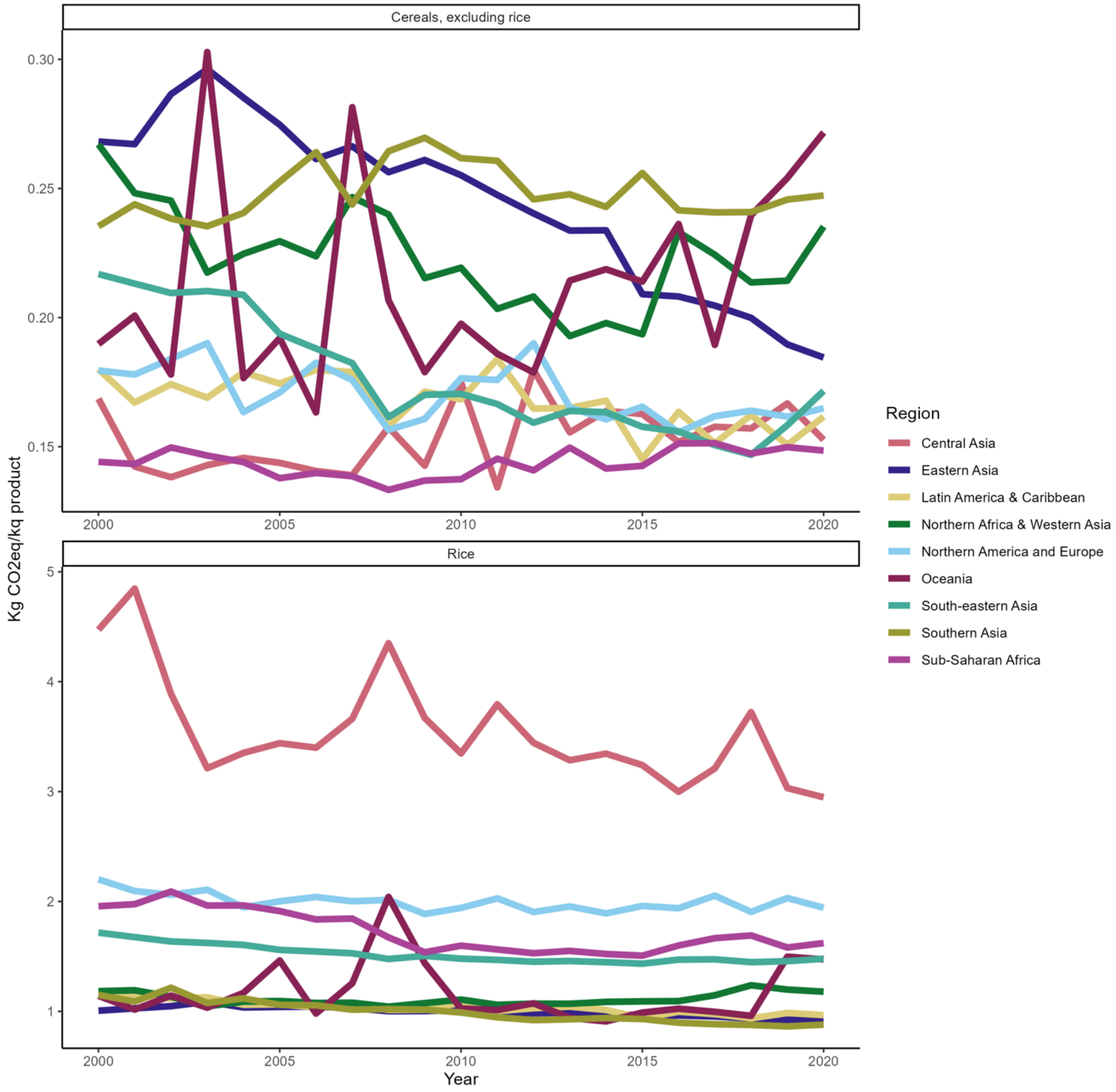

Production-weighted mean by region.

## S2.4 Emissions intensity (kg $CO_2$eq/kg product), Meats, 2000-2020, by region

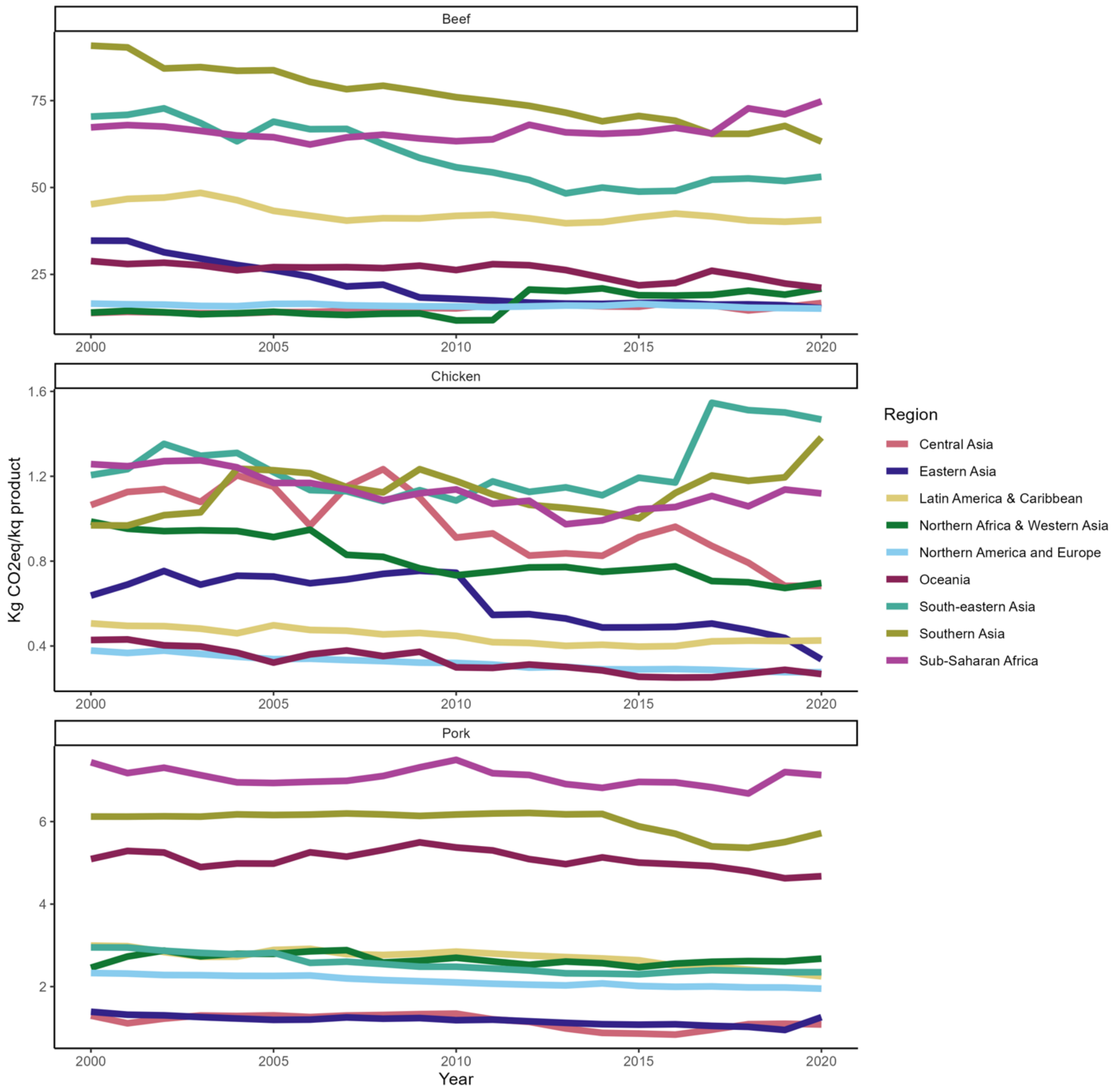

Production-weighted mean by region.

**S2.5 Emissions intensity (kg $CO_2$eq/kg product), Other animal-source foods, 2000-2020, by region**

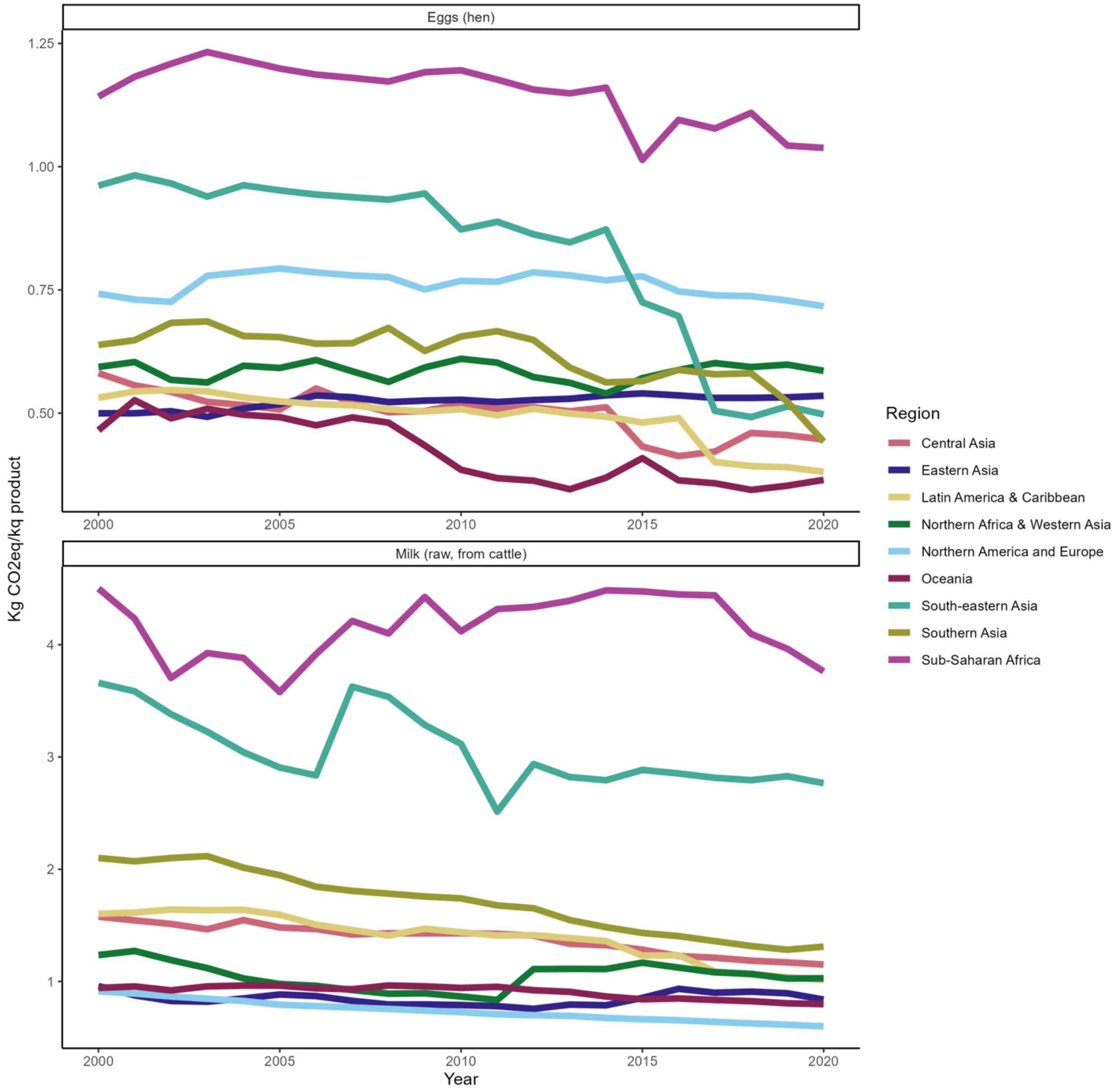

Production-weighted mean by region.

## S2.6 Yield, Staple foods, 2000-2020, by region

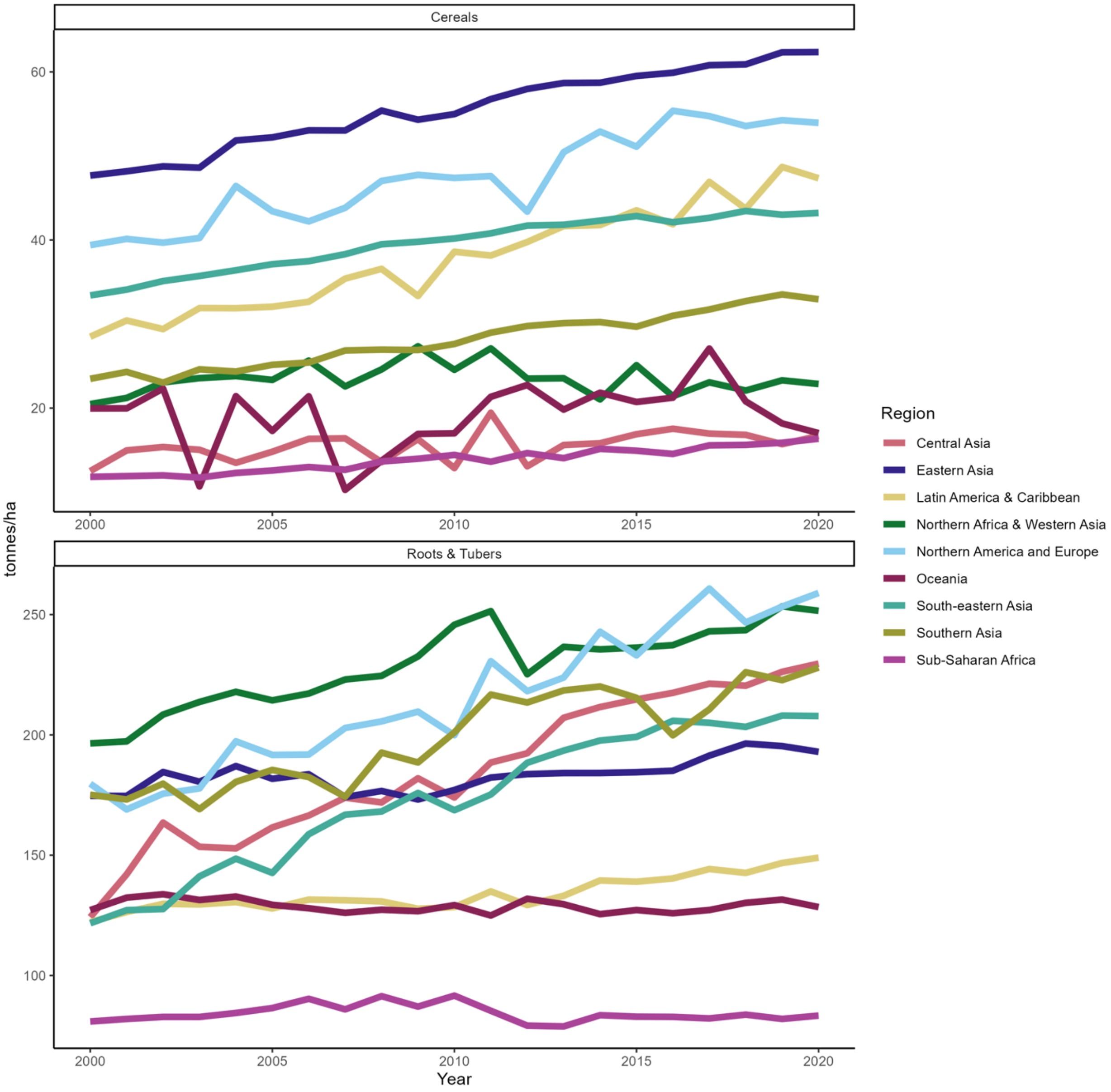

Area harvested-weighted mean by region.

## S2.7 Yield, Meat, 2000-2020, by region

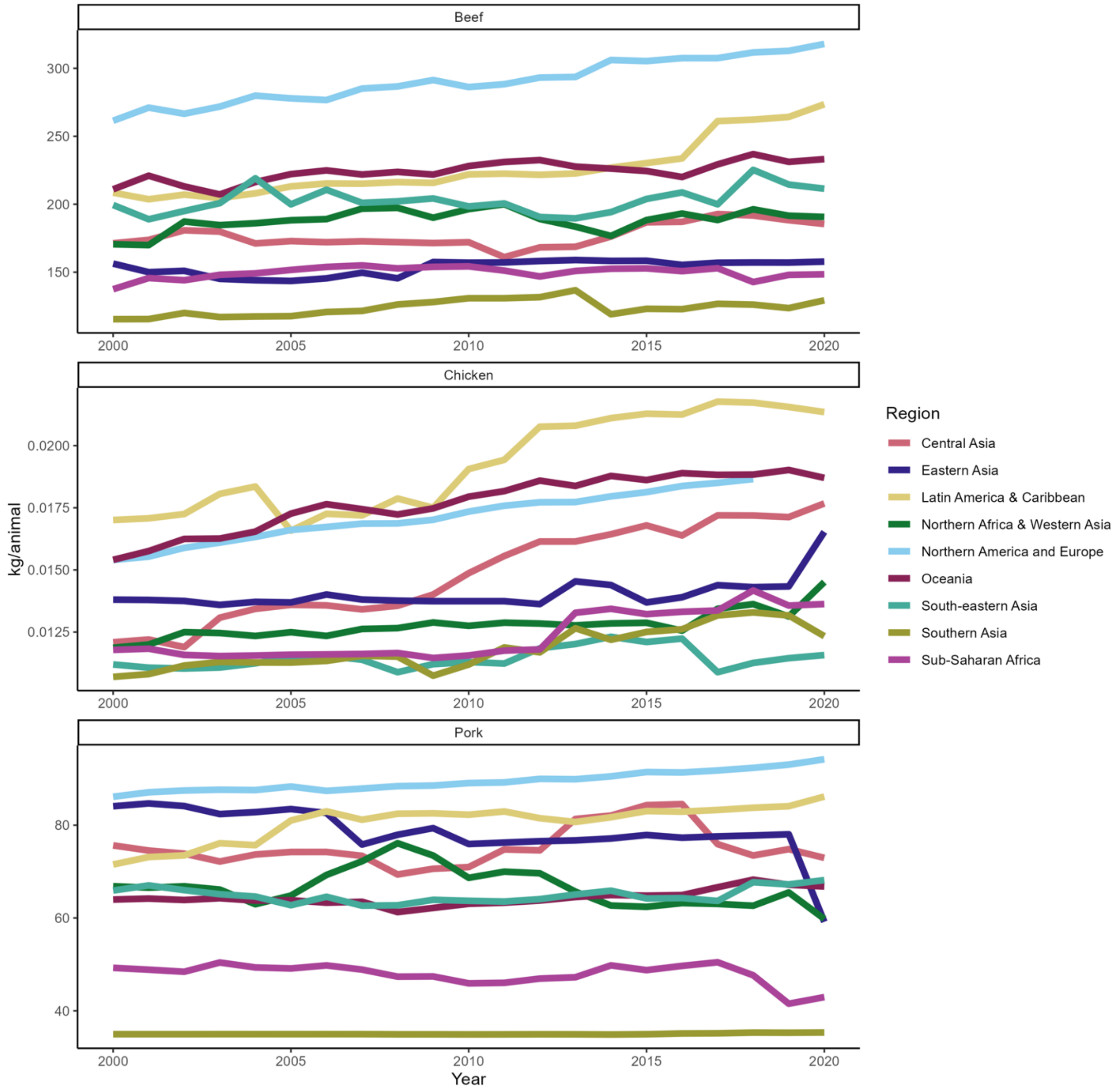

Producing animals-weighted mean by region.

## S2.8 Yield, Other animal-source foods, 2000-2020, by region

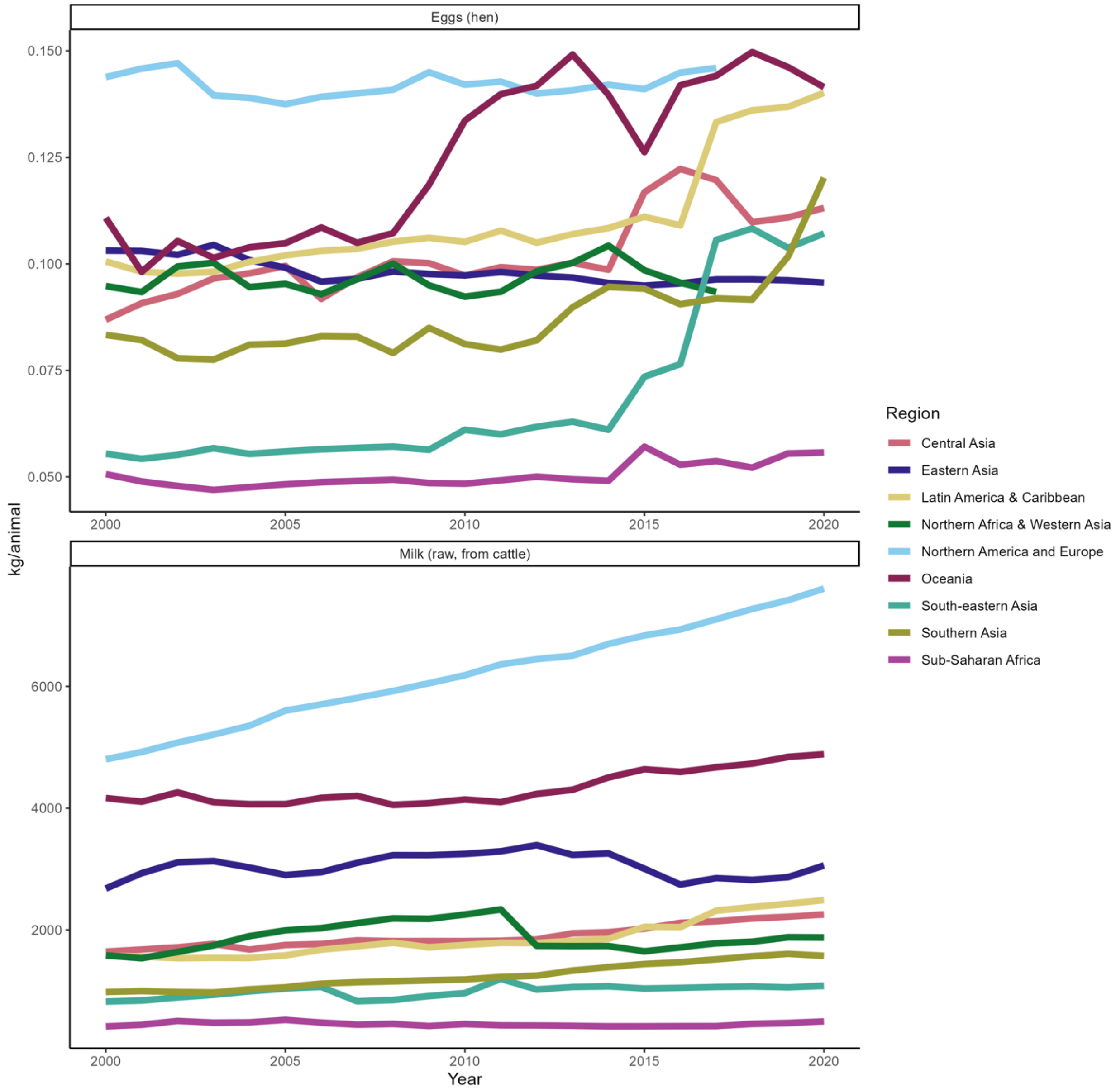

Producing animals-weighted mean by region.

## S2.9 Yield, Pulses & Nuts, 2000-2020, by region

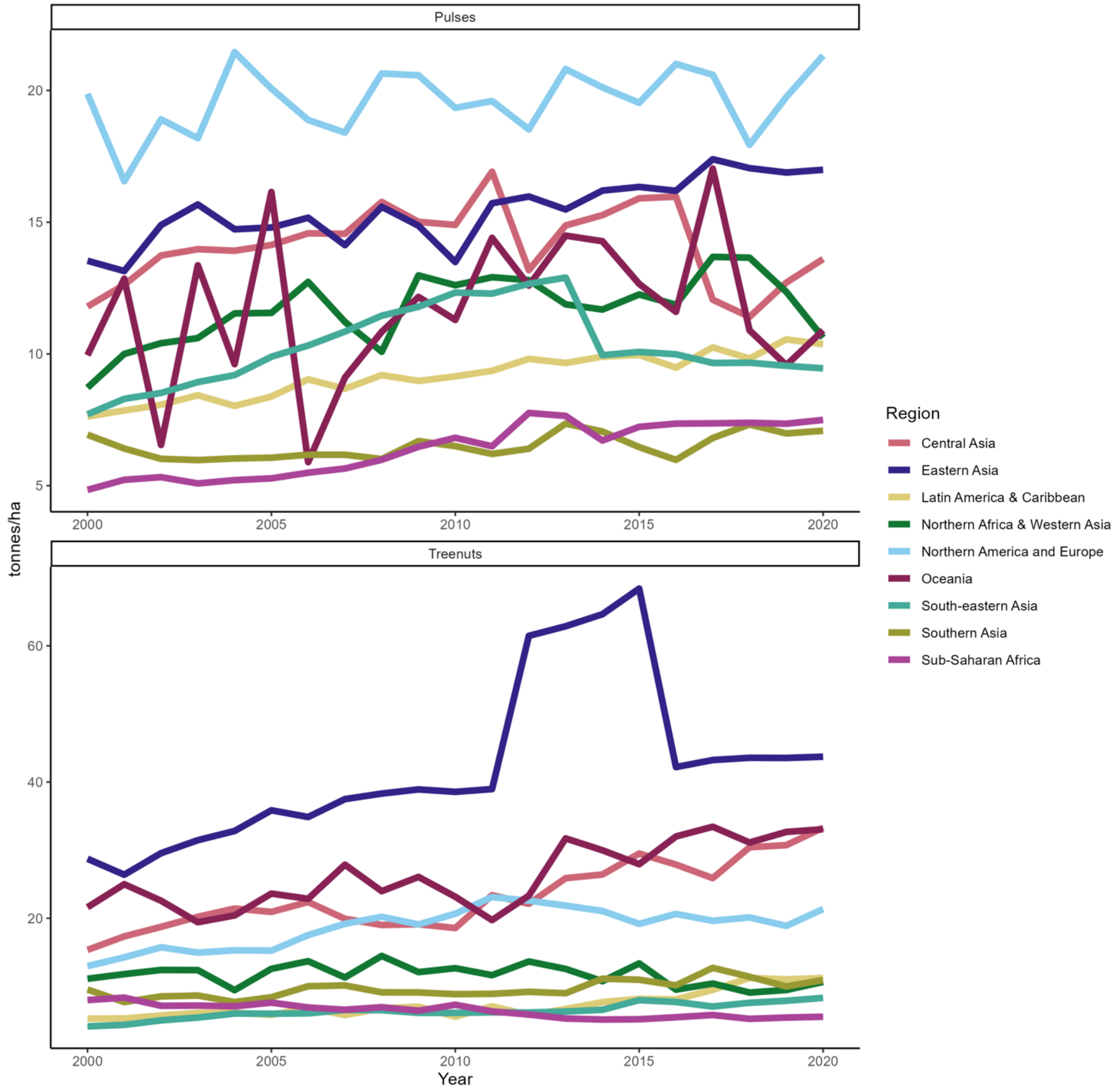

Area harvested-weighted mean by region.

## S2.10 Yield, Fruits & Vegetables, 2000-2020, by region

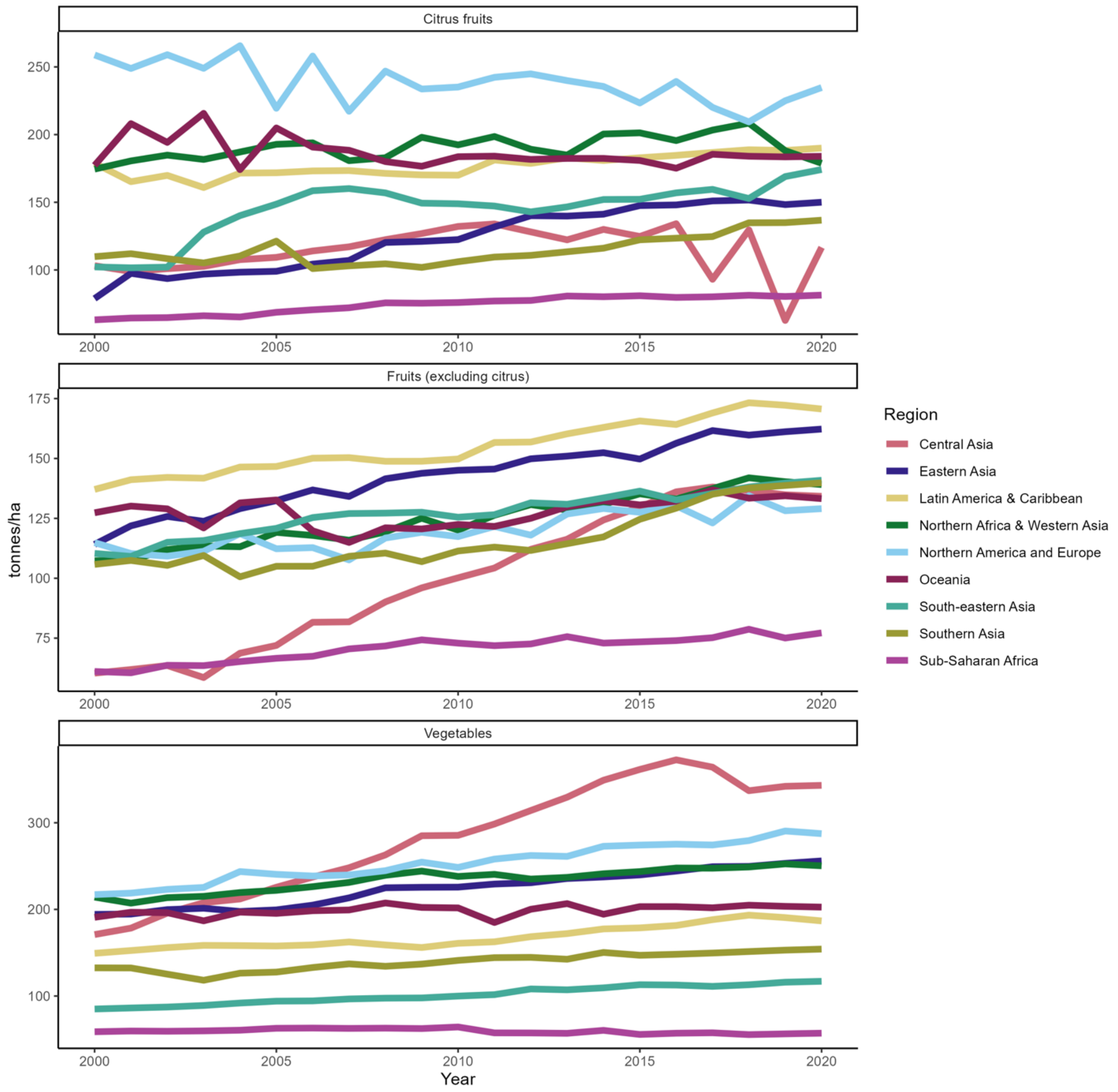

Area harvested-weighted mean by region.

### S2.11 Cropland expansion (relative change 2003-2019)

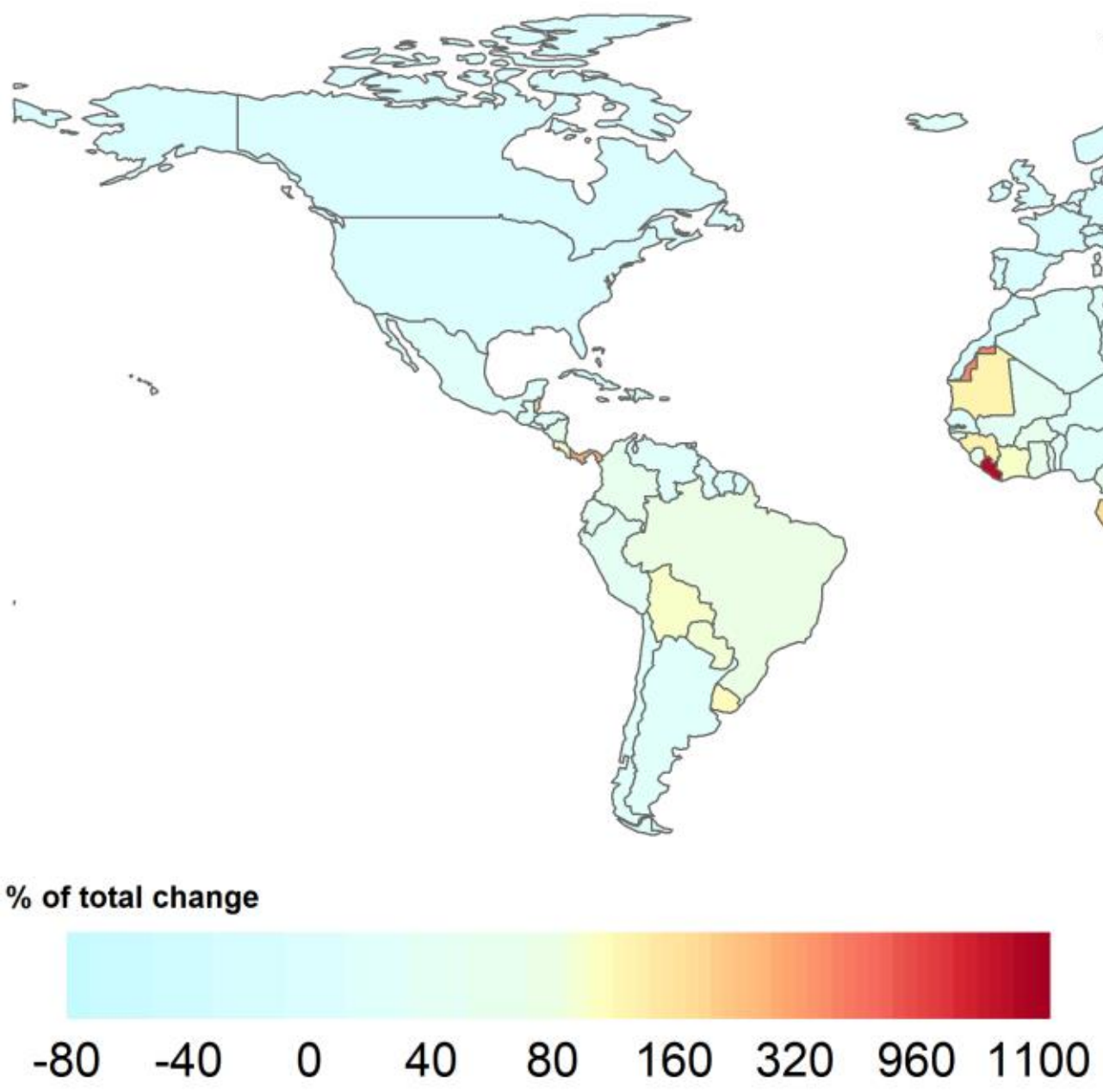

### S2.12 Agriculture water withdrawal as % of total renewable water resources, 2018

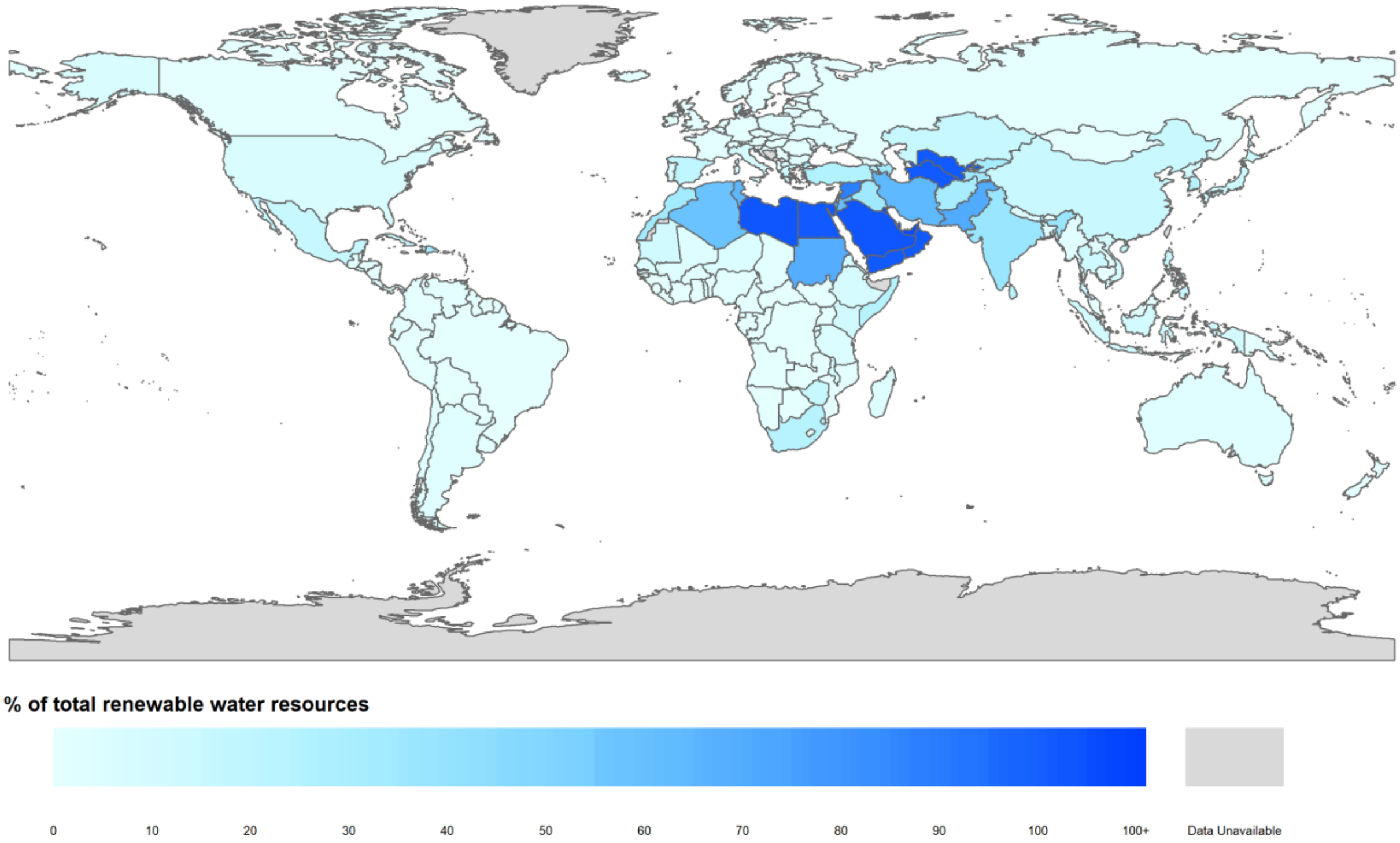

## S2.13 Agriculture water withdrawal as % of total renewable water resources, by region

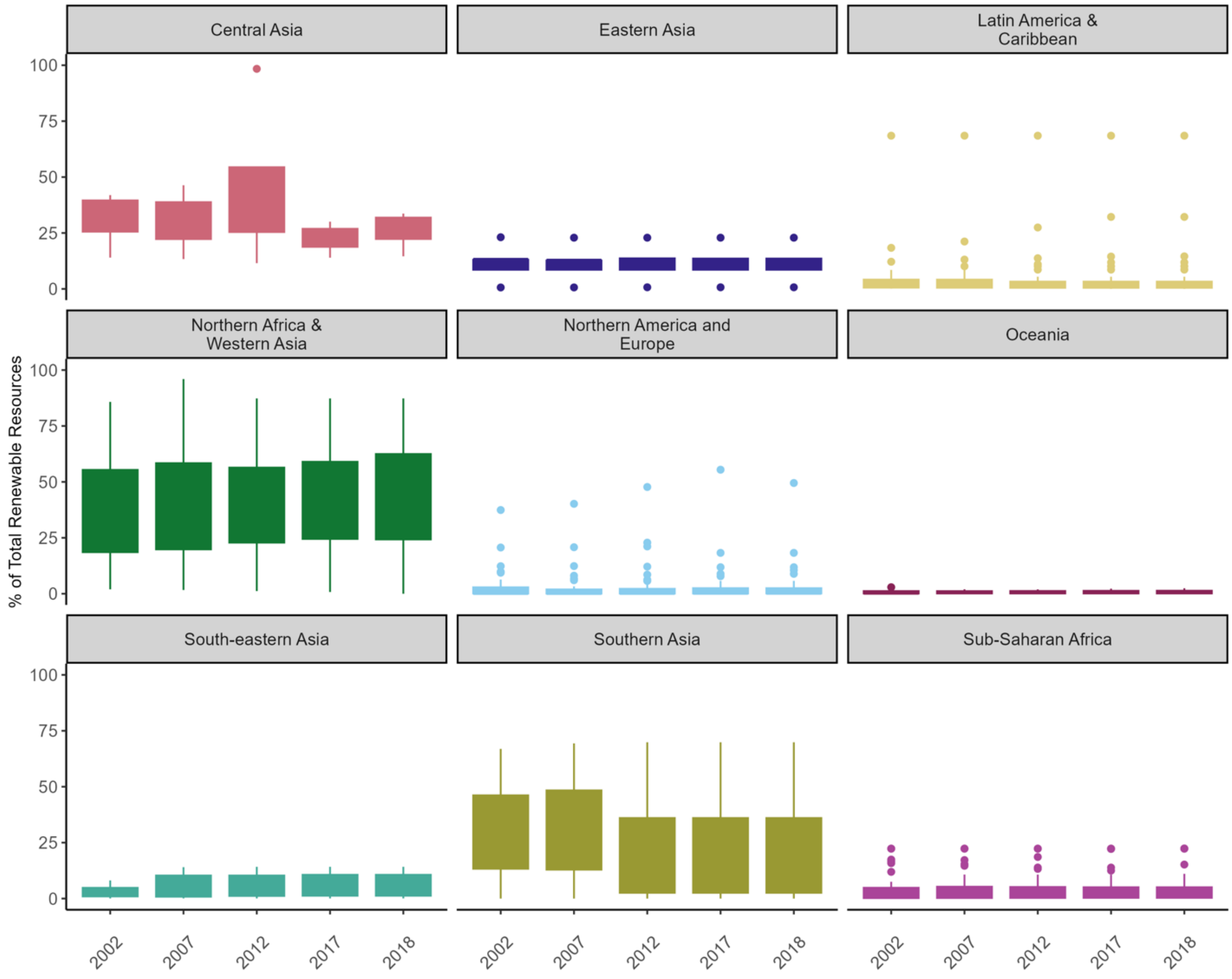

Unweighted mean by region.

## S2.14 Agriculture water withdrawal as % of total renewable water resources, by income group

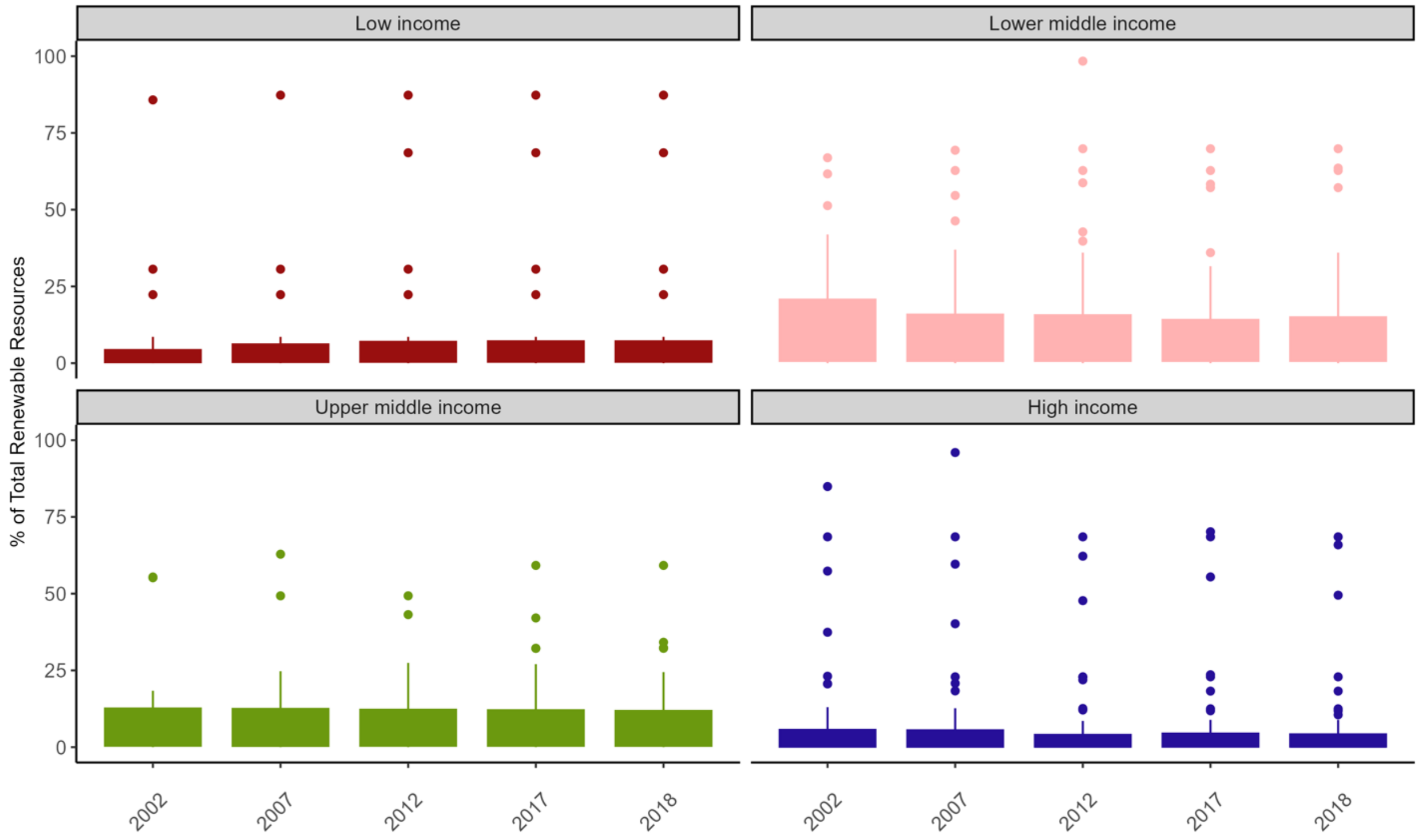

Unweighted mean by income group.

## S2.15 Functional integrity: Percent agricultural land with minimum level of natural habitat, 2015

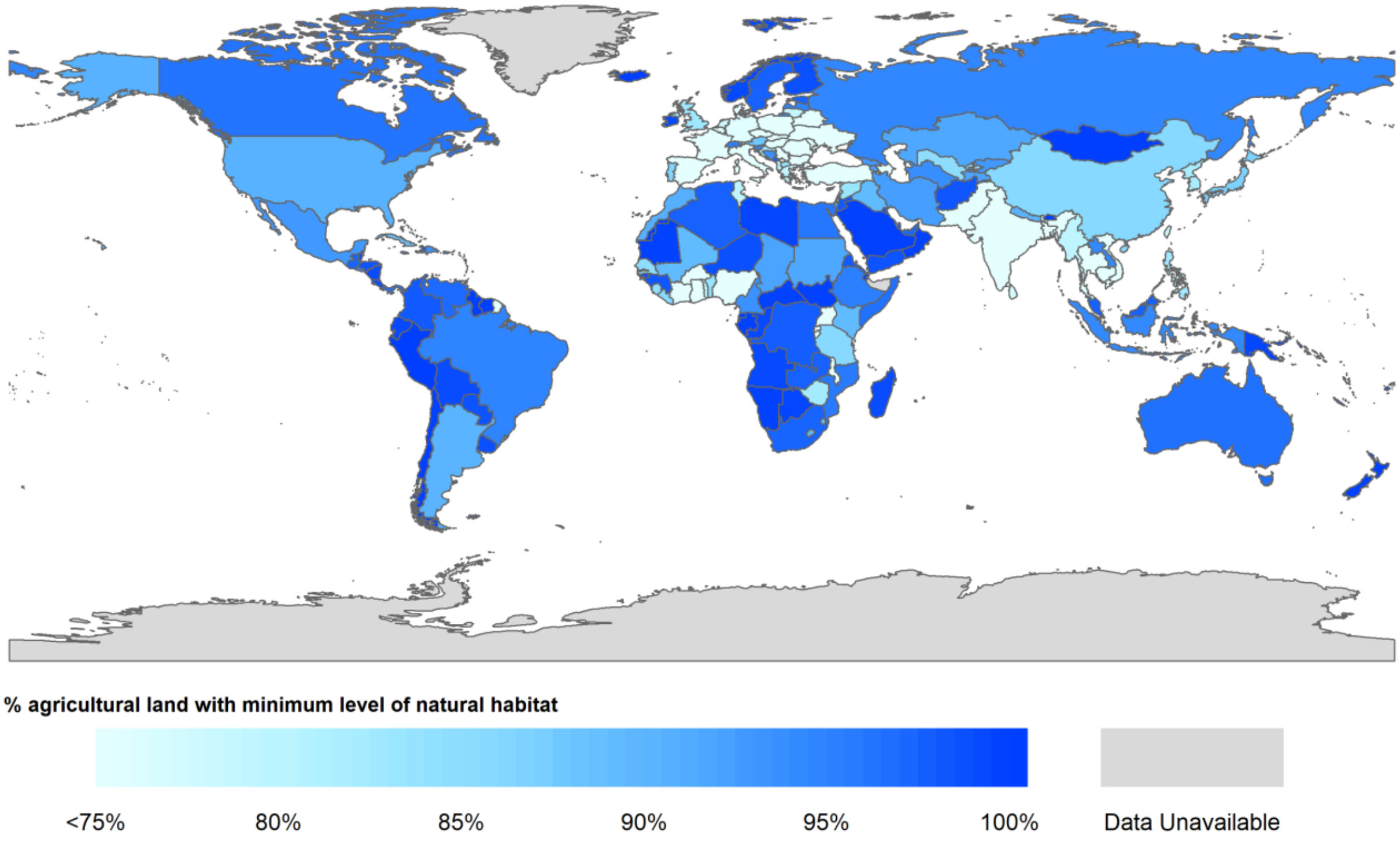

## S2.16 Fishery health index progress score, 2021

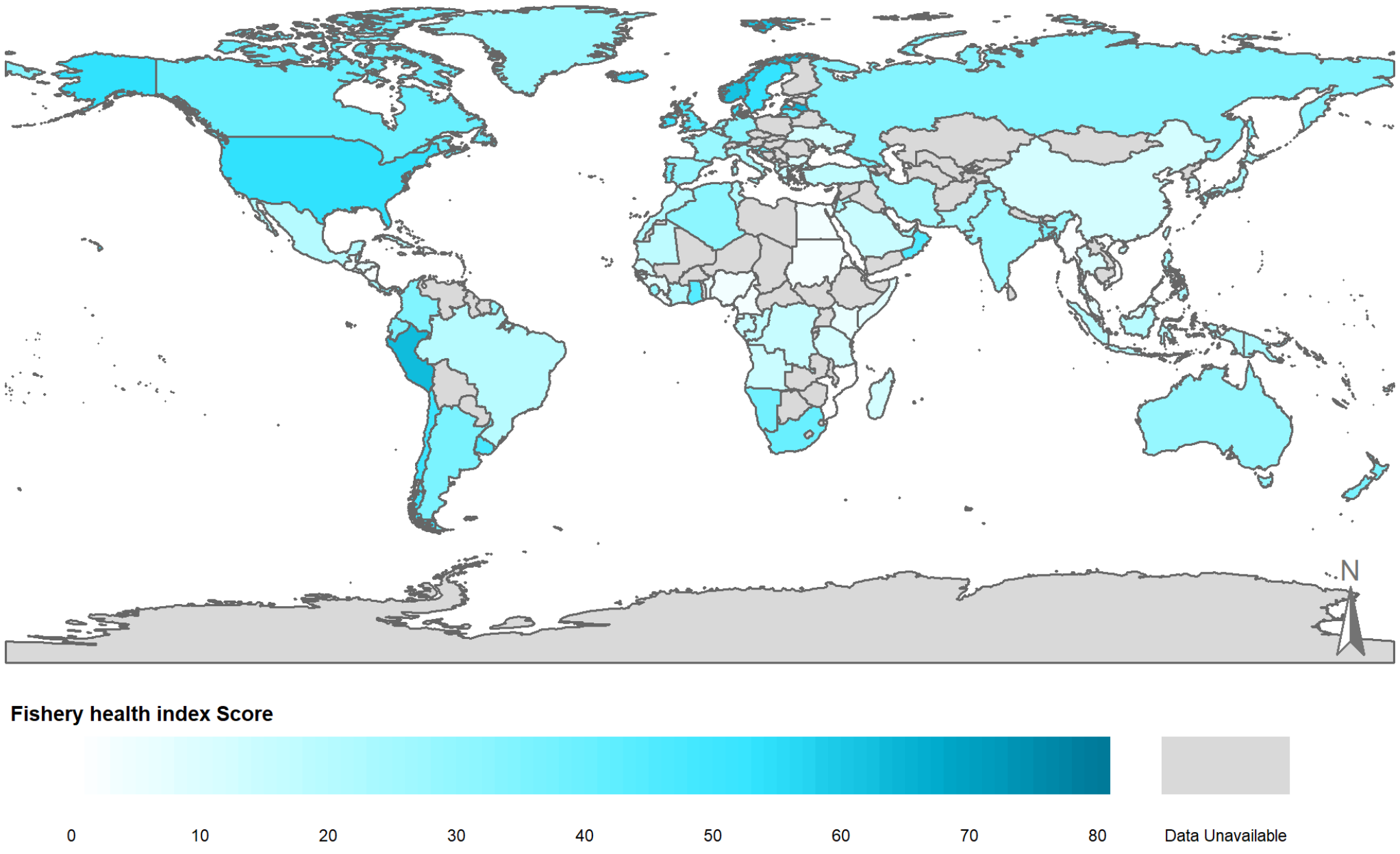

## S2.17 Total pesticides per unit of land (kg/ha), 2000-2019, by region

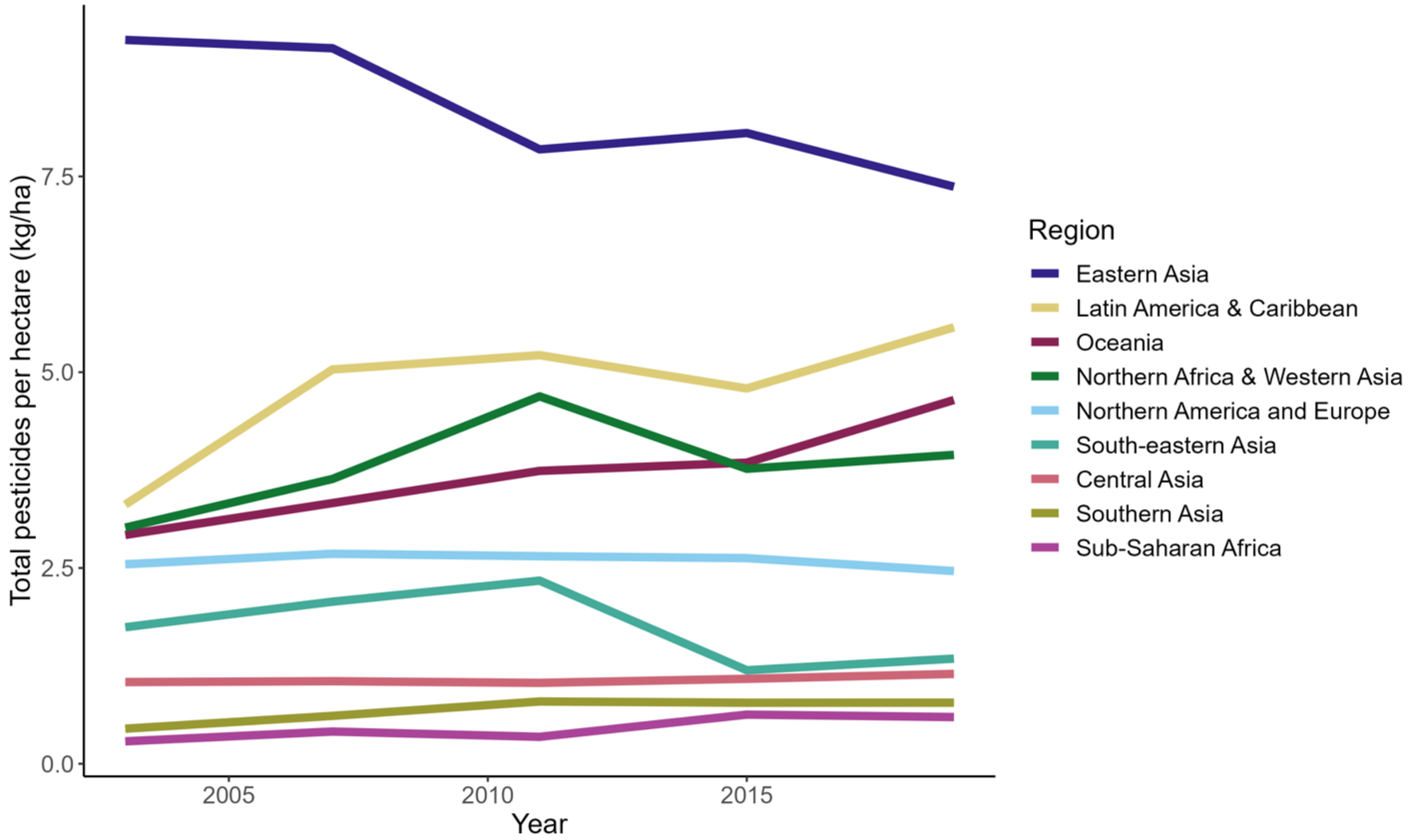

Cropland-weighted mean by region.

### S2.18 Total pesticides per unit of land (kg/ha), 2019

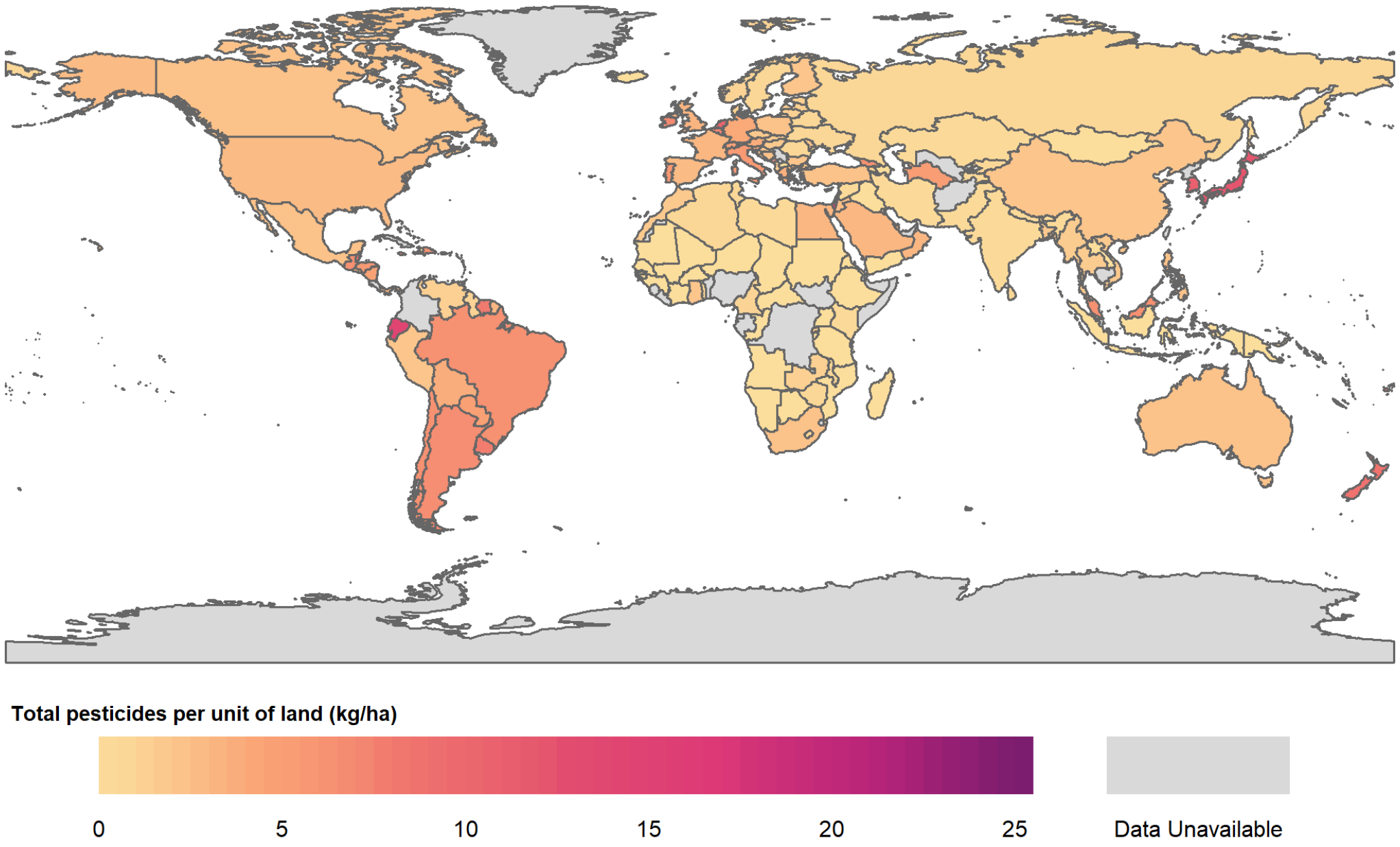

### S2.19 Sustainable nitrogen management index, 2000-2018, by region

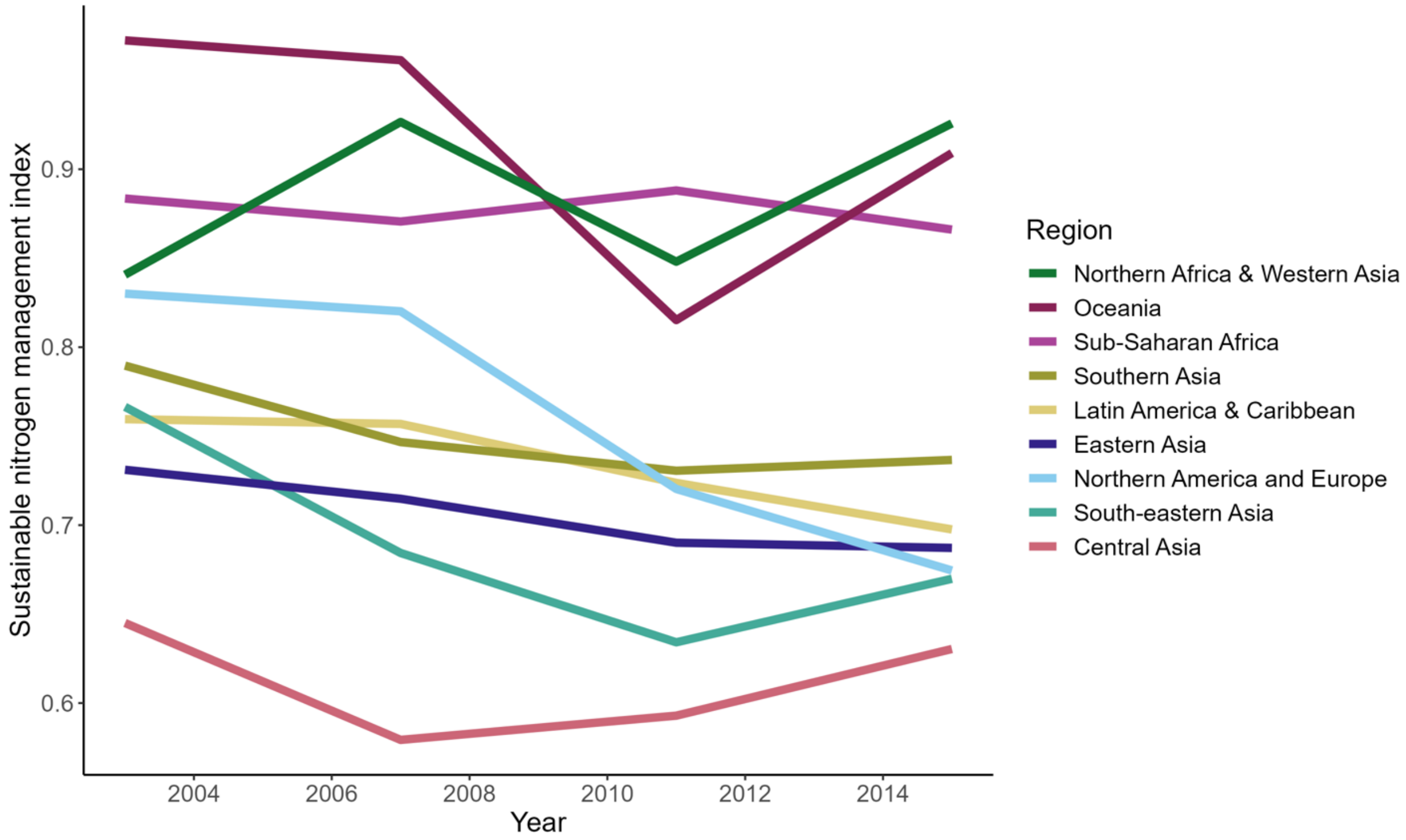

Cropland-weighted mean by region.

### S2.20 Sustainable nitrogen management index, 2018

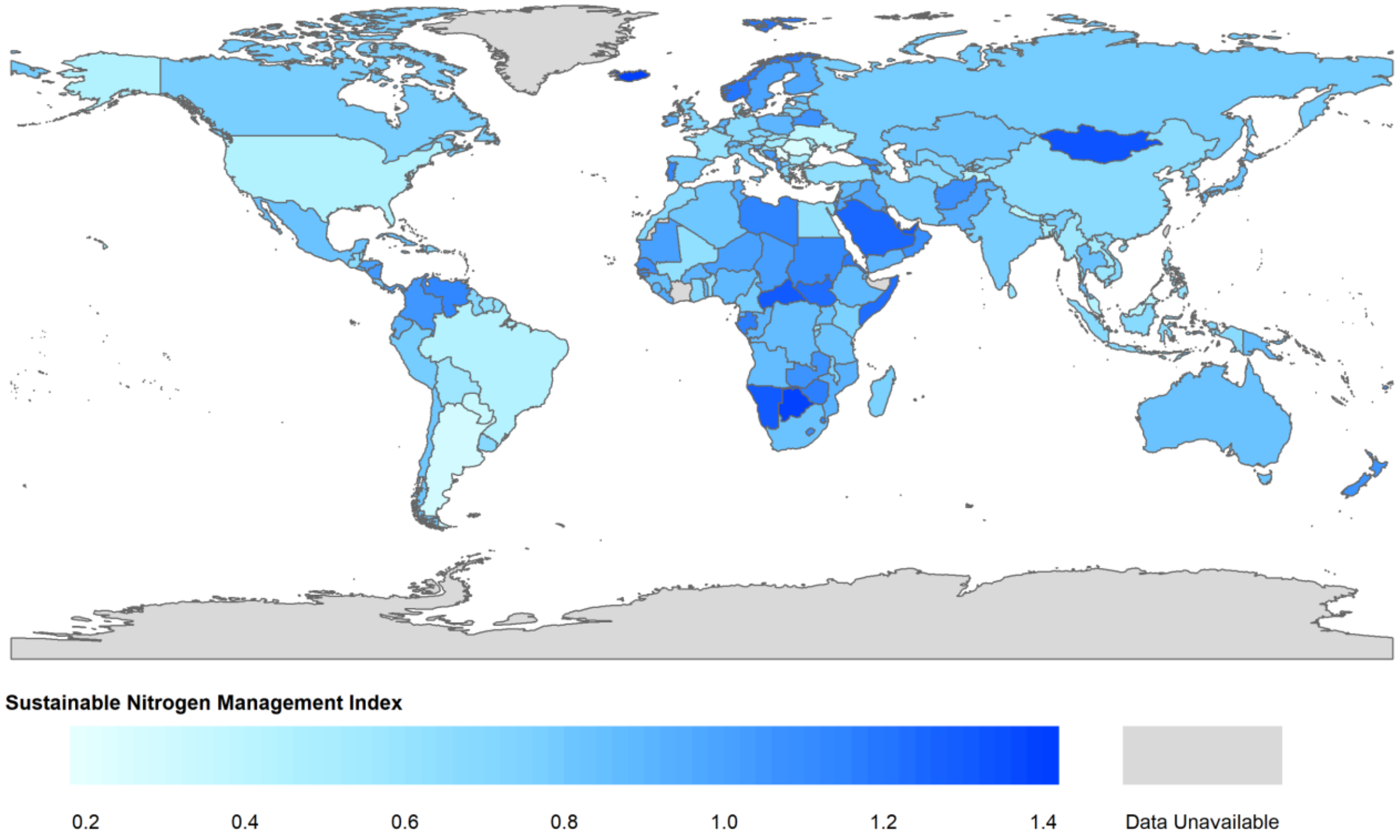

## S3.1 Share of agriculture in GDP, 2019

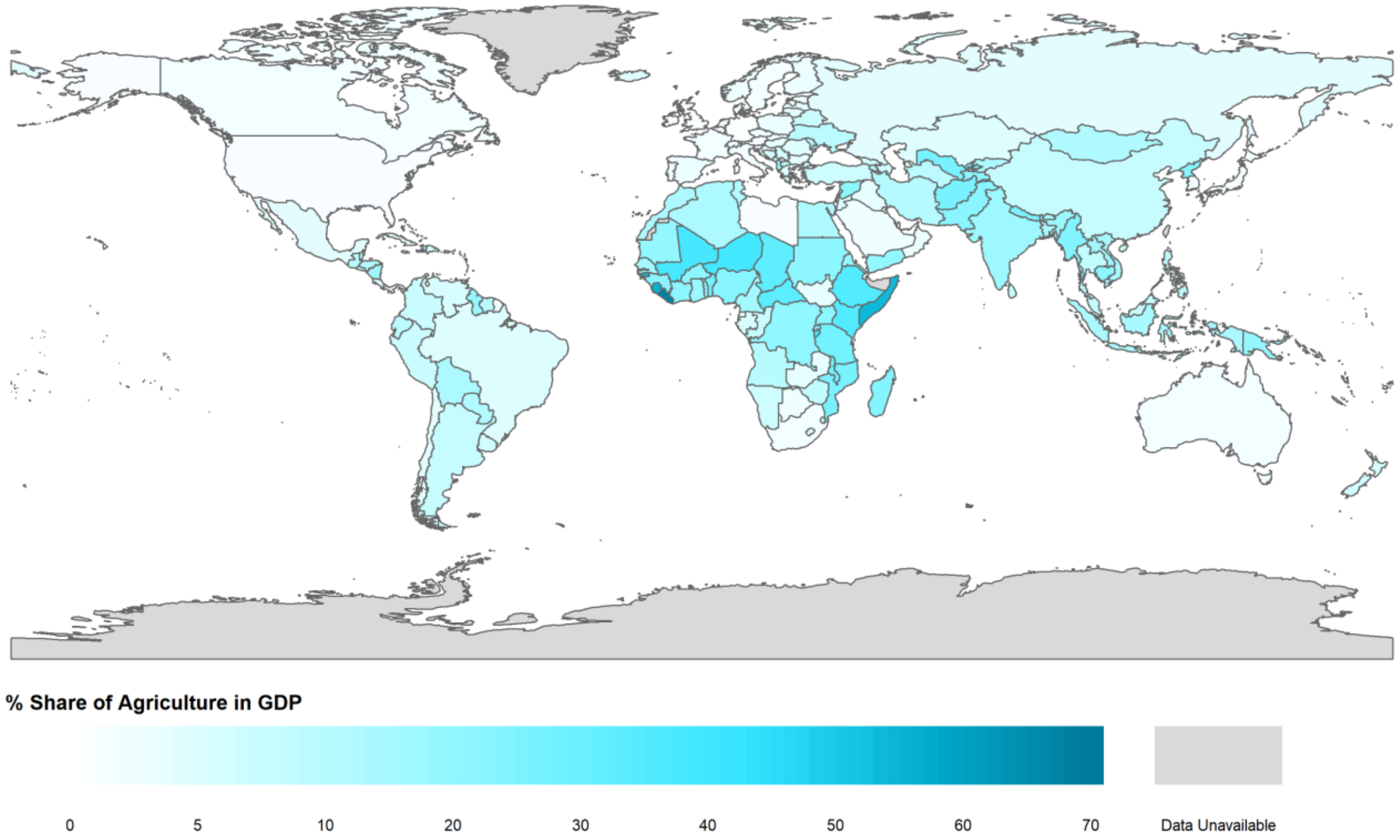

## S3.2 Share of agriculture in GDP, 2000-2019, by region

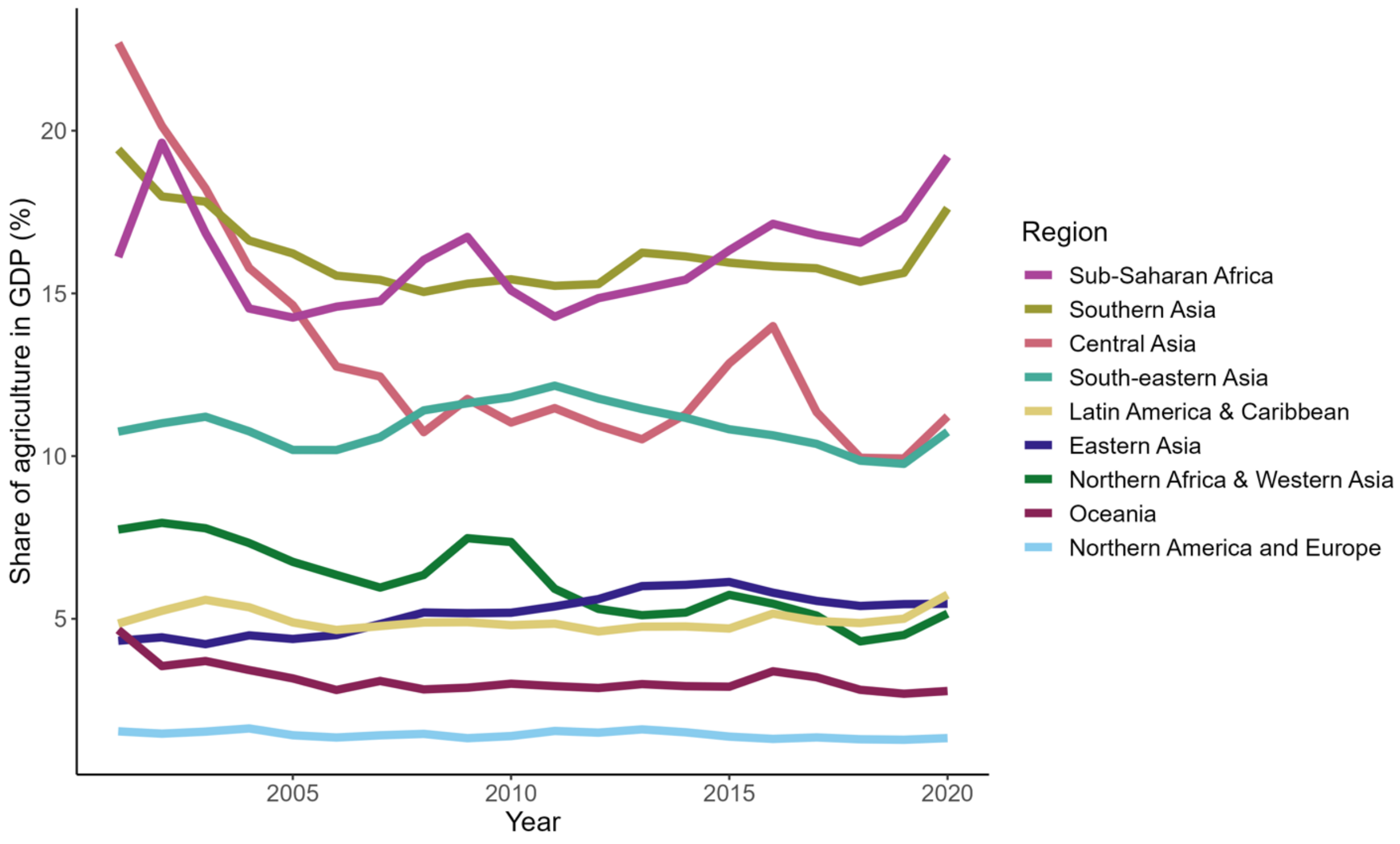

GDP-weighted regional means.

## S3.3 Unemployment and Underemployment rates by sex, age, and urban/rural, 2020

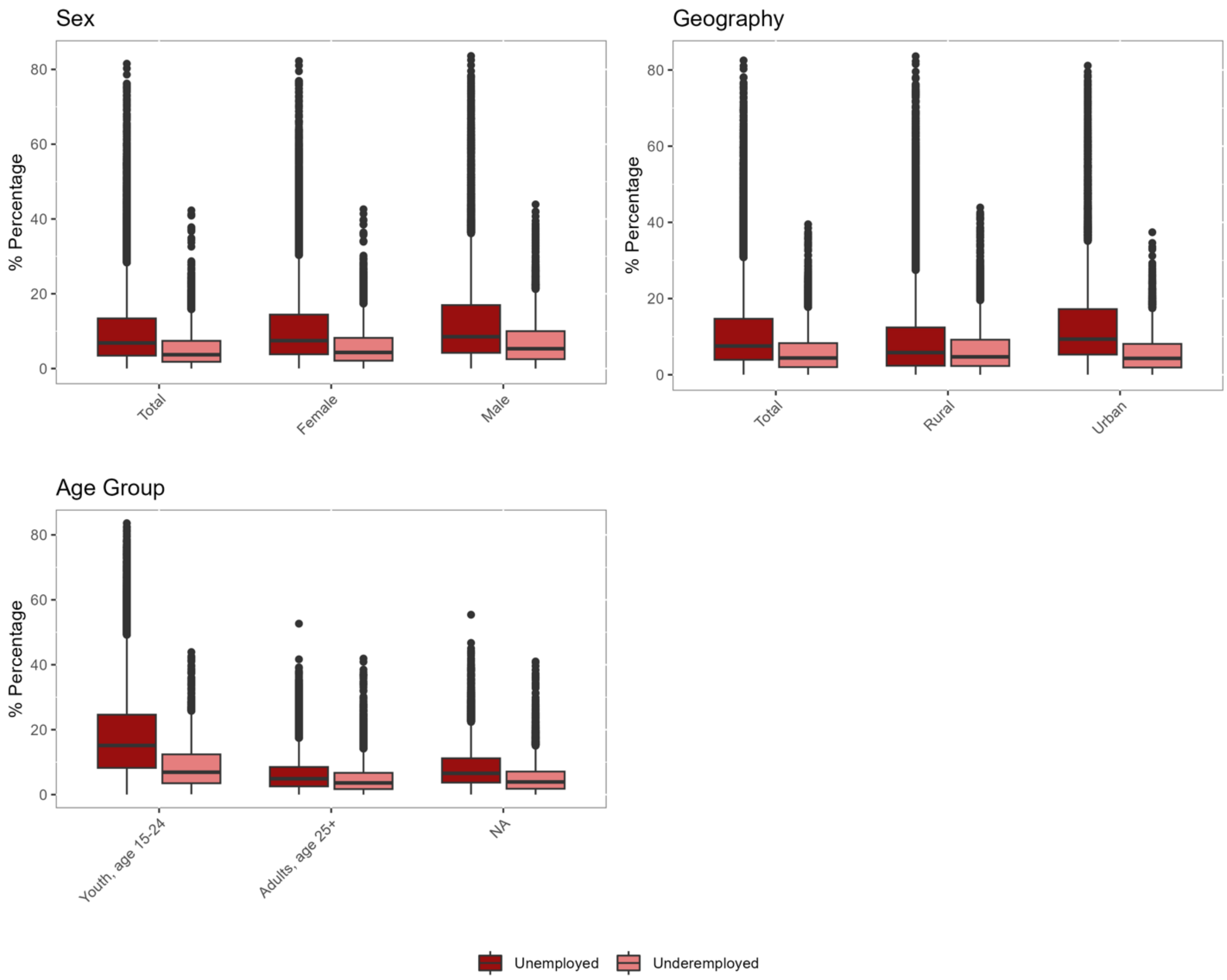

**S3.4 Unemployment rate (total), 2010-2020, by region**

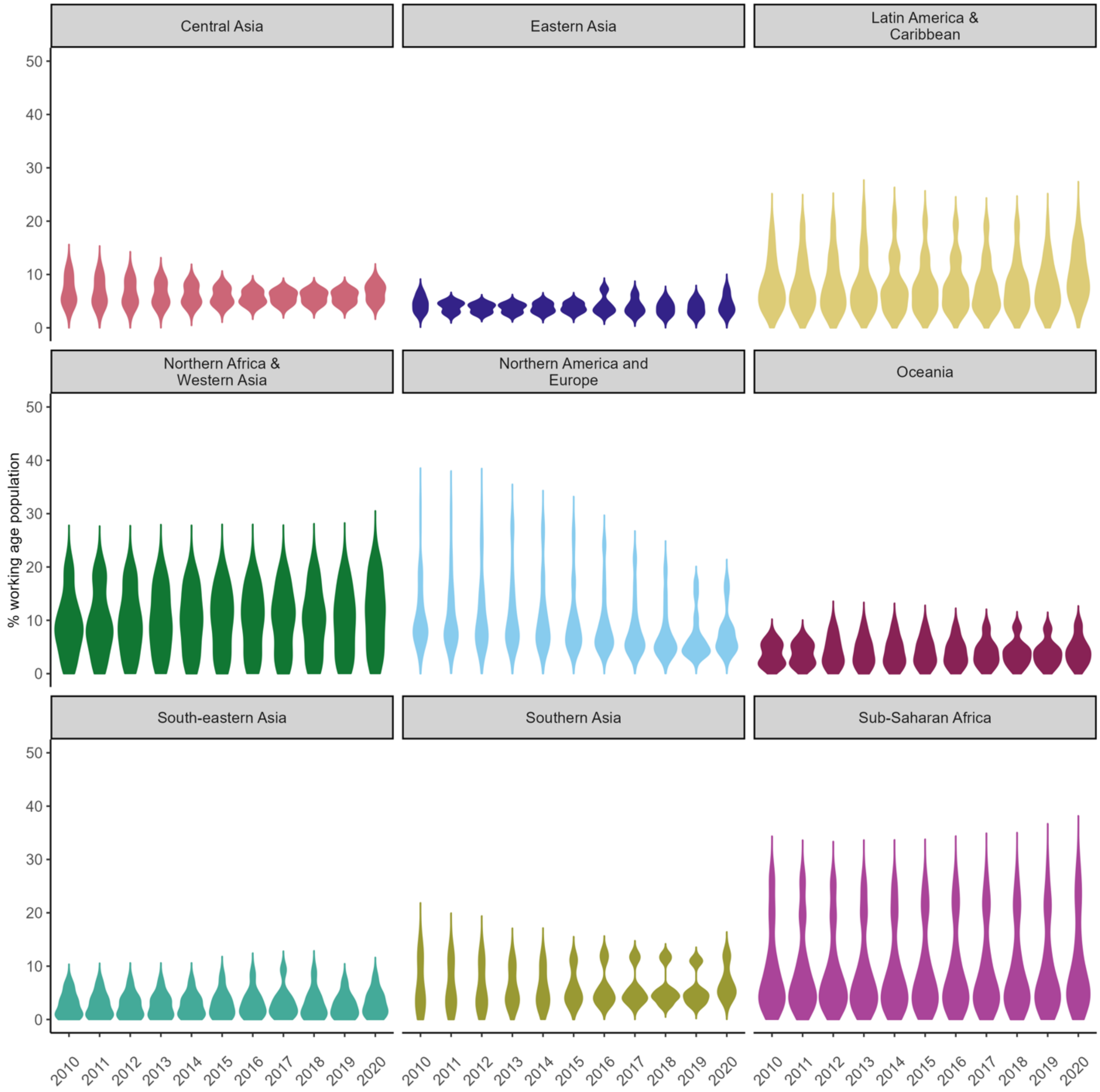

### S3.5 Underemployment rate (total), 2010-2020, by region

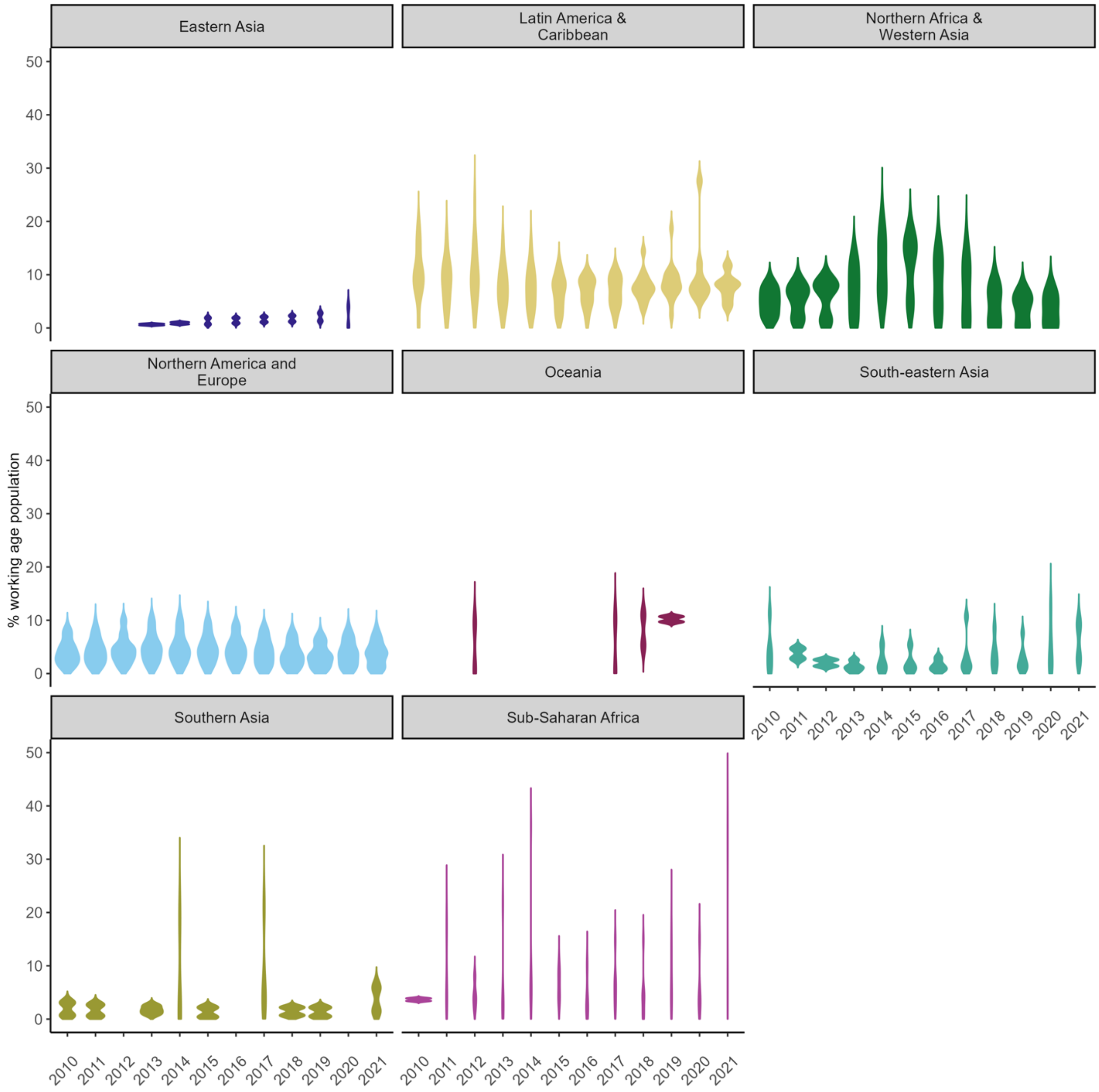

### S3.6 Unemployment rate (total), 2010-2020, by income group

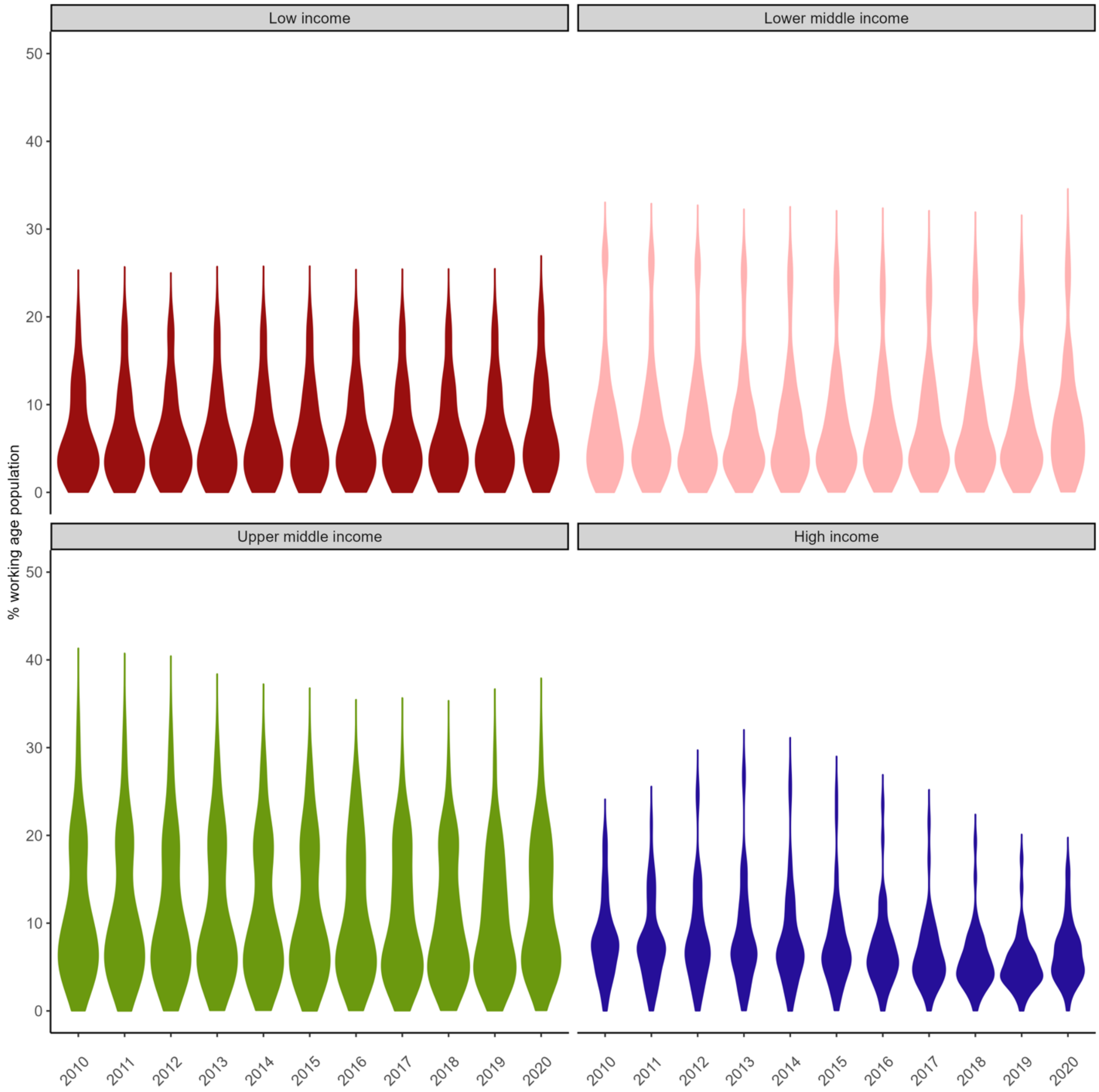

**S3.7 Underemployment rate (total), 2010-2020, by income group**

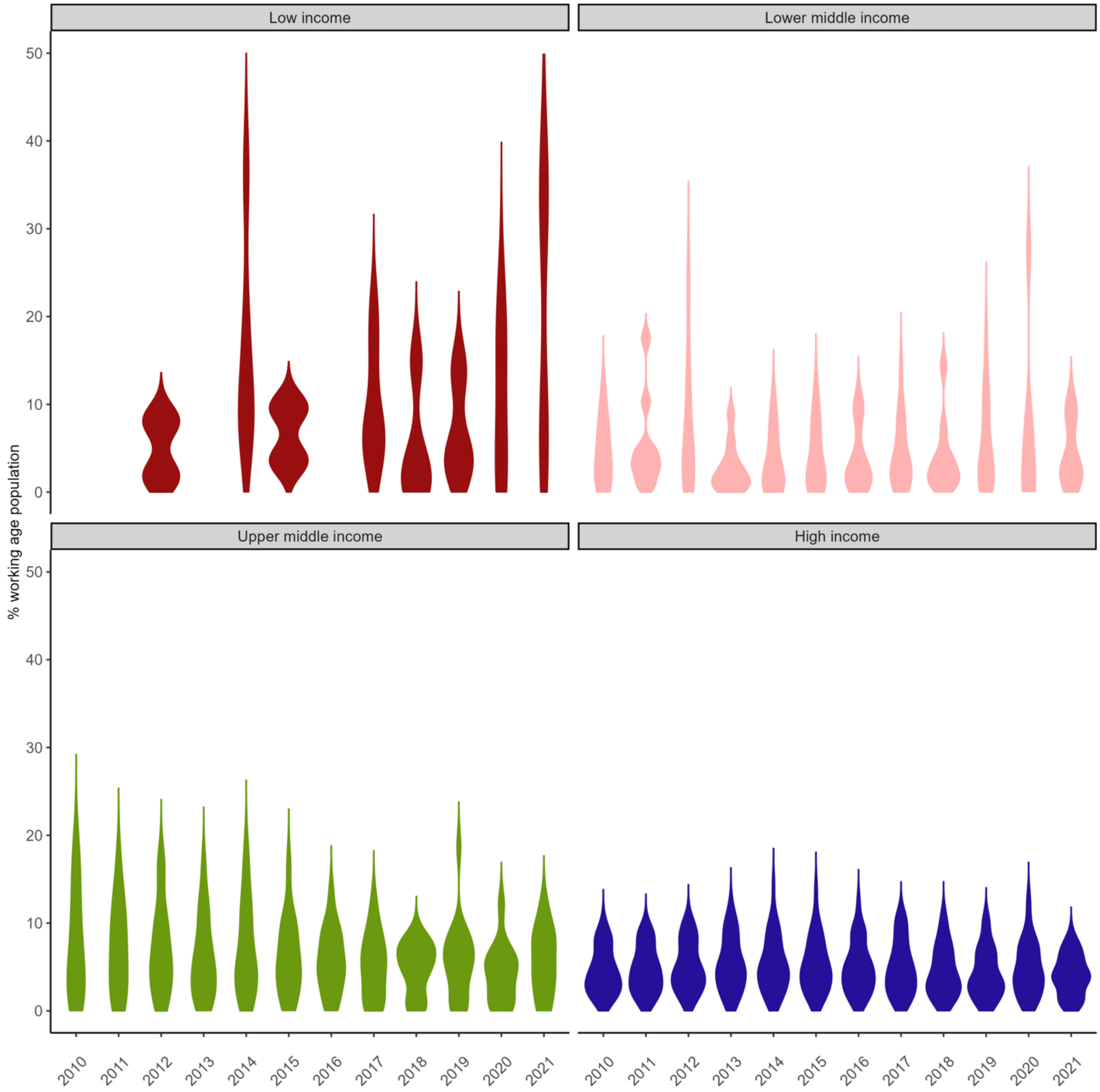

## S3.8 Unemployment and underemployment (total), 2005-2020, by region

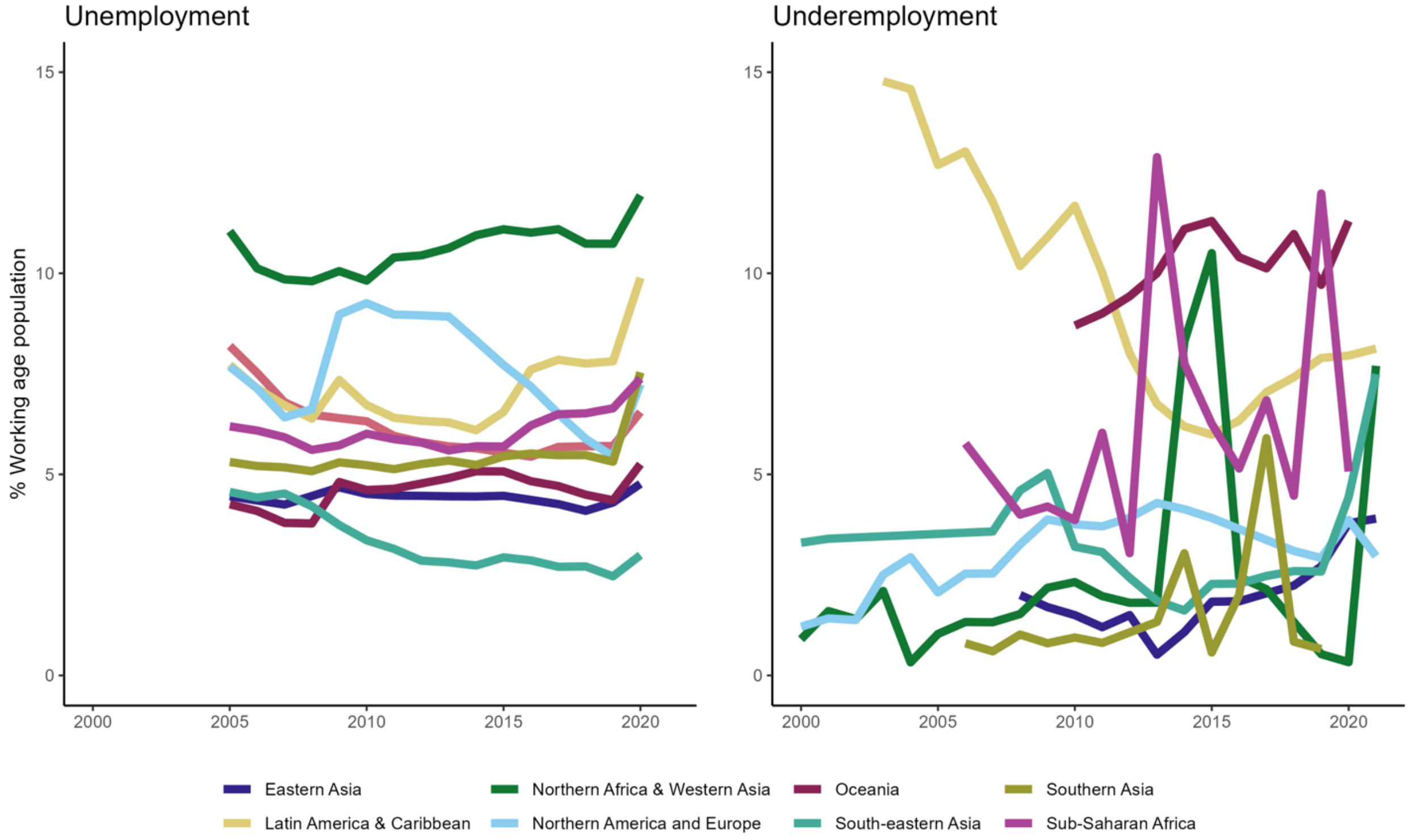

Population-weighted regional mean.

## S3.9 Unemployment and underemployment (total), 2005-2020, by income group

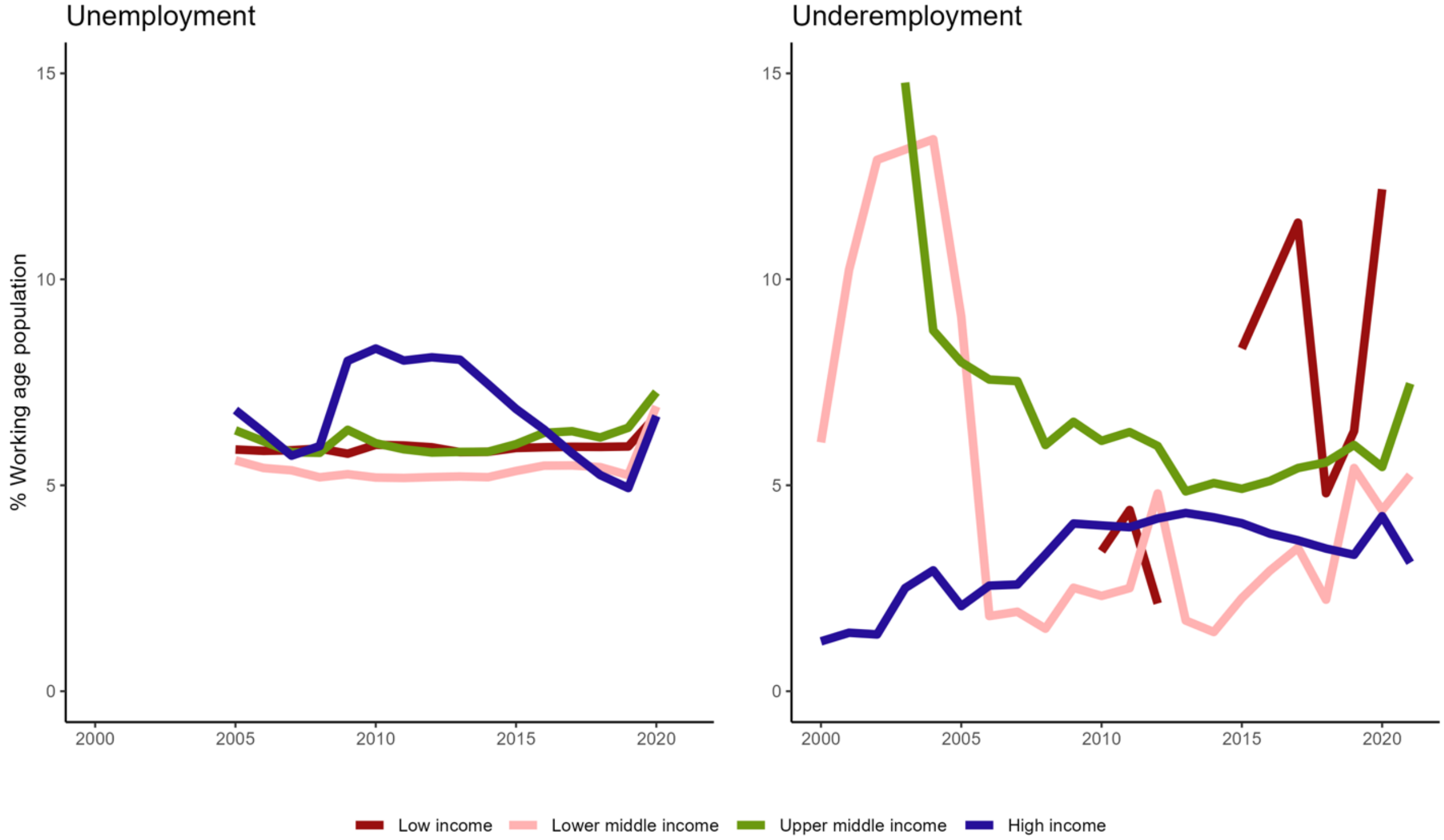

**Population-weighted regional mean.**

## S3.10 Social protection coverage and adequacy

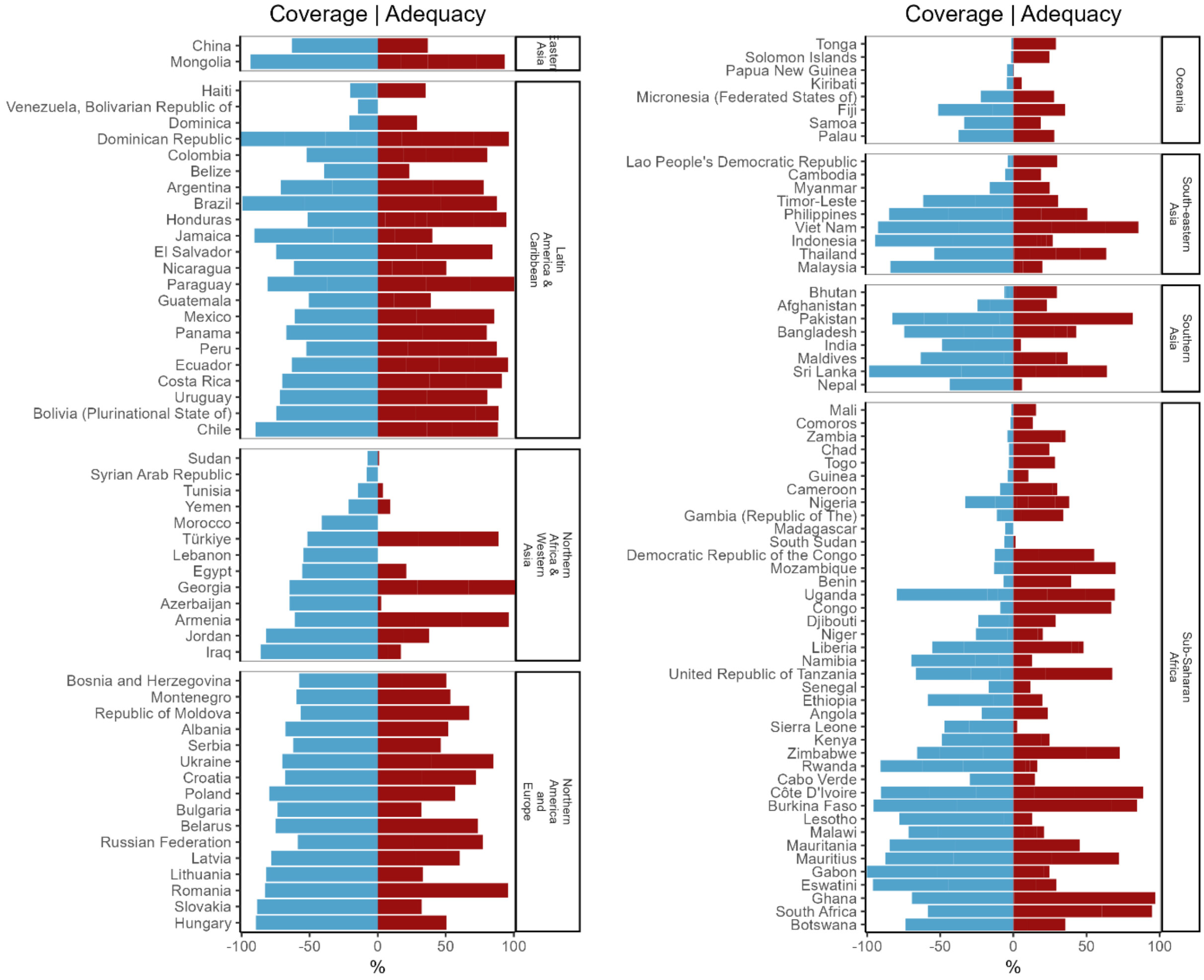

Data year differs by country. See table A1.2. The x-axis for coverage reflects the share of individuals in the total population from households where at least one member participates in a social protection and labor market program. The x-axis for adequacy reflects the total social protection benefit amount received by beneficiary households (direct and indirect beneficiaries) as a percentage of beneficiaries' post-transfer, household wealth.

**S3.11 Percent children 5-17 engaged in child labor, by sex, by region**

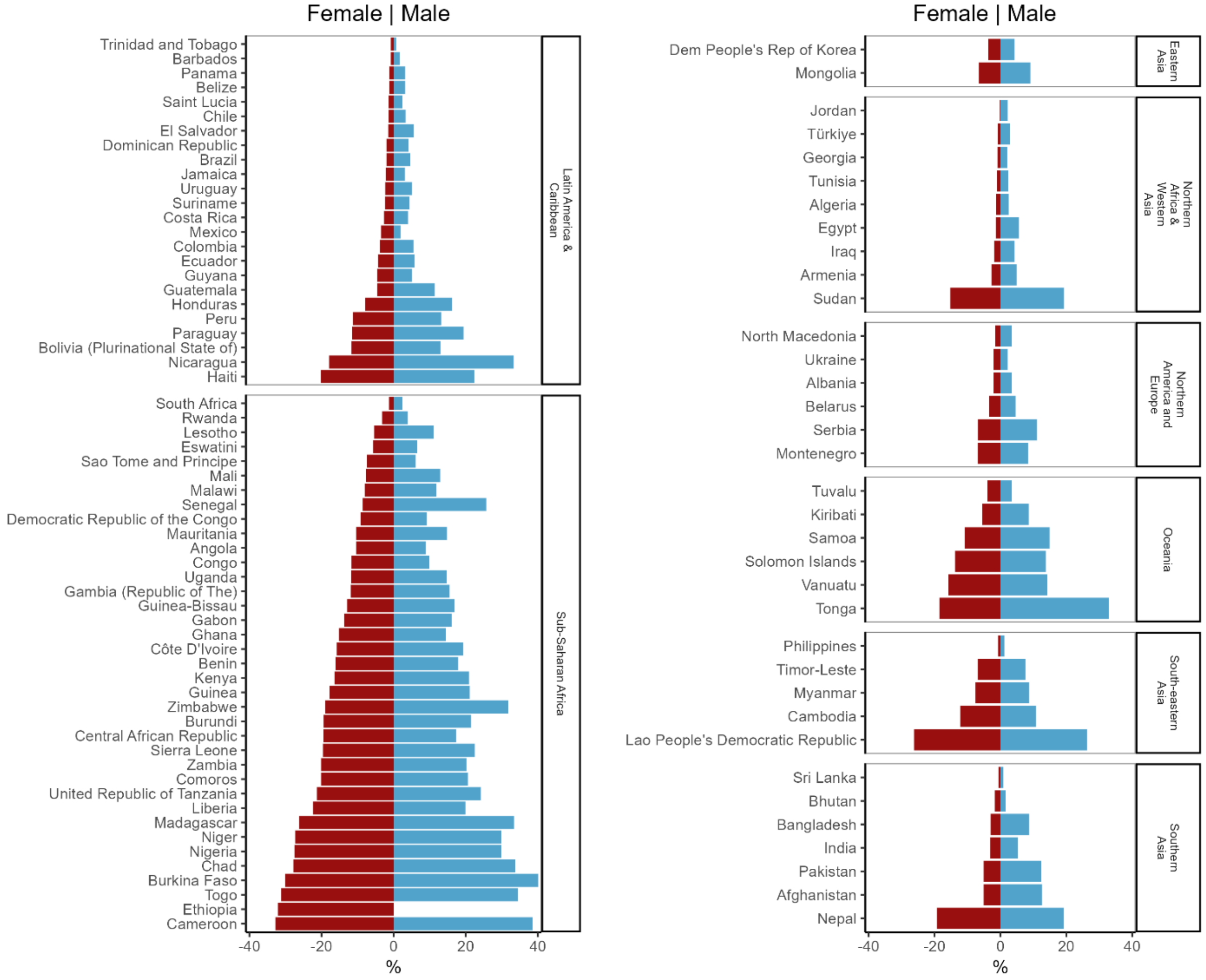

Unweighted regional mean. Data source year differs by country. See table A1.2.

## S3.12 Percent children 5-17 engaged in child labor, by sex, by income group

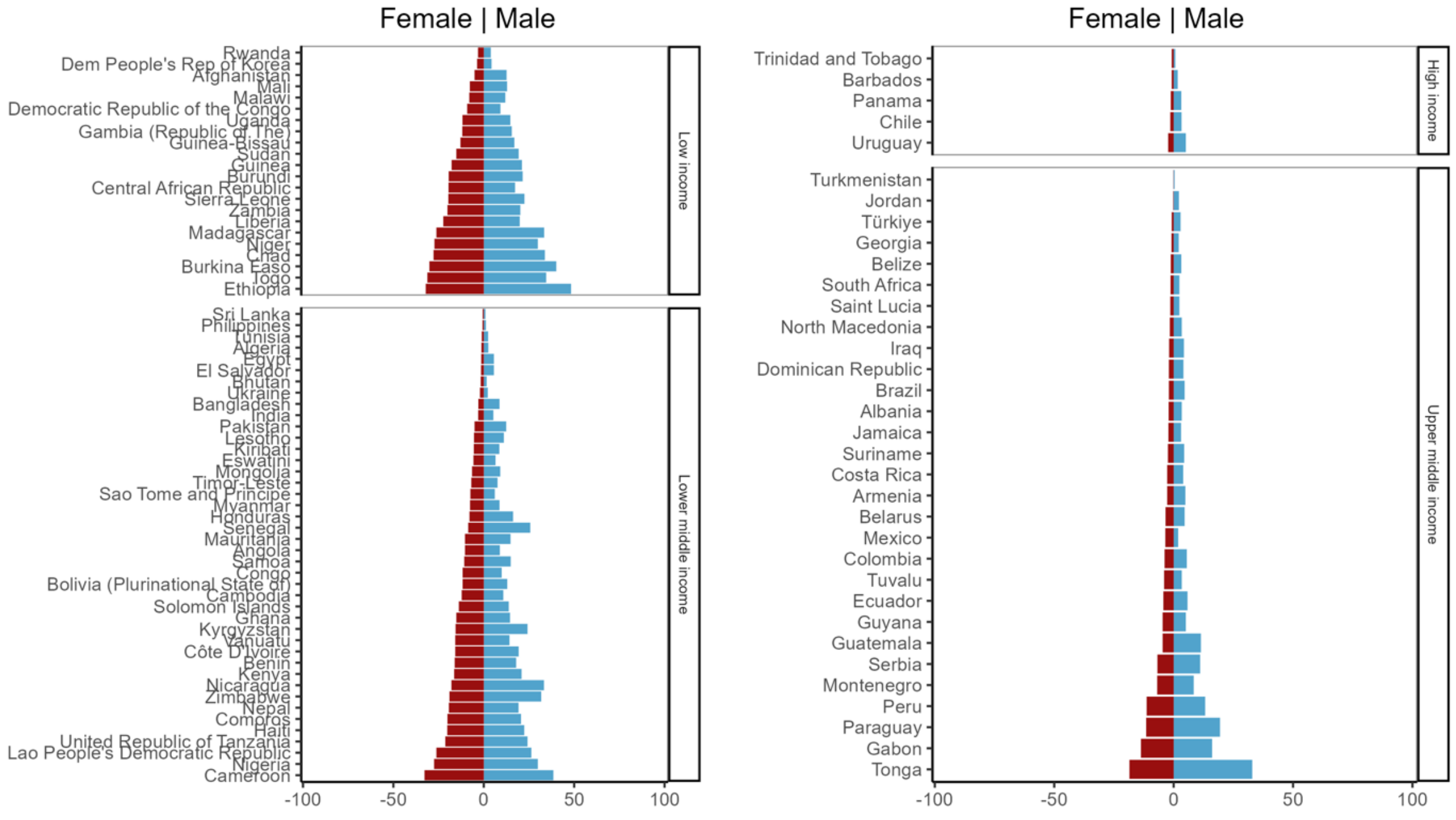

Unweighted income group mean. Data source year differs by country. See table A1.2.

## S3.13 Distribution of landholdings by sex

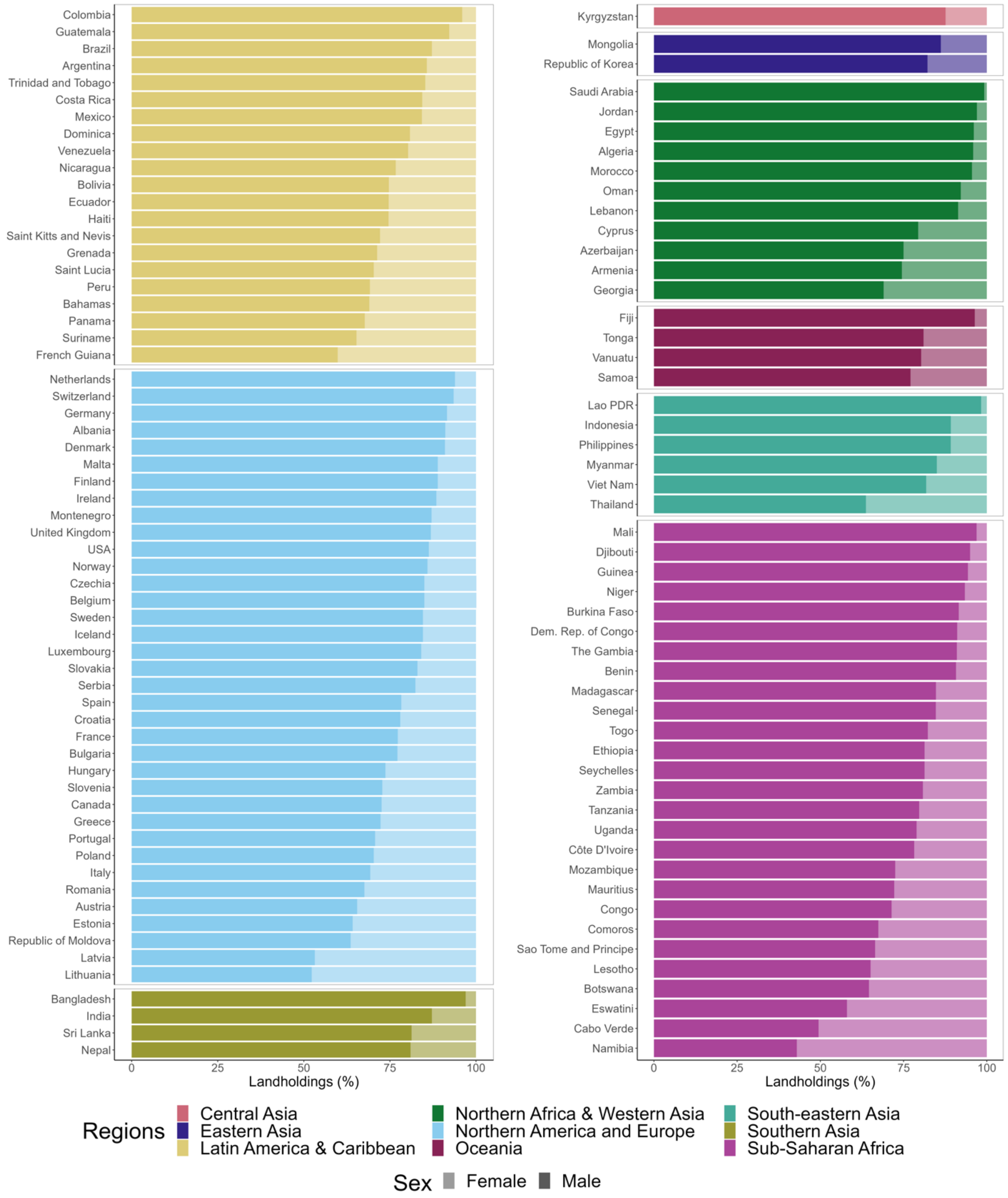

Data source year differs by country. See table A1.2.

## S4.1 Civil society participation index, 2021

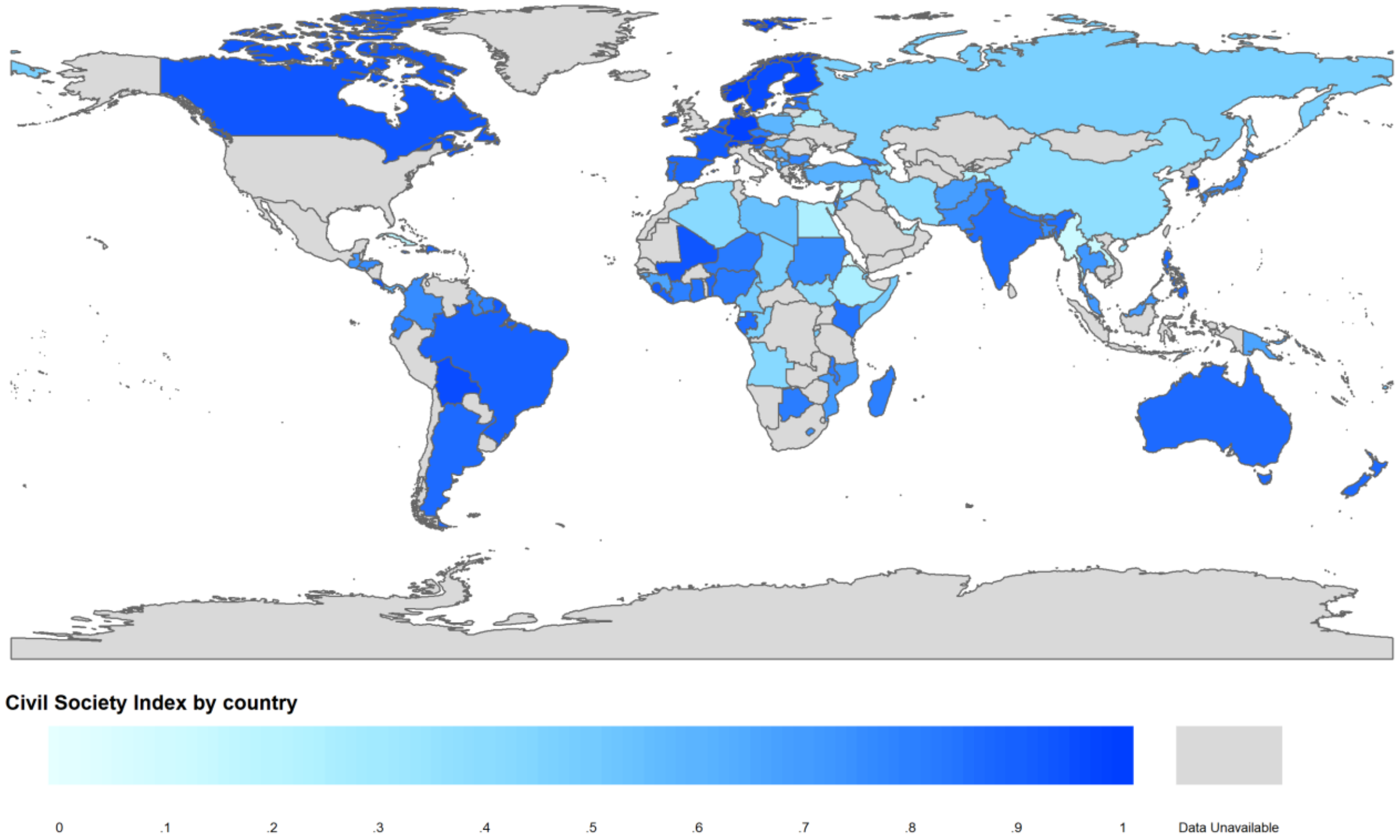

## S4.2 Civil society participation index, 2000-2021, by region

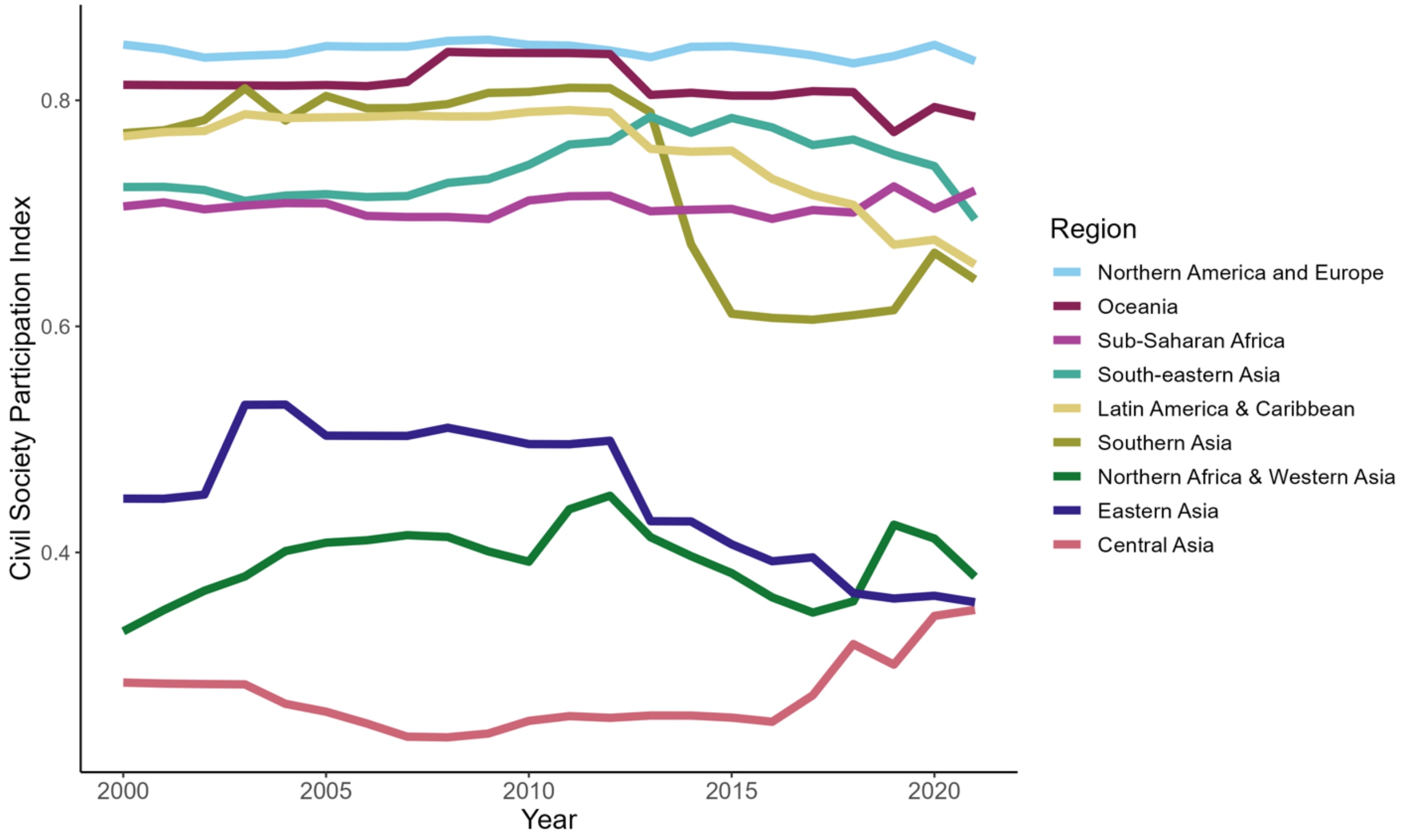

Population-weighted regional means.

## S4.3 Civil society participation index, 2000-2021, by income group

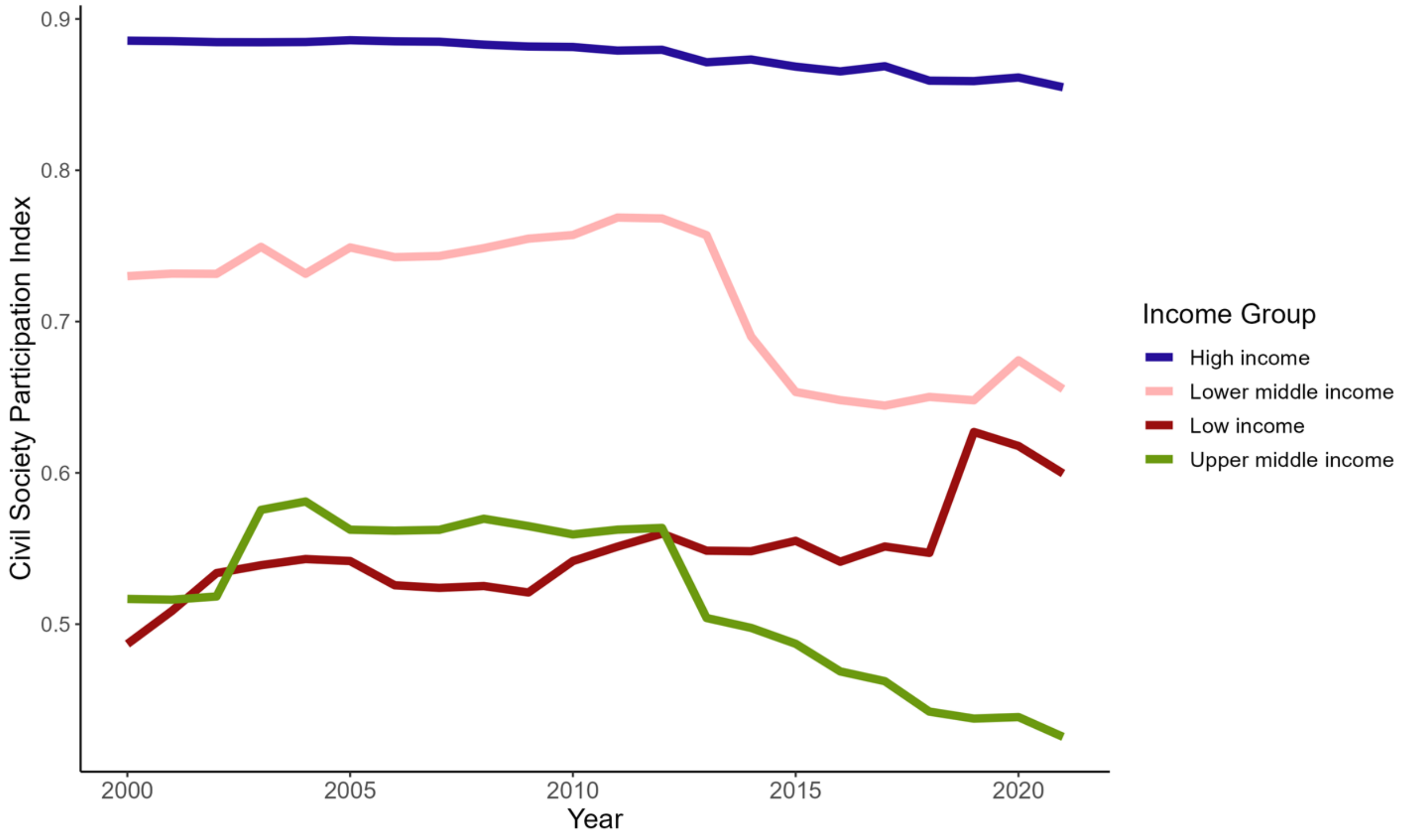

Population-weighted income group means.

## S4.4 Percent urban population living in cities signed onto the Milan Urban Food Policy Pact, 2020

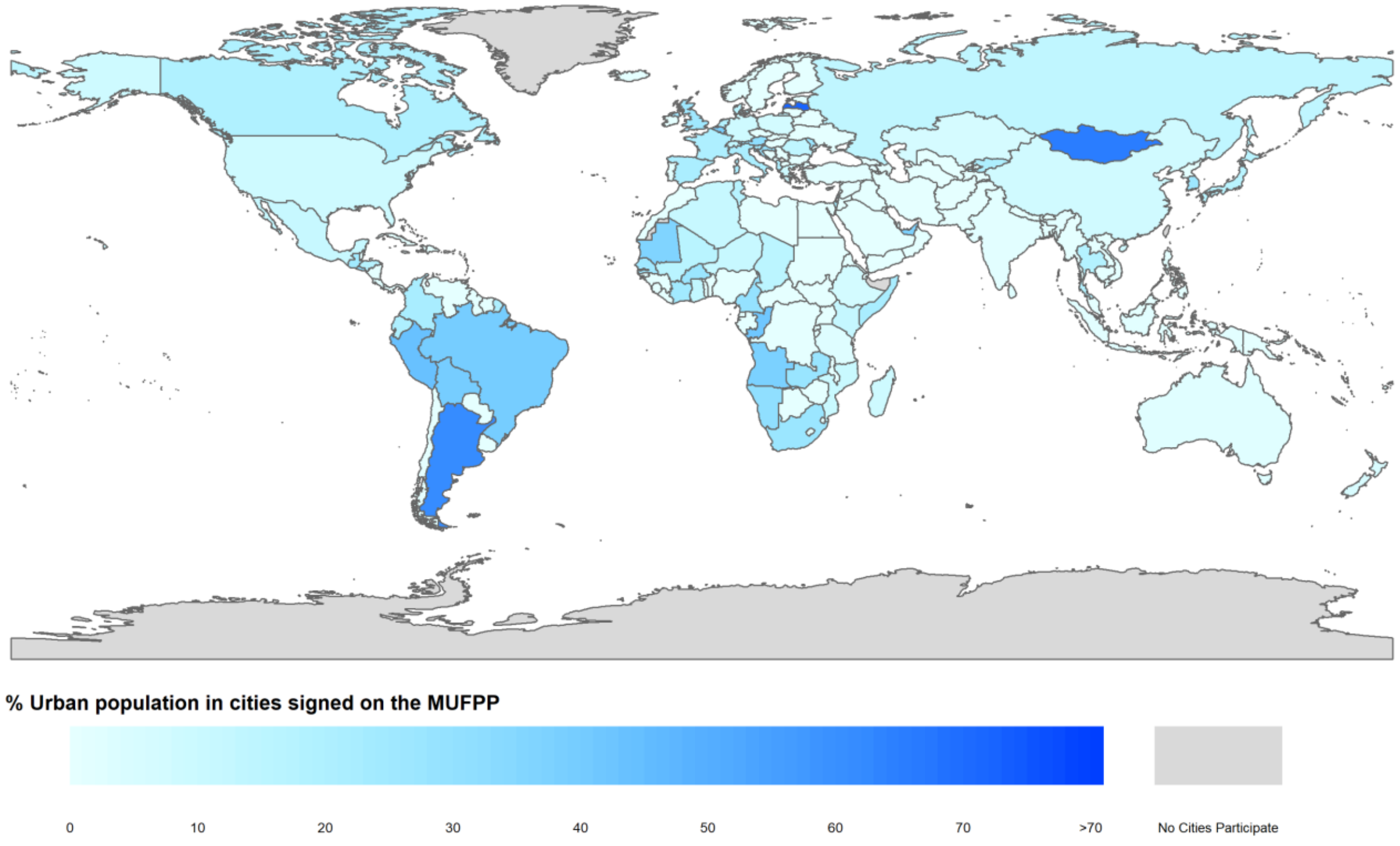

**S4.5 Percent urban population living in cities signed onto the Milan Urban Food Policy Pact, 2020, by country and region**

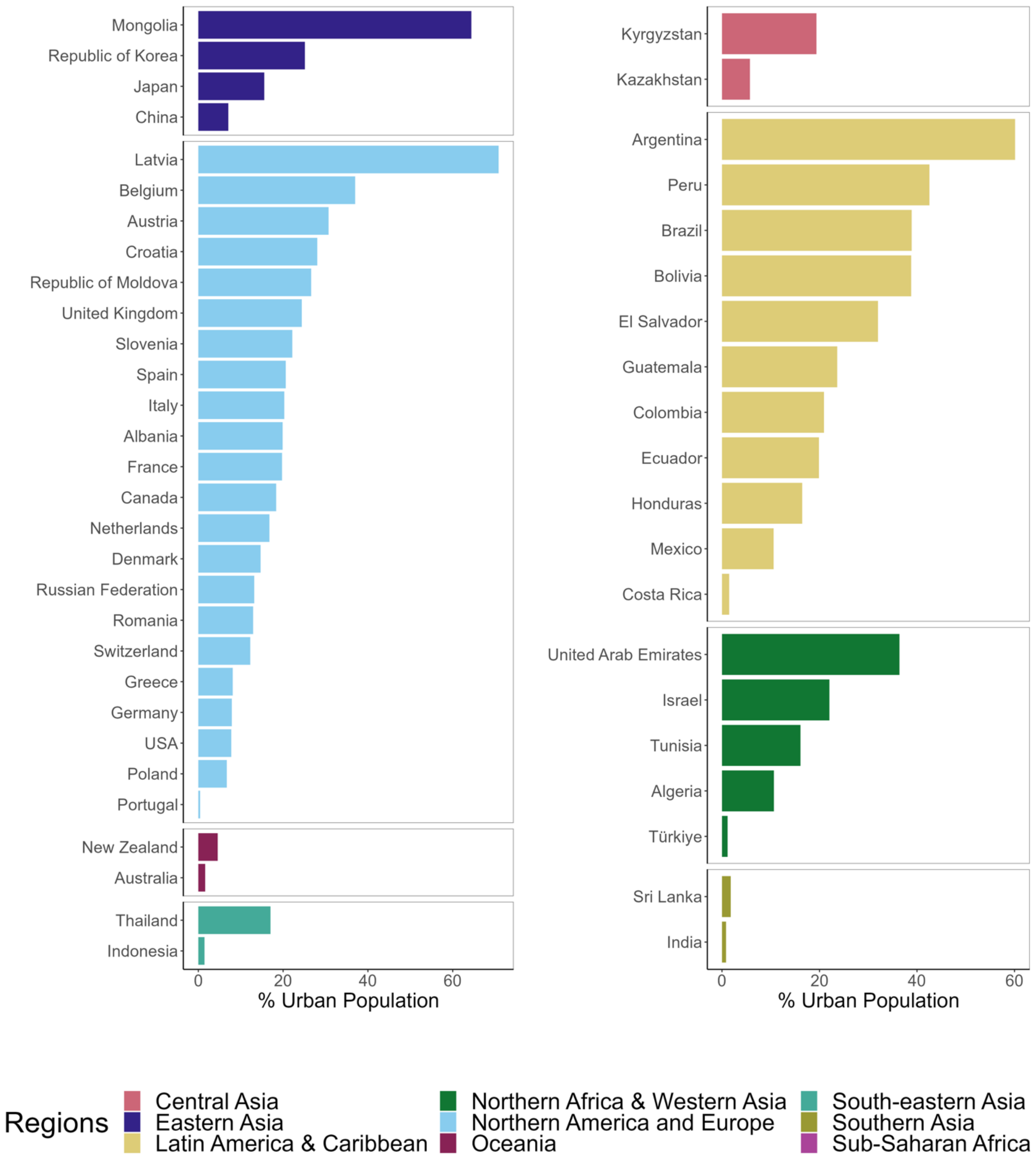

## S4.6 Degree of legal recognition of the Right to Food, 2021

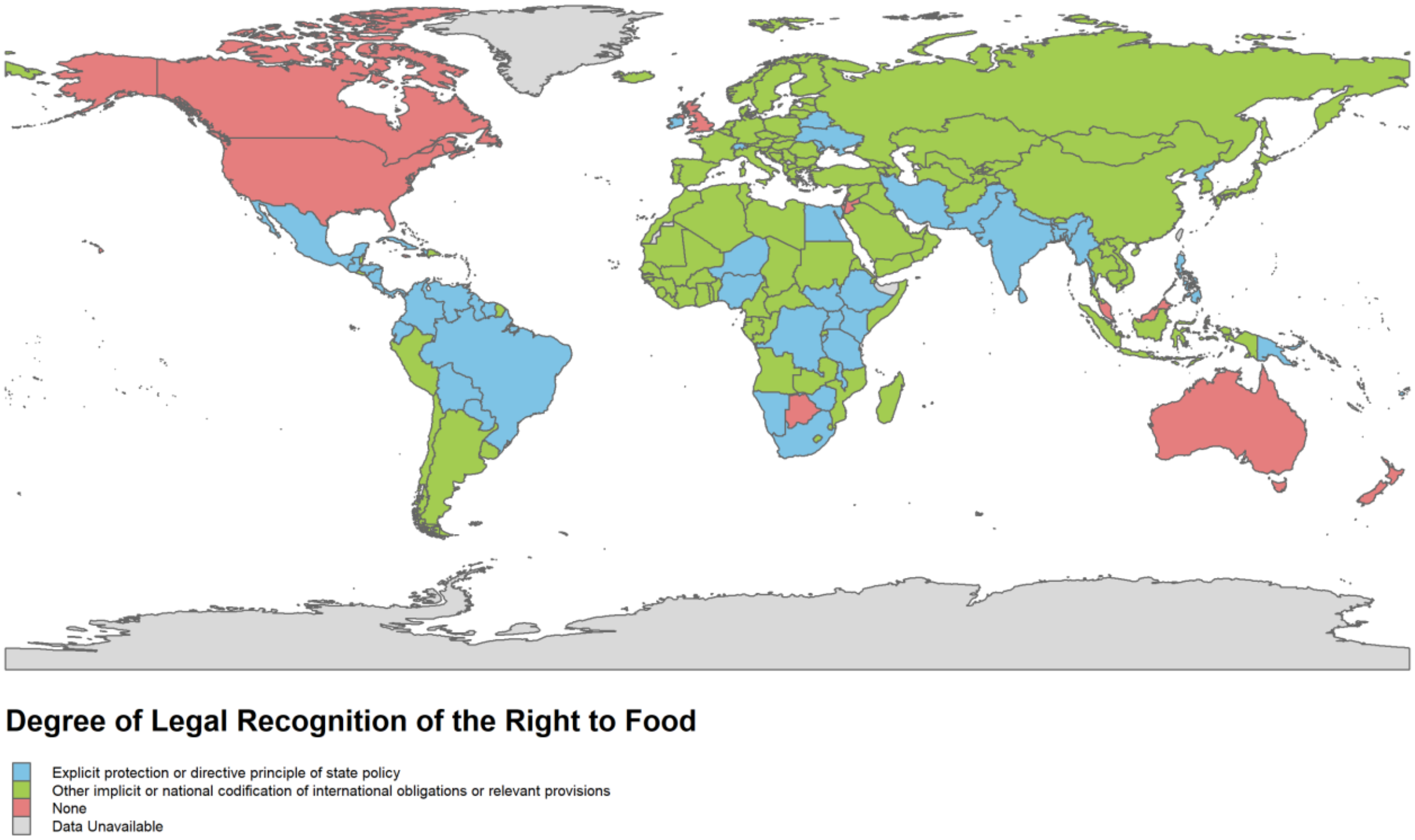

**Data as of 2021 update to the FAOLEX database.**

## S4.7 Presence of a food policy transformation pathway (from the UN Food Systems Summit processes), 2022

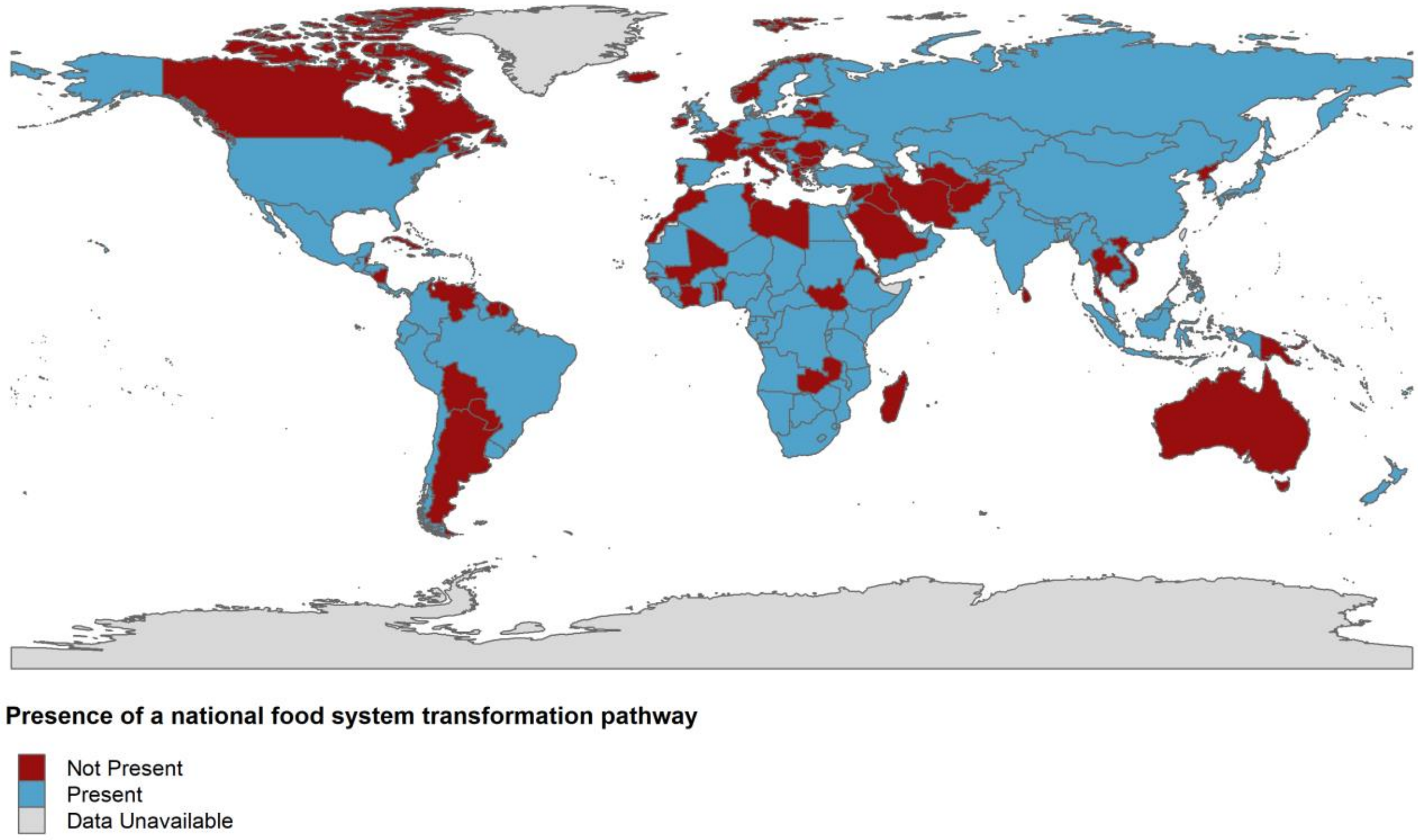

Data current as of October 2022.

## S4.8 Government Effectiveness Index, 2000-2020, by region

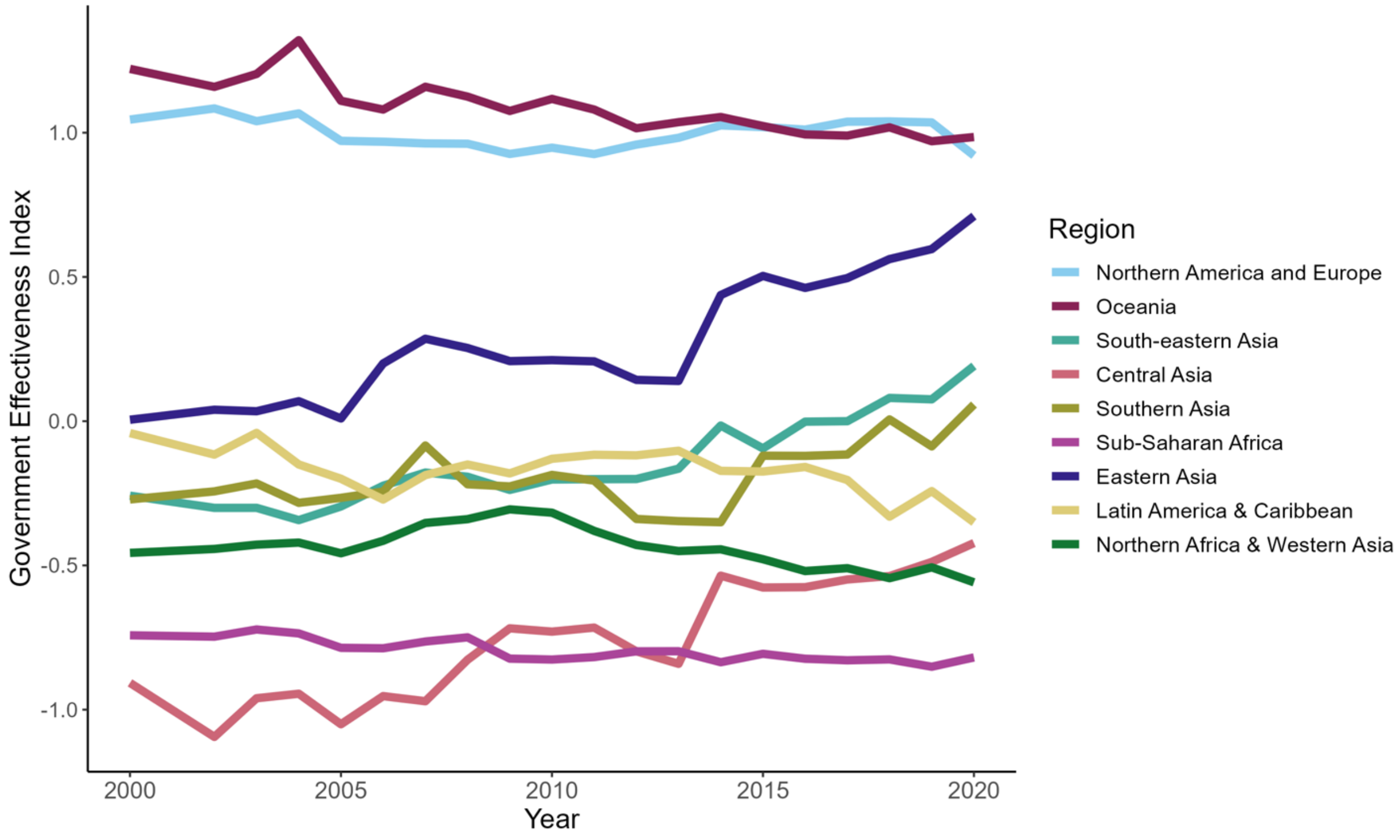

Population-weighted regional means.

## S4.9 Government Effectiveness Index, 2000-2020, by income group

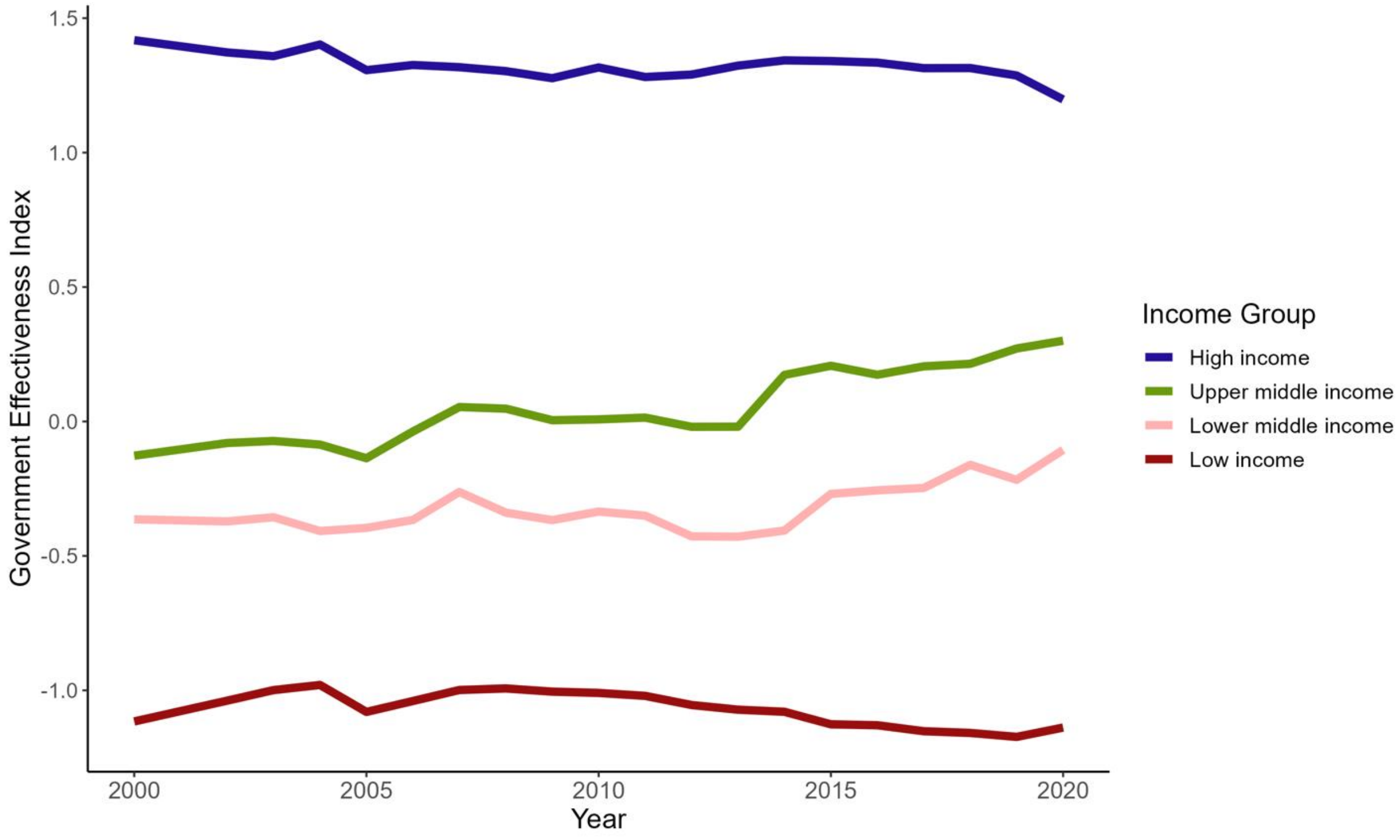

Population-weighted income group means.

### S4.10 International Health Regulations State Party Assessment report (IHR SPAR), Food safety capacity, 2020

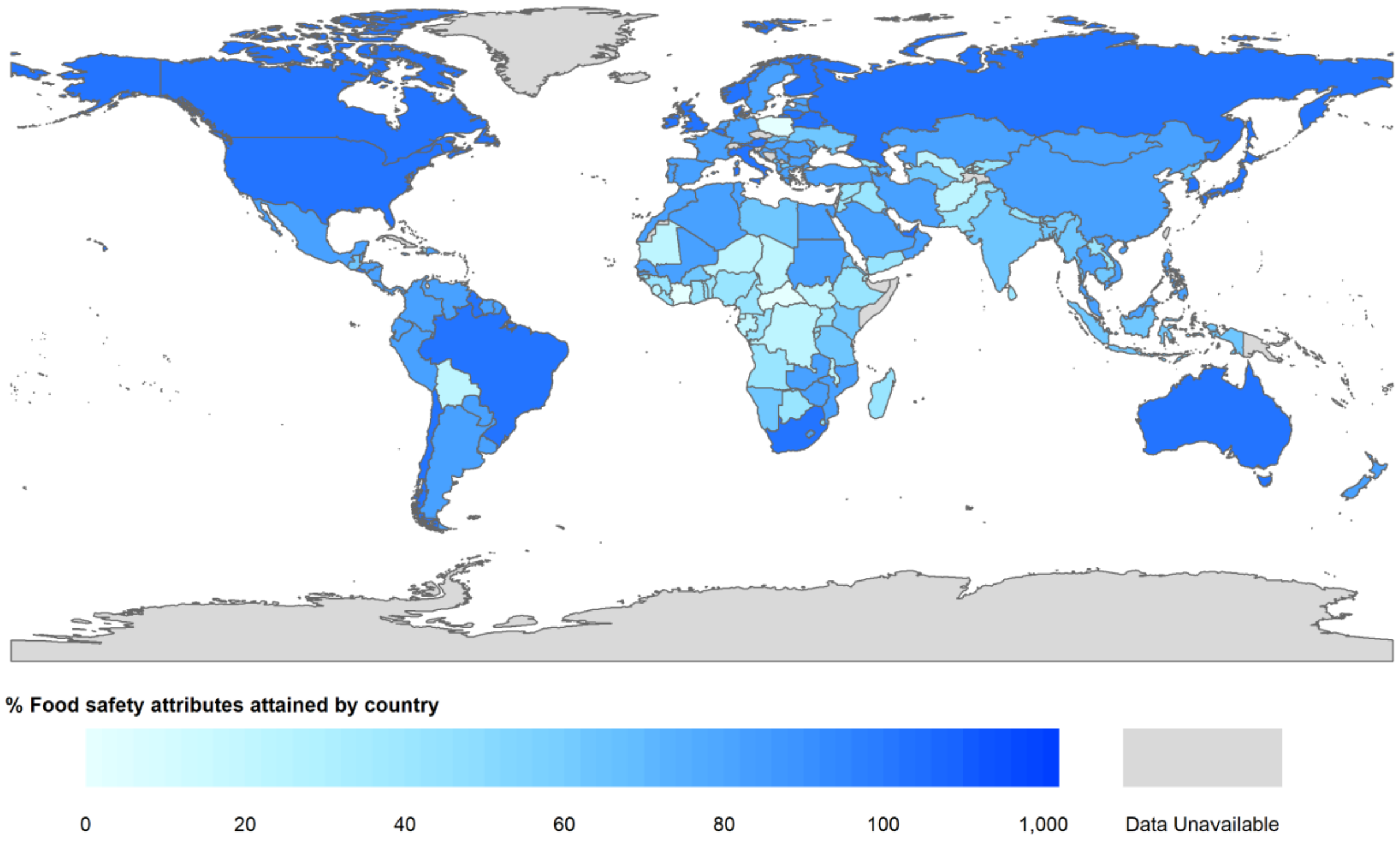

### S4.11 Presence of health-related food taxes, 2021

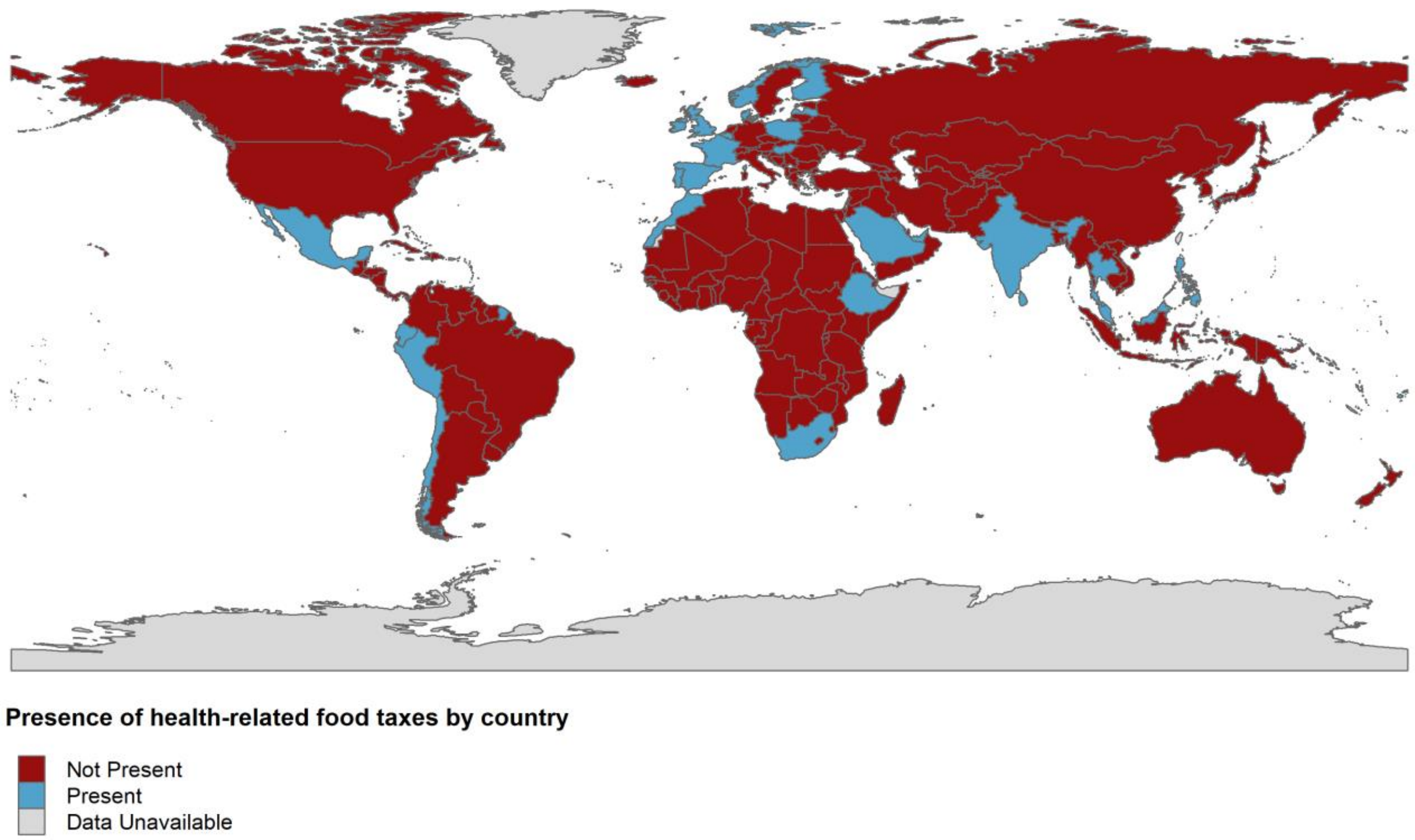

Data as of 2021 update to the World Cancer Research Fund NOURISHING database.

**S4.12 V-Dem Accountability index, 2021**

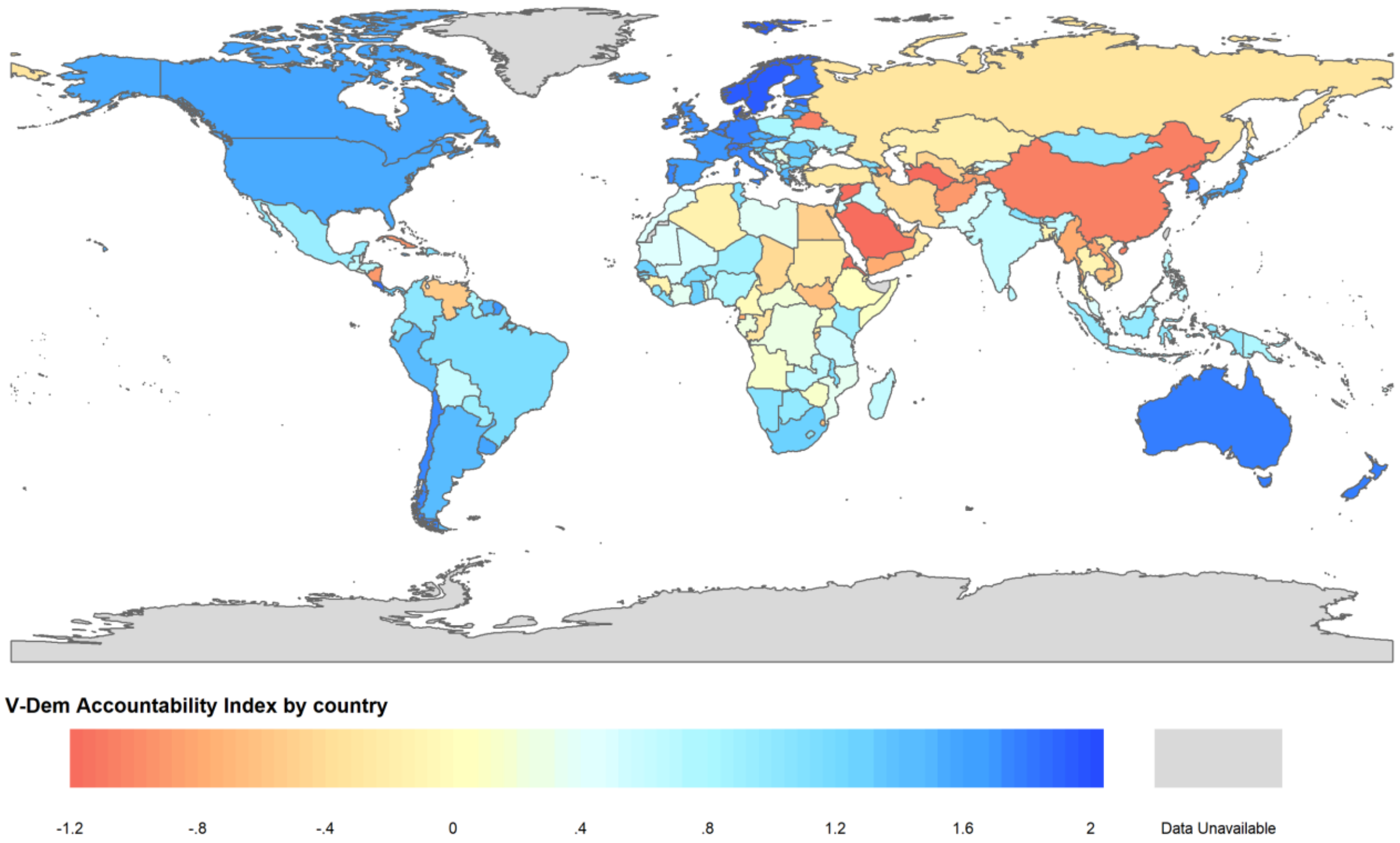

**S4.13 V-Dem Accountability index, 2000-2021, by region**

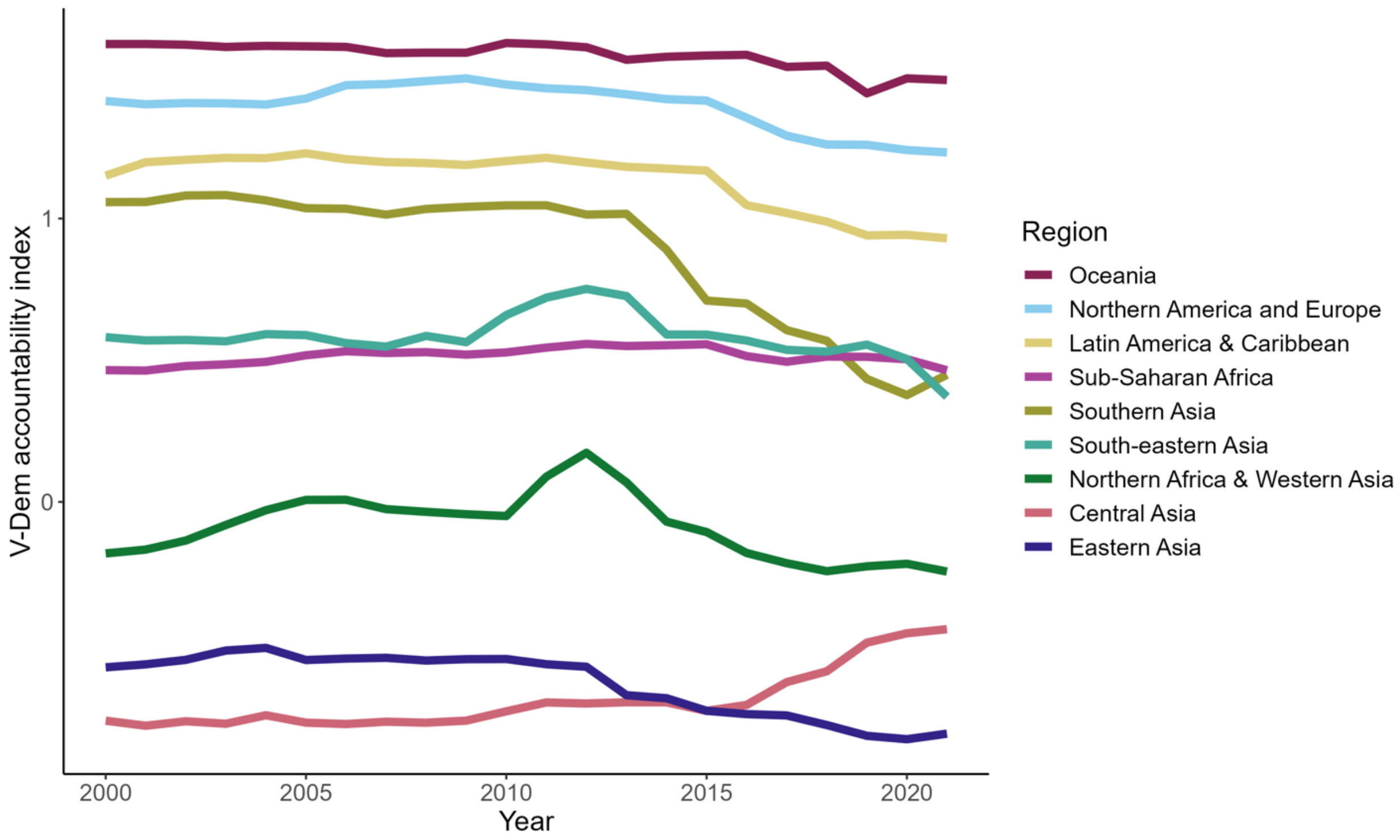

Population-weighted regional means.

## S4.14 V-Dem Accountability index, 2000-2021, by income group

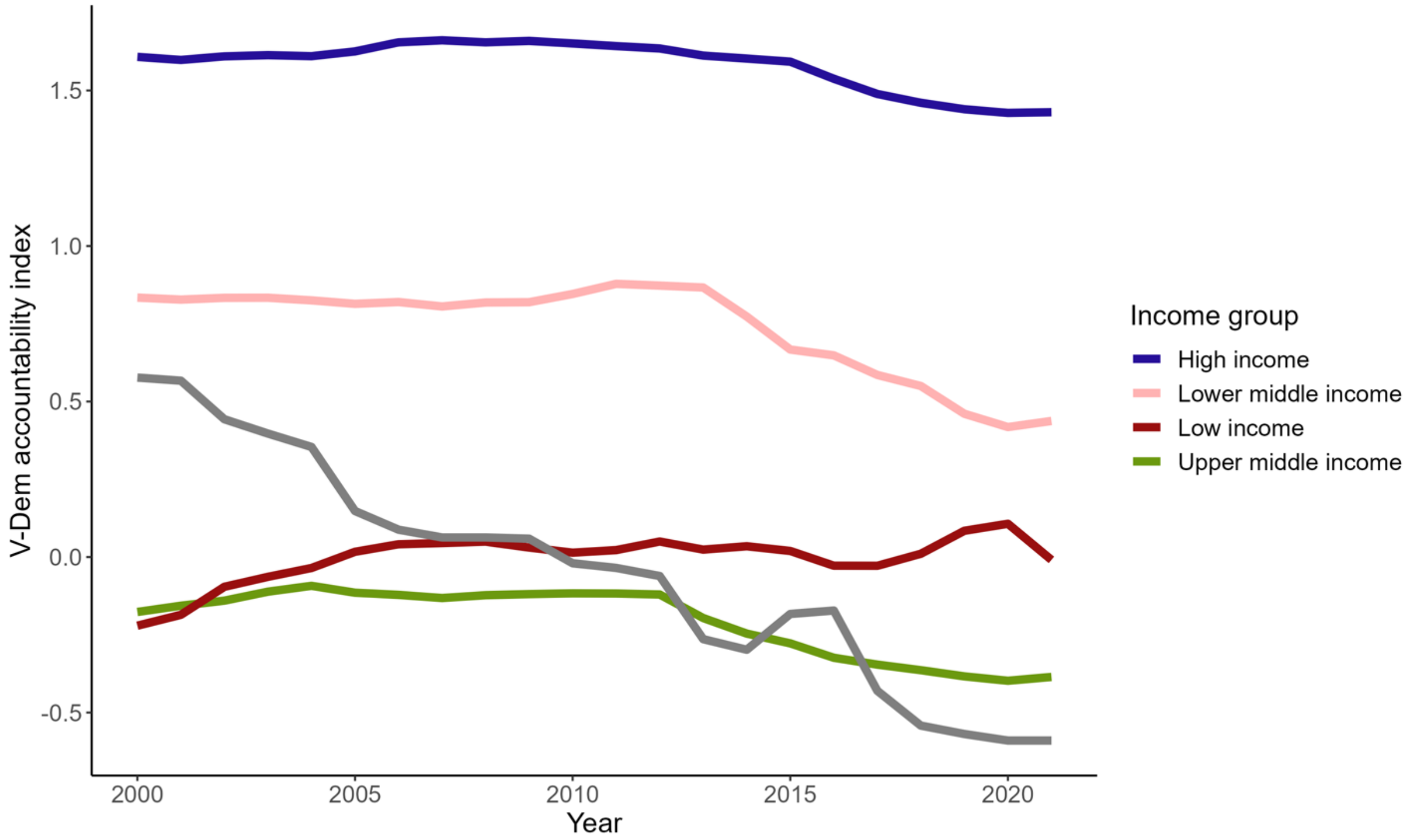

Population-weighted income group means.

## S4.15 Open Budget Index score, by income group

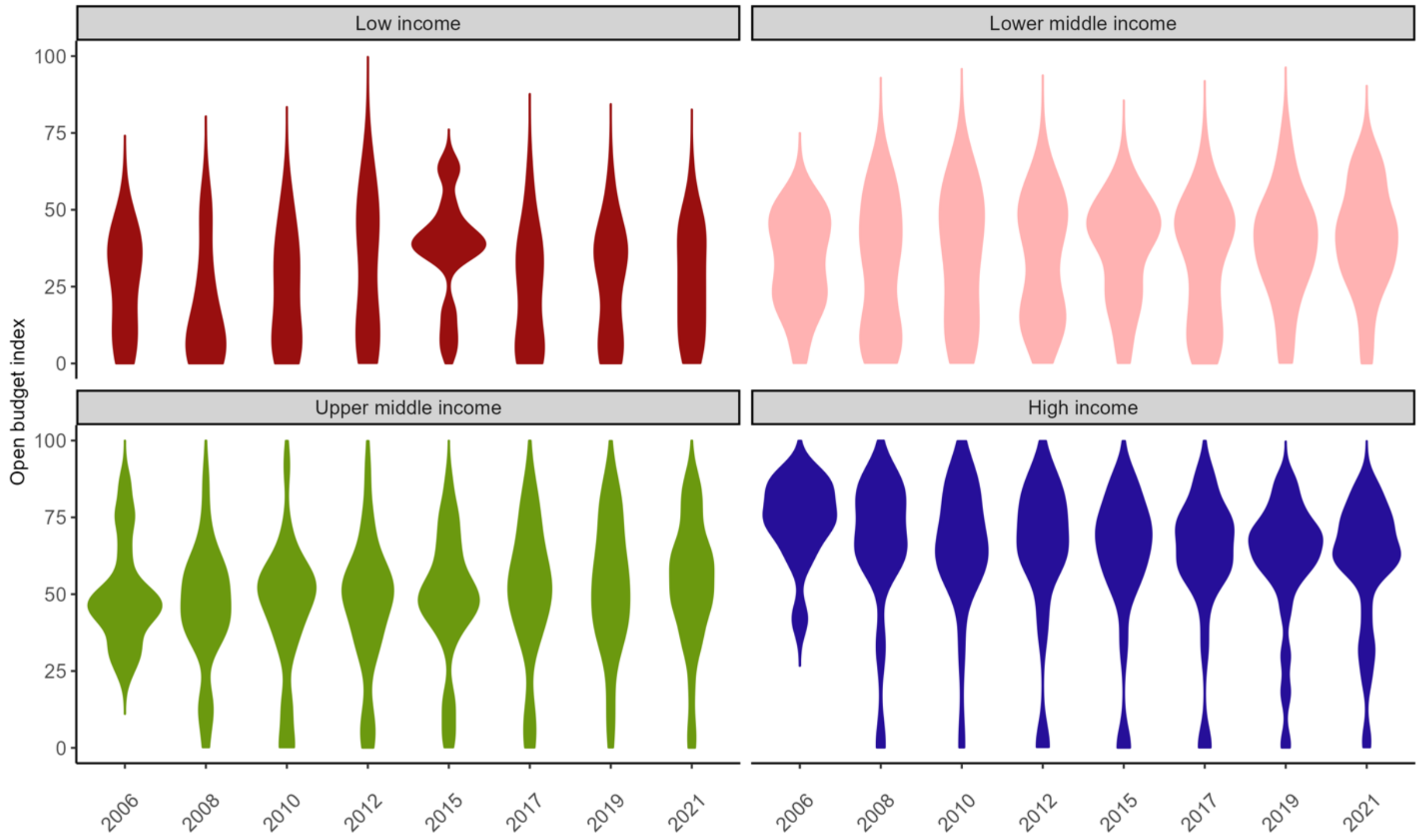

## S4.16 Guarantees for public access to information, by year of adoption

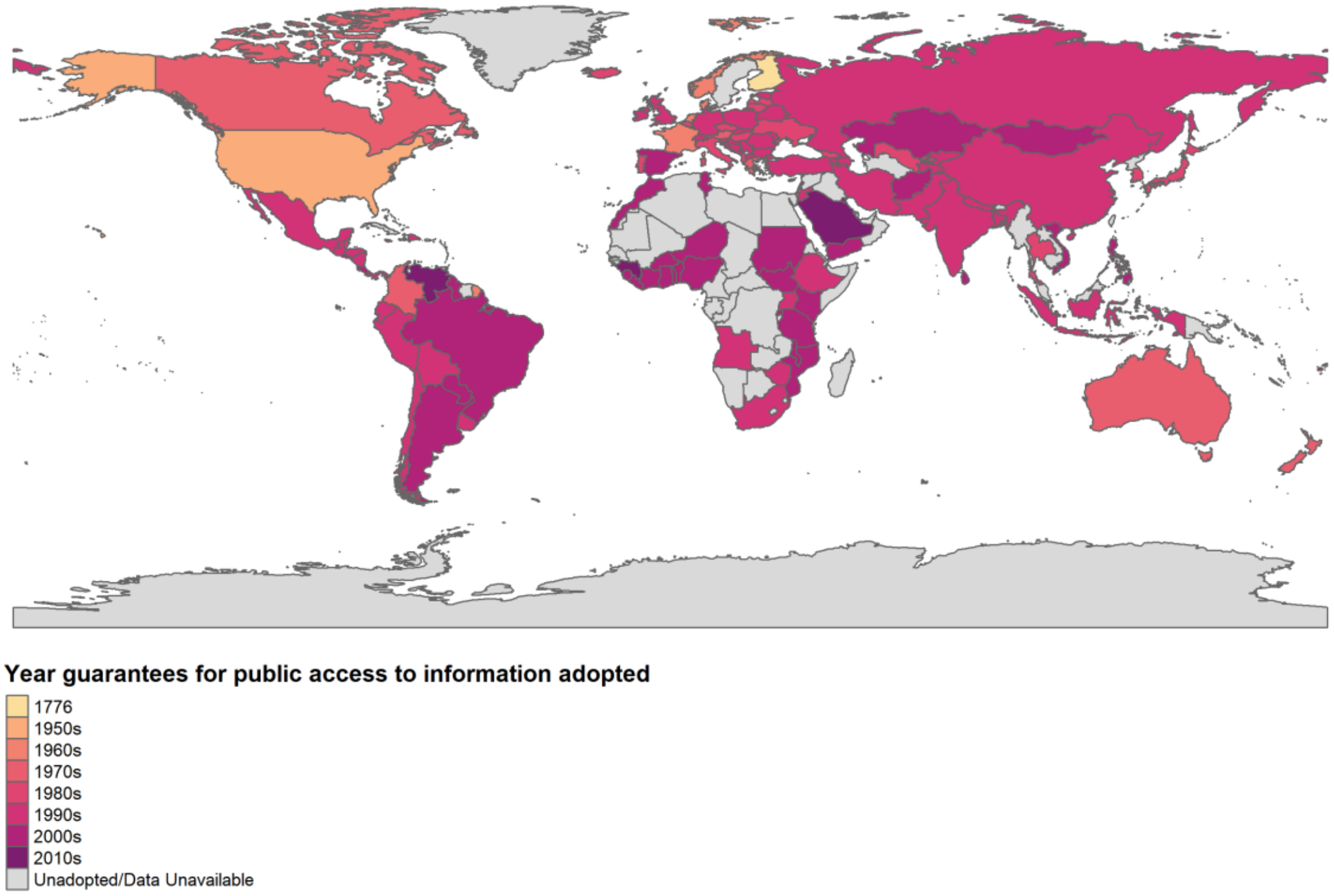

### S5.1 Exposure to shocks: Ratio of total damages of all disasters to GDP, 2012-2021

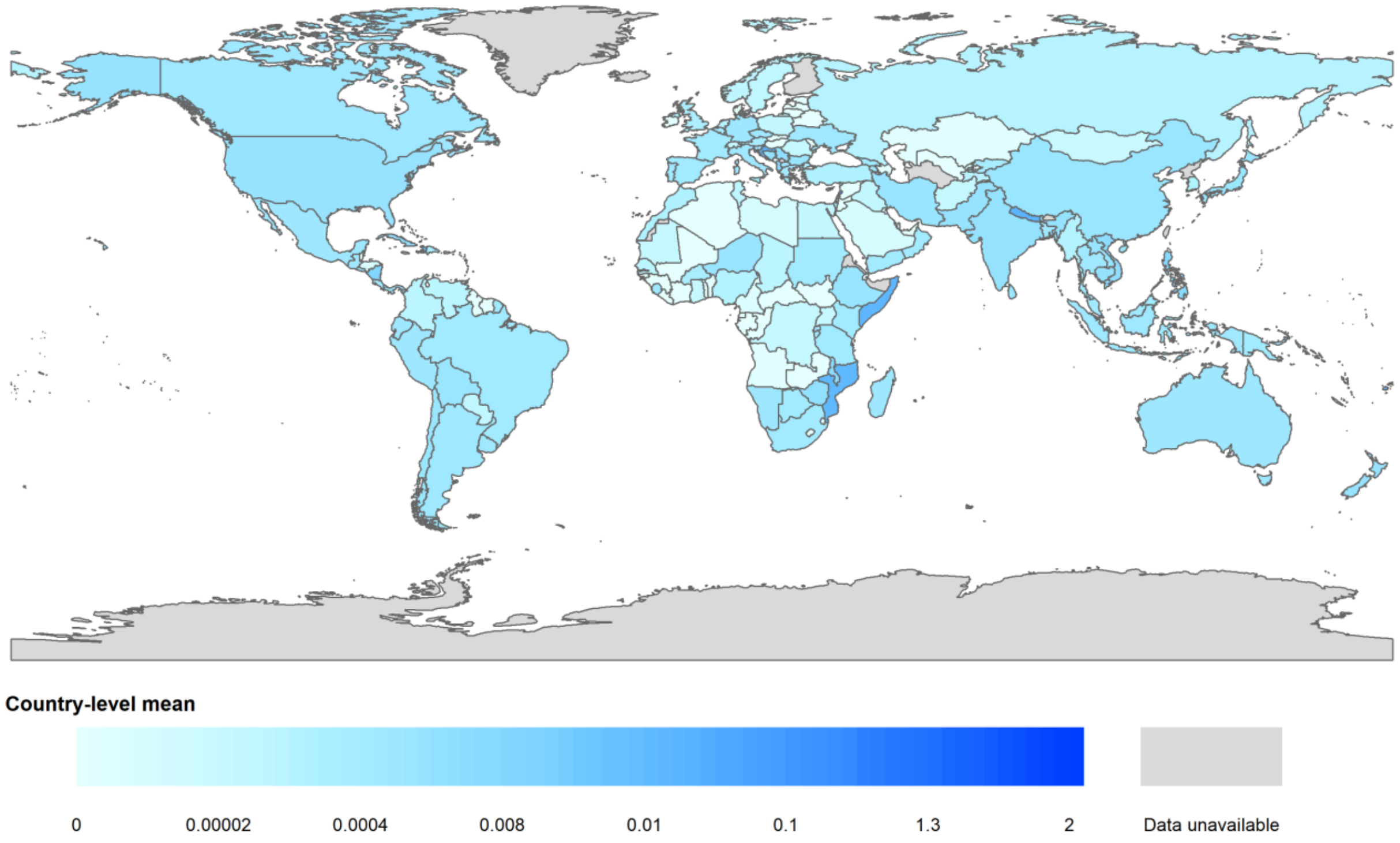

**S5.2 Dietary sourcing flexibility index, by nutrient / food group and country income level, 2018**

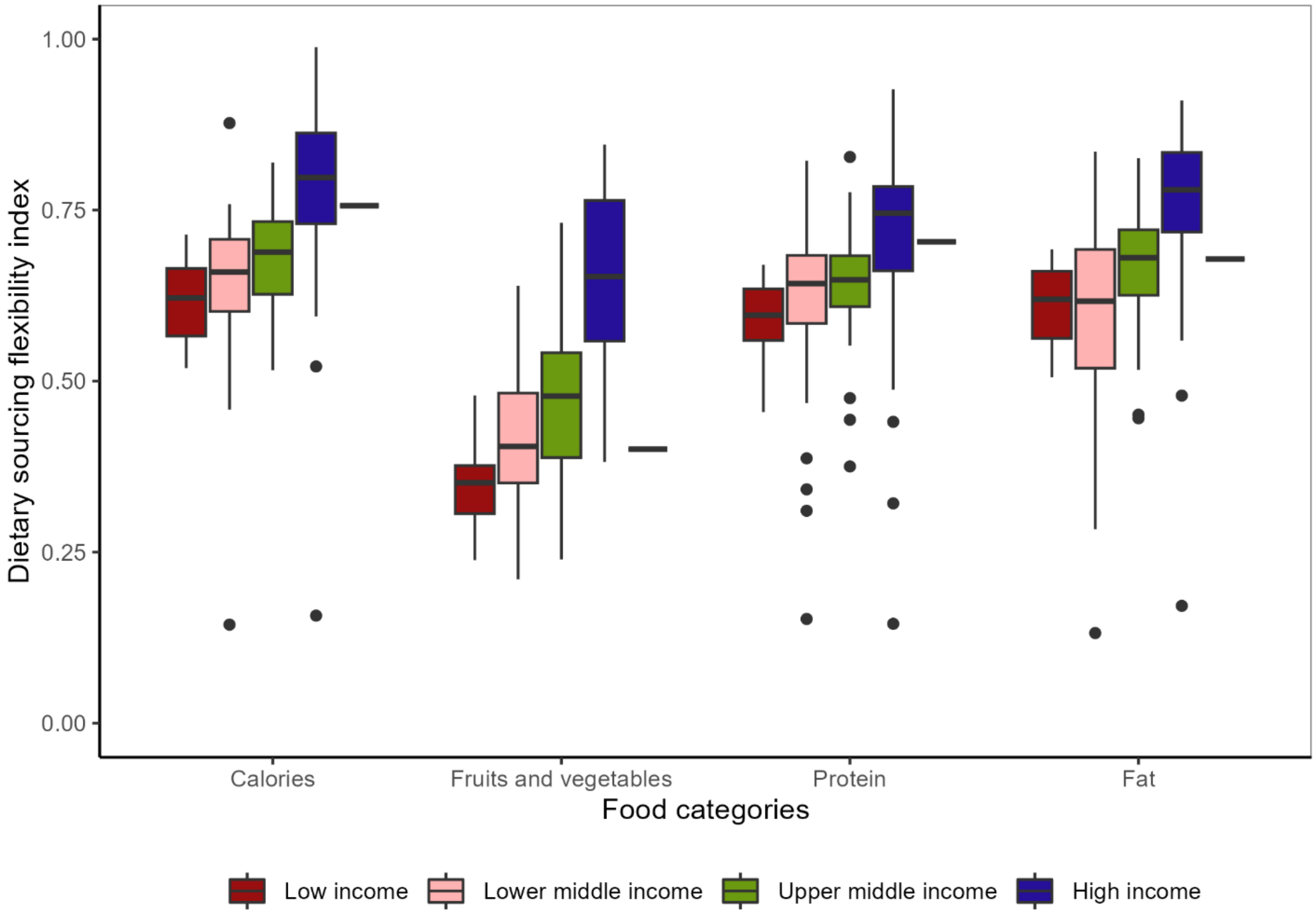

Calories, fruits and vegetables, protein, and fat are the total value from all sources.

### S5.3 Dietary sourcing flexibility index, Calories, 2018

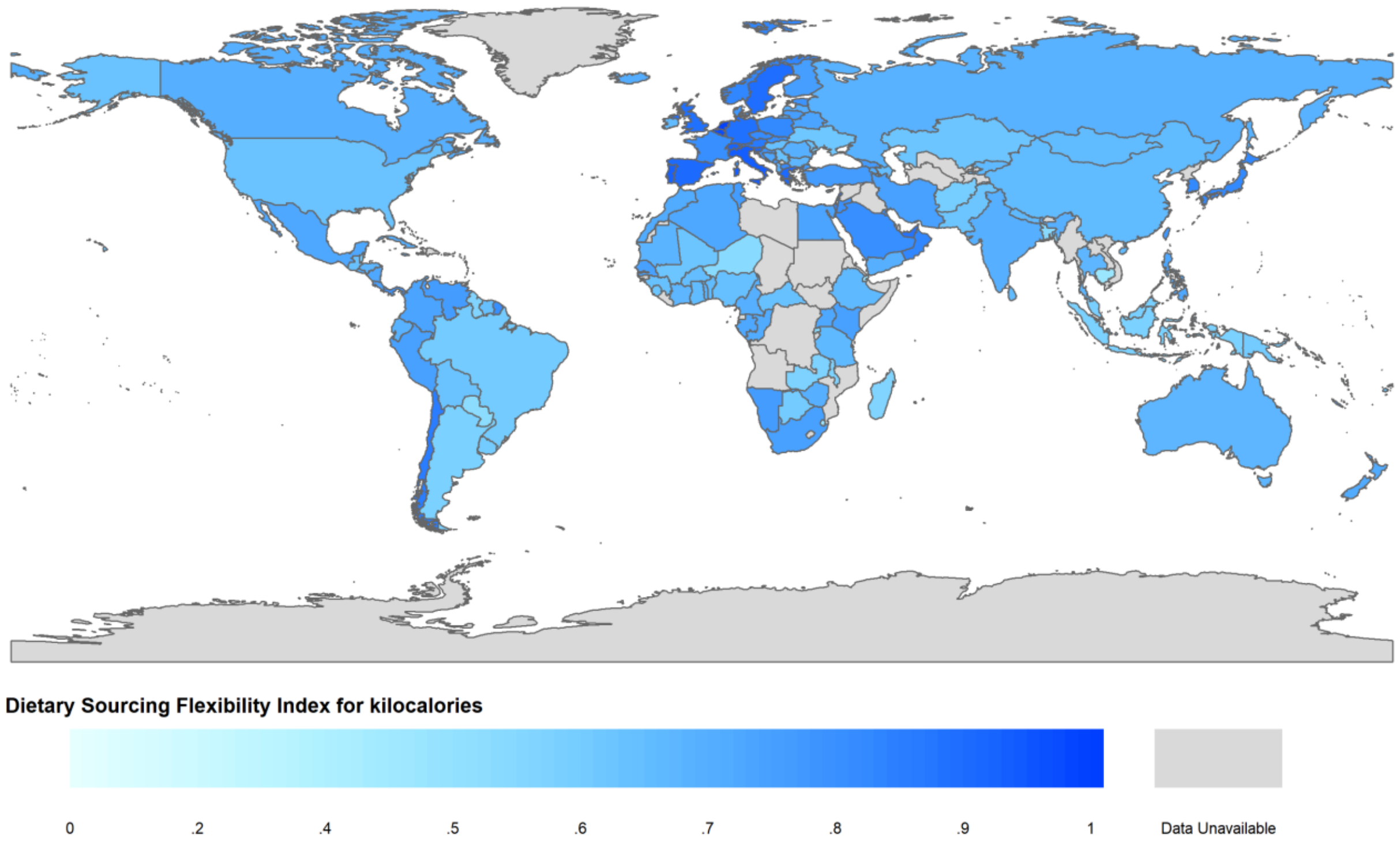

### S5.4 Mobile cellular subscriptions (per 100 people), 2000-2020, by region

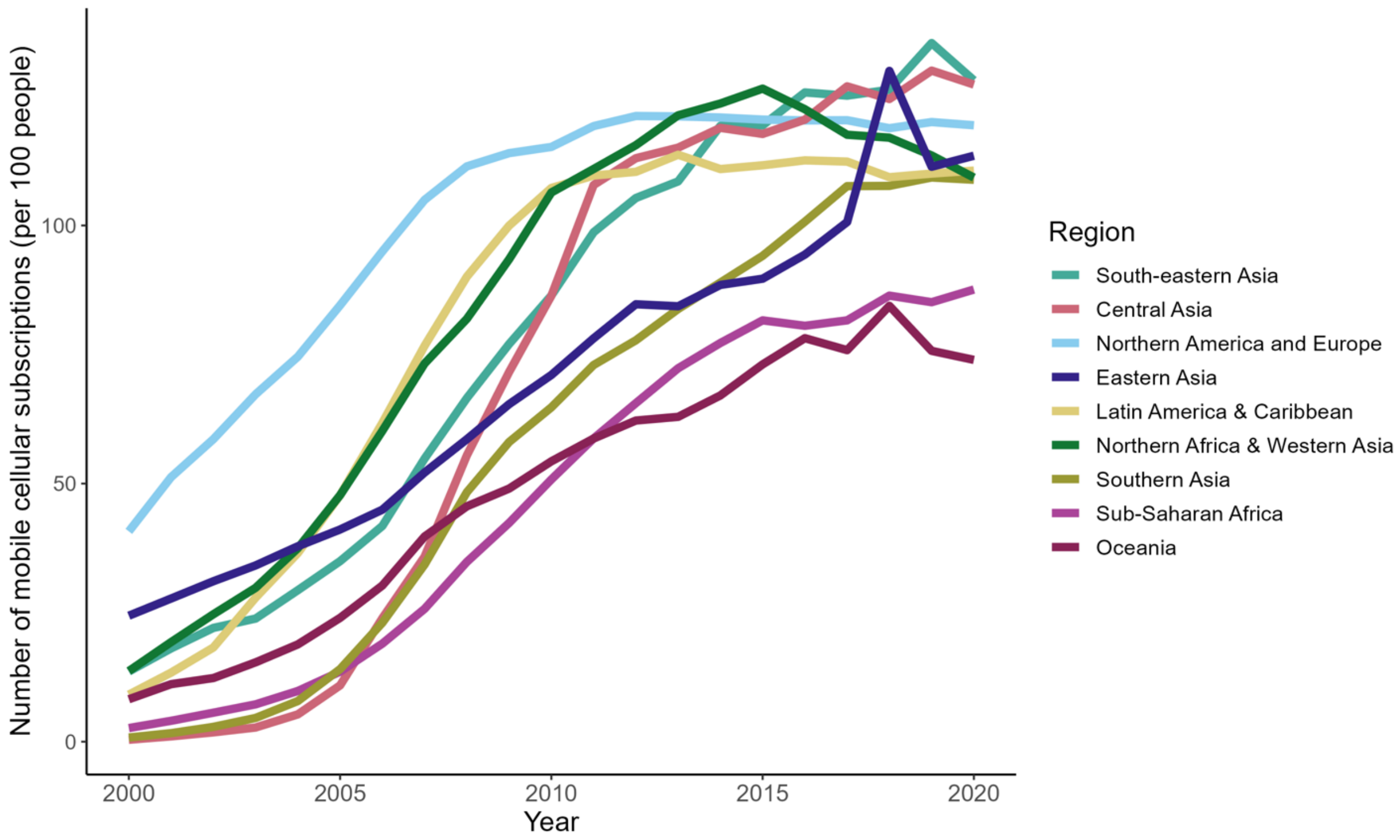

Unweighted mean by region.

## S5.5 Mobile cellular subscriptions (per 100 people), 2011-2020 average

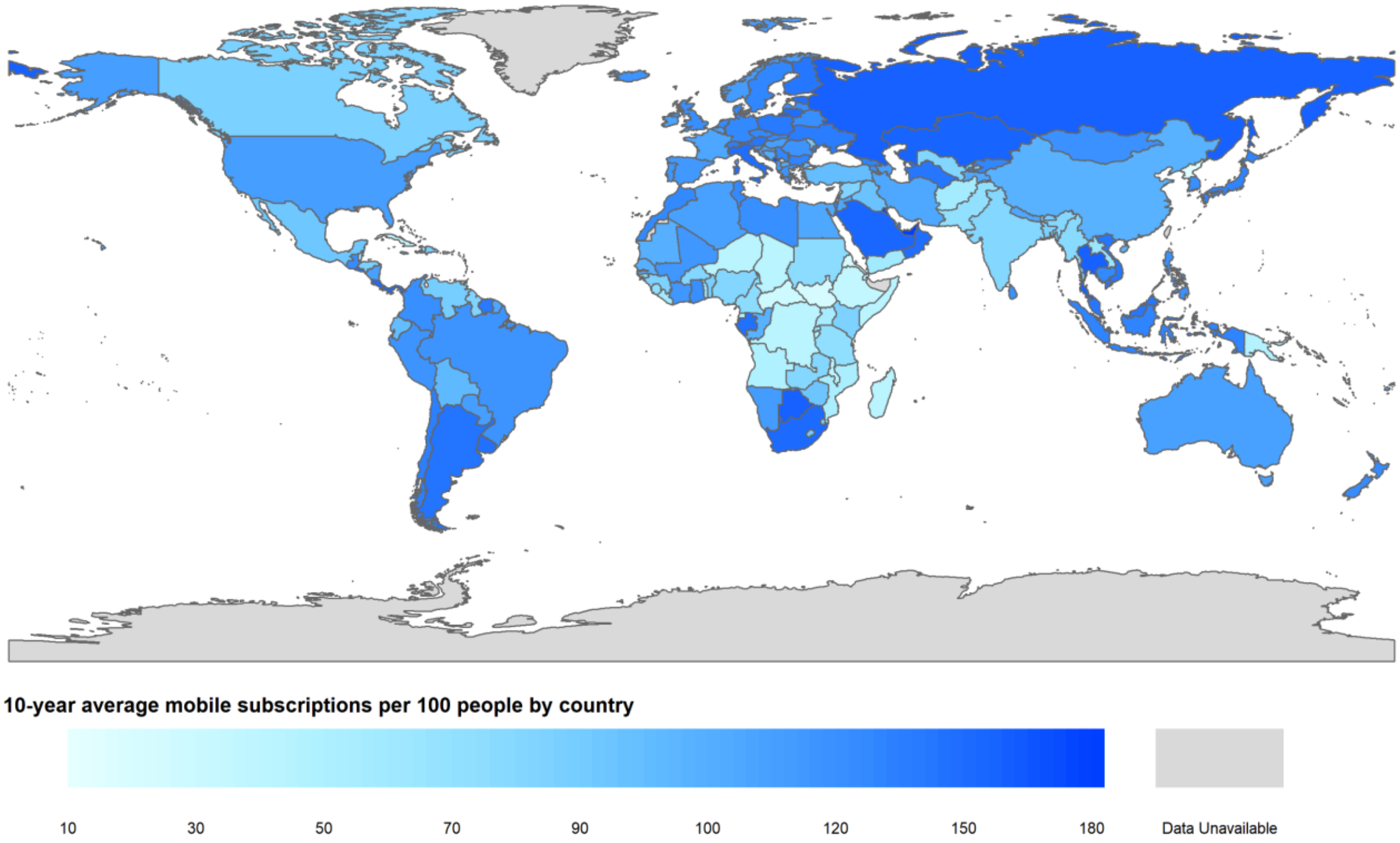

## S5.6 Social capital index, 2007-2021, by region

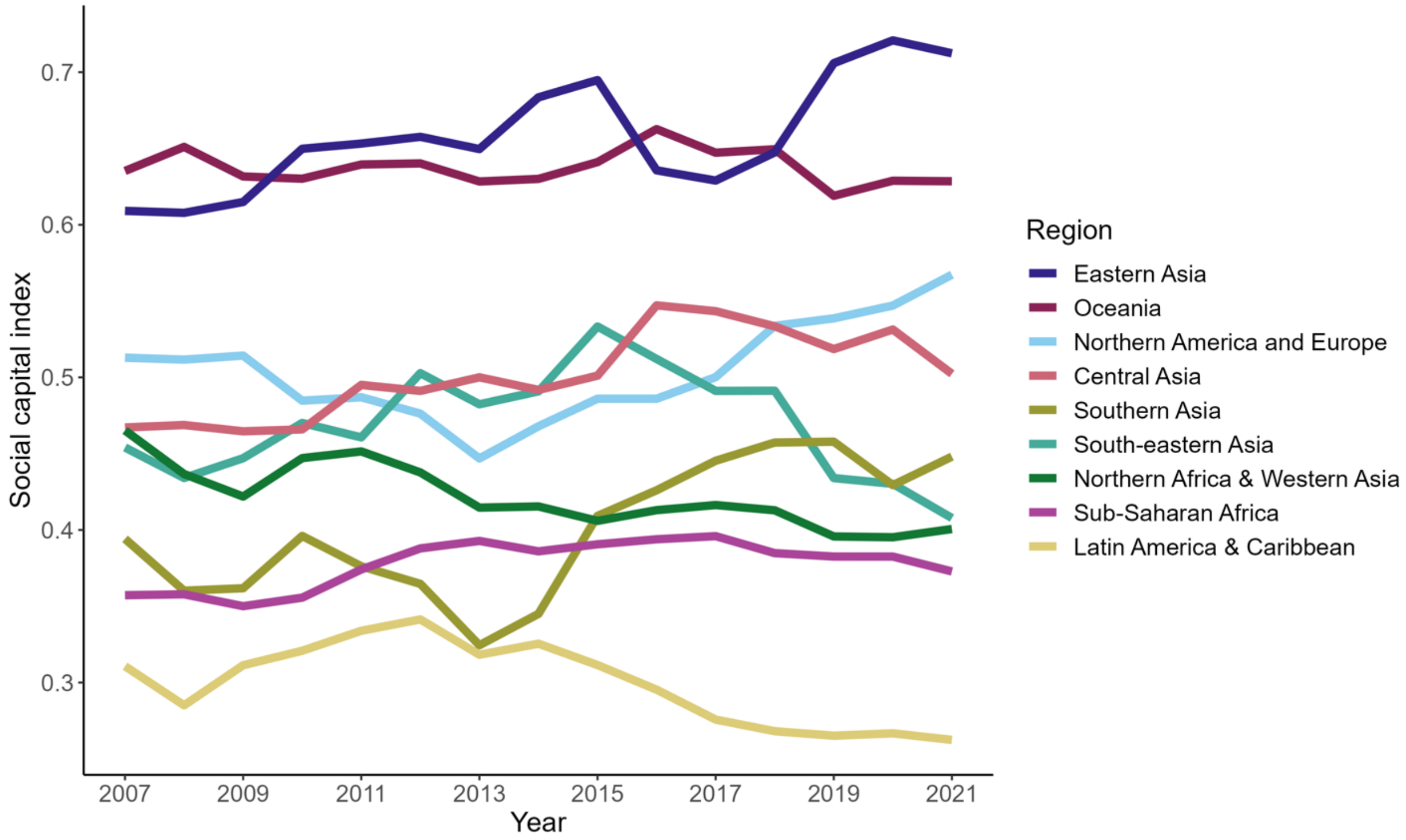

Population-weighted regional means.

## S5.7 Social capital index, 2007-2021, by income group

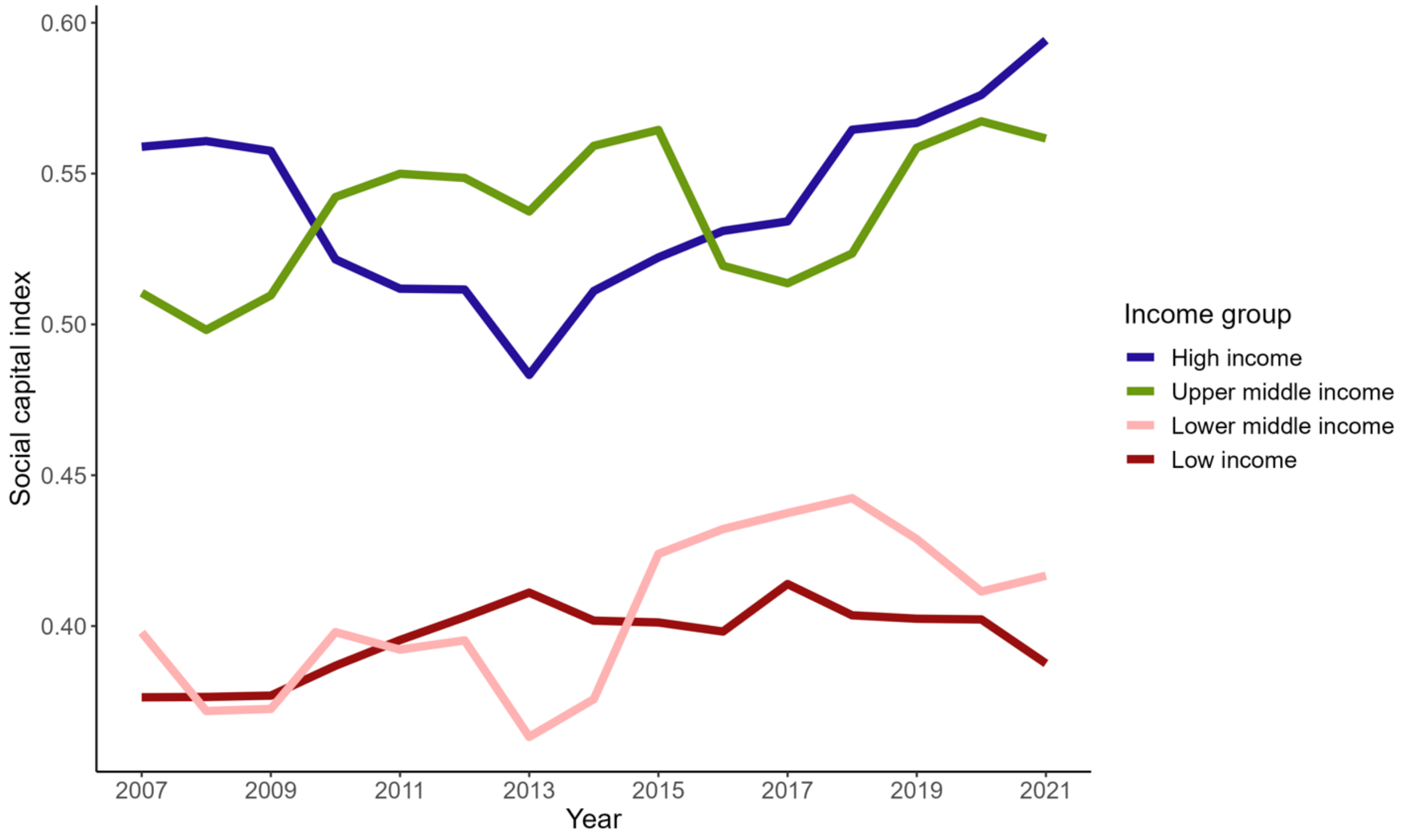

Population-weighted income group means

## S5.8 Social capital index, 2017-2021

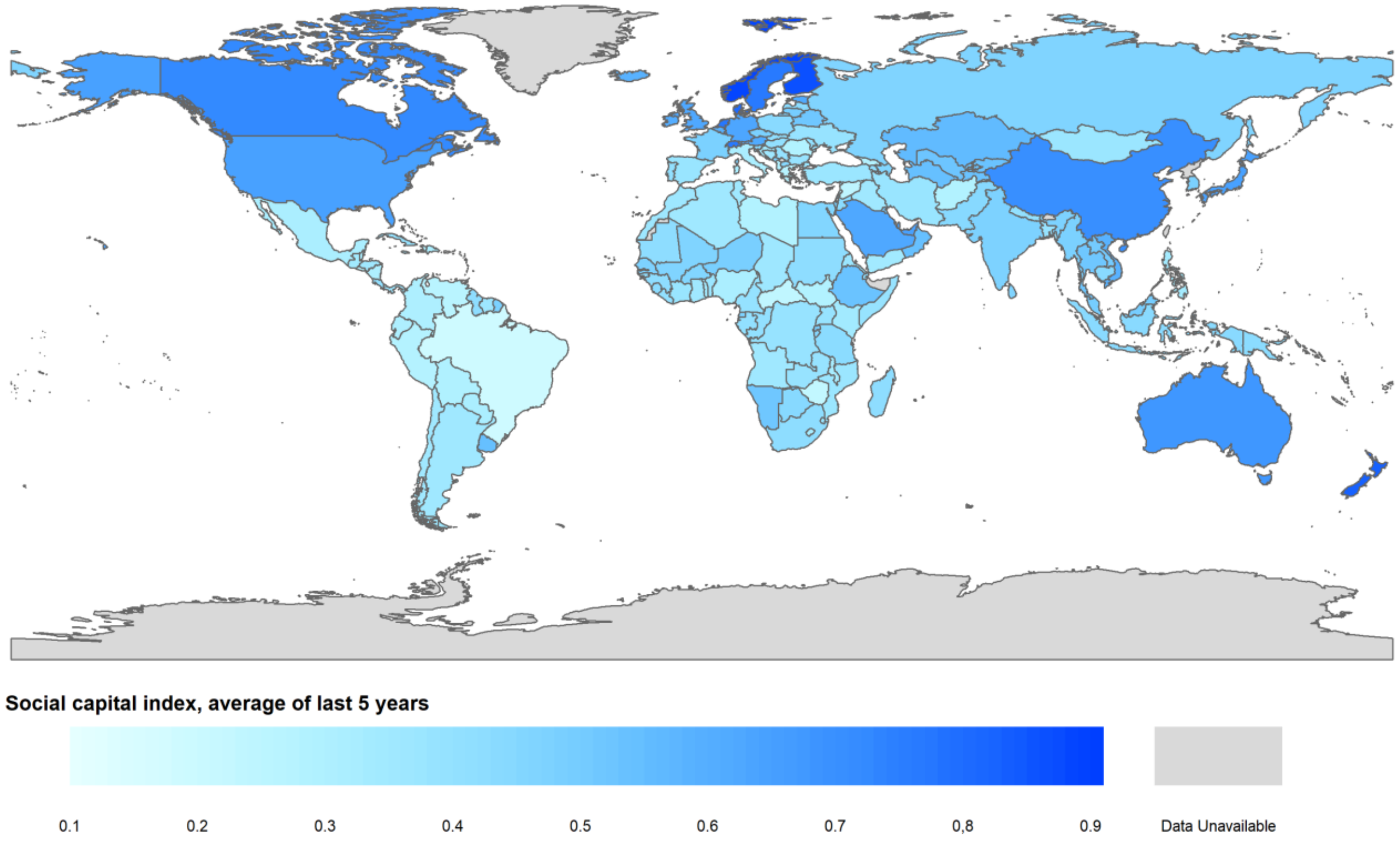

### S5.9 Percentage of agricultural land with minimum level of species diversity (crop and pasture), 2010

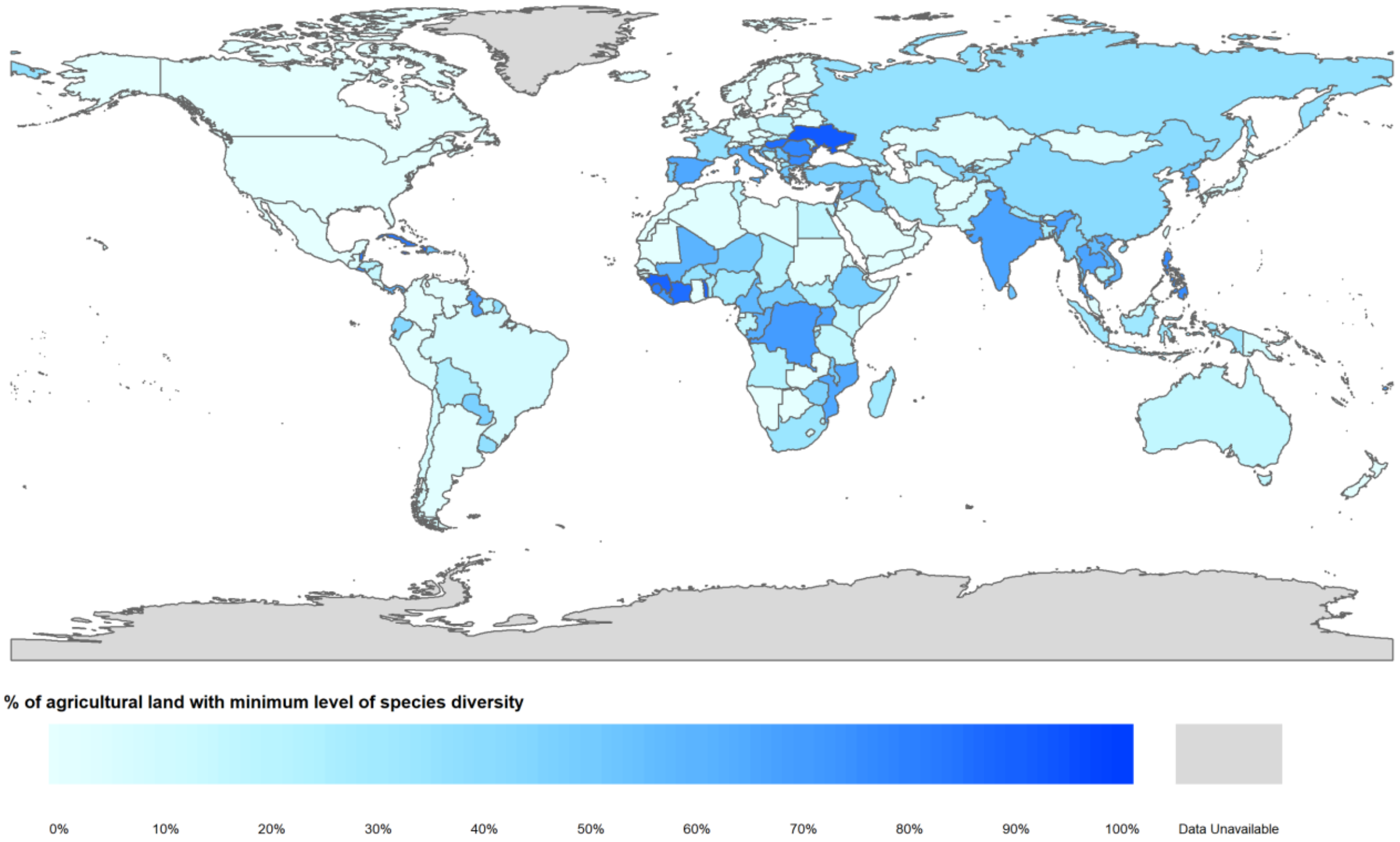

### S5.10 Number of wild useful plants for food and agriculture secured in conservation facilities (SDG 5.2.a), 2020

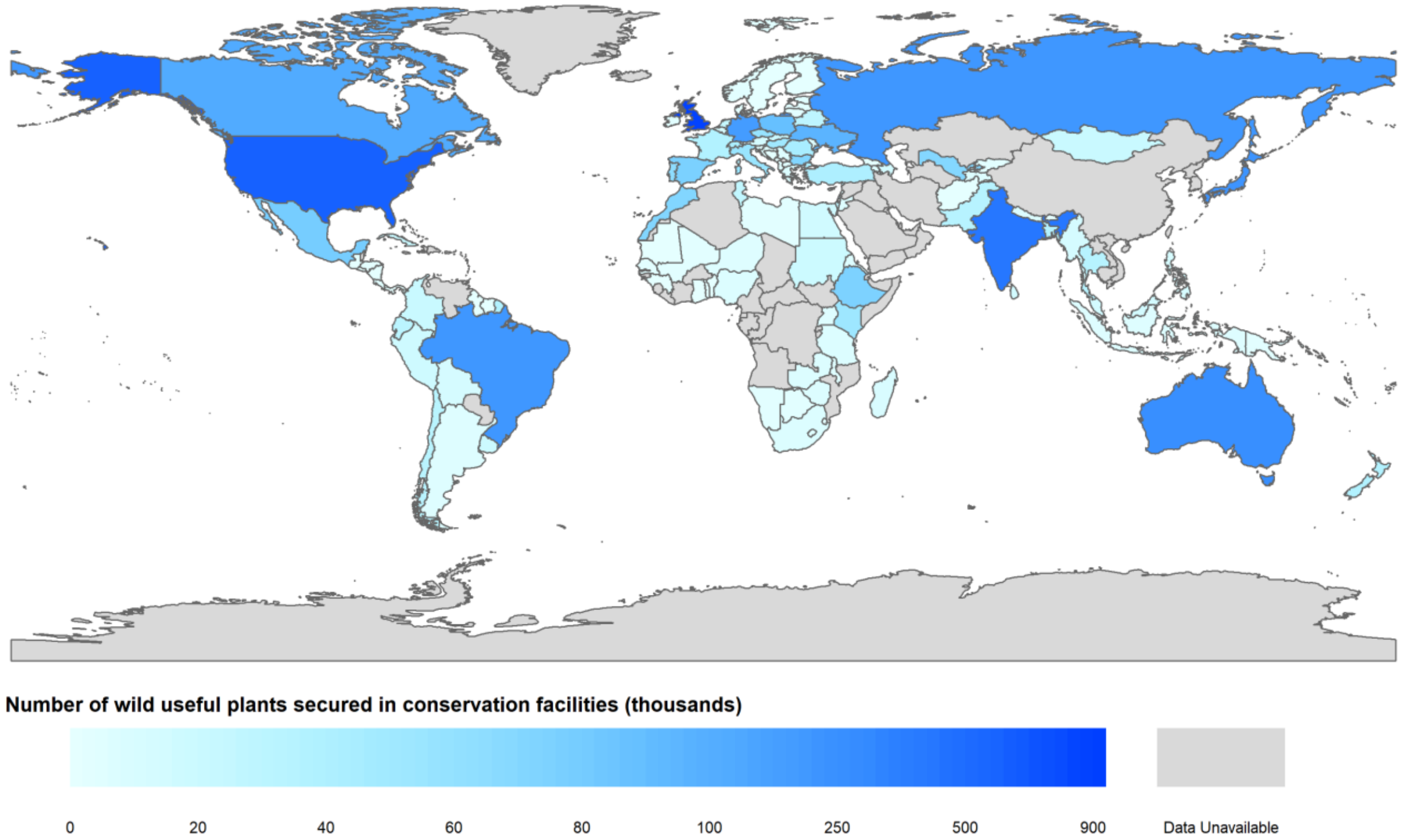

## S5.11 Number of animal genetic resources for food and agriculture secured in conservation facilities (SDG 5.2.b), 2021

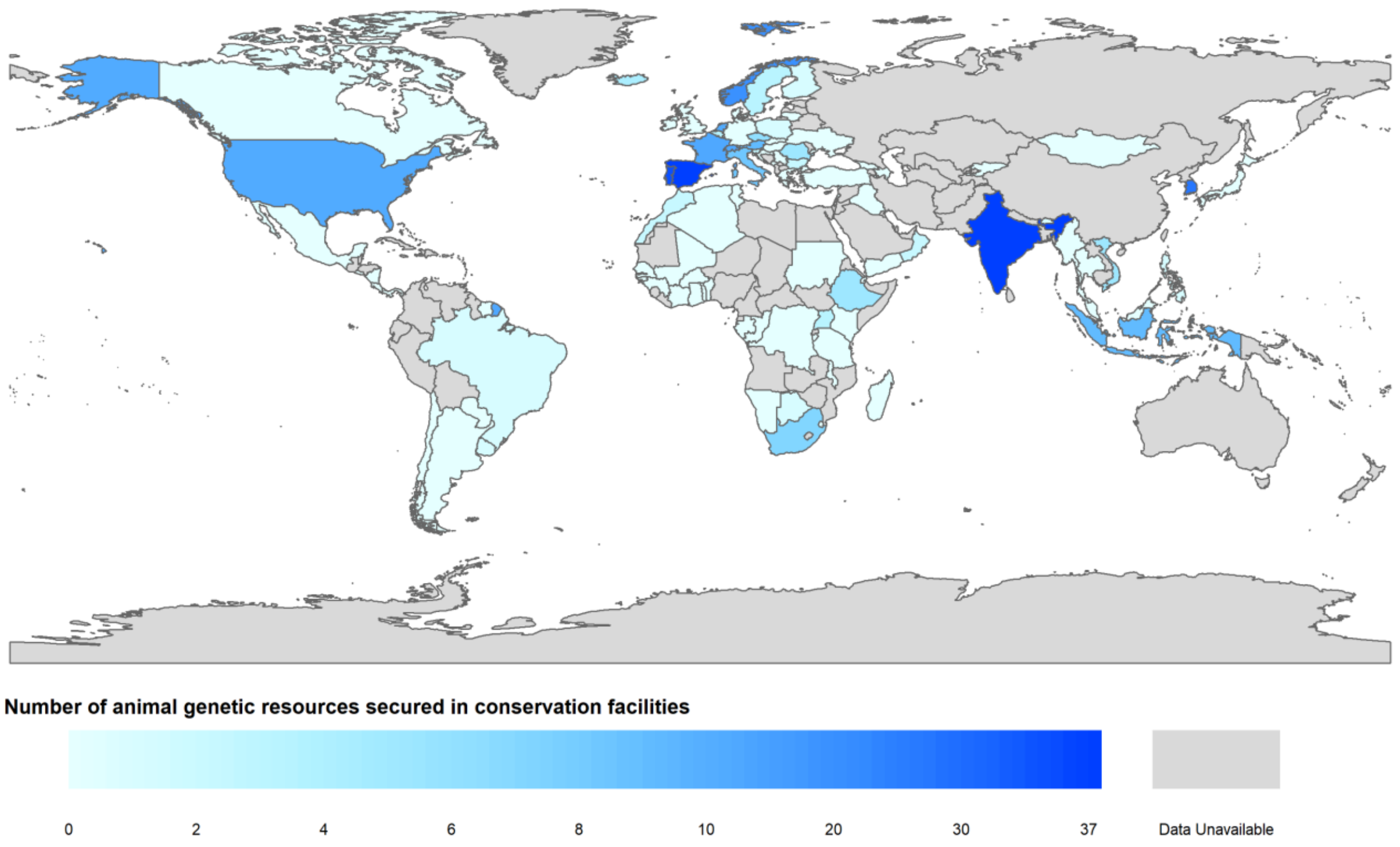

## S5.12 Prevalence of reduced coping strategies, 2021

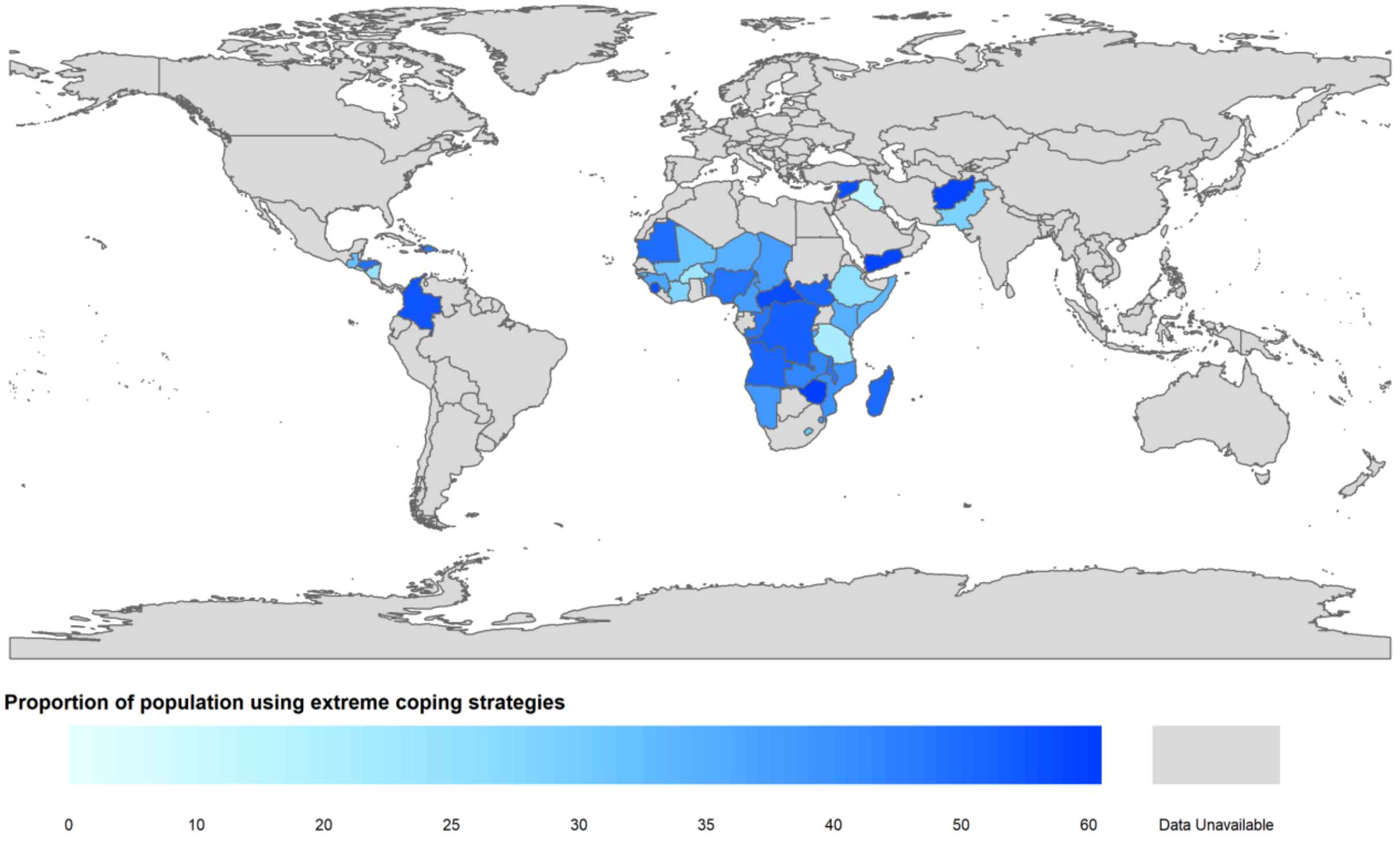

Dichotomous classification of reduced coping strategies uses a threshold of an rCSI score >=19. Based on daily data, the country-year prevalence is defined as the highest daily observation at any point during the year.

## S5.13 Food price volatility, 2019-2021 (average)

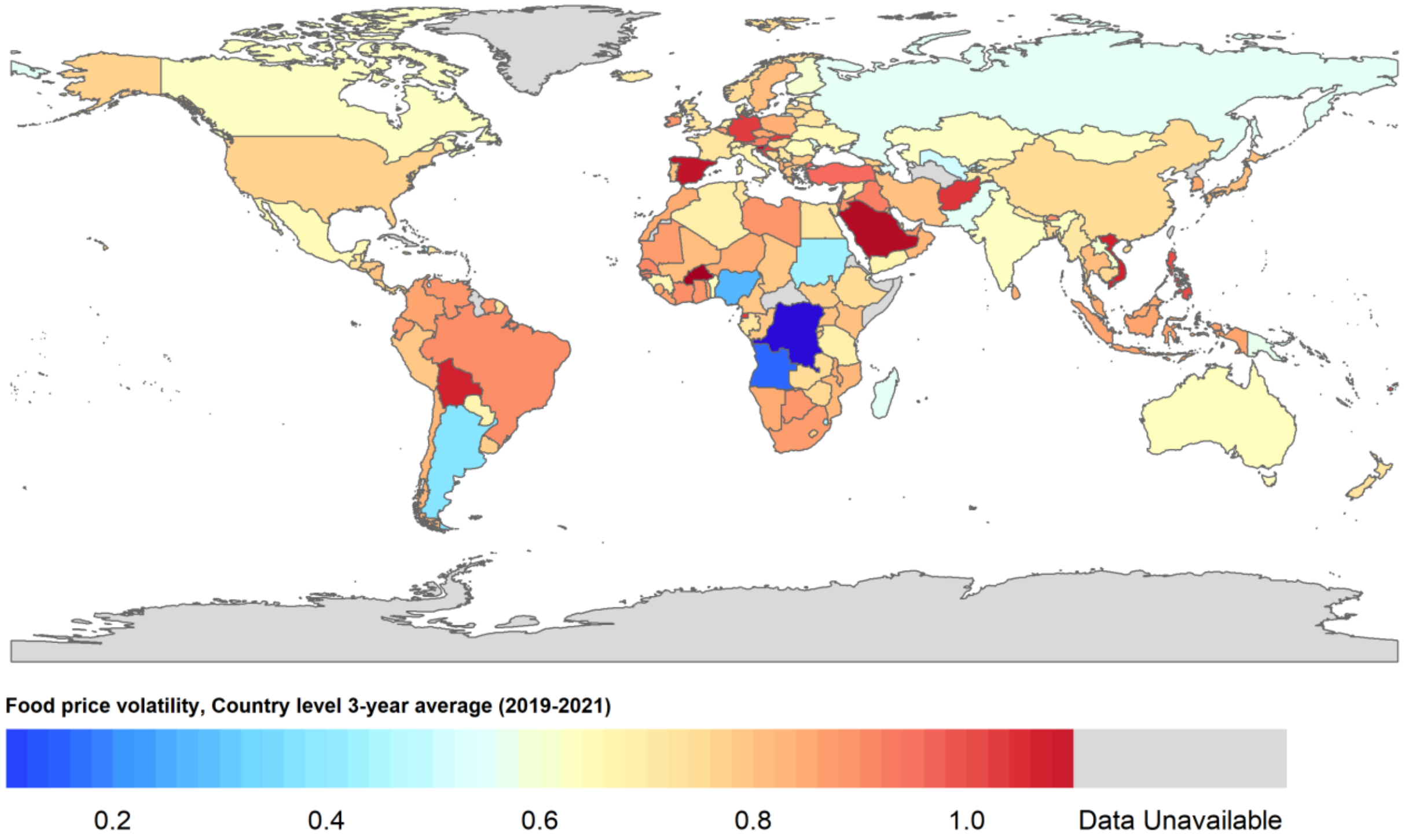

## S5.14 Food supply variability, 2000-2020, by region

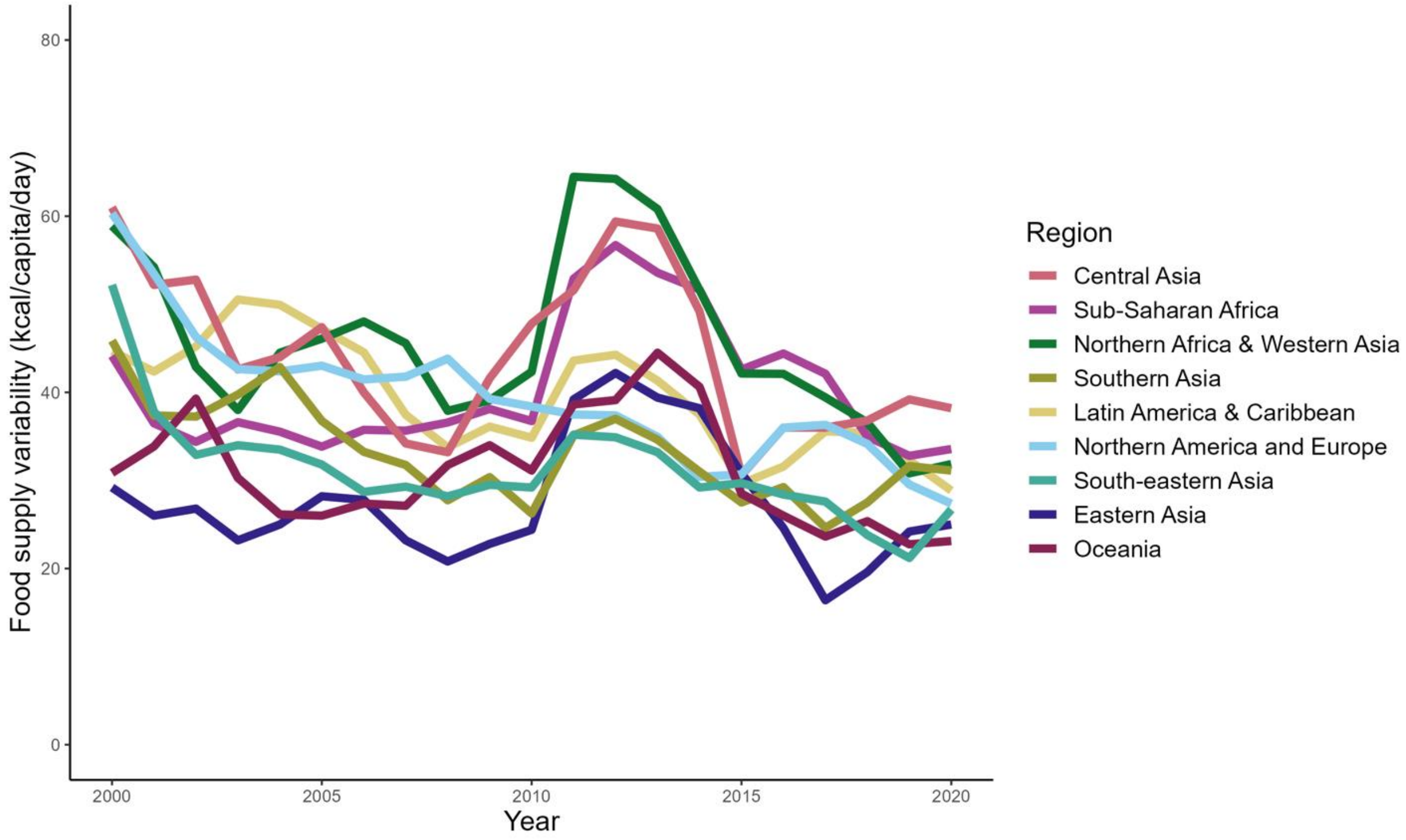

Unweighted regional means.

## S5.15 Food supply variability, 2000-2020, by income group

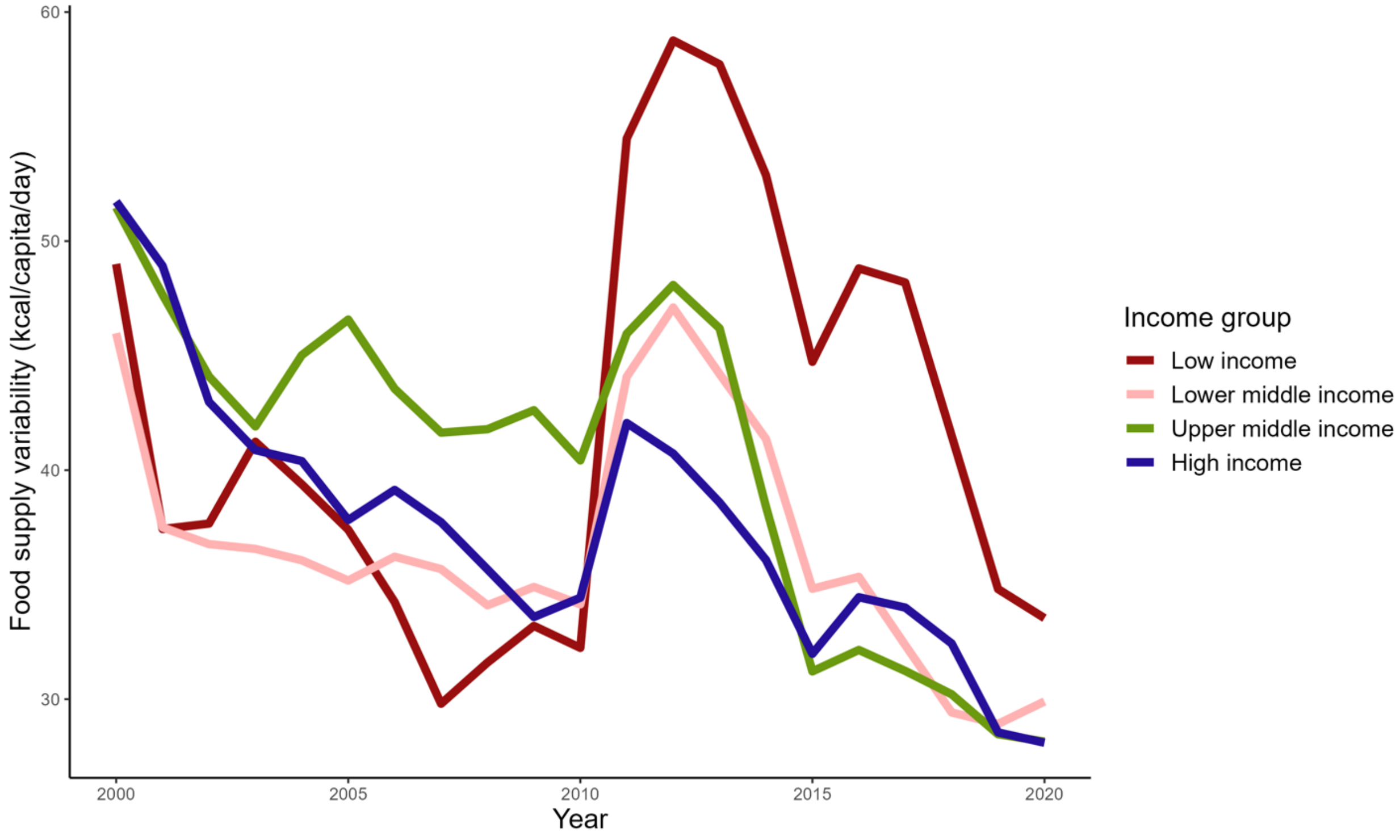

Unweighted income group means.

## S5.16 Food supply variability, 10-year average, (2011-2020)

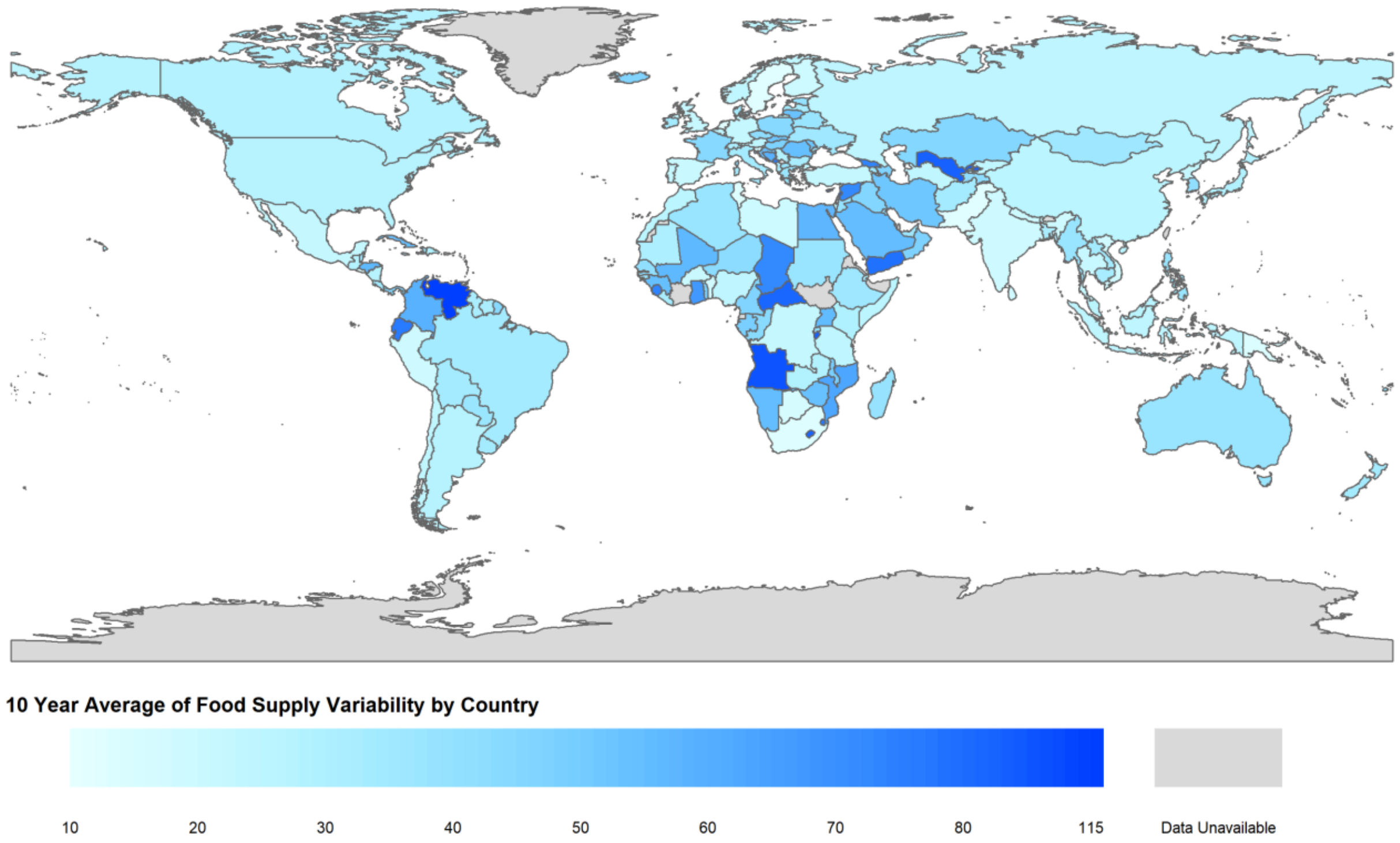

## S5.17 Food System Resilience

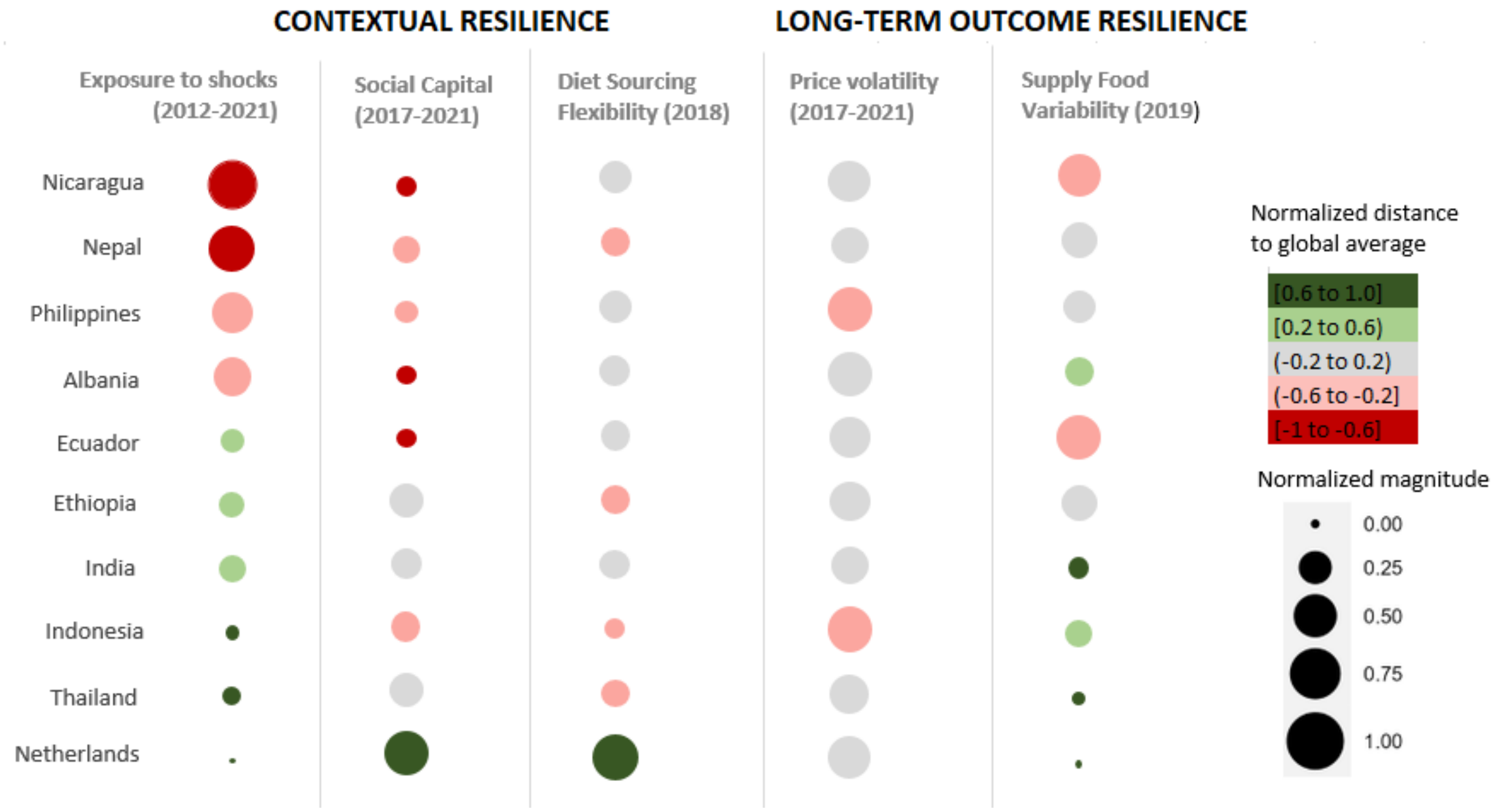

**Food system resilience.** Exploratory analysis based on five of the eight indicators initially proposed to measure food system resilience at country level. Three of the indicators capture the contextual dimension of resilience (exposure to shocks; social capital; and diet sourcing flexibility), while two indicators measure countries' food system resilience outcomes through the food price volatility and supply food variability. Ten countries were picked to illustrate the varied situation that characterizes the countries at present in relation to their food system resilience. To allow for comparison, the indicators have been normalized in two separate ways: their original value has been (i) normalized between 0 and +1 (thereafter represented by the size of the circle) and (ii) with respect to the distance to the world average (represented by the color of the circle, ranging from -1 (red) to +1 (green), while ensuring that the directionality of the gap with respect to the world average is appropriately accounted for.

Methodology note: The countries shown in this figure were selected by eliminating countries with fewer than 7 observations of exposure to shocks over the 10-year period (2012-2021) and where data were non-missing for all other indicators, resulting in a subset of 89 country-year observations. Exposure to shocks was calculated as the average of the highest three values per country over the 10-year period and the data were winsorized at the upper bound to the third-highest value for the purposes of visualization (Croatia and Nepal outliers, specifically). Normalization for the coloring was calculated using min-max scaling to the 89 observations in the subset dataset. Distance to the mean (mean of the 89 observations remaining in the subset dataset) was directionally adjusted to align with the desirable direction of change (see **Table 4**) but the average is not centered at zero. The normalization for the size of the marker was conducted with min-max scaling based on the 89 observations in the subset dataset.

# Supplementary Material, Appendix B

## Table B.1 Excluded indicators considered by expert and stakeholder consultations

** Denotes indicators added at the recommendation of respondents/participants of the survey and stakeholder consultation

| Indicators considered for survey and stakeholder consultation | Data sources(s) | Reason Excluded |
|---|---|---|
| ***Diets, Nutrition & Health*** | | |
| Egg and/or flesh food consumption among infants and young children(age 6-23 months) | UNICEF | This indicator is being tracked by UNICEF. Feedback from consultations did not rank this indicator as high as the others. |
| Sweet beverage consumption among infants and young children(age 6-23 months) | UNICEF | This indicator is now being collected in DHS-8 core module and MICS as of 2021. *Data are not yet available |
| Unhealthy food consumption among infants and young children(age 6-23 months) | UNICEF | This indicator is now being collected in DHS-8 core module and MICS as of 2021. *Data are not yet available |
| Consumption of all five recommended food groups among infants and young children(age 6-23 months) | UNICEF | This indicator is not currently reported by UNICEF but can be calculated from existing DHS and MICS source data (food group consumption). It mirrors the adult indicator. |
| Global Dietary Recommendations score (GDR score) | Gallup World Poll | It is the combined score of NCD-Protect and NCD-Risk, so duplicates information captured in those indicators. |
| Animal-source food consumption | Gallup World Poll | This indicator was considered because it is analogous to an IYCF indicator being tracked by UNICEF. Feedback from consultations did not rank this indicator as high as the others, in part because the directionality and policy relevance is less clear than some of the other diet indicators. |
| Pulses, nuts, and seeds consumption | Gallup World Poll | This indicator was considered because WHO recommends consumption of pulses, nuts and seeds as part of a healthy diet. Feedback from consultations did not rank this indicator as high as the others. |
| Whole grains consumption | Gallup World Poll | This indicator was considered because WHO recommends consumption of whole grains as part of a healthy diet. Feedback from consultations did not rank this indicator as high as the others. |
| Sweet beverage consumption | Gallup World Poll | This indicator includes soft drinks (sodas, energy drinks, and sports drinks), juice, and sweetened tea and coffee. Given that in the adult population, sweetened tea and coffee are highly prevalent traditional beverages, the indicator of sugar-sweetened soft drinks was selected because it is more policy relevant than this one. |
| Unhealthy food consumption (composed of sweet foods + salty or fried snacks) | Gallup World Poll | In the interest of having a smaller set of indicators, only one indicator for unhealthy food/beverage consumption is included. |
| Fast food or instant noodles consumption | Gallup World Poll | In the interest of having a smaller set of indicators, only one indicator for unhealthy food/beverage consumption is included. |
| Number of people who cannot afford a healthy diet | Food Prices for Nutrition Data Hub (soon FAOSTAT) | Duplicates information captured in the “proportion of people who cannot afford a healthy diet” |
| Cost of each food group (starchy staples, fats/oils, pulses/nuts/seeds, animal-source foods, fruits, vegetables) (per person per day) | Food Prices for Nutrition Data Hub (soon FAOSTAT) | Somewhat duplicative of the included Cost of a Healthy Diet indicator, so omitted in the interest of having a smaller set of indicators. |
| Cost of a Healthy Diet relative to average food expenditure | Food Prices for Nutrition DataHub | It may be duplicative of the other mainstreamed affordability indicators, the number and proportion of people who cannot afford a healthy diet. |
| Share of dietary energy from roots, cereals, tubers (%) | FAOSTAT (Food Systems Dashboard) | It may be difficult to interpret, because while a high share indicates low availability of diverse foods / imbalanced foods supply, a low share does not necessarily mean a nutritionally balanced food supply (could mean high sugars and fats). |
| Availability/supply (g/day/capita) - pulses | FAOSTAT | It could be compared to food group recommendations to determine adequacy of per capita supply, but it does not reflect disparities in access or consumption. |
| Availability/supply (kcal/day/capita) - animal-source foods | FAOSTAT | It could be compared to food group recommendations to determine adequacy of per capita supply, but it does not reflect disparities in access or consumption. |
| Availability (amount/day/capita) of macronutrients and micronutrients | FAOSTAT food and diet domain | It can be used to determine adequacy of per capita supply, but it does not reflect disparities in access or consumption. |
| Coverage of iodized salt (% of households) | UNICEF | The Working Group focused on the ends / outcomes, policies are not considered outcome, but rather a means to the end. |
| Fortification legislation (Any / Mandatory / None) | Food Systems Dashboard; Food Fortification Initiative, GAIN, Iodine Global Network, and the Micronutrient Forum, Global Fortification Data Exchange - legislation per food item | The Working Group focused on the ends / outcomes, policies are not considered outcome, but rather a means to the end. |
| Implementation of marketing of breast-milk substitutes restrictions (Fully achieved/ Partially achieved / Not achieved) | World Health Organization - Noncommunicable Diseases Progress Monitor | The Working Group focused on the ends / outcomes, policies are not considered outcome, but rather a means to the end. |
| Existence of policies on marketing junk foods to children (binary) | WHO (Food Systems Dashboard) | The Working Group focused on the ends / outcomes, policies are not considered outcome, but rather a means to the end. |
| Mandatory regulation of broadcast food advertising to children | World Cancer Research Fund International NOURISHING policy database | The Working Group focused on the ends / outcomes, policies are not considered outcome, but rather a means to the end. |
| Clearly visible "interpretative" labels and warning labels | World Cancer Research Fund International NOURISHING policy database | The Working Group focused on the ends / outcomes, policies are not considered outcome, but rather a means to the end. |
| Health-related food taxes | World Cancer Research Fund International NOURISHING policy database | The Working Group focused on the ends / outcomes, policies are not considered outcome, but rather a means to the end. |
| Best-practice policy implemented for industrially produced trans-fatty acids (TFA) (Y/N) | World Health Organization - Noncommunicable Diseases Progress Monitor Member State has adopted national policies to reduce population salt/sodium consumption | The Working Group focused on the ends / outcomes, policies are not considered outcome, but rather a means to the end. |
| Greenhouse gas emissions from food systems (farm to fork) | Crippa et al. (2021)[1] | Trade is yet adequately not accounted for in these aggregations, therefore it provides a rather misleading aggregated number. |
| Greenhouse gas emissions from food systems per capita (production-based emissions) | Crippa et al. (2021)[1] | Total greenhouse gas emissions were prioritized as per capita seems to imply that this also accounts for trade issues. |
| % change in soil organic carbon | Global Soil Organic Carbon Map | It is conceptually very important but data is not adequately there yet. |
| Biodiversity intactness | F. DeClerck et al. (2021) [2] | Biodiversity intactness is very important, but it is a society broad outcome indicator rather than food system specific. |
| Households with significant income from agriculture | LMICs: Harmonized Household Budget Surveys (LSMS type) - harmonized by FAO via RULiS<br>EU: Eurostat harmonized Labor Force Surveys<br>ILO: harmonized labor force surveys | RULiS data for LMICs had insufficient coverage. In addition, this indicator is less relevant outside LMICs. |
| % of rural population living below the poverty line | World Bank | Insufficient coverage |
| % population earning low pay | International labor force statistics - ILO | No urban/rural disaggregation available, so not very informative regarding agricultural workforce. |
| Informal employment rate in agriculture | International labor force statistics - ILO | Overlap with the underemployment rate. |
| Monthly wages for agricultural workers compared to the country's median monthly wage | World Bank | Lack of wage PPP data for inter-country comparability. |
| Coverage of school feeding programs | World Bank (2022). The Atlas of Social Protection: Indicators of Resilience and Equity (ASPIRE) Database. | Unclear definition of the indicator denominator, which appeared to include the total population rather than a population relevant for school-based programs (e.g. school-aged children). |
| ***Governance*** | | |
| Constitutional recognition of the right to adequate food | National policy documents (as collected and analyzed by FAOLEX and UNHCHR) and International Treaty documents | Constitutional recognition is not the only type of legal recognition that would facilitate the governance conditions to exercise the Right to Food. |
| Policy coordination | Bertelsmann Transformation Index | |
| SDG 16.7.2: Proportion of population who believe decision-making is inclusive and responsive, by sex, age, disability and population group** | UN Custodial Agency: UNDP | There is no data yet, perhaps forthcoming? |
| SDG 16.7.1: Proportions of positions (by sex, age, persons with disabilities and population groups) in public institutions (national and local legislatures, public service, and judiciary) compared to national distributions** | UN Custodial Agency: UNDP (Partnering Agencies: UN Women) | There is no data yet, perhaps forthcoming? |
| Policy implementation | Bertelsmann Transformation Index | |
| Implementation of marketing of breast-milk substitutes restrictions (Fully achieved/ Partially achieved / Not achieved) | World Health Organization - Noncommunicable Diseases Progress Monitor | Breastfeeding is a nutrition-relevant care practice and may not be considered directly tied to food systems. |
| Voice and Accountability, WGI | World Bank, Worldwide Governance Indicators | It incorporates data from V-Dem Index and Open Budget Index so it is less nuanced, and could lead to duplication of indicators. |
| ***Resilience & Sustainability*** | | |
| EM-DAT human impact | The database is made up of information from various sources, including UN agencies, governmental and non-governmental organizations, insurance companies, research institutes, and press agencies. | The ability of countries to better protect human life against shocks (e.g. cyclone, typhoons) has improved tremendously in the last 20 years thanks to investments in shelters and early warning systems, introducing a negative trend in long-term time series. If this indicator were to be used to proxy the exposure to shocks it would underestimate the actual level of exposure faced by countries in recent years. |
| SDG 16.1.2: Conflict-related deaths per 100,000 population, by sex, age, and cause. | Office of the High Commissioner for Human Rights (OHCHR) https://www.ohchr.org/en/instruments-and-mechanisms/human-rights-indicators/sdg-indicators-under-ohchrs-custodianship | Data is unavailable (only a very small number of countries seem to be collecting and making public this indicator). Some technical issues caused the Working Group to drop indicators that may be preferred by expert groups. |
| Economic value of violence | Institute for Economics and Peace | This indicator does not cover all types of shocks such as for instance natural disasters. Database is also not readily available. |
| People affected indicator | GCSI Monthly Dataset is uploaded to the ACAPS website and shared with INFORM partners on the last day of every month unless a new significant crisis occurs, in which case it may be updated ad hoc. The final scores and data are available in the ACAPS and the INFORM websites, uploaded in HDX, as well as through API. | Two indicators -- the economic impact and the people affected -- do not cover the same part of the world. The former covers low- and middle-income countries, while the latter covers high income countries. |
| FAO Relative detour cost (local impact) | Data available from University of Twente - ITC | New dataset that is tailored to food systems; however, there was uncertainty on how often it would be updated. |
| Road density (km of road per 100 sq. km of land area) | OpenStreetMap OSM, since road density indicator of the International Road Federation is under the private license. Open Street Map is open source and delivers better coverage. | The two global datasets available are not comparable due to difference in the methodology. |
| Poverty headcount ratio at $1.90 a day | World Bank | While poverty headcount could be considered as a possible resilience capacity indicator at household level, it was not considered a relevant indicator for food system resilience. |
| Access to electricity (% of population) | World Bank | With the exception of Sub-Sahara Africa and few countries in South Asia, most countries in the world now have electricity coverage over 85%, making this indicator not very useful or relevant at differentiating countries at a global level. |
| Renewable electricity output (% of total electricity output) | IEA statistics, World Bank | Renewable energy is not yet an appropriate global indicator as many countries still depend exclusively on fossil fuel-based energy. |
| Primary completion rate, total (% of relevant age group) | UNESCO, World Bank | While education could be considered a possible resilience capacity indicator at the household level, it was not considered a relevant indicator for food system resilience. |
| Social capital | BTI Transformation Index | A large number of Sub-Saharan Africa and all high income countries are not included in the BTI, making it less attractive as an indicator to track regional and global analyses than the Legatum Prosperity Index. |
| Global Innovation Index | Global Innovation Index | Although initially considered as a potential indicator of adaptation (one important dimension of resilience) the GII was not ranked as one of the top candidates by the different experts surveyed during the expert consultations. Additionally, the latest data available are from 2016. |
| Agrobiodiversity indicator 1: Species diversity in food supply | FAOSTAT food balance sheets | It is already included in the Dietary Sourcing and Flexibility Index, thus excluded to avoid double counting/bias. |
| Agrobiodiversity indicator 2: Species diversity in food production | FAOSTAT food balance sheets; | It is already included in the Dietary Sourcing and Flexibility Index, thus excluded to avoid double counting/bias. |
| Agrobiodiversity index | Jones et al. (2021)[3] | Combination of 22 indicators. Preference for non-aggregated indicators. |
| Stability of Food Consumption Score (FCS) | WFP curated Hunger Map https://hungermap.wfp.org | This indicator was excluded from the 'diets, food security and nutrition' domain as an insufficient measure of diet quality and therefore the stability thereof would not be an indicator of stability of diet quality. |
| Stability of Food Insecurity Experience Scale (FIES) - based indicators | FAO, MICS, DHS, World Bank high frequency phone surveys, country's own data | Was removed to avoid duplication or high cross-correlation with stability of the Food Consumption Score, however, that indicator was also ultimately excluded. The FIES (but not the stability of it) is included in the diet domain. |

# Table B.2 Indicators excluded prior to expert and stakeholder consultations, with rationale for exclusion

| Domain | Indicator name | Definition or description | Reason for elimination from consideration |
|---|---|---|---|
| ***Diets, Nutrition, & Health*** | | | |
| Diet Quality | Exclusive breastfeeding rate (children <6 months) | percentage of infants 0–5 months of age who were fed exclusively with breast milk during the previous day. | This is primarily a care practices indicator rather than a food systems indicator; outside the scope of the WG. Was excluded from Food Systems Dashboard diagnosis for this reason. |
| Diet Quality | Continued breastfeeding at 12-23 months | percentage of children 12–23 months of age who were fed breast milk during the previous day. | This is primarily a care practices indicator rather than a food systems indicator; outside the scope of the WG. Was excluded from FSD diagnosis for this reason. |
| Diet Quality | Minimum Meal Frequency for IYC (age 6-23 months) | percentage of children 6–23 months of age who consumed solid, semi-solid or soft foods (but also including milk feeds for non-breastfed children) at least the minimum number of times during the previous day. | This is primarily a care practices indicator rather than a food systems indicator; outside the scope of the WG. Was excluded from FSD diagnosis for this reason. |
| Diet Quality | Minimum Acceptable Diet for IYC (age 6-23 months) | percentage of children 6–23 months of age who consumed a minimum acceptable diet during the previous day. The minimum acceptable diet is defined as:<br>• for breastfed children: receiving at least the minimum dietary diversity and minimum meal frequency for their age during the previous day;<br>• for non-breastfed children: receiving at least the minimum dietary diversity and minimum meal frequency for their age during the previous day as well as at least two milk feeds. | This is primarily a care practices indicator rather than a food systems indicator; outside the scope of the WG. Was excluded from FSD diagnosis for this reason. |
| Diet Quality | Adolescent diet indicators | Various | Lack of geographic coverage. |
| Diet Quality | Healthy Eating Index | The Healthy Eating Index (HEI) is a measure of diet quality used to assess how well a set of foods aligns with key recommendations of the Dietary Guidelines for Americans. The Dietary Guidelines for Americans is designed for nutrition and health professionals to help individuals (ages 2 years and older) and families to consume a healthful and nutritionally adequate diet. | Not feasible; requires quantitative dietary intake data. |
| Diet Quality | Global Dietary Quality Score | The GDQS is a measure of diet quality with respect to both nutrient adequacy and diet-related NCD risk for use at the population level, with a range of 0-49. GDQS scores ≥23 are associated with a low risk of both nutrient adequacy and NCD risk, scores ≥15 and <23 | Not feasible; requires quantitative or semi-quantitative dietary intake data. |

| Domain | Indicator name | Definition or description | Reason for elimination from consideration |
|---|---|---|---|
| | | indicate moderate risk, and scores <15 indicate high risk. Points are assigned based on 3 or 4 categories of quantitative amounts (defined in g/d) specific to each group. | |
| Diet Quality | Percent of dietary energy from ultraprocessed foods | Percent of dietary energy from ultraprocessed foods, using the NOVA classification. | Not feasible; requires quantitative dietary intake data. |
| Diet Quality | Sweet foods consumption | The percent of the population age 15+ that consumed any sweet food in the previous day or night. This is a negative indicator. | Captured in the more aggregated indicator "Unhealthy food consumption". |
| Diet Quality | Salty or fried snacks consumption | The percent of the population age 15+ that consumed any salty or fried snack in the previous day or night. This is a negative indicator. | Captured in the more aggregated indicator "Unhealthy food consumption". |
| Diet Quality | Vegetable consumption | Prevalence of the population including vegetables as part of the diet, which is a global recommendation. | Captured in the more aggregated indicator "zero fruit and vegetable consumption". |
| Diet Quality | Fruit consumption | Prevalence of the population including fruits as part of the diet, which is a global recommendation. | Captured in the more aggregated indicator "zero fruit and vegetable consumption". |
| Diet Quality | WHO-FV | A score of 3 or more indicates likelihood of consuming at least 400g fruits and vegetables, which is a global dietary recommendation. Based on food group consumption data collected in the DQQ. *Provisional indicator; cutoff not yet globally validated. | Not yet globally validated. Possible to reconsider in future years; the indicator is constructed from the DQQ, which is collected in the Gallup World Poll. |
| Diet Quality | WHO-Sugar | A score of 2 or more indicates likelihood of exceeding 10% of dietary energy from free sugars, which is a negative indicator; limiting free sugar consumption to <10% of dietary energy is a global dietary recommendation. Based on food group consumption data collected in the DQQ. *Provisional indicator; cutoff not yet globally validated. | Not yet globally validated. Possible to reconsider in future years; the indicator is constructed from the DQQ, which is collected in the Gallup World Poll. |
| Diet Quality | WHO-Fiber | A score of 4 or more indicates likelihood of consuming at least 25g fiber, which is a global dietary recommendation. Based on food group consumption data collected in the DQQ. *Provisional indicator; cutoff not yet globally validated. | Not yet globally validated. Possible to reconsider in future years; the indicator is constructed from the DQQ, which is collected in the Gallup World Poll. |
| Diet Quality | WHO-SatFat | A score of 2 or more indicates likelihood of exceeding 10% of dietary energy from saturated fat, which is a negative indicator; limiting saturated fat consumption to <10% of dietary energy is a global dietary recommendation. Based on food group consumption data | Not yet globally validated. Possible to reconsider in future years; the indicator is constructed from the DQQ, which is collected in the Gallup World Poll. |

| Domain | Indicator name | Definition or description | Reason for elimination from consideration |
|---|---|---|---|
| | | collected in the DQQ. *Provisional indicator; cutoff not yet globally validated. | |
| Diet Quality | Global Burden of Disease - Estimated dietary intake for adults (g/day); available for: Fruit; Vegetables; Whole Grains; Legumes, Nuts, Seeds; Sugar-sweetened beverages; Milk; Red meat; Processed meat; Sodium; Calcium; Fiber; Poly-unsaturated fatty acids | Modeled estimate of food group intake per person, per day in each country. | Black box methodology. Beal et al. (2021)[4] showed that estimated intakes vary widely between estimation models, and as such may not be fit for advising policy. |
| Diet Quality | Global Dietary Database - Estimated dietary intake for adults (g/day); available for: Fruits, Non-starchy Vegetables, Potatoes, Other Starchy Vegetables, Beans and Legumes, Nuts and Seeds, Refined Grains, Whole Grains, Unprocessed Red Meats, Total Processed Meats, Total Seafoods, Eggs, Cheese, Yogurt (including fermented milk) | Modeled estimate of food group intake per person, per day in each country. | |
| Food security | Affordability of a healthy diet (as compared to Intl Poverty Line) | The ratio of the cost of an energy sufficient diet to the food poverty line (63% of the international poverty line of 1.90/day in 2017 USD). | Duplicative of other affordability indicators that are more informative; the indicator inherently critiques the denominator (i.e., highlights the limitations of food poverty line construction), suggesting that the denominator is not a solid standard for comparison. |
| Food security | Cost of nutrient adequacy (CoNA) relative to average food expenditure | The ratio of the cost of a nutrient adequate diet to observed per capita food expenditures from national accounts. CoNA is the cost of the least expensive locally available foods to meet average requirements for dietary energy and 23 essential macro- and micro-nutrients, per capita, per day (2017USD). | Duplicative/confusing alongside Cost and affordability of a Healthy Diet indicator. |
| Food environments | Relative cost of fruits and vegetables | The ratio of the cost of fruits and vegetables to the cost of starchy staples, each in the amount required to satisfy average daily recommendations from FBDG. | Overly disaggregated compared to the main indicators, cost and affordability of a healthy diet. |
| Food environments | Relative cost of legumes, nuts and seeds | The ratio of the cost of legumes, nuts and seeds to the cost of starchy staples, each in the amount required to satisfy average daily recommendations from FBDG. | Overly disaggregated compared to the main indicators, cost and affordability of a healthy diet. |
| Food environments | Relative cost of ASF | The ratio of the cost of animal-source foods to the cost of starchy staples, each in the amount required to satisfy average daily recommendations from FBDG. | Overly disaggregated compared to the main indicators, cost and affordability of a healthy diet. |
| Food environments | Relative cost of oils and fats | The ratio of the cost of oils and fats to the cost of starchy staples, each in the amount required to satisfy average daily recommendations from FBDG. | Overly disaggregated compared to the main indicators, cost and affordability of a healthy diet. |

| Domain | Indicator name | Definition or description | Reason for elimination from consideration |
|---|---|---|---|
| Food environments | Relative caloric price – available for the following food groups: Cereals, Fats and oils, Eggs, Nuts, Pulses, Milk, Fish, Meat, Dark green leafy vegetables, Vitamin A-rich fruits, Other vegetables, Other fruit, Salty snacks, Sugary snacks, SSBs | The ratio of a calorie of the specified food group to a calorie of cereals. | Not currently routinely calculated; feasibility may be limited. |
| Food environments | Per capita dietary energy supply (kcal/person/day) | kcal per person per day available in a country's foods supply. | Information is better captured in PoU; duplicative. |
| Food environments | Availability/supply (g/day/capita) – Available for: Meat, Fish, Eggs, and Milk | Grams per person per day of each meat, fish, eggs, and milk available in a country's food supply. | Information is better captured in aggregate indicator "Availability/supply of animal-source foods". There is no requirement or recommendations for this item specifically; rather recommendations are set for a more aggregated food group. Because of the absence of item-specific recommendations, this indicator is difficult to interpret. |
| Food environments | Food supply adequacy | This indicator estimates to what extent the national food supply is adequately meeting the daily requirements per person (g/capita/day) of the food groups recommended for a healthy diet (fruits, vegetables, pulses, animal-source foods). | Indicator not clearly defined / established. Possible to reconsider in future years. |
| Food environments | Household availability (unit/day/HH equivalent) of macro- and micronutrients | | Duplicative of availability at individual level (this indicator is the same indicator estimated at household level, which requires additional assumptions). |
| Food environments | Global demand for instant noodles | Interesting indicator of UFP in the food environment. Many but not all countries in the database. | Black box methodology. |
| Food environments | Retail value of packaged food sales per capita (also available as growth rate) | | Unclear interpretation. Packaged food may be healthy -- like canned beans -- or unhealthy. Unhealthy packaged food is better captured by retail value of UPF sales per capita. |
| Food environments | Modern grocery retailers per 100K people (also available as growth rate) | | Unclear interpretation. |
| Food environments | Supermarkets per 100K people (also available as growth rate) | | Unclear interpretation. |
| Food environments | Yields (tonnes/hectare) of cereals | | Outside the scope for the diets, nutrition and health theme. |
| Food environments | Yields (tonnes/hectare) of vegetables | | Outside the scope for the diets, nutrition and health theme. |
| Food environments | Losses of nutritious foods as pct of domestic supply; fruits | | Outside the scope for the diets, nutrition and health theme. Does not indicate where along supply chain loss occurs. |

| Domain | Indicator name | Definition or description | Reason for elimination from consideration |
|---|---|---|---|
| Food environments | Losses of nutritious foods as pct of domestic supply; vegetables | | Outside the scope for the diets, nutrition and health theme. Does not indicate where along supply chain loss occurs. |
| Food environments | Losses of nutritious foods as pct of domestic supply; pulses | | Outside the scope for the diets, nutrition and health theme. Does not indicate where along supply chain loss occurs. |
| Food environments | Losses of nutritious foods as pct of domestic supply; cereals | | Outside the scope for the diets, nutrition and health theme. Does not indicate where along supply chain loss occurs. |
| Policies affecting food environments | Food safety standards | Undefined | Indicators not defined. Possible to reconsider in future years. |
| Policies affecting food environments | Availability of food composition table (binary) | The country has a country-specific food composition table (1/0). | Unclear interpretation; unclear need for country specific food composition table for every country in the world. |
| Policies affecting food environments | Availability of food based dietary guidelines (binary) | The country has developed national food-based dietary guidelines (FBDG). | Unclear interpretation. |
| Policies affecting food environments | Fortification legislation, rice - Voluntary / Mandatory / None | The data may not exist in the form as disaggregated per food item (perhaps, salt is the most relevant one). May use the structure they exist in the database (fortification data site). | Overly disaggregated compared to the main indicator, presence of any fortification legislation. |
| Policies affecting food environments | Fortification legislation, maize flour - Voluntary / Mandatory / None | The data may not exist in the form as disaggregated per food item (perhaps, salt is the most relevant one). May use the structure they exist in the database (fortification data site). | Overly disaggregated compared to the main indicator, presence of any fortification legislation. |
| Policies affecting food environments | Fortification legislation, wheat flour - Voluntary / Mandatory / None | The data may not exist in the form as disaggregated per food item (perhaps, salt is the most relevant one). May use the structure they exist in the database (fortification data site). | Overly disaggregated compared to the main indicator, presence of any fortification legislation. |
| Policies affecting food environments | Fortification legislation, oil - Voluntary / Mandatory / None | The data may not exist in the form as disaggregated per food item (perhaps, salt is the most relevant one). May use the structure they exist in the database (fortification data site). | Overly disaggregated compared to the main indicator, presence of any fortification legislation. |
| Policies affecting food environments | Fortification legislation, salt- Voluntary / Mandatory / None | The data may not exist in the form as disaggregated per food item (perhaps, salt is the most relevant one). May use the structure they exist in the database (fortification data site). | Overly disaggregated compared to the main indicator, presence of any fortification legislation. |
| Policies affecting food environments | Biofortification policies and programs (binary: adopted/in process versus none) | Countries where biofortification is included in government legislation/programs/policies/strategies. | Not globally relevant; Lack of geographic coverage (additional note: conveys limited information and only 'yes' for a few countries in the Food Systems Dashboard: HarvestPlus made detailed policy (including information on which policies are in which country) and crops released databases available online. However, it does not meet the criteria to have at least 70 countries of coverage.) |

| Domain | Indicator name | Definition or description | Reason for elimination from consideration |
|---|---|---|---|
| Policies affecting food environments | Diet related (advertising, labelling, restrictions on trans fat, sodium etc.) | Unclear definition; topic rather than indicator. | Many indicators, but geographic coverage is unclear - also unclear if database is regularly updated. |
| Policies affecting food environments | Availability of legislation on nutrition labelling | | Precise indicator definition and Feasibility needs to be verified. |
| Policies affecting food environments | Availability of legislation on ingredient labelling on packaged foods | | Precise indicator definition and Feasibility needs to be verified. |
| Policies affecting food environments | Trans fat ban in place (binary? Voluntary / Mandatory / None?) | | Duplicative of other indicator on implementation of trans fat legislation. |
| Policies affecting food environments | COMP 1: Food composition targets/standards have been established for processed foods by the government for the content of the nutrients of concern in certain foods or food groups if they are major contributors to population intakes of these nutrients of concern (trans fats and added sugars in processed foods, salt in bread salt in snacks etc.)) | | Not feasible; Lack of geographic coverage (additional note: not monitorable across the countries). |
| Policies affecting food environments | COMP 2: Food composition targets/standards have been established by the government for out-of-home meals in food service outlets (such as fast food joints, food kiosks, check-check joints, restaurants, and other local food vendors) for the content of the nutrients of concern in certain foods or food groups if they are major contributors to population intakes of these nutrients of concern (e.g. trans fats, added sugars, salt, saturated fat, saturated fat in commercial frying fats/oils) | | Not feasible; Lack of geographic coverage (additional note: not monitorable across the countries). |
| Policies affecting food environments | LABEL1: Ingredient lists and nutrient declarations in line with Codex recommendations are present on the labels of packaged foods | | Not feasible; Lack of geographic coverage (additional note: not monitorable across the countries). |
| Policies affecting food environments | LABEL2: Robust, evidence-based regulatory systems are in place for approving/reviewing claims on foods, so that consumers are protected against unsubstantiated and misleading nutrition and health claims | | Not feasible; Lack of geographic coverage (additional note: not monitorable across the countries). |
| Policies affecting food environments | LABEL3: A single, consistent, interpretive, evidence-informed front-of-pack supplementary nutrition information system, which readily allows consumers to assess a product's healthiness, is applied to packaged foods | | Not feasible; Lack of geographic coverage (additional note: not monitorable across the countries). |
| Policies affecting food environments | LABEL4: A consistent, single, simple, clearly-visible system of labelling the menu boards of quick service restaurants (i.e. fast food chains) is applied by the government, which allows consumers to interpret the nutrient quality and energy content of foods and meals on sale | | Not feasible; Lack of geographic coverage (additional note: not monitorable across the countries). |
| Policies affecting food environments | PROMO1: Effective policies are implemented by the government to restrict exposure and power of promotion of unhealthy foods to or for children through broadcast media (TV, radio) | | Not feasible; Lack of geographic coverage (additional note: not monitorable across the countries). |
| Policies affecting food environments | PROMO2: Effective policies are implemented by the government to restrict exposure and power of promotion of unhealthy foods to or for children through non-broadcast media (e.g. Internet, social media, food packaging, sponsorship, religious events, outdoor advertising including around schools) | | Not feasible; Lack of geographic coverage (additional note: not monitorable across the countries). |
| Policies affecting food environments | PROMO3: Effective policies are implemented by the government to ensure that unhealthy foods are not commercially promoted to or for children in settings where children gather (e.g. preschools, schools, sport and cultural events) | | Not feasible; Lack of geographic coverage (additional note: not monitorable across the countries). |
| Policies affecting food environments | PROMO4: Effective policies are implemented by the government to restrict the marketing of breastmilk substitutes | | Not feasible; Lack of geographic coverage (additional note: not monitorable across the countries). |

| Domain | Indicator name | Definition or description | Reason for elimination from consideration |
|---|---|---|---|
| Policies affecting food environments | PRICES1: Taxes or levies on healthy foods are minimized to encourage healthy food choices where possible (e.g. low or no sales tax, excise, value-added or import duties on fruit and vegetables) | | Not feasible; Lack of geographic coverage (additional note: not monitorable across the countries). |
| Policies affecting food environments | PRICES2: Taxes or levies on unhealthy foods (e.g. sugar-sweetened beverages, foods high in nutrients of concern) are in place and increase the retail prices of these foods by at least 10% to discourage unhealthy food choices where possible, and these taxes are reinvested to improve population health | | Not feasible; Lack of geographic coverage (additional note: not monitorable across the countries). |
| Policies affecting food environments | PRICES3: The intent of existing subsidies on foods, including infrastructure funding support (e.g. research and development, supporting markets or transport systems), is to favour healthy rather than unhealthy foods | | Not feasible; Lack of geographic coverage (additional note: not monitorable across the countries). |
| Policies affecting food environments | PRICES4: The government ensures that food-related income support programs are for healthy foods | | Not feasible; Lack of geographic coverage (additional note: not monitorable across the countries). |
| Policies affecting food environments | PROV1: The government ensures that there are clear, consistent policies (including nutrition standards) implemented in schools and early childhood education services for food service activities (canteens, food at events, fundraising, promotions, vending machines etc.) to provide and promote healthy food choices | | Not feasible; Lack of geographic coverage (additional note: not monitorable across the countries). |
| Policies affecting food environments | PROV2: The government ensures that there are clear, consistent policies in other public sector settings for food service activities (canteens, hospitals, clinics, food at events, fundraising, promotions, vending machines, public procurement standards etc.) to provide and promote healthy food choices. | | Not feasible; Lack of geographic coverage (additional note: not monitorable across the countries). |
| Policies affecting food environments | PROV3: The Government ensures that there are good support and training systems to help schools and other public sector organisations and their caterers meet the healthy food service policies and guidelines | | Not feasible; Lack of geographic coverage (additional note: not monitorable across the countries). |
| Policies affecting food environments | RETAIL1: Zoning laws and policies are robust enough and are being used, where needed, by local governments to place limits on the density or placement of quick serve restaurants or other outlets selling mainly unhealthy foods in communities, and to encourage the availability of outlets selling healthy options such as fresh fruit and vegetables | | Not feasible; Lack of geographic coverage (additional note: not monitorable across the countries). |
| Policies affecting food environments | RETAIL2: The Government ensures existing support systems are in place to encourage food stores and food service outlets to promote the availability of healthy foods and to limit the promotion and availability of unhealthy foods | | Not feasible; Lack of geographic coverage (additional note: not monitorable across the countries). |
| Policies affecting food environments | RETAIL 3: Food hygiene policies are robust enough and are being enforced, where needed, by national and local government to protect human health and consumers' interests in relation to food. | | Not feasible; Lack of geographic coverage (additional note: not monitorable across the countries). |
| Policies affecting food environments | TRADE1: The Government undertakes risk impact assessments before and during the negotiation of trade and investment agreements, to identify, evaluate and minimize the direct and indirect negative impacts of such agreements on population nutrition and health | | Not feasible; Lack of geographic coverage (additional note: not monitorable across the countries). |
| Policies affecting food environments | TRADE2: The government adopts measures to manage investment and protect their regulatory capacity with respect to public health nutrition | | Not feasible; Lack of geographic coverage (additional note: not monitorable across the countries). |
| Policies affecting food environments | Implementation of salt/sodium policies (Fully achieved/ Partially achieved / Not achieved) | This indicator is about sodium restriction, and is not about salt iodization; that is covered by other indicators on fortification legislation and iodized salt coverage. Country has implemented | Lack of coverage - Could not find this indicator in the WHO Global Health Observatory database https://www.who.int/data/gho/data/indicators even though a TF team member indicated "World Health Organization - Noncommunicable |

| Domain | Indicator name | Definition or description | Reason for elimination from consideration |
|---|---|---|---|
| | | policies restricting salt/sodium in sold food products. | Diseases Progress Monitor Member State has adopted national policies to reduce population salt/sodium consumption". |
| ***Environment, Natural Resources, & Production*** | | | |
| GHG | Per capita biodiversity impact of food consumption (sp.yr*10^12) | The extinction rate expected under current food consumption patterns. These biodiversity losses result from the occupation of farmland and the effects of transformation of natural ecosystems into farmland. Food consumption is estimated using FAO Food Balance Sheet data as a proxy. | Too many parameter assumptions in calculating these per capita per country. |
| GHG | Per capita GHG emissions of food consumption (kg Co2Eq) | Greenhouse gas emissions (measured in carbon dioxide equivalent) related to current food consumption patterns. Food consumption is estimated using FAO Food Balance Sheet data as a proxy. | Too many parameter assumptions in calculating these per capita per country. |
| Water | Per capita eutrophication of food consumption (gPO43 eq) | The increase in phosphorous and nitrogen concentration (measured in phosphate equivalent) in water and soils as a result of current food consumption patterns. Food consumption is estimated using FAO Food Balance Sheet data as a proxy. | Too many parameter assumptions in calculating these per capita per country. |
| Water | Per capita water use linked to food consumption (L) | Freshwater withdrawals needed to produce food under current food consumption patterns. Includes irrigation water (for crops and livestock feed), animal drinking water, and water used during food processing. Withdrawals are weighted by local water scarcity. Food consumption is estimated using FAO Food Balance Sheets data as a proxy. | Too many parameter assumptions in calculating these per capita per country. |
| Water | Per capita water scarcity of food consumption (L eq) | Indicator undefined. | Too many parameter assumptions in calculating these per capita per country. |
| Biosphere integrity | Total Ecological Footprint of consumption [global ha] | The ecological footprint of production plus the ecological footprint of imports and minus the footprint of exports. | Too many parameter assumptions in calculating these per capita per country. |
| Biosphere integrity | Number of Earths required | Number of Earths required to support human's footprint if everyone lives like an average individual or country. | Too aggregated and too many assumptions. |
| Land & soils | Nitrogen fertilizer use per unit of land (tonnes ha-1) | Totals in nitrogen (N) for agriculture use of inorganic (chemical or mineral) fertilizers. Both straight and compound fertilizers are included. | Not normative - very context specific. |
| Land & soils | Phosphorous fertilizer use per unit of land (tonnes ha-1) | Totals in phosphorus (expressed as P2O5) for agriculture use of inorganic (chemical or mineral) fertilizers. Both straight and compound fertilizers are included. | Not normative - very context specific. |

| Domain | Indicator name | Definition or description | Reason for elimination from consideration |
|---|---|---|---|
| Land & soils | Agricultural land as % of arable land | Agricultural land includes arable land, land under permanent crops, and meadows or pasture. This indicator offers a quick snapshot of how much of the land area per country is occupied with agriculture. | Not normative - very context specific. |
| Land & soils | Percentage of intact area (% per country) | Intact land areas are those with low levels of human impact or disturbance, such that natural processes predominate. | Not necessarily normative and broader than food systems. |
| Land & soils | Soil Organic Content (tonnes ha-1) | Average soil organic carbon (SOC). Organic carbon is a major component of the organic material present in soils. It stabilizes soil structure and reduces erosion, improves soil fertility, and enhances its water-holding capacity. | Dependent on agroecological setting - replaced by change in SOC. |
| Biosphere integrity | % Natural vegetation in agricultural land | Cropland with a minimum level of proximate natural and seminatural vegetation to maintain ecosystem integrity. | Included a similar indicator “Functional Integrity (% agricultural land with minimum level of natural habitat)”. |
| Biosphere integrity | Crop richness (average number of crops per unit of land) | Crop richness in production systems. | Replaced by production diversity metric as part of resilience theme. |
| Biosphere integrity | Average Shannon diversity crops (dimensionless) | The Shannon diversity index reflects how many different types of foods (crops and livestock) are produced in a certain country, and how evenly these different types are distributed. This indicator identifies the diversity of crops in a country without regard for the nutrient content of each crop. | See resilience group. |
| Biosphere integrity | Calories diversity measured by Shannon index | How many different types of food items there are in a certain country, and how evenly these different types are distributed.[5] | See resilience group. |
| Biosphere integrity | Tree cover on agricultural land (%) | Average percentage tree cover on agricultural land in 2010. | Not normative - very context and crop specific. |
| Biosphere integrity | Benefits of biodiversity index (0= no biodiversity potential to 100 maximum) | | Too aggregated. |
| Biosphere integrity | Integrated plant nutrient management | Concept not an indicator. | Replaced by sustainable nitrogen management index. |
| Biosphere integrity | Total Ecological Footprint of production (global ha) | Combined land appropriation for producing agricultural, livestock, fishery and aquaculture, and timber products as well as the CO2 emissions and built-up surfaces (e.g. roads, factories, cities) linked to that production. | Too aggregated and too much a black box. |
| Biosphere integrity | Environmental Performance Index | Composite index of 37 indicators for measuring countries' environmental performance regarding environmental health (i.e. air quality, sanitation and drinking water, heavy metals and waste management) and ecosystem vitality (e.g. | Too aggregated. |

| Domain | Indicator name | Definition or description | Reason for elimination from consideration |
|---|---|---|---|
| | | biodiversity and habitat, ecosystem services, fisheries, water resources, climate change, pollution emissions, agriculture). | |
| GHG | Global Climate Risk Index | The Global Climate Risk Index assess the extent to which countries have been affected (in terms of fatalities and economic losses) by meteorological events such as tropical storms, winter storms, severe weather, hail, tornados, local storms; hydrological events such as storm surges, river floods, flash floods, mass movement (landslide); climatological events such as freezing, wildfires, droughts, and cold/heat waves. | Too aggregated and also a driver, not just an outcome. |
| GHG | Climate Risk fatalities per 100,000 (1999 – 2018) | | More a vulnerability indicator than an environmental indicator. |
| GHG | Climate Risk $ loss in million USD (1999 – 2018) (or per unit of GDP) | | More a vulnerability indicator than an environmental indicator. |
| Land & soils | Soil erosion | Concept not an indicator. | Not updated sufficiently and aggregated indicator. |
| Land & soils | Soil biodiversity threats index | Average levels of potential risk to soil biodiversity and life at a global scale. Considered threats include loss of above ground biodiversity, pollution and nutrient overloading, cropland percentage cover, cattle density, fire density between 1997-2010 , Water and Wind Erosion Vulnerability Indices, Desertification Vulnerability Index, and Global Aridity Index. | Not updated sufficiently and aggregated indicator. |
| Land & soils | Soil biodiversity potential index | Average Soil Biodiversity Potential Index describes the potential level of diversity (micro and macro fauna) living in soils on our planet. | Not updated sufficiently and aggregated indicator. |
| Pollution | (Change in) fertilizer use (tonnes per year) | | Not sufficiently normative - context specific. |
| Land & soils | Tree cover loss (%) | | Considered as part of agricultural land use change. |
| Land & soils | Wetland loss (%) | | Considered as part of agricultural land use change. |
| Land & soils | Grassland loss (%) | | Considered as part of agricultural land use change. |
| Land & Soil | % of agricultural land degraded | | No frequent updates. |
| Water | Water stress (SDG 6.4.2) | Level of water stress: freshwater withdrawal as a proportion of available freshwater resources. | Redundant with % withdrawal by agriculture. |
| Pollution | Trend in Chemical Production | Does not currently exist and would have to be created. | Not directly related to impact. |

| Domain | Indicator name | Definition or description | Reason for elimination from consideration |
|---|---|---|---|
| Biosphere integrity | Human trophic level | Trophic levels are critical for synthesizing species' diets, depicting energy pathways, understanding food web dynamics and ecosystem functioning, and monitoring ecosystem health. Specifically, trophic levels describe the position of species in a food web, from primary producers to apex predators (range, 1–5). Small differences in trophic level can reflect large differences in diet. Developed by Bonhommeau et al (2013)[6]. | Not validated. |
| ***Livelihoods, Poverty, & Equity*** | | | |
| Income and poverty | Share of agricultural income from fisheries | Agricultural income from fishery activities as a share of total income (%). | Disaggregated data by sub-sector to identify fishery-specific data are not available in harmonized global database. De novo calculation from microdata is not feasible at present. |
| Income and poverty | Agricultural income (livestock, crop, fishery, forestry, ag wage), share of total income (%) | Household income from all agricultural activities, as a share of total household income. | Data are only available for 64 low- and lower-middle income countries. There are no comparable data from countries at other income levels to combine with these to meet the prerequisite requirement for global coverage. |
| Income and poverty | Proportion of the rural population living below the international poverty line | The indicator "proportion of the population below the international poverty line" is defined as the percentage of the population living on less than $1.90 a day at 2011 international prices. | Data are not presently available. When the data for the SDG indicator 1.1.1 become available for urban/rural disaggregation, we can calculate the share of the rural population living below this line (by combining with population data). |
| Employment | Frequency rate of occupational injuries | | Insufficient disaggregation to link to food systems. |
| Employment | Farmer age | Average age of farmers per nation. | Closest proxy indicators are "Average age of self-employed or unpaid workers, aged 15-64, rural" and "Average age of wage workers, aged 15-64, rural", however these data have insufficient coverage across countries and country income levels. |
| Employment | Farm labour force - family labour | Family workers are persons who help another member of the family to run an agricultural holding or other business, provided they are not considered as employees. Persons working in a family business or on a family farm without pay should be living in the same household as the owner of the business or farm, or in a slightly broader interpretation, in a house located on the same plot of land and with common household interests. | Insufficient data coverage across countries and country income levels. |

| Domain | Indicator name | Definition or description | Reason for elimination from consideration |
|---|---|---|---|
| ***Governance*** | | | |
| Effective Implementation | Representation and influence of rural organizations and rural people, IFAD Rural Sector Performance Assessments | Representation and influence of rural organizations and rural people. Scale from 1 (worst) to 6 (best); Annual until 2014 and thereafter only conducted every 3 years; Targeted at LMICs and based on subjective assessments of IFAD country economists and then centrally reviewed. | The assessment approach seems to be a "black box"; data collected only every 3 years. |
| Accountability | Healthy Food Environment Policy Index (Food-EPI) developed by International Network for Food and Obesity/Non-communicable Diseases Research, Monitoring and Action Support (INFORMAS), Swinburn et al (2013).[7] | Food-EPI is comprised of 7 policy domains related to food environments and 6 infrastructure domains: leadership, governance, funding and resources, monitoring and intelligence, platforms for interaction and health-in-all-policies. Each domain is specified by several good practice indicators (47 in total) that encompass the directions necessary to improve the healthiness of food environments and to help prevent obesity and diet-related NCDs. | It only covers 49 countries. |
| Effective Implementation | World Benchmarking Alliance (WBA) Food and Agriculture Benchmark | | Data is at the level of companies rather than countries. |
| Effective Implementation | CAADP's Agricultural Transformation Scorecards (Department of Rural Economy and Agriculture, African Union, 2020) | | Only available for Africa. |
| Shared Vision | Hunger and Nutrition Commitment Index (HANCI) | | No longer being collected. |
| Strategic planning | WHO Landscape Analysis (WHO, 2012) | It is a tools package for in-depth country assessment that consist of stakeholder mapping tool, interview questionnaires, and analytical framework to do analysis of a country's capacities and resources, and identifies promising actions that could be scaled up to improve nutrition. | Seems to have only been done once (in 2012) and only covers a handful of countries. |
| Effective Implementation | WHO Global database on the Implementation of Nutrition Action (GINA) | GINA provides a repository of policies, actions and mechanisms related to nutrition. It is an interactive platform for sharing standardized information on nutrition policies and action, i.e. what are the commitments made and who is doing what, where, when, why and how (including lessons learnt). | Great resources on policy actions but it would need to be consolidated into quantifiable indicators. |
| Accountability | Access To Nutrition Index (ATNI) | The Global Access to Nutrition Index focusses on the role that food and beverage manufacturers play in making healthy food affordable and accessible to all consumers globally. | Focuses on companies, not countries. It does not seem to have been updated since 2018. |

| Domain | Indicator name | Definition or description | Reason for elimination from consideration |
|---|---|---|---|
| Accountability | Monitoring, Evaluation, Accountability, and Learning (MEAL), Scaling Up Nutrition (SUN) | Tracks progress towards overall SUN movement objectives and enforces mutual accountability. Relies on review of multi-stakeholder joint assessments, national budget analysis, national nutrition action plans, subnational action mapping, etc. | Not clear that this is still collected. Last update was 2018/2019. Also does not meet coverage requirement. |
| ***Resilience & Sustainability*** | | | |
| Resilience capacities | Positive peace index | The Positive Peace Index measures the level of societal resilience of a nation or region. | Black box, composite indicator. |
| Resilience capacities | Social trust | Based on responses to questions that include trust in various group, trust in other people confidence in police, courts, govt, etc. | Data available for only 57 countries. |
| Resilience responses/strategies | Agricultural irrigated land (% of total agricultural land) | Agricultural irrigated land refers to agricultural areas purposely provided with water, including land irrigated by controlled flooding. | Using irrigation for some countries does not make sense because they don't need it for their agriculture as much as others (e.g. North Europe vs South Europe), so irrigation is not a globally valid indicator. In Bene et al 2019[8] the following exclusion criterion was used: "Global validity. Were excluded indicators that refer to processes that are specific to some specific regions of the world and not to others. For instance, “Percentage of agricultural land lost yearly to desertification” is excluded as desertification is a phenomenon that by definition can only occur in some specific regions of the world" - similarly the use of irrigation is not globally valid. |

# Supplementary Material, Appendix C

This appendix contains the reports of the expert and stakeholder consultations.

# Expert Consultation Report

The Food Systems Countdown Initiative (FSCI) 2030 aims to track food systems transformation until 2030 and beyond. This tracking is done by means of a set of indicators applicable across various themes and geographies.

In order to develop a list of potential candidate indicator, the initiative undertook a process of consultations with various stakeholders.

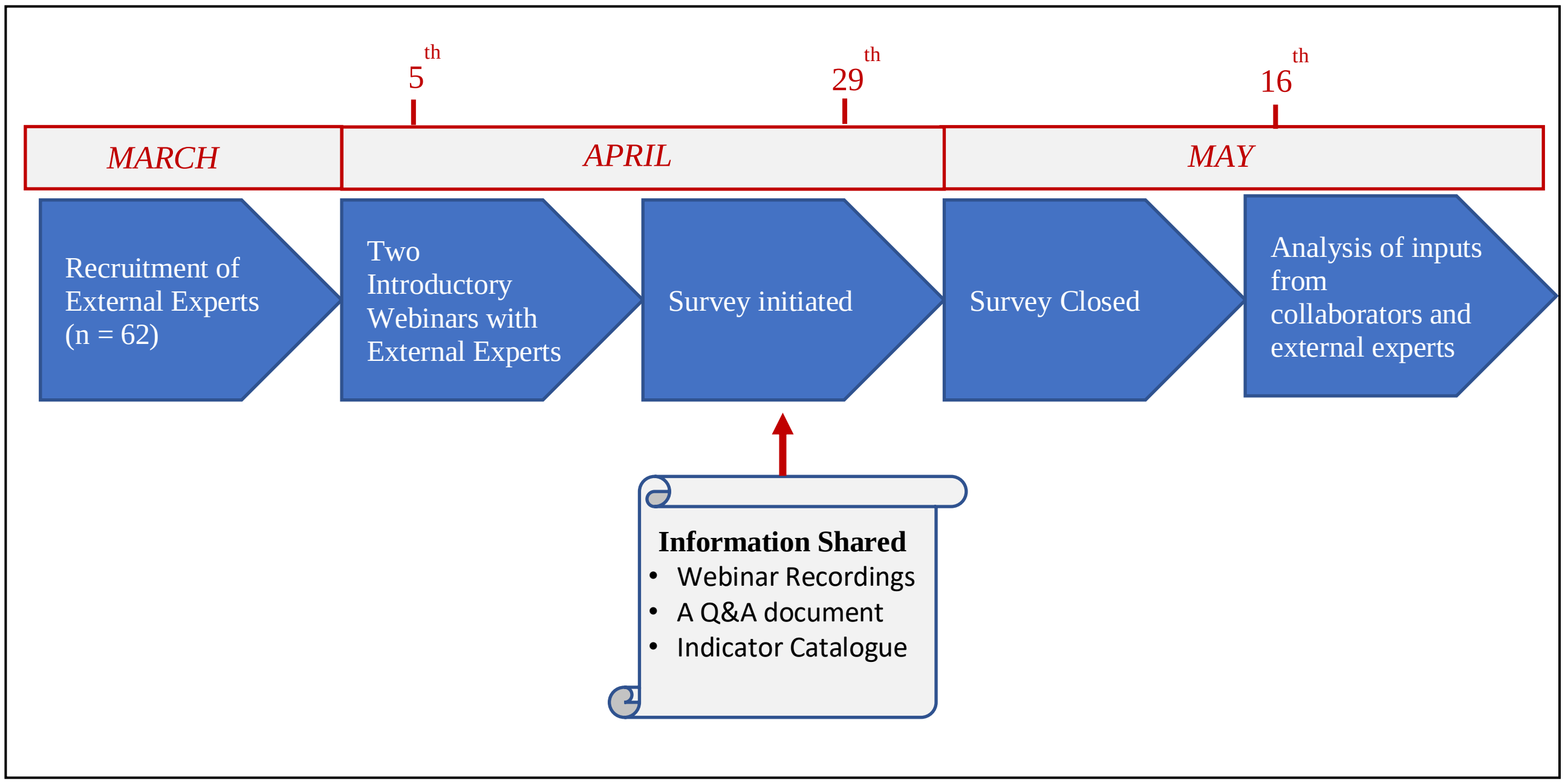

*Timeline for the expert consultation process*

The consultation process to finalize the list of indicators involved two groups of experts. One group comprised the collaborators—involved in the FSCI work from the beginning—who developed the long list of candidate indicators. The other group consisted of external experts who were recruited from a variety of networks and organizations. It was made sure that there was a diverse group of outside experts who represented various geographies, subject areas, and levels of experience.

During February and March 2022, 173 individuals were invited to join the external expert group; 62 of them replied and confirmed their attendance. They took part in the initial webinar and discussions on the FSCI indicator selection process. However, only 28 of the 62 external experts completed the indicator scoring survey. In addition, from the core collaborators group 39 completed the survey, thus a total of 67 individuals scored the FSCI indicators.

Two introductory webinars were held in the beginning of April 2022 to introduce the FSCI 2030 to external experts and describe the survey and indicator scoring criteria. After the webinar, the external experts received the scoring survey, webinar recordings, presentations, a document with questions and answers, and an indicator catalogue. The survey was closed Mid-May 2022, results were analyzed and presented to the FSCI core collaborator group.

**Composition of External Experts who confirmed their participation in the consultation.**

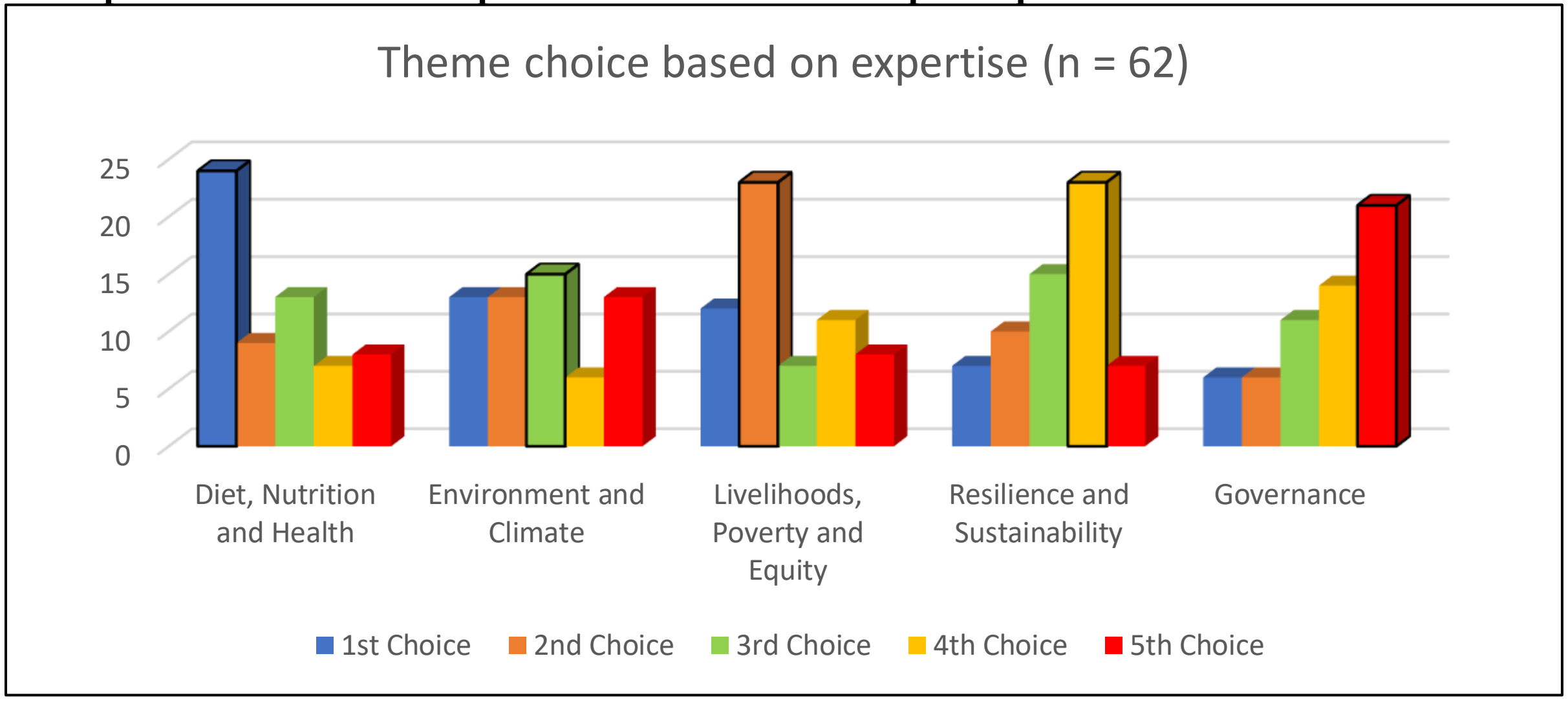

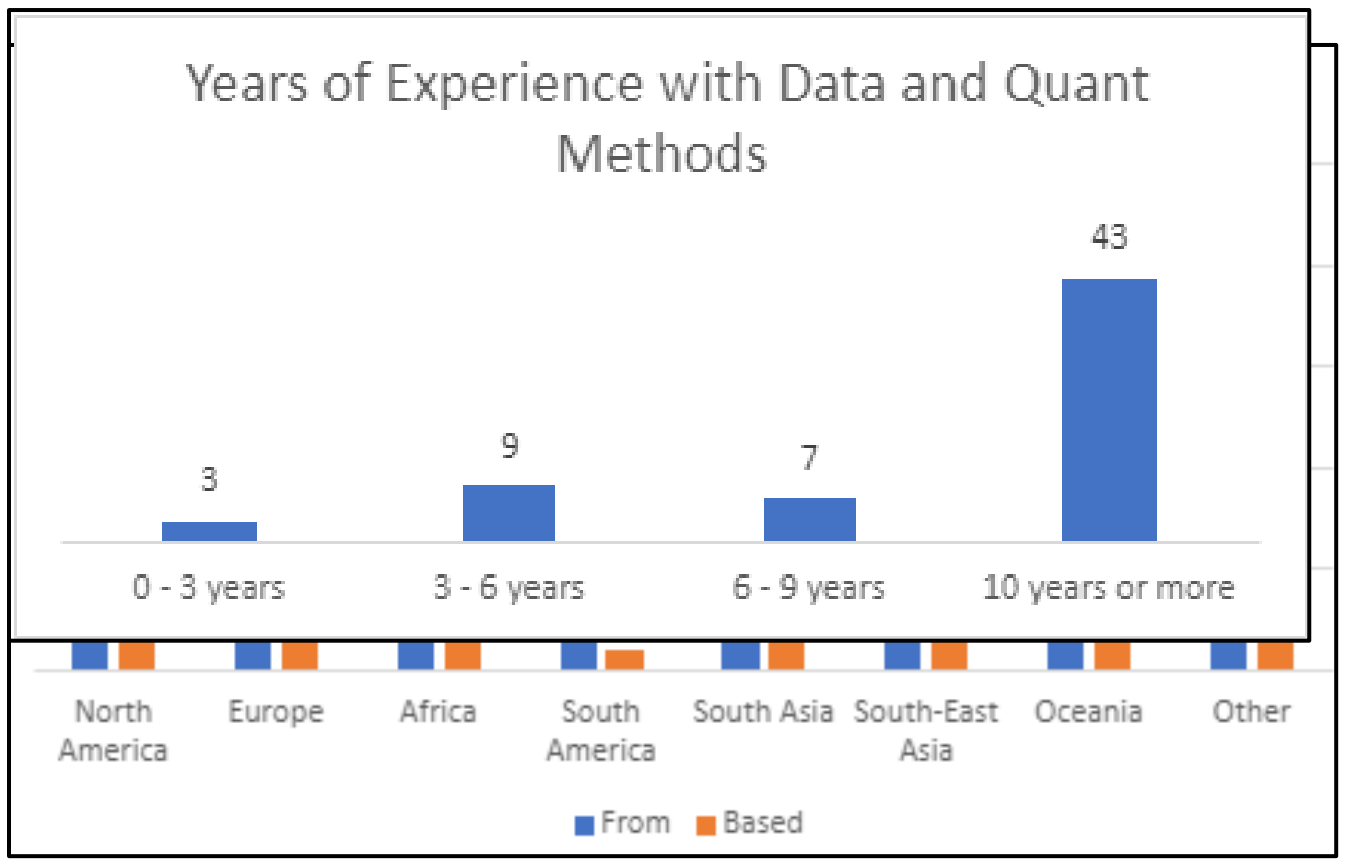

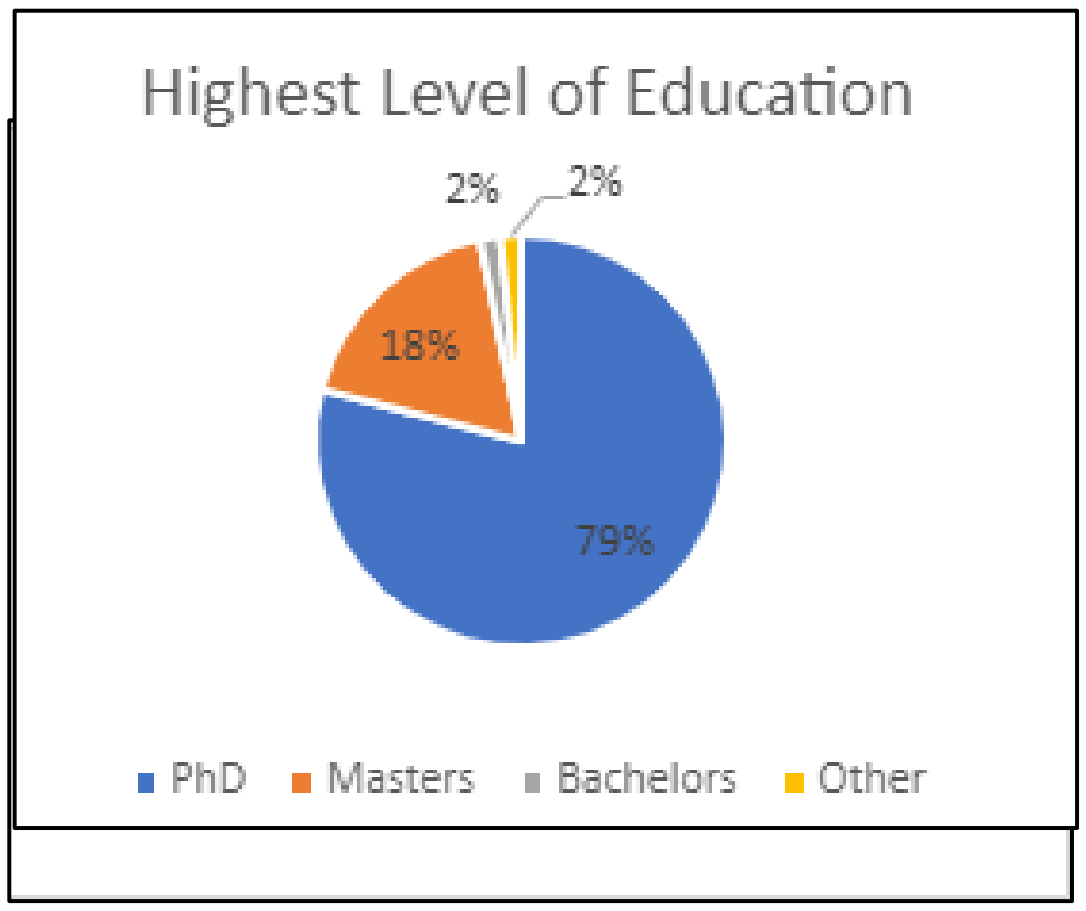

**The Indicator Scoring Survey Instrument**
The indicator scoring survey was created using the online platform KOBOToolbox. The survey included 98 indicators spread across five themes and 23 domains. The breakdown of the number of indicators in each domain under the five themes is shown in the table below.

| Theme and Domains | Number of Indicators |
|---|---|
| **DIETS, NUTRITION, AND HEALTH** | |
| Diet Quality | 19 |
| Food Security | 4 |
| Food Environments | 10 |
| Policies Affecting Food Environments | 8 |
| **ENVIRONMENT AND CLIMATE** | |
| Greenhouse Gas Emissions | 3 |
| Land and Soil | 2 |
| Pollution | 2 |
| Biosphere Integrity | 2 |
| Biosphere Intactness | 1 |
| Water | 1 |
| **LIVELIHOODS, POVERTY, AND EQUITY** | |
| Income and Poverty | 2 |
| Employment | 3 |
| Social Protection | 3 |
| Rights | 3 |
| **GOVERNANCE** | |
| Shared Vision | 2 |
| Strategic Planning | 1 |
| Effective Implementation | 3 |
| Accountability | 3 |
| **RESILIENCE** | |
| Exposure to Shocks | 5 |
| Resilience Capacities | 11 |
| Agro- and Food Diversity | 2 |
| Resilience Responses/Strategies | 2 |
| Long-term Outcomes | 6 |

For each indicator, the scorers were given a set of details that included the name of the indicator, its definition, relevant methodological details, data sources, links to the key data sources and methodology, and any additional information. Each indicator was evaluated using the following criteria:

**» Please indicate the extent to which you agree this indicator meets the criteria. (1 = least agreement, 5 = most agreement)**

**» » RELEVANCE**

| RELEVANCE | 1 (least) | 2 | 3 | 4 | 5 (most) | Opt Out |
|---|---|---|---|---|---|---|
| **Can be clearly mapped to the food systems framework** | ○ | ○ | ○ | ○ | ○ | ○ |
| **Observing change in the indicator is possible within a decade** | ○ | ○ | ○ | ○ | ○ | ○ |

**» » HIGH QUALITY**

| HIGH QUALITY | 1 (least) | 2 | 3 | 4 | 5 (most) | Opt Out |
|---|---|---|---|---|---|---|
| **Well-documented methodologies and metadata** | ○ | ○ | ○ | ○ | ○ | ○ |
| **Data are nationally representative** | ○ | ○ | ○ | ○ | ○ | ○ |

**» » INTERPRETABLE**

| INTERPRETABLE | 1 (least) | 2 | 3 | 4 | 5 (most) | Opt Out |
|---|---|---|---|---|---|---|
| **Change has a clear interpretation** | ○ | ○ | ○ | ○ | ○ | ○ |
| **Data are comparable across countries** | ○ | ○ | ○ | ○ | ○ | ○ |

**» » IMPORTANCE**

| IMPORTANCE | 1 (least) | 2 | 3 | 4 | 5 (most) | Opt Out |
|---|---|---|---|---|---|---|
| **Do you agree that this is an important indicator to assess desirable food system transformation?** | ○ | ○ | ○ | ○ | ○ | ○ |

**Composition of Survey Respondents by Domains**

There were more collaborators than external experts scoring indicators in general. The number of external experts exceeded the number of collaborators only in the Food Environments domain in the Diets, Nutrition and Health theme and were split 50/50 for Land and Soil domain in the Environment and Climate theme. The bar charts below illustrate the breakdown of survey respondents for each domain under the five themes.

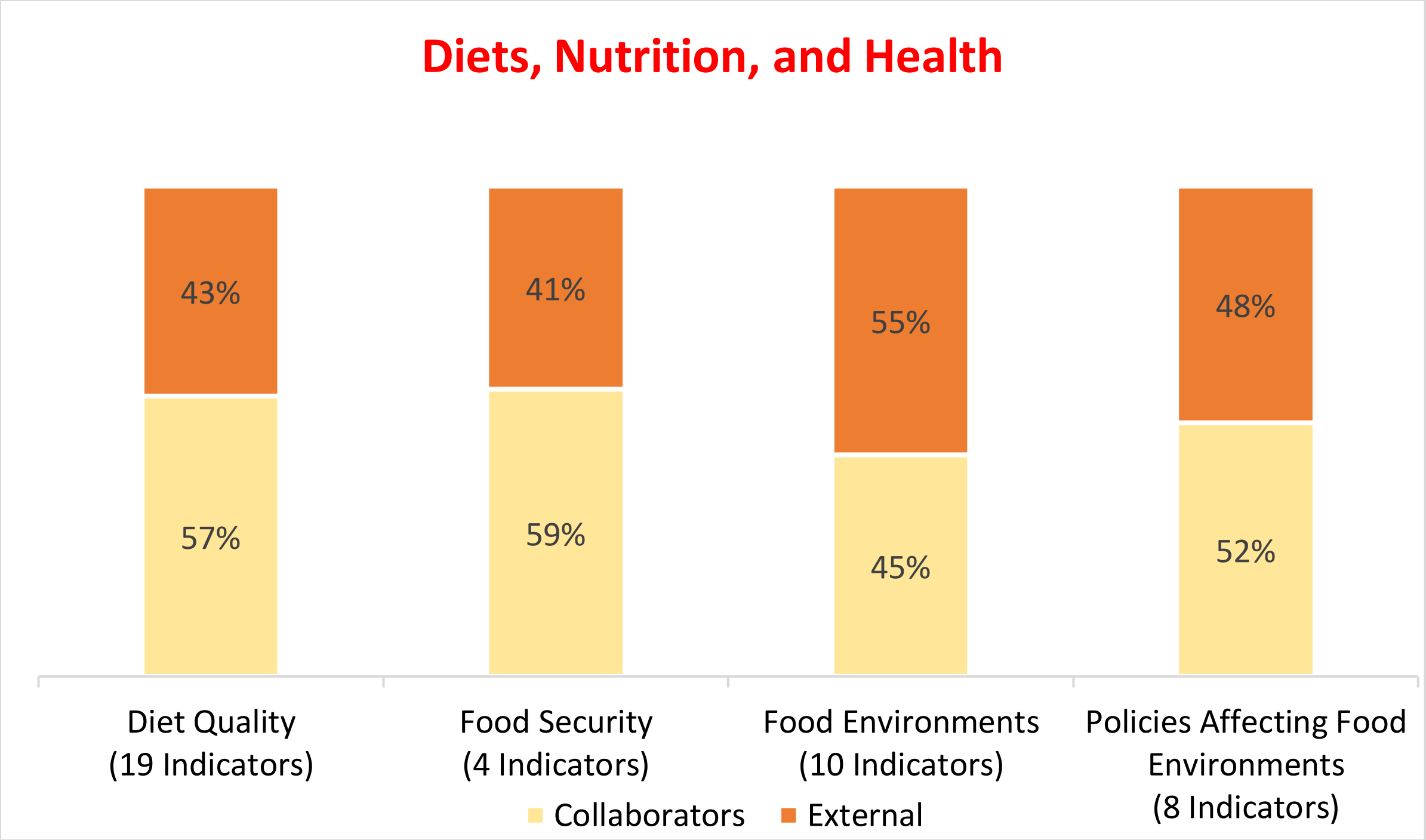

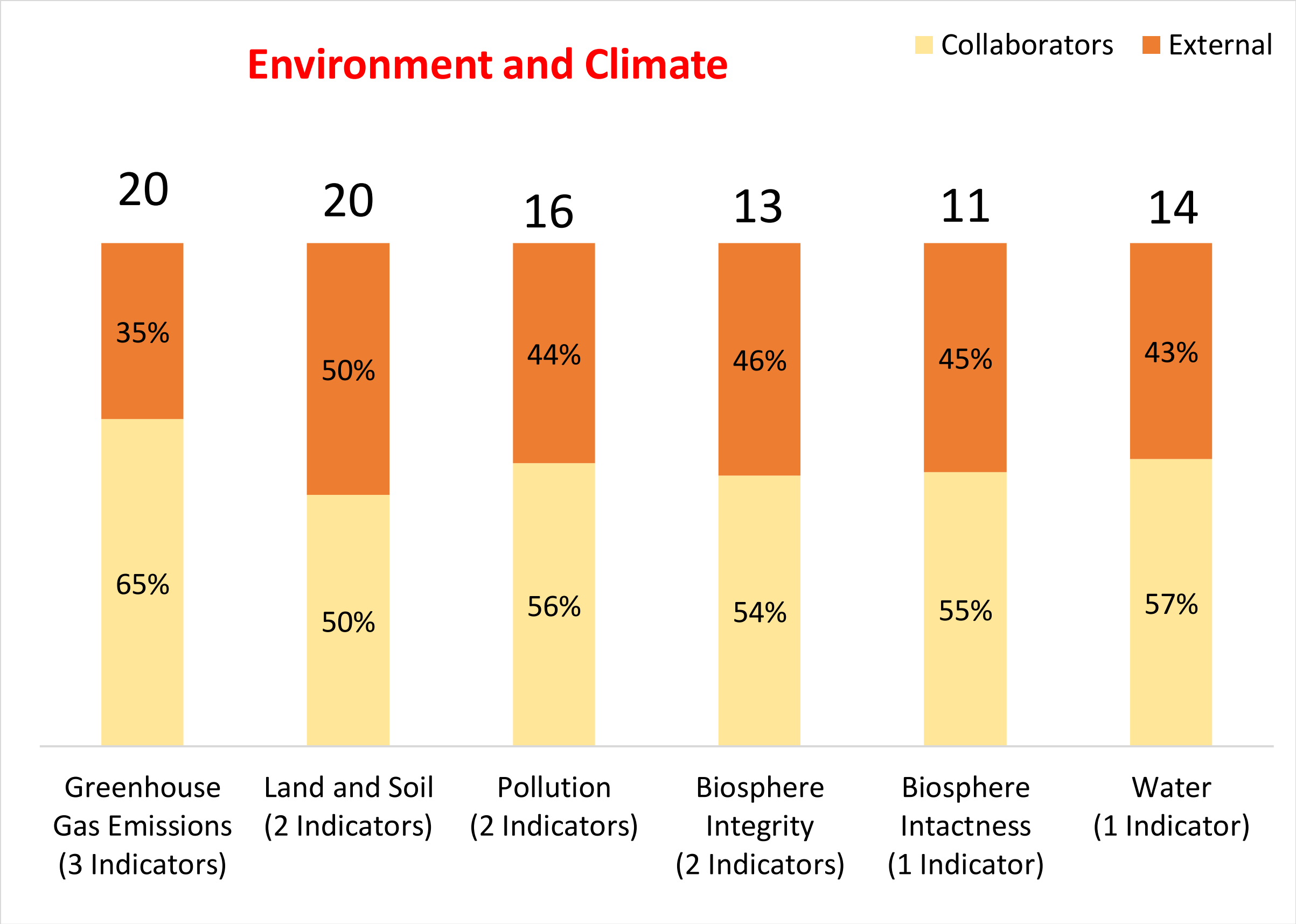

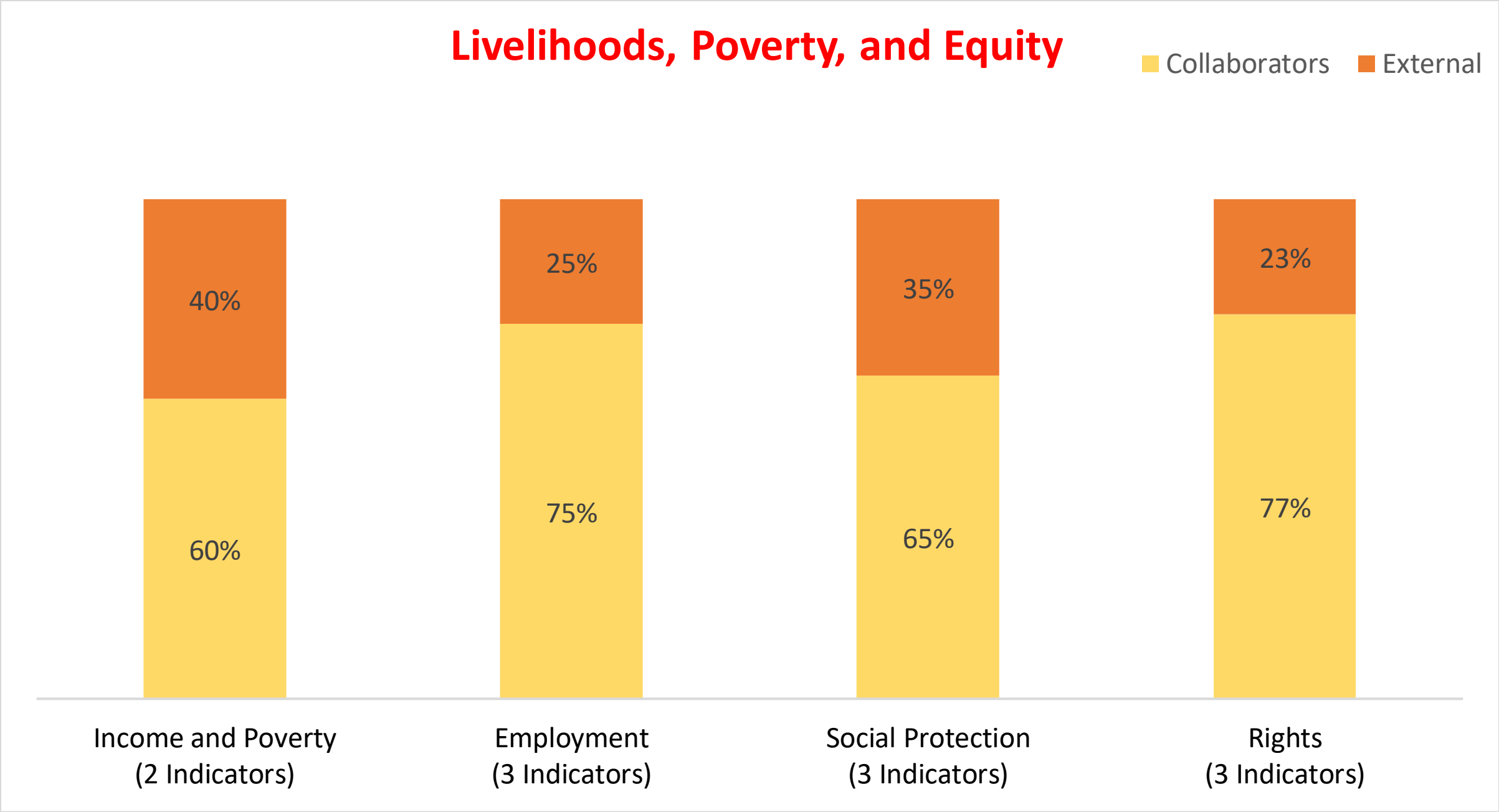

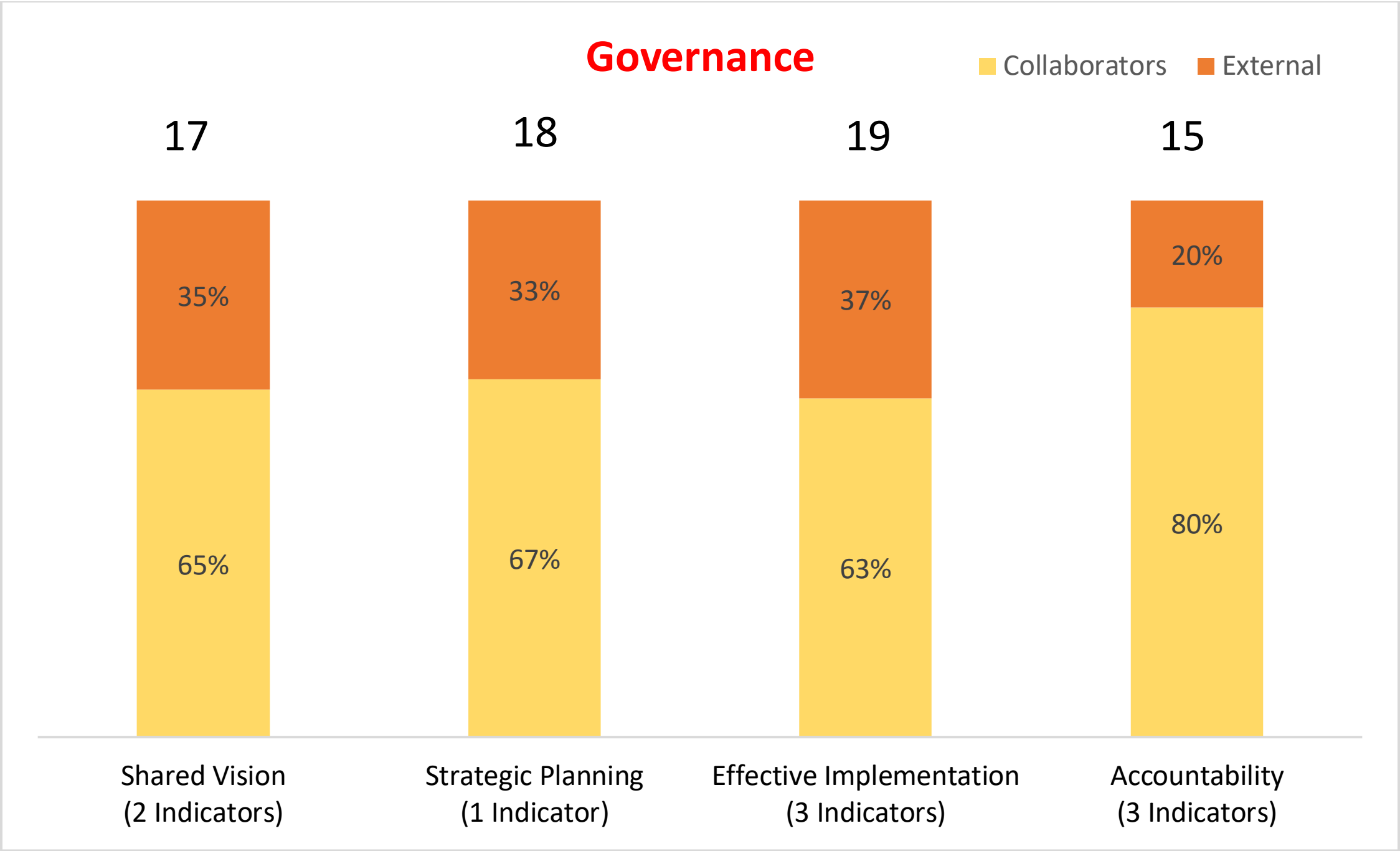

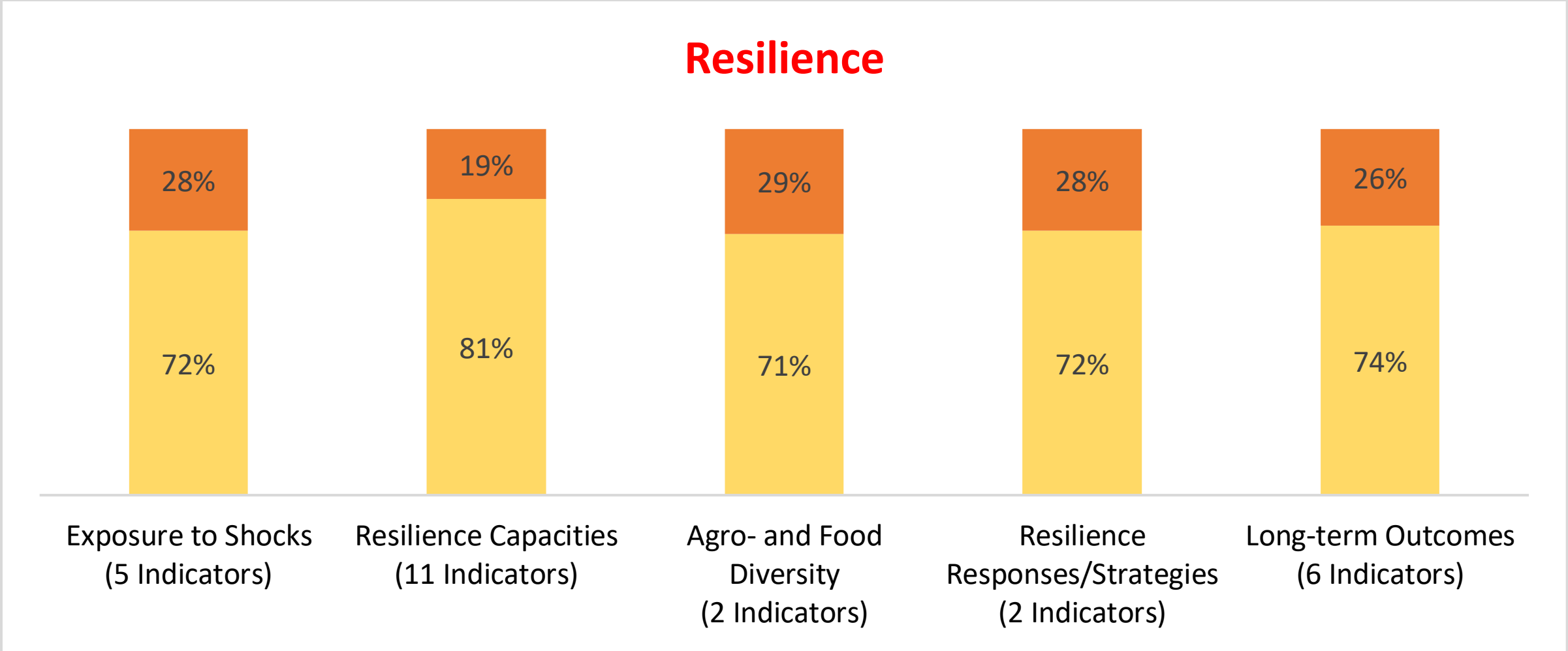

**Indicator Scoring Results**

Using the survey data, we calculated a summary score for each indicator. We began by calculating a weighted average score at the criterion level. Because different numbers of respondents scored each indicator, and there was an option to opt out of scoring any single criterion, the denominator of the weighted average is the number of respondents who selected a score (1 to 5) for that criterion. The criteria scores for each indicator were then added up to get an overall indicator score. The indicators within each domain were then ranked on the summary score to determine which were highest and lowest ranked. Scoring and ranking indicators helped the core collaborators determine which indicators to include for monitoring purposes.

The table below ranks the top two indicators for each domain. Some domains only have one or two indicators, so they are listed as is.

| Domain | Top 2 Indicators | Weighted Score | Rank within Domain | Rank within Theme |
|---|---|---|---|---|
| **DIET QUALITY** | Minimum Dietary Diversity for Women of Reproductive Age (MDD-W) | **30.64** | **1** | **2** |
| | Minimum Dietary Diversity (MDD) among IYC (age 6-23 months) | **30.11** | **2** | **5** |
| **Food Security** | Food Insecurity Experience Scale (FIES) - based indicators | **30.72** | **1** | **1** |
| | Proportion of people who cannot afford a healthy diet | **30.30** | **2** | **3** |
| **Food Environments** | Cost of a Healthy Diet | **29.65** | **1** | **6** |
| | Availability/supply (g/day/capita) - vegetables | **28.54** | **2** | **12** |
| **Policies Affecting Food Environments** | Best-practice policy implemented for industrially produced trans-fatty acids (TFA) (Y/N) | **28.86** | **1** | **9** |
| | Implementation of marketing of breast-milk substitutes restrictions (Fully achieved/ Partially achieved / Not achieved) | **28.06** | **2** | **17** |
| **Greenhouse Gas Emission** | Greenhouse gas emissions from food systems (farm to fork) | **31.22** | **1** | **1** |
| | GHG Emissions intensity | **29.74** | **2** | **3** |
| **Land and Soil** | Cropland expansion | **27.99** | **1** | **4** |
| | % change in soil organic carbon | **24.26** | **2** | **11** |
| **Biosphere Intactness** | biodiversity intactness | **25.11** | **1** | **8** |
| **Biosphere Integrity** | Functional Integrity (% agricultural land with minimum level of natural habitat) | **25.01** | **1** | **9** |
| | Fishery health index - Progress score | **24.57** | **2** | **10** |
| **WATER** | Agricultural water withdrawal as % of total renewable water resources | **30.15** | **1** | **2** |
| **Pollution** | Total pesticides per unit of land | **27.27** | **1** | **6** |
| | Sustainable Nitrogen Management index | **26.48** | **2** | **7** |

| Domain | Top 2 Indicators | Weighted Score | Rank within Domain | Rank within Theme |
|---|---|---|---|---|
| **Shared Vision** | Presence of a national food system transformation pathway | **25.01** | **1** | **5** |
| | Civil society index, Varieties of Democracy (V-Dem) | **24.31** | **2** | **9** |
| **Strategic Planning** | Policy coordination | **27.12** | **1** | **1** |
| **Effective Implementation** | Policy implementation | **27.05** | **1** | **2** |
| | Government Effectiveness | **25.61** | **2** | **3** |
| **Accountability** | Budget transparency score, Open Budget Initiative | **25.06** | 1 | **4** |
| | Voice and Accountability, WGI | **24.56** | 2 | **7** |
| **Income and Poverty** | Households with significant income from agriculture | **26.13** | 1 | **6** |
| | % population earning low pay | **25.68** | 2 | **8** |
| **Employment** | Unemployment, rural and urban | **28.14** | 1 | **2** |
| | Monthly wages for agricultural workers compared to the country's median monthly wage | **25.59** | 2 | **9** |
| **Social Protection** | Adequacy of benefits from social protection and labor programs | **27.94** | 1 | **3** |
| | Coverage of any social protection and labor program | **26.95** | 2 | **4** |
| **Rights** | Distribution of land holdings by sex (Female %) | **28.17** | 1 | **1** |
| | Constitutional recognition of the right to adequate food | **25.77** | 2 | **7** |
| **Exposure to Shocks** | SDG 16.1.2: Conflict-related deaths per 100,000 population, by sex, age, and cause. | **27.93** | 1 | **13** |
| | EM-DAT economic impact | **27.9** | 2 | **14** |
| **Resilience Capacities** | Poverty headcount ratio at $1.90 a day | **29.59** | 1 | **3** |
| | FAO Dietary Sourcing Flexibility Index | **29.43** | 2 | **6** |
| **Agro- and Food Diversity** | Agrobiodiversity indicator 2: Species diversity in food production | **29.13** | 1 | **10** |
| | Agrobiodiversity indicator 1: Species diversity in food supply | **29.08** | 2 | **11** |

| Domain | Top 2 Indicators | Weighted Score | Rank within Domain | Rank within Theme |
|---|---|---|---|---|
| **Resilience Responses/Strategies** | Consumption-based Coping Strategy Index (reduced CSI) | **29.92** | **1** | **2** |
| | Nature-based solution for adaptation | **26.73** | **2** | **22** |
| **Long-term Outcomes** | Domestic food price volatility index | **30.71** | **1** | **1** |
| | Stability of Food Insecurity Experience Scale (FIES) - based indicators | **29.57** | **2** | **4** |

Survey respondents were also asked to suggest alternative indicators that were not on the long list. These suggested indicators are as follows:

| Theme: **Domain** | Suggested Indicator | Suggested Indicator Data Source |
|---|---|---|
| Diet, Nutrition and Health: **Diet Quality** | Water and Sanitation (component of the Social Progress Index) | https://www.socialprogress.org/index/global |
| Diet, Nutrition and Health: **Policies Affecting Food Environments** | Food expenditure | LSMS/household surveys; https://inddex.nutrition.tufts.edu/data4diets/data-source/household-consumption-and-expenditure-surveys-hces |
| Diet, Nutrition and Health: **Policies Affecting Food Environments** | # of dieticians/public health nutritionist in a country | https://www.who.int/data/nutrition/nlis/info/nutrition-professionals-density |
| Diet, Nutrition and Health: **Diet Quality** | Hypertension | National surveys and WHO step surveys |
| Environment and Climate: **Pollution** | Food Loss and waste, N and P surplus | https://www.cell.com/one-earth/fulltext/S2590-3322(21)00473-5; https://www.fao.org/platform-food-loss-waste/flw-data/en/#:~:text=The%20Food%20Loss%20and%20Waste,grey%20literature%2C%20countries%20among%20others. |
| Environment and Climate: **Land and Soil** | deforestation emissions; peatland degradation; fires | FAOSTAT ET domains |

| Theme: **Domain** | Suggested Indicator | Suggested Indicator Data Source |
| --- | --- | --- |
| Environment and Climate: **Land and Soil** | organic agriculture; forest land | FAOSTAT RL and EL domains |
| Environment and Climate: **Water** | Water depletion | link |
| Governance: **Accountability** | I'd like to have seen indicators for discrimination under the "governance: accountability" aim. Additionally, and I appreciate such an off-the-shelf database doesn't exist, but indicators re: stigma for accessing food assistance would be important to know under governance and I would put safety net access and rights protections as governances issues (not under livelihoods) | |
| Governance: **Accountability** | Hanci - can we reinitiate it? | http://www.hancindex.org/the-index/ |
| Governance: **Effective Implementation** | Nutrition spending? | Check GNR |
| Governance: **Effective Implementation** | Regulatory Enforcement (component of the WJP Rule of Law Index) | https://worldjusticeproject.org/our-work/wjp-rule-law-index |
| Resilience: **Long-term Outcomes** | SDG 12.3.1.A Food Loss Index | https://www.fao.org/sustainable-development-goals/indicators/1231/en/ |
| Resilience: **Long-term Outcomes** | SDG 12.3.1.B Food Waste Index | https://www.unep.org/resources/report/unep-food-waste-index-report-2021 |
| Resilience: **Long-term Outcomes** | Gender Equality (component of the SDG Gender Index) | https://www.equalmeasures2030.org/wp-content/uploads/2021/06/SDG-Gender-Index-1.png |

| Theme: **Domain** | Suggested Indicator | Suggested Indicator Data Source |
|---|---|---|
| Resilience: **Long-term Outcomes** | Political Empowerment (component of the Global Gender Gap Index) | https://www.weforum.org/reports/gender-gap-2020-report-100-years-pay-equality |
| Resilience: **Long-term Outcomes** | Food Losses Index | FAO |
| Resilience: **Long-term Outcomes** | Food waste index | UNEP |

**Participants external to the co-authors:**

| **Name** | **Affiliation** |
|---|---|
| Carl Lachat | Department of Food Technology, Safety and Health, Ghent University, Belgium |
| Inge D. Brouwer | Division of Human Nutrition and Health, Wageningen University and Research, the Netherlands |
| Anna Lartey | |
| William A. Masters | Tufts University, Friedman School of Nutrition |
| Shauna Downs | Rutgers School of Public Health |
| Jamie Morrison | |
| Xin Zhang | Appalachian Laboratory, University of Maryland Center for Environmental Science |
| Chris Vogliano | USAID Advancing Nutrition |
| Elizabeth L. Fox | Department of Public and Ecosystem Health, Cornell University |
| Natalia Strigin | International Rescue Committee |
| Jinfeng Chang | Zhejiang University |
| Jonathan R.B. Fisher | The Pew Charitable Trusts |
| Frank Eyhorn | Biovision Foundation |
| Michelle Jurkovich | Assistant Professor of Political Science, University of Massachusetts Boston |
| Min Jung Cho | Leiden University |
| Jing Zhu | |
| Michelle Holdsworth | IRD (French National Research Institute for Sustainable Development) |
| Lijun Zuo | Aerospace Information Research Institute, Chinese Academy of Sciences |
| Mariana Rufino | Lancaster University, UK |
| Olutayo Adeyemi | Department of Human Nutrition and Dietetics, University of Ibadan, Nigeria |
| Rachel Nugent | RTI International Center for Global Noncommunicable Diseases |
| Ojwang A A | The Technical University of Kenya |
| Ramya Ambikapathi* | Purdue University |
| Mark Lawrence | IPAN, Deakin University |
| Julia Compton | |
| Dr. Isabel Madzorera | Department of Global Health and Population, Harvard School of Public Health |
| Jamie L. Thomas | Bread for the World |
| Roberto O. Valdivia | Applied Economics, Oregon State University, USA |

* Joined the FSCI collaborators after the execution of the survey.

# Regional Stakeholder Consultation Reports

### FAO *RAF* Regional Expert Consultation of the Food System Countdown Initiative's Indicator Framework

**Virtual, May *19th* 2022**

*This report was produced by the FAO Regional Office for Africa to summarize the results of the FAO RAF regional expert consultation held on May 19 2022.*

**Introduction:**

Accelerating action towards the achievement of SDGs is becoming more imperative that ever before. With only 8 years left to 2030, it is becoming imperative that concerted and collaborative actions be found to achieve the 2030 SDG Agenda. This is becoming especially urgent given the shocks in food systems that are taking the world off track in achieving the SDGs. For the Africa Region, recent shocks in the food systems are but not limited to Climate Change; Conflicts and terrorism; economic impact of COVID-19 and other pandemics; and the increased Cost of food (the 4C's). Climate shocks destroy lives, crops and livelihoods, undermine people's ability to feed themselves, and have displaced 30 million from their homes globally in 2020. Currently in the Horn of Africa region, Livestock are dying, crops are failing, and an estimated 13 million people wake up hungry every day across the region as it grapples with severe drought caused by the driest condition since 1981.[1]

Conflict are major threats to food security and nutrition and the leading cause of global food crises. Marked increases in the number and complexity of conflicts in the last ten years have eroded gains in food security and nutrition, leading several countries to the brink of famine. Of the more than 800 million chronically food insecure people in the world, 60 percent live in countries affected by conflict. In the Sahel, over 10.5 million people are facing Crisis (IPC 3) levels of hunger, including 1.1 million in Emergency (IPC 4), across five countries - Burkina Faso, Chad, Mali, Mauritania, and Niger. [2]
The COVID-19 pandemic has worsened the prevalence of multiple forms of malnutrition and could have lasting effects. More people slid into chronic hunger in 2020 than in the previous five years combined, and one in five children around the world are stunted. Children are paying the heaviest price, at the beginning of 2020, one in every two schoolchildren, or 388 million children, received school meals every day from national programmes in at least 161 countries from all income levels. The COVID-19 pandemic brought an end to this decade of global growth. At the height of the crisis in April 2020, 370 million children were suddenly deprived of what for many was their main meal of the day with the closure of schools.[3]

The rising food prices, resulting from the COVID-19 pandemic and other factors, including the recent conflict in Ukraine, risk of putting healthy diets out of reach of even more people, on top of the current 964.8 million people in Africa who cannot afford an adequate diet.[4] FAO has estimated that, globally, the number of undernourished people, which was around 817 million in 2021, could increase by 7.6 - 13.1 million people in 2022-2023 due to the conflict between Russia and Ukraine.[5]

---

[1] https://news.un.org/en/story/2022/02/1111472

[2] https://reliefweb.int/sites/reliefweb.int/files/resources/GHI%20February%202022%20Report%20-.pdf

[3] FAO, ECA and AUC. 2021. Africa – Regional Overview of Food Security and Nutrition 2021: Statistics and trends. Accra, FAO. https://doi.org/10.4060/cb7496en

[4] FAO, ECA and AUC. 2021. Africa – Regional Overview of Food Security and Nutrition 2021: Statistics and trends. Accra, FAO. https://doi.org/10.4060/cb7496en

[5] Technical Briefing to FAO Members on the impact of COVID-19 and the war in Ukraine on the outlook for food Security and nutrition, March 2022. https://www.fao.org/3/cb9241en/cb9241en.pdf

For the continent to be back on track to achieving the SDGs, the food systems must transform, in line with the commitments made during the UN Food Systems Summit.[6] Tracking can assess performance relative to established targets and goals and incentivize action. Doing so for food systems complements other global and regional monitoring and tracking initiatives focused on related outcomes, such as sustainable agriculture, nutrition and health. Tracking and assessment also offers food system actors and stakeholders (e.g. civil society, governments and international organizations) actionable evidence to hold governments, consumers (specifically, those with the privilege to choose), and the private sector accountable for food system transformation.

In this regard, The Food Systems Countdown Initiative has arrived at an opportune time for the continent. The Food Systems Countdown Initiative (the 'Initiative') is working to build a science-based observational system to track the performance and transformation of food systems globally. The 2021 UN Food Systems Summit presented a window of opportunity with food systems on the international political agenda, yet no rigorous mechanism currently exists to measure and track all aspects of food systems, their interactions and their changes over time. Deliberately changing complex systems that cut across sectors, jurisdictions and national borders calls for a comprehensive, ongoing programme of scientific measurement, tracking, and assessment of all aspects of the system to guide decision-makers and hold those in power to account for transformation. Led by John Hopkins University, GAIN and FAO, an unparalleled partnership has come together to implement the Initiative. Over 50 collaborating scientists have joined from nearly 30 organizations across academia, NGOs, and UN agencies, and from nearly all continents.

The principal goal of the Initiative is to provide independent tracking of a curated, parsimonious set of indicators that together cover all the important aspects of food systems. The architecture so far proposed covers the five thematic areas described below, each with three to five indicator domains (i.e. subtopics for which indicators must be found to capture that aspect of the thematic area). Fanzo et al. (2021) defined criteria for indicators to be included in the tracking system that will be used to guide the selection of indicators. The architecture covers five domains as follows:  1. Diets and nutrition; 2. Environment and climate domain 3. Livelihoods, poverty and equity Domain; 4. Governance Domain 5. Resilience and sustainability Domain

**Objectives of the RAF Regional Expert Consultation**

In line with The Initiative's commitment to an inclusive, consultative, and transparent process, the objective of the FAORAF consultation was to subject the set of propped indicators to a validation and peer review mechanism by carefully selected experts and scientist working with data and policy. The regional expert consultation is an opportunity to get inputs, comments, and suggestions on the monitoring framework proposed by the initiative. The framework is not mandatory. However, the consultations ensure that it has the capacity to be a useful tool for policy decision-making processes. The outcome of each regional consultation will be a public document that summarizes the inputs received in each thematic area.

For the Africa Region, the experts were drawn from those working at country level in policy development from the relevant Planning Units of the Ministries of Agriculture, National Statistical Offices and Ministries working on the thematic areas of the indicator framework; regional level experts, including non-state actors, academia and NGOs. The experts working and collaborating with the Comprehensive Africa Agriculture Development Programme (CAADP) were of particular value given that the CAADP initiative is an African led process sanctioned by Heads of Member States that has the objective of tracking collective performances in order to trigger continental, regional and national level action programmes to drive agricultural transformation in Africa.

[6] https://www.un.org/en/food-systems-summit

The CAADP process has strong political support that strengthens national and regional institutional capacity for agriculture data collection and knowledge management to inform actions that will support improved evidence-based planning, implementation, monitoring and evaluation, mutual learning and foster alignment, harmonization and coordination among multi-sectoral and multi-stakeholder efforts.

**Opening:**

Participants were welcomed by Mr Mphumuzi Sukati – Senior Nutrition and Food Systems Officer, who was stepping in for the Regional Programme Leader, Mr Ade Freeman. Following his welcoming of participant, Mr Jose Rosero Moncayo, Director of Statistics gave the opening remarks. He noted the importance of food systems transformation to deliver better on SDGs, and the need to track this transformation in line with the UN Food Systems Summit and the commitments made by partners and Member States to ensure sustainability of food systems. He then introduced briefly *The Food Systems Countdown Initiative* framework and what it entails in helping countries monitor their progress towards transforming food systems for the accelerated achievement of SDGs.

**Presentation of Themes**

**Methodology**

Following the opening remarks, the session started with keynote presentation that gave an overview of the role food systems play in meeting all 17 sustainable development goals. The overview highlighted the key role of The Food Systems Countdown Initiative (FSCI), which is to fill the gap on lack of rigorous mechanism to track food systems change, despite the need for food systems to transform. The overview highlighted the importance of actionable evidence to track progress, guide decision-makers, and inform transformation, and placed emphasis on complementing other monitoring and tracking of related goals at global and regional scales.

Following the overview, the FSCI was welcomed by the African experts, who felt that it has come at an opportune time. This was with the realization that the continent has its own framework to track progress towards transforming agriculture and food systems, which needs reinforcement. The greatest albatross of the CAADP initiative in particular, is the quality of data that is produced to track progress towards the Malabo targets of 2025 and challenges of aligning the Malabo Targets to the SDGs.

The consultation was conducted at a time when the African Union (AU) had convened a joint meeting of ministers in charge of agriculture, trade and finance from the AU Member States on the impacts of COVID-19 on African food systems, and the preparation and the presentation of the African common position to the UNFSS as a continent. In both instances, there was a recommitment to advance agricultural transformation of the continent through strengthening the implementation of CAADP and the Biennial Review process as the tool to assess progress on the implementation of the recommendations and the game changing solutions contained in the Africa common position to the UNFSS.

Following the plenary session, subsequent discussions took place in side events, under five sessions covering the themes of the indicator framework as follows: Session 1: Diets, Nutrition and Health; Session 2: Environment and climate Domain; Session 3: Livelihoods, poverty and equity Domain; Session 4: Governance Domain; and Session 5: Resilience and sustainability.

In each session, the discussions were motivated by a presentation on the proposed set of indicators. The discussions focused on the capacity of the proposed indicators to guide policy decisions and promote accountability mechanisms. In this sense, the focus was on the relevance, usefulness, and validity of the proposed set of indicators from a regional perspective.

The session on Diets, nutrition and Health need to monitor if:

- People have access to healthy diets? (*Food and nutrition security*)
- People consume healthy diets? (*Diet quality and adequate nutrients content*)
- Food environments support people to access and consume healthy diets? (*Access to Healthy Diets*)
- Policies contribute positively or negatively toward food availability, food access, product properties (e.g. food safety, labelling), and food messaging? (*Policies affecting food environments)*

The Session on Environment and Climate seeks to monitor:

- Food systems as a major source of environmental degradation
- Actions required achieving global environmental commitments
- Monitoring and accountability essential

The session on Livelihoods, poverty and equity aims to monitor;

- The number of people working as part of the food system, spanning rural and urban areas, high and low-income countries
- The diversity of livelihood supported by the diversity in food systems: farming, transport, processing, formal and informal retail, "gig economy", etc.
- The extent of poverty, vulnerability and exploitation across the food systems

The session on Governance seeks to monitor;

- Shared expectations for outcomes
- Relevant policy instruments to align efforts and bring about change
- Implementation capacity and resources
- Accountability for outcomes

The session on Resilience and sustainability seeks to monitor food systems based on the following issues:

- Food systems resilience being critical to food security and nutrition (e.g., fragile states = food insecurity, COVID-19).
- Food systems being critical for other functions (e.g., livelihoods, inclusion).
- Resilience being a pre-condition for sustainability.
- Sustainability having multi-dimensional interpretation
- Food system sustainability contributing to SDGs
- Normative element in the food systems transformation and resilience (transformation per se is not enough)
- Monitoring sustainability being necessary to capture food systems' holistic nature.

**Guiding questions to animate the discussions were the following:**

Do you consider the proposed indicators:

- Relevant, defined as their ability to measure something meaningful for food systems across a variety of settings, during relevant time periods?

- High quality, defined as using the best and most rigorous statistical methodologies and data available?

- Interpretable defined as having the ability to show a clear desirable direction of change, comparable across time and space, and easily communicated.

- Useful, defined as its ability to be used for policy and decision-making processes and by meeting actual information needs.

What are the data gaps that you can identify? Are these gaps structural? If not, are there some data or indicators that you know are available in your region and can be used for the purpose of this monitoring and assessment system?

What are the regional aspects we need to take into consideration now of selecting a set of indicators to monitor the state of food systems and its evolution?

**Key observations from all the sessions were as follows;**

- The Initiative was welcomed by participants who all appreciated the work done. It was noted that this Initiative was going to be an important addition to Africa's efforts t to strengthen tracking of progress towards achieving the SDGs and the Malabo Targets. It was noted that food systems play a major role in sustainable development and reversing the hunger and malnutrition statistics that continue to be an albatross weighing down on the continent.
- Having high quality data will help improve the statistical significance and reliability of estimated relationships between policies, investments, and outcomes. Hence, policymakers and investors can be more confident in using results of strategic analysis to make policies and investments that are more likely to lead to desirable outcomes.
- Capacity shortfalls to collect, analyze and report reliable statistics that will inform policy was a major observation.
- The indicators for tracking food systems transformation were many and at times confusing.
- Strengthening institutional capacity to track progress and putting in place strong M&E frameworks was key.
- The need to focus on demand driven indicators that respond to specific challenges in countries and regions.
- Gap in advocacy and wide dissemination of the indicators for national and regional buy in and adoption.
- Importance of political support to the initiative was highlighted and financing the data collection, following results based financing approaches.
- The need to analyze the interaction between the indicators was also discussed, considering the mechanisms in which these interactions progress over time – The "Law of Motion" of the indicators for food systems transformation in general.
- The need to bring in indicators outside the food systems that have an impact on food systems or influence food systems transformation.
- Advocacy for domestication of the indicators – based on international frameworks and conventions was also raised
- Analysis of the commonality of the indicators across countries and their divergence given countries and regional heterogeneities.
- Considerations on how the indicators are related to continental and regional initiatives like the African Continental Free Trade Area and Food Safety.
- There was a general concert to translate the indicators to French and other working languages in the continent where possible for ease of adoption of indicators by countries

**Discussion**

<u>Diets and nutrition</u>

*Do you consider the proposed indicators relevant, high quality, interpretable and useful?*

On this question, participants noted that the indicators proposed are relevant high quality, interpretable and useful. However, there is a need to define clearly the *"zero fruits and vegetable meaning"* for the indicator measuring the diet's quality. The indicators were many and were diverse, meaning it was important to aggregate some of them. Countries did not have the indicators, meaning that a lot of advocacy and disseminating the indicators was still needed.

The participants said that it might be difficult to interpret the indicator, as it should take into consideration the quantity, the frequency, the seasonality and safety. This will be complicated by the capacity at country level to collect, analyze and report on the data and the indicators.
Furthermore, the selection of the indicators in the category should be justified and it is important and relevant to compare the data according to gender, age and economical status of the respondent.

*What are the data gaps that you can identify? Are these gaps structural? If not, are there some data or indicators that you know are available in your region and can be used for the purpose of this monitoring and assessment system?*

The participants agreed that there is a lack of data in their countries and a lot of efforts should be done to ensure accurate and quality data to be analyzed.

Indicators on the percentage of locally produced and locally processed food components in the diet consumed, especially by urban consumers can be considered. Another indicator is the percentage of locally produced foods and commodities that are successfully integrated in urban food supply chains. Regarding the food policies, there is a need to have indicator(s) that track the enforcement of laws/policies, integrated into food safety standards and regulations for example.

*What are the regional aspects we need to take into consideration at the moment of selecting a set of indicators to monitor the state of food systems and its evolution?*

The data collection should be coordinated and harmonized building on existing initiatives. The CAADP M&E framework is seen as a key entry point.
The following points should be taken into consideration:

Growing population as the African population will double by 2050
Poor industrialization; high import of food commodities
Food safety
Cultures and food habits should be considered
Promotion of locally produced foods should be part of the strategies
Quality of water , as water play a key role in the health it is important to ensure that we have clean water

<u>Environment and climate domain</u>
*Do you consider the proposed indicators relevant, high quality, interpretable and useful?*
It was generally agreed that the data and indicators were of high quality and relevant.
However, it was felt that climate change was a complex phenomenon that requires specialized and highly technical data set and expertise to track its effect on food systems and importantly, the role of food systems in driving climate change itself.

*What are the data gaps that you can identify? Are these gaps structural? If not, are there some data or indicators that you know are available in your region and can be used for the purpose of this monitoring and assessment system?*

The climate change indicators must respond to global climate change frameworks, conversion, legislation and agreements.

Data and indicators to respond to environment and climate change must consider areas linked to climate mitigation, climate adaptation, and the linkages between climate change and the broader 2030 Agenda.
It was noted that Africa has lower carbon footprints yet the continents bears the brunt of the effects of climate change. In this regard, it was suggested that climate change and environment indicators for Africa must place strong emphasis on climate adaptation rather than mitigation, and make use of the global carbon trading schemes where possible.

The importance of coming up with regional data hubs for supporting the vetting of the indicators, data collection and coordination, reporting and mutual accountability was also raised. This will help in domestication of the indicators and identifying common indicators that will be easily comparable adaptable across countries.

It was noted that because of the complexity of the climate change subject and its wide scope and interlinkages, coming up with a set of smaller number of reliable indicators was not easy but this was a way to go, especially for countries with limited capacity to collect data.
Participants noted that data and indicators are becoming complex in line with the rapidly transforming food systems and the sporadic shocks in the system. For a complex subject like climate change, there is weak capacity to provide the system-wide coordination that is needed to provide greater access to data, identify and address data gaps or deliver data synthesis and assessment.

*What are the regional aspects we need to take into consideration at the moment of selecting a set of indicators to monitor the state of food systems and its evolution?*

Again, the importance of embedding the environment and climate change indicators to the CAADP process was raised, and linking the suggested indicators to the resilience indicators that already exist. The complexity of climate change and environmental systems and the rapidity in which they are progressing need special expertise and support in tracking these changes.

<u>Livelihoods, poverty and equity Domain</u>
*Do you consider the proposed indicators relevant, high quality, interpretable and useful?*
Participants agreed that, the indicators mentioned were very relevant and useful but still need other areas to be considered if we are indeed aiming at transforming food systems in Africa.
However, more collaborators should be included to enrich the existing information and be able to fill the missing gaps.

Participants also felt that reviewing of documents should be done to avoid duplication of ideas. To achieve sustainable agriculture and food systems, there should be more research and data metrics to fill knowledge gaps in terms of economic efficiency, environmental and social sustainability, and links to food security and nutrition for the transformation of food systems. The following indicators were considered important to have in the list:

- Post-harvest Losses
- Agric-business
- Agric value chain
- Involvement of the youth and integrating gender aspects
- Agriculture should be made attractive for all especially the youth
- Introduction of more user-technology / tools
- National Investment Programme (NIP) should be looked at again
- Identify the link between NIP and CAADP

*What are the data gaps that you can identify? Are these gaps structural? If not, are there some data or indicators that you know are available in your region and can be used for the purpose of this monitoring and assessment system?*
Data collection is always an issue; a lot of work still need to be done in that area. Lack of data makes it very difficult to move on. Without data, it will be very difficult to improve and transform the food systems in Africa and the world as a whole, and to have reliable baselines.
Other innovative indicators should be considered to achieving sustainable agriculture and food systems especially given the rapidity of the transformation in food systems.
*What are the regional aspects we need to take into consideration at the moment of selecting a set of indicators to monitor the state of food systems and its evolution?*

It was observed that there should be direct youth engagement and involvement in policymaking at local and global levels when it comes to agriculture and the need to make agriculture more attractive so many people could join especially the youth.

Governance Domain

*Do you consider the proposed indicators relevant, high quality, interpretable and useful?*

The indicators on Governance were welcomed by participants and were found to be very relevant. These indicators were observed to play a pivotal role in holding governments accountable to their commitments to transform food systems to deliver better on the SDGs. They also noted that the indicators were important in institutionalizing the national, regional and global commitments to transform food systems, grounded at country level.

*What are the data gaps that you can identify? Are these gaps structural? If not, are there some data or indicators that you know are available in your region and can be used for the purpose of this monitoring and assessment system?*

Data gaps on the governance domain were to do with lack of indicators to track public sector engagement in governance of food systems transformation, something that the initiative can consider. Balancing data from existing official structures with those from emerging or unofficial structures was again noted. Ensuring that the needs of data users are met and integrating this with country-level data support mechanisms was highlighted. Facilitating access to high quality, usable data and strengthening transparent and accountable governance structures, M&E and mutual accountability frameworks was raised.

*What are the regional aspects we need to take into consideration at the moment of selecting a set of indicators to monitor the state of food systems and its evolution?*

Indicators of governance of land rights and access to land and factors of production by marginalized groups and women was mentioned for consideration

It was again suggested that for the Initiative to gain the political mandate to hold governments accountable, it must be embedded into the CAADP accountability framework that has been sanction and adopted by Member States.

Resilience and sustainability Domain

*Do you consider the proposed indicators relevant, high quality, interpretable and useful?*

The indicators proposed are relevant, high quality, interpretable and useful. Additionally, participants stressed the fact that the indicators should be adapted to national realities as well as a difference should be made between individual, household, and community indicators. In fact each level has it own way to cope and it is crucial to capture them well for a better analysis.

*What are the data gaps that you can identify? Are these gaps structural? If not, are there some data or indicators that you know are available in your region and can be used for the purpose of this monitoring and assessment system?*

When it comes to measure resilience and sustainability, there are data gaps. This is because there is no consensus on the indicators to use. It leads to confusion i.e the Resilience Index Measurement and Analysis (RIMA) and other indicators.

*What are the regional aspects we need to take into consideration at the moment of selecting a set of indicators to monitor the state of food systems and its evolution?*

The CAADP M&E framework could be useful, and capacity building should be part of the plan. Furthermore, the selection of indicators should be demand driven and sub-national aspects should be considered.

In fact, the subnational aspect might pilot the indicators in certain countries to test their uptake and domestication.

**Overall recommendations**

- Strengthening institutional capacity to track progress and strong M&E frameworks.
- Coming up with regional and sub-regional data hubs for supporting data collection, analysis, M&E and peer review and mutual accountability mechanisms
- Mainstreaming the indicators into the CAADP framework was seen as a low hanging fruit for their adaptability and the political buy in.
- Strengthening advocacy for domestication of the indicators – based on international frameworks and conventions.
- The indicators should also respond to tools that monitor food and nutrition security. In particular, the issue of Food Based Dietary Guidelines was given as an example.
- There was a general concert to translate the indicators to French and other working languages in the continent where possible for ease of adoption of indicators by countries
- Gender and youth considerations in the data was recommended.

**Agenda of meeting**

*GMT or Accra Time*

| Time | Subject | Speakers |
|---|---|---|
| 9:00 – 9:30 | Plenary: Welcoming remarks and introduction to the Initiative | *FAO* |
| 9:30 -10:30 | Block 1:<br>Diet, Nutrition, and Health;<br>Environment and climate Domain;<br>Livelihoods, poverty and equity Domain | *FAO*<br>*Participants*<br>*Moderator* |
| 10:30 - 10:45 | Break | |
| 10:45 -11:45 | Block 2:<br>Governance Domain;<br>Resilience and sustainability Domain | *FAO*<br>*Participants*<br>*Moderator* |
| 11:45-12:00 | Break | |
| 12:00-13:00 | Wrap up and Conclusions | *Moderator* |

**Annex 2:** Invited Participants List

| Region | Country | Nom & Prénoms | Fonction | Organisation |
|---|---|---|---|---|
| ECCAS | Cameroon | **Ondoa Manga Tobie** | Point Focal PDDAA | Ministère de l'Agriculture et du Développement Rural |
| | | **Ntouda Betsogo julien** | Chef de Service du Traitement des Données | Ministère de l'Agriculture et du Développement Rural |
| | | **Amos Bassia Bassia** | | Ministère de l'Agriculture et du Développement Rural |
| | Central African Republic | **Francis Doui** | Point Focal PDDAA | Ministère de l'Agriculture et du Développement Rural |
| | | **Sylvie Solange NDODET BETIBANGUI** | Point Focal suivi de la Déclaration de Malabo | Ministère de l'Agriculture et du Développement Rural |
| | Chad | **AHOURDET DJAPANIA** | Point Focal PDDAA | Ministère de Production de l'Irrigation et des Equipements Agricoles |
| | | **Djimadoum Djeramian** | RB Point focal | Ministère de Production de l'Irrigation et des Equipements Agricoles |
| | | **EDMOND SOULEINGAR** | Statiscien agricole | CELLULE PERMANENTE |
| | | **Abdel Kerim Mahamat Issakha** | Expert Revue Biennale | Ministère de l'Agriculture |
| | | **Abakar Mahamat** | Expert Revue Biennale | Ministère de l'Agriculture |
| | Congo Republic | **Mobengue Cyprien** | Point Focal PDDAA | Ministère de l'Agriculture, de l'Evage et de la Pêche |
| | | **Pierre MPANDOU** | Planificateur en développement | Ministère de l'Agriculture, de l'Evage et de la Pêche |
| | | **HOMBESSA KIAKOUNDA Oriali Gaël Godlov** | Statisticien informatique et responsable de saisie des données RB | Ministère de l'Agriculture, de l'Evage et de la Pêche |
| | DR Congo | **Paul M'FINDA** | Chef de Division chargé des TIC et Point Focal RB | Ministère de l'Agriculture |
| | | **Alain Etango** | Chargé de saisie des données RB, Expert à la Direction d'Etudes et Planification et en charge de Programmes et Suivi | Ministère de l'Agriculture |
| | | **Jean LUFIMPU LUKOMBO** | Point Focal RB | Ministère de l'Agriculture |
| | | **Albine BUNGA** | Gestion des données | Ministère de l'Agriculture |
| | Equatorial Guinea | **José - Juan NDONG TOM** | Point Focal PDDAA | Ministerio de Agricultura, Ganadería y Alimentation (MAGA) |
| | | **Aquiles-Serafin CHONI BOLOPO** | Statiscien agricole | Ministerio de Agricultura, Ganadería y Alimentation (MAGA) |
| | | **Mba NGUI ANGUESOMO** | Responsable Adjoint de saisie des données RB | Ministerio de Agricultura, Ganadería y Alimentation (MAGA) |
| | | **Madame Estefania Isabel NGUEMA ANDEME** | Directeur Générale de vulgarisation, formation coopérative et de Mécanisation | Ministère de l'Agriculture, de l'Evage des Forêts et de l'Environnement |
| | Gabon | **Nguema Ndong Patrick** | Spécialiste en S&E et Revue Biennale | MAE et l'Alimentation chargé du Programme GRAINE |
| | | **Mesmin Ndong Biyo'o** | Point Focal PDDAA | MAE et l'Alimentation chargé du Programme GRAINE |
| | São Tomé and Príncipe | **Raimundo Gonsalves JB** | Technicien au Ministère | Ministère de l'Agriculture |
| | | **Idalécio Guadalupe Pereira Neto** | Technicien au Ministère | Ministère de l'Agriculture |
| **UMA** | Algeria | BENYAHIA-ZEGHLI | Point Focal PDDAA/Sous-directrice de la Coopération | |
| | | selsabil chouihat | ingénieur agronome | ministère de l'agriculture |
| | | Sabrina Ichou | Chargée d'études et de synthèse | Ministère de l'agriculture et du développement rural |
| | | Amel YESREF | | MADR |
| | Egypt | Ahmed Abdeen | Agriculture engineering/Animal production | Ministry of agreculture |
| | | Entessar El sayed Emam Desouky | Director General Department General for Coordinating the projects and Agricultural Affairs of the COMESA Countries | Ministry of Agriculture and Land Reclamation |
| | | hamdi swilam | Engineering of agriculture | Ministry of agriculture |
| | | DR. HOSSAM SALLAM | Agricultural engineer | Ministry of agriculture and Land reclamation |
| | Libya | Mohamed Ashour Adham | Point Focal PDDAA | |
| | Mauritania | Abdellahi zeyad | Point Focal PDDAA | |
| | Morocco | Zakaria Aharrouille | Point Focal PDDAA | |
| | | MUSTAPHA ABDERRAFIK | CAADP Biennial Review | |
| | | YAHYA FARESS | CAADP Biennial Review | |
| | Tunisia | Souhir Belaid | Point Focal PDDAA | |
| | | kamel zaidi | | ministère agriculture |
| | | Amina HICHRI | | Ministère de l'Agriculture, des Ressources Hydrauliques et de la Peche |
| **EAC** | Burundi | Mr Elias Ngendabanyikwa | CAADP Focal Person | Ministere de l'Agriculture, Elevage |
| | | Zénon Nsananikiye | CAADP Focal Person | Ministere de l'Agriculture, Elevage |
| | Rwanda | Bertrand Dushimayezu | CAADP Focal Person / Statistician | Ministry of Agriculture and Animal Resources |
| | | Joas Tugizimana | CAADP Focal Person / M&E | Ministry of Agriculture and Animal Resources |
| | Kenya | Josephine Love | CAADP Focal Person | Ministry of Agriculture |
| | | Isaiah Okeyo | CAADP Focal Person | Ministry of Agriculture |
| | Uganda | Stephen Kayongo | Commisioner/CAADP Focal Point | Ministry of Agriculture, Animal Industry and Fisheries |
| | | Fred Mayanja | Commisioner/CAADP Focal Point | Ministry of Agriculture and Animal Industry and Fisheries |
| | Tanzania | Daines P. Mtei | CAADP Focal Person | Ministry of Agriculture |
| | | DAINES MTEI | CAADP F.P | ministry of agriculture |
| | | Irene S. Lucas | M&E Officer | Ministry of Agriculture |
| | South Sudan | Loro George Leju Lugő | CAADP Focal Point | Ministry of Agriculture, Fisheries and Livestock |
| | | Stephen Wani Augustino | Statistician | Ministry of Agriculture and Forests |
| **IGAD, COMESA** | Djibouti | Ibrahim Djama Ismael | | Ministry of Agriculture, water , Livestock and Fisheries |
| | | Moussa Issak farah | Statistician | Ministre of Agriculture |
| | Ethiopia | Alemtsehay Hailegiorgis | food and nutrition coordination office head | Ministry of Agriculture |
| | | Seble Tilahun | Monitoring and Evaluation Expert | Ministry of Agriculture |
| | Somalia | Mohamud Artan | Director of Planning and Policy | Ministry of Agriculture and Irrigation |
| | | Mohamed Gure | Head of M & E | MOAI |
| | | Suleikha Ahmed Ali | Head of statistics | Ministry of agriculture and irrigation |
| | Sudan | Hamza Abdalla Siror Osman | Information and Analysis Unit Coordinator Food Technical Secretariat (FSTS) | Ministry of Agriculture & Forests |
| | | Mr. Babiker Hassan Adam Mohammed | CAADP M&E Officer | Ministry of Agriculture and Forests |
| | | Zeina Abbas | M&E officer | Ministry of Agriculture |
| | Eritrea | Bahlbi Goita | BR | Ministry of Agriculture |
| | | Bereket Tsehaye | BR | Ministry of Agriculture |
| **SADC** | Angola | Beto pinto | BR | Ministry of Agriculture |
| | | Gabriel Domingos | BR | minagrip |
| | Comoros | Issouf Miradji Ambadi | BR | Ministère de l'Agriculture, de la Pêche et de l'Environnement |
| | | Mahdy Youssouf | BR | Ministère de l'Agriculture, de la Pêche et de l'Environnement |
| | Botswana | GORATA BOIKANYO | AGRICULTURAL ECONOMIST | MINISTRY OF AGRICULTURE |
| | | Sandra Modise | Statistician | Statistics Botswana |
| | | Bueno Shanto Mokhutshwane | District Animal Production Officer | Ministry of Agricultural Development and Food Security; Department of Animal Production |
| | | Sandra Modise | Statistician | Statistics Botswana |
| | Eswatini | Howard mbuyisa | snr economist/ caadp focal point | ministry of agriculture |
| | | Kunene Sebenele | Planning officer | Ministry of agriculture |
| | Lesotho | Maoala Khesa | Senior Economic Planner and CAADP Focal Point | Ministry of Agriculture and Food Security |
| | | Leeto Semethe | Economic Planner | Ministry of Agriculture, Marketing and Food Security |
| | Malawi | Daudi Chinong'one | Principal Economist | Ministry of Agriculture |
| | | Charles Chinkhuntha | Chief Economist | Ministry of Agriculture |
| | | Readwell Musopole | | Ministry of Agriculture |
| | | Emmanuel Mwanaleza | Statistician | Ministry of Agriculture, Irrigation and Water Development |
| | Madagascar | Mr. RAMBOLARIMANANA Tahiana | Director of Planning and Monitoring and Evaluation | Ministry of Agriculture, Livestock and Fisheries (MAEP) |
| | | Mrs SOLONITOMPOARINONY Jocelyne | Head of Agricultural Statistics | Ministry of Agriculture, Livestock and Fisheries (MAEP) |
| | | Mr. RANDRIANAIVOMANANA Andritiana Luc | Directeur de la Planification et du Suivi-Evaluation | |
| | Mauritius | Krishna Chikhuri | Sappdo | Ministry of Agro-Industry & Food Security/Trade |
| | | Varuna Dreepaul | | Moafs |
| | Mozambique | Maria Manussa | head of Department do Agriculture Policy | Ministry of Agriculture and Rural Development |
| | | Duque Wilson | Hod | Ministry of Agriculture and Rural Development : Mozambique |
| | Namibia | Petrus Uushona | Policy Analyst, Policy Planning and Monitoring and Evaluation | Ministry of Agriculture, |
| | | Stephanus SANDA | Chief Statistician | Ministry of Agriculture, Water and Land Reform |
| | | Eric Kana Miaso | Acting Chief Development Planner-Monitoring & Evalutaion | Ministry of Agriculture, Water and Land Reform |
| | South Africa | Bongeka Mdleleni | Deputy Director | Department of Agriculture, Land Reform and Rural Development |
| | | Kelefetswe Seleka | Monitoring and Evaluation Specialist | Department of Agriculture, Land Reform and Rural Development |
| | | Heidi Phahlane | Agric Economist | DALRRD |
| | | Nico De Kock | Senior agric economist | Dept of Agriculture |
| | Zambia | Christopher Mbewe | caadp focal/agricultural economist | moa |
| | | Mweemba Chijoka | Senior Statistican | Ministry of Fisheries and Livestock |
| | | Mary Chilala | Principal Economist/Planning, Monitoring and Evaluation | Ministry of Agriculture |
| | | Yotam Nyirenda | Economists | Ministry of Agriculture |
| | Zimbabwe | Fredy Chinyavanhu | Deputy Director-Food Control | Ministry of Health and Child Care |
| | | Grace Nicholas | | |
| | Seychelles | Sylvie Larue | BR | Ministry of Agriculture |
| **ECOWAS** | Benin | Aguemon Dossa | Représentant du Point Focal PDDAA | Directeur de la Prospective et de la Programmation, Ministere de l/Agricultire, Elevage et peche |
| | | ACAKPO Charles | Représentant du Point Focal S&E du PDDAA | Chef Division Évaluation des Performances du Secteur et Capitalisation (CSE/DPP/MAEP), Ministere de l/Agricultire, Elevage et peche |
| | | BIAOU Samon Alexandre | Statisticien | Directeur de la Statistique Agricole, Ministere de l/Agricultire, Elevage et peche |
| | | Odilon Steeven ZOHOUN | | SAKSS Bénin |
| | Burkina Faso | SANA Souleimane | Spécialiste Suivi & Evaluation (S&E) PDDAA | Ministère de l’Agriculture et des Amengements Hydro-agricoles |
| | | SOME Gustave | Directeur Suivi & Evaluation (S&E) PDDAA | Ministère de l’Elevage et des Ressources Animales |
| | | NIKIEMA Adama | Statisticien/DGESS | Ministère de l’Agriculture et des Amengements Hydro-agricoles |
| | Cabo Verde | Clarimundo Pina Goncaives | Ponto Focal PNIA/PDDAA | Ministère de l'Agriculture |
| | | TEIXEIRA Arilde Emilia | Responsável Seguimento & Avaliação | Ministère de l'Agriculture |
| | | Emanuela.Santos | | Instituto Nacional de estatística |
| | | Emanuela Santos | Responsável pela componente Estatística (contas nacionais, dados macro - económicos, dados sectoriais etc.) | Ministère de l'Agriculture |
| | Côte d'Ivoire | Kouassi Adjoua Jeannine | Point Focal PDDAA | Ministère de l'Agriculture et du Développement Rural (MINADER) |
| | | Mme KOMENAN née Kouassi Mathilde Akissi | Spécialiste Suivi & Evaluation (S&E) PDDAA | MINADER |
| | | Gbahi Djoua Luc | Statisticien | MINADER |
| | Gambia | Momodou Sowe | CAADP Focal Person | Ministry of food and agriculture |
| | | Amet Sallah | CAADP Monitoring & Evaluation (M&E) Officer | Ministry of food and agriculture |
| | | Sabina K Mendy | Statistician | Ministry of food and agriculture |
| | Ghana | Mr. Faisal Munkaila | Senior Agricultural Economist (Policy, Planning Divsion) (Representing the CAADP Focal point ) | Ministry of Food and Agriculture |
| | | Mr. Patrick Ofori | Deputy Director (Monitoring &Evaluation Division) (Representing M&E) | Ministry of Food and Agriculture |
| | | Mr. Foster Azasi | Statistician (Representing the Statistics department) | Ministry of Food and Agriculture |
| | Guinea | Jean Luc Faber | Représentant le Point Focal PDDAA | Ministère de l'Agriculture |
| | | Atigou Balde | Spécialiste Suivi & Evaluation (S&E) PDDAA | Ministère de l'Agriculture |
| | | Mohamed Moustapha Camara | Chargé du rapport, en rapport avec les statistiques | Ministère de l'Agriculture |
| | Guinea-Bissau | Rui Jorge Alves da Fonseca | Point Focal PDDAA | Ministère de l'Agriculture et du Développement Rural |
| | | Ildo Afonso LOPES | Directeur des Statistiques Agricoles | Ministère de l'Agriculture et du Développement Rural |
| | | Suanda INFONDA | Technicien de la Direction de Suivi et Evaluation des Politiques et Programmes | Ministère de l'Agriculture et du Développement Rural |
| | Liberia | D. Musu B. Flomo Bendah | Director for Planning & Policy | Ministry of Agriculture |
| | | Jlopleh Dennis Wiagbe. Jr. | Director, Monitoring & Evaluation | Ministry of Agriculture |
| | | Peter W. Kun | Principal Data Analyst (Agriculture) | |
| | Mali | Mr. Abdoulaye Baba Arby | Point Focal PDDAA | Ministère de l'Agriculture |
| | | Mme Korotoumou TOURE | Spécialiste Suivi & Evaluation (S&E) PDDAA | Ministère de l'Agriculture |
| | | Mr. Aly KONE | Statisticien (Expert en statistiques agricoles) | Ministère de l'Agriculture |
| | Niger | Adamou Danguioua | Point Focal PDDAA | Ministre de l'Agriculture et de l'élevage |
| | | Arimi Mamadou Elh. Ousmane | Spécialiste Suivi & Evaluation (S&E) PDDAA | Ministre de l'Agriculture et de l'élevage |
| | | Neino Gondah | Chef de Division Gestion Base de Données, Informatiques et Archives | Ministre de l'Agriculture et de l'élevage |
| | Nigeria | Ibrahim Mohammed | Assistant Director (CAADP/NAIP Focal point) | Federal Ministry of Agriculture |
| | | Dare Olorunmola | Statistician | Fmard |
| | | muhammed adamu mele | ASST DIR (M & E) | PROJECT COORDINATING UNIT (PCU)OF THE FEDERAL MINISTRY OF AGRICULTURE AND RURAL DEVELO[PMENT |
| | Senegal | M. Ousmane SYLLA | Point Focal PDDAA | Ministère de l'Agriculture et de l'Equipement Rural |
| | | Mme Ndeye Dibor NGOM | Spécialiste Suivi & Evaluation (S&E) PDDAA | Ministère de l'Agriculture et de l'Equipement Rural |
| | | M. Moustapha NIANG | Statisticien | Ministère de l'Agriculture et de l'Equipement Rural |
| | Sierra Leone | Bai Bai Sesay | CAADP Focal Person | Ministry of Agriculture and Forestry |
| | | Umaru Sankoh | CAADP Monitoring & Evaluation (M&E) Officer | Ministry of Agriculture and Forestry |
| | | Mohamed Ajuba Sheriff (Dr) | Director of planning | Ministry of Agriculture and Forestry |
| | Togo | Alassani Ennardja | Point Focal PDDAA Directeur des Politiques, de la Planification et du suivi-évaluation (DPPSE) | Ministère de l'Agriculture, de la Production animale et Halieutique |
| | | TAWO Kodjovi Orénykuwua | Ingénieur Agronome à la Direction des Politiques, de la Planification et du suivi-évaluation (DPPSE), | Ministère de l’Agriculture, de la Production animale et Halieutique |
| | | AGOUDA Affèyitom | Statisticien à la Direction des Statistiques agricole, de l’informatique et de la documentation (DSID) | Ministère de l’Agriculture, de la Production animale et Halieutique |

**FAO RAP Regional Expert Consultation of the Food System Countdown Initiative´s Indicator Framework**
**Virtual Meeting, May 24**

This report was produced by the FAO Regional Office for Asia and the Pacific to summarize the results of the FAO RAP regional expert consultation held on May 24, 2022.

## 1. Introduction

Food systems play a role in meeting all 17 sustainable development goals (SDGs). With less than a decade to achieve the SDGs, the global community faces a critical juncture to transform food systems to be healthier, safer, more sustainable, more efficient, and more equitable. Lately, the UN Food Systems Summit has focused the global attention on food systems and set the stage of food system transformation. Country and independent dialogues catalyzed the development of shared visions for food systems that apply to different contexts and geographies.

It is widely recognized that to deliberately change food systems and, at the same time, cover all aspects of food systems and their interactions, a clear, rigorous and comprehensive set of metrics and indicators are required to guide decision-makers while, at the same time, to hold them accountable. However, no rigorous commonly agreed mechanism yet exists to track the state of food systems, their change and performance over time.

In much of 2021, most countries, including from Asia and the Pacific held multi-stakeholder national dialogues and prepared and submitted national papers expounding on their country's vision and plans to implement agri-food transformation pathways in line with national priorities, needs and capacity. In fact, the commitments reached at the UN Food Systems Summit, and the realization of each nation's food systems vision reflected in the national pathways need metrics to guide decisions and track progress. At the same time, food system actors and stakeholders (e.g., civil society, governments, and international organizations) require trustworthy, science-based metrics and assessment to ensure measurable progress and accountability.

Within this context and the articulated need to fill this gap, the Food Systems Countdown Initiative ("the Initiative") was formed in 2021 as a comprehensive, independent, inclusive, science-based mechanism to provide actionable evidence to track progress, guide decision-makers, and inform transformation. At the same time, it intends to complement other monitoring mechanisms and the tracking of related goals at global and regional scales (i.e., SDG agenda, CAADP). The main goal of the Initiative is to provide an independent tracking and assessment system based on a high-quality, curated, parsimonious set of indicators that cover all important aspects of food systems and measure food system performance. The Initiative expects to deliver an annual assessment of the state of global food systems and their transformation, published in a peer reviewed scientific paper. It is also envisioned that policy briefs will be delivered in parallel for a broader audience and to facilitate transformative action.

To implement the Initiative an unparalleled partnership and collaboration has been put together, led by FAO, GAIN and John Hopkins University and with the participation of more than 50 scientists from nearly 30 academic institutions, non-governmental organizations, and UN agencies from nearly all continents. The Initiative/FSCI has designed an architecture for such a system from a multidisciplinary point of view and is moving towards implementation.

The first milestone of the Initiative was the publication of the initial proposed architecture of the system and the description of an inclusive process to move from the concept to its execution[7]. Parallel to this, an initial framework inclusive of indicative set of indicators was produced and published in a paper to serve as a starting point. As a second step, the Initiative will aim to an agreed set of indicators in each of the five thematic areas identifies and to deliver a first assessment of the state of global food systems that will serve as a baseline for monitoring progress and performance.

**2. Objective of the Regional Expert Consultation**

The Initiative is committed to an inclusive, consultative, and transparent process that will allow for validation and peer review of the set of indicators that will be used for the assessments. As part of that process, a regional consultation to engage with national and regional stakeholders and receive their feedback and input on the FSCI was held in Asia and the Pacific region through a virtual meeting on May 24.

The objective of the regional consultation on FSCI was to receive inputs from policymakers and policy-adjacent users of data, on the **relevance, usefulness, and the validity** of the proposed set of indicators from a regional perspective. The consultation covered the proposed indicators in each of the five thematic areas that will be used for the first assessment of the state of global food systems and later for tracking progress and assessing performance.

The regional consultation for Asia and the Pacific brought together representatives from governments in Asia and the Pacific, as well as individual experts in food systems from development partners and FAO networks in the countries. Participants included senior staff engaged in policy development from the relevant Planning Units of the Ministry, as well as senior experts from across the five thematic areas. It is also extended to additional relevant staff from other Ministries working on the thematic areas of the indicator framework. Also invited were the individuals and institutions most directly involved in the national dialogue meetings in 2021 as part of the UN Food Systems Summit.

The meeting started with a short plenary introduction with welcoming remarks by Mr Takayuki Hagiwara, Regional Programme Leader, FAORAP followed by a short overview on FSCI by Jose Rosero Moncayo, Director, Statistics Division, FAO (see full agenda of the meeting in annex).

This was immediately followed by three parallel breakout sessions of 90 minutes each focusing on the three domains:
1. Diets, nutrition and health;
2. Environment and climate change; and
3. Livelihoods, poverty and social inclusion).

The participants are then reconvened in a plenary session for another 90 minutes to discuss the two cross cutting domains
Governance
Resilience and sustainability;

Followed by a wrap up and conclusion session.

The thematic sessions focused on the capacity of the indicators proposed to guide policy decisions and promote accountability mechanisms. Discussion also focused on the relevance, usefulness, and validity of

[7] Fanzo et al. Viewpoint: Rigorous monitoring is necessary to guide food system transformation in the countdown to the 2030 global goals. Food Policy 2021; 104.

the proposed set of indicators from a regional perspective. Each thematic discussion was guided by the following set of framing questions:

What are the regional aspects we need to take into consideration in selecting and prioritizing a set of indicators to monitor the state of food systems and its evolution?.

Evaluate the proposed indicators in terms of:
- Relevancy; defined as their ability to measure something meaningful for food systems across a variety of settings, during relevant time periods?
- High quality, defined as using the best and most rigorous statistical methodologies and data available?
- Interpretable defined as having the ability to show a clear desirable direction of change, comparable across time and space, and easily communicated.
- Useful, defined as its ability to be used for policy and decision-making processes and by meeting actual information needs

What are the data gaps that you can identify?. Are these gaps structural? If not, are there some data or indicators that you know are available in your region and can be used for the purpose of this monitoring and assessment system?

## 3. Key session outcomes

### 3.1 DOMAIN 1: Diet, Nutrition, and Health

**What are the regional aspects that should be taken into consideration when selecting the set of indicators to monitor the state of food systems and its evolution?**

The discussion did not emphasized specific regional aspects that should be considered, except for aspects that alluded to non-communicable diseases (NCD) indicators for the Pacific and food safety from Asia.

On the questions of proposed indicators relevancy, quality, ease of interpretation or usefulness, the group discussed yielded the following insights:

There were questions about these indicators including a question about the **decision scenarios** regarding the indicators at each level (global, national and even subnational) and about drivers of consumption that influence diets (e.g. fuel and water insecurity, prices, etc.).

Another concerned raised was the link between this initiative and **existing relevant frameworks** (complement/duplicate efforts).

While the set of indicators were acknowledged to be relevant, interpretable, etc., there was also a concern about government capacity and resources including budget availability for governments to be able to implement them.

Another question raised was the link between FSCI indicators and **SDGs** and whether the indicators can be used to benchmark the performance of countries in terms of Food Systems achievements.

Another issue raised is about the **usefulness/application of these indicators in emergency and humanitarian situations**. One issue raised about the type of indicators that could be useful to reflect or measure cross-countries collaboration and coaching to improve nutrition, food policies, etc. FSCI experts replied that the FSCI indicators retained left out issues like education or issues related to emergency and humanitarian situations.

One inquiry asked whether this set of indicators would be **built into existing surveys** like DHS or a new food systems survey.

Also pointed out was the need to think about **consumption indicators** that may not change over time, unless something drastic happens like disasters, or changes in food policies that would drastically affect the food system pathway. These indicators are important especially for children it was pointed out. While participants acknowledged the importance of these indicators, they raised the question of how to make these indicators as priority indicators at the country level.

**What are the data gaps that you can identify? Are these gaps structural? Which data or indicators listed are available in your region and can be used for the purpose of this monitoring and assessment system?**

<u>The discussion raised a number of points and issues:</u>

There were inquiries about the **monitoring of nutrition status**, including wasting of children, obesity and so on. These however were expletively excluded from the set of FSCI indicator, as they are already well established.

Also, participants asked about indicators to access adequacy of energy intake, macronutrients and selected micronutrients intake. In addition, it was emphasized the need to consider some **policy measure indicators** that can act as a driver in the transformation of food systems and influence health outcomes such as availability of policies/ legislation on unhealthy food marketing and taxation, among others.
A government representative inquired about the inclusion of **cognitive development in young children 0 - 16 years** that relates to diet, nutrition and health.

Another question was related to whether it would be possible to include **indicators to map out social, cultural norms and personal factors that influence diets and food environments**.
A government official inquired about how the proposed indicators would take into consideration the challenges brought about by the **pandemic**. The importance of the topic was acknowledged by FSCI, but the difficulties were noted.

Another intervention the need to think about some **indicators on consumption** that may not change over time unless something drastic happens like disasters, or changes in food policies that will drastically affect the food system pathway. These indicators are important especially those for children, and can be used as markers on the quality of food intake of children.

A participant underlined the possibility of taking into account **production, processing and logistics (post-harvest) factors** critical to ensuring food security and nutrition. Factors in the environment, including the involvement of NGOs and other facilitators were also pointed out. The FSCI team acknowledged the pertinence of these issues, particularly infrastructure [along the agrifood value chain].
An attendee raised the issue of food safety and asked whether there is an indicator that measures import and export of prohibited chemical fertilizers. This tends to happen in countries with weak policy enforcement environment.

A Government official recommended including a **specific indicator on food safety** as a crosscutting indicator. She also echoed the interventions of some participants on monitoring education in food systems including on safe food, better choice of food, etc. The difficulty with such indicator is capturing the data as these indicators are multiclausal and difficult to measure.

Another proposal from a Ministry official was to consider using the inclusion of an **indicator regarding to the percentage of dietary energy intake from ultra-processed food** using the NOVA classification, in light of the abundance of unhealthy ultra-processed food in the market.

A government official recommended including **NCD indicators**, noting that they are very common in the Pacific. Another recommendation was to consider the fact that different countries have different capacities to measure indicators and collect the data.

A participant recommended including under the food environment set of indicators, one related to the **volatility of food prices/food price regulation**, given their significant impact on food security and nutrition. She also mentioned the existence of data gaps in the Philippines regarding food losses and waste (FLW). In a later intervention, the same participant noted the importance of disaggregating the data on animal food consumption in terms of fish, seafood and others. FSCI team acknowledged the importance of taking into account food price regulations, understanding the trade-offs (farmers vs consumers), and taking account of their typical short notice implementation on food systems. Also FLW was acknowledged to be a major oversight in current FSCI indicators. However, data for the proposed indicators may not be available for all the countries.

**Additional questions raised by the session participants include:**

- The proposed indicators are mostly related to assess the practices related, what about the knowledge related indicators?
- Can we include cognitive development in young children 0 - 16 years that relates to diet, nutrition and health? Not the only reduction of malnutrition, stunting, and wasting but importantly on the total human development including the education and academic excellence in younger population?
- How will these indicators take into consideration the challenges brought about by the pandemic?
- How the FSCI indicators will be in line with SDGs (as some indicators also in SDGs)? Will the FSCI indicators be used to compare the country achievement on Food Systems?
- Would it be possible to include indicators to map out social, cultural norms and personal factors that influence diets and food environments.
- How about indicators to access adequacy of energy intake, macronutrients and selected micronutrients intake. Considering abundant of unhealthy ultra-processed food in the market, is recommended to access the % of dietary energy intake from ultra-processed food using the NOVA classification.
- Considering the effects of the pandemic on food supply and access (logistics particularly) it will really be helpful to know the extent and what can be done (decision scenarios) if this global situation persists
- Need to take into account countries different capacities to measure indicators and collect the data.
- Other possible indicator to access food environment - Percentage of monthly income spent on food.
- What kind of indicators can be useful to reflect or measure cross-countries collaboration and coaching to improve nutrition, food policies, and system?
- Monitoring education on food system including on safe food, better choice of food, etc.
- Will the FSCI made up of "parsimonious" set of indicators would be built into existing surveys like DHS? or a new food systems survey?
- Consider policies that measure indicators which can act as a driver in the transformation of food systems and influence health outcomes such as availability of policies/ legislation on unhealthy food marketing, taxation and etc.

### 3.2 Domain 2: Environment and Climate

This session covered a number of topics related to pollution, land and soil, water, biodiversity, climate and issues of implementation of indicators

**Pollution**

1. Pesticides – HHP
2. Risk-based approach and indicators – not just amount
3. Track hazardous pesticide use in terms of availability of such pesticides per unit. This unit can be crop/soil type/prevalent pests. Standard indicators:

a. Chronic Toxicity needs to be included

Acidification of land

The new indicator should be split into three different segments: X, Y, Z

Collapse of wild bees, and lady bird populations, would be good indicators of hazardous pesticides

Link with health - deaths from pesticide poisoning

4. Pollution and waste (no indicators defined for this, only pesticides) but it needs attention, otherwise it is not really a systemic indicators.
5. Crop residue management
6. Animal health - use of antibiotics is missing.

**Land – Soil**

Respondents indicated the importance to track not just crop land but also pasture land under different soil types

Overgrazing - temporary migration due to unavailability of grazing land for cattle - negatively affects the local ecosystem. It would be helpful to have an indicator that captures this interlinkage, perhaps in the cross cutting session

Climate induced migration – Look at WG3 and Cross cutting domain

Land productivity: Percentage of production or ag land under organic/sustainable cultivation (another difficult one but essential)

Soil losses and erosion - impact on land productivity (GLADA last updated in 2011 and may be out of date - what measures are regularly updated)

**Water**

Transboundary water dependence (there is some data on this, but may be difficult to separate the ag use from other)

Water use efficiency and water balance: extremes climatic events also include floods and sea-level rise that could impact on soil fertility and salinity wash-off surface soils.

How can we better measure irrigation performance for land productivity/food production per hectare

**Biodiversity**

Are the biodiversity intactness and fisheries health index indicators sufficient?

Consider ways to capture aquatic species loss

Monitoring of ecosystem diversity to better understand impacts on human health (Zoonoses)

**Climate**

GHG intensity indicators (if looking at intensity against Ag value added - needs careful consideration as when food prices go up dramatically, as they are, then intensities will go down)

"% change in soil organic carbon", is it anticipated to have synergies with the 4per1000 initiative?

Climate extremes and adaptation indicators – data and timescale for monitoring as climate extremes come up frequently – Look at Resilience domain

**Overall - Implementation**
It would be helpful to show clearer links between suggested indicators and relevant SDG target indicators and related commitments such under the UNFCCC Paris Agreement, UNCBD and others
Experts will share further ideas on possible indicators and data
Regional collaboration
Standardizing agreed measures of food system health to encourage learning and cooperation (but not government responses - no-one-size fits all approach)

**Additional comments from participants in the environment and climate session:**

- What would be the working modalities for this initiative be implemented at regional cooperation framework?
- A regionally useful set of indicators should look at 1. crop residue management 2. transboundary water dependence (there is some data on this, but may be difficult to separate the ag use from other) 3. pollution and waste (no indicators defined for this, only pesticides) but it needs attention, otherwise it is not really a systemic indicators. 4. a link with health - deaths from pesticide poisoning could be treated here, rather than in the diet and health rubric (?). 5. GHG intensity indicators (if looking at intensity against Ag value added - needs careful consideration as when food prices go up dramatically, as they are, then intensities will go down) 6. Animal health - use of antibiotics is missing. 7. percentage of production or ag land under organic/sustainable cultivation (another difficult one but essential) (ESCAP)
- On water use, extremes climatic events also include floods and sea-level rise that could impact on soil fertility salinity wash-off surface soils. Please consider also water use efficiency and water balance
- In regards to the indicator on "% change in soil organic carbon", is it anticipated to have synergies with the 4per1000 initiative for instance?

### 3.3 Domain 3: Livelihood, poverty and equity

What are the regional aspects that should be taken into consideration when selecting the set of indicators to monitor the state of food systems and its evolution?

It was generally agreed that the regional and country level aspects and differences need to be taken into account when considering the list of proposed indicators.

Countries are at different levels in terms of food systems (FS) monitoring, with some countries already with developed monitoring frameworks with a set indicators and others working towards the development of monitoring frameworks and collection of baselines.

The difference in country context was highlighted in the **Philippines** experience whereby it was noted that some of the indicators were already ones used in the country’s already established monitoring system, but that there were some indicators that were not included in their current framework which could be looked into for possible inclusion.

**Tonga** on the other hand is keen to review the indicators as part of its formulation of a monitoring framework for its national FS transformation.

Therefore, countries may use the list depending on specific country situations, and indicators can be adapted to the country’s own specific issues.

A participant pointed out that countries may find it challenging to produce reporting on these indicators, and that we need to assess the feasibility of reporting against these indicators.

Specific Recommendations

A participant asked whether the indicators covered '**off farm aspects of food systems'**. He noted that, across Asia, although the agriculture share of GDP is decreasing, the share of food enterprises (food processing, food selling and wholesale) is rising in GDP contribution. He noted that there were no **indicators covering employment and incomes of people in food enterprises**, and given this is an aspect of the food systems, indicators covering 'off farm' aspects should also be included. Data can probably be found in food enterprise surveys.

Also noted in areas such as South Asia, **gender aspects** are indeed important when looking at different indicators, not just in land ownership but we also need to consider **inequities such as the wage gap between men and women**. In addition to gender aspects, other areas of **inequalities** need to be captured such as **deprived communities**.

A participant proposed to consider an indicator to determine **youth participation** in Food Systems transformation.

A participant from FAO noted that indicators capturing **inclusiveness** are missing and that indicators on **access to finance and technology** should also be considered given the key role these two aspects have in food systems transformation.

A Government representative emphasized that **migration** plays a critical role in this process including on livelihoods. He mentioned that indicators should take into account the increasing number of migrant workers (e.g. seasonal, those in food insecurity), looking into rights and protection angles of these workers, as well as the laws and regulations concerning migration (immigration) and these workers (e.g. health, job security, etc). For specific indicators in identifying the baseline, there could be potential of migration in the relevant areas.

**Related the questions of indicators relevancy, quality, ease of interpretation and usefulness, the following discussion points were captured:**

- It was generally agreed that the list of indicators was relevant. However, working definition of each indicator is dependent on the country context.
- A government representative confirmed that these indicators are all relevant, but that, in terms of the definition of each indicator, the indicators need further elaboration. In particular:
- Income is relevant, as his country is doing the best on the national household savings as well as on household income expenditure survey (agricultural households).
- Population is not yet in the list, but this aspect can be looked at.
- In terms of other indicators related to income and poverty, they can consider other aspects and factors of poverty.
- Employment and social protection (such as school feeding programmes) are relevant, and can depend on the existing data (e.g. on agricultural workers).
- Indicators related to rights are currently not in the list of indicators of the country, but they can also look at this aspect. Rights to land tenure are included in the list and covered already.

**In relation to questions on data gaps, and data or indicators available in the region, the discussion noted the following points:**

It was confirmed that **data availability** is an important issue to be looked at. We also need to look at the feasibility and consistency in data availability across the countries. There is no one single solution to this issue.

A Ministry official, informed the meeting that right now the issue of data and monitoring is high on the agenda in his country. They are working to develop a baseline data to measure food systems transformation including measuring baseline for livelihood and employment in the food systems. He emphasized the need to improve the **capacity for collecting data**. It was noted that, with the indicators, there is a need to collect data and therefore capacity to collect data is important in considering the indicators to be adopted.

**Additional feedback from session on livelihood, poverty and equity:**
Would it be useful to highlight crucial aspects of food systems on which countries are not collecting data? The number of the proposed FSCI indicators for livelihoods poverty and equity which is 11. It would be a helpful menu from which countries can base their specific indicators and further apply these in their own ongoing monitoring and assessment and method/s of measuring and tracking. Certainly, there are some nuancing on country to country.

**3.4 CROSS CUTTING DOMAINS: GOVERNANCE AND SUSTAINABILITY**

**Key feedback from plenary session Governance and Resilience Domains:**

- For small countries, the set of monitoring data presented are quite complex and technically challenging requiring external technical assistance to collect and monitor relevant data.
- There are also institutional challenges in how to capture these data sets for some countries
- On resilience, consider integrating entire logistics system apart from road density when it comes to calamities/pandemic.
- Globalization is also posing a danger to decreasing resilience of villages
- The list of proposed indicators has their equivalencies in countries and the criteria used to list these down may also be applied differently across countries depending on those relevant to their situations.
- This FSCI Initiative is laudable but may find even more utility depending on what each country is already measuring and what else they could consider assessing and tracking, regarding their food systems, if not yet.
- The indicators for both resilience and governance should be directly linked to the component parts of the food system - these could be organized into two categories - those internal to the system (production, processing, supply chain, consumption...) and external to the system that affects the efficacy of food system.
- In the Pacific we need specific development to identify relevant data sets that we can use to monitor food systems. Similar issue to SDG where some of the indictors in the FSCI matrix are very complicated and are very hard for us (in the Pacific) to access and understand
- There is need for further discussion on datasets that we can use to monitor food system development in Pacific countries starting from baseline data where we are now in terms of food system development
- There is a need for closer collaboration between food system coordinators and National Statistics Office. Not all indices presented are applicable for all, each country can contextualize what is relevant. Not all indicators require quantitative datasets, some success stories and qualitative info may suffice.
- In South Asia, the most politically sensitive crops are Onion, potato and tomato and are important part of food system. All the three are prone to climate change and require dedicated analysis.
- Similarly the ratio between area, production, productivity should be a good indicator of which direction the fulcrum is moving

- There has to be a critical look at elasticity vs volatility of commodity prices under the influence of extreme events such as severe precipitation etc. This will be a measure of resilience vs threat analysis from Climate induced changes.

**AGENDA AND TIMELINE**

**The event took place virtually according to the following agenda and timeline:**

**Timeline:**

| Session | Pakistan | India | Bangkok | China/Mongolia | Japan | Australia | Apia, Samoa |
|---|---|---|---|---|---|---|---|
| Opening | 7:00 – 7:30 | 7:30 – 8:00 | **9:00 – 9:30** | 10:00 – 10:30 | 11:00 – 11:30 | 12:00 – 12:30 | 15:00 – 15:30 |
| BOS 1,2,3 | 7:30 – 9:00 | 8:00 – 9:30 | **9:30 – 11:00** | 10:30 – 12:00 | 11:30 – 13:00 | 12:30 – 14:00 | 15:30 – 17:00 |
| Break | 9:00- 9:15 | 9:30- 9:45 | **11:00- 11:15** | 12:00- 12:15 | 13:00- 13:15 | 14:00- 14:15 | 17:00- 17:15 |
| Plenary | 9:15 – 10:30 | 9:45 – 11:00 | **11:15 – 12:30** | 12:15 – 13:30 | 13:15 – 14:30 | 14:15 – 15:30 | 17:15 – 18:30 |
| Wrap up | 10:30 – 11:30 | 11:00 – 12:00 | **12:30 – 13:30** | 13:30 – 14:30 | 14:30 – 15:30 | 15:30 – 16:30 | 18:30 – 19:30 |

**AGENDA:**

| **Time** | **Subject** | **Speakers** |
|---|---|---|
| ***9:00 -9:30*** | Plenary: **Welcoming remarks and introduction to the Initiative**<br><br>Welcoming remarks, **Takayuki Hagiwara**, Regional Program leader, FAO RAP<br><br>Introduction to the Food Systems Countdown Initiative, **Jose Rosero Moncayo**, Director, Statistics Division, FAO<br><br>Opening | *FAO* |
| ***9:30-11:00*** | Break-out (BO) Sessions (parallel)<br><br>**Break out Session 1: Diet, Nutrition, and Health**<br>Moderator: **Joseph Nyemah**<br>Rapporteurs: **Eva GalvezNogales, Rosemary Kafa**<br><br>Discussion opener (**Kate Schneider**): Brief introduction to candidate indicators (10 minutes)<br><br>Open floor interventions- feedback from participants | *Moderator*<br><br>*Discussion starter*<br><br>*Participants* |
| | **Break out Session 2: Environment and climate Domain**<br>Moderator: **Hang Pham, Malia Talakai**<br>Rapporteur(s): **Beau Damen, Caroline Turner**<br><br>Discussion opener (**Mario Herrero**): Brief introduction to thematic indicators (10 minutes)<br><br>Open floor interventions- feedback from participants | *Moderator*<br><br>*Discussion starter*<br><br>*Participants* |
| | | |
| | **Break out Session 3:** **Livelihoods, poverty and equity Domain**<br>Moderator: **Aziz Elbehri, Fiasili Lam**<br>Rapporteur: **Kae Mihara**<br><br>Discussion opener (**Jikun Huang**): Brief introduction to thematic indicators (10 minutes)<br><br>Open floor interventions- feedback from participants | *Moderator*<br><br>*Discussion starter*<br><br>*Participants* |
| ***11:00-11:15*** | Break | |
| ***11:15-12:45*** | Plenary: **Governance and Resilience and sustainability Domains**<br><br>Moderator: **Aziz Elbehri**<br>Discussant opener (Governance – **Danielle Resnick**) (10 Min)<br>Discussant starter (Resilience **- Yuta Masuda**) (10 min)<br><br>Open floor for participants interventions, feedback and discussion | *Moderator*<br><br>*Discussion starters*<br><br>*Participants* |
| ***12:45-13:30*** | Plenary: **Summary and wrap up**<br><br>Moderators of BO session 1, 2 and 3 to present highlights of their sessions<br><br>15 minutes overall wrap up on what was discussed, key findings and follow up<br>**Lawrence Haddad**, GAIN<br><br><br>Closing remarks from FAO RAP | *Moderators*<br><br>*FSCI members*<br><br>*RAP-RPL* |

## List of participants

| Last Name | First Name | Organization | Country |
|---|---|---|---|
| Abbas | Touseef | UNICEF | PNG |
| Abella | Ellen Ruth | National Nutrition Council | Philippines |
| Agipar | Bakyei | Mongolian University of life Science | Mongolia |
| 'Aholelei | 'Isileli | Tonga Ministry of Agriculture, Food and Forest | Tonga |
| Ali | Pungkas | Bappenas | Indonesia |
| Arya | Aziz | FAO/UN | Thailand |
| Aye | Myint Myint | FAO | Myanmar |
| Azma | Nurul | Bappenas | Indonesia |
| Banura | Hitomi | Ministry of Agriculture, Forestry and Fisheries | Japan |
| Battulga | Altanshagai | UN FAO | Mongolia |
| Bicchieri | Marianna | FAO | Thailand |
| Bolima | Dean Marc | Department of Agriculture | Philippines |
| Boonchan | Surasak | FAO | Thailand |
| Chand | Ramesh | NITI Aayog | India |
| Chatrurvedi | Siddharth | Bill & Melinda Gates Foundation | India |
| Conda | Angel Mae | Department of Agriculture | Philippines |
| Da Cruz | Gil Rangel | Ministry of Agriculture and Fisheries | Timor-Leste |
| Damen | Beau | FAO | Thailand |
| Duran | Tamara | FAO | Philippines |
| Enkhtur | Anudari | FAO Mongolia | Mongolia |
| Estoesta | Alexander II | Department of Agriculture | Philippines |
| Fatmaningrum | Dewi | FAO | Indonesia |
| Felipe | Judi Anne | DA- Planning and Monitoring Service | Philippines |
| Ferrand | Pierre | FAO | Thailand |
| GalvezNogales | Eva | FAO | Thailand |
| Guterres | Maria Odete do Ce | Ministry of Agriculture and Fisheries | Timor-Leste |
| Haas | Linda | reu | Hungary |
| Haddad | Lawrence | GAIN | Switzerland |
| Hagiwara | Takayuki | FAO | Thailand |
| Huang | Jikun | Peking University | China |
| Hunter | David | Ministry of Agriculture and Fisheriess | Samoa |
| Iuvale | Soo Jr | MAF | Samoa |
| Jack | Randon | Ministry of Natural Resources and Commerce | Marshall Islanc |
| Jayanetti | Nandalal | Ministry of Agriculture | Sri Lanka |
| Kafa | Rosemary | United Nations FAO | Thailand |
| Kafle | Samjhana | Ministry of Agriculture and Livestock | Nepal |
| KAPUR | RAKESH | IFFCO | India |
| Khaine | Aye Aye | FAO | Myanmar |
| K-hasuwan | Prae-ravee | Ministry of Agriculture and Cooperatives | Thailand |
| kim | soojung | KOICA | Mongolia |
| Kishore | Avinash | IFPRI | India |
| Krishna | Gopal | ICAR CIFE | India |
| Kumwilai | Usakorn | The Queen Sirikit Department of Sericulture | Thailand |
| Lam | Fiasili | FAO | Samoa |
| Latifi | Muhebullah | FAO | Afghanistan |
| Lee | Nara | FAO | R.O. Korea |
| Lee Hang | Keyonce | Ministry of Agriculture and Fisheries | Samoa |
| Locke | Dan | FAO | Thailand |
| Magtibay | Jasmine | FAOPH | Philippines |
| Magyaya | Ariana | Department of Agriculture | Philippines |
| Mandal | Bir | FAO | PNG |
| Manu | Viliami | Ministry of Agriculture,  Food and Forests | Tonga |
| Marothia | Dinesh | Indian Society of Agricultural Economics | India |
| Marte | Karen | Department of Agriculture | Philippines |
| Masuda | Yuta | The Nature Conservancy | United States |
| Matatumua | Taimalietane | Ministry of Agriculture and Fisheries | Samoa |
| Menon | Purnima | IFPRI | India |
| Mihara | Kae | FAO | Thailand |
| Mishra | Rajendra | Government of Nepal,  MoALD | Nepal |
| Moon | Dooree | KOICA | Mongolia |
| Mumtaz | Amer | Pakistan Agricultural Research Council | Pakistan |
| Munir | Muhammad | FAO | Myanmar |
| Munkong | Tipaporn | The Queen Sirikit Department of Sericulture | Thailand |
| Munoz | Hernan | FAO | Italy |
| Myazoe | Walter | Ministry of Natural Resources and Commerce | Marshall Islanc |
| Nallur | Krishna Kumar | ICAR | India |
| Naz | Farrah | Global Alliance for Improved Nutrition (GAIN) | Pakistan |
| Neupane | Sumanta | International Food Policy Research Institute | United States |
| newman | scott | FAO | Thailand |
| Ngarmtab | Poranee | FAO | Thailand |
| Nkoroi | Alice | UNICEF | Philippines |
| Nyemah | Joseph | FAO | United States |
| On-ubol | Woranut | Ministry of Social Development and Human Se | Thailand |
| Panda | Ranjan | Water Initiatives | India |
| Pastores | Maria Cecilia | FAO | Philippines |
| Pechcho | Pachirarat | Department of Agricultural Extension | Thailand |
| Peixoto de Lima | Camilla Hellen | FAO | Thailand |
| Pham | HangThiThanh | FAO | Thailand |
| Pioneta | Joyce Mae | Department of Agriculture | Philippines |
| Piyawatpaphad | Pongpakorn | Cooperative promotion department | Thailand |
| Plianphanich | Ms. Phatthicha | Department of Agriculture | Thailand |
| Prasula | Paulina | FAO | Italy |
| Puana | Ilagi | NAQIA | PNG |
| Rahmani | Shahriar | FAO | null |
| Rana | Jai | Alliance of BI nd CIAT | India |
| Rankine | Hitomi | ESCAP | Thailand |
| Rasool | Faiz | Global Alliance for Improved Nutrition (GAIN) | Pakistan |
| Resnick | Danielle | Brookings | United States |
| Rosero Moncay | Jose | FAO | Italy |
| Rufus | Lajikit | Ministry of Natural Resources and Commerce | Marshall Islanc |
| Sainzaya | Batbayar/Johny | the world vision in Mongolia | Mongolia |
| SAMSON | MARIVIC | NATIONAL NUTRITION COUNCIL | Philippines |
| Schneider | Kate | Johns Hopkins University | United States |
| Scott | Samuel | IFPRI | India |
| Shaheen | Aslam | Planning Commission | Pakistan |
| Sharav | Nandinzaya | KOICA | Mongolia |
| Sharma | Inoshi | Food Safety & Standards Authority of India | India |
| Shivakoti | Sabnam | Ministry of Agriculture and Livestock Developm | Nepal |
| Shrestha | Ram | Ministry of Agriculture and Livestock Developm | Nepal |
| Sinclair | Chris | Ministry of Agriculture & Fisheries | Samoa |
| Singh | Nishmeet | IFPRI | India |
| Singh | Sunil | NPC | India |
| Singh | Lt.Col.(Dr) Baljit | National Cooperative Development Corporatio | India |
| Sinha | Alok | VillageNama | India |
| SJ | Balaji | Indian Council of Agricultural Research | India |
| Smektala | Isabelle | French Embassy | Mongolia |
| Sui | Hluan | FAO Myanmar | Myanmar |
| Talakai | Malia | FAO | Samoa |
| Tienpati | Supajit | FAO | Thailand |
| Tonga | Semisi Latu | Department of Agriculture | Tuvalu |
| Tripathi | Bhupendra Nath | Indian Council of Agricultural Research | India |
| Tupou | Siutoni | MAFF | Tonga |
| Turner | Caroline | FAO | Thailand |
| Vainikolo | Leody Eleutilde | Ministry of Agriculture, Food and Forests | Tonga |
| Varma | Anupam | Indian Agricultural Research Institute | India |
| Voratira | Yowanat | FAO | Thailand |
| Wani | Suhas | Formerly with ICRISAT, Current with FAO | India |
| Win | Le Le | FAO | Myanmar |
| Wongkanit | Napaporn | Agricultural Land Reform Office | PNG |
| Yoovatana | Dr. MARGARET C. | Department of Agriculture | Thailand |

**FAO REU Regional Expert Consultation of the Food System Countdown Initiative's Indicator Framework**
**Virtual, May 25, 2022**

This report was produced by the FAO Regional Office for Europe and Central Asia to summarize the results of the FAO REU regional expert consultation held on May 25, 2022.

**Introduction:**

*Food Systems Countdown Initiative – Regional Expert Consultation*
The Initiative is committed to an inclusive, consultative, and transparent process that will allow for validation and peer review of the set of indicators that will be used for the assessments. As part of that process, two efforts are envisioned. First, a consultation with expert scientists and, second, a series of regional expert consultations across the FAO regions.

*Objective and outcomes*
The objective of the FAO REU regional expert consultation was to bring an expert point of view from policymakers and policy-adjacent users of data, on the relevance, usefulness, and validity of the proposed set of indicators from a regional perspective and, finally, on the potential data gaps and resources. The consultation covered the proposed indicators in each of the five thematic areas. These will be used for the first assessment of the state of global food systems and later for tracking progress and assessing performance.

The regional expert consultation was an opportunity to provide inputs, comments, and suggestions on the proposed indicators both in terms of appropriateness as well as also feasibility related to the data availability. The Initiative proposes a monitoring framework. This framework is not mandatory, however, the consultations will ensure that it has the capacity to be a useful tool for policy decision-making processes.

The Initiative expects to deliver an annual assessment of the state of global food systems and their transformation, published in a peer-reviewed scientific paper (October 2021). It is also envisioned that policy briefs will be delivered in parallel for a broader audience and to facilitate transformative action.

The first milestone of the Initiative will be the publication of the initially proposed architecture of the system and the description of an inclusive process to move from the concept to its execution. The architecture covers the five thematic areas of diet, nutrition and health; environment and climate; livelihoods, poverty and equity; governance and resilience and sustainability (October 2022).
As a second step, the Initiative will aim to deliver the set of indicators in each of the five thematic areas above and to deliver a first assessment of the state of global food systems that will serve as a baseline for monitoring progress and performance. A report is expected to be published in October 2023 in conjunction with the UNFSS review mechanism to assess the status and performance of the food systems transformation path. In 2029, a Global Food Systems Conference will take place. "1 year left to achieve SDGs".

*Participant's general profile*
The regional experts' consultation for Europe and Central Asia brought together representatives from governments in the region, as well as individual food system experts.

Recognizing the complexity of the food systems challenges and actions, the participants were invited through a multidisciplinary, multi-stakeholder approach, while ensuring equitable geographical and gender-balanced representation. It enabled to capitalize on the expertise of regional and country representatives, who were nominated experts on the following aspects of food systems:

- Nutrition, food security, health;
- Environment and climate;
- Resilience and sustainability;
- Rural development, social protection and poverty alleviation.

Participants were mainly senior experts from the public sector, engaged in policy development, UNFSS National Convenors, statisticians, or data specialists.

The Regional Consultation also capitalized on the expertise of several UN agencies as an important element of the 2030 Agenda. In addition to FAO colleagues, some representatives of the Regional Issue-based Coalition on Sustainable Food Systems (IBC-SFS) - namely UNECE, UNICEF, WFP and WMO – as well as World Bank also joined the conversation.

In total, 83 participants joined the consultation from 18 countries. The number of participants in different breakout sessions was the following:

Nutrition and diets: 30
Environment: 22
Livelihoods: 26
Governance: 25
Resilience: 38

Regarding the geographical representation, most of the participants came from FAO program-countries in the region, but EU Member countries were also represented.

List of the countries participating in the FAO REU Regional Expert Consultation of the Food System Countdown Initiative's Indicator Framework:

Albania
Armenia
Belarus
Bosnia and Herzegovina
Finland
Georgia
Hungary
Iceland
Montenegro
North Macedonia
Republic of Moldova
Serbia
Tajikistan
Turkey
Ukraine

*Working arrangements*

The structure of the regional expert consultation was similar to other FAO regions.

The organization of the meeting was led by Raimund Jehle, Regional Programme Leader, Deputy Regional Representative – Programme, Regional Office for Europe and Central Asia, and Mary Kenny, Food Safety and Consumer Protection Officer and the Regional Initiative Coordinator on Transforming food systems and facilitating market access and integration. The FAO REU meeting Secretariat included

Aniko Nemeth, Food Safety and Nutrition Expert, Valeria Rocca, Regional SDGs Advisor, Klaudia Krizsan, Food Safety and Nutrition Junior Technical Officer, Valentina Gasbarri, Knowledge Management and Communication Specialist, and Linda Haas, Office Assistant.

The rationale and objectives of the consultations were introduced by Hernan Munoz, Statistician, FAO Statistics Division, and Lawrence Haddad, Executive Director, GAIN, Co-chair of the Food Systems Countdown Initiative while the closing sessions of the meeting was supported by José Rosero Moncayo, Director of the Statistics Division, FAO and Jessica Fanzo, PhD, Bloomberg Distinguished Professor of Global Food & Agricultural Policy and Ethics, as Co-Chairs of the Initiative.

*Methodology*

A short video introducing all the panelists was broadcasted at the beginning of the event.

A general overview on the food systems transformation in the Europe and Central Asia region and the brief introduction of the Food Systems Countdown Initiative were provided in the plenary.

The five breakout thematic sessions covered the domains of the indicator framework, namely:

Session 1: Diets, Nutrition and Health Domain;
Session 2: Environment and climate Domain;
Session 3: Livelihoods, poverty and equity Domain;
Session 4: Governance Domain;
and Session 5: Resilience and sustainability Domain.

In each session, the discussion was stimulated by an opening presentation of the experts of the Food Systems Countdown Initiative on the proposed set of indicators. The discussion was facilitated by Moderators from the FAO Regional Office in all breakout sessions. The Moderators were all technical experts on the topic of the session. The reporting process was supported by Note Takers, Rapporteurs and the Moderators (See Table 1).

Table 1. List of Working Groups, FSCI Experts, Moderators/Facilitators and Note Takers

| | Topic | FSCI Expert | Moderator/Facilitator FAO REU | Note Takers, FAO REU |
|---|---|---|---|---|
| **WG1** | Diet, nutrition and health | Musonda Mofu, Public Health Nutritionist and Director for the National Food and Nutrition Commission of Zambia | Keigo Obara, Food Security Officer | Klaudia Krizsan, Food Safety and Nutrition, Junior Technical Officer |
| **WG2** | Environment and climate | Fabrice DeClerck, Science Director of EAT and a Senior Scientist with One CGIAR | Anna Kanshieva, Biodiversity Expert | Virag Nagypal, Climate Change and Natural Resource Management Junior Technical Officer |
| **WG3** | Livelihoods, poverty and equity | Alejandro Guarin, Researcher at the International | Anna Jenderedjian, Gender and Social Protection Specialist | Anetta Szilagyi, Rural Development Consultant and Ildiko Buglyo, Assistant to |
| | | Institute for Environment and Development (IIED) | | the Forestry and Land Tenure Officers |
| **WG4** | Governance | Namukolo Covic, Nutritionist with the Health Professions Council of South Africa | Mary Kenny, Food Safety and Consumer Protection Officer and the Regional Initiative Coordinator | Gokce Akbalik, , Food Safety Consultant |
| **WG5** | Resilience and sustainability | Preet Lidder, Technical Adviser to the Chief Scientist, FAO | Pedro Arias, Economist | Fanni Zsilinszky, Agrifood Policy Junior Technical Officer |

The breakout sessions' discussions focused on the regional specificities and aspects that should be taken into consideration when selecting the set of indicators to monitor the state of food systems and its evolution capacity of the proposed indicators to guide policy decisions and promote accountability mechanisms.

**Opening:**
*Summary of Welcome Remarks*

**The opening remarks by Raimund Jehle**, Regional Programme Leader, Deputy Regional Representative, FAO Regional Office for Europe and Central Asia, highlighted the importance of keeping the momentum around sustainable food systems created by the UN Food Systems Summit in 2021 and its follow up mechanism at the global, regional national level. He reiterated that food systems related issues have also been put on the spotlight by the COVID-19 pandemic and, more recently, by the ongoing war in Ukraine, as reaffirmed during the 33rd Session of the Regional Conference for Europe in Łódź, Poland, where 50-plus FAO Members reaffirmed its commitments to ensuring food-secure, sustainable and inclusive agrifood systems, while, most importantly, pursuing peace. This is why food systems and their interactions require a clear, rigorous, and comprehensive set of metrics and indicators, to guide decision-makers and to hold them accountable, such as the Food Systems Countdown Initiative.

He highlighted that food system assessment is not an easy task, because of the complexity of the system and that this informal consultation is part of a range of work ongoing – such as a number of global, regional and national mechanisms that have been established to support the UN Food Systems follow- up. Such mechanisms include the UNFSS Coordination Hub, the UNFSS Coalitions, the national teams working with the national convenors on the implementation of national pathways and the work of the Issue-based Coalition on Sustainable Food Systems, a regional UN mechanism supporting the UN Country Teams in the region on the work on sustainable food systems.

*"Many decisions we could make, whether we are producers, transporters, retailers…could bring us toward the achievement of the SDGs outcomes or away from SDGs. Majority of decisions will bring us away from SDGs",* said **Lawrence Haddad,** Executive Director, GAIN; co-chair of the Food Systems Countdown Initiative during his keynote speech at the opening of the FAO REU regional Consultations.

Food systems play a role in meeting all 17 sustainable development goals (SDGs). With less than a decade to achieve the SDGs, the global community faces a critical juncture to transform food systems to be healthier, safer, more sustainable, more efficient, and more equitable.

He added: *"Food systems must be on the policy agenda for the coming years. Countries are thinking about food systems now.*
*We hope they will develop national monitoring mechanisms, in line with the Countdown Initiative".*

It is widely recognized that to enhance all aspects of food systems and their interactions, a clear, rigorous, and comprehensive set of metrics and indicators are required to guide decision-makers and to hold them accountable. However, no rigorous mechanism currently exists to track the state of food systems, their change, and performance over time.

**Presentation of Themes**
The themes were introduced by members of the Initiative.
Session 1 on Diet, Nutrition and Health Domain was presented by Musonda Mofu. In this theme, several indicators related to diet, nutrition and health were discussed by looking at the various areas which measure quality of diet, including nutrient adequacy and dietary risk factors for NCDs. As well as measuring the other factors affecting diet quality, namely the accessibility to healthy diets (food security); and a focus on the food environment (availability, affordability, messaging, and food & vendor properties). Finally, examining whether policies contribute positively or negatively towards food availability, food access, and product properties (policies affecting food environments).

Session 2 on Environment and climate Domain was presented by Fabrice DeClerck. This theme focused on the relation between food systems and the environment. The indicators focused on the main environmental systems and processes which interact with food systems: land use, climate, water use, biosphere integrity, and pollution (e.g., biogeochemical flows/novel entities). The indicators cover components and processes providing essential environmental services for the environment and humanity. As well as focusing on the areas where food systems can achieve the necessary change.

Session 3 on Livelihoods, poverty and equity Domain was presented by Alejandro Guarin. The indicators for this theme monitor the transformation created by food systems for the numerous people who work as part of the food system, in rural and urban areas, and in high and low-income countries. It focused on the areas of poverty and income, employment, social protection and rights. Cross-cutting issues were also pointed out: governance and resilience and sustainability.

Session 4 on Governance was presented by Namukolo Covic. The Governance Theme looks at how governance of the food systems domain can foster alignment and coherence across different food system actors, their activities, and progress toward results. It aims to monitor the shared vision of the outcomes, the relevant policy instruments to align efforts, the implementation of resources, and accountability for the outcomes.

Session 5 on Resilience and sustainability was presented by Preet Lidder. The theme of Resilience and sustainability was presented as a transversal aspect of the transformation of agri-food systems with a multi-dimensional interpretation. It covered the domains of exposure to shocks, resilience capacities, agrobiodiversity, food security stability, food system sustainability index. Resilience and sustainability are critical for food and nutritional security. They are also critical for other functions such as being a precondition for sustainability, and critical to tackling poverty and livelihoods issues.

**Discussion**

This section synthesizes the regional perspectives and results of the thematic discussions among the participants and contributors to the informal consultation.

Key general recommendations common to all the breakout groups:
Participants welcomed the Initiative. It was noted that this Initiative was going to be an important addition to the global, regional and country-led efforts to strengthen tracking of progress towards achieving the SDGs and transforming food systems (such as the UN Food Systems follow- up. i.e. the UNFSS Coordination Hub, the UNFSS Coalitions, the national teams working with the national convenors on the implementation of national pathways and the work of the Issue-based Coalition on Sustainable Food Systems).

It was noted that tracking food systems require a clear, rigorous and comprehensive set of metrics and indicators, to guide decision-makers and to hold them accountable. Having high-quality, timely, accessible, comparable and reliable data, which monitor and evaluate the whole agrifood value chain, will help improve the statistical significance and reliability of relationships between policies, investment, and outcomes.

Suggested indicators were generally found useful and high quality, but some region-specific measures were suggested for each thematic areas.

The debate highlighted the need to solve the issue of the lack of data through a global commitment to raise the bar on the international political agenda. At the EU level, there are many frameworks and approaches that could guide the work of FSCI.

The discussion provided an opportunity to explore demand-driven indicators that respond to specific challenges in countries and regions. It was also recommended to use a food systems approach, not only focus on the primary production phase (due to the lack of data available) as the majority of indicators included so far.

Importance of political support to the initiative was highlighted and financing the data collection. Advocacy for domestication and country-led implementation of the indicators was raised. Disaggregation of data based on geography, gender, youth and age and consideration of cross-sectorial indicators were also suggested.

There was a general concert to translate the indicators to Russian and other working languages in the region, where possible, for ease of adoption of indicators by countries.

1. Diets and nutrition
Summary

The impact of socio-economic and health-related shocks on food security and nutrition is a growing concern for the region, especially in import-dependent countries. In order to monitor these impacts and support evidence-based decision-making, it was suggested to collect more data related to vulnerable populations.

The prevalence of different forms of malnutrition and diet-related NCDs were also highlighted as prominent food system related issues in the ECA region. The monitoring and improvement of the food environment was recognized as an effective solution for these problems.

Finally, it was agreed that the availability of data should be improved through more strategic and predictable data collection. More advocacy is needed to help countries recognizing the value of data in supporting decision-making processes.

*Do you consider the proposed indicators relevant, high quality, interpretable and useful?*

The proposed indicators were considered necessary and useful to reach national and regional targets for all food system actors such as governments, civil society organizations, UN agencies and experts.

<u>Specific recommendations:</u>

Indicators measuring *processed foods*, foods high in sugar, salt etc. are useful when considering their contributions to *non-communicable diseases (*NCDs).
In order to complement the indicator of the coverage of iodized salt (% of households), an indicator to measure the *prevalence of iodine deficiency* among the population should also be considered.
Ensuring food safety is crucial for food security, and it is an integral aspect of sustainability of food systems. More *food safety indicators* in the framework.
*Food borne diseases and residues of pesticides and veterinary drugs* were highlighted as potential food safety risks in the region. Monitoring these issues would help to understand their health impacts.
Indicators mainly cover 6-23 months of age and adults. *Indicators for adolescents* should be also considered, for example the prevalence of anemia and other micronutrient deficiencies among adolescent girls could be a useful information for the ECA region.
In order to monitor the impacts of socio-economic, health-related and humanitarian crises, it was suggested to collect more data related to *vulnerable populations (e.g. refugees, internally displaced or lower income groups*). The changes in dietary patterns of vulnerable populations during crises could be interesting indicator for food security/resilience.
As food security and resilience becomes increasingly important in the region, it is recommended that indicators related to *self-sufficiency, strategic food stocks and food independence* could be also measured.

*What are the data gaps that you can identify? Are these gaps structural? If not, are there some data or indicators that you know are available in your region and can be used for the purpose of this monitoring and assessment system?*

It was agreed that data gaps exist, and the availability of data should be improved through more strategic and predictable data collection. In the ECA region, sometimes the national policies and strategic plans do not ensure effective and frequent data collection. One of the main reasons is the need for financial resources.

It was also noted that in many cases, the data is collected by external partners, such as academia and independent organizations. In this regard, mapping data availability would be very important.

In addition, more advocacy is needed to help countries to recognize the value of data in decision-making, as well as encouraging data sharing, to ensure that all available data is used by policy makers. Transparency should be also improved since in some countries data is considered a sensitive area, with a reluctance to share it.

*What are the regional aspects we need to take into consideration at the moment of selecting a set of indicators to monitor the state of food systems and its evolution?*

The prevalence of different forms of malnutrition and diet-related NCDs were highlighted as prominent food system related issues in the region. The prevalence of overweight and obesity causes concerns for both children and adults. NCDs represent a significant burden and contribute to a high percentage of the mortality in the region, therefore they should be addressed.

Improving the food environment is key in this process. Even when policies affecting food environments such as marketing of breast-milk substitutes, or marketing junk foods to children do exist, they are not well implemented. The implementation of large-scale nutrition related programmes targeting both children and adults are also necessary.

The impact of socio-economic and health-related shocks and the ongoing war in Ukraine on food security and nutrition is a growing concern for the region. In order to monitor these impacts and support evidence-based decision-making, it was suggested to collect more data related to vulnerable populations. For example, in Turkey, there are particular issues to be addressed among refugees' communities, including stunting, wasting and malnutrition under 5 years old, especially as the number of refugees are rising quickly.

It may be also needed to streamline the cultural aspect of diets in different countries, and analyze how food systems and technical groups can help countries and policies to adjust to this issue. There are certain countries in the region, where the traditional diet contains high proportion of animal-based products, while alternative protein sources, such as pulses are not commonly consumed. Raising consumer awareness on nutrition and sustainability are crucial in this context.

Disparity in terms of consumption patterns is significant in the region and it sometimes occurs within a country or within different population groups. Therefore, the indicators related to dietary patterns are particularly important for the region. However, the relevancy of the Indicator on "zero fruit and vegetable consumption in adults" was questioned.

2. Environment and climate domain
Summary

There was a shared, renewed understanding that food systems have an outstanding role regarding their impacts on nature and climate change.
Also, the ECA region was particularly affected by the negative impacts of COVID-19 and the ongoing war in Ukraine, resulting in rapidly rising food and energy prices.

In the region, a more conscious consumer behavior could have an impact on the climate-food nexus and the planetary boundaries. On one side, there is an increasing demand for animal food products, but, on the other side, consumers are becoming more aware of the challenges facing the food system, resulting in a growing number of them that are paying attention to how food is produced, processed, distributed and, finally, consumed and, sometimes, wasted, particularly in relation to the major challenges the ECA region is now facing.

There were additional indicators suggested particularly related to water scarcity and its sustainable management; soil degradation and land management; climate change adaptation and mitigation; pollution (air, water, land) and GHG emissions; food loss and food waste; energy efficiency at all stages of the food supply chain, and, finally, antimicrobial resistance (AMR).

Regarding the data sources and availability in the region, there were several suggestions at the international, regional, sub-regional and national level, with a common recommendations of all data sets to be gathered, collected and analysed in line with the EU environmental and climate framework of reference.

*Do you consider the proposed indicators relevant, high quality, interpretable and useful?*

A few participants pointed out that the *indicators are very much dedicated to the production*, whereas the other stages of the food supply chain are not well represented (processing, transformation, distribution, consumption and waste).

*Water availability issues and food loss and waste* were the topics that came out frequently in the discussion. Water systems decline due to improper construction of water reservoir, thus water vulnerability needs to be assessed in the region (Armenia). Drought was also highlighted as an increasing challenge, particularly in the Caucasus and some Central Asian countries that are downstream from the source of important natural water supplies, leading to significant decline in water systems.

*Land and soil pollution*, either by excessive input use and/or solid waste disposal:
land degradation and land management are very relevant to be considered at the regional level.
Discussion on the possibility to include indicators to monitor and measure crop rotations, not only to restore soil but also valuable for reversing biodiversity loss (i.e. the case of Albania)
The large extent of ploughed land and its growing trend was also considered a concern (with the consequent decrease in productivity and the increase in input use).

Reflection was made on the possibility to include within the *GHG domain, the reference to energy efficiency*, with particular attention to the renewable energy.

GHG domain:
Carbon capture is a relevant indicator, but there is a common understanding that it needs to be evaluated *throughout the whole food value chain* (raised by a participant from Ukraine)
There is a suggestion to incorporate evidence from the food loss and waste also at the GHG emissions level. There was also a specific suggestion that an indicator on food waste reduction could be included, and it was noted that countries have begun collecting this data in many instances (raised by a participant from Bosnia and Herzegovina).
Measurement, reporting and verification of greenhouse gas emissions and mitigation in agriculture are not common in the region and national emissions trade schemes (if exist) are not always in line with the European Union Emissions Trading Scheme - EU ETS.

Debate on the degree of localness of food value chains and GHG:
Suggestions were made to include an indicator of the "localness of food systems", although it would be difficult to define.
*However, concerns* were raised to avoid demonization of the food trade sector. While it might make sense intuitively, eating locally would only have a significant impact on reducing GHG if transport was responsible for a large share of food's final carbon footprint. For most foods, this is not the case, see the supporting paper (raised by UNECE).

*Consumption patterns*, particularly the change of diet and nutrition habits among the ECA population, and its positive environmental implications could represent a valuable evidence to connect the climate-related indicators. An example is the link between consumption of foods of animal origin, and the amount of land required for animal feed for livestock.

Linked to an earlier comment, there is a need to consider the different stages of the food supply chain and how they contribute to GHG emissions and environmental impacts. It was acknowledged that there is progress in the application of innovative solutions to reduce the environmental impacts of food processing through bioeconomy approaches. It was suggested that this is an area where progress will be made in the region in the next few years, and it can be a potential indicator for the future analysis.

Specific Recommendations:

- *Applying a food systems approach* to the selection of indicator, and not focusing only on the production stage of the food supply chain. There was a shared perception that the indicators suggested were only related to the primary production phase.[8]
- Use of *renewable energy* in the food system suggested as an indicator.
- The *percentage of countries with operational drought monitoring systems in place* is suggested as an additional indicator for the Land Use and Water use domains (this indicator could also be considered for the Resilience domain) (WMO); Drought and Flood Forecasting and Warning Systems and Services suggested as indicator (WMO).
- *Food loss and waste reduction* targets are suggested as additional indicators, from economic and environmental perspective at different levels, to identify potential causes and potential interventions. The indicator on food loss and waste is not only related to the environmental *Pollution domain* and also strictly related to the *GHG emissions* and natural inputs efficient management (water use)( raised by a participant from Ukraine). This requires assessing waste at the different stages of the food value chain, including prevention of food waste at the consumer stage.
- To expand indicators of Land Use domain and include *crops typology and crop rotation*.
- *Reduced use of pesticides and other relevant chemical inputs* in agrifood production can also be an indicator.
- *Air pollution and solid waste* should be addressed with specific indicators.
- *Meteorological data on climate change impact on different economic sectors* should be included
- The use of antibiotics in food – also associate with the One Health approach - is a missing aspect in the indicators and reveals much about the health of food system (in line with the EU Green Deal and in line with the national policies in the most countries in the region).
- *What are the data gaps that you can identify? Are these gaps structural? If not, are there some data or indicators that you know are available in your region and can be used for the purpose of this monitoring and assessment system?*

The European Green Deal, Farm to Fork Strategy, Biodiversity Strategy, Repower Strategy, EU Zero Pollution plan represent a good framework of reference for supporting the countries in the region to become more sustainable by turning climate and environmental challenges into opportunities, in line with international standards: UNFCCC; UN Convention on Biodiversity, United Nations Convention to Combat Desertification and SDGs.

Some data gaps exist in agricultural water withdrawal. The assessment becomes very difficult, when the irrigation largely relies on individual wells, for example in the case of Albania. Local estimation is possible, but it is dependent on local governance. It was advised to consider assessing underground aquifers with remote sensing. The majority of data indicators could be captured with remote sensing, including also burning residues.

Food loss and waste data are good quality and available in the region, and data collection is being invested in. FAO and UNEP are regularly publishing data on the food loss and food waste Indices.

On land use, statistics in the region are often available and reliable.

---

[8] According to latest data, the vast majority of the environmental footprint of food products (ca. 80%) is generated during the production stage (land use and farm-stage).
https://ourworldindata.org/food-choice-vs-eating-local

Bosnia and Herzegovina Council of Ministries recently approved a set of environmental indicators which might be useful for the initiative.

Serbia, Environmental Protection Agency, can provide data on GHG emissions, and on land use and land use changes. Whereas, common to all Western Balkan countries, it is more difficult to provide data on pollution indicators, there is the need to develop monitoring and data analysis systems on environmental pollution.

*What are the regional aspects we need to take into consideration at the moment of selecting a set of indicators to monitor the state of food systems and its evolution?*

The highly diverse region of Europe and Central Asia faces the same range of challenges due to the increasing vulnerability to climate change impacts, and environmental pollution.

One of the top policy priorities among the countries in the region is more rationale, sustainable and coordinated use of natural resources, namely water and land.
In addition to the threat posed by climate change to water availability in the region, wasteful and inefficient water use emerges as a common problem across the region and as one of the primary area of policy action, particularly for the Central Asia and Southern European countries, as well as for some countries in the North. But, cutting across discussions of the need for improved water resource management, there is an equally strong emphasis on the importance of strong water governance and cooperation (e.g. Azerbaijan and Turkey cooperation).
Land degradation, particularly the high-level mechanical degradation, and soil erosion are also common challenges across the region. Land use, and conservation approaches should be applied in a way that benefits both productivity and nature. Reference to heavy contamination of the soil from previous decades was made. Alongside the land management challenges, aspects of preventing environmental pollution through the responsible use of chemical inputs in food production, and importance to avoid the mismanagement of pesticides and fertilizers, in the region still remains a crucial issue, since these items generate soil and water pollution, contribute to the loss of biodiversity and diverse ecosystems in the region. The situation is further exacerbated in some countries due to the proliferation of obsolete pesticide stocks.

It was noted, that these actions are also important for human health. Furthermore, effective waste disposal, and biosecurity measures are also important to prevent environmental pollution, for example adequate waste disposal from *slaughterhouses and other livestock activities.* Also affected are the quality and cost of food, with impacts to the whole food system, food security, and human health. All of the components of food systems (production, processing, distribution and consumption) are generating huge amounts of externalities. These are primarily linked to damages to the environment throughout the whole food value chain and the costs of human health through consuming unhealthy food.

Plastic waste was also another major theme of concern in the region.

Antimicrobial resistance poses a major threat in the region and preventing the threat of AMR is an important priority. FAO is currently supporting data collection through surveys on the use of antimicrobials in the livestock sector in 14 programme countries.
In addition, relevant efforts are recorded in the EU countries to prevent food losses and food waste, and the topic is receiving increased attention in the Caucasus and Central Asian context.

3. Livelihoods, poverty and equity Domain
Summary

The importance of providing clear definitions for the indicators was noted, and they are suggested to be strengthened in some cases.
Concerns on the availability of data on agriculture and livelihoods at national level due to significant informal and unregistered information were raised several times during the discussion. Participants further stressed the need for sex-disaggregation of data and per urban and rural areas where possible as some of the issues could be very context specific.
The speaker explained the challenge on how to identify populations involved in the food systems, as some countries are collecting disaggregated data, while others do not have even the overall data. He referred to the background paper where ILO data which includes primary employment complimented with EUROSTAT. However, there are some countries that are excluded from ILO data, and informal, self-employed, migrant data is just estimated.
Participants suggested that indicators on the impact of forced or voluntary migration, remittances, EU indicators, and seasonal work could be also considered.
*Do you consider the proposed indicators relevant, high quality, interpretable and useful?*

Specific recommendations:

Countries (Turkey, Albania, Georgia) commented that the *agricultural income* is one of the key indicators, but the definition needs to be clarified by considering factors such as rural-urban households, percentage of the income, sources that are considered as agricultural income, and the part of the system this indicator considers (full food chain or only farm level). Additionally, the indicator "percentage of the population earning low pay" is important, since it can relate to the migration from rural areas due to low income. But the definition includes hourly earnings for agriculture; which is very difficult to obtain in many countries.

On the *employment indicators*, the group noted the importance of including informal employment as it is very common in the agricultural sector but data is difficult to obtain. While the number of smallholders and their contribution to the agricultural production is very significant in the region, they often remain in informal employment as not all are registered. The land holdings indicator is relevant, but measuring access to land is problematic due to the lack of registration.

"Households with significant income from agriculture" was highlighted as key indicator, but the definition needs to be clear on the target population whether they include full food chain or only agriculture. The indicators under employment (earnings/wages) are also noted as highly relevant as people are leaving rural areas because of low compensation and non-farming and shifting/seasonal sources; but data on hourly earnings is very limited or non-existent in the region. The indicators on human rights were found very relevant, with a note on access to land as well as resources.

*Access to resources* was also suggested as a potential indicator. The FSCI Expert noted that the rights and access to inputs and resources should be cross-examined closer because of their implications on the livelihoods.

*What are the data gaps that you can identify? Are these gaps structural? If not, are there some data or indicators that you know are available in your region and can be used for the purpose of this monitoring and assessment system?*
Participants noted the crucial need for accountable data sources. International organizations, such as World Bank, are reliable and accurate, but challenges exist at the national level, particularly for farm-relevant data. In certain countries, data on small- and medium scale food enterprises is also absent. At country level, informal data sources might provide the information that are not available in the official sources, or complement it. These could be considered, depending on their reliability.

Disaggregation of data (rural/urban, gender, age) wherever possible is strongly advised by the participants and their quality should be improved. Concerns were raised on the limited availability, informality, non-informality and access to data.

Bosnia and Herzegovina informed on their Ministry's available data source, which should be considered amongst the resources, as they had a recent survey on agriculture and rural development.

*What are the regional aspects we need to take into consideration at the moment of selecting a set of indicators to monitor the state of food systems and its evolution?*

For the EU candidate countries in the region, it is important that indicators are harmonized or taken into account with those also set by the EU, when agreeing on the Food Systems Countdown indicators.

Suggestion was made to include also an indicator on migration's impact on the food systems.
In addition, seasonal work is a significant factor in the region particularly for the agriculture sector and suggested to be considered.

4. Governance Domain
Summary

It was noted that in the Europe and Central Asia region, the governments have well-established systems of governance in place, which provide a good foundation and experience on agriculture, education, health, nutrition and trade. But it is important to take a critical view and to ensure that weaknesses in governance are addressed and to consider as part of the continuing evolution specific aspects to be improved as part of transforming agrifood systems. It was acknowledged that it is important to encourage monitoring, evaluation and continuous development. Selecting and tracking the right set of indicators will help countries to ensure that the transformation of food systems proceeds in the agreed direction, and the holistic food systems approach evolves in a continual momentum.

It was highlighted that UNFSS national pathways have been important to formulate policy documents and commitments related to sustainable food systems. This process of policy integration and committing the actions in the national pathways to strategic plans is on-going in many countries of the region.

*Do you consider the proposed indicators relevant, high quality, interpretable and useful?*

Discussing the indicators related to the governance thematic was interesting, but also a relatively new discussion for the audience.
It was challenging to identify specific, quantifiable indicators.

Specific recommendations:

In case of *food systems governance, multisectoral coordination* was highlighted as a key area for improvement, and a number of examples were given to demonstrate efforts both at national, and at local levels.

Under *coordination*, it was mentioned that it is important to look at the trade-offs between different policy goals and ensure they are balanced. The indicators would need to allow to look not only horizontal but also vertical levels of coordination. In addition, it would be important to capture the trends both on the national level and subnational level, including recognizing that is there are moves towards different government sectors working together, e.g. nutrition and agriculture, agriculture and environment – this is an essential prior step to full policy coherence (end objective). This information would help to identify specific interventions on the sub-national level and support the more efficient use of resources in food systems transformation. Responsible investment in food systems might also be an indicator.

There are several examples, where the formulation and implementation of national food system pathways are supported by national coordination platforms. However, the need for capturing nuances while looking at policy coordination and coherence was highlighted. Different countries may have different policy instruments or tools in place that can lead to the same transformation. Governance also needs to take place at many different levels and across many different aspects of the food system (value chain development, sustainable tourism, environmental protection, food safety, animal health, etc.) and the indicators need to ensure these are adequately measured. Or at least it would be helpful to understand at what level and how effective governance will be measured.

On *shared vision indicators*, it was suggested that it is important to look beyond what is reflected in the national food system transformation pathways. The candidate indicators could be expanded to capture the *cooperation and coordination level between the public and private sector; not only CSO level*.
For example, inter-branch organizations and similar public-private platforms contribute to the transformation process in terms of defining a national vision for specific sectors in Albania (fresh produce, dairy, meat etc.), similar to the EU model.

Under *implementation*, additional to the "marketing of breast-milk substitutes restrictions" indicator, it is suggested to look at breastfeeding in a more holistic manner and measure public health indicators, such as the *rate of exclusive breastfeeding*, as well as *the level of education provided to mothers related to breastfeeding*.

*What are the data gaps that you can identify? Are these gaps structural? If not, are there some data or indicators that you know are available in your region and can be used for the purpose of this monitoring and assessment system?*

The FSCI Expert explained that the expert team at the Initiative tried to identify the most relevant indicators and agreed that at least 70% of countries should have data on the selected indicator. However, in many cases, this may not be realistic. Due to this limitation, one recommendation is to choose potential indicators, even if data is not currently available and advocate for more data collection in the future.

Regarding data sources in the region, a recommendation is to make use of the regional platforms that are involved in food system analysis (e.g. Economic Coordination Organization - Regional Coordination Centre for Food Security (ECO-RCCFS), Black Sea Economic Cooperation - Regional Cooperation Centre for Sustainable Food Systems (BSEC-RCCSFS), etc.).

*What are the regional aspects we need to take into consideration at the moment of selecting a set of indicators to monitor the state of food systems and its evolution?*

As mentioned above, UNFSS national pathways are important tools to formulate holistic policy documents and commitments related to sustainable food systems in the region.

Some specific examples from Albania were shared that might be relevant for the whole region. These are the formulation and implementation of policies to promote healthy diets (school meals, nutrition education, marketing of food products, regulation of breastmilk substitutes) and food safety (e.g. on foodborne diseases and pesticide/veterinary residues). The One Health approach was also highlighted as an important example on policy coherence.

5. Resilience and sustainability Domain
Summary

- Overall, the participants found the candidate indicators important and relevant, although 26 indicators are quite a long list to be considered and evaluated.
- The importance of providing clear definitions for resilience considering the regional nuances for the indicators was noted.
- Resilience and sustainability indicators, as complex, cross-cutting themes, have to be identified avoiding overlaps with other indicators under the five domains.
- Resilience and sustainability are considered at all levels as a pre-condition for sustainability and its monitoring is necessary to capture food systems' holistic nature. Resilience is somehow a trade of short-term efficiency and longer-term survival.
- Identifying indicators in this area was very challenging, as it is very hard to find reliable, quantifiable indicators.
- It would be important to capture diversification as one of the potential paths to follow to transform food systems. If a country has diversified trade partners, food production, import and export, it will become more flexible to absorb shocks in the future.

*Do you consider the proposed indicators relevant, high quality, interpretable and useful?*
Specific recommendations:

- Indicators must capture *regional specificities*: sub-national and disaggregated.
- *Proposal of including renewable energy indicator*: already included into the following indicators: Mobile cellular subscriptions (per 100 people); Access to electricity (% of population); Renewable electricity output (% of total electricity output) (Ukraine).
- Suggestion to include *social protection indicators*: adequacy coverage, school feeding programs (also mentioned in the Livelihoods domain).
- Proposal of indicator to monitor and track the *exposure the dependency on international markets and trade, import dependency ratio* should be an aspect of resilience.
- *Proposal of inclusion of an indicator on the number of operational drought and early warming systems*.
- Indicators related to *higher levels of education (vocational education, and above)* would be interesting and should be included, too.
- *Corporate non-financial reporting – ESG & Corporate Social Responsibility* – are suggested as indicators.

*What are the data gaps that you can identify? Are these gaps structural? If not, are there some data or indicators that you know are available in your region and can be used for the purpose of this monitoring and assessment system?*

Gaps:
Environmental-related indicators: incidents related to adverse events – number of deaths (attributed only to agriculture) associated to the social impact of those events, considering increasing droughts.

Suggestions:
Global Innovation Index by WIPO and consortium of partners (also including private sector)
EM-DAT human impact
Combining also data from databases on Data conflict, armed conflict and IDPs.
Non financial reporting – CSR & ESG, in corporate reporting is increasingly gaining attention within EU countries and might represent a framework of reference for other countries in our region.

*What are the regional aspects we need to take into consideration at the moment of selecting a set of indicators to monitor the state of food systems and its evolution?*

Food systems resilience is critical to food security and nutrition (e.g., fragile states = food insecurity, COVID-19).
The theme is also critical for other functions (e.g., livelihoods, inclusion).
Within EU countries, we are witnessing an ever-lasting debate within the 2 different views on CAP: traditional family farming and market-oriented industrial agriculture efficiency view points, with key components including resilience.
Growing consensus on the diversification as one of the potential paths to follow to transform food systems. If a country has diversified trade partners, food production, import and export, it will become more flexible to absorb shocks in the future.
*Diversification – to capture the social, economic, environmental dimensions and make food systems intelligible.*
A relevant policy priority area for the region refers to migration, both forced and voluntary, (how migration reveals some aspects of the resilience ability of the food system of the region).
Green finance also is an indicator we may want to take into consideration and show the interest of governments in the region for investing responsibly in agrifood systems.

**Conclusions**
*"The need for food system transformation is undeniable. We need to transform the way that we produce, we need to do something about the way that we consume, we have to try to transform the way that we conceive food, if we want to be able to offer diets that are healthy but at the same time produced in a sustainable, resilient and equitable way"* said **José Rosero Moncayo,** Director of the Statistics Division, FAO, Co-Chair of the Initiative.

Food system transformation is urgent, requiring rigorous, science-based monitoring to guide public and private decisions and support those who hold decision-makers to account.

National and global governments are struggling to govern the increasingly complex crises and powerful forces in food systems, especially as the COVID-19 pandemic, climate change and the recent conflict in Ukraine have shown the fragility of food systems and the lack of sustainability to continue along the present course. These crises and challenges underscore the urgency to change the trajectory for food systems but also offer great potential to do so.

Monitoring food system transformation in this manner can aid governments in setting priorities and establishing incentives and regulations to align food systems in a transformative direction. High-quality evidence allows food system actors to undertake "course corrections" and make necessary changes.

This transformation is achievable. Rigorous monitoring is necessary to keep progress on track. It requires validation and to put indicators into reality of people that use this type of data to guide for the right policy.

Concerns about data availability:

- The initiative is aiming at solving a remarkable challenge: Using indicators based on the data available now.
- Lack of data could be solved through a global commitment to raise the bar on the international political agenda.
- Need to keep developing methods to compile indicators that are relevant to food systems and expand the list of indicators that we have now.

- The current set of proposed indicators – more bias towards production and agriculture in contrast to other aspects of the food system. The reason is the lack of data and methods.

**Jessica Fanzo,** PhD, Bloomberg Distinguished Professor of Global Food & Agricultural Policy and Ethics, Co-Chair of the Initiative highlighted the importance of the convening power of FAO: around 500 people joined to the discussions from all regions to provide their feedback.

**Raimund Jehle**, Regional Programme Leader, Deputy Regional Representative, FAO Regional Office for Europe and Central Asia confirmed that the there is an interest in measuring the progress to demonstrate the importance of the food system contributing to achieve the SDGs.

*"National ownership of sustainable food systems concept is the key to be successful*.
The challenge now is how to get these actions at country level and harmonize these indicators with different national sectorial strategies as in many countries a food systems strategy is not yet available. We need countries' input and support in order to work more coherently and to integrate into the national pathways developed as part of the UN Food Systems Summit", he added.

**Agenda of meeting**
**AGENDA**

| | |
|---|---|
| 08:30 -08:50 hours | *Welcoming remarks* |
| | **Raimund Jehle**<br>Regional Programme Leader, FAO Regional Office for Europe and Central Asia<br>**Hernán Daniel Muñoz**<br>Statistician, Statistics Division, FAO |
| 8:50 -09:15 hours | *Introduction to the initiative* |
| | **Lawrence Haddad**<br>Executive Director, GAIN |
| 09:15 -10.30 hours | *Breakout sessions, block 1:*<br>*• Diet, nutrition and health Domain;*<br>*• Environment and climate Domain;*<br>*• Livelihoods, poverty and equity Domain.* |
| | Moderators:<br>**Musonda Mofu**<br>Deputy Executive Director, National Food and Nutrition Commission, Zambia<br>**Fabrice DeClerck**<br>Science Director, EAT<br>Senior Scientist, One CGIAR<br>**Alejandro Guarin**<br>Researcher, International Institute for Environment and Development (IIED) |
| 10:30 -10.45 hours | *Summary of the breakout sessions* |

| | | |
|---|---|
| 10:45 -12.00 hours | *Breakout sessions, block 2:*<br>*• Governance Domain;*<br>*• Resilience and sustainability Domain.* |
| | Moderators:<br>**Namukolo Covic**<br>Registered Nutritionist, Health Professions Council of South Africa<br>Senior Research Coordinator at the International Food Policy Research Institute<br>**Dr. Preet Lidder**<br>Technical Adviser to the Chief Scientist, FAO |
| 12:00 -12.15 hours | *Summary of the breakout sessions* |
| 12:15 -13.00 hours | *Wrap up and conclusions* |

**Participants List**

| Country/ Organization | Name | Title |
|---|---|---|
| FAO - REU - RI1 | Ms Bianka Laskovics | Regional Initiative Coordination and Rural Development Associate |
| FAO - REU | Ms Valeria Rocca | Regional SDG Advisor |
| FAO - REU | Mr Giorgi Kvinikadze | Statisttician |
| FAO - REU | Anna Jenderedjian | Gender Mainstreaming and Social Protection Specialist |
| FAOTR | Ozcan Turkoglu | Senior Programme and Pipeline Coordinator |
| FAOTR | Yasemen Asli Karatas | Programme Associate |
| WMO | Natalia Berghi | UN Programme Officer |
| Regional DCO in Istanbul | Alper Almaz | |
| North Macedonia | Mr Aleksandar Musalevski | Head of sector for coordination of Minister’s Cabinet, Ministry of Agriculture, Forestry and Water Economy (also involved in Agriculture Policy Analysis Department) |
| North Macedonia | Ms Neda Gruevska | Rural Development Department, Ministry of Agriculture, Forestry and Water Economy |
| North Macedonia | Mr Jordan Chukaliev | Professor, University Ss. Cyril and Methodius, Faculty of Agricultural Sciences and Food – Skopje, Institute of natural resources, management and environment protection in agriculture |
| North Macedonia | Ms Aleksandra Martinovska Stojcheska | Professor, University Ss. Cyril and Methodius, Faculty of Agricultural Sciences and Food – Skopje, Institute of Agricultural Economics |
| North Macedonia | Mr Vladimir Petkovski | Institute for Economics |
| North Macedonia | Mr Igor Spirovski | Head of Department of Physiology and Monitoring of Nutrition, Public Health Institute |
| Finland National Convenor | Mr Jyri Ollila | Counsellor; World Food Systems Summit Ministry of Agriculture and Forestry, Finland |
| Albania | Ms Valbona Paluka | Advisor to minister of agriculture on food safety/security and Curator of National Food System Dialogues |
| Albania | Ms Renata Kongoli | Professor at Agriculture University of Tirana |
| Albania | Mr Drini Imami | Professor at Agriculture University of Tirana |
| Albania | Mr Engjell Skreli | Professor at Agriculture University of Tirana |
| Spain | Ms Patricia Villoch Carrión | Ministry of Agriculture, Fisheries and Food and the Spanish Representation to FAO |
| Republic of Moldova | Ms Maria Nagornii | Ministry of Environment, Head of the Division for Policies on Pollution Prevention |
| Republic of Moldova | Ms Galina Petrachi | Ministry of Agriculture and Food Industry, Department for Rural Development Policy and Programs |
| Republic of Moldova | Ms Marina Dintiu | National Food Safety Agency, Head of Department of Quality and Safety of food products of non-animal origin |
| Republic of Moldova | Mr Viaceslav Grigorita | Ministry of Agriculture and Food Industry, Head of Department Policies in Plant production and processing sector |
| Republic of Moldova (additional person) | Ms Liliana Martin | Ministry of Agriculture and Food Industry, Policy Analysis, Monitoring and Evaluation Department, Acting Head of Department |
| Republic of Moldova (additional person) | Ms Nicoleta Golovaci | Ministry of Agriculture and Food Industry, Policy Analysis, Monitoring and Evaluation Department, Senior Consultant |
| Kyrgyz Republic | Ms Meerim Esenkulova | Specialist, Food security Department Ministry of Agriculture |
| Kyrgyz Republic | Mr Abdymital Chyngojoev | Expert |
| Kyrgyz Republic | Mr Kanat Tilekeev | Research Fellow, UCA University |
| Kyrgyz Republic | Ms Miraida Naktaeva | Head of welfare department, Ministry of Labour, Social Protection and Migration |
| Georgia (Food System Summit Convenor) | Ms Ekaterine Zviadadze | Head of Primary Structural Unit of the Policy Coordination and Analysis Department at the Ministry of Environmental Protection and Agriculture of Georgia |
| Georgia (Thematic Expert) | Ms Maia Tskhvaradze | Head of the Climate Change Division at the Ministry of Environmental Protection and Agriculture of Georgia |
| Georgia | Mr Giorgi Machavariani (accompanies Ms Tskhvaradze) | Senior Specialist at the Climate Change Division at the Ministry of Environmental Protection and Agriculture of Georgia |
| Georgia (Data Expert) | Mr Giorgi Sanadze | Head of the Agricultural and Environment Statistics Department at Georgian National Statistics Office (GeoStat) |
| Georgia | Mr Irakli Tsikhelashvili (accomapies Mr Sanadze) | Head of Environment Statistics Division |
| Georgia | Mr Goga Talakhadze (accomapies Mr Sanadze) | Head of Agricultural Statistics Division |
| Hungary (Observer) | Ms Liliána Kaszás | Hungarian Ministry of Agriculture, FAO coordinator |
| Hungary (Observer) | Ms Dorka Mudrity | Hungarian Ministry of Agriculture, Project referee |
| Turkey | Pelin POLAT ÇAVUŞOĞLU | Engineer, General Directorate of Agricultural Reform (Ministry of Agriculture and Forestry) |
| Turkey | Dr. Aygül ÇAĞDAŞ | Engineer, Strategy Development Department (Ministry of Agriculture and Forestry) |
| Turkey | Arap Diri | Engineer, Turkish Statistical Institute |
| Turkey UNRC | Bulent Acikgoz | UNRC Office |
| Turkey | Mehmet Emre Uğur | Expert, Turkish Statistical Institute |
| Turkey UNRC | Bulent Acikgoz | Partnerships and Development Finance Officer, UNRC |
| Turkey FAO | Ozcan Turkoglu | FAO |
| Turkey UNICEF | Mehmet Ali Torunoglu | UNICEF |
| Serbia | Milos Stojanovic | FSS Curator, Ministry of Agriculture, Forestry and Water Management |
| Serbia | Dragana Vidojevic | Head of Section for Indicators and Reporting, Serbian Environmental Protection Agency, Ministry of Environmental Protection |
| Serbia | Milica Jevtic | Head of Unit for Analytics and Statistics, Ministry of Agriculture, Forestry and Water Management |
| Serbia | Jelena Perać | Group for agricultural statistics, Statistical Office of the Republic of Serbia |
| Tajikistan | Mr Sherali Rahmatulloev | Chief specialist of the Department of Maternal and Child Health and Family Planning of the Ministry of Health and Social Protection |
| Tajikistan | Mr. Nazarov Bahodur | Head of department on agrarian policy and food security monitoring of the Ministry of Agriculture of the Republic of Tajikistan |
| Tajikistan | Mr. Andamov Ismoil | Head of the Department on Veterinary and Breeding Supervision of the Committee for Food Security under the Government of the Republic of Tajikistan |
| Tajikistan | Mr. Murodjon Ergashev | Expert on Environment |
| Tajikistan | Mr. Firuz Saidov | National Consultant on Climate Change in Project Management Unit under Committee for Environmental Protection under the Government of the Republic of Tajikistan |
| Tajikistan | Mr. Jamshed Kamolov | Head of Main department on Population and Territory Protection of the Committee on Emergency Situations and Civil Defense under the Government of the Republic of Tajikistan |
| Tajikistan | Mr. Rustam Babadjanov | Professor, Institute of Economy and Demography of the Academy of Science of Tajikistan |
| Armenia World Bank | Artemis Ter Sargsyan | National Consultant to WB |
| Armenia World Bank | Irina Ghaplanyan | |
| Armenia World Bank | Arusyak Alaverdyan | |
| Armenia World Bank | Vahan Danielyan | |
| Armenia (recommendation from WFP) | Vardan Urutyan | Rector of Agrarian University |
| Armenia (recommendation from WFP) | Mr Gevorg V. Ghazaryan | |
| Armenia (recommendation from WFP) | Magda N. Hovhannesyan | |
| Ukraine | Olena Kovalova | Senior expert, IPRSA, Dr.Sc Econ., former DM of MAPF |
| Ukraine | Larysa Petriv | Head of Unit, National Scientific Center for the Climate-smart Agriculture, National Academy of Agrarian Science |
| Ukraine | Mykhailo Malkov | Environment Advisor, Office of the UN Resident Coordinator to Ukraine |
| Ukraine | Olena Safarova | Area Coordinator, Emergency Programme in Ukraine |
| Expert | Elisabeth Duban | |
| Expert | Valentina Bodrug-Lungu | |
| Expert | Eugenia Ganea | |
| WMO | Dr Elena Mateescu | |
| BiH | Mr Dalibor Vidacak | Expert Adviser, Food Safety Agency of BiH |
| BiH | Ms Vanda Medic | Senior Adviser, Ministry of Foreign Trade and Economic Relations BiH |
| BiH | Ms Sanida Sarajlić | Head of Department for Rural Development Policy Coordination, Ministry of Foreign trade and Economic relations of BiH |
| UNECE | Marit Nilses | |
| WMO | Milan Dacic | |
| WFP Cairo Regional Office | Khalid al-Qudsi | |
| WMO | Bob Stefanski | Chief, WMO Applied Climate Services Division |
| Georgia | Zaza Chelidze | Statistical Expert |
| Belgium | Louis Bout | Cabinet Member, Cabinet Hilde Crevits, Vice-Minister-President of the Flemish Government and Flemish Minister for Welfare, Public health and Family |
| Montenegro | Ms. Zorka Prljević | Deputy Director, Agency for Food Safety, Veterinary and Phitosanitary Affairs |
| Montenegro | Ms. Jelena Vračar Filipović | Advisor, Agency for Food Safety, Veterinary and Phitosanitary Affairs |
| Montenegro | Ms. Dubravka Radulović | Advisor in Directorate for Agriculture, MAFWM |
| Montenegro | Mr. Mirsad Spahić | Head of Department for Economy Analysis and Market, MAFWM |
| Montenegro | Ms. Kristina Radević | Advisor in Directorate for Rural Development, MAFWM |
| Belarus | Olga Lobanova | Consultant of the Agriculture and Environment Statistics Department, the National Statistical Committee of the Republic of Belarus |
| Belarus | Iryna Husakova | PhD, Head of the Food Market Sector, The Institute of System Research in Agro-Industrial Complex of National Academy of Sciences of Belarus |
| Belarus | Natallia Kireenko | DSc, Professor, Head of the Chair of Innovative Development in Agro-Industrial Complex, The Belarusian State Agrarian Technical University |
| Belarus | Irina Voitko | PhD, Associate professor of the Chair of Innovative Development in Agro-Industrial Complex, The Belarusian State Agrarian Technical University |
| Belarus FAO | Alesia Launikevich | FAO |
| JHU | Kate Schneider | JHU |
| Iceland | Dr. Matthías G. Pálsson | Permanent Representative of Iceland to FAO, WFP and IFAD |

**FAO RLC Regional Expert Consultation on the Food System Countdown Initiative's Indicator Framework**
**Virtual, May 17, 2022**

*This report was produced by RLC to summarize the results of the FAO RLC regional expert consultation held on* May 17, 2022.

**Introduction:**
The session took place normally on the Zoom platform, with opening words from the then-ADG Sr. Julio Berdegué and an introduction from the Statistics Division (ESS) Director, José Rosero. Subsequently, the audience was divided into working groups, to review in detail the different domains of the FSS Indicators. To close, a plenary session was held in which the moderators summarized what was discussed in each group.

Through a formal letter from the FAORs and RLC leaders, 270 people were invited to participate in the consultation, from more than 20 countries in the region. 244 of them were partners (especially government), while 26 were people from FAO and speakers. On the side of the invited UN institutions, there was ECLAC (with 7 guests) and IFAD (with 4). Before the event, 64 people confirmed their participation.

On the day of the event, there were more than 100 participants, and the range varied between 70 and 100 people, remaining stable among all blocks.

IT technical support, simultaneous interpretation and recording arrangements were provided. Each working group had a moderator and a note-taker from RLC, as well as a speaker chosen by ESS.

**Opening:**
Mr. Berdegué spoke about the challenges in relation to the concept and ways of measuring the transformation of agri-food systems. He pointed out that it is a broad concept, involving different types of actors, and all types of domains.

For Mr. Berdegué, the transformation of agri-food systems is a phenomenon that affects several dimensions, from the individual behavior of consumers to relationships with the environment, health institutions, the adoption of technologies, etc.

A key question was raised: "How to know if the actions that are being carried out – by civil society, governments, etc. – are contributing to the transformation of agri-food systems in the right direction and in a consistent way?" He concluded by pointing out that it is essential for the transformation of agri-food systems to have good analytical frameworks, to move from the conceptualization to the operationalization of these concepts.

**Presentation of Themes**

**Group 1: Diet, Nutrition, and Health Domain**
People were in the center. The objective of the exercise was to discuss several indicators related to diet, nutrition and health. To do this, the analysis needed to focus on different areas to measure people consumption (diet quality).
Nutrient adequacy and dietary risk factors for NCDs had to be measured, as well as the accessibility to healthy diets (food security). Other important issues were: food environment (availability, affordability, messaging, and food & vendor properties); and to answer whether policies contribute positively or negatively towards food availability, food access, and product properties (policies affecting food environments).
Finally, a little disclaimer was raised: "Why don't we have measurements like weight? Because those indicators are also affected by other contexts and factors".

**Group 2: Environment and climate Domain**

The "Environment and Climate Domain" group had 5 areas.
What is to be monitored? Food systems are a major source of environmental degradation, the actions required to achieve global environmental commitments, monitoring and accountability are essential.
The definition and the candidate indicators for each area of the "Environment and Climate" domain were presented: land use, greenhouse gas emissions, water use, pollution, and Integrity of the Biosphere.

**Group 3: Livelihoods, poverty and equity Domain**

The topics related to the group were introduced; these are: Poverty and income, employment, social protection and rights. Cross-cutting issues were also pointed out: governance, resilience and sustainability, emphasizing their importance.
The topics and problems that were sought to be measured were presented: the number of people who are part of the food system, in rural and urban areas, considering their diversity and also considering the poverty and vulnerability of those who work in these sectors. The definition of each one was given, and the questions were asked.

**Group 4: Governance Domain**

The objective of the meeting was to obtain feedback on the governance indicators' component, focused on the context of our region. Governance had four components: joint vision, policies and strategic planning, effective implementation, and accountability.
What is to be monitored? First, the common vision of the outcomes, relevant policy instruments to align efforts, the implementation of resources, and accountability for the outcomes.
The joint definition of governance is: inclusive and participatory processes that allow to identify priorities and to generate a guide with desired results. For this, different actors, civil society, relevant actors from the academy and decision-makers are involved.
To do this, it is necessary to measure power imbalances (market, influence asymmetries, concentration and market power).
The possible indicators are the following: Index of civil society, variants of democracy, and presence of a national food transformation system (strategic planning).

**Group 2: Resilience and sustainability Domain**

Resilience and sustainability were presented as a transversal aspect of the transformation of agri-food systems.
The subdomains are exposure to shocks, resilience capacities, agrobiodiversity, food security stability, and food system sustainability index. This issue is critical for food and nutritional security (for example, as states become more fragile, there is more insecurity, or the context of covid 19). It is critical for other functions: resilience is a precondition for sustainability; it has a multidimensional interpretation. The definition and candidate indicators were presented.

**Discussion**

- Some aspects to consider in the region are: retail trade, formality and informality, dependence on international trade, frequent natural disasters in certain areas (droughts, hurricanes, tornadoes, etc.), family farming, gender, generational change, among others.
- Cluster analysis idea is repeated in several groups.
- Considering demand factors, such as consumption habits, is highly relevant. The use of technologies and innovation in food systems should be considered. The financial support and inequality of countries should also be included, as well as the biodiversity of the production area, the trend in increasing food imports, crop productivity levels, and food losses and waste.

- Are indicators that are already on the SDG list being considered? That thinking of facilitating the follow-up of the countries to the indicators, and not repeating it...
- The indicators represent a great advance in the construction of measurement, but there is not going to be much correlation in an indicator that measures something in a construct (environment, or diets), and that again in another construct returns to exist in a similar way.
- Lack of data and availability of information in the region. For example, in the Dominican Republic, agricultural censuses have not been carried out since 1982. The absence of updated data is a problem for the formulation of good policies. There are NGOs that have good data, and good use could be made of it.
- An immediate task for LAC should be to make an inventory of available variables and indicators based on official information; multilateral organizations must draw on official information. Some national statistical systems are stronger than others in LAC. Many times, the measurements that are made are not published, due to political implications. The information exists.

1. Diets and nutrition

*Do you consider the proposed indicators relevant, high-quality, interpretable and useful?*

There are three repeated discussions: 1) among the proposed indicators, there are indicators of different levels and hierarchies (of processes, which seem to be inputs from others, of results, and whether or not impact indicators should be considered). 2) The aggregation capacity of these indicators at the sub-regional, regional and global levels, but also at local, territorial levels, to groups of certain ages, from different areas. 3) The data and data update frequency, the quality of the information.
With regard to food environment, governance indicators such as food supply in schools are often included, which is directly related to the food environment at school level. Therefore, it is suggested to adapt the definition, in such a way that it is not confused with governance aspects that directly aim to change the food environment.
The indicators of accessibility to healthy diets and the cost of healthy diets are very interesting and necessary, but it is important to consider countries that do not have data associated with this. On the other hand, the indicators are highly concentrated in food consumption and availability, but there are not many biometric or anthropometric indicators, which are very important having sustainable food systems.
It is pointed out that there is information associated with malnutrition due to excess or deficit; and this can be interesting to include, to think about medium and long-term indicators.
The healthy cost indicator seems very important; however, again the data gap is a big issue (for example, there is an absolute data gap in the case of Venezuela). On the other hand, the inclusion of food in social programs is very important and the use of instruments such as vouchers only for fruits and vegetables can make a big difference. The zero fruits indicator seems very relevant.
Finally, it would be necessary to make an ordering through the three elements of the food system (political, sectorial and individual), and to recognize which the process indicators that are required to achieve certain health are (and those of impact in each of these areas).

*What are the data gaps that you can identify? Are these gaps structural? If not, are there some data or indicators that you know are available in your region and can be used for the purpose of this monitoring and assessment system?*

It is suggested to incorporate the issue of policies, laws, strategies, regulations, which may be affecting the results and possibilities of transforming food systems, and also indicators that capture food loss, which are absent.

There are gaps in terms of physical access to food, the availability of food at nearby points. This is relevant and there are no indicators of this.

On the other hand, the type of place where food is purchased, countries that have a greater penetration of supermarkets, versus more informal fairs, etc.; this generates a substantially different food system, and it is interesting.

The main concern is the availability of data for the indicators. An example of this is the data on the costs of the diet for the Caribbean, how can this information gap be managed in the different countries of the region? On the other hand, it is suggested to think about how to involve the countries more in the process of collecting this data.

In addition, there are limitations for updating impact indicators, and there is also the issue of being able to disaggregate by region. There are big gaps in data and the time taken to collect this data is varies between countries.

It is very important to consider the food environment. There are two very different educational systems: the private and the public. The public system follows a menu for the school diet, but they are low-income children. It would be very good to monitor the effects of the policy changes that are needed to generate more agile ways of obtaining these nutrition indicators.
It could be interesting to address the measurement of food spending. On the other hand, it is also interesting to consider what happens with the eating behavior indicators.
Regarding child nutrition, it would be good to consider not only the child of breastfeeding age, but also beyond 23 months. It would be necessary to see other spaces to obtain more information to add these indicators.

*What are the regional aspects we need to take into consideration at the moment of selecting a set of indicators to monitor the state of food systems and its evolution?*

It is very important to consider the availability of data for indicators. In operational terms, the information needs to be disaggregated, not all regions have the same information. In terms of policy, it is necessary to consider the differences between provinces, departments, vulnerable groups; therefore, disaggregation is something that must be included in this proposal.

The cycles and food supply are different among countries; this should be something to take into account. Regarding food security policies, there is no a Central American level policy, based on the restructuring or relevance in data collection.

Caribbean countries are complaining that, over the years, the food system hasn’t been well-developed and that the import bills are getting more expensive every year. So, the questions are raised: "Is it possible to have indicators on this?", "Is it possible to have indicators related to imported food bills, so that countries can monitor this and take appropriate measures?"

2. Environment and climate domain

*Do you consider the proposed indicators relevant, high-quality, interpretable and useful?*

Specifically on the indicators, the following question arose: "what is the need or usefulness of differentiating an indicator of net emissions of greenhouse gases, and those of intensity?" It is considered that since there are different production systems throughout the region (with countries with high production intensity and others that are dispersed in small areas), the intensity indicator is one of the most relevant.
The indicators are not as rigorous as they would have liked.
Concern that selected indicators are not highly correlated.
Ease/pragmatism and follow-up skills are needed.

*What are the data gaps that you can identify? Are these gaps structural? If not, are there some data or indicators that you know are available in your region and can be used for the purpose of this monitoring and assessment system?*

- About the gaps: regarding land issues, there are two key indicators. One indicator is land degradation (which can be formulated as the % of land in relation to the total area), and this indicator can be divided into 3 indicators, which are those used in the convention to Combat Desertification and in SDG15: Vegetal Cover, productivity of the land, and soil organic carbon. That is essential, given the degradation of the land.
- In terms of gaps related to water: water stress (hydric stress), water availability (surface and groundwater), and water pollution. There is concern over soil contamination, but not about water.
- In the initial presentation, solid waste was discussed, but that was not being visualized in the dimension of contamination. On the issue of food waste, an indicator on the generation of this type of organic waste could be considered, which in any case is already contemplated in goal 12.3 of the SDGs, and some countries are already beginning to work on it. On the other hand, the entry of organic waste of this type to a sanitary landfill could be considered.
- In terms of pollution, as is the use of pesticides per unit of land, it is suggested to incorporate an indicator of the total bio-inputs on the land -natural, organic- (to think positively).
- It was asked why food loss and waste was not included in the indicators.
- There is concern regarding the Deforestation issue (It is not explicitly included in the proposed indicators, and it is one of the key elements of the expansion of the agricultural frontier in various countries, and especially in the Amazon region). Deforestation is reflected in 2 parts of the proposed indicators: 1) expansion of the cultivated area; 2) In relation to the issue of greenhouse gases, it is suggested to verify if once deforestation has been carried out in an area to expand crops, that remains in the proposal.
- Concern regarding the indicator "Nitrogen management index". It is much more interesting to include an index of sustainable carbon management in the soil, since nitrogen is part of organic matter.
- It is suggested to include an indicator related to agroecology, which is very important for production.
- On the subject of climate variability, it is suggested to include indicators regarding Climate Change and production: changes in rainfall patterns, changes in seasonality (there are areas where the dry season was short and now it has extended), intensity of drought, to name a few.
- Functional integrity within food systems can go further. Biologists and conservationist work on landscape fragmentation indexes, which are possible to calculate with secondary information, and give a much better estimate than just the % of cover.
- Diversity of ecosystems should also be included.
- There are no indicators of the increase in pests and diseases as a result of climate variability that leads to an increase in the use of pesticides.
- It would be good to include an indicator on Circular Bioproduction in relation to proposing alternatives for sustainable agriculture.

*What are the regional aspects we need to take into consideration at the moment of selecting a set of indicators to monitor the state of food systems and its evolution?*

- Regarding the data needs in the region, there are biodiversity characteristics that must be considered in the construction of the indicators. Climate variability and heterogeneity and their impact on agriculture were also mentioned.
- One participant points out that small farmers are not prepared for the weather effects that occur in the region. They do not have a contingency plan to prevent these effects. In the Dominican Republic, for example, a donation plan is beginning according to weather stations, to give them the appropriate support. The importance of considering the issue of vulnerability due to the CC effect within the indicators is highlighted.
- About carbon, a worry is that its behavior is very variable depending on where it is. In Costa Rica, for example it is very difficult to see increases because the organic matter materializes a lot due to the tropical soils.
- On agroecology, there is a recent proposal that emerged at the Food Systems Forum. It is suggested to take into account what experts on this subject recommended there.
- When there are extreme events, food security disappears. It is worrying that the proposed indicators focus on FF emissions, and not on the production systems of the countries that emit the most.
- On issues of biodiversity and ecosystem integrity, there is a comment that has to do with the % of natural area indicator: there are ways to measure the integrity of landscapes, measuring the patches, the distances between patches, and this can give a better vision of the areas that are involved in productive sectors. Another issue is the security of these areas. It is not enough for the areas to exist, but for there to be certainty about it, a management agreement with the communities, for example, is needed.

*Others:*

- The need to see if there are indicators of the total or partial use of agricultural inputs (fertilizers, pesticides) arises.
- The management of fruit trees stands out as a good indicator; that is directly related to peatlands, and these are directly related to CC mitigation.
- It is marked that FAO has a new methodology developed together on the Multidimensional Assessment of Agroecological Systems, based on 10 indicators. It is suggested to analyze and include them.
- There is a concern, which is the efficient use of water, or re-use of water. There are areas of our country with a lot of water deficiency. Using water in several productive units can be very good (for example, the experience with fish farming and vegetable production.)
- There is a voluntary code for the issue of food loss and waste, and it is pointed out that some elements of monitoring could be considered beyond strictly the FLI or FWI of table 12.3 of the SDGs. Some more qualitative elements may emerge, but they could add up in terms of initiatives or preferred actions as an indicator https://www.fao.org/3/nf393en/nf393en.pdf
- Some key indicators on water issues: water availability, rainwater changes, groundwater availability, water pollution, water stress.
- There is a new framework developed by the United Nations Statistics Division: link.

3. Livelihoods, poverty and equity Domain

From the nature and regional reality, some definitional issues are raised in relation to the role of family farming, the role of food systems. The contribution that agriculture makes to the economy is also mentioned; issues of inflation, GDP, expanded agricultural participation are discussed.

*Do you consider the proposed indicators relevant, high-quality, interpretable and useful?*

- The distinction between more or fewer indicators arises, with the risk that if there are more indicators, they may be contradictory to each other. The idea of a brief index or of forming clusters arises. On the other hand, assess whether there are indicators that are more relevant than others.
- The availability of information, censuses, statistical institutes, incorporate studies and data from NGOs, academic institutions, and other United Nations organizations were discussed.
- The dilemma about what can and cannot be done in terms of measurements and available information emerges.
- It is pointed out that it is better to use simple indicators than indices (indices become complex and feel unclear).
- It is stated that including many indicators at all levels (impact, results, outputs and inputs; according to theory of change), several can cancel each other out. The cluster exercise is key.
- In the same way, it is not recommended to mix so many different levels of indicators: there are final objectives, intermediate objectives, immediate results and inputs, and, in addition, policies (it could be very confusing, and cannot be distinguished where the growth, the production, the trade is).
- It is suggested to know what the expenses for agriculture and expenses for food systems are.
- It is recommended to reduce the number of indicators. A cluster analysis should be considered.
- The existence of external macroeconomic factors is mentioned. It would be a mistake to associate poverty with agriculture or with agri-food systems that are still developing.
- The question arises: "How is the issue of double causality going to be addressed?" (Lower levels of poverty contribute to further development of FSS, and more developed FSS also contribute to poverty alleviation).
- It is noted that the characterization of agri-food systems (those that are less developed, with less added value, etc.) was not contemplated.
- The indicator "Recognition in the Constitution of the Right to Food" could be included in the group of indicators for the topic of Governance.
- A dialogue was developed around including the indicator of participation of agricultural GDP in GDP. This could generate incorrect interpretations, because participation falls with the level of development of the countries and is sometimes associated with less importance of the agricultural sector, and higher participation is associated with higher levels of poverty.
- Based on the above, the question arises as to what indicator would be proposed and what it would include. "If it were included (with the problem that data is not available for all countries), the indicator should be the share of extended agriculture (agriculture + sectors linked to agribusiness, agribusiness and food). The traditional agricultural GDP only captures production at farm level."
- Interpretability criteria is not clear in terms of direction and context on the indicator "households of significant income from agriculture."
- Another issue is the distribution of land ownership by sects (or sectors?). Currently, the definition of the indicator does not represent all land holdings (only agricultural land). Therefore, it is noted that the title should be adjusted. In addition, for the data sources for this indicator, there is more than just the agricultural census because agricultural surveys also collect information on this, and in some cases, household surveys with agricultural modules collect data on this, too; there, data sources could be expanded. For the coverage of school feeding programs, there is no clarity on the coverage of this indicator in the region.

*What are the data gaps that you can identify? Are these gaps structural? If not, are there some data or indicators that you know are available in your region and can be used for the purpose of this monitoring and assessment system?*

- It is indicated that there is nothing about the contribution of food systems to economic growth, and that is the main "trade off" that can be carried out for transformation policies and processes. “We need to know if the proposed political trends are going to have an impact on economic growth. Employment is measured, but that is not direct enough.”
- There should be indicators on the budget allocation that countries have, specifically for the issue of sustainable agri-food systems. “You have to look at the other groups of indicators, as well as the government budget allocations. See how to link government development programs and public policies to country roadmaps.”
- In relation to the income indicator, the direction is not so clear; only agricultural ownership is considered. Agricultural censuses should be considered, and data sources expanded.
- Regarding data gaps, it was proposed to include the contribution of agriculture to GDP as one of the indicators. Also, to include some productivity indicator, to see if the agricultural sector is becoming more productive.
- The agricultural orientation index is available for all countries using national accounts’ data, and this would be a good indication for how much of government expenditures are going to agricultural programs.
- Lack of explicit indicators on access and appropriation of technology. This access must be generated in an inclusive manner, so that there is no greater gap between rural inhabitants.
- It is proposed to create a funnel system and to have sets to order and find out which the most common indicators in the countries are and which could have the desired comparability. Making a chain of results to organize and have the indicators that are most relevant is also mentioned.
- “It would be very important for the FAO to have economic, productive, and social indicators for all the countries. As an investment for agriculture, a budget for agricultural innovation, nutrition, per capita consumption, etc.”
- It is proposed to delve into the gender gap within the indicators: “especially how employment is defined within rural women; many statistics do not consider all the activities carried out by them as ‘employment’, and they are in the inactive category. In Colombia, rural women are the ones who work the most hours without any income and when it is paid, the gender gaps are larger than in urban areas.”
- It is suggested to incorporate ECLAC indicators of poverty and extreme poverty, which is calculated for rural and urban areas, and the gaps could be observed.
- It is suggested to build specific indicators for the group that account for the sensitivities of the different households.
- The issue of income has a lot to do with distribution within the agri-food chain; for example, rural and agricultural producers are being affected by the prices of production. One could try to see this problem within the indicators: how is value distributed within the food chain? Try to value, for example, the natural resources within each of the chains. How to deal with exports? Exporting ecosystem services; water is being exported, natural resources that in many cases are not going to be renewed.

*What are the regional aspects we need to take into consideration at the moment of selecting a set of indicators to monitor the state of food systems and its evolution?*

- The needs of the region must be taken into account, highlighting that family farming is not included in the indicators presented. "From the Dominican Republic it is mandatory that, when talking about food systems, family farming is considered."
- Family farming: the idea that there is nothing about the contribution of food systems to economic growth is reinforced, and this is the main challenge for many policies and transformation processes. “We need to be able to know if certain political trends are going to have a certain impact on economic growth. Employment is being measured, but that is not enough. It should include the economic consequences.”

- It is pointed out that within the measurement of the indicators that are being proposed, there is an invisibility of subsistence family farming in rural areas. They are part of the negative effect of this system, since they are the most malnourished and those with the least possibility of production and the most difficulty in obtaining accessibility. Difficulties arise when using indicators that are not linked to the issue of poverty. The importance of making visible what governments are doing to include families in these areas is highlighted.
- Food security for the migrant and refugee population is considered an immediate issue. “Many of these people are excluded from formal systems and are also often reluctant to participate in surveys because of fear of persecution/prosecution in places where they do not have adequate social protections or permissions to work legally. Given their lack of formal inclusion, is there a framework in place to ensure their conditions are considered? From our perspective, food security is of paramount concern for this community, especially facing rising inflation and recessionary economic conditions.”
- Information on urban population should be incorporated. There is a fairly strong bias towards agriculture, “food systems are beyond agriculture, and an important part of the economy is in the postproduction of food systems in many Latin American countries.”
- How to capture the income, jobs, employment conditions in the part of retail, wholesale, formal and informal trade, and exports? “As a region, a lot of food comes out, how will that be considered?”
- "In countries like Colombia, the countryside is becoming uninhabited and that puts food supply at high risk at the internal level." There is an issue of generational change and the legal occupation of the territory. There are people who are being displaced from their natural habitat.

*Others:*

- One participant suggests that in the specific case of poverty, equity and life systems, development indicators go beyond food. As an example, a classic indicator of development is "infant mortality rates, which is not directly related to food systems.” It is suggested to learn from UNDP. However, it is known that there is incomplete data for many countries (for example, multidimensional poverty, many countries do not report it). National statistical systems are very weak and incomplete, and that is the reason why non-governmental organizations are emerging. It is recommended to see to what extent there is a possibility, in some countries, of taking reliable indicators that are constantly measured by NGOs as an alternative.
- It is talked about the topic of crime in the rural sector: “sectors in rural development, stressed by climate, economic problems, etc. This ends up generating migration, which creates empty spaces and that's where crime comes in, which, in return, feeds back the problems of stress and migration. This does not only happen in LAC, but also in Africa, and if we are talking about livelihoods, one of the most serious problems they have is the lack of crime and violence control systems in rural sectors.”

4. Governance Domain

*Do you consider the proposed indicators relevant, high-quality, interpretable and useful?*

The indicators are not evidently clear to the group.
A participant points out a reflection: from a conceptual point of view, the indicators seem fine. The issue is that it is logically ordered, but it is not very operational; it does not refer specifically to the agri-food sector. In addition, the indicators are based on people’s opinion; there is no hard data. Therefore, there is no comparison. So, it is important to generate indicators that can be used as hard data. One solution to this can be to think about how to formulate the question; for example, a specific question can be: did you or someone you know have to pay a bribe to the police? It is a yes-or-no answer. This type of question could be hard data. What would be useful then? An institutional analysis of the entire system that is around the agri-food system, which does not exist right now, because they are all separated into different ministries. It is suggested to consider, when using these indicators, that this is a sensitive issue for governments, especially when there is no hard data, because all the indicators on the institutional level at the international level are perceptions. It is important to build the most objective data possible.

The importance of focusing on the institutional framework in general is highlighted (what the coordination mechanisms are, the plans and the budget). A vision of budget analysis with a food system perspective is lacking, and at the other level of effective implementation, it is necessary to check if what has to be done is being done. Perhaps it would be necessary to go a little further and review the institutional system, which is what is needed to give long-term sustainability to the desired changes in agri-food systems. It is suggested to review the intersectoral plan in the countries, clearly indicating what each institution contributes and does and how this is translated into an institutional plan.

Finally, there are doubts about the interpretability of the indicators, about whether it will be a faithful reflection of situations with better governance for the transformation of agri-food systems.

*What are the data gaps that you can identify? Are these gaps structural? If not, are there some data or indicators that you know are available in your region and can be used for the purpose of this monitoring and assessment system?*

- It is proposed to pay attention to consultation mechanisms for consumers, producers and family farmers, and also for vulnerable groups that go unnoticed (native peoples). Many countries have many mechanisms, but they are not effective, they do not translate into dialogue processes, or coordination between actions of public sectors themselves. This should be taken into account.
- It is recommended to accompany the indicators with an institutional analysis (graph what is the institutionality that exists in the countries for the government of agri-food systems).
- It is desirable to incorporate public spending variables and an analysis of effective policy implementation regulations.
- The lack of data for several countries in the region and the periodicity of this data collection (regarding the inclusion of indigenous peoples or consumers, there is not enough international data to have them as indicators) is pointed out as an important gap.
- It is pointed out that governance indicators are a challenge for many countries in the region, since the institutions and laws are not always applied, for example, Venezuela.
- It is indicated that the region has a problem at the level of dialogue in agri-food systems and chains, and there are no laws that stimulate dialogue between the actors of the system and the agri-food value chains.

*What are the regional aspects we need to take into consideration at the moment of selecting a set of indicators to monitor the state of food systems and its evolution?*

- Regional peculiarities were identified: corruption, informality, limited capacity of public sectors to implement decisions, and policies that are made.
- A participant indicates that in the region, the generation of dominant actors is usually sought (in all institutional relations and among the various actors of the society itself). In practice, these actors subordinate the true peasant and productive representation; this creates an environment where small producers are no longer heard. Therefore, it is suggested to plan actions so that this power does not end up submerging those small producers, who are the ones who generate the production that guarantees the food consumed in the country.
- It is suggested to consider that the existing indicators generate controversy.

*Others:*

- A suggestion is made: the indicators should be comparable across countries and over time, but that does not mean that discrete indicators are no longer important. Proxies must be used. For example, regarding the budget issue, the countries have defined agricultural spending for their countries; this expense is a % of the budget, of the GDP. This can be important information, hard data. There is also information on agricultural financing and financing for small producers that can be measured annually -this is to mention some indicators that are measurable over time.
- It is suggested to design an indicator that measures the relevance and functionality of the institutional offer.
- It is also suggested to have data on the sustainability of projects implemented by governments ("when a project is installed, after the management is over, the project is lost; and this is worrying"). It is recommended that one of the indicators focus on government effectiveness. Related to this, it is also proposed to design an indicator that measures the relevance and functionality of the institutional offer.
- It is pointed out that one of the missing indicators for the region is the issue of corruption, and the voice of consumers and users when it comes to accessing quality food, because when implementing the laws, it is important to measure consumer satisfaction. Generally, international NGOs work to improve production, with different tools, but it is not seen until the end of the process (once the customer buys the product, knowing if there are problems, where to complain, to whom they can express their satisfaction or dissatisfaction).
- It is mentioned that, at this time, the countries compile the information based on the thematic classification of the monetary fund in which agriculture appears, but not what is being spent on agri-food systems in general. It is suggested that countries have satellite accounts based on this approach, which will allow for something to define what is important. It is also pointed out that the same thing happens with the regulations (for example, the USA has the code of federal regulations, where is possible to see all the federal regulations that exist, and this would be very useful for the agri-food sector).
- In another topic, it is pointed out that part of the problem of non-effectiveness is the non-continuity of operations and the fragmentation of operations. It is highlighted that there are many projects that cover very few people, leave large gaps or often compete for the same beneficiaries, but since no one does a global mapping, there is duplication and little coverage. In addition to that, there is discontinuity; projects are often completed because there are contracts with entities that are later rescinded. One proposal that is presented is that, in addition to the indicators mentioned, more institutional data and more qualitative data with identification recommendations be generated, with alternatives (satellite accounts, for example). With this there will be no comparative tables between countries, but it will be possible to have a vision of their governance schemes and the impacts on agri-food systems.
- A participant wonders if, taking into account that the article talks about systems theory, there is an interrelation between the components of the system.

<u>5. Resilience and sustainability Domain</u>

*Do you consider the proposed indicators relevant, high-quality, interpretable and useful?*

- It is pointed out that something important to begin with is to define what is meant by resilience within agri-food systems. That is a starting point to understand the proposed indicators.
- It is mentioned that the definition of food system resilience is the capacity of the different individual and institutional actors of the food system to maintain, protect or quickly recover the key functions of that system despite the impacts of disturbances. Link.

- One participant comment that resilience is easy to describe and pronounce, but difficult to measure; its interpretation is highly variable.
- A participant asks about the methodology: having indicators at different levels, how is this synthesis exercise going to be carried out? Are you thinking about cluster analysis? When there are many indicators, they can cancel each other out.
- There is concern regarding the indicators. In the case of Colombia, the armed conflict and migration determine the transformation of agri-food systems.
- The concern that the concepts do not overlap (resilience capacities and resilience strategies) is presented and repeated. Information is lost when it is not easily understood. It is necessary to avoid redundancy.
- It is recognized that the major limitation to this process is the availability of data for most countries.
- One participant note that he agrees with the text's definition: “resilience is the capacity to ensure that stressors and disturbances do not have lasting adverse consequences for development”.

*What are the data gaps that you can identify? Are these gaps structural? If not, are there some data or indicators that you know are available in your region and can be used for the purpose of this monitoring and assessment system?*

- It is recognized that indicators referring to food agrodiversity, produced and consumed, are taken into account, but the importance of talking about food biodiversity produced at the country level is recalled, because there are countries that depend a lot on imports, but little is said about efforts made by the country for biodiversity.
- Given that the Sustainable Food Systems Program exists, and that these consultations are taking place in other regions of the world, it is suggested that the co-leaders of the Program be included in this invitation.
- The conceptual proposal is recognized as good, but some gaps are identified: 1) despite how complicated it is to define resilience, whenever you want to measure something, it is essential to define and agree on what is meant by each concept, 2) there is no consideration of economic shocks (current context of Eastern Europe, which has involved export restrictions). It is indicated that measures are now being tested to interrupt the normal flow of trade, which has a severe impact on the agri-food system. The idea of diverse agri-food systems, which is related to local trade versus exports, is valued, but it is suggested to add indicators that measure a) diversification of consumption, b) diversification of production (diversifying more will help resilience), and c) trade diversification.
- Along with the volatility of domestic food prices, there could be another similar indicator for inputs (fertilizers, insecticides, labor, etc.).
- It is highlighted that the volatility of food prices is an aspect of affordability, but also of income, savings, etc.
- The importance of considering the climatic shocks that are often experienced in this particular region of the Central American Caribbean is highlighted.
- For purposes of resilience in the purchasing economic sphere, and for decision-making, the importance of considering indicators that exist in El Salvador is highlighted: Indicator of the purchasing power of wages (if purchasing power drops, wages must be adjusted, so that the poorest families have greater purchasing power). This indicator is not visible in the proposal.
- The indicator of the degree of dependence on imports of basic grains is not observed. It is emphasized that it is important to know which foods are depended on... Another indicator of foreign trade is the % of total exports dedicated to importing food for the basic basket, which was not on the list either.
- There are gaps in terms of contamination, it is important to define it much better. The question is: What are the practices that are based on nature that are going to be identified as useful for the sustainability and resilience of agri-food systems? How the implementation of topics such as agroecology could help the nitrogen concentration of certain cultivars? The idea would be to make a filter thinking about the ecosystem services that these solutions provide.

- About diversity: the importance of identifying which are non-native or non-local species is highlighted. Introduced species can help a territory economically, but their impact can be significant, killing some local species. It is key that this biodiversity indicator can be adjusted taking into account the non-promotion of the introduction of non-native species.
- It is important to have an indicator to know when an economy goes from one equilibrium to another, an indicator that shows: if "x" amount of food was produced, and a hurricane comes, "x" less is produced, showing statistical changes relative to an initial reference point. In this way, the concept of resilience could be better understood.
- It is commented that **resilience is not only individual, but collective,** and there is no indicator that shows this: the formation of new organizations, for example.
- The following question arises: what will be the weighting of each of the indicators, which together add up to resilience?
- A participant states that he does not see indicators of resilience. He understands that resilience is a product of native biodiversity. He works in Peru, where it is important to work with Amazonian fish. They do not depend on imports, but on the technology that has been developed in the area. He suggests incorporating an indicator that reflects the use of local species (fish, plants or animals).
- On the gaps: measure of dependency or integration with global markets. It may be affected if there is an international introduction.
- There is an issue that has to do with the volume of food that does not go through informal commercial relations. For some populations it is highly significant to have more informal exchanges: for example, the family that sends cheese or beans. The following question arises: "To what extent are we clear about how important this form of provisioning is? And if it is important, how can this be captured?"

*What are the regional aspects we need to take into consideration at the moment of selecting a set of indicators to monitor the state of food systems and its evolution?*

- From the regional point of view, there are issues that are inherent in the Caribbean zone, which have to do with vulnerability to dependency on food imports and inputs for agriculture and, at the same time, vulnerability to climatic shocks (hurricanes).
- Dependence on food and imports is an issue that concerns the entire region.
- A mention is made about the SICA area: it is a large staple food market that should be promoted. The resilience of the region lies here. It is recommended to consider all this. In addition, it was pointed out that agricultural variability and livestock activities, which sustain the countries' food, should also be considered.
- It is pointed out that it is important to consider the definition of food sustainability from the nutritional point of view. How, from the region, is it being produced healthily? (With fewer transgenics and reduction of agrochemicals).

*Others:*

- Discussion around the purchasing power of money: to what extent domestic and international shocks have a local impact through the economic part, purchasing power and the ability to ensure the basic food basket.
- It is commented that in Peru there is a great commitment to the indicators, and to the SDGs. However, the concern arises as to how to ground the theoretical framework to more specific tools. It is requested to illustrate with practical examples (Excel sheets, etc.), and exemplify the sources of information, in order to comply better.

- The INTA (Nicaragua), with the Agrarian University and the Movement of Agroecological Producers developed a resilience assessment methodology, based on the Stoplight method, which could be shared. The Methodology is called REDAGRES.
- It is recommended to have up-to-date and quality data to identify and address the problems faced by food systems, such as distortions at the level of production, the marketing chain and consumption.
- The question is raised: Are the policies or market and regulatory incentives that countries have to encourage the adoption of environmentally sustainable practices that are profitable at the same time contemplated?

**sAgenda of meeting**

**Agenda:**

| | | |
|---|---|---|
| **10.00-10.05** | Opening | *SDG and RR of RLC Julio Berdegué.* |
| **10.05-11.15** | Introduction to the Initiative | *José Rosero, Statistics Division (ESS) Director.* |
| **11.15-12.15** | Work sessions in 3 breakout rooms:<br>1. Diet, Nutrition, and Health Domain.<br>2. Environment and climate Domain.<br>3. Livelihoods, poverty and equity Domain. | *1. Nancy Aburto.*<br>*2. Mario Herrero.*<br>*3. Rashid Sumaila.* |
| **12.15-12.30** | ***Break*** | |
| **12.30-13.30** | Work sessions in 2 breakout rooms:<br>1. Governance Domain.<br>2. Resilience and sustainability Domain. | *1. Simón Barquera.*<br>*2. Ty Beal.* |
| **13.30-13.45** | Closing and next steps. | *José Rosero.* |

**FAO RNE Regional Expert Consultation of the Food System Countdown Initiative's Indicator Framework**
**Virtual, May 18 2022**

*This report was produced by the FAO Regional Office for the Near East and North Africa to summarize the results of the FAO RNE regional expert consultation held on* May *18* 2022.

## Introduction

Food systems play a role in meeting all 17 sustainable development goals (SDGs). With less than a decade to achieve the SDGs, the global community faces a critical juncture to transform food systems to be healthier, safer, more sustainable, more efficient, and more equitable. Lately, the UN Food Systems Summit has focused global attention on food systems and set the stage for food system transformation. Country and independent dialogues catalyzed the development of shared visions for food systems that apply to different contexts and geographies.

It is widely recognized that to enhance all aspects of food systems and their interactions, a clear, rigorous, and comprehensive set of metrics and indicators are required to guide decision-makers and to hold them accountable. However, no rigorous mechanism currently exists to track the state of food systems, their change, and performance over time. In fact, the commitments reached at the UN Food Systems Summit, and the realization of the visions reflected in the national food systems pathways need metrics to guide decisions and track progress. At the same time, food system actors and stakeholders (e.g., civil society, governments, and international organizations) require trustworthy, science-based metrics and assessment.

*Food Systems Countdown Initiative*

With the ultimate objective to fill this gap, the Food Systems Countdown Initiative ("the Initiative") was formed in 2021 as a comprehensive, independent, inclusive, science-based mechanism to provide actionable evidence to track progress, guide decision-makers, and inform transformation. At the same time, it intends to complement other monitoring mechanisms and the tracking of related goals at global and regional scales (i.e., SDG agenda CAADP).

To implement the Initiative an unparalleled partnership and collaboration have been put together, led by FAO, GAIN, and John Hopkins University and with the participation of more than 50 scientists from nearly 30 academic institutions, non-governmental organizations, and UN agencies from nearly all continents.

*Independent tracking and assessment system*

The main goal of the Initiative is to provide an independent tracking and assessment system based on a high-quality, curated, parsimonious set of indicators that cover all important aspects of food systems and measure food system performance. The Initiative/FSCI has designed an architecture for such a system from a multidisciplinary point of view and is moving towards implementation.

The Initiative expects to deliver an annual assessment of the state of global food systems and their transformation, published in a peer-reviewed scientific paper. It is also envisioned that policy briefs will be delivered in parallel for a broader audience and to facilitate transformative action.
The first milestone of the Initiative was the publication of the initially proposed architecture of the system and the description of an inclusive process to move from the concept to its execution.[9] The architecture covers the five thematic areas of diet, nutrition and health; environment and climate; livelihoods, poverty and equity; governance and resilience and sustainability.

[9] Fanzo et al (2021).

As a second step, the Initiative will aim to deliver the set of indicators in each of the five thematic areas above and to deliver a first assessment of the state of global food systems that will serve as a baseline for monitoring progress and performance.

The proposed tracking and assessment system are an independent effort which do not represent any obligation or commitment for reporting from countries.

*Objectives of the RNE Regional Expert Consultation*
The Initiative is committed to an inclusive, consultative, and transparent process that will allow for validation and peer review of the set of indicators that will be used for the assessments. Two sets of consultations have taken place. The first was a consultation with expert scientists. The second, a series of regional expert consultations across the FAO regions, including this Regional Expert Consultation for the Near East and North Africa Region.

The Regional Expert Consultation for the Near East and North Africa Region brought an expert point of view from policymakers and policy-adjacent users of data, on the relevance, usefulness, and validity of the proposed set of indicators from a regional perspective. The consultation covered the proposed indicators in each of the five thematic areas.

It provided an opportunity to get inputs, comments, and suggestions on the monitoring framework proposed by the initiative. The framework is voluntary, however, the consultations will ensure that it has the capacity to be a useful tool for policy decision-making processes. The results will be used for the first assessment of the state of global food systems and later for tracking progress and assessing performance.

The FAO RNE Regional Expert Consultation of the Food Systems Countdown Initiative's Indicator Framework was held virtually on May 18 2022. It brought together representatives from governments in the Near East and North Africa region, as well as individual experts in food systems from the FAO network in countries. Participants were drawn from senior staff engaged in policy development from the relevant Planning Units of the Ministries of Agriculture, National Statistical Offices and Ministries working on the thematic areas of the indicator framework. As well as senior experts from international, regional and national agencies in the region from across the five thematic areas.

**Opening**
The opening remarks by FAO Regional Office Assistant Director General (ADG-RNE), Mr. Abdulhakim Elwaer, highlighted the importance placed on monitoring of food systems at the UN Food Systems Summit held in New York in September 2021. He highlighted the importance of food systems transformation in achieving sustainable development and the SDGs. It was underlined that transformation is needed across four key areas of food systems transformation highlighted from the Food Systems Summit for the which the Near East and North Africa Region. Namely, hunger, food security and nutrition; nature based solutions; advancing equitable livelihoods, decent work and empowered communities; and building resilience to vulnerabilities, shocks and stresses. He highlighted the work FAO is doing across the region linked to food systems transformation and reinforced that success requires enhanced national capacity building, investment, science based solutions; and strong governance along the food supply chain and regional cooperation and integration. The importance of the Food Systems Countdown Initiative framework and monitoring system in leading the way forward to track progress, guide decision-makers, and inform transformation was stressed and welcomed.

**Presentation of Themes**
The themes were introduced by members of the Initiative.

Session 1 on Diet, Nutrition and Health Domain was presented by Stella Nordhagen. In this theme people are at the center, and it discusses several indicators related to diet, nutrition and health by looking at the various areas which measure quality of diet, including nutrient adequacy and dietary risk factors for NCDs. As well as measuring the other factors affecting diet quality, namely the accessibility to healthy diets (food security); and a focus on on the food environment (availability, affordability, messaging, and food & vendor properties). Finally, examining whether policies contribute positively or negatively towards food availability, food access, and product properties (policies affecting food environments).

Session 2 on Environment and climate Domain was presented by Rosaline Remans. This theme focused on the relation between food systems and the environment. The indicators focused on the main environmental systems and processes which interact with food systems: land use, climate, water use, biosphere integrity, and pollution (e.g., biogeochemical flows/novel entities). The indicators cover components and processes providing essential environmental services for the environment and humanity. As well as focusing on the areas where food systems can achieve the necessary change.

Session 3 on Livelihoods, poverty and equity Domain was presented by Alejandro Guarin. The indicators for this theme monitor the transformation created by food systems for the numerous people who work as part of the food system, in rural and urban areas, and in high and low-income countries. It focused on the areas of poverty and income, employment, social protection and rights. Cross-cutting issues were also pointed out: governance and resilience and sustainability.

Session 4 on Governance was presented by Sheryl L. Hendriks. The Governance Theme looks at how governance of the food systems domain can foster alignment and coherence across different food system actors, their activities, and progress toward results. It aims to monitor the shared vision of the outcomes, the relevant policy instruments to align efforts, the implementation of resources, and accountability for the outcomes.
Session 5 on Resilience and sustainability was presented by Paulina Bizzotto Molina. The theme of Resilience and sustainability was presented as a transversal aspect of the transformation of agri-food systems with a multi-dimensional interpretation. It covered the domains of exposure to shocks, resilience capacities, agrobiodiversity, food security stability, food system sustainability index. Resilience and sustainability are critical for food and nutritional security. They are also critical for other functions such as being a precondition for sustainability, and critical to tackling poverty and livelihoods issues.

**Discussion**

**Methodology**

Discussions took place in plenary, under five sessions covering the themes of the indicator framework, namely: Session 1: Diets, Nutrition and Health; Session 2: Environment and climate Domain; Session 3: Livelihoods, poverty and equity Domain; Session 4: Governance Domain; and Session 5: Resilience and sustainability

In each session, the discussion was motivated by a presentation on the proposed set of indicators. The discussions focused on the capacity of the proposed indicators to guide policy decisions and promote accountability mechanisms. In this sense, the focus is on the relevance, usefulness, and validity of the proposed set of indicators from a regional perspective.

This was followed by discussions based on the three guiding questions listed below.

Do you consider the proposed indicators **(this question examines the indicators suggested by the Initiative):**

*Relevant*, defined as their ability to measure something meaningful for food systems across a variety of settings, during relevant time periods?
*High quality,* defined as using the best and most rigorous statistical methodologies and data available?
*Interpretable* defined as having the ability to show a clear desirable direction of change, comparable across time and space, and easily communicated.
*Useful*, defined as its ability to be used for policy and decision-making processes and by meeting actual information needs.

What are the data gaps that you can identify? **(this question examines data issues).** Are these gaps structural? If not, are there some data or indicators that you know are available in your region and can be used for the purpose of this monitoring and assessment system?

What are the regional aspects we need to take into consideration at the moment of selecting a set of indicators to monitor the state of food systems and its evolution? **(this question identifies any new indicators needed to cover relevant issues)**

**Discussion points relevant to all Domains**
This section discusses points which are cross-cutting across all Domains.

The indicator sets are presented in silos. However, it was noted that there are links between Domains which are not reflected. Particularly for Diets, Nutrition and Health; Livelihoods, Poverty and Equity; and also for Environment. The details are discussed under each Domain.

It was also highlighted, that for several issues relevant to the region, disaggregation of indicators is needed: by gender, urban/rural location, and youth. In addition, sub-national reporting of internally displaced persons (IDPs) and refugees is needed as these groups are of particular importance to food systems monitoring in the region.

Participants emphasized the lack of data availability in the region. In particular, the region lacks recent data for many of the indicators included in the catalogue. Even where data is available, this is often out of date or lacks a complete time series. Which makes monitoring and examining trends problematic. Dependence on global data may not be very useful as they do not capture country/ regional specificities.

Participants raised concerns about the cost of such data collection as well as the local capacity to do data collection/ analysis/ reporting. Despite numerous capacity building initiatives held in the region, data collection remains an institutional as well as a political problem. These problems need to be addressed to facilitate availability of data from this region. Capacity building alone is not sufficient to ensure data availability.

Key issues which are important to the region, and cut across all domains were water, trade, food import dependency, conflict and refugees/IDPs. These were not well reflected in the set of indicators.
It was also noted that indicators of agricultural production and productivity were not explicitly mentioned in the indicator set, despite being a core element of food systems transformation. Such indicators should be included as it is important from the point of view of availability and affordability of nutritious food at national level in the context of high import dependency.

## 4.1 Diets and nutrition

***Do you consider the proposed indicators relevant, high quality, interpretable and useful?***
Overall, indicators selected for the theme are relevant and they are of high quality depending on the source they are coming from. However, in some cases data interpretation may be difficult. It will be important to think about the ways of communicating the indicators to non-experts on the theme, as some of these indicators are not easy to interpret. For example, the indicator on egg consumption is not very clear – e.g. consumption of how many eggs is considered to be good for (or detrimental to) health.

Participants have raised a number of other questions such as the number of indicators included in the catalogue; the ways of selecting a limited number of indicators to convey clear messages to policy makers; and how to construct a clear dashboard with fewer indicators which clearly indicates those which need improvement.

The issue of the baseline information for monitoring was mentioned. Participants suggested to use internationally sourced data for the baseline as national data might not always be available. The question was raised whether countries would be ready to share national data for reporting.

Participants have inquired whether the countries would need to collect these data. They have expressed their concerns about the cost of such data collection as well as the local capacity to do data collection/ analysis/ reporting.

There was a question regarding a geographic level of reporting on the indicators – whether this should be done at global, national and/or regional levels. The selection of indicators should ensure that they are useful for national and regional policy making. Sub-national data is of particular importance for the countries in conflict as national data may not cover the issues of internally displaced persons (IDPs) and refugees.

Participants pointed out that the proposed indicators were biased towards high-income countries, where more data was available compared to low-income countries. The data situation is also different across countries of the region.

Specific Recommendations

*Indicators must capture regional specificities*: sub-national and disaggregated indicators should be included to capture specific issues related to the region such as the situation of refugees and IDPs. Not only are these important for national policy making, but disaggregated data would also be useful for FAO-RNE programming.

*Number of indicators should be limited*: this will be helpful for better understanding and interpretation by policymakers and non-experts. It is also important to flag the indicators that need to be addressed urgently, which is crucial for policy guidance.

*Developing a data manual is important*: the manual will document the details of data requirements for the monitoring framework, definitions of proxy indicators used, methods of data computation, and a guide on the interpretation of the results and relevance to policies.

***What are the data gaps that you can identify? Are these gaps structural? If not, are there some data or indicators that you know are available in your region and can be used for the purpose of this monitoring and assessment system?***

Participants emphasized the data availability issue in the region. In particular, the region lacks up to date data for many of the indicators included in the catalogue (this applies to all themes discussed during the meeting). Due to existing gaps in the data series, monitoring and examining trends of different indicators will be problematic. Dependence on global data may not be very useful as they do not capture country/ regional specificities.

There are also issues with timeliness and completeness of the data series, which are important data quality characteristics and are required for interpreting the change. For example, UNICEF has flagged the issues about the availability, quality and timeliness of data related to children's diet quality (6-23 months). In most cases, countries may collect data for the Minimum Dietary Diversity (MDD) indicator, but less so for other diet-related indicators. It also depends if you use national surveys or other surveys (e.g. SMART surveys) for this information. There are also challenges with the existence of relevant indicators around the food environment and related policies across countries.

Participants mentioned the issues related to data interpretability and comparability among different national data sources. For example, data on healthy diets is collected by different entities and organizations inside the same country and data comparability is a systemic issue. A consortium from all the concerned parties or any other form of cooperation is needed to collect data on the same theme.

Participants have pointed out that data availability problems were endemic and long term. Despite numerous capacity building initiatives held in the region, data collection remains an institutional as well as a political problem. The statistics institutions lack resources to collect data on a regular basis as well as the mandate to disseminate data. These problems need to be addressed to facilitate availability of data from this region. Capacity building alone is not sufficient to ensure data availability.

Specific Recommendations

When selecting indicators for monitoring, it is important to consider not only data availability, but also timeliness of data. Even if some data is available it is likely to be out of date. There are also many gaps in the data series so considering completeness of the time series is important.

***What are the regional aspects we need to take into consideration at the moment of selecting a set of indicators to monitor the state of food systems and its evolution?***

There is a need for indicators to capture the situation of countries affected by conflicts, considering that conflict is one of the major driving factors of food insecurity and malnutrition in the region. Namely, conflicts cause disruptions to livelihoods (loss of assets and incomes), food production (crop and livestock) and food systems, which affect affordability and accessibility of nutritious foods for healthy diets. Conflicts also affect access to clean water, which is key to ensure food safety and food utilization. It is also crucial to think about the rapid changes and implications caused by conflicts and their impact on the extent towards which data can depict the current situation.

The situation of IDPs and refugees from countries affected by conflicts, including the countries of the region that host refugees (e.g., Lebanon, Jordan) needs to be captured. National data do not always depict situation of these population groups.

Participants reemphasized throughout the discussion the region's dependence on food imports and the need to incorporate indicators of food trade into the catalogue. It is important to distinguish between imported and locally (country or region) produced food and agriculture products.

A binary indicator whether the country has any tax policy levied on foods for health reasons may not be very efficient since there might be several such policies for different food products. Currently, the catalogue of indicators does not make differentiation between having one policy or several policies.

The Economic Commission for Western Asia and Africa (ESCWA) has suggested to examine the Arab Food Security Monitoring Framework (developed jointly by ESCWA, FAO and AOAD[10]) for indicators that potentially could be included in the indicator catalogue for monitoring food systems transformation. https://www.unescwa.org/publications/tracking-food-security-arab-region.

Specific recommendations for additional indicators

Import dependency ratio was suggested to be included in the indicator catalogue. This indicator would reflect food systems transformation as well as development of the agriculture sector. This would help to understand their impacts on the costs and affordability of different food products. It will be important to highlight import dependency on strategic food products such as wheat, oils and sugar among others; monitoring of imports of unhealthy food items (highly processed food, sugar-sweetened beverages, etc.) is also needed.

Indicators of agricultural production and productivity related to specific or strategic crops and animal products relevant to food habits in the region should be included. This is important from the point of view of availability and affordability of nutritious food at national level in the context of high import dependency.

Indicators measuring the efficiency of marketing infrastructure in the countries and percentage of food loss of agriculture products (wheat, fruits and vegetables, animal products) need to be included.

A logistics indicator such as the food cooling indicator can be used to assess food quality and availability.

A number of enacted legislations/policies promoting/limiting the import of unhealthy food items would be more useful indicator rather than a binary indicator of having a policy to tax foods for health reasons.

## 4.2 Environment and climate domain

***Do you consider the proposed indicators relevant, high quality, interpretable and useful?***

Participants expressed concerns about the growing number of indicators being developed. It was suggested that there should be some consideration to reduce the number of indicators, and to select those which have a broader relevance and are applicable to multiple country and regional contexts. Some highlighted the need for indicators which are applicable across different countries. One proposal for a more compact set of indicators was for the creation of a matrix to indicate an indicator's relevance to multiple processes and states, including soil health, agricultural productivity, or consumption. Other participants highlighted the need to build capacity in countries, as well as to account for the cost of collecting data related to environment and climate.

***What are the data gaps that you can identify? Are these gaps structural? If not, are there some data or indicators that you know are available in your region and can be used for the purpose of this monitoring and assessment system?***

Concerns were raised about the complexity of measuring GHG emissions, biodiversity loss, land encroachment, and soil carbon. This challenge was noted during the preparation of the upcoming report on the *State of Land and Water Resources SOLAW in the Near East and North Africa.* Remote sensing is a good tool for monitoring and data collection, but there is a need to ensure that the results produced are triangulated with realities on the ground. In order to better support informed decision making, the methodologies used should be robust.

[10] https://www.unescwa.org/sites/default/files/pubs/pdf/manual-monitoring-food-security-arab-region-english_1.pdf

Participants highlighted that one aspect which is missing from the set of environment and climate indicators is the negative impacts of encroachment on agricultural land by the expansion of urban and industrial uses. The expansion of these uses are reducing the area available for crop production and livestock. The threat of encroachment was highlighted by representatives of at least three countries, including Egypt and Libya. In one case (Egypt), data on this indicator was available.

Regarding the indicator on pesticides, participants suggested adding an indicator on countries' uptake of biological fertilizers containing bacteria, fungi and microorganisms. Participants agreed there could be the potential also to capture the transition to sustainable agricultural practices. Furthermore, in terms of soil biodiversity, there is an indicator called 'soil biodiversity potential' which is included in a global atlas, but there are currently no plans to update this. Instead, data on changes in soil organic carbon is available. Another participant noted that soil data collection is ongoing in several countries within the context of a FAO-Technical Cooperation Project, and maps of soil properties and soil organic carbon are expected to be produced by the end of the year.

Water is critical to the region and the production of cash crops has implications on water use. Currently there is only one indicator for water. One suggestion was for an indicator on virtual water, measuring the water use through exports and imports of agricultural and other products, that may be helpful in understanding the transition to a sustainable food system in water-scarce contexts.

***What are the regional aspects we need to take into consideration at the moment of selecting a set of indicators to monitor the state of food systems and its evolution?***
Participants highlighted the links between the environment/climate and food security themes for the region. For example, the ongoing conflict in Ukraine poses a threat to global and regional food security, highlighting the vulnerability of countries and food systems to such threats.

While participants were reminded that the indicators came from already-established secondary sources and datasets, several participants drew attention to relevant indicators that already exist in countries, and encouraged the narrative developed to include also these indicators. This could assist countries in identifying gaps and investment needs.

There was a need for an evaluation of the results and lessons learned from past programmes, as well as achievements so far, to better understand the regional context for food systems monitoring.

### 4.3 Livelihoods, poverty and equity Domain

***Do you consider the proposed indicators relevant, high quality, interpretable and useful?***
The indicators on livelihoods, poverty and equality the indicators are generally relevant but some points were raised regarding their quality and interpretability.

One drawback is that several of the proposed indicators cannot be measured for the specific population that is engaged in the agrifood system, but only for the entire national population. Without this specificity, the usefulness of indicators, such as national rates of unemployment, rate of informality or the percent of the population earning low pay is limited in terms of assessing progress in agrifood system transformation. As they reflect the general population, any increase or decrease in these indicators cannot be attributed to changes within the agrifood sector.

Another key issue is that disaggregation by gender and space, are key elements of equality in the region, and are thus needed to adequately assess the theme. Gender gaps and geospatial gaps can only be reflected if the data is disaggregated by sex and rural versus urban geography. Although this data is not readily available, ignoring disaggregation would be detrimental to the quality and usefulness of the proposed indicators. Age and migrant status are two additional disaggregations that could be important when assessing equality.

In terms of interpretability, the indicators of households with significant income from agriculture and coverage of any social protection programme presented difficulties. It is not immediately clear if an increase in the share of households with significant income from agriculture would be considered a positive or negative outcome. Such a conclusion would depend on the context and drivers of this change. Moreover, off-farm activities including post-harvest handling, aggregation, processing and distribution are often key additional sources of income in rural areas that should not be ignored.

Similarly, social protection coverage could mean different things in different contexts. For example, an increase in social protection coverage could be positive if the number of those in poverty has not changed and if the programmes are adequately targeting those in need. Indicators could be added regarding the availability, efficiency and coverage of social safety nets, and the availability of official social solidarity systems that encompass dimensions related to food security such as food subsidies.

Questions were raised on the usefulness of the indicator of population earning low pay, and whether this is an adequate measure of poverty. Poverty headcount is included in the indictors under resilience and may be better placed under livelihoods. Another indicator for poverty is the Multi-dimensional Poverty Index which is also disaggregated by rural and urban for select Arab countries.

***What are the data gaps that you can identify? Are these gaps structural? If not, are there some data or indicators that you know are available in your region and can be used for the purpose of this monitoring and assessment system?***
Data on the distribution of land holdings is collected through the agriculture census in Egypt as in many countries. This census is conducted every 10 years on average meaning the data would not allow for bi-annual monitoring. To address the issue of data availability more broadly, one suggestion is to also examine other sources of data such as those generated by the private sector.

Informal employment, especially in agriculture will be difficult to measure without a specific study in each country. Most of the agriculture in the region is family agriculture with family labour that is unpaid. This indicator may not fit the rural reality in the region.

In terms of methodology, the indicators need to have clear definitions of poverty, informal employment and significant income as well as indicate clearly how national discrepancies in these definitions are overcome.

***What are the regional aspects we need to take into consideration at the moment of selecting a set of indicators to monitor the state of food systems and its evolution?***
There are very high levels of gender inequality in the NENA region which affect agrifood system development. While the indicator on female ownership of landholdings attempts to reflect gender disparities, this issue is insufficiently captured by the proposed indicators. The share of women in agriculture through paid or unpaid work, inclusion in rural institutions, exposure to gender based violence, discriminatory laws and practices are critical elements that should be included to the extent possible.

The NENA region has the highest youth unemployment rates in the world. Unemployment for young women is significantly higher than young men, reaching 40%. This concern would only be reflected if the unemployment indicator was disaggregated by sex and age. The region is also witnessing an unprecedented level of youth in the population where, in several NENA countries, more than half the population is below the age of 25. At the same time young people are less engaged in agriculture production and are often migrating out of rural areas to escape poverty.

A closely related issue is that employment in rural areas is often higher than in urban areas but the quality and decency of work is lower due to its informal and physical nature. The low wages, unhealthy and unsafe working conditions, and irregularity are among the reasons that young people are reluctant to work in agriculture. Possible indicators which can reflect not only the quantity of jobs but their quality include: work related injury in agriculture sector, prevalence of unpaid work, applicability of labour laws to the agriculture sector (including minimum wage).

Another critical aspect to be taken into consideration is the refugee crisis faced by so many NENA countries. Forced migrants tend to settle in urban areas where services are more available, however a significant share of refugees and IDPs (up to 40%) end up in rural areas where they work in food production and processing. These workers are especially vulnerable to abuse and exploitation. An indicator on Share of migrants in the agriculture workforce, could help indicate the presence of vulnerable labour. In addition, international and rural migration for economic reasons may also have an important bearing on agrifood systems through remittances and labour movement, although data looking at the intersection of migration and agrifood sectors is limited.

The discussion highlighted a number of assets and resources that influence livelihood, poverty and equality. Principle among these are access to financial services, access to land and water (of particular importance in the NENA) and access to digital technologies. The digital divide and asymmetry in the use of digital tools can in many cases exacerbate inequalities across the agrifood sector. However, digital technologies can also be leveraged to close gaps. For example mobile payments can be especially empowering for female producers who face restrictions on their mobility. Smartphone penetration is included under the indicators for resilience, but can fall under the livelihoods and poverty domain as well along with indicators like internet penetration and computer use. It is important to capture how digital technologies are transforming the sector and to whose benefit.

Finally, the indicators focus mainly on the agricultural producers, but for the region a focus on the consumers (which reflect the consumption and demand side) is important for poverty. To track livelihoods of consumers an indicator on price volatility could be useful as it affects consumer purchasing power and many trends can be covered with this one indicator. In addition, agricultural productivity impacts resilience, livelihoods and environmental sustainability yet is not included as an indicator in any of the domains.

### 4.4 Governance Domain

***Do you consider the proposed indicators relevant, high quality, interpretable and useful?***
It was noted that food systems are complex and delivers on a range of outcomes from nutrition, environmental to socio economic. Therefore, in order to capture governance it may be more useful to focus on one major outcome rather than looking at the food system as a whole, as that may not be manageable through broad indicators.

In terms of interpretability, it was noted that many of the indicators proposed are composite indicators. The point was made that these will need to be explained and unpacked to policy makers to make the governance aspect of the framework useful for them.

Regarding the proposed indicators, discussion reflected the view that these provide only a limited view of governance and indicators reflecting quality of food systems governance is also needed. For instance, Policy Coordination can be measured looking at existence of multi stakeholder platforms but the quality in terms of inclusivity and linkage to centres of decision making is also important. Some of the indicators were too specific while some others were too broad due to composite indices. The narrow or specific indicators may need to be reviewed as these may not be applicable universally, at least in the context of data and reporting.

The need to consider quality was also highlighted for the indicator titled "Presence of a national food systems pathways". The development of a pathway is just a start, and it should reflect the quality of the pathway, with complementary indicators which measure 1)whether there is an associated investment or resource mobilization plan (key to its success in implementation), 2) is it sufficiently gender and age responsive and 3) does it have clear and measurable targets. The power of the food systems pathways for decision makers could also be indicated by the presence of communication or dialogue mechanisms.

Discussion was raised around specific indicators, namely:
The point was made that many of the governance indicators are not specific to food systems, while the theme should focus more directly on food systems governance than trying to monitor governance in general. The point applies to the indicators: Civil Society Index, Varieties of Democracy; VDEM Accountability Index, and Voice and Accountability, WGI. The role of an international food systems framework is to measure food systems transformation rather than wider governance transformation, and in addition, the wording of the indicators in terms of accountability of governance does not reflect the ethos of international frameworks. A suggestion was made that for the Civil Society Index a more useful indicator would be a measure of public engagement and would allow measurement of the voices of the ordinary public in food systems transformations.

The indicator on Government Effectiveness describes it as including " the degree of independence of the civil service from political pressures". Clarification is needed on the interpretation of this indicator. The civil service develops and implement policies required by government. Its role is also to advise, not to decide policy which is the prerogative of the Minister, therefore by its nature, it is not independent from politics.

Participants noted that the indicator on implementation of marketing of breast-milk substitutes restrictions may be more relevant to Theme 1 and also qualifies as too specific/narrow for the governance theme. In addition, for the NENA region it may be more of social than governance issue, as was the case in countries which faced resistance from industry to legislative change.

***What are the data gaps that you can identify? Are these gaps structural? If not, are there some data or indicators that you know are available in your region and can be used for the purpose of this monitoring and assessment system?***
The indicators are drawn from non-national sources. However, the general issues of data gaps also applies to the governance theme. This theme has a particular requirement for timely and accurate data which may not be available, at least in many of the NENA regional countries

***What are the regional aspects we need to take into consideration at the moment of selecting a set of indicators to monitor the state of food systems and its evolution?***
Relevant governance indicators can be found under SDG Goal 16 "Promote peaceful and inclusive societies for sustainable development, provide access to justice for all and build effective, accountable and inclusive institutions at all levels". In particular, Target 16.6: Develop effective, accountable and transparent institutions at all levels with Indicator 16.6.1: Primary government expenditures as a proportion of original approved budget, by sector (or by budget codes or similar) and Indicator 16.6.2: Proportion of population satisfied with their last experience of public services

In the region conflict is an issue affecting governance of food systems as it weakens the structure and increases informality.

Another issue which is important for food systems governance is asymmetry of power and power dynamics. Some area of interest are the driving power of cooperate and retail actors and their level of influence on food systems; and the conflict of interest for the transformation of food systems. It was suggested that levels of private and public sector partnership might be useful to monitor

There are many indicators by city region networks, for example, the Milan Networks. It may be useful to examine city-region governance networks as urban food governance also provide suggestions for indicators at national level. An indicator is needed which covers the existence of urban food system policies/plans/strategies, and a possible source is the Milan Food Policy site and the City Region Networks

Examination of the transition from centralized to decentralized governance is a useful aspect of food systems transformation. For example, the presence of municipal food system policies/plans/strategies. However, it was noted that this would have to be adjusted for countries where there is no strong urban system.

Another suggested indicator is whether a country has implemented the Voluntary Guidelines on Food Security and Nutrition; it was suggested that relevant information is available from the Committee on Food Security.

### 4.5 Resilience and sustainability Domain

***Do you consider the proposed indicators relevant, high quality, interpretable and useful?***
A number of the indicators this domain are also relevant to the Diets, Nutrition and Health, and Livelihoods, Poverty and Equity, but also to the Environment domain. It was suggested that Resilience could focus on indicators which reduce the impact of shocks and help systems to better sustain or recover from those shocks.

It was also noted that composite indexes such as the Dietary Sourcing Flexibility Index and the Global Innovation Index can be difficult to interpret by policy makers, and can hide issues revealed by their individual components. It would also be useful to highlight the causal pathways between specific interventions and sustainability outcomes.

Discussion focused on specific indicators, namely:
Regarding the Stability of Food Insecurity Experience Scale (FIES) - based indicators and Stability of Prevalence of Undernourishment, stability is not the direction of interest. A stable POU and FIES indicator can be high, but to show improvement the trend should be decreasing – not stable.

The relevance of mobile cellular subscriptions (per 100 people) and renewable electricity output (% of total electricity output) for policy makers was not clear.

It was also questioned whether road density and distance to food are both needed as they provide a similar measure.

Specific Recommendations

Resilience should focus on resilience and capacities according to its definition, as well as presence of instruments and tools used to reduce the impact of or recover from shocks, such as the different insurance tools.
The number of resilience indicators can be more limited, with triage led by examination of use from the policy perspective, giving priority to those that are simple, easy to understand and can convey key messages to policymakers.
Comprehension of averaged composite indexes is difficult for policy makers; and an unpacking of the sub-indicators within the composite index would be useful.

***What are the data gaps that you can identify? Are these gaps structural? If not, are there some data or indicators that you know are available in your region and can be used for the purpose of this monitoring and assessment system?***
The issues discussed in Diets, Nutrition and Health are also relevant to this theme.

It was mentioned that the data on economic impact of disasters in the region is incomplete and could lead to underestimation of the impact.

***What are the regional aspects we need to take into consideration at the moment of selecting a set of indicators to monitor the state of food systems and its evolution?***
The discussion opened highlighted the high vulnerability of food systems in a region, particularly as there is a high dependence on imports for the majority of food staples. Vulnerability was seen as being determined by three generic characteristics: (1) the wealth available in the system, (2) how connected is the system, and (3) how much diversity exists in the system.

It was also suggested that consideration be given to food system vulnerability to <u>future</u> shocks (not only current shocks) is important and strongly related to food system resilience.

This is also linked to the need to balance efficiency and resilience (which is is one of the largest trade-offs in food systems). For example, vulnerability increases with the interconnectedness of food systems and trade patterns. Importing countries which rely on a single source have a lower diversity in terms of trade partners. This increases the vulnerability and exposure to future shocks.

A second issue relevant to the resilience of the region was that of economic migration and refugees as a coping strategy.

<u>Specific Recommendations</u>
It is important to include indicators that link issues of trade and migration as a coping mechanism to resilience of food systems.
Indicators on how to build resilience that consider both domestic production and transportation flows between countries within the region.
Diversity of the food systems in the region and the extent of concentration of imports of major food and agriculture products, such as wheat, should be considered in the indicators used.
Indicators must assess the vulnerability of the food systems in the region to withstand future shocks. For example, what are the commodities imported for major food staples, and the level of vulnerability. There has also been debate on future shocks around the FSTI dietary sources index.
Migration can be considered as a coping strategy in the región. For example, relevant indicators are remittances and how they are used in consumption and investment.

**Annex 1: Agenda of meeting**

**Wednesday 18 May 2022**

*Cairo time*

| Time | Subject | Speaker |
|---|---|---|
| | | *Master of Ceremonies, Jean Marc Faurés, FAO* |
| 10:00-10:30 | Welcoming remarks and introduction to the Initiative | *RNE ADG*<br>*José Rosero Moncayo, FAO*<br>*Kate Schneider, the Initiative* |
| 10:30-11:45 | Session 1: Diet, Nutrition and Health Domain<br>*Motivational presentation*<br>*Moderated discussion*<br>*Summary* | *Stella Nordhagen, the Initiative*<br>*Moderator: Tamara Nanitashvili, FAO* |
| 11:45-12:00 | Break | |
| 12:00-13:15 | Session 2: Environment and climate Domain<br>*Motivational presentation*<br>*Moderated discussion*<br>*Summary* | *Roseline Remans, the Initiative*<br>*Moderator: Theresa Wong, FAO* |
| 13:15-14:30 | Session 3: Livelihoods, poverty and equity Domain<br>*Motivational presentation*<br>*Moderated discussion*<br>*Summary* | *Alejandro Guarin, the Initiative*<br>*Moderator: Dalia Abulfotuh, FAO* |
| 14:30-14:45 | Break | |
| 14:45-16:00 | Session 4: Governance Domain<br>*Motivational presentation*<br>*Moderated discussion*<br>*Summary* | *Sheryl L. Hendriks, the Initiative*<br>*Moderator: Ahmad Mukhtar, FAO* |
| 16:00-17:15 | Session 5: Resilience and sustainability<br>*Motivational presentation*<br>*Moderated discussion*<br>*Summary* | *Paulina Bizzotto Molina, the Initiative*<br>*Moderator: Ahmad Mukhtar, FAO* |
| 17:15-17:30 | Wrap up and Conclusions | *Jean Marc Faurés*<br>*Kate Schneider, the Initiative* |

**Participants List**

**<u>Invited Experts:</u>**

**AHL ZINE Brahim**
Chef Division
Le Haut Commissariat au Plan HCP
Morocco

**AITKADI Mohamed**
President
General Council of Agricultural Development
Morocco

**ALAJMI Asma**
Senior Economic Analyst
Public Authority of Food & Nutrition
Kuwait

**AL DOMOR Lamia**
Manager
Ministry of Agriculture
Jordan

**ALESTAD Jeehan**
First Permanent Representation of Kuwait to Food and Agriculture Organization of the United Nations FAO & World Food Programme WFP
Secretary
Kuwait

**ALFRIHAT Nada**
Head of Organizations Division
Ministry of Agriculture
Jordan

**ALHAMMAD Mohammad**
Food Quality Regulations Manager
The Saudi Grains Organization SAGO
Saudi Arabia

**AL JAILANI Fatma**
Head Section of Studies and Research
Ministry of Agriculture, Fisheries Wealth and Water Resources
Oman

**ALKOSTABAN Khaled**
Agriculture Value Chain Officer
Food and Agriculture Organization of the United Nations FAO
Yemen

**ALLAN Mohamed**
Deputy Minister
Ministry Agriculture Irrigation and Fishing Wealth
Yemen

**AL-OTHAMN Wajd**
Food Security Acting Director
Public Authority for Food and Nutrition
Kuwait

**ALRAWAHIA Nargis**
Head of Sectors Accounts Department
National Centre of Statistics and Information (NCSI)
Oman

**AL SAADI Anwar**
Director of Statistics and Planning
Ministry of Fish Wealth
Yemen

**ALSABAH Manar**
Attaché
Permanent Representation of Kuwait to Food and Agriculture Organization of the United Nations FAO & World Food Programme WFP
Kuwait

**ALSULAIMAN Mohammed**
Consultant
Ministry of Environment, Water and Agriculture MEWA
Saudi Arabia

**ALWAN Abdulla**
Chairman
Ministry of Agriculture, Irrigation and Fisheries
Yemen

**BAKAR Lutfi**
Expert
Libya

**BALUSHI Sultan**
Head of Consumer Price Statistics Section
National Centre for Statistical Information NCSI

Oman

**BOUCHAMA Khalid**
Conseiller
Conseil Général du Développement Agricole
Morocco

**DAKHAKHNI Sumaiya**
Engineer
Ministry of Agriculture
Libya

**EID Mariam**
Head of Agro Industries Department
Ministry of Agriculture
Lebanon

**ELMEZDAWI Walid Khalid Omar**
Representative of the International Cooperation Office
Ministry of Agriculture and Livestock
Libya

**FADLAOUI Aziz**
Chercheur
Institut National de la Recherche Agronomique
Morocco

**FARESS Yahya**
Chef de Service
Département de l'Agriculture
Morocco

**HATTER Raed**
Deputy General Manager
Arab Organization for Agricultural Development
Egypt AOAD
Sudan

**HOBIESHAN Ahmed Saeed**
Manager of Coffee Department
Ministry of Agriculture and Irrigation & Fish Wealth
Yemen

**KAMOUKA Sadeg**
Chairman
National Committee for Desert Locust Control
Libya

**KHAMIS Shokry**
General Manager
Ministry of Agriculture
Yemen

**KORDASS Hassan**
Chargé de Mission
Conseil Général du Développement Agricole (CGDA)
Morocco

**MAAKOUL Siham**
Département de l'Agriculture
Chargée de Dossiers de Coopération Multilatérale
Morocco

**MANSOURI Amal**
Economist
Le Haut Commissariat au Plan HCP
Morocco

**MHANNA Maya**
Head of Rural Engineering Service
Ministry of Agriculture
Lebanon

**QTANANNI Iqbal**
Head of Studies of Value Chain Department
Ministry of Agriculture MOA
Jordan

**RASHID Hanaa**
Director
General Authority for Marine and Aquatic Sciences Research
Yemen

**RIZK Reda**
PGR Expert
Biodiversity expert
Egypt

**SALIBI Amal**
Head of Economic Studies and Statistics Service
Lebanese Ministry of Agriculture
Lebanon

**TALY Ahmed**
Assistant of General Director
Ministry of Agriculture MOA
Iraq

**YASSIN Dalia**
Prof. Economist,
Director of Food Loss and Waste
Monitoring and Evaluation Unit
Agriculture Research Center,
Agriculture Economic Research Institute Egypt

**ZAROUALI Said**
Le Haut Commissariat au Plan HCP
Morocco

**International and Regional Agencies and Other Institutions:**

**ABD-ALLAH Eman**
Expert
Agricultural Economics
Arab Organization for Agricultural Development
AOAD
Egypt

**ABDELGADIR Mohamed**
Country Director
International Fund for Agricultural Development
IFAD
Egypt

**ABOUARAB Julie**
United Nations Economic and Social
Commission for Western Asia UN ESCWA
Associate Coordination Officer
Lebanon

**AL-QUDSI Khalid**
Regional Programme Advisor
World Food Programme WFP
Egypt

**AL-SAMAWI Ahmed**
Agricultural Expert
Arab Organization for Agricultural Development
AOAD
Sudan

**BARIGOU Sabah**
Regional Head
Nutrition, HIV and School Based Programmes
World Food Programme
Egypt

**BERTINI Raffaele**
Data and Statistics Officer
International Organization for Migration
IOM
Egypt

**FAISAL Walaa**
Rural Women Development Expert
Arab Organization for Agricultural Development
AOAD
Sudan

**GARG Aashima**
Senior Nutrition Specialist
United Nations Children's Fund UNICEF
Middle East and North Africa Regional Office
Jordan

**GUARIN Alejandro**
Researcher
International Institute for Environment and
Development IIED
United Kingdom of Great Britain and Northern
Ireland (the)

**HENDRIKS Sheryl**
Professor
University of Pretoria
South Africa

**MADANI Wigdan**
United Nations Children's Fund UNICEF
Nutrition specialist
Jordan

**MOLINA Paulina Bizzotto**
Policy Officer
European Centre for Development Policy
Management (ECDPM)
Netherlands

**NEJDAWI Reem**
Chief of Food and Environment Section
United Nations Economic and Social
Commission for Western Asia UN ESCWA
Lebanon

**NORDHAGEN Stella**
Senior Technical Specialist
Global Alliance for Improved Nutrition
Switzerland

**OUMID Bedreddine**
Directeur Régional
L'Organisation Arabe du Développement de l'Agriculture OADA
Algeria

**RAWABDEH Fida'a Ali**
Head of Eastern Region Office
Arab Organization for Agriculture Development AOAD
Jordan

**REMANS Roseline**
Visiting Researcher
Alliance of Biodiversity and International Centre for Tropical Agriculture CIAT, CGIAR
Switzerland

**SCHNEIDER Kate**
Researcher
Johns Hopkins University
United States of America (the)

**<u>FAO ESS</u>**

**ROSERO MONCAYO José**
Director
Food and Agriculture Organization of the United Nations, Statistics Division, FAO-ESS
Rome, Italy

**MUNOZ Hernan Daniel**
Statistician
Food and Agriculture Organization of the United Nations, Statistics Division, FAO-ESS
Rome, Italy

**<u>FAO RNE</u>**

**ELWAER Abdulhakim**
Assistant Director General
Regional Representative
Food and Agriculture Organization of the United Nations FAO Regional Office for the Near East and North Africa
Cairo, Egypt

**FAURÈS JeanMarc**
Regional Programme Leader
Food and Agriculture Organization of the United Nations FAO Regional Office for the Near East and North Africa
Cairo, Egypt

**ABDELMAGEED Samar**
Data Analyst
Food and Agriculture Organization of the United Nations FAO Regional Office for the Near East and North Africa
Cairo, Egypt

**ABULFOTUH Dalia**
Economist
Food and Agriculture Organization of the United Nations FAO Regional Office for the Near East and North Africa
Cairo, Egypt

**AHMED Mohamed**
Policy Officer
Food and Agriculture Organization of the United Nations FAO Regional Office for the Near East and North Africa
Cairo, Egypt

**ALAOUI Hicham El Mhamdi El**
Monitoring and Evaluation M&E expert
Food and Agriculture Organization of the United Nations FAO Regional Office for the Near East and North Africa
Cairo, Egypt

**AMIN Karim**
Innovation Consultant
Food and Agriculture Organization of the United Nations FAO Regional Office for the Near East and North Africa
Cairo, Egypt

**CHIN Nancy**
Regional Statistician
Food and Agriculture Organization of the United Nations FAO Regional Office for the Near East and North Africa
Cairo, Egypt

**FETSI Theodora**
Consultant

Food and Agriculture Organization of the United Nations FAO Regional Office for the Near East and North Africa
Cairo, Egypt

**MUKHTAR Ahmad**
Senior Economist
Food and Agriculture Organization of the United Nations FAO Regional Office for the Near East and North Africa
Cairo, Egypt

**NANITASHVILI Tamara**
Nutrition and Food Systems Officer
Food and Agriculture Organization of the United Nations FAO Regional Office for the Near East and North Africa
Cairo, Egypt

**REZAEI Maryam**
Agroindustry officer
Food and Agriculture Organization of the United Nations FAO Regional Office for the Near East and North Africa
Cairo, Egypt

**WONG Theresa**
Natural Resources Officer
Food and Agriculture Organization of the United Nations FAO Regional Office for the Near East and North Africa

**<u>FAO Iraq</u>**

**ALI Hamid**
Sr. Tech. Dialogues Officer
Food and Agriculture Organization of the United Nations FAO Iraq
Iraq

**Fadhil Ismael**
National Dialogue Supporter
Food and Agriculture Organization of the United Nations FAO Iraq
Iraq

**SHLASH Khalid**
Assistant FAO Representative AFAOR (Programme)
Food and Agriculture Organization of the United Nations FAO Iraq
Iraq

**<u>FAO Lebanon</u>**

**SAADE Solange Matta-**
Assistant FAO Representative AFAOR (Programme)
Food and Agriculture Organization of the United Nations FAO Lebanon
Lebanon

**<u>FAO</u>**

**AQUILA Dell**
Policy advisor for Palestine
Food and Agriculture Organization of the United Nations FAO

# Supplementary Material, Appendix D

## Table D.1 Related and other relevant initiatives

| Initiative Name | Description | Distinction from FSCI |
|---|---|---|
| 50x2030 | 50x2030 is a 10-year, ~US$500 million initiative that aims to increase the capacity of 50 low and lower middle-income countries to produce, analyze, interpret, and apply data to decisions in the agricultural sector that support rural development and food security. It is Implemented through a unique partnership between the World Bank, Food and Agriculture Organization of the United Nations (FAO) and the International Fund for Agricultural Development (IFAD). | 50x2030 focuses on improving country-level data systems, particularly for collecting new survey data. In contrast, FSCI will not collect new data or build country capacity to do so. The two are complementary in the FSCI can identify key indicators for 50x2030 to collect data on, helping with their priority setting, and flag gaps where more data is needed, which 50x2030 can then help fill. |
| Access to Nutrition Initiative (ATNI) - Global Index | ATNI is hosted by the Access to Nutrition Foundation, an independent not-for-profit organization based in the Netherlands that works internationally.<br>ATNI focuses on developing tools and initiatives that track and drive the contribution made by the food and beverage sector to addressing the world's global nutrition challenges.<br>ATNI actively seeks partnerships with other organizations taking a multi-stakeholder, holistic approach to everything it does.<br>Funding comes from foundations, governments and fees. ATNI does not take any funding from — nor undertake projects commissioned by — food and beverage companies or industry associations. | ATNI focuses primarily on tracking performance of major food and beverage firms vis-a-vis nutrition outcomes. It does not generally look beyond these firms, nor does it look beyond nutrition, and its unit of analysis is the firm. In contrast, FSCI considers country-level data and looks well beyond nutrition, with private sector companies being only one stakeholder of interest among many. |
| Accountability Pact | A statement to set out a collaborative agenda for measurement, monitoring, and accountability following the UNFSS | The Pact focuses on galvanizing action from researchers and others to hold leaders accountable for food systems transformation; it does not endorse any specific indicators or curate, analyse, or present any actual data, unlike the FSCI. The two are complementary in that FSCI will be useful source of curated data that can be used in accountability efforts and flags data gaps where particular work on measurement and monitoring are needed and that, if the Pact leads to better monitoring and measurement of food systems transformation by signatories, those data could feed into FSCI in later years, helping to fill data gaps. |
| Agrifood Systems Technologies and Innovations Outlook (ATIO) | Agrifood system transformation to achieve the Sustainable Development Goals requires increased attention to developing, adapting and diffusing impactful science, technology and innovation (STI). Current levels and patterns of STI uptake are inadequate to facilitate needed agrifood system transformations, especially in today's low- and middle-income countries. Moreover, the descriptive and evaluative evidence on current and emergent STI is also insufficiently well understood to permit intentional management of STI to meet the multiple objectives of future agrifood systems: efficient, inclusive, resilient and sustainable. This report introduces the vision, rationale, scope and methods for new knowledge products FAO will launch as part of a new Agrifood System Technologies and Innovations Outlook (ATIO). ATIO's objective is to curate existing information on the current, measurable state of STI and upcoming changes, as well as their transformative potential, to inform evidence-based policy dialogue and decisions, including on investments. | ATIO focuses specifically on science, technology and innovation within agrifood systems, tracking progress on discrete technologies and innovations that can be used within those systems. In contrast, FSCI is tracking outcomes and key characteristics of the agrifood systems themselves. The two are complementary in that science, technology and innovation will be key to achieving desires shifts in the outcomes that FSCI tracks, and that FSCI can be used to pinpoint outcomes or areas of the food system for which additional technology and innovation are particularly needed. |
| Ceres 2030 | Ceres2030 has brought together economic modelling, machine learning, and evidence-based synthesis into one initiative, helping fill a major knowledge gap in the field of agricultural and food policy. Ceres2030 connects this knowledge back to the donor community, making sure decision makers have the cost figures and evidence they need when deciding where and how to make their investments. The partnership brought together Cornell University, the International Food Policy Research Institute (IFPRI) and the International Institute for Sustainable Development (IISD).<br>Funding support came from Germany's Federal Ministry of Economic Cooperation and Development (BMZ) and the Bill & Melinda Gates Foundation (BMGF). | Ceres2030 focused on identifying impactful interventions and investments that can support reductions in hunger and food insecurity, with its analysis focusing primarily in agricultural production and at the level of the intervention/investment. In contrast, FSCI is tracking outcomes and key characteristics at the level of the overall food system, looking beyond hunger as an outcome and agriculture as a sector and without a focus on specific interventions or investments. The two are complementary in that investments and interventions will be key to achieving desires shifts in the outcomes that FSCI tracks, and FSCI emphasizes continued undernourishment and food insecurity as a central challenge of food system transformation. |
| EDGAR-food | EDGAR is a multipurpose, independent, global database of anthropogenic emissions of greenhouse gases and air pollution on Earth. EDGAR provides independent emission estimates compared to what reported by European Member States or by Parties under the United Nations Framework Convention on Climate Change (UNFCCC), using international statistics and a consistent IPCC methodology.<br>EDGAR provides both emissions as national totals and gridmaps at 0.1 x 0.1 degree resolution at global level, with yearly, monthly and up to hourly data. | EDGAR's focus is narrower than the FSCI's, looking only at GHG emissions and air pollution; at the same time, its level of data resolution for this topic is much higher than that of the FSCI. The two are complementary in that FSCI highlights the importance of GHG emissions as one food system outcome among many, and EDGAR offers considerably more detail on how these vary across space and time. |
| FACT Alliance | The FACT Alliance embarked on a project with USAID, co-led by J-WAFS and D-Lab, that uses a combination of machine learning and traditional systematic review to identify evidence gaps in USAID's evaluation metrics in the areas of agriculture, nutrition, water, and resilience. The mission of the FACT Alliance is to transform the sustainability of food systems through collaborative, actionable research.<br>Central to the work of the FACT Alliance is the development of new methodologies for aligning data across scales and food systems components, improving data access, integrating research across the diverse disciplines that address aspects of food systems, making stakeholders partners in the research process, and assessing impact in the context of complex and interconnected food and climate systems. | FACT focuses primarily on climate impacts of the food system and on the development of new methodologies and conducting primary research. In contrast, the FSCI takes a broader perspective across food systems domains and focuses on indicator prioritisation and curation and analysis of existing data. |
| Food Action Alliance | The Food Action Alliance was catalysed by the World Economic Forum, International Fund for Agricultural Development (IFAD) and Rabobank, and today engages over 35 strategic and affiliate partners, in addition to a vast network of over 700 global and regional organisations including government, business, international organisations, civil society and farmer organisations. This allows the FAA to provide strong access to a wide range of implementing, commercial and funding partners across the food system with the ability to design and execute complex systemic solutions. | The FAA aims to focus on stakeholder coordination and collaboration in support of implementing food systems transformation pathways. Unlike FSCI, it does not have a focus on data or monitoring of food systems transformation. The two are complementary in that FSCI highlights the need for transformation and priorities for transformative policies, and tracks their ultimate outcomes, while the FAA could support cooperation around implementing those transformative policies. |
| Food Sustainability Index | The Food Sustainability Index (FSI), developed by the Economist Intelligence Unit (EIU) with the Barilla Center for Food & Nutrition (BCFN), measures the sustainability of food systems in 67 countries around three key issues outlined in the 2015 BCFN Milan Protocol and designed around the Sustainable Development Goals (SDGs): nutrition, sustainable agriculture and food loss and waste. The index looks at policies and outcomes around sustainable food systems and diets through a series of key performance indicators that consider environmental, social and economic sustainability. | The FSCI takes a broader scope, looking across a wider set of domains and indicators within those domains, as well as having global coverage. |
| Food system sustainability metric (Bene et. al.) | Compiles a metric of **12 key drivers** of food system from a globally-representative set of low, middle, and high-income countries and analyze the relationships between these drivers and a composite index that integrates the four key dimensions of food system sustainability, namely: food security & nutrition, environment, social, and economic dimensions | Bene et al is a one-off paper, not an ongoing tracking initiative and aims to analyse correlations and relationships between four dimensions of food system sustainability and specific drivers. While there is overlap between the domains and indicators examined by Bene et al and those of the FSCI, the FSCI includes some additional areas (e.g., governance) and aims to be an ongoing tracking initiative. |
| Food Systems Initiative on Shaping the Future of Food (WEF) | The mission of the World Economic Forum's System Initiative on Shaping the Future of Food is to build inclusive, sustainable, efficient, and nutritious food systems through leadership-driven, market-based action and collaboration, informed by insights and innovation, in alignment with the Sustainable Development Goals.<br>The System Initiative on Shaping the Future of Food aims to:<br>Strengthen global food systems by developing new insights; facilitating collaboration on priority action areas, including leveraging technology and innovation for food systems change; and mobilizing leadership and expertise at the global level.<br>Supporting regional and country platforms by achieving the New Vision for Agriculture through the Food Action Alliance - strengthening multistakeholder collaboration at the country and regional level, and by mobilizing new investments, partnerships, and best practices to achieve concrete results.<br>Harness the power of technological innovations to transform the food system through Innovation with a Purpose, as a large-scale partnership aggregator and project accelerator. | This Initiative aims to focus on stakeholder collaboration, mobilizing leadership and expertise, and one-off analysis on specific topics. Unlike FSCI, it does not have a focus on data curation, analysis, or monitoring of food systems transformation. The two are complementary in that FSCI highlights the need for transformation and priorities for transformative policies, and tracks their ultimate outcomes, while this Initiative could support cooperation around implementing those transformative policies and add new insights on specific topics related to them. |
| Food System Development Pathways (Gaupp et al.) | "Food system development pathways (FSDPs)… elicit the biophysical and technical feasibility of food systems transformation and potential trade-offs among multiple food systems objectives, notably between health, environmental and inclusion goals. These pathways are meant to provide decision-makers with possible combinations of policy options to achieve an inclusive food systems transformation." | Gaupp et al (2021) is similar to the FSCI in that they present a set of indicators that can be used to capture food systems transformation, spanning health, environment, and inclusion. However, their intention is not to track these indicators over time or to compare across countries but rather to use them to illustrate a framework and analytical approach to prioritising policies and interventions. They thus present limited quantitative data on the indicators. The FSCI also includes additional domains (resilience and governance) that have little overlap with those of Gaupp et al. |
| Global Strategy to Improve Agricultural and Rural Statistics | The purpose of the global strategy is to provide a framework for national and international statistical systems that enables them to produce and to apply the basic data and information needed to guide decision making in the twenty-first century. The Global Strategy is the result of an extensive consultation process with national and international statistical organizations as well as with agriculture ministries and other governmental institutions represented in Food and Agricultural Organization of the United Nations (FAO) governing bodies. | The Global Strategy is focused on guiding governments and national statistics agencies to collect more accurate and efficient data. It concentrates on data collection and the collection of many of the indicators from which the FSCI draws. |
| INFORMAS | INFORMAS (International Network for Food and Obesity / Non-communicable Diseases (NCDs) Research, Monitoring and Action Support) is a global network of public-interest organisations and researchers that aims to monitor, benchmark and support public and private sector actions to increase healthy food environments and reduce obesity and NCDs and their related inequalities. INFORMAS supports the WHO's Global Action Plan for the Prevention and Control of Non-Communicable Diseases (2013-2020) and the World Cancer Research Fund International NOURISHING framework. | INFORMAS has a primary focus on NCD/obesity prevention (though expanding into other issues) and looks particularly at policy-related indicators. It offers very detailed analysis and indicators on these topics, but for only 42 countries and without the coverage of larger food system issues (e.g., undernutrition, livelihoods, resilience) that the FSCI provides. |
| CAADP Results Framework (Malabo Declaration) | The Twenty-Third ordinary session of the African Union Assembly held in Malabo, Equatorial Guinea recommitted to the Comprehensive African Agriculture Development Programme (CAADP) principles and goals and defined a set of targets and goals, referred to as the Accelerated Agricultural Growth and Transformation Goals 2025. In their Declaration in Malabo, the Heads of State recalled the progress made and noted the need for monitoring, tracking and reporting on the implementation of the Declaration using the CAADP Results Framework.<br>Progress tracked by the CAADP Biennial Review Report (latest available: 2015-2021) | The CAADP Results Framework primarily tracks indicators related to agricultural production and productivity, and is focused only on Africa. FSCI considers a broader set of food system domains and indicators, with a global focus. |
| Regional Strategic Analysis and Knowledge Support System (ReSAKSS) | The Regional Strategic Analysis and Knowledge Support System (ReSAKSS) compiles and analyzes national indicators to help monitor the progress of the CAADP, Africa's policy framework for agricultural transformation. | As ReSAKSS aims to support monitoring progress according to the CAADP Results Framework, it also primarily tracks indicators related to agricultural production and productivity, and is focused only on Africa. FSCI considers a broader set of food system domains and indicators, with a global focus. |
| Africa Trends and Outlooks Report (ATOR) | A product of ReSAKSS. Latest ATOR 2022 on Agrifood processing strategies for successful food systems transformation in Africa<br>https://www.resakss.org/sites/default/files/ReSAKSS_AW_ATOR_2022.pdf | As a product of ReSAKSS aiming to support monitoring progress according to the CAADP Results Framework, ATOR's main focus is on agricultural production and productivity, and it is focused only on Africa. FSCI considers a broader set of food system domains and indicators, with a global focus. |
| Nutrition for Growth (N4G) (Global Nutrition Report) & Nutrition Accountability Framework | Nutrition for Growth (N4G) is a global pledging moment to drive greater action toward ending malnutrition and helping ensure everyone, everywhere can reach their full potential.<br><br>Over the past decade, the governments of the United Kingdom, Brazil, and Japan each stepped up to mobilize N4G nutrition commitments against the backdrop of the Olympics—a symbol of health, strength, and human potential. With the support of governments, donors, civil society, and the private sector, the result has been unprecedented, coordinated, and impactful commitments to improve global nutrition.<br><br>The Nutrition Accountability Framework (NAF) creates the world's first independent and comprehensive platform for registering SMART nutrition commitments and monitoring nutrition action. It has been endorsed by the government of Japan, the SUN Movement, the World Health Organization, UNICEF, USAID and many others, and will hold all data on commitments made for the Tokyo Nutrition for Growth (N4G) Summit 2021 and progress made against them over time. | The NAF focuses on tracking commitments (primarily from N4G) and progress on them, with a narrow focus on nutrition; its analysis is at the level of the commitment. In contrast, FSCI takes a holistic view across food system domains and considers data at the country level, with no focus on commitment tracking. |
| OmniAction | OmniAction takes existing metrics and methodologies from across the food system, supports the development of those that are less established, and harmonises the data into one global and unifying framework. We are a not-for-profit making this resource available for the public good | While this initiative has not yet published its framework or data, its intended focus appears to be on rating individual foods or food products, as opposed to tracking transformation at the level of the food system -- the goal of the FSCI. |
| Progress towards Sustainable Agriculture (PROSA) (FAO) | A methodological approach aimed at measuring progress towards sustainable agriculture in countries and across agri-food systems typologies, by measuring socio-economic and environmental dimensions with available national statistics, with sixteen indicators defined and constructed from FAOSTAT data | PROSA primarily considers indicators related to agricultural production and sustainability, drawing on FAO data. FSCI considers a broader set of food system domains and indicators as well as broader set of data sources. |
| Scaling Up Nutrition (SUN) | Since 2010, the SUN Movement has inspired a new way of working collaboratively to end malnutrition, in all its forms. With the governments of SUN Countries in the lead, it unites people—from civil society, the United Nations, donors, businesses and researchers—in a collective effort to improve nutrition.<br>The Scaling Up Nutrition Movement Strategy SUN 3.0 (2021–2025) continues to highlight the importance of nutrition as a universal agenda – and one which is integral to achieving the Sustainable Development Goals (SDGs).<br>To realize the vision of a world without hunger and malnutrition, the SUN Movement Principles of Engagement guide members as they work in a multi-sectoral and multi-stakeholder space to effectively work together to end malnutrition, in all its forms. These principles ensure that the SUN Movement is flexible while maintaining a common purpose and mutual accountability. | SUN is a global movement aimed primarily at increasing focus on malnutrition and supporting collaboration among actors to end it. It does not have a major focus on data curation, analysis, or monitoring and focuses only on nutrition, as opposed to the other food system domains the FSCI considers. The two are complementary in that FSCI flags areas where action on nutrition is particularly needed, and tracks progress on that at the level of the food system, whereas SUN can support ensuring that action happens within member countries. |
| Sustainable Agriculture Matrix (SAM) | SAM aims to serve as a platform to engage conversations among stakeholders involved in agriculture and to forge positive changes towards sustainability while avoiding unintended consequences. SAM reports indicators by country and year so that end-users can track a country's progress along time and make comparison across countries among different dimensions of sustainability. However, SAM does not only rely on available data reported on the national level but also synthesizes data in the literature and in other public domains on various spatial and temporal scales.<br>There are mainly three goals for the development and publication of SAM:<br>Provide a consistent and transparent measurement of countries' performance in sustainable agricultural production<br>Investigate the socioeconomic and ecological drivers for a country achieving sustainability.<br>Quantify and visualize the impacts of current agricultural production on its future sustainability. | SAM focuses on agriculture and the sustainability of agricultural production; FSCI considers a broader set of food system domains and indicators, without as much detail on the topic of sustainable agriculture. |
| Sustainable Nutrition Security (SNS) food system metrics (Gustafson et al.) | New methodology based on the concept of "sustainable nutrition security" (SNS). This novel assessment methodology is intended to remedy both kinds of deficiencies in the previous work by defining seven metrics, each based on a combination of multiple indicators, for use in characterizing sustainable nutrition outcomes of food systems: **(1) food nutrient adequacy; (2) ecosystem stability; (3) food affordability and availability; (4) sociocultural wellbeing; (5) food safety; (6) resilience; and (7) waste and loss reduction**. Each of the metrics comprises multiple indicators that are combined to derive an overall score (0–100). | Gustafson et al is a one-off paper, not an ongoing tracking initiative, and aimed to develop new composite metrics. While there is overlap between the domains and indicators examined by Gustafson et al and those of the FSCI, the FSCI: does not aim to create composite metrics; includes some additional areas (e.g., governance, more details on diets) and omits some of those included by Gustafson; and aims to be an ongoing tracking initiative. |
| The Paris Declaration on Aid Effectiveness and the Accra Agenda for Action | At the Second High Level Forum on Aid Effectiveness (2005) it was recognised that aid could - and should - be producing better impacts. The Paris Declaration was endorsed in order to base development efforts on first-hand experience of what works and does not work with aid. It is formulated around five central pillars: Ownership, Alignment, Harmonisation, Managing for Results and Mutual Accountability.<br><br>In 2008 at the Third High Level Forum on Aid Effectiveness all OECD donors, more than 80 developing countries and some 3 000 civil society organisations from around the world joined representatives of emerging economies, United Nations and multilateral institutions and global funds in the negotiations leading up to and taking place during the Accra meeting. The Accra Agenda for Action (AAA) was endorsed. The AAA both reaffirms commitment to the Paris Declaration and calls for greater partnership between different parties working on aid and development. | The Accra Agenda / Paris Declaration is process oriented: it focuses on the process of development and the provision of development aid. As such, its monitoring also takes this focus. Moreover, it is not specific to food systems, considering multiple areas of development. In contrast, FSCI looks at outcomes of food systems without a focus on the provision of development aid of the process of setting development agendas. The two are complementary in that effective use of development aid and appropraite setting of development agendas are critical for achieving the food system transformation advocated for, and tracked by, FSCI. |
| World Benchmarking Alliance | To achieve the SDGs by 2030, we need transformational change from farm to fork. Food systems transformation requires large-scale and fundamental action led by those who drive environmental, health and social pressures in the system.<br>The benchmark's aim is to stimulate the most influential food and agriculture companies to apply sustainable business practices throughout their operations as well as use their influence to encourage value chain partners to do the same.<br>Developed a framework that set out the critical areas and topics where private sector action is needed and where companies must step up their efforst to collectively transform the food system. | The benchmark focuses on tracking performance of major food and beverage companies. It does not look beyond these companies, and its unit of analysis is the company. In contrast, FSCI considers country-level data, with private sector companies being only one stakeholder of interest among many. |
| Global Nutrition Report and Country Nutrition Profiles | The Global Nutrition Report provides a concise data-focused update on the state of diets and nutrition in the world. Its Country Profiles enable users to explore the latest data on nutrition at global, regional and country level. | The GNR focuses just on diets and nutrition, without considering the other domains included in the FSCI (i.e., sustainability, resilience, livelihoods, and governance). The two are complementary in that FSCI highlights the importance of nutrition outcomes (among other food system outcomes), while GNR provides a more in-depth perspective on diet and nutrition indicators. |
| Food Systems Dashboard | The Food Systems Dashboard combines data from multiple sources to give users a complete view of food systems. Users can compare components of food systems across countries and regions. They can also identify and prioritize ways to sustainably improve diets and nutrition in their food systems. | The Dashboard aims at a complete or comprehensive view of food systems, including some that are not useful for tracking progress towards transformation but have other purposes, while FSCI focuses on identifying a smaller set of trackable key indicators. The Dashboard also does not currently include indicators for livelihoods or governance. The two are complementary in that many FSCI indicators are included in the Dashboard, which provides an easy platform for visualization and comparison. |
| The State of Food Security and Nutrition in the World | The State of Food Security and Nutrition in the World (SOFI) is an annual flagship report jointly prepared by FAO, IFAD, UNICEF, WFP and WHO to inform on progress towards ending hunger, achieving food security and improving nutrition and to provide in depth analysis on key challenges for achieving this goal in the context of the 2030 Agenda for Sustainable Development. The report targets a wide audience, including policy-makers, international organizations, academic institutions and the general public. | These annual reports focus on food security and nutrition, with in-depth analysis of a given topic (which varies by the year), without considering the other domains included in the FSCI (i.e., sustainability, resilience, livelihoods, and governance). The two are complementary in that FSCI highlights the importance of nutrition outcomes (among other food system outcomes), while SOFI provides a more in-depth perspective on diet and nutrition indicators, and the indicators developed and reported by SOFI are among those included in the FSCI |
| The State of Food and Agriculture | The State of Food and Agriculture, one of FAO's major annual flagship publications, aims at bringing to a wider audience balanced science-based assessments of important issues in the field of food and agriculture. Each edition of the report contains a comprehensive, yet easily accessible, overview of a selected topic of major relevance for rural and agriculture development and for global food security. | These annual reports focus on agriculture, with in-depth analysis of a given topic (which varies by the year). It does not aim to comprehensively track food systems indicators and its focus is on agriculture. The two are complementary in that indicators developed and reported by SOFA play a key role in plugging food system data gaps and could feed into those included in the FSCI. |
| Healthy Diets Monitoring Initiative (WHO, UNICEF, FAO) managed by TEAM | The overall objective of the Healthy Diets Monitoring Initiative is to enable national and global monitoring of the diet quality to inform policies and programmes across a wide range of sectors. Specifically, the Initiative will determine how best to measure healthy diets for different purposes, build consensus, and promote uptake and use of healthy diet measures and indicators among national and global stakeholders. It is currently in planning stages: The work of the Initiative will be phased over three to five years, with the initial planning and priority-setting occurring during summer 2022 and a first technical consultation in late 2022, with further work to follow in 2023-24. | This initiative is exclusively focused on healthy diets, whereas FSCI includes healthy diets as part of a broader agenda to monitor the whole of food systems. When this cross-UN initiative produces recommendations, and subsequently when data become available, the indicators will be considered for inclusion in the FSCI. |